\documentclass[english,11pt]{article}
\usepackage[latin9]{inputenc}
\usepackage{geometry}
\usepackage{color}
\usepackage{babel}
\usepackage{amsmath}
\usepackage{amsthm}
\usepackage{amssymb}
\usepackage{mleftright}
\usepackage[unicode=true,pdfusetitle,
 bookmarks=true,bookmarksnumbered=false,bookmarksopen=false,
 breaklinks=false,pdfborder={0 0 0},pdfborderstyle={},backref=false,colorlinks=true]
 {hyperref}
\hypersetup{
 linkcolor=vblue, citecolor=vblue}

\makeatletter

\numberwithin{equation}{section}
\theoremstyle{plain}
\newtheorem{thm}{\protect\theoremname}[section]
\theoremstyle{plain}
\newtheorem{prop}[thm]{\protect\propositionname}
\theoremstyle{plain}
\newtheorem{lem}[thm]{\protect\lemmaname}
\theoremstyle{plain}
\newtheorem{cor}[thm]{\protect\corollaryname}
\theoremstyle{plain}
\newtheorem*{prop*}{\protect\propositionname}
\theoremstyle{plain}
\newtheorem*{thm*}{\protect\theoremname}
\theoremstyle{plain}
\newtheorem*{lem*}{\protect\lemmaname}

\definecolor{vblue}{RGB}{0,0,255}

\allowdisplaybreaks

\makeatother

\providecommand{\corollaryname}{Corollary}
\providecommand{\lemmaname}{Lemma}
\providecommand{\propositionname}{Proposition}
\providecommand{\theoremname}{Theorem}

\begin{document}
\title{Emergent Quasi-Bosonicity in Interacting Fermi Gases}
\author{Martin Ravn Christiansen\\
\\
{\small{}Ph.D. advisors: Christian Hainzl and Phan Th\`anh Nam}}
\maketitle
\begin{abstract}
This thesis concerns the correlation structure of interacting Fermi
gases on a torus in the mean-field regime. A bosonization method in
the spirit of Sawada\cite{Sawada-57} is developed to analyze the
system, and is applied to obtain an upper bound for the correlation
energy of the system for a wide class of repulsive interaction potentials,
including the Coulomb potential.

This upper bound includes both a bosonic contribution, as found in
the bosonic model of Sawada, and an exchange contribution, as was
found by Gell-Mann and Brueckner\cite{GellMannBrueckner-57} but which
was missed by Sawada's model.

An extension to weakly attractive potentials is also presented, as
is an outline of the derivation of an effective Hamiltonian for regular
interaction potentials, and the construction of plasmon states for
this outside of the mean-field setting.

This thesis is based on the papers \cite{ChrHaiNam-21,ChrHaiNam-22a,ChrHaiNam-22b}.
\end{abstract}
\tableofcontents{}

\section{Introduction}

A Fermi gas is a quantum system described by a Hamiltonian of the
form
\[
H=-\sum_{i=1}^{N}\Delta_{i}+\sum_{1\leq i<j\leq N}V\mleft(x_{i}-x_{j}\mright)
\]
on a fermionic $N$-particle space. Here the first term represents
the kinetic energy of the fermions (in units where $\frac{\hbar^{2}}{2m}=1$)
while the second term represents pair interactions through a potential
$V$.

The potential of greatest physical interest is the (background-subtracted)
Coulomb potential, in which case the system is referred to as \textit{jellium}.
Jellium is the simplest model of electrons in a metal which still
includes all electron-electron interactions.

\medskip{}

In the 1930-40's, theoretical calculations based on applying the Hartree-Fock
approximation to the jellium model exhibited a large discrepancy when
compared to experimental values. Furthermore, pertubative methods
broke down already at second-order, presenting the physicists of the
time with the puzzle of how to model an interacting many-body system
without being able to apply perturbative methods.

As the Hartree-Fock approximation amounts to neglecting particle correlations,
the question was how to include these in the computation. The first
steps toward this was taken in the early 1950's by Bohm and Pines\cite{BohmPines-51,BohmPines-52,BohmPines-53,Pines-53},
who argued that the correlations at play were of an essentially bosonic
nature, which would manifest itself as quantized collective electron
oscillations, which they dubbed \textit{plasmons}.

Adding plasmon modes to the jellium model by hand, they argued that
these served to regularize the electron-electron interaction to the
point that second-order perturbation could be applied - provided that
certain terms appearing in their analysis could be neglected, the
assumption of which was referred to as the ``Random Phase Approximation''
(RPA).

\medskip{}

The validity of the RPA and the manner in which the plasmons were
introduced was a somewhat controversial issue, but they were effectively
justified by two later works: The first was by Gell-Mann and Brueckner\cite{GellMannBrueckner-57},
who were able to derive the \textit{correlation energy} - the difference
between the ground state and Fermi state energies - of the jellium
model directly, by performing a formal resummation of the divergent
perturbation series for this, and finding agreement with Pines' calculation.

The second work was by Sawada\cite{Sawada-57} (and expanded on by
Sawada-Brueckner-Fukuda-Brout\cite{SawBruFukBro-57}). He observed
that certain terms of the Hamiltonian could, when expressed in the
second-quantized picture, be interpreted as quadratic operators with
respect to almost-bosonic operators. By studying this corresponding
bosonic Hamiltonian, he was also able to derive the correlation energy
- with the exception of one term, which was explicitly fermionic in
nature.

\medskip{}

With these works, the correlation energy was thought to be well-understood
by the physics community, but presenting a mathematically rigorous
derivation of this remains a major open problem in mathematical physics
to this day. Recently there has however been much progress on the
corresponding mean-field problem, in which the potential is scaled
by a factor proportional to $N^{-\frac{1}{3}}$.

The first results on this problem were by Benedikter-Nam-Porta-Schlein-Seiringer\cite{BenNamPorSchSei-20,BenNamPorSchSei-21}
(see also \cite{BenPorSchSei-21}), who were able to prove an asymptotic
formula for the correlation energy for highly regular potentials $V$
by employing a bosonization method, albeit in a manner different from
Sawada's original observation, to define an analog of a bosonic Bogolubov
transformation which could be applied to analyze the system.

Subsequently I and my Ph.D. advisors extended this result significantly
in \cite{ChrHaiNam-21}, in which we both proved an asymptotic formula
for the correlation energy for more general potentials and additionally
derived an effective quasi-bosonic Hamiltonian governing the low-lying
eigenstates of the Fermi gas. We accomplished this by developing a
bosonization method different from that of \cite{BenNamPorSchSei-20,BenNamPorSchSei-21}
and more in the spirit of Sawada.

\medskip{}

The aim of this thesis is to present this method and the results we
have obtained by it.

\subsection{Main Results}

Before stating the main results, let us introduce the setting properly
and define some notation: We consider for a given Fermi momentum $k_{F}>0$
the mean-field Hamiltonian
\begin{equation}
H_{N}=-\sum_{i=1}^{N}\Delta_{i}+k_{F}^{-1}\sum_{1\leq i<j\leq N}V\mleft(x_{i}-x_{j}\mright)
\end{equation}
on $\mathcal{H}_{N}=\bigwedge^{N}L^{2}\mleft(\mathbb{T}^{3};\mathbb{C}^{s}\mright)$,
where $\mathbb{T}^{3}$ is the $3$-torus of sidelength $2\pi$ and
$s\in\mathbb{N}$ is the number of spin states of the system. The
number of particles, $N$, is determined by $k_{F}$ through the relation
$N=s\left|\overline{B}\mleft(0,k_{F}\mright)\cap\mathbb{Z}^{3}\right|$.

We take the interaction potential $V$ to admit the Fourier decomposition
\begin{equation}
V\mleft(x\mright)=\frac{1}{\mleft(2\pi\mright)^{3}}\sum_{k\in\mathbb{Z}^{3}}\hat{V}_{k}e^{ik\cdot x}
\end{equation}
and assume that the Fourier coefficients obey (with $\mathbb{Z}_{\ast}^{3}=\mathbb{Z}^{3}\backslash\{0\}$)
\begin{equation}
\hat{V}_{k}=\hat{V}_{-k}\quad\text{and}\quad\hat{V}_{k}\geq0,\quad k\in\mathbb{Z}_{\ast}^{3},
\end{equation}
in other words we consider a symmetric and repulsive interaction potential.

We define for $k\in\mathbb{Z}_{\ast}^{3}$ the \textit{lune} $L_{k}$
by
\begin{equation}
L_{k}=\left\{ p\in\mathbb{Z}^{3}\mid\left|p-k\right|\leq k_{F}<\left|p\right|\right\} 
\end{equation}
and let further $\lambda_{k,p}=\frac{1}{2}\mleft(\left|p\right|^{2}-\left|p-k\right|^{2}\mright)$
for $p\in L_{k}$.

\medskip{}

The main focus of the thesis is the derivation of the following:
\begin{thm}
\label{them:MainTheorem}Let $\sum_{k\in\mathbb{Z}^{3}}\hat{V}_{k}^{2}<\infty$.
Then it holds that
\[
\mathrm{inf}\mleft(\sigma\mleft(H_{N}\mright)\mright)\leq E_{F}+E_{\mathrm{corr},\mathrm{bos}}+E_{\mathrm{corr},\mathrm{ex}}+C\sqrt{\sum_{k\in\mathbb{Z}_{\ast}^{3}}\hat{V}_{k}^{2}\min\left\{ \left|k\right|,k_{F}\right\} },\quad k_{F}\rightarrow\infty,
\]
where $E_{F}=\left\langle \psi_{F},H_{N}\psi_{F}\right\rangle $ is
the energy of the Fermi state,
\[
E_{\mathrm{corr},\mathrm{bos}}=\frac{1}{\pi}\sum_{k\in\mathbb{Z}_{\ast}^{3}}\int_{0}^{\infty}F\mleft(\frac{s\hat{V}_{k}k_{F}^{-1}}{\mleft(2\pi\mright)^{3}}\sum_{p\in L_{k}}\frac{\lambda_{k,p}}{\lambda_{k,p}^{2}+t^{2}}\mright)dt,\quad F\mleft(x\mright)=\log\mleft(1+x\mright)-x,
\]
is the bosonic contribution (to the correlation energy) and
\[
E_{\mathrm{corr},\mathrm{ex}}=\frac{sk_{F}^{-2}}{4\,\mleft(2\pi\mright)^{6}}\sum_{k,l\in\mathbb{Z}_{\ast}^{3}}\hat{V}_{k}\hat{V}_{l}\sum_{p,q\in L_{k}\cap L_{l}}\frac{\delta_{p+q,k+l}}{\lambda_{k,p}+\lambda_{k,q}}
\]
is the exchange contribution, for a constant $C>0$ depending only
on $\sum_{k\in\mathbb{Z}_{\ast}^{3}}\hat{V}_{k}^{2}$ and $s$.
\end{thm}

This result was originally presented in \cite{ChrHaiNam-22b} (for
$s=1$). Although we have so far only been able to prove this asymptotic
statement as an upper bound, it constitutes a major improvement over
the corresponding one of \cite{ChrHaiNam-21}: Not only does it apply
to singular potentials (including the Coulomb potential), it also
includes the ``exchange contribution'' $E_{\mathrm{corr},\mathrm{ex}}$,
which is the term that was missing from Sawada's purely bosonic model,
and which was also lacking in the previously proved results for non-singular
potentials (for which $E_{\mathrm{corr},\mathrm{ex}}$ is of much
lower order than the rest).

In the case of the Coulomb potential, i.e. $\hat{V}_{k}\sim\left|k\right|^{-2}$,
$E_{\mathrm{corr},\mathrm{bos}}$ is of order $k_{F}\log\mleft(k_{F}\mright)$
and $E_{\mathrm{corr},\mathrm{ex}}$ is of order $k_{F}$, while the
error term of the theorem is of order $\sqrt{\log\mleft(k_{F}\mright)}$.
The precision of the result is thus almost an entire order of magnitude.
Furthermore, we may observe that for any potential with $\sum_{k\in\mathbb{Z}^{3}}\hat{V}_{k}^{2}<\infty$,
the error term is at most of order $\sqrt{k_{F}}$ whereas $E_{\mathrm{corr},\mathrm{bos}}$
is at least order $k_{F}$, so there is always a sharp distinction
between the correlation energy and the error term.

\medskip{}

After concluding this theorem we will make the observation that our
proof in fact allows us to generalize this result to slightly \textit{attractive}
potentials, proving the following:
\begin{thm}
\label{them:AttractiveGeneralization}Assuming the weaker condition
that $\hat{V}_{k}\geq-\mleft(1-\epsilon\mright)\frac{4\pi^{2}}{s}$
for some $\epsilon>0$ and all $k\in\mathbb{Z}_{\ast}^{3}$, it continues
to hold that
\[
\mathrm{inf}\mleft(\sigma\mleft(H_{N}\mright)\mright)\leq E_{F}+E_{\mathrm{corr},\mathrm{bos}}+E_{\mathrm{corr},\mathrm{ex}}+C\sqrt{\sum_{k\in\mathbb{Z}_{\ast}^{3}}\hat{V}_{k}^{2}\min\left\{ \left|k\right|,k_{F}\right\} },\quad k_{F}\rightarrow\infty,
\]
where now $C>0$ depends on $\sum_{k\in\mathbb{Z}_{\ast}^{3}}\hat{V}_{k}^{2}$,
$s$ and $\epsilon$.
\end{thm}

This result has not been presented before. We remark that the condition
on $\hat{V}_{k}$ is nearly optimal, in the sense that if $\hat{V}_{k}<-\frac{4\pi^{2}}{s}$
for some $k\in\mathbb{Z}_{\ast}^{3}$ then the corresponding term
of $E_{\mathrm{corr},\mathrm{bos}}$ is not even well-defined, as
the argument of the logarithm of the integrand is then strictly negative
near $t=0$.

\medskip{}

These results only concern upper bounds for the ground state energy
of $H_{N}$. In \cite{ChrHaiNam-21} we also proved the following
stronger operator-level result regarding $H_{N}$, albeit only under
high regularity assumptions on $V$:
\begin{thm}
\label{them:OperatorStatement}Let $\sum_{k\in\mathbb{Z}_{\ast}^{3}}\hat{V}_{k}\left|k\right|<\infty$.
Then there exists a unitary transformation $\mathcal{U}:\mathcal{H}_{N}\rightarrow\mathcal{H}_{N}$,
depending implicitly upon $k_{F}$, such that
\[
\mathcal{U}H_{N}\mathcal{U}^{\ast}=E_{F}+E_{\mathrm{corr},\mathrm{bos}}+H_{\mathrm{eff}}+\mathcal{E}
\]
where
\[
H_{\mathrm{eff}}=H_{\mathrm{kin}}^{\prime}+2\sum_{k\in\mathbb{Z}_{\ast}^{3}}\sum_{p,q\in L_{k}}\langle e_{p},(\tilde{E}_{k}-h_{k})e_{q}\rangle b_{k,p}^{\ast}b_{k,q}
\]
for $\tilde{E}_{k}=\mleft(h_{k}^{\frac{1}{2}}\mleft(h_{k}+2P_{k}\mright)h_{k}^{\frac{1}{2}}\mright)^{\frac{1}{2}}$.

Furthermore, it holds for every normalized eigenstate $\Psi$ of $H_{N}$
with $\left\langle \Psi,H_{N}\Psi\right\rangle \leq E_{F}+\kappa k_{F}$,
$\kappa>0$, that the error operator $\mathcal{E}$ obeys
\[
\left|\left\langle \Psi,\mathcal{E}\Psi\right\rangle \right|,\,\left|\left\langle \mathcal{U}\Psi,\mathcal{E}\mathcal{U}\Psi\right\rangle \right|\leq Ck_{F}^{1-\frac{1}{94}+\varepsilon},\quad k_{F}\rightarrow\infty,
\]
for any $\varepsilon>0$, the constant $C>0$ depending only on $V$,
$\kappa$ and $\varepsilon$.
\end{thm}

In words, the theorem states that the Hamiltonian $H_{N}$ is, with
respect to the \textit{low-lying} eigenstates (as demarked by the
condition $\left\langle \Psi,H_{N}\Psi\right\rangle \leq E_{F}+\kappa k_{F}$),
up to the constant terms $E_{F}+E_{\mathrm{corr},\mathrm{bos}}$ unitarily
equivalent with the effective Hamiltonian $H_{\mathrm{eff}}$, to
leading order in $k_{F}$.

Here the effective Hamiltonian consists of two parts: The \textit{localized
kinetic operator} $H_{\mathrm{kin}}^{\prime}$, which appears naturally
during the extraction of $E_{F}$, and a \textit{quasi-bosonic} term
involving the \textit{excitation operators} (for $s=1$)
\begin{equation}
b_{k,p}=c_{p-k}^{\ast}c_{p},\quad b_{k,p}^{\ast}=c_{p}^{\ast}c_{p-k},\quad k\in\mathbb{Z}_{\ast}^{3},\,p\in L_{k},
\end{equation}
where $(c_{p}^{\ast})_{p\in\mathbb{Z}_{\ast}^{3}}$ and $(c_{p})_{p\in\mathbb{Z}_{\ast}^{3}}$
denote the fermionic creation and annihilation operators associated
with the plane-wave states. In the definition of the quasi-bosonic
term also appears certain ``one-body operators'' $h_{k},P_{k}:\ell^{2}\mleft(L_{k}\mright)\rightarrow\ell^{2}\mleft(L_{k}\mright)$
which naturally appear during the diagonalization process which extracts
$E_{\mathrm{corr},\mathrm{bos}}$.

From the fact that $\tilde{E}_{k}\geq h_{k}$ it follows that $H_{\mathrm{eff}}\geq0$,
so (as the ground state certainly is low-lying) the theorem in particular
implies that
\begin{equation}
\mathrm{inf}\mleft(\sigma\mleft(H_{N}\mright)\mright)=E_{F}+E_{\mathrm{corr},\mathrm{bos}}+O\mleft(k_{F}^{1-\frac{1}{94}+\varepsilon}\mright),\quad k_{F}\rightarrow\infty,
\end{equation}
i.e. the ground-state energy is indeed $E_{F}+E_{\mathrm{corr},\mathrm{bos}}$
to leading order, provided $\sum_{k\in\mathbb{Z}_{\ast}^{3}}\hat{V}_{k}\left|k\right|<\infty$.
Note that $E_{\mathrm{corr},\mathrm{ex}}$ is absent, which is a consequence
of the assumed regularity - even just assuming boundedness of $V$,
i.e. that $\sum_{k\in\mathbb{Z}_{\ast}^{3}}\hat{V}_{k}<\infty$, it
holds that $E_{\mathrm{corr},\mathrm{ex}}\leq Ck_{F}^{-1}$.

We will not give a full proof of Theorem \ref{them:OperatorStatement}
in this thesis, but in Section \ref{sec:OverviewoftheOperatorResult}
we present the main ideas and techniques that lead to its conclusion.\medskip{}

What is particularly noteworthy about Theorem \ref{them:OperatorStatement}
is that it not only yields a lower bound on the correlation energy,
but also identifies the operator which should govern the low-lying
excitations of the system - in the physical case this would include
the plasmon states. Unfortunately the mean-field scaling suppresses
these states, making it difficult to say much about $H_{\mathrm{eff}}$
in this setting.

Given the physical importance of plasmons it is however interesting
to extrapolate this result and consider $H_{\mathrm{eff}}$ by itself
without imposing the mean-field scaling or strict regularity assumptions
on the potential, which is what we did in \cite{ChrHaiNam-22a}, obtaining
a result of the following form:
\begin{thm}
\label{them:PlasmonStates}In the non-mean-field scaled setting the
following holds: Let $\hat{V}_{k}=g\left|k\right|^{-2}$, $k\in\mathbb{Z}_{\ast}^{3}$,
for some $g>0$. Then for any $\delta\in\mleft(0,\frac{1}{2}\mright)$
and $\varepsilon\in\mleft(0,2\mright)$ there exists for all $k\in\overline{B}\mleft(0,k_{F}^{\delta}\mright)\cap\mathbb{Z}_{\ast}^{3}$
and $M\leq k_{F}^{\varepsilon}$ a normalized state $\Psi\in\mathcal{H}_{N}$
such that
\[
\left\Vert \mleft(H_{\mathrm{eff}}-M\epsilon_{k}\mright)\Psi\right\Vert \leq C\left|k\right|^{-1}\sqrt{k_{F}}M^{\frac{5}{2}},\quad k_{F}\rightarrow\infty,
\]
where $\epsilon_{k}$ denotes the greatest eigenvalue of $2\tilde{E}_{k}$.
$\epsilon_{k}$ obeys $\epsilon_{k}\geq ck_{F}^{\frac{3}{2}}$ and
\[
0\leq\epsilon_{k}-2\sqrt{\frac{s}{\mleft(2\pi\mright)^{3}}\frac{g}{\left|k\right|^{2}}\sum_{p\in L_{k}}\lambda_{k,p}+\frac{\sum_{p\in L_{k}}\lambda_{k,p}^{3}}{\sum_{p\in L_{k}}\lambda_{k,p}}}\leq Ck_{F}^{-\frac{1}{2}}\left|k\right|^{4},\quad k_{F}\rightarrow\infty,
\]
for constants $c,C>0$ depending only on $g$.
\end{thm}

We present a proof of this in Section \ref{sec:PlasmonModesoftheEffectiveHamiltonian}.

The theorem states that for $k$ and $M$ in certain ranges, there
exists an ``approximate eigenvector'' $\Psi$ for $H_{\mathrm{eff}}$
with approximate eigenvalue $M\epsilon_{k}$ - in fact $\Psi$ is
explicitly given as the normalization of
\begin{equation}
b^{\ast}\mleft(\phi\mright)^{M}\psi_{F},\quad b_{k}^{\ast}\mleft(\phi\mright)=\sum_{p\in L_{k}}\left\langle e_{p},\phi\right\rangle b_{k,p}^{\ast},
\end{equation}
where $\phi$ is the normalized eigenstate of $2\tilde{E}$ with eigenvalue
$\epsilon_{k}$, which mimics the definition of a bosonic state with
$M$ ``$\phi$ excitations''.

Calling $\Psi$ an approximate eigenvector is justified by Markov's
inequality in the operator form $1_{\mathbb{R}\backslash\left[E-\delta,E+\delta\right]}\mleft(H\mright)\leq\delta^{-1}\left|H-E\right|$,
as it implies that
\begin{equation}
\left\Vert 1_{\mathbb{R}\backslash\left[M\varepsilon_{k}-\delta,M\varepsilon_{k}+\delta\right]}\mleft(H_{\mathrm{eff}}\mright)\Psi\right\Vert \ll1,\quad\left|k\right|^{-1}\sqrt{k_{F}}M^{\frac{5}{2}}\ll\delta,
\end{equation}
i.e. $\Psi$ is spectrally localized at $E=M\varepsilon_{k}$ on the
scale $\left|k\right|^{-1}\sqrt{k_{F}}M^{\frac{5}{2}}$. As $M\varepsilon_{k}\sim Mk_{F}^{\frac{3}{2}}$
this is a nontrivial statement for $M\ll\mleft(k_{F}\left|k\right|\mright)^{\frac{2}{3}}$.
One can also view this in a dynamical setting: By the time evolution
estimate $\left\Vert \mleft(e^{-itH}-e^{-itE}\mright)\psi\right\Vert \leq\left\Vert \mleft(H-E\mright)\psi\right\Vert t$
the theorem implies that
\begin{equation}
\left\Vert \mleft(e^{-itH_{\mathrm{eff}}}-e^{iM\epsilon_{k}t}\mright)\Psi\right\Vert \ll1,\quad M\epsilon_{k}t\ll k_{F}\left|k\right|M^{-\frac{3}{2}};
\end{equation}
as $\mleft(M\epsilon_{k}\mright)^{-1}$ is the characteristic timescale
of oscillation of $\Psi$ this is again non-trivial for $M\ll\mleft(k_{F}\left|k\right|\mright)^{\frac{2}{3}}$.

The formula for $\epsilon_{k}$ is also interesting - if one formally
replaces the Riemann sums by their corresponding integrals, one finds
that (to leading order)
\begin{equation}
\epsilon_{k}\sim\sqrt{2gn+\frac{12}{5}k_{F}^{2}\left|k\right|^{2}}
\end{equation}
where $n=\frac{N}{\mleft(2\pi\mright)^{3}}=\frac{s\left|B_{F}\right|}{\mleft(2\pi\mright)^{3}}\sim\frac{1}{\mleft(2\pi\mright)^{3}}\frac{4\pi s}{3}k_{F}^{3}$
is the number density of the system. In the physical case $g=4\pi e^{2}$
($e$ being the elementary charge), so recalling that $\frac{\hbar^{2}}{2m}=1$
we find
\begin{equation}
\epsilon_{k}\sim\sqrt{8\pi e^{2}n\frac{\hbar^{2}}{2m}+\frac{12}{5}k_{F}^{2}\left|k\right|^{2}}=\hbar\sqrt{\frac{4\pi ne^{2}}{m}+\frac{12}{5}\frac{k_{F}^{2}\left|k\right|^{2}}{\hbar^{2}}}=\hbar\sqrt{\omega_{0}^{2}+\frac{3}{5}v_{F}^{2}\left|k\right|^{2}}
\end{equation}
where $\omega_{0}=\sqrt{\frac{4\pi ne^{2}}{m}}$ is the famous plasmon
frequency (in CGS units) and $v_{F}=m^{-1}\hbar k_{F}=2\hbar^{-1}k_{F}$
is the Fermi velocity, corresponding to the well-known plasmon frequency
dispersion relation
\begin{equation}
\omega_{k}^{2}\approx\omega_{0}^{2}+\frac{3}{5}v_{F}^{2}\left|k\right|^{2}.
\end{equation}
This shows that if Theorem \ref{them:OperatorStatement} could be
generalized to the full physical setting, it would not only account
for the correlation energy but also for the plasmons predicted by
Bohm and Pines in the 1950's.

Although proving such a result would be an extremely challenging task,
it is our hope that the work covered by this thesis will be useful
in this endeavor.

\subsection{Outline of the Thesis}

We begin our analysis of the Hamiltonian $H_{N}$ in Section \ref{sec:LocalizationoftheHamiltonian}
by extracting the leading order contribution to the ground state energy
of $H_{N}$, which is the energy of the Fermi state $\psi_{F}$. We
do this by normal-ordering $H_{N}$ (in its second-quantized form)
``with respect to $\psi_{F}$''. After doing so we observe that
the resulting terms which violate the separation between states inside
and outside the Fermi ball are quasi-bosonic, in that they obey commutation
relations reminiscent of the canonical commutation relations of a
bosonic system.

\smallskip{}

In Section \ref{sec:OverviewofBosonicBogolubovTransformations} we
review the theory of bosonic Bogolubov transformations, originally
introduced in \cite{Bogolubov-47} to explain the phenomenon of superfluidity,
to prepare for the analysis of the quasi-bosonic operators. In particular
we describe how one may explicitly define a Bogolubov transformation
which diagonalizes a given positive-definite quadratic Hamiltonian.

\smallskip{}

We then apply the bosonic theory to our study of the Fermi gas in
Section \ref{sec:DiagonalizationoftheBosonizableTerms} wherein we
implement the diagonalization procedure in the quasi-bosonic setting.
This is done by mimicking the bosonic case to define a quasi-bosonic
Bogolubov transformation $e^{\mathcal{K}}$ which diagonalizes the
bosonizable terms of $H_{N}$ up to \textit{exchange terms} - terms
which arise due to the deviation from the exact CCR.

\smallskip{}

In Section \ref{sec:ControllingtheTransformationKernel} we justify
that the transformation $e^{\mathcal{K}}$ is well-defined by establishing
that the generating kernel $\mathcal{K}$ is in fact a bounded operator
under the condition $\sum_{k\in\mathbb{Z}_{\ast}^{3}}\hat{V}_{k}^{2}<\infty$.
More generally we establish a bound on $\mathcal{K}$ in terms of
the \textit{excitation number operator} $\mathcal{N}_{E}$ which will
also allow us to control error terms later on by a Gronwall-type argument.

\smallskip{}

In order to analyze the exchange terms which appeared during the diagonalization
procedure we require detailed information on the one-body operators
of the corresponding bosonic problem. We analyze these in section
\ref{sec:AnalysisofOne-BodyOperators}, obtaining asymptotically optimal
elementwise estimates of the main operators.

\smallskip{}

We then turn to the exchange terms themselves in Section \ref{sec:AnalysisofExchangeTerms}.
By performing a detailed analysis of all of the possible kinds of
terms which emerge from these upon normal-ordering with respect to
$\psi_{F}$, we extract the exchange contribution $E_{\mathrm{corr},\mathrm{ex}}$
and bound the remaining terms using $\mathcal{N}_{E}$.

\smallskip{}

In Section \ref{sec:EstimationoftheNon-BosonizableTermsandGronwallEstimates}
we bring all our work together. After deriving bounds on the \textit{non-bosonizable
terms} - the terms of $H_{N}$ which do not fit into the quasi-bosonic
setting - we apply our prior results to estimate the energy of the
trial state $e^{\mathcal{K}}\psi_{F}$, which results in the proof
of Theorem \ref{them:MainTheorem}.

\smallskip{}

This is followed by Section \ref{sec:ExtensiontoAttractivePotentials}
wherein we describe the modifications necessary to extend Theorem
\ref{them:MainTheorem} to weakly attractive potentials in order to
conclude Theorem \ref{them:AttractiveGeneralization}.\smallskip{}

In Section \ref{sec:OverviewoftheOperatorResult} we first present
a general outline of the approach that leads to Theorem \ref{them:OperatorStatement},
followed by a more detalied examination of the key ideas which leads
to its conclusion.\smallskip{}

Finally, in Section \ref{sec:PlasmonModesoftheEffectiveHamiltonian},
we consider plasmon states for the effective operator of Theorem \ref{them:OperatorStatement}
in the non-mean-field setting, proving a generalization of Theorem
\ref{them:PlasmonStates} valid for arbitrary repulsive potentials
$V$.

\subsubsection*{Acknowledgements}

I thank my Ph.D. advisors Christian Hainzl and Phan Th\`anh Nam for
guidance and assistance throughout the project, and for encouraging
me to approach and understand problems in my own way.

I also thank my father for the copious amounts of proofreading and
encouragement he has supported me with.

\section{\label{sec:LocalizationoftheHamiltonian}Localization of the Hamiltonian
at the Fermi State}

In this section we begin our study of the interacting Fermi gas by
extracting the energy of the Fermi state $\psi_{F}$ from the Hamiltonian
operator $H_{N}$. We do this by normal-ordering $H_{N}$ ``with
respect to $\psi_{F}$'', a procedure which we refer to as localization
since it serves to fix $\psi_{F}$ as our point of reference, making
it analogous to the vacuum state of a field theory.

The result of this procedure is summarized in the following:
\begin{prop}
\label{prop:LocalizedHamiltonian}It holds that
\[
H_{N}=E_{F}+H_{\mathrm{kin}}^{\prime}+\sum_{k\in\mathbb{Z}_{\ast}^{3}}\frac{\hat{V}_{k}k_{F}^{-1}}{2\,\mleft(2\pi\mright)^{3}}\mleft(2B_{k}^{\ast}B_{k}+B_{k}B_{-k}+B_{-k}^{\ast}B_{k}^{\ast}\mright)+\mathcal{C}+\mathcal{Q}
\]
where $E_{F}=\left\langle \psi_{F},H_{N}\psi_{F}\right\rangle $ is
the energy of the Fermi state,
\[
H_{\mathrm{kin}}^{\prime}=\sum_{p\in B_{F}^{c}}^{\sigma}\left|p\right|^{2}c_{p,\sigma}^{\ast}c_{p,\sigma}-\sum_{p\in B_{F}}^{\sigma}\left|p\right|^{2}c_{p,\sigma}c_{p,\sigma}^{\ast},\quad B_{k}=\sum_{p\in L_{k}}^{\sigma}c_{p-k,\sigma}^{\ast}c_{p,\sigma},
\]
and
\begin{align*}
\mathcal{C} & =\frac{k_{F}^{-1}}{\mleft(2\pi\mright)^{3}}\sum_{k\in\mathbb{Z}_{\ast}^{3}}\hat{V}_{k}\,\mathrm{Re}\mleft(\mleft(B_{k}+B_{-k}^{\ast}\mright)^{\ast}D_{k}\mright)\\
\mathcal{Q} & =\frac{k_{F}^{-1}}{2\,\mleft(2\pi\mright)^{3}}\sum_{k\in\mathbb{Z}_{\ast}^{3}}\hat{V}_{k}\mleft(D_{k}^{\ast}D_{k}-\sum_{p\in L_{k}}^{\sigma}\mleft(c_{p,\sigma}^{\ast}c_{p,\sigma}+c_{p-k,\sigma}c_{p-k,\sigma}^{\ast}\mright)\mright)
\end{align*}
for $D_{k}=\mathrm{d}\Gamma\mleft(P_{B_{F}}e^{-ik\cdot x}P_{B_{F}}\mright)+\mathrm{d}\Gamma\mleft(P_{B_{F}^{c}}e^{-ik\cdot x}P_{B_{F}^{c}}\mright)$.
\end{prop}

After carrying out this procedure we will see how the concept of quasi-bosonicity
emerges: The operators $B_{k}$ of the above representation obey commutation
relations which are analogous to the canonical commutation relations
of a bosonic system. We end the section by exploring this phenomenon,
in particular showing how the kinetic operator $H_{\mathrm{kin}}^{\prime}$
can be made to fit into such a bosonic picture by considering the
\textit{excitation operators}
\begin{equation}
b_{k,p}=\frac{1}{\sqrt{s}}\sum_{\sigma=1}^{s}c_{p-k,\sigma}^{\ast}c_{p,\sigma},\quad b_{k,p}^{\ast}=\frac{1}{\sqrt{s}}\sum_{\sigma=1}^{s}c_{p,\sigma}^{\ast}c_{p-k,\sigma}.
\end{equation}

\subsection{Notation and Conventions}

Before we begin the analysis proper we review the notation which we
will use throughout the paper.

We consider the one-particle space $\mathfrak{h}=L^{2}\mleft(\mathbb{T}^{3};\mathbb{C}^{s}\mright)$,
where $\mathbb{T}^{3}=\left[0,2\pi\right]^{3}$ with periodic boundary
conditions and $s\in\mathbb{N}$ is the number of spin states of the
system. We denote by $\mathcal{H}_{N}=\bigwedge^{N}\mathfrak{h}$
the associated fermionic $N$-particle space.

$\mathfrak{h}$ is spanned by the orthonormal basis of plane wave
states $\mleft(u_{p,\sigma}\mright)_{p\in\mathbb{Z}^{3}}^{1\leq\sigma\leq s}$,
given by
\begin{equation}
u_{p,\sigma}\mleft(x\mright)=\mleft(2\pi\mright)^{-\frac{3}{2}}e^{ip\cdot x}v_{\sigma},\quad p\in\mathbb{Z}^{3},
\end{equation}
where $v_{\sigma}$ denotes the $\sigma$-th standard basis vector
of $\mathbb{C}^{s}$.

We denote by $c_{p,\sigma}^{\ast}$, $c_{p,\sigma}$ the creation
and annihilation operators associated to the plane wave states, which
obey the canonical anticommutation relations (CAR)
\begin{equation}
\left\{ c_{p,\sigma},c_{q,\tau}^{\ast}\right\} =\delta_{p,q}\delta_{\sigma,\tau},\quad\left\{ c_{p,\sigma},c_{q,\tau}\right\} =0=\left\{ c_{p,\sigma}^{\ast},c_{q,\tau}^{\ast}\right\} ,
\end{equation}
for all $p,q\in\mathbb{Z}^{3}$ and $1\leq\sigma,\tau\leq s$.

Sums involving the creation and annihilation operators will generally
run over all spin states. To reduce clutter we will denote this by
writing the summed indices over the sum signs, leaving the summation
range implicit, e.g. for the number operators $\mathcal{N}$ we simply
write
\begin{equation}
\mathcal{N}=\sum_{p\in\mathbb{Z}^{3}}\sum_{\sigma=1}^{s}c_{p,\sigma}^{\ast}c_{p,\sigma}=\sum_{p\in\mathbb{Z}^{3}}^{\sigma}c_{p,\sigma}^{\ast}c_{p,\sigma}.
\end{equation}
For a given Fermi momentum $k_{F}>0$ we denote by $B_{F}$ the (closed)
Fermi ball
\begin{equation}
B_{F}=\overline{B}\mleft(0,k_{F}\mright)\cap\mathbb{Z}^{3}
\end{equation}
and write $B_{F}^{c}$ for the complement of $B_{F}$ with respect
to $\mathbb{Z}^{3}$. We define $\psi_{F}$ to be the Fermi state
\begin{equation}
\psi_{F}=\bigwedge_{p\in B_{F}}^{\sigma}u_{p,\sigma}\in\mathcal{H}_{N},\quad N=s\left|B_{F}\right|.
\end{equation}
For the sake of brevity we define $\mathbb{Z}_{\ast}^{3}=\mathbb{Z}^{3}\backslash\{0\}$
and for $k\in\mathbb{Z}_{\ast}^{3}$ define the lune $L_{k}$ by
\begin{equation}
L_{k}=\mleft(B_{F}+k\mright)\backslash B_{F}=\left\{ p\in\mathbb{Z}^{3}\mid\left|p-k\right|\leq k_{F}<\left|p\right|\right\} .
\end{equation}

\subsubsection*{The Hamiltonian Operator $H_{N}$}

We consider for given $k_{F}>0$ the mean-field Hamiltonian
\begin{equation}
H_{N}=H_{\mathrm{kin}}+k_{F}^{-1}H_{\mathrm{int}},\quad D\mleft(H_{N}\mright)=D\mleft(H_{\mathrm{kin}}\mright),
\end{equation}
on $\mathcal{H}_{N}$, where $N=s\left|B_{F}\right|$. $H_{\mathrm{kin}}$
is the standard kinetic operator
\begin{equation}
H_{\mathrm{kin}}=\mathrm{d}\Gamma\mleft(-\Delta\mright)=-\sum_{i=1}^{N}\Delta_{i},\quad D\mleft(H_{\mathrm{kin}}\mright)=\bigwedge^{N}H^{2}\mleft(\mathbb{T}^{3};\mathbb{C}^{s}\mright),
\end{equation}
and $H_{\mathrm{int}}$ describes the pairwise interaction between
$N$ particles through a potential $V:\mathbb{T}^{3}\rightarrow\mathbb{R}$,
\begin{equation}
H_{\mathrm{int}}=\sum_{1\leq i<j\leq N}V\mleft(x_{i}-x_{j}\mright).
\end{equation}
We will take $V\in L^{2}\mleft(\mathbb{T}^{3}\mright)$, in which case
$H_{N}$ is a self-adjoint operator on $\mathcal{H}_{N}$. Letting
the Fourier decomposition of $V$ be given by
\begin{equation}
V\mleft(x\mright)=\frac{1}{\mleft(2\pi\mright)^{3}}\sum_{k\in\mathbb{Z}^{3}}\hat{V}_{k}e^{ik\cdot x}
\end{equation}
we furthermore assume that $\hat{V}_{k}=\hat{V}_{-k}$ and $\hat{V}_{k}\geq0$
for all $k\in\mathbb{Z}_{\ast}^{3}$, i.e. that $V$ is \textit{repulsive}.

For the remainder of the thesis we will work in the second-quantized
picture, in which it is well-known that $H_{\mathrm{kin}}$ and $H_{\mathrm{int}}$
can be expressed as
\begin{equation}
H_{\mathrm{kin}}=\sum_{p\in\mathbb{Z}^{3}}^{\sigma}\left|p\right|^{2}c_{p,\sigma}^{\ast}c_{p,\sigma},\quad H_{\mathrm{int}}=\frac{1}{2\,\mleft(2\pi\mright)^{3}}\sum_{k\in\mathbb{Z}^{3}}\hat{V}_{k}\sum_{p,q\in\mathbb{Z}^{3}}^{\sigma,\tau}c_{p+k,\sigma}^{\ast}c_{q-k,\tau}^{\ast}c_{q,\tau}c_{p,\sigma}.\label{eq:SecondQuantizedOperators}
\end{equation}

\subsection{Extraction of the Fermi State Energy}

It is well-known that the Fermi state $\psi_{F}$ is characterized
by the conditions
\begin{equation}
c_{p}\psi_{F}=0=c_{q}^{\ast}\psi_{F},\quad p\in B_{F}^{c},\,q\in B_{F},
\end{equation}
and so the Fermi state energy $E_{F}=\left\langle \psi_{F},H_{N}\psi_{F}\right\rangle $
can be extracted from $H_{N}$ by normal-ordering this ``with respect
to $\psi_{F}$'', in the sense that the creation and annihilation
operators of equation (\ref{eq:SecondQuantizedOperators}) are normal-ordered
as if $c_{p,\sigma}^{\ast}$ were an annihilation operator for $p\in B_{F}$.

Consider first the kinetic operator: By the CAR we can write $H_{\mathrm{kin}}$
in the form
\begin{align}
H_{\mathrm{kin}} & =\sum_{p\in B_{F}^{c}}^{\sigma}\left|p\right|^{2}c_{p,\sigma}^{\ast}c_{p,\sigma}+\sum_{p\in B_{F}}^{\sigma}\left|p\right|^{2}c_{p,\sigma}^{\ast}c_{p,\sigma}=\sum_{p\in B_{F}^{c}}^{\sigma}\left|p\right|^{2}c_{p,\sigma}^{\ast}c_{p,\sigma}+\sum_{p\in B_{F}}^{\sigma}\left|p\right|^{2}-\sum_{p\in B_{F}}^{\sigma}\left|p\right|^{2}c_{p,\sigma}c_{p,\sigma}^{\ast}\\
 & =s\sum_{p\in B_{F}}\left|p\right|^{2}+H_{\mathrm{kin}}^{\prime}\nonumber 
\end{align}
where we define the \textit{localized kinetic operator} $H_{\mathrm{kin}}^{\prime}:D\mleft(H_{\mathrm{kin}}\mright)\subset\mathcal{H}_{N}\rightarrow\mathcal{H}_{N}$
by
\begin{equation}
H_{\mathrm{kin}}^{\prime}=\sum_{p\in B_{F}^{c}}^{\sigma}\left|p\right|^{2}c_{p,\sigma}^{\ast}c_{p,\sigma}-\sum_{p\in B_{F}}^{\sigma}\left|p\right|^{2}c_{p,\sigma}c_{p,\sigma}^{\ast}.
\end{equation}
$H_{\mathrm{kin}}^{\prime}$ is clearly normal-ordered with respect
to $\psi_{F}$, and so the quantity $s\sum_{p\in B_{F}}\left|p\right|^{2}$
is simply the kinetic energy of $\psi_{F}$, whence we can write the
relation between $H_{\mathrm{kin}}$ and $H_{\mathrm{kin}}^{\prime}$
as
\begin{equation}
H_{\mathrm{kin}}=\left\langle \psi_{F},H_{\mathrm{kin}}\psi_{F}\right\rangle +H_{\mathrm{kin}}^{\prime}.\label{eq:LocalizedKineticOperator}
\end{equation}
To normal-order $H_{\mathrm{int}}$ we first rewrite this in a factorized
form: By the CAR we can write
\begin{align}
H_{\mathrm{int}} & =\frac{1}{2\,\mleft(2\pi\mright)^{3}}\sum_{k\in\mathbb{Z}^{3}}\hat{V}_{k}\sum_{p,q\in\mathbb{Z}^{3}}^{\sigma,\tau}c_{p+k,\sigma}^{\ast}\mleft(c_{p,\sigma}c_{q-k,\tau}^{\ast}-\delta_{p,q-k}\delta_{\sigma,\tau}\mright)c_{q,\tau}\nonumber \\
 & =\frac{1}{2\,\mleft(2\pi\mright)^{3}}\sum_{k\in\mathbb{Z}^{3}}\hat{V}_{k}\mleft(\mleft(\sum_{p\in\mathbb{Z}^{3}}^{\sigma}c_{p+k,\sigma}^{\ast}c_{p,\sigma}\mright)\mleft(\sum_{q\in\mathbb{Z}^{3}}^{\tau}c_{q-k,\tau}^{\ast}c_{q,\tau}\mright)-\sum_{q\in\mathbb{Z}^{3}}^{\sigma}c_{q,\sigma}^{\ast}c_{q,\sigma}\mright)\label{eq:HintFactorizedForm}\\
 & =\frac{1}{2\,\mleft(2\pi\mright)^{3}}\sum_{k\in\mathbb{Z}^{3}}\hat{V}_{k}\mleft(\mathrm{d}\Gamma\mleft(e^{-ik\cdot x}\mright)^{\ast}\mathrm{d}\Gamma\mleft(e^{-ik\cdot x}\mright)-\mathcal{N}\mright)\nonumber \\
 & =\frac{N\mleft(N-1\mright)}{2\,\mleft(2\pi\mright)^{3}}\hat{V}_{0}+\frac{1}{2\,\mleft(2\pi\mright)^{3}}\sum_{k\in\mathbb{Z}_{\ast}^{3}}\hat{V}_{k}\mleft(\mathrm{d}\Gamma\mleft(e^{-ik\cdot x}\mright)^{\ast}\mathrm{d}\Gamma\mleft(e^{-ik\cdot x}\mright)-N\mright)\nonumber 
\end{align}
where we recognized the operator $\mathrm{d}\Gamma\mleft(e^{-ik\cdot x}\mright)$
as
\begin{equation}
\mathrm{d}\Gamma\mleft(e^{-ik\cdot x}\mright)=\sum_{p,q\in\mathbb{Z}^{3}}^{\sigma,\tau}\left\langle u_{p,\sigma},e^{-ik\cdot x}u_{q,\tau}\right\rangle c_{p,\sigma}^{\ast}c_{q,\tau}=\sum_{p,q\in\mathbb{Z}^{3}}^{\sigma,\tau}\delta_{p,q-k}\delta_{\sigma,\tau}c_{p,\sigma}^{\ast}c_{q,\tau}=\sum_{q\in\mathbb{Z}^{3}}^{\tau}c_{q-k,\tau}^{\ast}c_{q,\tau}
\end{equation}
and used that $\mathrm{d}\Gamma\mleft(e^{-i\mleft(0\cdot x\mright)}\mright)=\mathrm{d}\Gamma\mleft(1\mright)=\mathcal{N}=N$
on $\mathcal{H}_{N}$. Now, with $P_{B_{F}}:\mathfrak{h}\rightarrow\mathfrak{h}$
denoting the orthogonal projection onto $\mathrm{span}\mleft(u_{p,\sigma}\mright)_{p\in\mathbb{Z}^{3}}^{1\leq\sigma\leq s}$
and $P_{B_{F}^{c}}=1-P_{B_{F}}$ denoting its complement, we can decompose
$\mathrm{d}\Gamma\mleft(e^{-ik\cdot x}\mright)$ as
\begin{equation}
\mathrm{d}\Gamma\mleft(e^{-ik\cdot x}\mright)=\mathrm{d}\Gamma\mleft(\mleft(P_{B_{F}}+P_{B_{F}^{c}}\mright)e^{-ik\cdot x}\mleft(P_{B_{F}}+P_{B_{F}^{c}}\mright)\mright)=B_{k}+B_{-k}^{\ast}+D_{k}
\end{equation}
where the operator $B_{k}$ is given by
\begin{equation}
B_{k}=\mathrm{d}\Gamma\mleft(P_{B_{F}}e^{-ik\cdot x}P_{B_{F}^{c}}\mright)=\sum_{p\in B_{F}}^{\sigma}\sum_{q\in B_{F}^{c}}^{\tau}\delta_{p,q-k}\delta_{\sigma,\tau}c_{p,\sigma}^{\ast}c_{q,\tau}=\sum_{q\in L_{k}}^{\tau}c_{q-k,\tau}^{\ast}c_{q,\tau}
\end{equation}
as the Kronecker delta $\delta_{p,q-k}$ precisely restrict the summation
to $q\in L_{k}$, and the operator $D_{k}$ is simply
\begin{equation}
D_{k}=\mathrm{d}\Gamma\mleft(P_{B_{F}}e^{-ik\cdot x}P_{B_{F}}\mright)+\mathrm{d}\Gamma\mleft(P_{B_{F}^{c}}e^{-ik\cdot x}P_{B_{F}^{c}}\mright).
\end{equation}
We can thus write $H_{\mathrm{int}}$ as
\begin{align}
H_{\mathrm{int}} & =\frac{N\mleft(N-1\mright)}{2\,\mleft(2\pi\mright)^{3}}\hat{V}_{0}+\frac{1}{2\,\mleft(2\pi\mright)^{3}}\sum_{k\in\mathbb{Z}_{\ast}^{3}}\hat{V}_{k}\mleft(\mleft(\mleft(B_{k}+B_{-k}^{\ast}\mright)^{\ast}\mleft(B_{k}+B_{-k}^{\ast}\mright)-N\mright.\mright)\\
 & \qquad\qquad\qquad\qquad\qquad\qquad\qquad\qquad\qquad\;\,\mleft.+2\,\mathrm{Re}\mleft(\mleft(B_{k}+B_{-k}^{\ast}\mright)^{\ast}D_{k}\mright)+D_{k}^{\ast}D_{k}\mright).\nonumber 
\end{align}
Now, it is easily verified that $B_{k}\psi_{F}=D_{k}\psi_{F}=D_{k}^{\ast}\psi_{F}=0$
for any $k\in\mathbb{Z}_{\ast}^{3}$, and so the terms on the last
line are effectively normal-ordered, and it only remains to normal-order
the terms of the first sum. For this we calculate the commutator $\left[B_{k},B_{k}^{\ast}\right]$:
By the CAR and basic commutator identities, we find that
\begin{align}
\left[B_{k},B_{k}^{\ast}\right] & =\sum_{p\in L_{k}}^{\sigma}\sum_{q\in L_{k}}^{\tau}\left[c_{p-k,\sigma}^{\ast}c_{p,\sigma},c_{q,\tau}^{\ast}c_{q-k,\tau}\right]=\sum_{p\in L_{k}}^{\sigma}\sum_{q\in L_{k}}^{\tau}\mleft(c_{p-k,\sigma}^{\ast}\left[c_{p,\sigma},c_{q,\tau}^{\ast}c_{q-k,\tau}\right]+\left[c_{p-k,\sigma}^{\ast},c_{q,\tau}^{\ast}c_{q-k,\tau}\right]c_{p,\sigma}\mright)\nonumber \\
 & =\sum_{p\in L_{k}}^{\sigma}\sum_{q\in L_{k}}^{\tau}c_{p-k,\sigma}^{\ast}\mleft(\left\{ c_{p,\sigma},c_{q,\tau}^{\ast}\right\} c_{q-k,\tau}-c_{q,\tau}^{\ast}\left\{ c_{p,\sigma},c_{q-k,\tau}\right\} \mright)\nonumber \\
 & +\sum_{p\in L_{k}}^{\sigma}\sum_{q\in L_{k}}^{\tau}\mleft(\left\{ c_{p-k,\sigma}^{\ast},c_{q,\tau}^{\ast}\right\} c_{q-k,\tau}-c_{q,\tau}^{\ast}\left\{ c_{p-k,\sigma}^{\ast},c_{q-k,\tau}\right\} \mright)c_{p,\sigma}\label{eq:BkBkastCommutator}\\
 & =\sum_{p\in L_{k}}^{\sigma}\sum_{q\in L_{k}}^{\tau}\delta_{p,q}\delta_{\sigma,\tau}c_{p-k,\sigma}^{\ast}c_{q-k,\tau}-\sum_{p\in L_{k}}^{\sigma}\sum_{q\in L_{k}}^{\tau}\delta_{p-k,q-k}\delta_{\sigma,\tau}c_{q,\tau}^{\ast}c_{p,\sigma}\nonumber \\
 & =\sum_{p\in L_{k}}^{\sigma}c_{p-k,\sigma}^{\ast}c_{p-k,\sigma}-\sum_{p\in L_{k}}^{\sigma}c_{p,\sigma}^{\ast}c_{p,\sigma}=s\left|L_{k}\right|-\sum_{p\in L_{k}}^{\sigma}\mleft(c_{p,\sigma}^{\ast}c_{p,\sigma}+c_{p-k,\sigma}c_{p-k,\sigma}^{\ast}\mright)\nonumber 
\end{align}
and using also that $\hat{V}_{k}=\hat{V}_{-k}$ we may then write
$H_{\mathrm{int}}$ as
\begin{align}
H_{\mathrm{int}} & =\frac{N\mleft(N-1\mright)}{2\,\mleft(2\pi\mright)^{3}}\hat{V}_{0}-\frac{1}{2\,\mleft(2\pi\mright)^{3}}\sum_{k\in\mathbb{Z}_{\ast}^{3}}\hat{V}_{k}\mleft(N-s\left|L_{k}\right|\mright)+\frac{1}{2\,\mleft(2\pi\mright)^{3}}\sum_{k\in\mathbb{Z}_{\ast}^{3}}\hat{V}_{k}\mleft(2B_{k}^{\ast}B_{k}+B_{k}^{\ast}B_{-k}^{\ast}+B_{-k}B_{k}\mright)\nonumber \\
 & +\frac{1}{2\,\mleft(2\pi\mright)^{3}}\sum_{k\in\mathbb{Z}_{\ast}^{3}}\hat{V}_{k}\mleft(2\,\mathrm{Re}\mleft(\mleft(B_{k}+B_{-k}^{\ast}\mright)^{\ast}D_{k}\mright)+D_{k}^{\ast}D_{k}-\sum_{p\in L_{k}}^{\sigma}\mleft(c_{p,\sigma}^{\ast}c_{p,\sigma}+c_{p-k,\sigma}c_{p-k,\sigma}^{\ast}\mright)\mright).
\end{align}
Note that the sum $\sum_{k\in\mathbb{Z}_{\ast}^{3}}\hat{V}_{k}\mleft(N-s\left|L_{k}\right|\mright)$
is actually finite, as $s\left|L_{k}\right|=s\left|B_{F}\right|=N$
when $\left|k\right|>2k_{F}$.

The terms on the right-hand side of this equation are now normal-ordered,
and in particular we see that
\begin{equation}
\left\langle \psi_{F},H_{\mathrm{int}}\psi_{F}\right\rangle =\frac{N\mleft(N-1\mright)}{2\,\mleft(2\pi\mright)^{3}}\hat{V}_{0}-\frac{1}{2\,\mleft(2\pi\mright)^{3}}\sum_{k\in\mathbb{Z}_{\ast}^{3}}\hat{V}_{k}\mleft(N-s\left|L_{k}\right|\mright)\label{eq:FSInteractionEnergy}
\end{equation}
whence we can write
\begin{equation}
k_{F}^{-1}H_{\mathrm{int}}=\left\langle \psi_{F},k_{F}^{-1}H_{\mathrm{int}}\psi_{F}\right\rangle +\sum_{k\in\mathbb{Z}_{\ast}^{3}}\frac{\hat{V}_{k}k_{F}^{-1}}{2\,\mleft(2\pi\mright)^{3}}\mleft(2B_{k}^{\ast}B_{k}+B_{k}^{\ast}B_{-k}^{\ast}+B_{-k}B_{k}\mright)+\mathcal{C}+\mathcal{Q}\label{eq:LocalizedInterationOperator}
\end{equation}
where the \textit{cubic} and \textit{quartic} terms $\mathcal{C}$
and $\mathcal{Q}$ are defined by
\begin{align}
\mathcal{C} & =\frac{k_{F}^{-1}}{\mleft(2\pi\mright)^{3}}\sum_{k\in\mathbb{Z}_{\ast}^{3}}\hat{V}_{k}\,\mathrm{Re}\mleft(\mleft(B_{k}+B_{-k}^{\ast}\mright)^{\ast}D_{k}\mright)\\
\mathcal{Q} & =\frac{k_{F}^{-1}}{2\,\mleft(2\pi\mright)^{3}}\sum_{k\in\mathbb{Z}_{\ast}^{3}}\hat{V}_{k}\mleft(D_{k}^{\ast}D_{k}-\sum_{p\in L_{k}}^{\sigma}\mleft(c_{p,\sigma}^{\ast}c_{p,\sigma}+c_{p-k,\sigma}c_{p-k,\sigma}^{\ast}\mright)\mright).\nonumber 
\end{align}
The terms $\mathcal{C}$ and $\mathcal{Q}$ constitute the \textit{non-bosonizable
terms}: They fall outside the quasi-bosonic approach we will introduce
below, and so we consider them as error terms to be analyzed separately
at the end.

Combining the equations (\ref{eq:LocalizedKineticOperator}) and (\ref{eq:LocalizedInterationOperator})
now yields Proposition \ref{prop:LocalizedHamiltonian}.

\subsection{Remarks on the Localization Procedure}

Before continuing with our analysis we must comment on some subtle
details of the localization procedure.

Consider the localized kinetic operator $H_{\mathrm{kin}}^{\prime}$,
which we defined by
\begin{equation}
H_{\mathrm{kin}}^{\prime}=\sum_{p\in B_{F}^{c}}^{\sigma}\left|p\right|^{2}c_{p,\sigma}^{\ast}c_{p,\sigma}-\sum_{p\in B_{F}}^{\sigma}\left|p\right|^{2}c_{p,\sigma}c_{p,\sigma}^{\ast}.
\end{equation}
This expression is a sum of two terms, one manifestly positive and
one manifestly negative. As the creation and annihilation operator
for orthogonal states are (algebraically) independent, one would therefore
not expect $H_{\mathrm{kin}}^{\prime}$ to have a definite sign. But
this is not the case, as we can argue that
\begin{equation}
H_{\mathrm{kin}}^{\prime}=H_{\mathrm{kin}}-\left\langle \psi_{F},H_{\mathrm{kin}}\psi_{F}\right\rangle \geq0
\end{equation}
since $\left\langle \psi_{F},H_{\mathrm{kin}}\psi_{F}\right\rangle $
is the ground state energy of $H_{\mathrm{kin}}$.

The resolution of this apparent paradox lies in the domains of definition:
The argument for non-definiteness of $H_{\mathrm{kin}}^{\prime}$
is valid\textit{ when viewed as an operator on the full Fock space
$\mathcal{F}^{-}\mleft(\mathfrak{h}\mright)$}, where the assertion
that $\left\langle \psi_{F},H_{\mathrm{kin}}\psi_{F}\right\rangle $
is the ground state energy of $H_{\mathrm{kin}}$ is wrong.

That $H_{\mathrm{kin}}^{\prime}\geq0$ is nonetheless correct when
viewed as an operator on $\mathcal{H}_{N}$, precisely by the second
observation. The first argument fails in this case because the creation
and annihilation operators (or more precisely, the products $c_{p,\sigma}^{\ast}c_{p,\sigma}$)
are \textit{not} independent on $\mathcal{H}_{N}$: Normal-ordering
$\mathcal{N}$ with respect to $\psi_{F}$, we see that
\begin{align}
N & =\mathcal{N}=\sum_{p\in\mathbb{Z}^{3}}^{\sigma}c_{p,\sigma}^{\ast}c_{p,\sigma}=\sum_{p\in B_{F}^{c}}^{\sigma}c_{p,\sigma}^{\ast}c_{p,\sigma}+\sum_{p\in B_{F}}^{\sigma}1-\sum_{p\in B_{F}}^{\sigma}c_{p,\sigma}c_{p,\sigma}^{\ast}\\
 & =s\left|B_{F}\right|+\sum_{p\in B_{F}^{c}}^{\sigma}c_{p,\sigma}^{\ast}c_{p,\sigma}-\sum_{p\in B_{F}}^{\sigma}c_{p,\sigma}c_{p,\sigma}^{\ast},\nonumber 
\end{align}
so as $N=s\left|B_{F}\right|$ we conclude the identity
\begin{equation}
\sum_{p\in B_{F}^{c}}^{\sigma}c_{p,\sigma}^{\ast}c_{p,\sigma}=\sum_{p\in B_{F}}^{\sigma}c_{p,\sigma}c_{p,\sigma}^{\ast}\quad\text{on}\,\mathcal{H}_{N}.\label{eq:ParticleHoleSymmetry}
\end{equation}
This is the statement of \textit{particle-hole symmetry}: The expression
on the left-hand side is appropriately labeled the \textit{excitation
number operator} $\mathcal{N}_{E}$, since just as $\mathcal{N}$
``counts'' the number of particles in a state of the full Fock space,
$\mathcal{N}_{E}$ ``counts'' the number of states lying outside
$B_{F}$ in a state on $\mathcal{H}_{N}$, which is to say the number
of excitations relative to $\psi_{F}$.

The expression on the right-hand side may be similarly thought of
as a ``hole number operator'', as it similarly counts the number
of states lying inside $B_{F}$ that a given state is \textit{lacking}.
Equation (\ref{eq:ParticleHoleSymmetry}) thus makes explicit the
observation that any excitation relative to $\psi_{F}$ must be accompanied
by a ``hole''.

This also explains why $H_{\mathrm{kin}}^{\prime}$, despite being
the difference of two positive operators, remains positive: To take
advantage of the negative part, one must create a hole in the Fermi
ball. But particle number conservation then demands that one must
create an excitation outside this, and as $\left|p\right|>k_{F}\geq\left|q\right|$
for all $p\in B_{F}^{c}$, $q\in B_{F}$, this procedure will always
lead to an increase in energy.

In fact we can use equation (\ref{eq:ParticleHoleSymmetry}) to make
this argument precise, since it implies that
\begin{equation}
H_{\mathrm{kin}}^{\prime}=H_{\mathrm{kin}}^{\prime}-k_{F}^{2}\mathcal{N}_{E}+k_{F}^{2}\mathcal{N}_{E}=\sum_{p\in B_{F}^{c}}^{\sigma}\mleft(\left|p\right|^{2}-k_{F}^{2}\mright)c_{p,\sigma}^{\ast}c_{p,\sigma}+\sum_{p\in B_{F}}^{\sigma}\mleft(k_{F}^{2}-\left|p\right|^{2}\mright)c_{p,\sigma}c_{p,\sigma}^{\ast}
\end{equation}
and now both of the sums on the right-hand side are manifestly non-negative.

\subsection{The Quasi-Bosonic Excitation Operators}

Now we consider the structure of the terms
\begin{equation}
\sum_{k\in\mathbb{Z}_{\ast}^{3}}\frac{\hat{V}_{k}k_{F}^{-1}}{2\,\mleft(2\pi\mright)^{3}}\mleft(2B_{k}^{\ast}B_{k}+B_{k}^{\ast}B_{-k}^{\ast}+B_{-k}B_{k}\mright),
\end{equation}
which appear in the decomposition of $H_{N}$ of Proposition \ref{prop:LocalizedHamiltonian},
further. Consider the operators $B_{k}=\sum_{p\in L_{k}}^{\sigma}c_{p-k,\sigma}^{\ast}c_{p,\sigma}$:
It is easily seen that for any $k,l\in\mathbb{Z}_{\ast}^{3}$ it holds
that $\left[B_{k},B_{l}\right]=\left[B_{k}^{\ast},B_{l}^{\ast}\right]=0$,
while a slight modification of the calculation of equation (\ref{eq:BkBkastCommutator})
shows that
\begin{equation}
\left[B_{k},B_{l}^{\ast}\right]=s\left|L_{k}\right|\delta_{k,l}-\sum_{p\in L_{k}}^{\sigma}\sum_{q\in L_{l}}\mleft(\delta_{p-k,q-l}c_{q,\sigma}^{\ast}c_{p,\sigma}+\delta_{p,q}c_{q-l,\sigma}c_{p-k,\sigma}^{\ast}\mright).\label{eq:BkBlastCommutator}
\end{equation}
Consider the sum on the right: By the Cauchy-Schwarz and triangle
inequalities we can bound the first part as
\begin{align}
 & \quad\;\;\;\left|\sum_{p\in L_{k}}^{\sigma}\sum_{q\in L_{l}}\left\langle \Psi,\delta_{p-k,q-l}c_{q,\sigma}^{\ast}c_{p,\sigma}\Psi\right\rangle \right|\leq\sum_{p\in L_{k}}^{\sigma}\sum_{q\in L_{l}}\delta_{p-k,q-l}\left\Vert c_{q,\sigma}\Psi\right\Vert \left\Vert c_{p,\sigma}\Psi\right\Vert \\
 & \leq\sqrt{\sum_{q\in L_{l}}^{\sigma}\left\Vert c_{q,\sigma}\Psi\right\Vert ^{2}}\sqrt{\sum_{p\in L_{k}}^{\sigma}\left\Vert c_{p,\sigma}\Psi\right\Vert ^{2}}\leq\left\langle \Psi,\mathcal{N}_{E}\Psi\right\rangle \nonumber 
\end{align}
for any $\Psi\in\mathcal{H}_{N}$, and likewise for the second part
of the sum. If one now defines the rescaled operators $B_{k}^{\prime}=\mleft(s\left|L_{k}\right|\mright)^{-\frac{1}{2}}B_{k}$,
one sees that these obey commutation relations of the form
\begin{equation}
\left[B_{k}^{\prime},\mleft(B_{l}^{\prime}\mright)^{\ast}\right]=\delta_{k,l}+O\mleft(k_{F}^{-2}\mathcal{N}_{E}\mright),\quad\left[B_{k}^{\prime},B_{l}^{\prime}\right]=0=\left[\mleft(B_{k}^{\prime}\mright)^{\ast},\mleft(B_{l}^{\prime}\mright)^{\ast}\right],
\end{equation}
since (as we will see) $\left|L_{k}\right|\geq ck_{F}^{2}$. With
respect to states for which $\left\langle \Psi,\mathcal{N}_{E}\Psi\right\rangle $
is small, these relations approximate the canonical commutation relations
for \textit{bosonic} creation and annihilation operators $a_{k}^{\ast}$,
$a_{k}$, which are
\begin{equation}
\left[a_{k},a_{l}^{\ast}\right]=\delta_{k,l},\quad\left[a_{k},a_{l}\right]=0=\left[a_{k}^{\ast},a_{l}^{\ast}\right].
\end{equation}
This motivates describing the $B_{k}$ as being \textit{quasi-bosonic}
operators. In view of this, it is tempting to view the terms
\begin{equation}
\sum_{k\in\mathbb{Z}_{\ast}^{3}}\frac{\hat{V}_{k}k_{F}^{-1}}{2\,\mleft(2\pi\mright)^{3}}\mleft(2B_{k}^{\ast}B_{k}+B_{k}^{\ast}B_{-k}^{\ast}+B_{-k}B_{k}\mright)
\end{equation}
as analogous to a quadratic Hamiltonian in the bosonic setting, to
which the theory of Bogolubov transformations applies. This is the
spirit of what we will do, but there is a catch: The kinetic operator
$H_{\mathrm{kin}}^{\prime}$ is not of a similar form, and the operators
$B_{k}$ do not behave bosonically with respect to it.

The solution to this problem is to further decompose the operators
$B_{k}$: We define for $k\in\mathbb{Z}_{\ast}^{3}$, $p\in L_{k}$,
the \textit{excitation operators} $b_{k,p}^{\ast}$, $b_{k,p}$ by
\begin{equation}
b_{k,p}=\frac{1}{\sqrt{s}}\sum_{\sigma=1}^{s}c_{p-k,\sigma}^{\ast}c_{p,\sigma},\quad b_{k,p}^{\ast}=\frac{1}{\sqrt{s}}\sum_{\sigma=1}^{s}c_{p,\sigma}^{\ast}c_{p-k,\sigma}.
\end{equation}
The name is due to the fact that the action of $b_{k,p}^{\ast}$ is
to annihilate a state at momentum $p-k\in B_{F}$ and create a state
at momentum $p\in B_{F}^{c}$ (irrespective of spin), which is to
say excite the state $p-k$ to $p$.

Note that the $b_{k,p}$ and $B_{k}$ operators are simply related
as $B_{k}=\sqrt{s}\sum_{p\in L_{k}}b_{k,p}$. Furthermore, the excitation
operators also obey quasi-bosonic commutation relations:
\begin{lem}
For any $k,l\in\mathbb{Z}_{\ast}^{3}$, $p\in L_{k}$ and $q\in L_{l}$
it holds that
\[
\left[b_{k,p},b_{l,q}^{\ast}\right]=\delta_{k,l}\delta_{p,q}+\varepsilon_{k,l}\mleft(p;q\mright),\quad\left[b_{k,p},b_{l,q}\right]=0=\left[b_{k,p}^{\ast},b_{l,q}^{\ast}\right],
\]
where $\varepsilon_{k,l}\mleft(p;q\mright)=-s^{-1}\sum_{\sigma=1}^{s}\mleft(\delta_{p,q}c_{q-l,\sigma}c_{p-k,\sigma}^{\ast}+\delta_{p-k,q-l}c_{q,\sigma}^{\ast}c_{p,\sigma}\mright)$.
\end{lem}

\textbf{Proof:} By the CAR and commutator identities we calculate
that
\begin{align}
\left[b_{k,p},b_{l,q}\right] & =\frac{1}{s}\sum_{\sigma,\tau=1}^{s}\left[c_{p-k,\sigma}^{\ast}c_{p,\sigma},c_{q-l,\tau}^{\ast}c_{q,\tau}\right]=\frac{1}{s}\sum_{\sigma,\tau=1}^{s}\mleft(c_{p-k,\sigma}^{\ast}\left[c_{p,\sigma},c_{q-l,\tau}^{\ast}c_{q,\tau}\right]+\left[c_{p-k,\sigma}^{\ast},c_{q-l,\tau}^{\ast}c_{q,\tau}\right]c_{p,\sigma}\mright)\\
 & =\frac{1}{s}\sum_{\sigma,\tau=1}^{s}\mleft(c_{p-k,\sigma}^{\ast}\left\{ c_{p,\sigma},c_{q-l,\tau}^{\ast}\right\} c_{q,\tau}-c_{q-l,\tau}^{\ast}\left\{ c_{p-k,\sigma}^{\ast},c_{q,\tau}\right\} c_{p,\sigma}\mright)=0\nonumber 
\end{align}
as the anticommutators vanish by disjointness of $B_{F}$ and $B_{F}^{c}$.
$[b_{k,p}^{\ast},b_{l,q}^{\ast}]$ then likewise vanishes, while for
$[b_{k,p},b_{l,q}^{\ast}]$
\begin{align}
\left[b_{k,p},b_{l,q}^{\ast}\right] & =\frac{1}{s}\sum_{\sigma,\tau=1}^{s}\left[c_{p-k,\sigma}^{\ast}c_{p,\sigma},c_{q,\tau}^{\ast}c_{q-l,\tau}\right]=\frac{1}{s}\sum_{\sigma,\tau=1}^{s}\mleft(c_{p-k,\sigma}^{\ast}\left\{ c_{p,\sigma},c_{q,\tau}^{\ast}\right\} c_{q-l,\tau}-c_{q,\tau}^{\ast}\left\{ c_{p-k,\sigma}^{\ast},c_{q-l,\tau}\right\} c_{p,\sigma}\mright)\\
 & =\frac{1}{s}\sum_{\sigma=1}^{s}\mleft(\delta_{p,q}c_{p-k,\sigma}^{\ast}c_{q-l,\sigma}-\delta_{p-k,q-l}c_{q,\sigma}^{\ast}c_{p,\sigma}\mright)=\delta_{k,l}\delta_{p,q}+\varepsilon_{k,l}\mleft(p;q\mright).\nonumber 
\end{align}
$\hfill\square$

Again these commutation relations are similar to those of bosonic
operators, now indexed by $k\in\mathbb{Z}_{\ast}^{3}$ and $p\in L_{k}$,
but differing by the appearance of the \textit{exchange correction}
$\varepsilon_{k,l}\mleft(p;q\mright)$, which evidently acts by exchanging
the hole states with momenta $p-k$ and $q-l$ if $p=q$, i.e. if
the excited states match, or swaps the states with momenta $p$ and
$q$ if $p-k=q-l$, i.e. if the hole states match.

The presence of $\varepsilon_{k,l}\mleft(p;q\mright)$ can be considered
a consequence of the fact that holes and excited states are not uniquely
associated with one another - indeed, for any $p\in B_{F}^{c}$, \textit{every}
hole state can be excited into this state, so there is a kind of ``overlap''
between the excitation operators, which the exchange correction accounts
for.

Unlike what was the case for the $B_{k}^{\prime}$ operators, these
correction terms can however not be expected to be ``small'' individually.
They can however still be considered small ``on average'', as the
sum $\sum_{p\in L_{k}}\sum_{q\in L_{l}}\varepsilon_{k,l}\mleft(p;q\mright)$
simply reproduces the correction term of equation (\ref{eq:BkBlastCommutator})
(up to a spin factor).

This is generally an unavoidable point: As we will see in Section
\ref{sec:AnalysisofExchangeTerms}, the exchange contribution of Theorem
\ref{them:MainTheorem} in fact originates from these exchange corrections,
so an attempt at treating these as simple error terms (as was done
in the works \cite{BenNamPorSchSei-20,BenNamPorSchSei-21,BenPorSchSei-21})
is bound to miss this.

Now, the reason that the excitation operators are preferable to the
$B_{k}$ operators is that these do in fact behave bosonically with
respect to $H_{\mathrm{kin}}^{\prime}$:
\begin{lem}
\label{lemma:HkinBkpCommutator}For any $k\in\mathbb{Z}_{\ast}^{3}$
and $p\in L_{k}$ it holds that
\[
\left[H_{\mathrm{kin}}^{\prime},b_{k,p}^{\ast}\right]=\mleft(\left|p\right|^{2}-\left|p-k\right|^{2}\mright)b_{k,p}^{\ast}.
\]
\end{lem}

\textbf{Proof:} As $H_{\mathrm{kin}}^{\prime}=\sum_{q\in B_{F}^{c}}^{\tau}\left|q\right|^{2}c_{q,\tau}^{\ast}c_{q,\tau}-\sum_{q\in B_{F}}^{\tau}\left|q\right|^{2}c_{q,\tau}c_{q,\tau}^{\ast}$
we calculate the commutator with each sum: First is
\begin{align}
\sum_{q\in B_{F}^{c}}^{\tau}\left[\left|q\right|^{2}c_{q,\tau}^{\ast}c_{q,\tau},b_{k,p}^{\ast}\right] & =\frac{1}{\sqrt{s}}\sum_{q\in B_{F}^{c}}^{\sigma,\tau}\left|q\right|^{2}\left[c_{q,\tau}^{\ast}c_{q,\tau},c_{p,\sigma}^{\ast}c_{p-k,\sigma}\right]\nonumber \\
 & =\frac{1}{\sqrt{s}}\sum_{q\in B_{F}^{c}}^{\sigma,\tau}\left|q\right|^{2}\mleft(c_{q,\tau}^{\ast}\left\{ c_{q,\tau},c_{p,\sigma}^{\ast}\right\} c_{p-k,\sigma}-c_{p,\sigma}^{\ast}\left\{ c_{q,\tau}^{\ast},c_{p-k,\sigma}\right\} c_{q,\tau}\mright)\\
 & =\frac{1}{\sqrt{s}}\sum_{q\in B_{F}^{c}}^{\sigma,\tau}\delta_{p,q}\delta_{\sigma,\tau}\left|q\right|^{2}c_{q,\tau}^{\ast}c_{p-k,\sigma}=\frac{\left|p\right|^{2}}{\sqrt{s}}\sum_{\sigma=1}^{s}c_{p,\sigma}^{\ast}c_{p-k,\sigma}=\left|p\right|^{2}b_{k,p}^{\ast}\nonumber 
\end{align}
as the second anticommutator vanishes by disjointness of $B_{F}$
and $B_{F}^{c}$. Similarly, for the second sum
\begin{align}
\sum_{q\in B_{F}}^{\tau}\left[\left|q\right|^{2}c_{q,\tau}c_{q,\tau}^{\ast},b_{k,p}^{\ast}\right] & =\frac{1}{\sqrt{s}}\sum_{q\in B_{F}}^{\sigma,\tau}\left|q\right|^{2}\left[c_{q,\tau}c_{q,\tau}^{\ast},c_{p,\sigma}^{\ast}c_{p-k,\sigma}\right]\nonumber \\
 & =\frac{1}{\sqrt{s}}\sum_{q\in B_{F}}^{\sigma,\tau}\left|q\right|^{2}\mleft(-c_{q,\tau}c_{p,\sigma}^{\ast}\left\{ c_{q,\tau}^{\ast},c_{p-k,\sigma}\right\} +\left\{ c_{q,\tau},c_{p,\sigma}^{\ast}\right\} c_{p-k,\sigma}c_{q,\tau}^{\ast}\mright)\\
 & =-\frac{1}{\sqrt{s}}\sum_{q\in B_{F}}^{\sigma,\tau}\left|q\right|^{2}\delta_{q,p-k}\delta_{\sigma,\tau}c_{q,\tau}c_{p,\sigma}^{\ast}=\frac{1}{\sqrt{s}}\sum_{\sigma=1}^{s}\left|p-k\right|^{2}c_{p,\sigma}^{\ast}c_{p-k,\sigma}=\left|p-k\right|^{2}b_{k,p}^{\ast}\nonumber 
\end{align}
and the claim follows.

$\hfill\square$

This commutation relation mimicks that of a diagonal bosonic quadratic
operator, which is
\begin{equation}
\left[\sum\epsilon_{l,q}a_{l,q}^{\ast}a_{l,q},a_{k,p}^{\ast}\right]=\epsilon_{k,p}a_{k,p}^{\ast}
\end{equation}
whence we may informally think of $H_{\mathrm{kin}}^{\prime}$ as
\begin{equation}
H_{\mathrm{kin}}^{\prime}\sim\sum_{k\in\mathbb{Z}^{3}}\sum_{p\in L_{k}}\mleft(\left|p\right|^{2}-\left|p-k\right|^{2}\mright)b_{k,p}^{\ast}b_{k,p}.
\end{equation}
In fact the lemma tells us that $H_{\mathrm{kin}}^{\prime}$ is much
better behaved than the expression on the right-hand side: Unlike
that, the commutator $[H_{\mathrm{kin}}^{\prime},b_{k,p}^{\ast}]$
behaves \textit{exactly} bosonically, without any additional error
terms. In the subsequent sections we will see that it is precisely
through such commutators that $H_{\mathrm{kin}}^{\prime}$ will enter
our analysis. For this reason, working with the excitation operators
$b_{k,p}$ will prove to be extremely advantageous.

\section{\label{sec:OverviewofBosonicBogolubovTransformations}Overview of
Bosonic Bogolubov Transformations}

In this section we review some of the general theory of Bogolubov
transformations in the bosonic setting. Although the object of study
of this thesis is a fermionic system, our approach to this will be
through a quasi-bosonic analysis of the fermionic Hamiltonian, and
while this of course differs from the exact bosonic case, we will
carry out the quasi-bosonic analysis by imitating the exact bosonic
setting. For this reason we find it best to review this first so that
we may focus on the implementation of the analysis and the discrepancies
arising from the quasi-bosonicity in the remainder of the thesis.

This is particularly important as our treatment of Bogolubov transformations
will differ from the ``usual'' one, in that we will view \textit{quadratic
operators}, formed by pairs of creation and annihilation operators,
as the fundamental object of study, rather than the creation and annihilation
operators themselves.

Before we begin the review we must remark on the level of rigor of
this section: Bosonic creation and annihilation operators are inherently
unbounded operators, and so a full account of this subject would necessitate
discussing domains of definition and other subtle details. As the
purpose of this section is only to motivate our approach to the fermionic
problem later on we will however not address these here.

We will employ the following notation: $V$ denotes a \textit{real}
$n$-dimensional Hilbert space, to which is associated the bosonic
Fock space $\mathcal{F}^{+}\mleft(V\mright)=\bigoplus_{N=0}^{\infty}\bigotimes_{\mathrm{sym}}^{N}V$.
To any element $\varphi\in V$ there corresponds the creation and
annihilation operators $a^{\ast}\mleft(\varphi\mright)$ and $a\mleft(\varphi\mright)$,
which act on $\mathcal{F}^{+}\mleft(V\mright)$. These are (formal)
adjoints of one another and obey the canonical commutation relations
(CCR): For any $\varphi,\psi\in V$ it holds that
\begin{equation}
\left[a\mleft(\varphi\mright),a^{\ast}\mleft(\psi\mright)\right]=\left\langle \varphi,\psi\right\rangle ,\quad\left[a\mleft(\varphi\mright),a\mleft(\psi\mright)\right]=0=\left[a^{\ast}\mleft(\varphi\mright),a^{\ast}\mleft(\psi\mright)\right].
\end{equation}
Furthermore, the mappings $\varphi\mapsto a\mleft(\varphi\mright),a^{\ast}\mleft(\varphi\mright)$
are linear.

\subsection{Quadratic Hamiltonians and Bogolubov Transformations}

Similarly to how we can to any $\varphi\in V$ associate the two operators
$a\mleft(\varphi\mright)$ and $a^{\ast}\mleft(\varphi\mright)$, we can
to any symmetric operators $A,B:V\rightarrow V$ associate two kinds
of \textit{quadratic operators} acting on $\mathcal{F}^{+}\mleft(V\mright)$:
The first kind is the usual second-quantization, given by
\begin{equation}
\mathrm{d}\Gamma\mleft(A\mright)=\sum_{i,j=1}^{n}\left\langle e_{i},Ae_{j}\right\rangle a^{\ast}\mleft(e_{i}\mright)a\mleft(e_{j}\mright)=\sum_{i=1}^{n}a^{\ast}\mleft(Ae_{i}\mright)a\mleft(e_{i}\mright)\label{eq:BosonicQuadraticOperator1}
\end{equation}
where $\mleft(e_{i}\mright)_{i=1}^{n}$ denotes any orthonormal basis
of $V$ (the operator is independent of this choice, as guaranteed
by Lemma \ref{lemma:TraceFormLemma} below). The second kind is of
the form
\begin{align}
Q\mleft(B\mright) & =\sum_{i,j=1}^{n}\left\langle e_{i},Be_{j}\right\rangle \mleft(a\mleft(e_{i}\mright)a\mleft(e_{j}\mright)+a^{\ast}\mleft(e_{j}\mright)a^{\ast}\mleft(e_{i}\mright)\mright)\label{eq:BosonicQuadraticOperator2}\\
 & =\sum_{i=1}^{n}\mleft(a\mleft(Be_{i}\mright)a\mleft(e_{i}\mright)+a^{\ast}\mleft(e_{i}\mright)a^{\ast}\mleft(Be_{i}\mright)\mright).\nonumber 
\end{align}
We define a \textit{quadratic Hamiltonian} to be an operator $H$,
acting on $\mathcal{F}^{+}\mleft(V\mright)$, of the form
\begin{equation}
H=2\,\mathrm{d}\Gamma\mleft(A\mright)+Q\mleft(B\mright).
\end{equation}
(The factor of $2$ will be convenient below.)

The importance of quadratic Hamiltonians lies in the fact that they
can (under suitable assumptions) be \textit{diagonalized}, in the
sense that there exists a unitary transformation $\mathcal{U}:\mathcal{F}^{+}\mleft(V\mright)\rightarrow\mathcal{F}^{+}\mleft(V\mright)$
such that
\begin{equation}
\mathcal{U}H\mathcal{U}^{\ast}=2\,\mathrm{d}\Gamma\mleft(E\mright)+E_{0}
\end{equation}
for a symmetric operator $E:V\rightarrow V$ and $E_{0}\in\mathbb{R}$,
i.e. a quadratic Hamiltonian is unitarily equivalent to a second-quantized
one-body operator plus a constant. As second-quantized operators are
simple objects, the properties of quadratic Hamiltonians are thus
in principle also simple, provided one can describe $\mathcal{U}$
explicitly enough to relate the operators $A$ and $B$ to $E$.

In this section we review the explicit construction of such \textit{Bogolubov
transformations} $\mathcal{U}$. More precisely, we will consider
the Bogolubov transformations which can be written as $\mathcal{U}=e^{\mathcal{K}}$
where $\mathcal{K}$ is of the form
\begin{align}
\mathcal{K} & =\frac{1}{2}\sum_{i,j=1}^{n}\left\langle e_{i},Ke_{j}\right\rangle \mleft(a\mleft(e_{i}\mright)a\mleft(e_{j}\mright)-a^{\ast}\mleft(e_{j}\mright)a^{\ast}\mleft(e_{i}\mright)\mright)\label{eq:BosonicBogolubovKernel}\\
 & =\frac{1}{2}\sum_{i=1}^{n}\mleft(a\mleft(Ke_{i}\mright)a\mleft(e_{i}\mright)-a^{\ast}\mleft(e_{i}\mright)a^{\ast}\mleft(Ke_{i}\mright)\mright)\nonumber 
\end{align}
for a symmetric operator $K:V\rightarrow V$ (the \textit{transformation
kernel}). Note that from the second line it is clear that $\mathcal{K}^{\ast}=-\mathcal{K}$,
so such a $\mathcal{K}$ will indeed generate a unitary transformation.

The action of $e^{\mathcal{K}}$ on creation and annihilation operators
can be determined as follows: By the CCR we compute that
\begin{align}
\left[\mathcal{K},a\mleft(\varphi\mright)\right] & =\frac{1}{2}\sum_{i=1}^{n}\left[a\mleft(Ke_{i}\mright)a\mleft(e_{i}\mright)-a^{\ast}\mleft(e_{i}\mright)a^{\ast}\mleft(Ke_{i}\mright),a\mleft(\varphi\mright)\right]=\frac{1}{2}\sum_{i=1}^{n}\left[a\mleft(\varphi\mright),a^{\ast}\mleft(e_{i}\mright)a^{\ast}\mleft(Ke_{i}\mright)\right]\nonumber \\
 & =\frac{1}{2}\sum_{i=1}^{n}\mleft(a^{\ast}\mleft(e_{i}\mright)\left[a\mleft(\varphi\mright),a^{\ast}\mleft(Ke_{i}\mright)\right]+\left[a\mleft(\varphi\mright),a^{\ast}\mleft(e_{i}\mright)\right]a^{\ast}\mleft(Ke_{i}\mright)\mright)\\
 & =\frac{1}{2}\sum_{i=1}^{n}\mleft(a^{\ast}\mleft(e_{i}\mright)\left\langle \varphi,Ke_{i}\right\rangle +\left\langle \varphi,e_{i}\right\rangle a^{\ast}\mleft(Ke_{i}\mright)\mright)\nonumber \\
 & =\frac{1}{2}\mleft(a^{\ast}\mleft(\sum_{i=1}^{n}\left\langle e_{i},K\varphi\right\rangle e_{i}\mright)+a^{\ast}\mleft(K\sum_{i=1}^{n}\left\langle e_{i},\varphi\right\rangle e_{i}\mright)\mright)=a^{\ast}\mleft(K\varphi\mright)\nonumber 
\end{align}
and taking the adjoint likewise shows that $\left[\mathcal{K},a^{\ast}\mleft(\varphi\mright)\right]=a\mleft(K\varphi\mright)$,
so
\begin{align}
\left[\mathcal{K},a\mleft(\varphi\mright)\right] & =a^{\ast}\mleft(K\varphi\mright)\label{eq:ActionofcalKonCAOperators}\\
\left[\mathcal{K},a^{\ast}\mleft(\varphi\mright)\right] & =a\mleft(K\varphi\mright).\nonumber 
\end{align}
$\left[\mathcal{K},\cdot\right]$ thus acts on creation and annihilation
operators by ``swapping'' each type into the other and applying
the operator $K$ to their arguments. From this one can now deduce
that
\begin{align}
e^{\mathcal{K}}a\mleft(\varphi\mright)e^{-\mathcal{K}} & =a\mleft(\cosh\mleft(K\mright)\mright)+a^{\ast}\mleft(\sinh\mleft(K\mright)\mright)\\
e^{\mathcal{K}}a^{\ast}\mleft(\varphi\mright)e^{-\mathcal{K}} & =a^{\ast}\mleft(\cosh\mleft(K\mright)\mright)+a\mleft(\sinh\mleft(K\mright)\mright)\nonumber 
\end{align}
since by the Baker-Campbell-Hausdorff formula
\begin{align}
e^{\mathcal{K}}a\mleft(\varphi\mright)e^{-\mathcal{K}} & =a\mleft(\varphi\mright)+\left[\mathcal{K},a\mleft(\varphi\mright)\right]+\frac{1}{2!}\left[\mathcal{K},\left[\mathcal{K},a\mleft(\varphi\mright)\right]\right]+\frac{1}{3!}\left[\mathcal{K},\left[\mathcal{K},\left[\mathcal{K},a\mleft(\varphi\mright)\right]\right]\right]+\cdots\nonumber \\
 & =a\mleft(\varphi\mright)+a^{\ast}\mleft(K\varphi\mright)+\frac{1}{2!}a\mleft(K^{2}\varphi\mright)+\frac{1}{3!}a^{\ast}\mleft(K^{3}\varphi\mright)+\cdots\\
 & =a\mleft(\mleft(1+\frac{1}{2!}K^{2}+\cdots\mright)\varphi\mright)+a^{\ast}\mleft(\mleft(K+\frac{1}{3!}K^{3}+\cdots\mright)\varphi\mright)=a\mleft(\cosh\mleft(K\mright)\mright)+a^{\ast}\mleft(\sinh\mleft(K\mright)\mright)\nonumber 
\end{align}
and likewise for $e^{\mathcal{K}}a^{\ast}\mleft(\varphi\mright)e^{-\mathcal{K}}$.

\subsection{The Action of $e^{\mathcal{K}}$ on Quadratic Operators}

As our interest in Bogolubov transformations lie in their diagonalization
of quadratic Hamiltonians it is however not the transformation of
$a\mleft(\cdot\mright)$ and $a^{\ast}\mleft(\cdot\mright)$ that will
interest us, but rather the transformation of $\mathrm{d}\Gamma\mleft(\cdot\mright)$
and $Q\mleft(\cdot\mright)$. The latter can of course be deduced from
the former, but this approach is disadvantageous in the quasi-bosonic
setting, which is why we will proceed differently.

First, let us make an observation on the structure of the quadratic
operators which will simplify calculation significantly: The operators
\begin{align}
\mathrm{d}\Gamma\mleft(A\mright) & =\sum_{i=1}^{n}a^{\ast}\mleft(Ae_{i}\mright)a\mleft(e_{i}\mright)\\
Q\mleft(B\mright) & =\sum_{i=1}^{n}\mleft(a\mleft(Be_{i}\mright)a\mleft(e_{i}\mright)+a^{\ast}\mleft(e_{i}\mright)a^{\ast}\mleft(Be_{i}\mright)\mright)\nonumber 
\end{align}
are both of a ``trace-form'', in the sense that we can write $\mathrm{d}\Gamma\mleft(A\mright)$
(say) in the form $\mathrm{d}\Gamma\mleft(A\mright)=\sum_{i=1}^{n}q\mleft(e_{i},e_{i}\mright)$
where
\begin{equation}
q\mleft(x,y\mright)=a^{\ast}\mleft(Ax\mright)a\mleft(y\mright),\quad x,y\in V,
\end{equation}
defines a bilinear mapping from $V\times V$ into the space of operators
on $\mathcal{F}^{+}\mleft(V\mright)$, similar to how $\mathrm{tr}\mleft(T\mright)=\sum_{i=1}^{n}q\mleft(e_{i},e_{i}\mright)$
for $q\mleft(x,y\mright)=\left\langle x,Ty\right\rangle $. This is
worth noting since all such expressions are both basis-independent
and obey an additional property, which for the trace is the familiar
cyclicity property.

As we will encounter such trace-form sums repeatedly throughout this
paper, we state this property in full generality:
\begin{lem}
\label{lemma:TraceFormLemma}Let $\left\langle V,\left\langle \cdot,\cdot\right\rangle \right\rangle $
be an $n$-dimensional Hilbert space and let $q:V\times V\rightarrow W$
be a sesquilinear mapping into a vector space $W$. Let $\mleft(e_{i}\mright)_{i=1}^{n}$
be an orthonormal basis for $V$. Then for any linear operators $S,T:V\rightarrow V$
it holds that
\[
\sum_{i=1}^{n}q\mleft(Se_{i},Te_{i}\mright)=\sum_{i=1}^{n}q\mleft(ST^{\ast}e_{i},e_{i}\mright).
\]
As a particular consequence, the expression $\sum_{i=1}^{n}q\mleft(e_{i},e_{i}\mright)$
is independent of the basis chosen.
\end{lem}

\textbf{Proof:} By orthonormal expansion we find that
\begin{align}
\sum_{i=1}^{n}q\mleft(Se_{i},Te_{i}\mright) & =\sum_{i=1}^{n}q\mleft(Se_{i},\sum_{j=1}^{n}\left\langle e_{j},Te_{i}\right\rangle e_{j}\mright)=\sum_{i,j=1}^{n}\left\langle T^{\ast}e_{j},e_{i}\right\rangle q\mleft(Se_{i},e_{j}\mright)\\
 & =\sum_{j=1}^{n}q\mleft(S\sum_{i=1}^{n}\left\langle e_{i},T^{\ast}e_{j}\right\rangle e_{i},e_{j}\mright)=\sum_{i=1}^{n}q\mleft(ST^{\ast}e_{i},e_{i}\mright).\nonumber 
\end{align}
The basis-independence follows from this by noting that if $\mleft(e_{i}^{\prime}\mright)_{i=1}^{n}$
is any other orthonormal basis, then with $U:V\rightarrow V$ denoting
the unitary transformation defined by $Ue_{i}=e_{i}^{\prime}$, $1\leq i\leq n$,
we see that
\begin{equation}
\sum_{i=1}^{n}q\mleft(e_{i}^{\prime},e_{i}^{\prime}\mright)=\sum_{i=1}^{n}q\mleft(Ue_{i},Ue_{i}\mright)=\sum_{i=1}^{n}q\mleft(UU^{\ast}e_{i},e_{i}\mright)=\sum_{i=1}^{n}q\mleft(e_{i},e_{i}\mright).
\end{equation}
$\hfill\square$

(In the present real case sesquilinearity is of course just bilinearity.)

The lemma thus allows us to move operators from one argument to the
other when under a sum, which will be immensely useful when simplifying
expressions. This can indeed be seen as a generalization of the cyclicity
property of the trace, since the lemma can be applied to see that
\begin{equation}
\mathrm{tr}\mleft(ST\mright)=\sum_{i=1}^{n}\left\langle e_{i},STe_{i}\right\rangle =\sum_{i=1}^{n}\left\langle S^{\ast}e_{i},Te_{i}\right\rangle =\sum_{i=1}^{n}\left\langle S^{\ast}T^{\ast}e_{i},e_{i}\right\rangle =\sum_{i=1}^{n}\left\langle e_{i},TSe_{i}\right\rangle =\mathrm{tr}\mleft(TS\mright),
\end{equation}
but it should be noted that cyclicity in this sense is not a general
property of trace-form sums.

With this lemma we can easily calculate the commutators of $\mathcal{K}$
with $\mathrm{d}\Gamma\mleft(\cdot\mright)$ and $Q\mleft(\cdot\mright)$:
\begin{prop}
\label{prop:BosonicQuadraticCommutators}For any symmetric operators
$A,B:V\rightarrow V$ it holds that
\begin{align*}
\left[\mathcal{K},2\,\mathrm{d}\Gamma\mleft(A\mright)\right] & =Q\mleft(\left\{ K,A\right\} \mright)\\
\left[\mathcal{K},Q\mleft(B\mright)\right] & =2\,\mathrm{d}\Gamma\mleft(\left\{ K,B\right\} \mright)+\mathrm{tr}\mleft(\left\{ K,B\right\} \mright).
\end{align*}
\end{prop}

\textbf{Proof:} Using equation (\ref{eq:ActionofcalKonCAOperators})
and the lemma we compute
\begin{align}
\left[\mathcal{K},\mathrm{d}\Gamma\mleft(A\mright)\right] & =\sum_{i=1}^{n}\left[\mathcal{K},a^{\ast}\mleft(Ae_{i}\mright)a\mleft(e_{i}\mright)\right]=\sum_{i=1}^{n}\mleft(a^{\ast}\mleft(Ae_{i}\mright)\left[\mathcal{K},a\mleft(e_{i}\mright)\right]+\left[\mathcal{K},a^{\ast}\mleft(Ae_{i}\mright)\right]a\mleft(e_{i}\mright)\mright)\\
 & =\sum_{i=1}^{n}\mleft(a^{\ast}\mleft(Ae_{i}\mright)a^{\ast}\mleft(Ke_{i}\mright)+a\mleft(KAe_{i}\mright)a\mleft(e_{i}\mright)\mright)=\sum_{i=1}^{n}\mleft(a\mleft(KAe_{i}\mright)a\mleft(e_{i}\mright)+a^{\ast}\mleft(e_{i}\mright)a^{\ast}\mleft(KAe_{i}\mright)\mright),\nonumber 
\end{align}
and since the annihilation operators commute there holds the identity
\begin{equation}
\sum_{i=1}^{n}a\mleft(KAe_{i}\mright)a\mleft(e_{i}\mright)=\sum_{i=1}^{n}a\mleft(e_{i}\mright)a\mleft(\mleft(KA\mright)^{\ast}e_{i}\mright)=\sum_{i=1}^{n}a\mleft(AKe_{i}\mright)a\mleft(e_{i}\mright)
\end{equation}
and likewise for the second term, so including a factor of $2$ we
can write
\begin{equation}
\left[\mathcal{K},2\,\mathrm{d}\Gamma\mleft(A\mright)\right]=\sum_{i=1}^{n}\mleft(a\mleft(\left\{ K,A\right\} e_{i}\mright)a\mleft(e_{i}\mright)+a^{\ast}\mleft(e_{i}\mright)a^{\ast}\mleft(\left\{ K,A\right\} e_{i}\mright)\mright)=Q\mleft(\left\{ K,B\right\} \mright)
\end{equation}
as claimed. For $Q\mleft(B\mright)$ we note that $Q\mleft(B\mright)=2\,\mathrm{Re}\mleft(\sum_{i=1}^{n}a\mleft(Be_{i}\mright)a\mleft(e_{i}\mright)\mright)$
and calculate as above that
\begin{align}
\left[\mathcal{K},Q\mleft(B\mright)\right] & =2\,\mathrm{Re}\mleft(\sum_{i=1}^{n}\left[\mathcal{K},a\mleft(Be_{i}\mright)a\mleft(e_{i}\mright)\right]\mright)=2\,\mathrm{Re}\mleft(\sum_{i=1}^{n}\mleft(a\mleft(Be_{i}\mright)\left[\mathcal{K},a\mleft(e_{i}\mright)\right]+\left[\mathcal{K},a\mleft(Be_{i}\mright)\right]a\mleft(e_{i}\mright)\mright)\mright)\nonumber \\
 & =2\,\mathrm{Re}\mleft(\sum_{i=1}^{n}\mleft(a\mleft(Be_{i}\mright)a^{\ast}\mleft(Ke_{i}\mright)+a^{\ast}\mleft(KBe_{i}\mright)a\mleft(e_{i}\mright)\mright)\mright)\\
 & =2\,\mathrm{Re}\mleft(\sum_{i=1}^{n}\mleft(a\mleft(BKe_{i}\mright)a^{\ast}\mleft(e_{i}\mright)+a^{\ast}\mleft(KBe_{i}\mright)a\mleft(e_{i}\mright)\mright)\mright).\nonumber 
\end{align}
By the lemma we see that
\begin{align}
2\,\mathrm{Re}\mleft(\sum_{i=1}^{n}a\mleft(BKe_{i}\mright)a^{\ast}\mleft(e_{i}\mright)\mright) & =\sum_{i=1}^{n}\mleft(a\mleft(BKe_{i}\mright)a^{\ast}\mleft(e_{i}\mright)+a\mleft(e_{i}\mright)a^{\ast}\mleft(BKe_{i}\mright)\mright)\nonumber \\
 & =\sum_{i=1}^{n}\mleft(a\mleft(e_{i}\mright)a^{\ast}\mleft(\mleft(BK\mright)^{\ast}e_{i}\mright)+a\mleft(e_{i}\mright)a^{\ast}\mleft(BKe_{i}\mright)\mright)\\
 & =\sum_{i=1}^{n}a\mleft(e_{i}\mright)a^{\ast}\mleft(\left\{ K,B\right\} e_{i}\mright)\nonumber 
\end{align}
and likewise
\begin{equation}
2\,\mathrm{Re}\mleft(\sum_{i=1}^{n}a^{\ast}\mleft(KBe_{i}\mright)a\mleft(e_{i}\mright)\mright)=\sum_{i=1}^{n}a^{\ast}\mleft(\left\{ K,B\right\} e_{i}\mright)a\mleft(e_{i}\mright),
\end{equation}
so by the CCR
\begin{align}
\left[\mathcal{K},Q\mleft(B\mright)\right] & =\sum_{i=1}^{n}\mleft(a\mleft(e_{i}\mright)a^{\ast}\mleft(\left\{ K,B\right\} e_{i}\mright)+a^{\ast}\mleft(\left\{ K,B\right\} e_{i}\mright)a\mleft(e_{i}\mright)\mright)\nonumber \\
 & =2\sum_{i=1}^{n}a^{\ast}\mleft(\left\{ K,B\right\} e_{i}\mright)a\mleft(e_{i}\mright)+\sum_{i=1}^{n}\left\langle e_{i},\left\{ K,B\right\} e_{i}\right\rangle \\
 & =2\,\mathrm{d}\Gamma\mleft(\left\{ K,B\right\} \mright)+\mathrm{tr}\mleft(\left\{ K,B\right\} \mright).\nonumber 
\end{align}
$\hfill\square$

Note the similarity between these commutators and those of equation
(\ref{eq:ActionofcalKonCAOperators}) - again $\left[\mathcal{K},\cdot\right]$
acts by ``swapping the types and applying $K$ to the argument'',
although now the types are those of the quadratic operators and the
application of $K$ is taking the anticommutator.

Although the action of $e^{\mathcal{K}}$ on the quadratic operators
can again be deduced from the Baker-Campbell-Hausdorff formula, we
now derive this by an ``ODE-style'' argument, as this will generalize
better to the quasi-bosonic setting of the next section:
\begin{prop}
\label{prop:BosonicQuadraticTransformationLaw}For any symmetric operator
$T:V\rightarrow V$ it holds that
\begin{align*}
e^{\mathcal{K}}\mleft(2\,\mathrm{d}\Gamma\mleft(T\mright)\mright)e^{-\mathcal{K}} & =2\,\mathrm{d}\Gamma\mleft(T_{1}\mright)+Q\mleft(T_{2}\mright)+\mathrm{tr}\mleft(T_{1}-T\mright)\\
e^{\mathcal{K}}Q\mleft(T\mright)e^{-\mathcal{K}} & =2\,\mathrm{d}\Gamma\mleft(T_{2}\mright)+Q\mleft(T_{1}\mright)+\mathrm{tr}\mleft(T_{2}\mright)
\end{align*}
where $T_{1},T_{2}:V\rightarrow V$ are given by
\[
T_{1}=\frac{1}{2}\mleft(e^{K}Te^{K}+e^{-K}Te^{-K}\mright),\quad T_{2}=\frac{1}{2}\mleft(e^{K}Te^{K}-e^{-K}Te^{-K}\mright).
\]
\end{prop}

\textbf{Proof:} We prove the first identity, the second following
similarly.

Consider an expression of the form $e^{-t\mathcal{K}}\mleft(2\,\mathrm{d}\Gamma\mleft(A\mleft(t\mright)\mright)+Q\mleft(B\mleft(t\mright)\mright)\mright)e^{t\mathcal{K}}$
where $A\mleft(t\mright),B\mleft(t\mright):V\rightarrow V$ are any symmetric
operators with $t\mapsto A\mleft(t\mright),B\mleft(t\mright)$ differentiable.
Taking the derivative, we find by Proposition \ref{prop:BosonicQuadraticCommutators}
that
\begin{align}
 & \,\frac{d}{dt}e^{-t\mathcal{K}}\mleft(2\,\mathrm{d}\Gamma\mleft(A\mleft(t\mright)\mright)+Q\mleft(B\mleft(t\mright)\mright)\mright)e^{t\mathcal{K}}\nonumber \\
 & =e^{-t\mathcal{K}}\mleft(2\,\mathrm{d}\Gamma\mleft(A'\mleft(t\mright)\mright)+Q\mleft(B'\mleft(t\mright)\mright)-\left[\mathcal{K},2\,\mathrm{d}\Gamma\mleft(A\mleft(t\mright)\mright)+Q\mleft(B\mleft(t\mright)\mright)\right]\mright)e^{t\mathcal{K}}\\
 & =e^{-t\mathcal{K}}\mleft(2\,\mathrm{d}\Gamma\mleft(A'\mleft(t\mright)-\left\{ K,B\mleft(t\mright)\right\} \mright)\mright)e^{t\mathcal{K}}+e^{-t\mathcal{K}}Q\mleft(B'\mleft(t\mright)-\left\{ K,A\mleft(t\mright)\right\} \mright)e^{t\mathcal{K}}-\mathrm{tr}\mleft(\left\{ K,B\mleft(t\mright)\right\} \mright).\nonumber 
\end{align}
Consequently, if $A\mleft(t\mright)$ and $B\mleft(t\mright)$ are solutions
of the system
\begin{equation}
A'\mleft(t\mright)=\left\{ K,B\mleft(t\mright)\right\} ,\quad B'\mleft(t\mright)=\left\{ K,A\mleft(t\mright)\right\} ,
\end{equation}
then the first two terms vanish, i.e.
\begin{equation}
\frac{d}{dt}e^{-t\mathcal{K}}\mleft(2\,\mathrm{d}\Gamma\mleft(A\mleft(t\mright)\mright)+Q\mleft(B\mleft(t\mright)\mright)\mright)e^{t\mathcal{K}}=-\,\mathrm{tr}\mleft(\left\{ K,B\mleft(t\mright)\right\} \mright).
\end{equation}
The fundamental theorem of calculus thus implies that
\begin{align}
e^{-\mathcal{K}}\mleft(2\,\mathrm{d}\Gamma\mleft(A\mleft(1\mright)\mright)+Q\mleft(B\mleft(1\mright)\mright)\mright)e^{\mathcal{K}} & =2\,\mathrm{d}\Gamma\mleft(A\mleft(0\mright)\mright)+Q\mleft(B\mleft(0\mright)\mright)-\int_{0}^{1}\mathrm{tr}\mleft(\left\{ K,B\mleft(t\mright)\right\} \mright)\,dt\\
 & =2\,\mathrm{d}\Gamma\mleft(A\mleft(0\mright)\mright)+Q\mleft(B\mleft(0\mright)\mright)-\mathrm{tr}\mleft(A\mleft(1\mright)-A\mleft(0\mright)\mright),\nonumber 
\end{align}
and imposing also the initial conditions
\begin{equation}
A\mleft(0\mright)=T,\quad B\mleft(0\mright)=0,
\end{equation}
this can be rearranged to
\begin{equation}
e^{\mathcal{K}}\mleft(2\,\mathrm{d}\Gamma\mleft(T\mright)\mright)e^{-\mathcal{K}}=2\,\mathrm{d}\Gamma\mleft(A\mleft(1\mright)\mright)+Q\mleft(B\mleft(1\mright)\mright)+\mathrm{tr}\mleft(A\mleft(1\mright)-T\mright).
\end{equation}
The claim now follows by the observation that
\begin{align}
A\mleft(t\mright) & =\frac{1}{2}\mleft(e^{tK}Te^{tK}+e^{-tK}Te^{-tK}\mright)\\
B\mleft(t\mright) & =\frac{1}{2}\mleft(e^{tK}Te^{tK}-e^{-tK}Te^{-tK}\mright)\nonumber 
\end{align}
are precisely the solutions of this system: The initial conditions
are clearly satisfied, as is the ODE since
\begin{equation}
\frac{d}{dt}\mleft(e^{tK}Te^{tK}\pm e^{-tK}Te^{-tK}\mright)=e^{tK}\left\{ K,T\right\} e^{tK}\pm e^{-tK}\left\{ -K,T\right\} e^{-tK}=\left\{ K,e^{tK}Te^{tK}\mp e^{-tK}Te^{-tK}\right\} .
\end{equation}
$\hfill\square$

\subsubsection*{Diagonalization of Quadratic Hamiltonians}

Having derived the transformation laws we can now describe how to
diagonalize the quadratic Hamiltonian $H=2\,\mathrm{d}\mathrm{\Gamma}\mleft(A\mright)+Q\mleft(B\mright)$:
By Proposition \ref{prop:BosonicQuadraticTransformationLaw}, this
transforms as
\begin{align}
e^{\mathcal{K}}He^{-\mathcal{K}} & =2\,\mathrm{d}\Gamma\mleft(\frac{1}{2}\mleft(e^{K}Ae^{K}+e^{-K}Ae^{-K}\mright)\mright)+Q\mleft(\frac{1}{2}\mleft(e^{K}Ae^{K}-e^{-K}Ae^{-K}\mright)\mright)\nonumber \\
 & +2\,\mathrm{d}\Gamma\mleft(\frac{1}{2}\mleft(e^{K}Be^{K}-e^{-K}Be^{-K}\mright)\mright)+Q\mleft(\frac{1}{2}\mleft(e^{K}Be^{K}+e^{-K}Be^{-K}\mright)\mright)\nonumber \\
 & +\mathrm{tr}\mleft(\frac{1}{2}\mleft(e^{K}Ae^{K}+e^{-K}Ae^{-K}\mright)-A\mright)+\mathrm{tr}\mleft(\frac{1}{2}\mleft(e^{K}Be^{K}-e^{-K}Be^{-K}\mright)\mright)\\
 & =2\,\mathrm{d}\Gamma\mleft(\frac{1}{2}\mleft(e^{K}\mleft(A+B\mright)e^{K}+e^{-K}\mleft(A-B\mright)e^{-K}\mright)\mright)+Q\mleft(\frac{1}{2}\mleft(e^{K}\mleft(A+B\mright)e^{K}-e^{-K}\mleft(A-B\mright)e^{-K}\mright)\mright)\nonumber \\
 & +\mathrm{tr}\mleft(\frac{1}{2}\mleft(e^{K}\mleft(A+B\mright)e^{K}+e^{-K}\mleft(A-B\mright)e^{-K}\mright)-A\mright).\nonumber 
\end{align}
As the diagonalization of $H$ is the statement that the $Q\mleft(\cdot\mright)$
term vanishes, we see that the \textit{diagonalization condition}
is that $K$ obeys
\begin{equation}
e^{K}\mleft(A+B\mright)e^{K}=e^{-K}\mleft(A-B\mright)e^{-K}.\label{eq:DiagonalizationCondition}
\end{equation}
Indeed, if this holds then we evidently have that
\begin{equation}
e^{\mathcal{K}}He^{-\mathcal{K}}=2\,\mathrm{d}\Gamma\mleft(E\mright)+\mathrm{tr}\mleft(E-A\mright)
\end{equation}
for $E=e^{K}\mleft(A+B\mright)e^{K}=e^{-K}\mleft(A-B\mright)e^{-K}$.

There remains the question of when such a kernel $K$ exists. For
this it holds that the condition $A\pm B>0$ not only suffices, but
in this case a diagonalizing $K$ can be explicitly defined, which
is furthermore unique (the following is a generalization and simplification
of the arguments used in \cite{BenNamPorSchSei-20,GrechSeiringer-13}):
\begin{prop}
\label{prop:UniqueDiagonalizingKernel}Let $A,B:V\rightarrow V$ be
symmetric operators such that $A\pm B>0$. Then
\[
K=-\frac{1}{2}\log\mleft(\mleft(A-B\mright)^{-\frac{1}{2}}\mleft(\mleft(A-B\mright)^{\frac{1}{2}}\mleft(A+B\mright)\mleft(A-B\mright)^{\frac{1}{2}}\mright)^{\frac{1}{2}}\mleft(A-B\mright)^{-\frac{1}{2}}\mright)
\]
is the unique symmetric solution of
\[
e^{K}\mleft(A+B\mright)e^{K}=e^{-K}\mleft(A-B\mright)e^{-K}.
\]
\end{prop}

\textbf{Proof:} Write $A_{\pm}=A\pm B$ for brevity. Then we can write
the diagonalization condition as
\begin{equation}
A_{+}=e^{-2K}A_{-}e^{-2K}.
\end{equation}
Multiplying by $A_{-}^{-\frac{1}{2}}$ on both sides yields
\begin{equation}
A_{-}^{\frac{1}{2}}A_{+}A_{-}^{\frac{1}{2}}=A_{-}^{\frac{1}{2}}e^{-2K}A_{-}e^{-2K}A_{-}^{\frac{1}{2}}=\mleft(A_{-}^{\frac{1}{2}}e^{-2K}A_{-}^{\frac{1}{2}}\mright)^{2},
\end{equation}
so as both $A_{-}^{\frac{1}{2}}A_{+}A_{-}^{\frac{1}{2}}$ and $A_{-}^{\frac{1}{2}}e^{-2K}A_{-}^{\frac{1}{2}}$
are positive operators it must be the case that
\begin{equation}
A_{-}^{\frac{1}{2}}e^{-2K}A_{-}^{\frac{1}{2}}=\mleft(A_{-}^{\frac{1}{2}}A_{+}A_{-}^{\frac{1}{2}}\mright)^{\frac{1}{2}}
\end{equation}
whence
\begin{equation}
-2K=\log\mleft(A_{-}^{-\frac{1}{2}}\mleft(A_{-}^{\frac{1}{2}}A_{+}A_{-}^{\frac{1}{2}}\mright)^{\frac{1}{2}}A_{-}^{-\frac{1}{2}}\mright)
\end{equation}
which is the claim.

$\hfill\square$

\section{\label{sec:DiagonalizationoftheBosonizableTerms}Diagonalization
of the Bosonizable Terms}

In this section we diagonalize the \textit{bosonizable terms}, which
is to say the expression
\begin{equation}
H_{\mathrm{kin}}^{\prime}+\sum_{k\in\mathbb{Z}_{\ast}^{3}}\frac{\hat{V}_{k}k_{F}^{-1}}{2\,\mleft(2\pi\mright)^{3}}\mleft(2B_{k}^{\ast}B_{k}+B_{k}B_{-k}+B_{-k}^{\ast}B_{k}^{\ast}\mright).
\end{equation}
In Section \ref{sec:LocalizationoftheHamiltonian} we saw that these
behave in a quasi-bosonic fashion, and this ``diagonalization''
is indeed in the sense of Bogolubov transformations. To this end we
start by casting the bosonizable terms into a form which more closely
mirrors that of the quadratic operators which we considered in the
previous section.

Once this is done it will be clear how to define a quasi-bosonic Bogolubov
transformation $e^{\mathcal{K}}$ which emulates the properties of
the transformation in the exact bosonic setting. We can then repeat
the calculations of the previous section - keeping also in mind the
additional terms which arise from the exchange correction - to determine
the action of this transformation on the bosonizable terms.

With this established we then specify a particular generator $\mathcal{K}$
which will diagonalize these terms, and in the process extract the
bosonic contribution to the correlation energy. The main result of
this section is summarized in the following (in notation defined below):
\begin{thm}
\label{thm:DiagonalizationoftheBosonizableTerms}Let $\sum_{k\in\mathbb{Z}_{\ast}^{3}}\hat{V}_{k}^{2}<\infty$.
Then there exists a unitary transformation $e^{\mathcal{K}}:\mathcal{H}_{N}\rightarrow\mathcal{H}_{N}$
such that
\begin{align*}
 & \;\,e^{\mathcal{K}}\mleft(H_{\mathrm{kin}}^{\prime}+\sum_{k\in\mathbb{Z}_{\ast}^{3}}\frac{\hat{V}_{k}k_{F}^{-1}}{2\,\mleft(2\pi\mright)^{3}}\mleft(2B_{k}^{\ast}B_{k}+B_{k}B_{-k}+B_{-k}^{\ast}B_{k}^{\ast}\mright)\mright)e^{-\mathcal{K}}\\
 & =\sum_{k\in\mathbb{Z}_{\ast}^{3}}\mathrm{tr}\mleft(e^{-K_{k}}h_{k}e^{-K_{k}}-h_{k}-P_{k}\mright)+H_{\mathrm{kin}}^{\prime}+2\sum_{k\in\mathbb{Z}_{\ast}^{3}}Q_{1}^{k}\mleft(e^{-K_{k}}h_{k}e^{-K_{k}}-h_{k}\mright)\\
 & +\sum_{k\in\mathbb{Z}_{\ast}^{3}}\int_{0}^{1}e^{\mleft(1-t\mright)\mathcal{K}}\mleft(\varepsilon_{k}\mleft(\left\{ K_{k},B_{k}\mleft(t\mright)\right\} \mright)+2\,\mathrm{Re}\mleft(\mathcal{E}_{k}^{1}\mleft(A_{k}\mleft(t\mright)\mright)\mright)+2\,\mathrm{Re}\mleft(\mathcal{E}_{k}^{2}\mleft(B_{k}\mleft(t\mright)\mright)\mright)\mright)e^{-\mleft(1-t\mright)\mathcal{K}}dt
\end{align*}
where for any $k\in\mathbb{Z}_{\ast}^{3}$ the operators $h_{k},P_{k}:\ell^{2}\mleft(L_{k}\mright)\rightarrow\ell^{2}\mleft(L_{k}\mright)$
are defined by
\[
\begin{array}{ccccccc}
h_{k}e_{p} & = & \lambda_{k,p}e_{p} &  & \lambda_{k,p} & = & \frac{1}{2}\mleft(\left|p\right|^{2}-\left|p-k\right|^{2}\mright)\\
P_{k}\mleft(\cdot\mright) & = & \left\langle v_{k},\cdot\right\rangle v_{k} &  & v_{k} & = & \sqrt{\frac{s\hat{V}_{k}k_{F}^{-1}}{2\,\mleft(2\pi\mright)^{3}}}\sum_{p\in L_{k}}e_{p},
\end{array}
\]
the operator $K_{k}:\ell^{2}\mleft(L_{k}\mright)\rightarrow\ell^{2}\mleft(L_{k}\mright)$
is defined by
\[
K_{k}=-\frac{1}{2}\log\mleft(h_{k}^{-\frac{1}{2}}\mleft(h_{k}^{\frac{1}{2}}\mleft(h_{k}+2P_{k}\mright)h_{k}^{\frac{1}{2}}\mright)^{\frac{1}{2}}h_{k}^{-\frac{1}{2}}\mright)
\]
and for $t\in\left[0,1\right]$ the operators $A_{k}\mleft(t\mright),B_{k}\mleft(t\mright):\ell^{2}\mleft(L_{k}\mright)\rightarrow\ell^{2}\mleft(L_{k}\mright)$
are given by
\begin{align*}
A_{k}\mleft(t\mright) & =\frac{1}{2}\mleft(e^{tK_{k}}\mleft(h_{k}+2P_{k}\mright)e^{tK_{k}}+e^{-tK_{k}}h_{k}e^{-tK_{k}}\mright)-h_{k}\\
B_{k}\mleft(t\mright) & =\frac{1}{2}\mleft(e^{tK_{k}}\mleft(h_{k}+2P_{k}\mright)e^{tK_{k}}-e^{-tK_{k}}h_{k}e^{-tK_{k}}\mright).
\end{align*}
\end{thm}

The condition that $\sum_{k\in\mathbb{Z}_{\ast}^{3}}\hat{V}_{k}^{2}<\infty$
arises to ensure that the diagonalizing generator $\mathcal{K}$ is
a well-defined (and even bounded) operator. We will however postpone
the proof of this until the next section, to focus on the diagonalization
procedure first.

(Even though $\mathcal{K}$ is bounded, there are still some subleties
to address due to the unboundedness of the transformed operators.
We have included these considerations in appendix section \ref{sec:CarefulJustificationoftheTransformationFormulas}
for the interested reader.)

\subsection{Formalizing the Bosonic Analogy}

Recall that we defined the quasi-bosonic excitation operators by
\begin{equation}
b_{k,p}=\frac{1}{\sqrt{s}}\sum_{\sigma=1}^{s}c_{p-k,\sigma}^{\ast}c_{p,\sigma},\quad b_{k,p}^{\ast}=\frac{1}{\sqrt{s}}\sum_{\sigma=1}^{s}c_{p,\sigma}^{\ast}c_{p-k,\sigma},\quad k\in\mathbb{Z}_{\ast}^{3},\,p\in L_{k},
\end{equation}
which obey the commutation relations
\begin{equation}
\left[b_{k,p},b_{l,q}^{\ast}\right]=\delta_{k,l}\delta_{p,q}+\varepsilon_{k,l}\mleft(p;q\mright),\quad\left[b_{k,p},b_{l,q}\right]=0=\left[b_{k,p}^{\ast},b_{l,q}^{\ast}\right],\label{eq:QuasiBosonicCommutationRelationReminder}
\end{equation}
for $\varepsilon_{k,l}=-s^{-1}\sum_{\sigma=1}^{s}\mleft(\delta_{p,q}c_{q-l,\sigma}c_{p-k,\sigma}^{\ast}+\delta_{p-k,q-l}c_{q,\sigma}^{\ast}c_{p,\sigma}\mright)$.
The relation between these and the $B_{k}$ operators is simply $B_{k}=\sqrt{s}\sum_{p\in L_{k}}b_{k,p}$,
so we can express the non-kinetic part of the bosonizable terms as
\begin{align}
 & \quad\,\sum_{k\in\mathbb{Z}_{\ast}^{3}}\frac{\hat{V}_{k}k_{F}^{-1}}{2\,\mleft(2\pi\mright)^{3}}\mleft(2B_{k}^{\ast}B_{k}+B_{k}B_{-k}+B_{-k}^{\ast}B_{k}^{\ast}\mright)\label{eq:ExpandedBosonizableTerms}\\
 & =\sum_{k\in\mathbb{Z}_{\ast}^{3}}\mleft(2\sum_{p,q\in L_{k}}\frac{s\hat{V}_{k}k_{F}^{-1}}{2\,\mleft(2\pi\mright)^{3}}b_{k,p}^{\ast}b_{k,p}+\sum_{p,q\in L_{k}}\frac{s\hat{V}_{k}k_{F}^{-1}}{2\,\mleft(2\pi\mright)^{3}}\mleft(b_{k,p}b_{-k,-p}+b_{-k,-p}^{\ast}b_{k,p}^{\ast}\mright)\mright).\nonumber 
\end{align}
The expressions inside the parenthesis are similar to the quadratic
operators we considered in the previous section, and to exploit this
similarly we define for any operators $A,B:\ell^{2}\mleft(L_{k}\mright)\rightarrow\ell^{2}\mleft(L_{k}\mright)$
quasi-bosonic quadratic operators $Q_{1}^{k}\mleft(A\mright)$ and $Q_{2}^{k}\mleft(B\mright)$
by
\begin{align}
Q_{1}^{k}\mleft(A\mright) & =\sum_{p,q\in L_{k}}\left\langle e_{p},Ae_{q}\right\rangle b_{k,p}^{\ast}b_{k,q}\\
Q_{2}^{k}\mleft(B\mright) & =\sum_{p,q\in L_{k}}\left\langle e_{p},Be_{q}\right\rangle \mleft(b_{k,p}b_{-k,-q}+b_{-k,-q}^{\ast}b_{k,p}^{\ast}\mright),\nonumber 
\end{align}
where $\mleft(e_{p}\mright)_{p\in L_{k}}$ is the standard orthonormal
basis of $\ell^{2}\mleft(L_{k}\mright)$.

Note that the spaces $\ell^{2}\mleft(L_{k}\mright)$ play the role of
the one-body space $V$ of the previous section\footnote{As in that case we will only consider $\ell^{2}\mleft(L_{k}\mright)$
as a \textit{real} vector space.}, and that $Q_{1}^{k}\mleft(A\mright)$ and $Q_{2}^{k}\mleft(B\mright)$
are analogous to $\mathrm{d}\Gamma\mleft(A\mright)$ and $Q\mleft(B\mright)$
of the equations (\ref{eq:BosonicQuadraticOperator1}) and (\ref{eq:BosonicQuadraticOperator2})
(since we already use $\mathrm{d}\Gamma\mleft(\cdot\mright)$ to denote
the fermionic second-quantization on $\mathcal{H}_{N}$, we deviate
slightly from that notation for the quasi-bosonic operators).

Note also that the $Q_{2}^{k}\mleft(\cdot\mright)$ terms involve excitation
operators of both momentum $k$ and $-k$. For this reason we will
have to treat operators corresponding to the lunes $L_{k}$ and $L_{-k}$
simultaneously when deriving the transformation identities below.

To write the right-hand side of equation (\ref{eq:ExpandedBosonizableTerms})
in this notation, define a vector $v_{k}\in\ell^{2}\mleft(L_{k}\mright)$
by
\begin{equation}
v_{k}=\sqrt{\frac{s\hat{V}_{k}k_{F}^{-1}}{2\,\mleft(2\pi\mright)^{3}}}\sum_{p\in L_{k}}e_{p}
\end{equation}
and consider the operator $P_{k}:\ell^{2}\mleft(L_{k}\mright)\rightarrow\ell^{2}\mleft(L_{k}\mright)$
which acts according to $P_{k}\mleft(\cdot\mright)=\left\langle v_{k},\cdot\right\rangle v_{k}$.
Then
\begin{equation}
\left\langle e_{p},P_{k}e_{q}\right\rangle =\left\langle e_{p},v_{k}\right\rangle \left\langle v_{k},e_{q}\right\rangle =\frac{s\hat{V}_{k}k_{F}^{-1}}{2\,\mleft(2\pi\mright)^{3}},\quad p,q\in L_{k},
\end{equation}
so we simply have
\begin{equation}
\sum_{k\in\mathbb{Z}_{\ast}^{3}}\frac{\hat{V}_{k}k_{F}^{-1}}{2\,\mleft(2\pi\mright)^{3}}\mleft(2B_{k}^{\ast}B_{k}+B_{k}B_{-k}+B_{-k}^{\ast}B_{k}^{\ast}\mright)=\sum_{k\in\mathbb{Z}_{\ast}^{3}}\mleft(2\,Q_{1}^{k}\mleft(P_{k}\mright)+Q_{2}^{k}\mleft(P_{k}\mright)\mright).
\end{equation}

\subsubsection*{Generalized Excitation Operators}

For the purpose of computation (in particular so that we can exploit
Lemma \ref{lemma:TraceFormLemma} to the fullest) it is convenient
to also introduce a basis-independent notation for the quasi-bosonic
operators. We thus define, for any $k\in\mathbb{Z}_{\ast}^{3}$ and
$\varphi\in\ell^{2}\mleft(L_{k}\mright)$, the \textit{generalized excitation
operators} $b_{k}\mleft(\varphi\mright)$ and $b_{k}^{\ast}\mleft(\varphi\mright)$
by
\begin{equation}
b_{k}\mleft(\varphi\mright)=\sum_{p\in L_{k}}\left\langle \varphi,e_{p}\right\rangle b_{k,p},\quad b_{k}^{\ast}\mleft(\varphi\mright)=\sum_{p\in L_{k}}\left\langle e_{p},\varphi\right\rangle b_{k,p}^{\ast}.
\end{equation}
The assignments $\varphi\mapsto b_{k}\mleft(\varphi\mright),b_{k}^{\ast}\mleft(\varphi\mright)$
are then linear, and so it follows from equation (\ref{eq:QuasiBosonicCommutationRelationReminder})
that the generalized excitation operators obey the commutation relations
\begin{align}
\left[b_{k}\mleft(\varphi\mright),b_{l}\mleft(\psi\mright)\right] & =\left[b_{k}^{\ast}\mleft(\varphi\mright),b_{l}^{\ast}\mleft(\psi\mright)\right]=0\label{eq:GeneralizedQBCR}\\
\left[b_{k}\mleft(\varphi\mright),b_{l}^{\ast}\mleft(\psi\mright)\right] & =\delta_{k,l}\left\langle \varphi,\psi\right\rangle +\varepsilon_{k,l}\mleft(\varphi;\psi\mright)\nonumber 
\end{align}
for all $k,l\in\mathbb{Z}^{3}$, $\varphi\in\ell^{2}\mleft(L_{k}\mright)$
and $\psi\in\ell^{2}\mleft(L_{l}\mright)$, where the exchange correction
$\varepsilon_{k,l}\mleft(\varphi;\psi\mright)$ is given by
\begin{equation}
\varepsilon_{k,l}\mleft(\varphi;\psi\mright)=-\frac{1}{s}\sum_{p\in L_{k}}^{\sigma}\sum_{q\in L_{l}}\left\langle \varphi,e_{p}\right\rangle \left\langle e_{q},\psi\right\rangle \mleft(\delta_{p,q}c_{q-l,\sigma}c_{p-k,\sigma}^{\ast}+\delta_{p-k,q-l}c_{q,\sigma}^{\ast}c_{p,\sigma}\mright).
\end{equation}
In terms of these the quadratic operators $Q_{1}^{k}\mleft(A\mright)$
and $Q_{2}^{k}\mleft(B\mright)$ are expressed as
\begin{align}
Q_{1}^{k}\mleft(A\mright) & =\sum_{p\in L_{k}}b_{k}^{\ast}\mleft(Ae_{p}\mright)b_{k,p}\\
Q_{2}^{k}\mleft(B\mright) & =\sum_{p\in L_{k}}\mleft(b_{k}\mleft(Be_{p}\mright)b_{-k,-p}+b_{-k,-p}^{\ast}b_{k}^{\ast}\mleft(Be_{p}\mright)\mright).\nonumber 
\end{align}
It will also be useful to express the relation
\begin{equation}
\left[H_{\mathrm{kin}}^{\prime},b_{k,p}^{\ast}\right]=\mleft(\left|p\right|^{2}-\left|p-k\right|^{2}\mright)b_{k,p}^{\ast}
\end{equation}
of Lemma \ref{lemma:HkinBkpCommutator} in a basis-independent way:
Defining operators $h_{k}:\ell^{2}\mleft(L_{k}\mright)\rightarrow\ell^{2}\mleft(L_{k}\mright)$
by
\begin{equation}
h_{k}e_{p}=\lambda_{k,p}e_{p},\quad\lambda_{k,p}=\frac{1}{2}\mleft(\left|p\right|^{2}-\left|p-k\right|^{2}\mright),
\end{equation}
linearity yields the general commutator
\begin{equation}
\left[H_{\mathrm{kin}}^{\prime},b_{k}^{\ast}\mleft(\varphi\mright)\right]=\sum_{p\in L_{k}}\mleft(\left|p\right|^{2}-\left|p-k\right|^{2}\mright)\left\langle e_{p},\varphi\right\rangle b_{k,p}^{\ast}=2\,b_{k}^{\ast}\mleft(h_{k}\varphi\mright).\label{eq:HkinGeneralizedCommutator}
\end{equation}

\subsection{The Quasi-Bosonic Bogolubov Transformation}

Let a collection of symmetric operators $K_{l}:\ell^{2}\mleft(L_{l}\mright)\rightarrow\ell^{2}\mleft(L_{l}\mright)$,
$l\in\mathbb{Z}_{\ast}^{3}$, be given. Then we define the associated
quasi-bosonic Bogolubov kernel $\mathcal{K}:\mathcal{H}_{N}\rightarrow\mathcal{H}_{N}$
by
\begin{align}
\mathcal{K} & =\frac{1}{2}\sum_{l\in\mathbb{Z}_{\ast}^{3}}\sum_{p,q\in L_{l}}\left\langle e_{p},K_{l}e_{q}\right\rangle \mleft(b_{l,p}b_{-l,-q}-b_{-l,-q}^{\ast}b_{l,p}^{\ast}\mright)\label{eq:calKDefinition}\\
 & =\frac{1}{2}\sum_{l\in\mathbb{Z}_{\ast}^{3}}\sum_{q\in L_{l}}\mleft(b_{l}\mleft(K_{l}e_{q}\mright)b_{-l,-q}-b_{-l,-q}^{\ast}b_{l}^{\ast}\mleft(K_{l}e_{q}\mright)\mright),\nonumber 
\end{align}
in analogy with equation (\ref{eq:BosonicBogolubovKernel}). It is
clear from the second equation that $\mathcal{K}^{\ast}=-\mathcal{K}$,
and so $\mathcal{K}$ generates a unitary transformation $e^{\mathcal{K}}$.

This of course depends on $\mathcal{K}$ being well-defined - as it
is an infinite sum, this is not obvious. As mentioned at the beginning
of this section, we will consider this issue in the next section,
in which we establish that $\mathcal{K}$ is in fact a bounded operator
provided $\sum_{l\in\mathbb{Z}_{\ast}^{3}}\left\Vert K_{l}\right\Vert _{\mathrm{HS}}^{2}<\infty$.

We will make the additional assumption about the operators $K_{k}$
that they are symmetric under the negation $k\rightarrow-k$, in the
sense that
\begin{equation}
\left\langle e_{p},K_{k}e_{q}\right\rangle =\left\langle e_{-p},K_{-k}e_{-q}\right\rangle ,\quad k\in\mathbb{Z}_{\ast}^{3},\,p,q\in L_{k}.
\end{equation}
Letting $I_{k}:\ell^{2}\mleft(L_{k}\mright)\rightarrow\ell^{2}\mleft(L_{-k}\mright)$
denote the unitary mapping acting according to $I_{k}e_{p}=e_{-p}$,
$p\in L_{k}$, this condition is expressed in terms of operators as
\begin{equation}
I_{k}K_{k}=K_{-k}I_{k}.
\end{equation}
It is easily seen that the operators $h_{k}$ and $P_{k}$ defined
above also satisfy this relation. The reason for imposing this condition
is to ensure that Lemma \ref{lemma:TraceFormLemma} allows us to move
operators between arguments also for $Q_{2}^{k}\mleft(\cdot\mright)$-type
terms, since e.g.
\begin{align}
\sum_{q\in L_{l}}b_{l}\mleft(K_{l}e_{q}\mright)b_{-l,-q} & =\sum_{q\in L_{l}}b_{l}\mleft(K_{l}e_{q}\mright)b_{-l}\mleft(e_{-q}\mright)=\sum_{q\in L_{l}}b_{l}\mleft(K_{l}e_{q}\mright)b_{-l}\mleft(I_{l}e_{q}\mright)\\
 & =\sum_{q\in L_{l}}b_{l}\mleft(e_{q}\mright)b_{-l}\mleft(I_{l}K_{l}^{\ast}e_{q}\mright)=\sum_{q\in L_{l}}b_{l,q}b_{-l}\mleft(I_{l}K_{l}e_{q}\mright)=\sum_{q\in L_{l}}b_{l,q}b_{-l}\mleft(K_{-l}e_{-q}\mright).\nonumber 
\end{align}

\subsubsection*{$\mathcal{K}$ Commutators}

As in the previous section we must calculate several commutators involving
$\mathcal{K}$ before we can determine the action of $e^{\mathcal{K}}$
on the bosonizable terms. We start by computing the commutator of
$\mathcal{K}$ with an excitation operator:
\begin{prop}
\label{prop:calKExcitationCommutator}For any $k\in\mathbb{Z}_{\ast}^{3}$
and $\varphi\in\ell^{2}\mleft(L_{k}\mright)$ it holds that
\begin{align*}
\left[\mathcal{K},b_{k}\mleft(\varphi\mright)\right] & =b_{-k}^{\ast}\mleft(I_{k}K_{k}\varphi\mright)+\mathcal{E}_{k}\mleft(\varphi\mright)\\
\left[\mathcal{K},b_{k}^{\ast}\mleft(\varphi\mright)\right] & =b_{-k}\mleft(I_{k}K_{k}\varphi\mright)+\mathcal{E}_{k}\mleft(\varphi\mright)^{\ast}
\end{align*}
where
\[
\mathcal{E}_{k}\mleft(\varphi\mright)=\frac{1}{2}\sum_{l\in\mathbb{Z}_{\ast}^{3}}\sum_{q\in L_{l}}\left\{ \varepsilon_{k,l}\mleft(\varphi;e_{q}\mright),b_{-l}^{\ast}\mleft(K_{-l}e_{-q}\mright)\right\} .
\]
\end{prop}

\textbf{Proof:} It suffices to determine $\left[\mathcal{K},b_{k}\mleft(\varphi\mright)\right]$.
Using Lemma \ref{lemma:TraceFormLemma} we calculate that
\begin{align}
\left[\mathcal{K},b_{k}\mleft(\varphi\mright)\right] & =\frac{1}{2}\sum_{l\in\mathbb{Z}_{\ast}^{3}}\sum_{q\in L_{l}}\mleft(\left[b_{l}\mleft(K_{l}e_{q}\mright)b_{-l}\mleft(e_{-q}\mright)-b_{-l}^{\ast}\mleft(e_{-q}\mright)b_{l}^{\ast}\mleft(K_{l}e_{q}\mright),b_{k}\mleft(\varphi\mright)\right]\mright)\nonumber \\
 & =\frac{1}{2}\sum_{l\in\mathbb{Z}_{\ast}^{3}}\sum_{q\in L_{l}}\mleft(b_{-l}^{\ast}\mleft(e_{-q}\mright)\left[b_{k}\mleft(\varphi\mright),b_{l}^{\ast}\mleft(K_{l}e_{q}\mright)\right]+\left[b_{k}\mleft(\varphi\mright),b_{-l}^{\ast}\mleft(e_{-q}\mright)\right]b_{l}^{\ast}\mleft(K_{l}e_{q}\mright)\mright)\\
 & =\frac{1}{2}\sum_{l\in\mathbb{Z}_{\ast}^{3}}\sum_{q\in L_{l}}\mleft(b_{-l}^{\ast}\mleft(K_{-l}e_{-q}\mright)\left[b_{k}\mleft(\varphi\mright),b_{l}^{\ast}\mleft(e_{q}\mright)\right]+\left[b_{k}\mleft(\varphi\mright),b_{-l}^{\ast}\mleft(e_{-q}\mright)\right]b_{l}^{\ast}\mleft(K_{l}e_{q}\mright)\mright)\nonumber \\
 & =\frac{1}{2}\sum_{l\in\mathbb{Z}_{\ast}^{3}}\sum_{q\in L_{l}}\left\{ \left[b_{k}\mleft(\varphi\mright),b_{-l}^{\ast}\mleft(e_{-q}\mright)\right],b_{l}^{\ast}\mleft(K_{l}e_{q}\mright)\right\} \nonumber 
\end{align}
where we lastly substituted $l\rightarrow-l$, $q\rightarrow-q$ in
the first term. Using the commutation relations of equation (\ref{eq:GeneralizedQBCR})
we then find that
\begin{align}
\left[\mathcal{K},b_{k}\mleft(\varphi\mright)\right] & =\frac{1}{2}\sum_{l\in\mathbb{Z}_{\ast}^{3}}\sum_{q\in L_{l}}\left\{ \delta_{k,-l}\left\langle \varphi,e_{-q}\right\rangle +\varepsilon_{k,-l}\mleft(\varphi;e_{-q}\mright),b_{l}^{\ast}\mleft(K_{l}e_{q}\mright)\right\} \nonumber \\
 & =\sum_{q\in L_{-k}}\left\langle \varphi,e_{-q}\right\rangle b_{-k}^{\ast}\mleft(K_{-k}e_{q}\mright)+\frac{1}{2}\sum_{l\in\mathbb{Z}_{\ast}^{3}}\sum_{q\in L_{l}}\left\{ \varepsilon_{k,-l}\mleft(\varphi;e_{-q}\mright),b_{l}^{\ast}\mleft(K_{l}e_{q}\mright)\right\} \\
 & =b_{-k}^{\ast}\mleft(K_{-k}I_{k}\sum_{q\in L_{k}}\left\langle \varphi,e_{q}\right\rangle e_{q}\mright)+\frac{1}{2}\sum_{l\in\mathbb{Z}_{\ast}^{3}}\sum_{q\in L_{l}}\left\{ \varepsilon_{k,l}\mleft(\varphi;e_{q}\mright),b_{-l}^{\ast}\mleft(K_{-l}e_{-q}\mright)\right\} \nonumber \\
 & =b_{-k}^{\ast}\mleft(I_{k}K_{k}\varphi\mright)+\mathcal{E}_{k}\mleft(\varphi\mright).\nonumber 
\end{align}
$\hfill\square$

Note how these commutators compare to those of equation (\ref{eq:ActionofcalKonCAOperators})
- again $\mathcal{K}$ ``swaps the type and applies $K$'', but
now there is also a reflection from $L_{k}$ to $L_{-k}$, as well
as an additional term involving the exchange correction.

Using this relation we can now determine the commutator with $Q_{1}^{k}$
terms:
\begin{prop}
\label{prop:calKQ1Commutator}For any $k\in\mathbb{Z}_{\ast}^{3}$
and symmetric operators $A_{\pm k}:\ell^{2}\mleft(L_{\pm k}\mright)\rightarrow\ell^{2}\mleft(L_{\pm k}\mright)$
such that $I_{k}A_{k}=A_{-k}I_{k}$, it holds that
\[
\left[\mathcal{K},2\,Q_{1}^{k}\mleft(A_{k}\mright)+2\,Q_{1}^{-k}\mleft(A_{-k}\mright)\right]=Q_{2}^{k}\mleft(\left\{ K_{k},A_{k}\right\} \mright)+2\,\mathrm{Re}\mleft(\mathcal{E}_{k}^{1}\mleft(A_{k}\mright)\mright)+\mleft(k\rightarrow-k\mright)
\]
where
\[
\mathcal{E}_{k}^{1}\mleft(A_{k}\mright)=\sum_{l\in\mathbb{Z}_{\ast}^{3}}\sum_{p\in L_{k}}\sum_{q\in L_{l}}b_{k}^{\ast}\mleft(A_{k}e_{p}\mright)\left\{ \varepsilon_{k,l}\mleft(e_{p};e_{q}\mright),b_{-l}^{\ast}\mleft(K_{-l}e_{-q}\mright)\right\} .
\]
\end{prop}

\textbf{Proof:} Using Proposition \ref{prop:calKExcitationCommutator}
(and Lemma \ref{lemma:TraceFormLemma} together with symmetry of $A_{k}$)
we find that
\begin{align}
\left[\mathcal{K},Q_{1}^{k}\mleft(A_{k}\mright)\right] & =\sum_{p\in L_{k}}\left[\mathcal{K},b_{k}^{\ast}\mleft(A_{k}e_{p}\mright)b_{k}\mleft(e_{p}\mright)\right]=\sum_{p\in L_{k}}\mleft(b_{k}^{\ast}\mleft(A_{k}e_{p}\mright)\left[\mathcal{K},b_{k}\mleft(e_{p}\mright)\right]+\left[\mathcal{K},b_{k}^{\ast}\mleft(A_{k}e_{p}\mright)\right]b_{k}\mleft(e_{p}\mright)\mright)\nonumber \\
 & =\sum_{p\in L_{k}}\mleft(b_{k}^{\ast}\mleft(A_{k}e_{p}\mright)b_{-k}^{\ast}\mleft(I_{k}K_{k}e_{p}\mright)+b_{-k}\mleft(I_{k}K_{k}A_{k}e_{p}\mright)b_{k}\mleft(e_{p}\mright)\mright)\nonumber \\
 & +\sum_{p\in L_{k}}\mleft(b_{k}^{\ast}\mleft(A_{k}e_{p}\mright)\mathcal{E}_{k}\mleft(e_{p}\mright)+\mathcal{E}_{k}\mleft(A_{k}e_{p}\mright)^{\ast}b_{k}\mleft(e_{p}\mright)\mright)\\
 & =\sum_{p\in L_{k}}\mleft(b_{k}^{\ast}\mleft(A_{k}K_{k}e_{p}\mright)b_{-k,-p}^{\ast}+b_{-k,-p}b_{k}\mleft(A_{k}K_{k}e_{p}\mright)\mright)+2\,\mathrm{Re}\mleft(\sum_{p\in L_{k}}b_{k}^{\ast}\mleft(A_{k}e_{p}\mright)\mathcal{E}_{k}\mleft(e_{p}\mright)\mright)\nonumber \\
 & =Q_{2}^{k}\mleft(A_{k}K_{k}\mright)+2\,\mathrm{Re}\mleft(\sum_{p\in L_{k}}b_{k}^{\ast}\mleft(A_{k}e_{p}\mright)\mathcal{E}_{k}\mleft(e_{p}\mright)\mright).\nonumber 
\end{align}
Now, the assumption that $I_{k}A_{k}=A_{-k}I_{k}$ yields
\begin{align}
\sum_{p\in L_{k}}b_{k}\mleft(A_{k}K_{k}e_{p}\mright)b_{-k,-p} & =\sum_{p\in L_{k}}b_{k}\mleft(I_{k}A_{-k}K_{-k}e_{-p}\mright)b_{-k}\mleft(e_{-p}\mright)=\sum_{p\in L_{k}}b_{k}\mleft(e_{p}\mright)b_{-k}\mleft(K_{-k}A_{-k}e_{-p}\mright)\nonumber \\
 & =\sum_{p\in L_{-k}}b_{-k}\mleft(K_{-k}A_{-k}e_{p}\mright)b_{k,-p}
\end{align}
and likewise $\sum_{p\in L_{k}}b_{-k,-p}^{\ast}b_{k}^{\ast}\mleft(A_{k}K_{k}e_{p}\mright)=\sum_{p\in L_{-k}}b_{k,-p}^{\ast}b_{-k}^{\ast}\mleft(K_{-k}A_{-k}e_{p}\mright)$,
whence
\begin{equation}
Q_{2}^{k}\mleft(A_{k}K_{k}\mright)=Q_{2}^{-k}\mleft(K_{-k}A_{-k}\mright).
\end{equation}
Summing over both $k$ and $-k$, and introducing a factor of $2$,
we thus find
\begin{align}
\left[\mathcal{K},2\,Q_{1}^{k}\mleft(A_{k}\mright)+2\,Q_{1}^{-k}\mleft(A_{-k}\mright)\right] & =2\,Q_{2}^{k}\mleft(A_{k}K_{k}\mright)+2\,\mathrm{Re}\mleft(2\sum_{p\in L_{k}}b_{k}^{\ast}\mleft(A_{k}e_{p}\mright)\mathcal{E}_{k}\mleft(e_{p}\mright)\mright)\\
 & +2\,Q_{2}^{-k}\mleft(A_{-k}K_{-k}\mright)+2\,\mathrm{Re}\mleft(2\sum_{p\in L_{-k}}b_{-k}^{\ast}\mleft(A_{-k}e_{p}\mright)\mathcal{E}_{-k}\mleft(e_{p}\mright)\mright)\nonumber \\
 & =Q_{2}^{k}\mleft(\left\{ K_{k},A_{k}\right\} \mright)+2\,\mathrm{Re}\mleft(\mathcal{E}_{k}^{1}\mleft(A_{k}\mright)\mright)+\mleft(k\rightarrow-k\mright)\nonumber 
\end{align}
where $\mathcal{E}_{k}^{1}\mleft(A_{k}\mright)=2\sum_{p\in L_{k}}b_{k}^{\ast}\mleft(A_{k}e_{p}\mright)\mathcal{E}_{k}\mleft(e_{p}\mright)$
follows simply by expansion.

$\hfill\square$

To state the commutator of $\mathcal{K}$ with $Q_{2}^{k}$-type terms,
we first note the identity
\begin{align}
\sum_{p\in L_{k}}b_{k}\mleft(e_{p}\mright)b_{k}^{\ast}\mleft(A_{k}e_{p}\mright) & =\sum_{p\in L_{k}}b_{k}^{\ast}\mleft(A_{k}e_{p}\mright)b_{k}\mleft(e_{p}\mright)+\sum_{p\in L_{k}}\left[b_{k}\mleft(e_{p}\mright),b_{k}^{\ast}\mleft(A_{k}e_{p}\mright)\right]\nonumber \\
 & =\sum_{p\in L_{k}}b_{k}^{\ast}\mleft(A_{k}e_{p}\mright)b_{k}\mleft(e_{p}\mright)+\sum_{p\in L_{k}}\left\langle e_{p},A_{k}e_{p}\right\rangle +\sum_{p\in L_{k}}\varepsilon_{k,k}\mleft(e_{p};A_{k}e_{p}\mright)\\
 & =Q_{1}^{k}\mleft(A_{k}\mright)+\mathrm{tr}\mleft(A_{k}\mright)+\varepsilon_{k}\mleft(A_{k}\mright)\nonumber 
\end{align}
where we introduced the convenient notation
\begin{align}
\varepsilon_{k}\mleft(A_{k}\mright) & =\sum_{p\in L_{k}}\varepsilon_{k,k}\mleft(e_{p};A_{k}e_{p}\mright)=-\frac{1}{s}\sum_{p,q\in L_{k}}^{\sigma}\left\langle e_{q},A_{k}e_{p}\right\rangle \mleft(\delta_{p,q}c_{q-k,\sigma}c_{p-k,\sigma}^{\ast}+\delta_{p-k,q-k}c_{q,\sigma}^{\ast}c_{p,\sigma}\mright)\\
 & =-\frac{1}{s}\sum_{p\in L_{k}}^{\sigma}\left\langle e_{p},A_{k}e_{p}\right\rangle \mleft(c_{p,\sigma}^{\ast}c_{p,\sigma}+c_{p-k,\sigma}c_{p-k,\sigma}^{\ast}\mright).\nonumber 
\end{align}
The commutator is then given by the following:
\begin{prop}
\label{prop:calKQ2Commutator}For any $k\in\mathbb{Z}_{\ast}^{3}$
and symmetric operators $B_{\pm k}:\ell^{2}\mleft(L_{\pm k}\mright)\rightarrow\ell^{2}\mleft(L_{\pm k}\mright)$
such that $I_{k}B_{k}=B_{-k}I_{k}$, it holds that
\begin{align*}
\left[\mathcal{K},Q_{2}^{k}\mleft(B_{k}\mright)+Q_{2}^{-k}\mleft(B_{-k}\mright)\right] & =2\,Q_{1}^{k}\mleft(\left\{ K_{k},B_{k}\right\} \mright)+\mathrm{tr}\mleft(\left\{ K_{k},B_{k}\right\} \mright)+\varepsilon_{k}\mleft(\left\{ K_{k},B_{k}\right\} \mright)\\
 & +2\,\mathrm{Re}\mleft(\mathcal{E}_{k}^{2}\mleft(B_{k}\mright)\mright)+\mleft(k\rightarrow-k\mright)
\end{align*}
where
\[
\mathcal{E}_{k}^{2}\mleft(B_{k}\mright)=\frac{1}{2}\sum_{l\in\mathbb{Z}_{\ast}^{3}}\sum_{p\in L_{k}}\sum_{q\in L_{l}}\left\{ b_{k}\mleft(B_{k}e_{p}\mright),\left\{ \varepsilon_{-k,-l}\mleft(e_{-p};e_{-q}\mright),b_{l}^{\ast}\mleft(K_{l}e_{q}\mright)\right\} \right\} .
\]
\end{prop}

\textbf{Proof:} Writing $Q_{2}^{k}\mleft(B_{k}\mright)$ as $Q_{2}^{k}\mleft(B_{k}\mright)=2\,\mathrm{Re}\mleft(\sum_{p\in L_{k}}b_{k}\mleft(B_{k}e_{p}\mright)b_{-k}\mleft(e_{-p}\mright)\mright)$,
we calculate
\begin{align}
\left[\mathcal{K},Q_{2}^{k}\mleft(B_{k}\mright)\right] & =2\,\mathrm{Re}\mleft(\sum_{p\in L_{k}}\mleft(b_{k}\mleft(B_{k}e_{p}\mright)\left[\mathcal{K},b_{-k}\mleft(e_{-p}\mright)\right]+\left[\mathcal{K},b_{k}\mleft(B_{k}e_{p}\mright)\right]b_{-k}\mleft(e_{-p}\mright)\mright)\mright)\nonumber \\
 & =2\,\mathrm{Re}\mleft(\sum_{p\in L_{k}}\mleft(b_{k}\mleft(B_{k}e_{p}\mright)\left[\mathcal{K},b_{-k}\mleft(e_{-p}\mright)\right]+\left[\mathcal{K},b_{k}\mleft(e_{p}\mright)\right]b_{-k}\mleft(B_{-k}e_{-p}\mright)\mright)\mright)\nonumber \\
 & =2\,\mathrm{Re}\mleft(\sum_{p\in L_{k}}\mleft(b_{k}\mleft(B_{k}e_{p}\mright)b_{k}^{\ast}\mleft(I_{-k}K_{-k}e_{-p}\mright)+b_{-k}^{\ast}\mleft(I_{k}K_{k}e_{p}\mright)b_{-k}\mleft(B_{-k}e_{-p}\mright)\mright)\mright)\\
 & +2\,\mathrm{Re}\mleft(\sum_{p\in L_{k}}\mleft(b_{k}\mleft(B_{k}e_{p}\mright)\mathcal{E}_{-k}\mleft(e_{-p}\mright)+\mathcal{E}_{k}\mleft(e_{p}\mright)b_{-k}\mleft(B_{-k}e_{-p}\mright)\mright)\mright)\nonumber \\
 & =2\,\mathrm{Re}\mleft(\sum_{p\in L_{k}}\mleft(b_{k,p}b_{k}^{\ast}\mleft(K_{k}B_{k}e_{p}\mright)+b_{-k}^{\ast}\mleft(K_{-k}B_{-k}e_{-p}\mright)b_{-k,-p}\mright)\mright)\nonumber \\
 & +2\,\mathrm{Re}\mleft(\sum_{p\in L_{k}}\mleft(b_{k}\mleft(B_{k}e_{p}\mright)\mathcal{E}_{-k}\mleft(e_{-p}\mright)+\mathcal{E}_{k}\mleft(e_{p}\mright)b_{-k}\mleft(B_{-k}e_{-p}\mright)\mright)\mright).\nonumber 
\end{align}
Now
\begin{align}
2\,\mathrm{Re}\mleft(\sum_{p\in L_{k}}b_{k,p}b_{k}^{\ast}\mleft(K_{k}B_{k}e_{p}\mright)\mright) & =\sum_{p\in L_{k}}b_{k,p}b_{k}^{\ast}\mleft(K_{k}B_{k}e_{p}\mright)+\sum_{p\in L_{k}}b_{k}\mleft(K_{k}B_{k}e_{p}\mright)b_{k,p}^{\ast}\nonumber \\
 & =\sum_{p\in L_{k}}b_{k,p}b_{k}^{\ast}\mleft(K_{k}B_{k}e_{p}\mright)+\sum_{p\in L_{k}}b_{k,p}b_{k}^{\ast}\mleft(B_{k}K_{k}e_{p}\mright)\\
 & =\sum_{p\in L_{k}}b_{k,p}b_{k}^{\ast}\mleft(\left\{ K_{k},B_{k}\right\} e_{p}\mright)\nonumber 
\end{align}
and likewise $2\,\mathrm{Re}\mleft(\sum_{p\in L_{k}}b_{-k}^{\ast}\mleft(K_{-k}B_{-k}e_{-p}\mright)b_{-k,-p}\mright)=\sum_{p\in L_{k}}b_{-k}^{\ast}\mleft(\left\{ K_{-k},B_{-k}\right\} e_{-p}\mright)$,
so
\begin{align}
 & \;\,2\,\mathrm{Re}\mleft(\sum_{p\in L_{k}}\mleft(b_{k,p}b_{k}^{\ast}\mleft(K_{k}B_{k}e_{p}\mright)+b_{-k}^{\ast}\mleft(K_{-k}B_{-k}e_{-p}\mright)b_{-k,-p}\mright)\mright)\nonumber \\
 & =\sum_{p\in L_{k}}b_{k,p}b_{k}^{\ast}\mleft(\left\{ K_{k},B_{k}\right\} e_{p}\mright)+\sum_{p\in L_{k}}b_{-k}^{\ast}\mleft(\left\{ K_{-k},B_{-k}\right\} e_{-p}\mright)b_{-k,-p}\\
 & =Q_{1}^{k}\mleft(\left\{ K_{k},B_{k}\right\} \mright)+\mathrm{tr}\mleft(\left\{ K_{k},B_{k}\right\} \mright)+\varepsilon_{k}\mleft(\left\{ K_{k},B_{k}\right\} \mright)+Q_{1}^{-k}\mleft(\left\{ K_{-k},B_{-k}\right\} \mright),\nonumber 
\end{align}
whence summing over $k$ and $-k$ yields
\begin{align}
\left[\mathcal{K},Q_{2}^{k}\mleft(B_{k}\mright)+Q_{2}^{-k}\mleft(B_{-k}\mright)\right] & =2\,Q_{1}^{k}\mleft(\left\{ K_{k},B_{k}\right\} \mright)+\mathrm{tr}\mleft(\left\{ K_{k},B_{k}\right\} \mright)+\varepsilon_{k}\mleft(\left\{ K_{k},B_{k}\right\} \mright)\\
 & +2\,\mathrm{Re}\mleft(\sum_{p\in L_{k}}\left\{ b_{k}\mleft(B_{k}e_{p}\mright),\mathcal{E}_{-k}\mleft(e_{-p}\mright)\right\} \mright)+\mleft(k\rightarrow-k\mright)\nonumber 
\end{align}
and $\mathcal{E}_{k}^{2}\mleft(B_{k}\mright)=\sum_{p\in L_{k}}\left\{ b_{k}\mleft(B_{k}e_{p}\mright),\mathcal{E}_{-k}\mleft(e_{-p}\mright)\right\} $
follows by expansion, yielding the claim.

$\hfill\square$

Finally, for the transformation of $H_{\mathrm{kin}}^{\prime}$, we
also calculate the commutator $\left[\mathcal{K},H_{\mathrm{kin}}^{\prime}\right]$:
\begin{prop}
\label{prop:calKHkinCommutator}It holds that
\[
\left[\mathcal{K},H_{\mathrm{kin}}^{\prime}\right]=\sum_{k\in\mathbb{Z}_{\ast}^{3}}Q_{2}^{k}\mleft(\left\{ K_{k},h_{k}\right\} \mright).
\]
\end{prop}

\textbf{Proof:} By equation (\ref{eq:HkinGeneralizedCommutator})
we have
\begin{equation}
\left[H_{\mathrm{kin}}^{\prime},b_{k}\mleft(\varphi\mright)\right]=-2\,b_{k}\mleft(h_{k}\varphi\mright),\quad\left[H_{\mathrm{kin}}^{\prime},b_{k}^{\ast}\mleft(\varphi\mright)\right]=2\,b_{k}^{\ast}\mleft(h_{k}\varphi\mright),
\end{equation}
so using that $I_{k}h_{k}=h_{-k}I_{k}$ we find
\begin{align}
\left[\mathcal{K},H_{\mathrm{kin}}^{\prime}\right] & =\frac{1}{2}\sum_{k\in\mathbb{Z}_{\ast}^{3}}\sum_{q\in L_{k}}\mleft(\left[b_{k}\mleft(K_{k}e_{q}\mright)b_{-k}\mleft(e_{-q}\mright),H_{\mathrm{kin}}^{\prime}\right]-\left[b_{-k}^{\ast}\mleft(e_{-q}\mright)b_{k}^{\ast}\mleft(K_{k}e_{q}\mright),H_{\mathrm{kin}}^{\prime}\right]\mright)\nonumber \\
 & =-\frac{1}{2}\sum_{k\in\mathbb{Z}_{\ast}^{3}}\sum_{q\in L_{k}}\mleft(b_{k}\mleft(K_{k}e_{q}\mright)\left[H_{\mathrm{kin}}^{\prime},b_{-k}\mleft(e_{-q}\mright)\right]+\left[H_{\mathrm{kin}}^{\prime},b_{k}\mleft(K_{k}e_{q}\mright)\right]b_{-k}\mleft(e_{-q}\mright)\mright)\nonumber \\
 & +\frac{1}{2}\sum_{k\in\mathbb{Z}_{\ast}^{3}}\sum_{q\in L_{k}}\mleft(b_{-k}^{\ast}\mleft(e_{-q}\mright)\left[H_{\mathrm{kin}}^{\prime},b_{k}^{\ast}\mleft(K_{k}e_{q}\mright)\right]+\left[H_{\mathrm{kin}}^{\prime},b_{-k}^{\ast}\mleft(e_{-q}\mright)\right]b_{k}^{\ast}\mleft(K_{k}e_{q}\mright)\mright)\nonumber \\
 & =\sum_{k\in\mathbb{Z}_{\ast}^{3}}\sum_{q\in L_{k}}\mleft(b_{k}\mleft(K_{k}e_{q}\mright)b_{-k}\mleft(h_{-k}e_{-q}\mright)+b_{k}\mleft(h_{k}K_{k}e_{q}\mright)b_{-k}\mleft(e_{-q}\mright)\mright)\\
 & +\sum_{k\in\mathbb{Z}_{\ast}^{3}}\sum_{q\in L_{k}}\mleft(b_{-k}^{\ast}\mleft(e_{-q}\mright)b_{k}^{\ast}\mleft(h_{k}K_{k}e_{q}\mright)+b_{-k}^{\ast}\mleft(h_{-k}e_{-q}\mright)b_{k}^{\ast}\mleft(K_{k}e_{q}\mright)\mright)\nonumber \\
 & =\sum_{k\in\mathbb{Z}_{\ast}^{3}}\sum_{q\in L_{k}}\mleft(b_{k}\mleft(\left\{ K_{k},h_{k}\right\} e_{q}\mright)b_{-k}\mleft(e_{-q}\mright)+b_{-k}^{\ast}\mleft(e_{-q}\mright)b_{k}^{\ast}\mleft(\left\{ K_{k},h_{k}\right\} e_{q}\mright)\mright)\nonumber \\
 & =\sum_{k\in\mathbb{Z}_{\ast}^{3}}Q_{2}^{k}\mleft(\left\{ K_{k},h_{k}\right\} \mright).\nonumber 
\end{align}
$\hfill\square$

\subsection{Transformation of the Bosonizable Terms}

With all the commutators calculated we can now determine the action
of $e^{\mathcal{K}}$ on quadratic operators:
\begin{prop}
\label{prop:TransformationofQuadraticTerms}For any $k\in\mathbb{Z}_{\ast}^{3}$
and symmetric operators $T_{\pm k}:\ell^{2}\mleft(L_{\pm k}\mright)\rightarrow\ell^{2}\mleft(L_{\pm k}\mright)$
such that $I_{k}T_{k}=T_{-k}I_{k}$ it holds that
\begin{align*}
 & \;\,e^{\mathcal{K}}\mleft(2\,Q_{1}^{k}\mleft(T_{k}\mright)+2\,Q_{1}^{-k}\mleft(T_{-k}\mright)\mright)e^{-\mathcal{K}}=\mathrm{tr}\mleft(T_{k}^{1}\mleft(1\mright)-T_{k}\mright)+2\,Q_{1}^{k}\mleft(T_{k}^{1}\mleft(1\mright)\mright)+Q_{2}^{k}\mleft(T_{k}^{2}\mleft(1\mright)\mright)\\
 & +\int_{0}^{1}e^{\mleft(1-t\mright)\mathcal{K}}\mleft(\varepsilon_{k}\mleft(\left\{ K_{k},T_{k}^{2}\mleft(t\mright)\right\} \mright)+2\,\mathrm{Re}\mleft(\mathcal{E}_{k}^{1}\mleft(T_{k}^{1}\mleft(t\mright)\mright)\mright)+2\,\mathrm{Re}\mleft(\mathcal{E}_{k}^{2}\mleft(T_{k}^{2}\mleft(t\mright)\mright)\mright)\mright)e^{-\mleft(1-t\mright)\mathcal{K}}dt+\mleft(k\rightarrow-k\mright)
\end{align*}
and
\begin{align*}
 & \;\,e^{\mathcal{K}}\mleft(Q_{2}^{k}\mleft(T_{k}\mright)+Q_{2}^{-k}\mleft(T_{-k}\mright)\mright)e^{-\mathcal{K}}=\mathrm{tr}\mleft(T_{k}^{2}\mleft(1\mright)\mright)+2\,Q_{1}^{k}\mleft(T_{k}^{2}\mleft(1\mright)\mright)+Q_{2}^{k}\mleft(T_{k}^{1}\mleft(1\mright)\mright)\\
 & +\int_{0}^{1}e^{\mleft(1-t\mright)\mathcal{K}}\mleft(\varepsilon_{k}\mleft(\left\{ K_{k},T_{k}^{1}\mleft(t\mright)\right\} \mright)+2\,\mathrm{Re}\mleft(\mathcal{E}_{k}^{1}\mleft(T_{k}^{2}\mleft(t\mright)\mright)\mright)+2\,\mathrm{Re}\mleft(\mathcal{E}_{k}^{2}\mleft(T_{k}^{1}\mleft(t\mright)\mright)\mright)\mright)e^{-\mleft(1-t\mright)\mathcal{K}}dt+\mleft(k\rightarrow-k\mright)
\end{align*}
where for $t\in\left[0,1\right]$
\begin{align*}
T_{k}^{1}\mleft(t\mright) & =\frac{1}{2}\mleft(e^{tK_{k}}T_{k}e^{tK_{k}}+e^{-tK_{k}}T_{k}e^{-tK_{k}}\mright)\\
T_{k}^{2}\mleft(t\mright) & =\frac{1}{2}\mleft(e^{tK_{k}}T_{k}e^{tK_{k}}-e^{-tK_{k}}T_{k}e^{-tK_{k}}\mright).
\end{align*}
\end{prop}

\textbf{Proof:} We prove the first identity, the second following
by a similar argument.

As in the proof of Proposition \ref{prop:BosonicQuadraticTransformationLaw}
we consider the expression $e^{-t\mathcal{K}}\mleft(2\,Q_{1}^{k}\mleft(T_{k}^{1}\mleft(t\mright)\mright)+Q_{2}^{k}\mleft(T_{k}^{2}\mleft(t\mright)\mright)\mright)e^{t\mathcal{K}}$,
where $T_{k}^{1}\mleft(t\mright)$ and $T_{k}^{2}\mleft(t\mright)$ are
the solutions of the system
\begin{equation}
\mleft(T_{k}^{1}\mright)'\mleft(t\mright)=\left\{ K_{k},T_{k}^{2}\mleft(t\mright)\right\} ,\quad\mleft(T_{k}^{2}\mright)'\mleft(t\mright)=\left\{ K_{k},T_{k}^{1}\mleft(t\mright)\right\} ,
\end{equation}
with initial conditions $T_{k}^{1}\mleft(0\mright)=T_{k}$, $T_{k}^{2}\mleft(0\mright)=0$.

By the Propositions \ref{prop:calKQ1Commutator} and \ref{prop:calKQ2Commutator}
the derivative of such an expression is
\begin{align}
 & \,\frac{d}{dt}e^{-t\mathcal{K}}\mleft(2\,Q_{1}^{k}\mleft(T_{k}^{1}\mleft(t\mright)\mright)+Q_{2}^{k}\mleft(T_{k}^{2}\mleft(t\mright)\mright)\mright)e^{t\mathcal{K}}+\mleft(k\rightarrow-k\mright)\nonumber \\
 & =e^{-t\mathcal{K}}\mleft(2\,Q_{1}^{k}\mleft(\mleft(T_{k}^{1}\mright)^{\prime}\mleft(t\mright)\mright)+Q_{2}^{k}\mleft(\mleft(T_{k}^{2}\mright)^{\prime}\mleft(t\mright)\mright)-\left[\mathcal{K},2\,Q_{1}^{k}\mleft(T_{k}^{1}\mleft(t\mright)\mright)+Q_{2}^{k}\mleft(T_{k}^{2}\mleft(t\mright)\mright)\right]\mright)e^{t\mathcal{K}}+\mleft(k\rightarrow-k\mright)\\
 & =e^{-t\mathcal{K}}\mleft(2\,Q_{1}^{k}\mleft(\mleft(T_{k}^{1}\mright)^{\prime}\mleft(t\mright)\mright)-2\,Q_{1}^{k}\mleft(\left\{ K_{k},T_{k}^{2}\mleft(t\mright)\right\} \mright)-\mathrm{tr}\mleft(\left\{ K_{k},T_{k}^{2}\mleft(t\mright)\right\} \mright)-\varepsilon_{k}\mleft(\left\{ K_{k},T_{k}^{2}\mleft(t\mright)\right\} \mright)-2\,\mathrm{Re}\mleft(\mathcal{E}_{k}^{2}\mleft(T_{k}^{2}\mleft(t\mright)\mright)\mright)\mright)e^{t\mathcal{K}}\nonumber \\
 & +e^{-t\mathcal{K}}\mleft(Q_{2}^{k}\mleft(\mleft(T_{k}^{2}\mright)^{\prime}\mleft(t\mright)\mright)-Q_{2}^{k}\mleft(\left\{ K_{k},T_{k}^{1}\mleft(t\mright)\right\} \mright)-2\,\mathrm{Re}\mleft(\mathcal{E}_{k}^{1}\mleft(T_{k}^{1}\mleft(t\mright)\mright)\mright)\mright)e^{t\mathcal{K}}+\mleft(k\rightarrow-k\mright)\nonumber \\
 & =-\mathrm{tr}\mleft(\mleft(T_{k}^{1}\mright)^{\prime}\mleft(t\mright)\mright)-e^{-t\mathcal{K}}\mleft(\varepsilon_{k}\mleft(\left\{ K_{k},T_{k}^{2}\mleft(t\mright)\right\} \mright)+2\,\mathrm{Re}\mleft(\mathcal{E}_{k}^{1}\mleft(T_{k}^{1}\mleft(t\mright)\mright)\mright)+2\,\mathrm{Re}\mleft(\mathcal{E}_{k}^{2}\mleft(T_{k}^{2}\mleft(t\mright)\mright)\mright)\mright)e^{t\mathcal{K}}+\mleft(k\rightarrow-k\mright),\nonumber 
\end{align}
so by the fundamental theorem of calculus
\begin{align}
 & e^{-\mathcal{K}}\mleft(2\,Q_{1}^{k}\mleft(T_{k}^{1}\mleft(1\mright)\mright)+Q_{2}^{k}\mleft(T_{k}^{2}\mleft(1\mright)\mright)\mright)e^{\mathcal{K}}+\mleft(k\rightarrow-k\mright)=2\,Q_{1}^{k}\mleft(T_{k}\mright)-\mathrm{tr}\mleft(T_{k}^{1}\mleft(1\mright)-T_{k}\mright)\\
 & \qquad\qquad\qquad\qquad\;\;\;-\int_{0}^{1}e^{-t\mathcal{K}}\mleft(\varepsilon_{k}\mleft(\left\{ K_{k},T_{k}^{2}\mleft(t\mright)\right\} \mright)+2\,\mathrm{Re}\mleft(\mathcal{E}_{k}^{1}\mleft(T_{k}^{1}\mleft(t\mright)\mright)\mright)+2\,\mathrm{Re}\mleft(\mathcal{E}_{k}^{2}\mleft(T_{k}^{2}\mleft(t\mright)\mright)\mright)\mright)e^{t\mathcal{K}}dt+\mleft(k\rightarrow-k\mright)\nonumber 
\end{align}
whence conjugation by $e^{\mathcal{K}}$ and rearrangement yields
\begin{align}
 & \;\,e^{\mathcal{K}}\mleft(2\,Q_{1}^{k}\mleft(T_{k}\mright)+2\,Q_{1}^{-k}\mleft(T_{-k}\mright)\mright)e^{-\mathcal{K}}=\mathrm{tr}\mleft(T_{k}^{1}\mleft(1\mright)-T_{k}\mright)+2\,Q_{1}^{k}\mleft(T_{k}^{1}\mleft(1\mright)\mright)+Q_{2}^{k}\mleft(T_{k}^{2}\mleft(1\mright)\mright)\\
 & +\int_{0}^{1}e^{\mleft(1-t\mright)\mathcal{K}}\mleft(\varepsilon_{k}\mleft(\left\{ K_{k},T_{k}^{2}\mleft(t\mright)\right\} \mright)+2\,\mathrm{Re}\mleft(\mathcal{E}_{k}^{1}\mleft(T_{k}^{1}\mleft(t\mright)\mright)\mright)+2\,\mathrm{Re}\mleft(\mathcal{E}_{k}^{2}\mleft(T_{k}^{2}\mleft(t\mright)\mright)\mright)\mright)e^{-\mleft(1-t\mright)\mathcal{K}}dt+\mleft(k\rightarrow-k\mright)\nonumber 
\end{align}
which is the claim.

$\hfill\square$

With the transformation of quadratic operators determined we can also
derive the transformation of $H_{\mathrm{kin}}^{\prime}$:
\begin{prop}
\label{prop:HKinTransformation}It holds that
\begin{align*}
e^{\mathcal{K}}H_{\mathrm{kin}}^{\prime}e^{-\mathcal{K}} & =\sum_{k\in\mathbb{Z}_{\ast}^{3}}\mathrm{tr}\mleft(h_{k}^{1}\mleft(1\mright)-h_{k}\mright)+H_{\mathrm{kin}}^{\prime}+\sum_{k\in\mathbb{Z}_{\ast}^{3}}\mleft(2\,Q_{1}^{k}\mleft(h_{k}^{1}\mleft(1\mright)-h_{k}\mright)+Q_{2}^{k}\mleft(h_{k}^{2}\mleft(1\mright)\mright)\mright)\\
 & +\sum_{k\in\mathbb{Z}_{\ast}^{3}}\int_{0}^{1}e^{\mleft(1-t\mright)\mathcal{K}}\mleft(\varepsilon_{k}\mleft(\left\{ K_{k},h_{k}^{2}\mleft(t\mright)\right\} \mright)+2\,\mathrm{Re}\mleft(\mathcal{E}_{k}^{1}\mleft(h_{k}^{1}\mleft(t\mright)-h_{k}\mright)\mright)+2\,\mathrm{Re}\mleft(\mathcal{E}_{k}^{2}\mleft(h_{k}^{2}\mleft(t\mright)\mright)\mright)\mright)e^{-\mleft(1-t\mright)\mathcal{K}}dt
\end{align*}
where for $t\in\left[0,1\right]$
\begin{align*}
h_{k}^{1}\mleft(t\mright) & =\frac{1}{2}\mleft(e^{tK_{k}}h_{k}e^{tK_{k}}+e^{-tK_{k}}h_{k}e^{-tK_{k}}\mright)\\
h_{k}^{2}\mleft(t\mright) & =\frac{1}{2}\mleft(e^{tK_{k}}h_{k}e^{tK_{k}}-e^{-tK_{k}}h_{k}e^{-tK_{k}}\mright).
\end{align*}
\end{prop}

\textbf{Proof:} By the Propositions \ref{prop:calKQ1Commutator} and
\ref{prop:calKHkinCommutator} we see that
\begin{equation}
\left[\mathcal{K},H_{\mathrm{kin}}^{\prime}-\sum_{k\in\mathbb{Z}_{\ast}^{3}}2\,Q_{1}^{k}\mleft(h_{k}\mright)\right]=-\sum_{k\in\mathbb{Z}_{\ast}^{3}}2\,\mathrm{Re}\mleft(\mathcal{E}_{k}^{1}\mleft(h_{k}\mright)\mright)
\end{equation}
whence by the fundamental theorem of calculus
\begin{equation}
e^{\mathcal{K}}\mleft(H_{\mathrm{kin}}^{\prime}-\sum_{k\in\mathbb{Z}_{\ast}^{3}}2\,Q_{1}^{k}\mleft(h_{k}\mright)\mright)e^{-\mathcal{K}}=H_{\mathrm{kin}}^{\prime}-\sum_{k\in\mathbb{Z}_{\ast}^{3}}2\,Q_{1}^{k}\mleft(h_{k}\mright)-\sum_{k\in\mathbb{Z}_{\ast}^{3}}\int_{0}^{1}e^{t\mathcal{K}}\mleft(2\,\mathrm{Re}\mleft(\mathcal{E}_{k}^{1}\mleft(h_{k}\mright)\mright)\mright)e^{-t\mathcal{K}}dt
\end{equation}
or
\begin{align}
e^{\mathcal{K}}H_{\mathrm{kin}}^{\prime}e^{-\mathcal{K}} & =H_{\mathrm{kin}}^{\prime}+\sum_{k\in\mathbb{Z}_{\ast}^{3}}\mleft(e^{\mathcal{K}}\mleft(2\,Q_{1}^{k}\mleft(h_{k}\mright)\mright)e^{-\mathcal{K}}-2\,Q_{1}^{k}\mleft(h_{k}\mright)\mright)\\
 & -\sum_{k\in\mathbb{Z}_{\ast}^{3}}\int_{0}^{1}e^{\mleft(1-t\mright)\mathcal{K}}\mleft(2\,\mathrm{Re}\mleft(\mathcal{E}_{k}^{1}\mleft(h_{k}\mright)\mright)\mright)e^{-\mleft(1-t\mright)\mathcal{K}}dt.\nonumber 
\end{align}
Applying Proposition \ref{prop:TransformationofQuadraticTerms} now
yields the claim.

$\hfill\square$

With the transformation formulas derived we can now conclude the main
part of Theorem \ref{thm:DiagonalizationoftheBosonizableTerms}: By
the two previous propositions, we see that
\begin{align}
 & \;e^{\mathcal{K}}\mleft(H_{\mathrm{kin}}^{\prime}+\sum_{k\in\mathbb{Z}_{\ast}^{3}}\mleft(2\,Q_{1}^{k}\mleft(P_{k}\mright)+Q_{2}^{k}\mleft(P_{k}\mright)\mright)\mright)e^{-\mathcal{K}}=\sum_{k\in\mathbb{Z}_{\ast}^{3}}\mathrm{tr}\mleft(h_{k}^{1}\mleft(1\mright)-h_{k}+P_{k}^{1}\mleft(1\mright)-P_{k}+P_{k}^{2}\mleft(1\mright)\mright)\nonumber \\
 & +H_{\mathrm{kin}}^{\prime}+\sum_{k\in\mathbb{Z}_{\ast}^{3}}\mleft(2\,Q_{1}^{k}\mleft(h_{k}^{1}\mleft(1\mright)-h_{k}+P_{k}^{1}\mleft(1\mright)+P_{k}^{2}\mleft(1\mright)\mright)+Q_{2}^{k}\mleft(h_{k}^{2}\mleft(1\mright)+P_{k}^{2}\mleft(1\mright)+P_{k}^{1}\mleft(1\mright)\mright)\mright)\\
 & +\sum_{k\in\mathbb{Z}_{\ast}^{3}}\int_{0}^{1}e^{\mleft(1-t\mright)\mathcal{K}}\mleft(\varepsilon_{k}\mleft(\left\{ K_{k},h_{k}^{2}\mleft(t\mright)+P_{k}^{2}\mleft(t\mright)+P_{k}^{1}\mleft(t\mright)\right\} \mright)+2\,\mathrm{Re}\mleft(\mathcal{E}_{k}^{1}\mleft(h_{k}^{1}\mleft(t\mright)-h_{k}+P_{k}^{1}\mleft(t\mright)+P_{k}^{2}\mleft(t\mright)\mright)\mright)\mright.\nonumber \\
 & \qquad\qquad\qquad\qquad\qquad\qquad\qquad\qquad\qquad\qquad\qquad\quad\,+\mleft.2\,\mathrm{Re}\mleft(\mathcal{E}_{k}^{2}\mleft(h_{k}^{2}\mleft(t\mright)+P_{k}^{2}\mleft(t\mright)+P_{k}^{1}\mleft(t\mright)\mright)\mright)\mright)e^{-\mleft(1-t\mright)\mathcal{K}}dt,\nonumber 
\end{align}
which is to say
\begin{align}
 & \;\,e^{\mathcal{K}}\mleft(H_{\mathrm{kin}}^{\prime}+\sum_{k\in\mathbb{Z}_{\ast}^{3}}\mleft(2\,Q_{1}^{k}\mleft(P_{k}\mright)+Q_{2}^{k}\mleft(P_{k}\mright)\mright)\mright)e^{-\mathcal{K}}\nonumber \\
 & =\sum_{k\in\mathbb{Z}_{\ast}^{3}}\mathrm{tr}\mleft(A_{k}\mleft(1\mright)-P_{k}\mright)+H_{\mathrm{kin}}^{\prime}+\sum_{k\in\mathbb{Z}_{\ast}^{3}}\mleft(2\,Q_{1}^{k}\mleft(A_{k}\mleft(1\mright)\mright)+Q_{2}^{k}\mleft(B_{k}\mleft(1\mright)\mright)\mright)\\
 & +\sum_{k\in\mathbb{Z}_{\ast}^{3}}\int_{0}^{1}e^{\mleft(1-t\mright)\mathcal{K}}\mleft(\varepsilon_{k}\mleft(\left\{ K_{k},B_{k}\mleft(t\mright)\right\} \mright)+2\,\mathrm{Re}\mleft(\mathcal{E}_{k}^{1}\mleft(A_{k}\mleft(t\mright)\mright)\mright)+2\,\mathrm{Re}\mleft(\mathcal{E}_{k}^{2}\mleft(B_{k}\mleft(t\mright)\mright)\mright)\mright)e^{-\mleft(1-t\mright)\mathcal{K}}dt\nonumber 
\end{align}
where the operators $A_{k}\mleft(t\mright),B_{k}\mleft(t\mright):\ell^{2}\mleft(L_{k}\mright)\rightarrow\ell^{2}\mleft(L_{k}\mright)$
are given by
\begin{align}
A_{k}\mleft(t\mright) & =\frac{1}{2}\mleft(e^{tK_{k}}\mleft(h_{k}+2P_{k}\mright)e^{tK_{k}}+e^{-tK_{k}}h_{k}e^{-tK_{k}}\mright)-h_{k}\\
B_{k}\mleft(t\mright) & =\frac{1}{2}\mleft(e^{tK_{k}}\mleft(h_{k}+2P_{k}\mright)e^{tK_{k}}-e^{-tK_{k}}h_{k}e^{-tK_{k}}\mright).\nonumber 
\end{align}
We can now choose the kernels $K_{k}$ such that this expression is
diagonalized, i.e. such that the $Q_{2}^{k}\mleft(\cdot\mright)$ terms
vanish. Evidently this is saying that $B_{k}\mleft(1\mright)=0$, so
we arrive at the diagonalization condition
\begin{equation}
e^{K_{k}}\mleft(h_{k}+2P_{k}\mright)e^{K_{k}}=e^{-K_{k}}h_{k}e^{-K_{k}}.
\end{equation}
Note that this is really the condition of equation (\ref{eq:DiagonalizationCondition})
of the previous section, with $A=h_{k}+P_{k}$ and $B_{k}=P_{k}$.
As such we see by Proposition \ref{prop:UniqueDiagonalizingKernel}
that we must choose
\begin{equation}
K_{k}=-\frac{1}{2}\log\mleft(h_{k}^{-\frac{1}{2}}\mleft(h_{k}^{\frac{1}{2}}\mleft(h_{k}+2P_{k}\mright)h_{k}^{\frac{1}{2}}\mright)^{\frac{1}{2}}h_{k}^{-\frac{1}{2}}\mright).\label{eq:KkDefinition}
\end{equation}
Since the diagonalization condition is then fulfilled, it follows
that also
\begin{equation}
A_{k}\mleft(1\mright)=e^{-K_{k}}h_{k}e^{-K_{k}}-h_{k}
\end{equation}
and the formula of Theorem \ref{thm:DiagonalizationoftheBosonizableTerms}
is proved.

\section{\label{sec:ControllingtheTransformationKernel}Controlling the Transformation
Kernel}

In this section we prove that under the condition $\sum_{l\in\mathbb{Z}_{\ast}^{3}}\left\Vert K_{l}\right\Vert _{\mathrm{HS}}^{2}<\infty$,
the operator defined by
\begin{equation}
\mathcal{K}=\frac{1}{2}\sum_{l\in\mathbb{Z}_{\ast}^{3}}\sum_{p,q\in L_{l}}\left\langle e_{p},K_{l}e_{q}\right\rangle \mleft(b_{l,p}b_{-l,-q}-b_{-l,-q}^{\ast}b_{l,p}^{\ast}\mright)
\end{equation}
is bounded. More precisely, we prove the following estimate:
\begin{prop}
\label{prop:calKNumberBound}For all $\Phi,\Psi\in\mathcal{H}_{N}$
it holds that
\[
\left|\left\langle \Phi,\mathcal{K}\Psi\right\rangle \right|\leq\sqrt{5}\sqrt{\sum_{l\in\mathbb{Z}_{\ast}^{3}}\left\Vert K_{l}\right\Vert _{\mathrm{HS}}^{2}}\sqrt{\left\langle \Phi,\mleft(\mathcal{N}_{E}+1\mright)\Phi\right\rangle \left\langle \Psi,\mleft(\mathcal{N}_{E}+1\mright)\Psi\right\rangle }.
\]
\end{prop}

Recalling that
\begin{equation}
\mathcal{N}_{E}=\sum_{p\in B_{F}^{c}}^{\sigma}c_{p,\sigma}^{\ast}c_{p,\sigma}=\sum_{p\in B_{F}}^{\sigma}c_{p,\sigma}c_{p,\sigma}^{\ast}
\end{equation}
we have the trivial bound $\mathcal{N}_{E}=\sum_{p\in B_{F}}^{\sigma}c_{p,\sigma}c_{p,\sigma}^{\ast}\leq s\left|B_{F}\right|=N$,
whence the proposition indeed implies boundedness, as an estimate
of the form $\left|\left\langle \Phi,\mathcal{K}\Psi\right\rangle \right|\leq C\left\Vert \Phi\right\Vert \left\Vert \Psi\right\Vert $
follows. Additionally, we will see in the next section that the kernels
of equation (\ref{eq:KkDefinition}) obey
\begin{equation}
\left\Vert K_{k}\right\Vert _{\mathrm{HS}}\leq C\hat{V}_{k},\quad k\in\mathbb{Z}_{\ast}^{3},
\end{equation}
for a constant $C>0$ independent of $k$, so the criterion $\sum_{k\in\mathbb{Z}_{\ast}^{3}}\hat{V}_{k}^{2}<\infty$
does indeed imply boundedness of our diagonalizing kernel $\mathcal{K}$
hence existence of the unitary transformation $e^{\mathcal{K}}$ asserted
by Theorem \ref{thm:DiagonalizationoftheBosonizableTerms}.

\subsubsection*{Preliminary Analysis}

Define
\begin{equation}
\tilde{\mathcal{K}}=\frac{1}{2}\sum_{l\in\mathbb{Z}_{\ast}^{3}}\sum_{p,q\in L_{l}}\left\langle e_{p},K_{l}e_{q}\right\rangle b_{l,p}b_{-l,-q}
\end{equation}
so that $\mathcal{K}=\tilde{\mathcal{K}}-\tilde{\mathcal{K}}^{\ast}$.
Then for any $\Phi,\Psi\in\mathcal{H}_{N}$
\begin{equation}
\left|\left\langle \Phi,\mathcal{K}\Psi\right\rangle \right|\leq\vert\langle\Phi,\tilde{\mathcal{K}}\Psi\rangle\vert+\vert\langle\Psi,\tilde{\mathcal{K}}\Phi\rangle\vert
\end{equation}
so we need only bound a quantity of the form $\vert\langle\Phi,\tilde{\mathcal{K}}\Psi\rangle\vert$.

Note that by expanding $b_{-l,-q}=s^{-\frac{1}{2}}\sum_{\sigma=1}^{s}c_{-q+l,\sigma}^{\ast}c_{-q,\sigma}$
we can write $\tilde{\mathcal{K}}$ as
\begin{equation}
\tilde{\mathcal{K}}=\frac{1}{2\sqrt{s}}\sum_{l\in\mathbb{Z}_{\ast}^{3}}\sum_{p,q\in L_{l}}^{\sigma}\left\langle e_{p},K_{l}e_{q}\right\rangle b_{l,p}c_{-q+l,\sigma}^{\ast}c_{-q,\sigma}=\frac{1}{2\sqrt{s}}\sum_{q\in B_{F}^{c}}^{\sigma}\mleft(\sum_{l\in\mathbb{Z}_{\ast}^{3}}\sum_{p\in L_{l}}1_{L_{l}}\mleft(q\mright)\left\langle e_{p},K_{l}e_{q}\right\rangle b_{l,p}c_{-q+l,\sigma}^{\ast}\mright)c_{-q,\sigma}
\end{equation}
whence we may estimate
\begin{align}
\vert\langle\Phi,\tilde{\mathcal{K}}\Psi\rangle\vert & =\frac{1}{2\sqrt{s}}\left|\sum_{q\in B_{F}^{c}}^{\sigma}\left\langle \sum_{l\in\mathbb{Z}_{\ast}^{3}}\sum_{p\in L_{l}}1_{L_{l}}\mleft(q\mright)\left\langle K_{l}e_{q},e_{p}\right\rangle c_{-q+l,\sigma}b_{l,p}^{\ast}\Phi,c_{-q,\sigma}\Psi\right\rangle \right|\nonumber \\
 & \leq\frac{1}{2\sqrt{s}}\sum_{q\in B_{F}^{c}}^{\sigma}\left\Vert \sum_{l\in\mathbb{Z}_{\ast}^{3}}\sum_{p\in L_{l}}1_{L_{l}}\mleft(q\mright)\left\langle K_{l}e_{q},e_{p}\right\rangle c_{-q+l,\sigma}b_{l,p}^{\ast}\Phi\right\Vert \left\Vert c_{-q,\sigma}\Psi\right\Vert \label{eq:calKtildeEstimate1}\\
 & \leq\frac{1}{2\sqrt{s}}\sqrt{\sum_{q\in B_{F}^{c}}^{\sigma}\left\Vert \sum_{l\in\mathbb{Z}_{\ast}^{3}}\sum_{p\in L_{l}}1_{L_{l}}\mleft(q\mright)\left\langle K_{l}e_{q},e_{p}\right\rangle c_{-q+l,\sigma}b_{l,p}^{\ast}\Phi\right\Vert ^{2}}\sqrt{\sum_{q\in B_{F}^{c}}^{\sigma}\left\Vert c_{-q,\sigma}\Psi\right\Vert ^{2}}\nonumber \\
 & =\frac{1}{2}\sqrt{\frac{1}{s}\sum_{q\in B_{F}^{c}}^{\sigma}\left\Vert \sum_{l\in\mathbb{Z}_{\ast}^{3}}\sum_{p\in L_{l}}1_{L_{l}}\mleft(q\mright)\left\langle K_{l}e_{q},e_{p}\right\rangle c_{-q+l,\sigma}b_{l,p}^{\ast}\Phi\right\Vert ^{2}}\sqrt{\left\langle \Psi,\mathcal{N}_{E}\Psi\right\rangle }.\nonumber 
\end{align}
Now, the operator appearing under the root can be written as
\begin{align}
 & \quad\;\sum_{l\in\mathbb{Z}_{\ast}^{3}}\sum_{p\in L_{l}}1_{L_{l}}\mleft(q\mright)\left\langle K_{l}e_{q},e_{p}\right\rangle c_{-q+l,\sigma}b_{l,p}^{\ast}=\frac{1}{\sqrt{s}}\sum_{l\in\mathbb{Z}_{\ast}^{3}}\sum_{p\in L_{l}}^{\tau}1_{L_{l}}\mleft(q\mright)\left\langle K_{l}e_{q},e_{p}\right\rangle c_{p,\tau}^{\ast}c_{p-l,\tau}c_{-q+l,\sigma}\label{eq:calKtildeEstimate2}\\
 & =\frac{1}{\sqrt{s}}\sum_{p'\in B_{F}^{c}}^{\tau}\sum_{q',r'\in B_{F}}\mleft(\sum_{l\in\mathbb{Z}_{\ast}^{3}}\sum_{p\in L_{l}}\delta_{p',p}\delta_{q',p-l}\delta_{r',-q+l}1_{L_{l}}\mleft(q\mright)\left\langle K_{l}e_{q},e_{p}\right\rangle \mright)c_{p',\tau}^{\ast}c_{q',\tau}c_{r',\sigma}.\nonumber 
\end{align}
The introduction of these Kronecker $\delta$'s has no effect by itself,
but it highlights that this operator can be written simply in the
form
\begin{equation}
\frac{1}{\sqrt{s}}\sum_{p\in B_{F}^{c}}^{\tau}\sum_{q,r\in B_{F}}A_{p,q,r}c_{p,\tau}^{\ast}c_{q,\tau}c_{r,\sigma}
\end{equation}
for some coefficients $A_{p,q,r}$. We will now derive a general estimate
for such an expression.

\subsection{A Higher Order Fermionic Estimate}

Recall that the ``standard fermionic estimate'' can be stated as
\begin{equation}
\left\Vert \sum A_{k}c_{k}\Psi\right\Vert ,\,\left\Vert \sum A_{k}c_{k}^{\ast}\Psi\right\Vert \leq\sqrt{\sum\left|A_{k}\right|^{2}}\left\Vert \Psi\right\Vert ,
\end{equation}
which can be proved by appealing to the CAR as follows: Trivially
\begin{align}
\left\Vert \sum A_{k}c_{k}\Psi\right\Vert ^{2} & =\left\langle \sum A_{k}c_{k}\Psi,\sum A_{l}c_{l}\Psi\right\rangle =\left\langle \Psi,\mleft(\sum A_{k}c_{k}\mright)^{\ast}\mleft(\sum A_{l}c_{l}\mright)\Psi\right\rangle \\
 & \leq\left\langle \Psi,\mleft(\sum A_{k}c_{k}\mright)^{\ast}\mleft(\sum A_{l}c_{l}\mright)\Psi\right\rangle +\left\langle \Psi,\mleft(\sum A_{l}c_{l}\mright)\mleft(\sum A_{k}c_{k}\mright)^{\ast}\Psi\right\rangle \nonumber \\
 & =\left\langle \Psi,\left\{ \mleft(\sum A_{k}c_{k}\mright)^{\ast},\mleft(\sum A_{l}c_{l}\mright)\right\} \Psi\right\rangle \nonumber 
\end{align}
since all that was done was the addition of a non-negative term. By
the CAR, however,
\begin{equation}
\left\{ \mleft(\sum A_{k}c_{k}\mright)^{\ast},\mleft(\sum A_{l}c_{l}\mright)\right\} =\sum\overline{A_{k}}A_{l}\left\{ c_{k}^{\ast},c_{l}\right\} =\sum\overline{A_{k}}A_{l}\delta_{k,l}=\sum\left|A_{k}\right|^{2}
\end{equation}
whence the bound immediately follows. This establishes the uniquely
fermionic property that sums of creation and annihilation operators
can be estimated independently of the number operator, unlike in the
bosonic case.

One can imagine generalizing this to quadratic expressions of the
form $\sum_{k,l}A_{k,l}c_{k}c_{l}$, but this fails: The issue is
that the CAR only yields a commutation relation for such expressions,
and not an anticommutation relation, whence the argument above can
not be applied.

We may however make the observation that for cubic expressions, such
as $\sum_{k,l,m}A_{k,l,m}c_{k}^{\ast}c_{l}c_{m}$, the CAR does yield
an anticommutation relation, allowing the trick to be applied. The
anticommutator is of course not constant, but rather a combination
of quadratic, linear and constant expressions, but this still yields
a reduction in ``number operator order'', which will be crucial
for our estimation of $e^{\mathcal{K}}\mathcal{N}_{E}^{m}e^{-\mathcal{K}}$
later on.

To derive such an estimate we first calculate the following basic
anticommutator:
\begin{lem}
For any $p,p'\in B_{F}^{c}$, $q,q',r,r'\in B_{F}$ and $1\leq\sigma,\tau,\tau'\leq s$
it holds that
\begin{align*}
\left\{ \mleft(c_{p,\tau}^{\ast}c_{q,\tau}c_{r,\sigma}\mright)^{\ast},c_{p',\tau'}^{\ast}c_{q',\tau'}c_{r',\sigma}\right\}  & =\delta_{p,p'}^{\tau,\tau'}c_{q',\tau'}c_{r',\sigma}c_{r,\sigma}^{\ast}c_{q,\tau}^{\ast}+\delta_{q,q'}^{\tau,\tau'}c_{p',\tau'}^{\ast}c_{r',\sigma}c_{r,\sigma}^{\ast}c_{p,\tau}+\delta_{r,r'}c_{p',\tau'}^{\ast}c_{q',\tau'}c_{q,\tau}^{\ast}c_{p,\tau}\\
 & -\delta_{r,q'}^{\sigma,\tau'}c_{p',\tau'}^{\ast}c_{r',\sigma}c_{q,\tau}^{\ast}c_{p,\tau}-\delta_{q,r'}^{\tau,\sigma}c_{p',\tau'}^{\ast}c_{q',\tau'}c_{r,\sigma}^{\ast}c_{p,\tau}\\
 & -\delta_{q,q'}^{\tau,\tau'}\delta_{r,r'}c_{p',\tau'}^{\ast}c_{p,\tau}-\delta_{p,p'}^{\tau,\tau'}\delta_{q,q'}^{\tau,\tau'}c_{r',\sigma}c_{r,\sigma}^{\ast}-\delta_{p,p'}^{\tau,\tau'}\delta_{r,r'}c_{q',\tau'}c_{q,\tau}^{\ast}\\
 & +\delta_{q,r'}^{\tau,\sigma}\delta_{r,q'}^{\sigma,\tau'}c_{p',\tau'}^{\ast}c_{p,\tau}+\delta_{p,p'}^{\tau,\tau'}\delta_{r,q'}^{\sigma,\tau'}c_{r',\sigma}c_{q,\tau}^{\ast}+\delta_{p,p'}^{\tau,\tau'}\delta_{q,r'}^{\tau,\sigma}c_{q',\tau'}c_{r,\sigma}^{\ast}\\
 & +\delta_{p,p'}^{\tau,\tau'}\delta_{q,q'}^{\tau,\tau'}\delta_{r,r'}-\delta_{p,p'}^{\tau,\tau'}\delta_{q,r'}^{\tau,\sigma}\delta_{r,q'}^{\sigma,\tau'}.
\end{align*}
\end{lem}

\textbf{Proof:} The proof is a straightforward but lengthy calculation
using the CAR: First we note
\begin{align}
\mleft(c_{p,\tau}^{\ast}c_{q,\tau}c_{r,\sigma}\mright)^{\ast}c_{p',\tau'}^{\ast}c_{q',\tau'}c_{r',\sigma} & =c_{r,\sigma}^{\ast}c_{q,\tau}^{\ast}c_{p,\tau}c_{p',\tau'}^{\ast}c_{q',\tau'}c_{r',\sigma}=-c_{r,\sigma}^{\ast}c_{q,\tau}^{\ast}c_{p',\tau'}^{\ast}c_{p,\tau}c_{q',\tau'}c_{r',\sigma}+\delta_{p,p'}^{\tau,\tau'}c_{r,\sigma}^{\ast}c_{q,\tau}^{\ast}c_{q',\tau'}c_{r',\sigma}\nonumber \\
 & =-c_{p',\tau'}^{\ast}c_{r,\sigma}^{\ast}c_{q,\tau}^{\ast}c_{q',\tau'}c_{r',\sigma}c_{p,\tau}+\delta_{p,p'}^{\tau,\tau'}c_{r,\sigma}^{\ast}c_{q,\tau}^{\ast}c_{q',\tau'}c_{r',\sigma}\label{eq:CARLemma1}
\end{align}
and
\begin{align}
c_{r,\sigma}^{\ast}c_{q,\tau}^{\ast}c_{q',\tau'}c_{r',\sigma} & =-c_{r,\sigma}^{\ast}c_{q',\tau'}c_{q,\tau}^{\ast}c_{r',\sigma}+\delta_{q,q'}^{\tau,\tau'}c_{r,\sigma}^{\ast}c_{r',\sigma}=c_{r,\sigma}^{\ast}c_{q',\tau'}c_{r',\sigma}c_{q,\tau}^{\ast}-\delta_{q,r'}^{\tau,\sigma}c_{r,\sigma}^{\ast}c_{q',\tau'}+\delta_{q,q'}^{\tau,\tau'}c_{r,\sigma}^{\ast}c_{r',\sigma}\nonumber \\
 & =-c_{q',\tau'}c_{r,\sigma}^{\ast}c_{r',\sigma}c_{q,\tau}^{\ast}+\delta_{r,q'}^{\sigma,\tau'}c_{r',\sigma}c_{q,\tau}^{\ast}-\delta_{q,r'}^{\tau,\sigma}c_{r,\sigma}^{\ast}c_{q',\tau'}+\delta_{q,q'}^{\tau,\tau'}c_{r,\sigma}^{\ast}c_{r',\sigma}\nonumber \\
 & =c_{q',\tau'}c_{r',\sigma}c_{r,\sigma}^{\ast}c_{q,\tau}^{\ast}-\delta_{r,r'}c_{q',\tau'}c_{q,\tau}^{\ast}+\delta_{r,q'}^{\sigma,\tau'}c_{r',\sigma}c_{q,\tau}^{\ast}-\delta_{q,r'}^{\tau,\sigma}c_{r,\sigma}^{\ast}c_{q',\tau'}+\delta_{q,q'}^{\tau,\tau'}c_{r,\sigma}^{\ast}c_{r',\sigma}\\
 & =c_{q',\tau'}c_{r',\sigma}c_{r,\sigma}^{\ast}c_{q,\tau}^{\ast}-\delta_{q,q'}^{\tau,\tau'}c_{r',\sigma}c_{r,\sigma}^{\ast}-\delta_{r,r'}c_{q',\tau'}c_{q,\tau}^{\ast}+\delta_{r,q'}^{\sigma,\tau'}c_{r',\sigma}c_{q,\tau}^{\ast}+\delta_{q,r'}^{\tau,\sigma}c_{q',\tau'}c_{r,\sigma}^{\ast}\nonumber \\
 & +\delta_{q,q'}^{\tau,\tau'}\delta_{r,r'}-\delta_{q,r'}^{\tau,\sigma}\delta_{r,q'}^{\sigma,\tau'}.\nonumber 
\end{align}
Consequently
\begin{align}
-c_{p',\tau'}^{\ast}c_{r,\sigma}^{\ast}c_{q,\tau}^{\ast}c_{q',\tau'}c_{r',\sigma}c_{p,\tau} & =-c_{p',\tau'}^{\ast}c_{q',\tau'}c_{r',\sigma}c_{r,\sigma}^{\ast}c_{q,\tau}^{\ast}c_{p,\tau}+c_{p',\tau'}^{\ast}\mleft(\delta_{q,q'}^{\tau,\tau'}c_{r',\sigma}c_{r,\sigma}^{\ast}+\delta_{r,r'}c_{q',\tau'}c_{q,\tau}^{\ast}\mright)c_{p,\tau}\nonumber \\
 & -c_{p',\tau'}^{\ast}\mleft(\delta_{r,q'}^{\sigma,\tau'}c_{r',\sigma}c_{q,\tau}^{\ast}+\delta_{q,r'}^{\tau,\sigma}c_{q',\tau'}c_{r,\sigma}^{\ast}\mright)c_{p,\tau}-c_{p',\tau'}^{\ast}\mleft(\delta_{q,q'}^{\tau,\tau'}\delta_{r,r'}-\delta_{q,r'}^{\tau,\sigma}\delta_{r,q'}^{\sigma,\tau'}\mright)c_{p,\tau}\\
 & =-c_{p',\tau'}^{\ast}c_{q',\tau'}c_{r',\sigma}\mleft(c_{p,\tau}^{\ast}c_{q,\tau}c_{r,\sigma}\mright)^{\ast}+\delta_{q,q'}^{\tau,\tau'}c_{p',\tau'}^{\ast}c_{r',\sigma}c_{r,\sigma}^{\ast}c_{p,\tau}+\delta_{r,r'}c_{p',\tau'}^{\ast}c_{q',\tau'}c_{q,\tau}^{\ast}c_{p,\tau}\nonumber \\
 & -\delta_{r,q'}^{\sigma,\tau'}c_{p',\tau'}^{\ast}c_{r',\sigma}c_{q,\tau}^{\ast}c_{p,\tau}-\delta_{q,r'}^{\tau,\sigma}c_{p',\tau'}^{\ast}c_{q',\tau'}c_{r,\sigma}^{\ast}c_{p,\tau}\nonumber \\
 & -\delta_{q,q'}^{\tau,\tau'}\delta_{r,r'}c_{p',\tau'}^{\ast}c_{p,\tau}+\delta_{q,r'}^{\tau,\sigma}\delta_{r,q'}^{\sigma,\tau'}c_{p',\tau'}^{\ast}c_{p,\tau}\nonumber 
\end{align}
and
\begin{align}
\delta_{p,p'}^{\tau,\tau'}c_{r,\sigma}^{\ast}c_{q,\tau}^{\ast}c_{q',\tau'}c_{r',\sigma} & =\delta_{p,p'}^{\tau,\tau'}c_{q',\tau'}c_{r',\sigma}c_{r,\sigma}^{\ast}c_{q,\tau}^{\ast}-\delta_{p,p'}^{\tau,\tau'}\mleft(\delta_{q,q'}^{\tau,\tau'}c_{r',\sigma}c_{r,\sigma}^{\ast}+\delta_{r,r'}c_{q',\tau'}c_{q,\tau}^{\ast}\mright)\nonumber \\
 & +\delta_{p,p'}^{\tau,\tau'}\mleft(\delta_{r,q'}^{\sigma,\tau'}c_{r',\sigma}c_{q,\tau}^{\ast}+\delta_{q,r'}^{\tau,\sigma}c_{q',\tau'}c_{r,\sigma}^{\ast}\mright)+\delta_{p,p'}^{\tau,\tau'}\mleft(\delta_{q,q'}^{\tau,\tau'}\delta_{r,r'}-\delta_{q,r'}^{\tau,\sigma}\delta_{r,q'}^{\sigma,\tau'}\mright)\\
 & =\delta_{p,p'}^{\tau,\tau'}c_{q',\tau'}c_{r',\sigma}c_{r,\sigma}^{\ast}c_{q,\tau}^{\ast}-\delta_{p,p'}^{\tau,\tau'}\delta_{q,q'}^{\tau,\tau'}c_{r',\sigma}c_{r,\sigma}^{\ast}-\delta_{p,p'}^{\tau,\tau'}\delta_{r,r'}c_{q',\tau'}c_{q,\tau}^{\ast}\nonumber \\
 & +\delta_{p,p'}^{\tau,\tau'}\delta_{r,q'}^{\sigma,\tau'}c_{r',\sigma}c_{q,\tau}^{\ast}+\delta_{p,p'}^{\tau,\tau'}\delta_{q,r'}^{\tau,\sigma}c_{q',\tau'}c_{r,\sigma}^{\ast}+\delta_{p,p'}^{\tau,\tau'}\delta_{q,q'}^{\tau,\tau'}\delta_{r,r'}-\delta_{p,p'}^{\tau,\tau'}\delta_{q,r'}^{\tau,\sigma}\delta_{r,q'}^{\sigma,\tau'}.\nonumber 
\end{align}
Insertion of these two identities into equation (\ref{eq:CARLemma1})
yields the claim.

$\hfill\square$

We can now conclude the desired bound:
\begin{prop}
\label{prop:CubicSumBound}Let $A_{p,q,r}\in\mathbb{C}$ for $p\in B_{F}^{c}$
and $q,r\in B_{F}$ with $\sum_{p\in B_{F}^{c}}\sum_{q,r\in B_{F}}\left|A_{p,q,r}\right|^{2}<\infty$
be given. Then for any $\Psi\in\mathcal{H}_{N}$
\[
\frac{1}{s}\sum_{\sigma=1}^{s}\left\Vert \sum_{p\in B_{F}^{c}}^{\tau}\sum_{q,r\in B_{F}}A_{p,q,r}c_{p,\tau}^{\ast}c_{q,\tau}c_{r,\sigma}\Psi\right\Vert ^{2}\leq5s\sum_{p\in B_{F}^{c}}\sum_{q,r\in B_{F}}\left|A_{p,q,r}\right|^{2}\left\langle \Psi,\mleft(\mathcal{N}_{E}+1\mright)\Psi\right\rangle .
\]
\end{prop}

\textbf{Proof:} As in the proof of the standard fermionic estimate,
we have
\begin{align*}
\left\Vert \sum_{p\in B_{F}^{c}}^{\tau}\sum_{q,r\in B_{F}}A_{p,q,r}c_{p,\tau}^{\ast}c_{q,\tau}c_{r,\sigma}\Psi\right\Vert ^{2} & =\left\langle \sum_{p\in B_{F}^{c}}^{\tau}\sum_{q,r\in B_{F}}A_{p,q,r}c_{p,\tau}^{\ast}c_{q,\tau}c_{r,\sigma}\Psi,\sum_{p'\in B_{F}^{c}}^{\tau'}\sum_{q',r'\in B_{F}}A_{p',q',r'}c_{p',\tau'}^{\ast}c_{q',\tau'}c_{r',\sigma}\Psi\right\rangle \\
 & \leq\sum_{p,p'\in B_{F}^{c}}^{\tau,\tau'}\sum_{q,q',r,r'\in B_{F}}\overline{A_{p,q,r}}A_{p',q',r'}\left\langle \Psi,\left\{ \mleft(c_{p,\tau}^{\ast}c_{q,\tau}c_{r,\sigma}\mright)^{\ast},c_{p',\tau'}^{\ast}c_{q',\tau'}c_{r',\sigma}\right\} \Psi\right\rangle 
\end{align*}
so by the identity of the preceding lemma
\begin{align}
 & \quad\;\;\;\sum_{\sigma=1}^{s}\left\Vert \sum_{p\in B_{F}^{c}}\sum_{q,r\in B_{F}}A_{p,q,r}c_{p}^{\ast}c_{q}c_{r}\Psi\right\Vert ^{2}\nonumber \\
 & \leq\sum_{\sigma,\tau,\tau'=1}^{s}\sum_{p,p'\in B_{F}^{c}}\sum_{q,q',r,r'\in B_{F}}\overline{A_{p,q,r}}A_{p',q',r'}\left\langle \Psi,\mleft(\delta_{p,p'}^{\tau,\tau'}c_{q',\tau'}c_{r',\sigma}c_{r,\sigma}^{\ast}c_{q,\tau}^{\ast}+\delta_{q,q'}^{\tau,\tau'}c_{p',\tau'}^{\ast}c_{r',\sigma}c_{r,\sigma}^{\ast}c_{p,\tau}+\delta_{r,r'}c_{p',\tau'}^{\ast}c_{q',\tau'}c_{q,\tau}^{\ast}c_{p,\tau}\mright)\Psi\right\rangle \nonumber \\
 & -\sum_{\sigma,\tau,\tau'=1}^{s}\sum_{p,p'\in B_{F}^{c}}\sum_{q,q',r,r'\in B_{F}}\overline{A_{p,q,r}}A_{p',q',r'}\left\langle \Psi,\mleft(\delta_{r,q'}^{\sigma,\tau'}c_{p',\tau'}^{\ast}c_{r',\sigma}c_{q,\tau}^{\ast}c_{p,\tau}+\delta_{q,r'}^{\tau,\sigma}c_{p',\tau'}^{\ast}c_{q',\tau'}c_{r,\sigma}^{\ast}c_{p,\tau}\mright)\Psi\right\rangle \label{eq:HigherFermionicEstimateTerms}\\
 & -\sum_{\sigma,\tau,\tau'=1}^{s}\sum_{p,p'\in B_{F}^{c}}\sum_{q,q',r,r'\in B_{F}}\overline{A_{p,q,r}}A_{p',q',r'}\left\langle \Psi,\mleft(\delta_{q,q'}^{\tau,\tau'}\delta_{r,r'}c_{p',\tau'}^{\ast}c_{p,\tau}+\delta_{p,p'}^{\tau,\tau'}\delta_{q,q'}^{\tau,\tau'}c_{r',\sigma}c_{r,\sigma}^{\ast}+\delta_{p,p'}^{\tau,\tau'}\delta_{r,r'}c_{q',\tau'}c_{q,\tau}^{\ast}\mright)\Psi\right\rangle \nonumber \\
 & +\sum_{\sigma,\tau,\tau'=1}^{s}\sum_{p,p'\in B_{F}^{c}}\sum_{q,q',r,r'\in B_{F}}\overline{A_{p,q,r}}A_{p',q',r'}\left\langle \Psi,\mleft(\delta_{q,r'}^{\tau,\sigma}\delta_{r,q'}^{\sigma,\tau'}c_{p',\tau'}^{\ast}c_{p,\tau}+\delta_{p,p'}^{\tau,\tau'}\delta_{r,q'}^{\sigma,\tau'}c_{r',\sigma}c_{q,\tau}^{\ast}+\delta_{p,p'}^{\tau,\tau'}\delta_{q,r'}^{\tau,\sigma}c_{q',\tau'}c_{r,\sigma}^{\ast}\mright)\Psi\right\rangle \nonumber \\
 & +\sum_{\sigma,\tau,\tau'=1}^{s}\sum_{p,p'\in B_{F}^{c}}\sum_{q,q',r,r'\in B_{F}}\overline{A_{p,q,r}}A_{p',q',r'}\left\langle \Psi,\mleft(\delta_{p,p'}^{\tau,\tau'}\delta_{q,q'}^{\tau,\tau'}\delta_{r,r'}-\delta_{p,p'}^{\tau,\tau'}\delta_{q,r'}^{\tau,\sigma}\delta_{r,q'}^{\sigma,\tau'}\mright)\Psi\right\rangle .\nonumber 
\end{align}
We estimate the different types of expressions appearing above. Firstly,
by the standard fermionic estimate,
\begin{align}
 & \;\;\;\sum_{\sigma,\tau,\tau'=1}^{s}\sum_{p,p'\in B_{F}^{c}}\sum_{q,q',r,r'\in B_{F}}\overline{A_{p,q,r}}A_{p',q',r'}\left\langle \Psi,\delta_{p,p'}^{\tau,\tau'}c_{q',\tau'}c_{r',\sigma}c_{r,\sigma}^{\ast}c_{q,\tau}^{\ast}\Psi\right\rangle \nonumber \\
 & =\sum_{p\in B_{F}^{c}}^{\sigma,\tau}\left\langle \sum_{q',r'\in B_{F}}\overline{A_{p,q',r'}}c_{r',\sigma}^{\ast}c_{q',\tau}^{\ast}\Psi,\sum_{q,r\in B_{F}}\overline{A_{p,q,r}}c_{r,\sigma}^{\ast}c_{q,\tau}^{\ast}\Psi\right\rangle =\sum_{p\in B_{F}^{c}}^{\sigma,\tau}\left\Vert \sum_{q,r\in B_{F}}\overline{A_{p,q,r}}c_{r,\sigma}^{\ast}c_{q,\tau}^{\ast}\Psi\right\Vert ^{2}\\
 & \leq\sum_{p\in B_{F}^{c}}^{\sigma,\tau}\mleft(\sum_{q\in B_{F}}\left\Vert \sum_{r\in B_{F}}\overline{A_{p,q,r}}c_{r,\sigma}^{\ast}c_{q,\tau}^{\ast}\Psi\right\Vert \mright)^{2}\leq\sum_{p\in B_{F}^{c}}^{\sigma,\tau}\mleft(\sum_{q\in B_{F}}\sqrt{\sum_{r\in B_{F}}\left|A_{p,q,r}\right|^{2}}\left\Vert c_{q,\tau}^{\ast}\Psi\right\Vert \mright)^{2}\nonumber \\
 & \leq\sum_{p\in B_{F}^{c}}^{\sigma,\tau}\mleft(\sum_{q,r\in B_{F}}\left|A_{p,q,r}\right|^{2}\mright)\mleft(\sum_{q\in B_{F}}\left\Vert c_{q,\tau}^{\ast}\Psi\right\Vert ^{2}\mright)=s\sum_{p\in B_{F}^{c}}\sum_{q,r\in B_{F}}\left|A_{p,q,r}\right|^{2}\left\langle \Psi,\mathcal{N}_{E}\Psi\right\rangle \nonumber 
\end{align}
and likewise for the other two terms on the first line of equation
(\ref{eq:HigherFermionicEstimateTerms}). For the terms on the second
line we similarly estimate
\begin{align}
 & \quad\,\left|\sum_{\sigma,\tau,\tau'=1}^{s}\sum_{p,p'\in B_{F}^{c}}\sum_{q,q',r,r'\in B_{F}}\overline{A_{p,q,r}}A_{p',q',r'}\left\langle \Psi,\delta_{r,q'}^{\sigma,\tau'}c_{p',\tau'}^{\ast}c_{r',\sigma}c_{q,\tau}^{\ast}c_{p,\tau}\Psi\right\rangle \right|\nonumber \\
 & =\left|\sum_{r\in B_{F}}^{\sigma,\tau}\left\langle \sum_{p'\in B_{F}^{c}}\sum_{r'\in B_{F}}\overline{A_{p',r,r'}}c_{r',\sigma}^{\ast}c_{p',\sigma}\Psi,\sum_{p\in B_{F}^{c}}\sum_{q\in B_{F}}\overline{A_{p,q,r}}c_{q,\tau}^{\ast}c_{p,\tau}\Psi\right\rangle \right|\nonumber \\
 & \leq\sum_{r\in B_{F}}^{\sigma,\tau}\left\Vert \sum_{p'\in B_{F}^{c}}\sum_{r'\in B_{F}}\overline{A_{p',r,r'}}c_{r',\sigma}^{\ast}c_{p',\sigma}\Psi\right\Vert \left\Vert \sum_{p\in B_{F}^{c}}\sum_{q\in B_{F}}\overline{A_{p,q,r}}c_{q,\tau}^{\ast}c_{p,\tau}\Psi\right\Vert \\
 & \leq\sum_{r\in B_{F}}^{\sigma,\tau}\sum_{p,p'\in B_{F}^{c}}\sqrt{\sum_{r'\in B_{F}}\left|A_{p',r,r'}\right|^{2}}\left\Vert c_{p',\sigma}\Psi\right\Vert \sqrt{\sum_{q\in B_{F}}\left|A_{p,q,r}\right|^{2}}\left\Vert c_{p,\tau}\Psi\right\Vert \nonumber \\
 & \leq\sum_{r\in B_{F}}\sqrt{\sum_{p'\in B_{F}^{c}}^{\sigma}\sum_{r'\in B_{F}}\left|A_{p',r,r'}\right|^{2}}\sqrt{\sum_{p'\in B_{F}^{c}}^{\sigma}\left\Vert c_{p',\sigma}\Psi\right\Vert ^{2}}\sqrt{\sum_{p\in B_{F}^{c}}^{\tau}\sum_{q\in B_{F}}\left|A_{p,q,r}\right|^{2}}\sqrt{\sum_{p\in B_{F}^{c}}^{\tau}\left\Vert c_{p,\tau}\Psi\right\Vert ^{2}}\nonumber \\
 & \leq s\sum_{p\in B_{F}^{c}}\sum_{q,r\in B_{F}}\left|A_{p,q,r}\right|^{2}\left\langle \Psi,\mathcal{N}_{E}\Psi\right\rangle .\nonumber 
\end{align}
The terms on the third line of equation (\ref{eq:HigherFermionicEstimateTerms})
all factorize in a manifestly non-positive fashion, and so can be
dropped, while for the fourth line
\begin{align}
 & \quad\,\left|\sum_{\sigma,\tau,\tau'=1}^{s}\sum_{p,p'\in B_{F}^{c}}\sum_{q,q',r,r'\in B_{F}}\overline{A_{p,q,r}}A_{p',q',r'}\left\langle \Psi,\delta_{q,r'}^{\tau,\sigma}\delta_{r,q'}^{\sigma,\tau'}c_{p',\tau'}^{\ast}c_{p,\tau}\Psi\right\rangle \right|\\
 & =\left|\sum_{q,r\in B_{F}}^{\sigma}\left\langle \sum_{p'\in B_{F}^{c}}\overline{A_{p',r,q}}c_{p',\sigma}\Psi,\sum_{p\in B_{F}^{c}}\overline{A_{p,q,r}}c_{p,\sigma}\Psi\right\rangle \right|\leq\sum_{q,r\in B_{F}}^{\sigma}\left\Vert \sum_{p'\in B_{F}^{c}}\overline{A_{p',r,q}}c_{p',\sigma}\Psi\right\Vert \left\Vert \sum_{p\in B_{F}^{c}}\overline{A_{p,q,r}}c_{p,\sigma}\Psi\right\Vert \nonumber \\
 & \leq\sum_{q,r\in B_{F}}^{\sigma}\sqrt{\sum_{p\in B_{F}^{c}}\left|A_{p',r,q}\right|^{2}}\sqrt{\sum_{p\in B_{F}^{c}}\left|A_{p,q,r}\right|^{2}}\left\Vert \Psi\right\Vert ^{2}\leq s\sum_{p\in B_{F}^{c}}\sum_{q,r\in B_{F}}\left|A_{p,q,r}\right|^{2}\left\Vert \Psi\right\Vert ^{2}.\nonumber 
\end{align}
Lastly, the terms on the fifth line are seen to simply be constant
and easily bounded by $s^{2}\sum_{p\in B_{F}^{c}}\sum_{q,r\in B_{F}}\left|A_{p,q,r}\right|^{2}\left\Vert \Psi\right\Vert ^{2}$,
whence the proposition follows.

$\hfill\square$

We can now conclude the following bound for $\tilde{\mathcal{K}}$:
\begin{prop}
\label{prop:calKTildeBound}For any $\Phi,\Psi\in\mathcal{H}_{N}$
it holds that
\[
\vert\langle\Phi,\tilde{\mathcal{K}}\Psi\rangle\vert\leq\frac{\sqrt{5}}{2}\sqrt{\sum_{l\in\mathbb{Z}_{\ast}^{3}}\left\Vert K_{l}\right\Vert _{\mathrm{HS}}^{2}}\sqrt{\left\langle \Phi,\mleft(\mathcal{N}_{E}+1\mright)\Phi\right\rangle \left\langle \Psi,\mathcal{N}_{E}\Psi\right\rangle }.
\]
\end{prop}

\textbf{Proof:} By the equations (\ref{eq:calKtildeEstimate1}) and
(\ref{eq:calKtildeEstimate2}), combined with the estimate of the
previous proposition, we can estimate
\begin{align}
\vert\langle\Phi,\tilde{\mathcal{K}}\Psi\rangle\vert & \leq\frac{\sqrt{5}}{2}\sqrt{\sum_{q\in B_{F}^{c}}\sum_{p'\in B_{F}^{c}}\sum_{q',r'\in B_{F}}\left|\sum_{l\in\mathbb{Z}_{\ast}^{3}}\sum_{p\in L_{l}}\delta_{p',p}\delta_{q',p-l}\delta_{r',-q+l}1_{L_{l}}\mleft(q\mright)\left\langle K_{l}e_{q},e_{p}\right\rangle \right|^{2}}\\
 & \qquad\qquad\qquad\qquad\qquad\qquad\qquad\qquad\qquad\qquad\cdot\sqrt{\left\langle \Phi,\mleft(\mathcal{N}_{E}+1\mright)\Phi\right\rangle \left\langle \Psi,\mathcal{N}_{E}\Psi\right\rangle },\nonumber 
\end{align}
and by repeated elimination of the Kronecker $\delta$'s the sum reduces
to
\begin{align}
 & \quad\;\sum_{q\in B_{F}^{c}}\sum_{p'\in B_{F}^{c}}\sum_{q',r'\in B_{F}}\left|\sum_{l\in\mathbb{Z}_{\ast}^{3}}\sum_{p\in L_{l}}\delta_{p',p}\delta_{q',p-l}\delta_{r',-q+l}1_{L_{l}}\mleft(q\mright)\left\langle K_{l}e_{q},e_{p}\right\rangle \right|^{2}\nonumber \\
 & =\sum_{q\in B_{F}^{c}}\sum_{p'\in B_{F}^{c}}\sum_{q'\in B_{F}}\sum_{l\in\mathbb{Z}_{\ast}^{3}}\left|\sum_{p\in L_{l}}\delta_{p',p}\delta_{q',p-l}1_{L_{l}}\mleft(q\mright)\left\langle K_{l}e_{q},e_{p}\right\rangle \right|^{2}\\
 & =\sum_{q\in B_{F}^{c}}\sum_{q'\in B_{F}}\sum_{l\in\mathbb{Z}_{\ast}^{3}}\sum_{p\in L_{l}}\left|\delta_{q',p-l}1_{L_{l}}\mleft(q\mright)\left\langle K_{l}e_{q},e_{p}\right\rangle \right|^{2}=\sum_{l\in\mathbb{Z}_{\ast}^{3}}\sum_{p,q\in L_{l}}\left|\left\langle K_{l}e_{q},e_{p}\right\rangle \right|^{2}=\sum_{l\in\mathbb{Z}_{\ast}^{3}}\left\Vert K_{l}\right\Vert _{\mathrm{HS}}^{2}.\nonumber 
\end{align}
$\hfill\square$

The bound of Proposition \ref{prop:calKNumberBound} now follows by
the observation that $\left|\left\langle \Phi,\mathcal{K}\Psi\right\rangle \right|\leq\vert\langle\Phi,\tilde{\mathcal{K}}\Psi\rangle\vert+\vert\langle\Psi,\tilde{\mathcal{K}}\Phi\rangle\vert$.

\section{\label{sec:AnalysisofOne-BodyOperators}Analysis of One-Body Operators}

In this section we analyze the operators $K_{k}$, $A_{k}\mleft(t\mright)$
and $B_{k}\mleft(t\mright)$ which appeared during the diagonalization
process of Section \ref{sec:DiagonalizationoftheBosonizableTerms}.

We first consider operators of the form $e^{-2K_{k}}$ and $e^{2K_{k}}$
in detail, obtaining asymptotically optimal matrix element estimates
for these. We then extend these estimates to $K_{k}$ itself, as well
as $\sinh\mleft(-tK_{k}\mright)$ and $\cosh\mleft(-tK_{k}\mright)$ for
any $t\in\left[0,1\right]$. With these we then turn to $A_{k}\mleft(t\mright)$
and $B_{k}\mleft(t\mright)$.

We end the analysis with the integral $\int_{0}^{1}B_{k}\mleft(t\mright)\,dt$,
which will appear in the next section during our extraction of the
exchange contribution.

In all, we prove the following:
\begin{thm}
\label{them:OneBodyEstimates}It holds for any $k\in\mathbb{Z}_{\ast}^{3}$
that
\[
\mathrm{tr}\mleft(e^{-K_{k}}h_{k}e^{-K_{k}}-h_{k}-P_{k}\mright)=\frac{1}{\pi}\int_{0}^{\infty}F\mleft(\frac{s\hat{V}_{k}k_{F}^{-1}}{\mleft(2\pi\mright)^{3}}\sum_{p\in L_{k}}\frac{\lambda_{k,p}}{\lambda_{k,p}^{2}+t^{2}}\mright)dt,
\]
where $F\mleft(x\mright)=\log\mleft(1+x\mright)-x$. Furthermore, as $k_{F}\rightarrow\infty$,
\[
\left\Vert K_{k}\right\Vert _{\mathrm{HS}}\leq C\hat{V}_{k}\min\,\{1,k_{F}^{2}\left|k\right|^{-2}\}
\]
and for all $p,q\in L_{k}$ and $t\in\left[0,1\right]$
\begin{align*}
\left|\left\langle e_{p},K_{k}e_{q}\right\rangle \right| & \leq C\frac{\hat{V}_{k}k_{F}^{-1}}{\lambda_{k,p}+\lambda_{k,q}}\\
\left|\left\langle e_{p},\mleft(-K_{k}\mright)e_{q}\right\rangle -\frac{s\hat{V}_{k}k_{F}^{-1}}{2\,\mleft(2\pi\mright)^{3}}\frac{1}{\lambda_{k,p}+\lambda_{k,q}}\right| & \leq C\frac{\hat{V}_{k}^{2}k_{F}^{-1}}{\lambda_{k,p}+\lambda_{k,q}}\\
\left|\left\langle e_{p},A_{k}\mleft(t\mright)e_{q}\right\rangle \right|,\left|\left\langle e_{p},B_{k}\mleft(t\mright)e_{q}\right\rangle \right| & \leq C\mleft(1+\hat{V}_{k}^{2}\mright)\hat{V}_{k}k_{F}^{-1}\\
\left|\left\langle e_{p},\mleft(\int_{0}^{1}B_{k}\mleft(t\mright)\,dt\mright)e_{q}\right\rangle -\frac{s\hat{V}_{k}k_{F}^{-1}}{4\,\mleft(2\pi\mright)^{3}}\right| & \leq C\mleft(1+\hat{V}_{k}\mright)\hat{V}_{k}^{2}k_{F}^{-1}\\
\left|\left\langle e_{p},\left\{ K_{k},B_{k}\mleft(t\mright)\right\} e_{q}\right\rangle \right| & \leq C\mleft(1+\hat{V}_{k}^{2}\mright)\hat{V}_{k}^{2}k_{F}^{-1}
\end{align*}
for a constant $C>0$ depending only on $s$.
\end{thm}

\subsection{Matrix Element Estimates for $K$-Quantities}

To ease the notation we will abstract the problem slightly: Instead
of $\ell^{2}\mleft(L_{k}\mright)$ we consider a general $n$-dimensional
Hilbert space $\mleft(V,\left\langle \cdot,\cdot\right\rangle \mright)$,
let $h:V\rightarrow V$ be a positive self-adjoint operator on $V$
with eigenbasis $\mleft(x_{i}\mright)_{i=1}^{n}$ and eigenvalues $\mleft(\lambda_{i}\mright)_{i=1}^{n}$,
and let $v\in V$ be any vector such that $\left\langle x_{i},v\right\rangle \geq0$
for all $1\leq i\leq n$. Theorem \ref{them:OneBodyEstimates} will
then be obtained at the end by insertion of the particular operators
$h_{k}$ and $P_{k}$.

Throughout this section we will also write $P_{w}:V\rightarrow V$,
$w\in V$, to denote the operator
\begin{equation}
P_{w}\mleft(\cdot\mright)=\left\langle w,\cdot\right\rangle w.
\end{equation}
We define $K:V\rightarrow V$ by
\begin{equation}
K=-\frac{1}{2}\log\mleft(h^{-\frac{1}{2}}\mleft(h^{\frac{1}{2}}\mleft(h+2P_{v}\mright)h^{\frac{1}{2}}\mright)^{\frac{1}{2}}h^{-\frac{1}{2}}\mright)=-\frac{1}{2}\log\mleft(h^{-\frac{1}{2}}\mleft(h^{2}+2P_{h^{\frac{1}{2}}v}\mright)^{\frac{1}{2}}h^{-\frac{1}{2}}\mright).
\end{equation}
Then $e^{-2K}$ is given by
\begin{equation}
e^{-2K}=h^{-\frac{1}{2}}\mleft(h^{2}+2P_{h^{\frac{1}{2}}v}\mright)^{\frac{1}{2}}h^{-\frac{1}{2}}
\end{equation}
while $e^{2K}$ takes the form
\begin{equation}
e^{2K}=h^{\frac{1}{2}}\mleft(h^{2}+2P_{h^{\frac{1}{2}}v}\mright)^{-\frac{1}{2}}h^{\frac{1}{2}}=h^{\frac{1}{2}}\mleft(\mleft(h^{2}+2P_{h^{\frac{1}{2}}v}\mright)^{-1}\mright)^{\frac{1}{2}}h^{\frac{1}{2}}.
\end{equation}
We can rewrite the inverse of $h^{2}+2P_{h^{\frac{1}{2}}v}$ using
the Sherman-Morrison formula:
\begin{lem}[The Sherman-Morrison Formula]
\label{lemma:ShermanMorrison}Let $A:V\rightarrow V$ be an invertible
self-adjoint operator. Then for any $w\in V$ and $g\in\mathbb{C}$,
the operator $A+gP_{w}$ is invertible if and only if $\left\langle w,A^{-1}w\right\rangle \neq-g^{-1}$,
with inverse
\[
\mleft(A+gP_{w}\mright)^{-1}=A^{-1}-\frac{g}{1+g\left\langle w,A^{-1}w\right\rangle }P_{A^{-1}w}.
\]
\end{lem}

Applying the Sherman-Morrison formula with $A=h^{2}$, $w=h^{\frac{1}{2}}v$
and $g=2$ we obtain
\begin{equation}
\mleft(h^{2}+2P_{h^{\frac{1}{2}}v}\mright)^{-1}=h^{-2}-\frac{2}{1+2\left\langle v,h^{-1}v\right\rangle }P_{h^{-\frac{3}{2}}v}
\end{equation}
so $e^{-2K}$ and $e^{2K}$ are given by
\begin{align}
e^{-2K} & =h^{-\frac{1}{2}}\mleft(h^{2}+2P_{h^{\frac{1}{2}}v}\mright)^{\frac{1}{2}}h^{-\frac{1}{2}}\label{eq:e-2Ke2KIdentities}\\
e^{2K} & =h^{\frac{1}{2}}\mleft(h^{-2}-\frac{2}{1+2\left\langle v,h^{-1}v\right\rangle }P_{h^{-\frac{3}{2}}v}\mright)^{\frac{1}{2}}h^{\frac{1}{2}}.\nonumber 
\end{align}
To proceed further we apply the following integral representation
of the square root of a one-dimensional perturbation, first presented
in \cite{BenNamPorSchSei-20}:
\begin{prop}
\label{prop:SquareRootofAOneDimensionalPerturbation}Let $A:V\rightarrow V$
be a positive self-adjoint operator. Then for any $w\in V$ and $g\in\mathbb{R}$
such that $A+gP_{w}>0$ it holds that
\[
\mleft(A+gP_{w}\mright)^{\frac{1}{2}}=A^{\frac{1}{2}}+\frac{2g}{\pi}\int_{0}^{\infty}\frac{t^{2}}{1+g\left\langle w,\mleft(A+t^{2}\mright)^{-1}w\right\rangle }P_{\mleft(A+t^{2}\mright)^{-1}w}dt
\]
and
\[
\mathrm{tr}\mleft(\mleft(A+gP_{w}\mright)^{\frac{1}{2}}\mright)=\mathrm{tr}\mleft(A^{\frac{1}{2}}\mright)+\frac{1}{\pi}\int_{0}^{\infty}\log\mleft(1+g\left\langle w,\mleft(A+t^{2}\mright)^{-1}w\right\rangle \mright)dt.
\]
\end{prop}

We have included a proof of this in appendix section \ref{subsec:TheSquareRootofaRankOnePerturbtation}.

By the trace formula we can immediately deduce the following identity:
\begin{prop}
\label{prop:GeneralBosonizableContribution}It holds that
\[
\mathrm{tr}\mleft(e^{-K}he^{-K}-h-P_{v}\mright)=\frac{1}{\pi}\int_{0}^{\infty}F\mleft(2\left\langle v,h\mleft(h^{2}+t^{2}\mright)^{-1}v\right\rangle \mright)dt
\]
where $F\mleft(x\mright)=\log\mleft(1+x\mright)-x.$
\end{prop}

\textbf{Proof:} By cyclicity of the trace and the previous proposition
\begin{align}
\mathrm{tr}\mleft(e^{-K}he^{-K}-h-P_{v}\mright) & =\mathrm{tr}\mleft(h^{\frac{1}{2}}e^{-2K}h^{\frac{1}{2}}\mright)-\mathrm{tr}\mleft(h\mright)-\mathrm{tr}\mleft(P_{v}\mright)=\mathrm{tr}\mleft(\mleft(h^{2}+2P_{h^{\frac{1}{2}}v}\mright)^{\frac{1}{2}}\mright)-\mathrm{tr}\mleft(h\mright)-\left\Vert v\right\Vert ^{2}\nonumber \\
 & =\frac{1}{\pi}\int_{0}^{\infty}\log\mleft(1+2\left\langle v,h\mleft(h^{2}+t^{2}\mright)^{-1}v\right\rangle \mright)\,dt-\left\Vert v\right\Vert ^{2},
\end{align}
so noting that the integral identity $\int_{0}^{\infty}\frac{a}{a^{2}+t^{2}}dt=\frac{\pi}{2}$,
$a>0$, implies that
\begin{equation}
\frac{1}{\pi}\int_{0}^{\infty}2\left\langle v,h\mleft(h^{2}+t^{2}\mright)^{-1}v\right\rangle dt=\frac{2}{\pi}\sum_{i=1}^{n}\left|\left\langle e_{i},v\right\rangle \right|^{2}\int_{0}^{\infty}\frac{\lambda_{i}}{\lambda_{i}^{2}+t^{2}}dt=\sum_{i=1}^{n}\left|\left\langle e_{i},v\right\rangle \right|^{2}=\left\Vert v\right\Vert ^{2}
\end{equation}
we can absorb the term $-\left\Vert v\right\Vert ^{2}$ into the integral
for the claim.

$\hfill\square$

\subsubsection*{Estimation of $e^{-2K}$ and $e^{2K}$}

Using the square-root formula we now derive elementwise estimates
for $e^{-2K}$ and $e^{2K}$:
\begin{prop}
\label{prop:e-2Ke2KElementEstimates}For all $1\leq i,j\leq n$ it
holds that
\[
\frac{2}{1+2\left\langle v,h^{-1}v\right\rangle }\frac{\left\langle x_{i},v\right\rangle \left\langle v,x_{j}\right\rangle }{\lambda_{i}+\lambda_{j}}\leq\left\langle x_{i},\mleft(e^{-2K}-1\mright)x_{j}\right\rangle ,\left\langle x_{i},\mleft(1-e^{2K}\mright)x_{j}\right\rangle \leq2\frac{\left\langle x_{i},v\right\rangle \left\langle v,x_{j}\right\rangle }{\lambda_{i}+\lambda_{j}}.
\]
\end{prop}

\textbf{Proof:} From the first equality of equation (\ref{eq:e-2Ke2KIdentities})
we can apply the identity of Proposition \ref{prop:SquareRootofAOneDimensionalPerturbation}
with $A=h^{2}$, $w=h^{\frac{1}{2}}v$ and $g=2$ to see that
\begin{align}
e^{-2K} & =h^{-\frac{1}{2}}\mleft(h+\frac{4}{\pi}\int_{0}^{\infty}\frac{t^{2}}{1+2\left\langle h^{\frac{1}{2}}v,\mleft(h^{2}+t^{2}\mright)^{-1}h^{\frac{1}{2}}v\right\rangle }P_{\mleft(h^{2}+t^{2}\mright)^{-1}h^{\frac{1}{2}}v}dt\mright)h^{-\frac{1}{2}}\\
 & =1+\frac{4}{\pi}\int_{0}^{\infty}\frac{t^{2}}{1+2\left\langle v,h\mleft(h^{2}+t^{2}\mright)^{-1}v\right\rangle }P_{\mleft(h^{2}+t^{2}\mright)^{-1}v}dt\nonumber 
\end{align}
whence for any $1\leq i,j\leq n$
\begin{align}
\left\langle x_{i},\mleft(e^{-2K}-1\mright)x_{j}\right\rangle  & =\frac{4}{\pi}\int_{0}^{\infty}\frac{t^{2}}{1+2\left\langle v,h\mleft(h^{2}+t^{2}\mright)^{-1}v\right\rangle }\frac{\left\langle x_{i},v\right\rangle }{\lambda_{i}^{2}+t^{2}}\frac{\left\langle v,x_{j}\right\rangle }{\lambda_{j}^{2}+t^{2}}dt\\
 & =\frac{4}{\pi}\left\langle x_{i},v\right\rangle \left\langle v,x_{j}\right\rangle \int_{0}^{\infty}\frac{1}{1+2\left\langle v,h\mleft(h^{2}+t^{2}\mright)^{-1}v\right\rangle }\frac{t}{\lambda_{i}^{2}+t^{2}}\frac{t}{\lambda_{j}^{2}+t^{2}}dt.\nonumber 
\end{align}
Noting that
\begin{equation}
\frac{1}{1+2\left\langle v,h^{-1}v\right\rangle }\leq\frac{1}{1+2\left\langle v,h\mleft(h^{2}+t^{2}\mright)^{-1}v\right\rangle }\leq1,\quad t\geq0,
\end{equation}
and recalling that $\left\langle x_{i},v\right\rangle \geq0$ by assumption,
we conclude
\[
\frac{4}{\pi}\frac{\left\langle x_{i},v\right\rangle \left\langle v,x_{j}\right\rangle }{1+2\left\langle v,h^{-1}v\right\rangle }\int_{0}^{\infty}\frac{t}{\lambda_{i}^{2}+t^{2}}\frac{t}{\lambda_{j}^{2}+t^{2}}dt\leq\left\langle x_{i},\mleft(e^{-2K}-1\mright)x_{j}\right\rangle \leq\frac{4}{\pi}\left\langle x_{i},v\right\rangle \left\langle v,x_{j}\right\rangle \int_{0}^{\infty}\frac{t}{\lambda_{i}^{2}+t^{2}}\frac{t}{\lambda_{j}^{2}+t^{2}}dt
\]
from which the claim follows by an application of the integral identity
\begin{equation}
\int_{0}^{\infty}\frac{t}{a^{2}+t^{2}}\frac{t}{b^{2}+t^{2}}\,dt=\frac{\pi}{2}\frac{1}{a+b},\quad a,b>0.
\end{equation}
Similarly, for $e^{2K}$, we have by equation (\ref{eq:e-2Ke2KIdentities})
that applying Proposition \ref{prop:SquareRootofAOneDimensionalPerturbation}
with $A=h^{-2}$, $w=h^{-\frac{3}{2}}v$ and $g=-2\mleft(1+2\left\langle v,h^{-1}v\right\rangle \mright)^{-1}$
yields
\begin{align}
e^{2K} & =h^{\frac{1}{2}}\mleft(h^{-1}-\frac{4}{\pi}\int_{0}^{\infty}\frac{t^{2}}{1+2\left\langle v,h^{-1}v\right\rangle -2\left\langle h^{-\frac{3}{2}}v,\mleft(h^{-2}+t^{2}\mright)^{-1}h^{-\frac{3}{2}}v\right\rangle }P_{\mleft(h^{-2}+t^{2}\mright)^{-1}h^{-\frac{3}{2}}v}dt\mright)h^{\frac{1}{2}}\nonumber \\
 & =1-\frac{4}{\pi}\int_{0}^{\infty}\frac{t^{2}}{1+2\left\langle v,h^{-1}\mleft(h^{-2}+t^{2}\mright)^{-1}v\right\rangle t^{2}}P_{\mleft(h^{-2}+t^{2}\mright)^{-1}h^{-1}v}dt
\end{align}
from which the claimed inequality follows as before by the observation
that
\begin{equation}
\frac{1}{1+2\left\langle v,h^{-1}v\right\rangle }\leq\frac{1}{1+2\left\langle v,h^{-1}\mleft(h^{-2}+t^{2}\mright)^{-1}v\right\rangle t^{2}}\leq1,\quad t\geq0,
\end{equation}
as well as the integral identity
\begin{equation}
\frac{1}{ab}\int_{0}^{\infty}\frac{t}{a^{-2}+t^{2}}\frac{t}{b^{-2}+t^{2}}\,dt=\frac{\pi}{2}\frac{1}{a+b},\quad a,b>0.
\end{equation}
$\hfill\square$

Note that these estimates are asymptotically optimal, in the sense
that the left-hand side reduces to the right-hand side as $\left\langle v,h^{-1}v\right\rangle \rightarrow0$.
In our case we will see that $\left\langle v_{k},h_{k}^{-1}v_{k}\right\rangle \sim\hat{V}_{k}$,
so this amounts to optimal estimates for ``small'' $\hat{V}_{k}$.

Below it will be more convenient to consider the hyperbolic functions
$\sinh\mleft(-2K\mright)$ and $\cosh\mleft(-2K\mright)$ rather than
$e^{-2K}$ and $e^{2K}$. The previous proposition implies the following
for these operators:
\begin{cor}
\label{coro:HyperbolicBounds}For any $1\leq i,j\leq n$ it holds
that
\begin{align*}
\left\langle x_{i},\sinh\mleft(-2K\mright)x_{j}\right\rangle  & \leq2\frac{\left\langle x_{i},v\right\rangle \left\langle v,x_{j}\right\rangle }{\lambda_{i}+\lambda_{j}}\\
\left\langle x_{i},\mleft(\cosh\mleft(-2K\mright)-1\mright)x_{j}\right\rangle  & \leq\frac{2\left\langle v,h^{-1}v\right\rangle }{1+2\left\langle v,h^{-1}v\right\rangle }\frac{\left\langle x_{i},v\right\rangle \left\langle v,x_{j}\right\rangle }{\lambda_{i}+\lambda_{j}}.
\end{align*}
\end{cor}

\textbf{Proof:} As $\sinh\mleft(-2K\mright)=\frac{1}{2}\mleft(\mleft(e^{-2K}-1\mright)+\mleft(1-e^{2K}\mright)\mright)$
we can bound
\begin{equation}
\left\langle x_{i},\sinh\mleft(-2K\mright)x_{j}\right\rangle =\frac{1}{2}\mleft(\left\langle x_{i},\mleft(e^{-2K}-1\mright)x_{j}\right\rangle +\left\langle x_{i},\mleft(1-e^{2K}\mright)x_{j}\right\rangle \mright)\leq2\frac{\left\langle x_{i},v\right\rangle \left\langle v,x_{j}\right\rangle }{\lambda_{i}+\lambda_{j}}
\end{equation}
and as similarly $\cosh\mleft(-2K\mright)-1=\frac{1}{2}\mleft(\mleft(e^{-2K}-1\mright)-\mleft(1-e^{2K}\mright)\mright)$
also
\begin{align}
\left\langle x_{i},\mleft(\cosh\mleft(-2K\mright)-1\mright)x_{j}\right\rangle  & =\frac{1}{2}\mleft(\left\langle x_{i},\mleft(e^{-2K}-1\mright)x_{j}\right\rangle -\left\langle x_{i},\mleft(1-e^{2K}\mright)x_{j}\right\rangle \mright)\nonumber \\
 & \leq\frac{1}{2}\mleft(2\frac{\left\langle x_{i},v\right\rangle \left\langle v,x_{j}\right\rangle }{\lambda_{i}+\lambda_{j}}-\frac{2}{1+2\left\langle v,h^{-1}v\right\rangle }\frac{\left\langle x_{i},v\right\rangle \left\langle v,x_{j}\right\rangle }{\lambda_{i}+\lambda_{j}}\mright)\\
 & =\frac{2\left\langle v,h^{-1}v\right\rangle }{1+2\left\langle v,h^{-1}v\right\rangle }\frac{\left\langle x_{i},v\right\rangle \left\langle v,x_{j}\right\rangle }{\lambda_{i}+\lambda_{j}}.\nonumber 
\end{align}
$\hfill\square$

\subsubsection*{General Estimates}

Now we extend our elementwise estimates to more general operators.
First we consider $K$ itself:
\begin{prop}
\label{prop:KElementEstimates}For any $1\leq i,j\leq n$ it holds
that
\[
\frac{1}{1+2\left\langle v,h^{-1}v\right\rangle }\frac{\left\langle x_{i},v\right\rangle \left\langle v,x_{j}\right\rangle }{\lambda_{i}+\lambda_{j}}\leq\left\langle x_{i},\mleft(-K\mright)x_{j}\right\rangle \leq\frac{\left\langle x_{i},v\right\rangle \left\langle v,x_{j}\right\rangle }{\lambda_{i}+\lambda_{j}}.
\]
\end{prop}

\textbf{Proof:} As $K=-\frac{1}{2}\log\mleft(h^{-\frac{1}{2}}\mleft(h^{2}+2P_{h^{\frac{1}{2}}v}\mright)^{\frac{1}{2}}h^{-\frac{1}{2}}\mright)$
and
\begin{equation}
h^{-\frac{1}{2}}\mleft(h^{2}+2P_{h^{\frac{1}{2}}v}\mright)^{\frac{1}{2}}h^{-\frac{1}{2}}\geq h^{-\frac{1}{2}}hh^{-\frac{1}{2}}=1
\end{equation}
we see that $K\leq0$. From the identity
\begin{equation}
-x=\frac{1}{2}\sum_{m=1}^{\infty}\frac{1}{m}\mleft(1-e^{2x}\mright)^{m},\quad x\leq0,
\end{equation}
which follows by the Mercator series, we thus have that $-K=\frac{1}{2}\sum_{m=1}^{\infty}\frac{1}{m}\mleft(1-e^{2K}\mright)^{m}$.
Noting that Proposition \ref{prop:e-2Ke2KElementEstimates} in particular
implies that $\left\langle x_{i},\mleft(1-e^{2K}\mright)x_{j}\right\rangle \geq0$
for all $1\leq i,j\leq n$, whence also $\left\langle x_{i},\mleft(1-e^{2K}\mright)^{m}x_{j}\right\rangle \geq0$
for any $m\in\mathbb{N}$, we may estimate
\begin{align}
\left\langle x_{i},\mleft(-K\mright)x_{j}\right\rangle  & =\frac{1}{2}\sum_{m=1}^{\infty}\frac{1}{m}\left\langle x_{i},\mleft(1-e^{2K}\mright)^{m}x_{j}\right\rangle \geq\frac{1}{2}\left\langle x_{i},\mleft(1-e^{2K}\mright)x_{j}\right\rangle \\
 & \geq\frac{1}{1+2\left\langle v,h^{-1}v\right\rangle }\frac{\left\langle x_{i},v\right\rangle \left\langle v,x_{j}\right\rangle }{\lambda_{i}+\lambda_{j}}\nonumber 
\end{align}
which is the lower bound. This similarly implies that $\left\langle x_{i},\mleft(-K\mright)^{m}x_{j}\right\rangle \geq0$
for all $1\leq i,j\leq n$, $m\in\mathbb{N}$, so the upper bound
now also follows from Proposition \ref{prop:e-2Ke2KElementEstimates}
by noting that
\begin{equation}
\frac{\left\langle x_{i},v\right\rangle \left\langle v,x_{j}\right\rangle }{\lambda_{i}+\lambda_{j}}\geq\frac{1}{2}\left\langle x_{i},\mleft(e^{-2K}-1\mright)x_{j}\right\rangle =\frac{1}{2}\sum_{m=1}^{\infty}\frac{1}{m!}\left\langle x_{i},\mleft(-2K\mright)^{m}x_{j}\right\rangle \geq\left\langle x_{i},\mleft(-K\mright)x_{j}\right\rangle .
\end{equation}
$\hfill\square$

The fact that $\left\langle x_{i},\mleft(-K\mright)^{m}x_{j}\right\rangle \geq0$
for all $1\leq i,j\leq n$, $m\in\mathbb{N}$, has the important consequence
that for any such $i$ and $j$, the functions
\begin{equation}
t\mapsto\left\langle x_{i},\sinh\mleft(-tK\mright)x_{j}\right\rangle ,\,\left\langle x_{i},\mleft(\sinh\mleft(-tK\mright)+tK\mright)x_{j}\right\rangle ,\,\left\langle x_{i},\mleft(\cosh\mleft(-tK\mright)-1\mright)x_{j}\right\rangle 
\end{equation}
are non-negative and convex for $t\in\mleft[0,\infty\mright)$, as follows
by considering the Taylor expansions of the operators involved. This
allows us to extend the bounds of Corollary \ref{coro:HyperbolicBounds}
to arbitrary $t\in\left[0,1\right]$:
\begin{prop}
\label{prop:tDependentElementEstimates}For all $1\leq i,j\leq n$
and $t\in\left[0,1\right]$ it holds that
\begin{align*}
\frac{1}{1+2\left\langle v,h^{-1}v\right\rangle }\frac{\left\langle x_{i},v\right\rangle \left\langle v,x_{j}\right\rangle }{\lambda_{i}+\lambda_{j}}t\leq\left\langle x_{i},\sinh\mleft(-tK\mright)x_{j}\right\rangle  & \leq\frac{\left\langle x_{i},v\right\rangle \left\langle v,x_{j}\right\rangle }{\lambda_{i}+\lambda_{j}}t\\
0\leq\left\langle x_{i},\mleft(\cosh\mleft(-tK\mright)-1\mright)x_{j}\right\rangle  & \leq\frac{\left\langle v,h^{-1}v\right\rangle }{1+2\left\langle v,h^{-1}v\right\rangle }\frac{\left\langle x_{i},v\right\rangle \left\langle v,x_{j}\right\rangle }{\lambda_{i}+\lambda_{j}}\\
\left|\left\langle x_{i},\mleft(e^{tK}-1\mright)x_{j}\right\rangle \right| & \leq\frac{\left\langle x_{i},v\right\rangle \left\langle v,x_{j}\right\rangle }{\lambda_{i}+\lambda_{j}}.
\end{align*}
\end{prop}

\textbf{Proof:} By the noted convexity we immediately conclude the
upper bounds
\begin{align}
\left\langle x_{i},\sinh\mleft(-tK\mright)x_{j}\right\rangle  & \leq\frac{t}{2}\left\langle x_{i},\sinh\mleft(-2K\mright)x_{j}\right\rangle \leq\frac{\left\langle x_{i},v\right\rangle \left\langle v,x_{j}\right\rangle }{\lambda_{i}+\lambda_{j}}t\\
\left\langle x_{i},\mleft(\cosh\mleft(-tK\mright)-1\mright)x_{j}\right\rangle  & \leq\frac{t}{2}\left\langle x_{i},\mleft(\cosh\mleft(-2K\mright)-1\mright)x_{j}\right\rangle \leq\frac{\left\langle v,h^{-1}v\right\rangle }{1+2\left\langle v,h^{-1}v\right\rangle }\frac{\left\langle x_{i},v\right\rangle \left\langle v,x_{j}\right\rangle }{\lambda_{i}+\lambda_{j}}\nonumber 
\end{align}
and by non-negativity of $\left\langle x_{i},\mleft(\sinh\mleft(-tK\mright)+tK\mright)x_{j}\right\rangle $
and Proposition \ref{prop:KElementEstimates}, the lower bound
\begin{equation}
\left\langle x_{i},\sinh\mleft(-tK\mright)x_{j}\right\rangle \geq\left\langle x_{i},\mleft(-tK\mright)x_{j}\right\rangle \geq\frac{1}{1+2\left\langle v,h^{-1}v\right\rangle }\frac{\left\langle x_{i},v\right\rangle \left\langle v,x_{j}\right\rangle }{\lambda_{i}+\lambda_{j}}t.
\end{equation}
Lastly we can apply the non-negativity of the hyperbolic operators
to conclude the bound for $e^{tK}-1$ as
\begin{align}
\left|\left\langle x_{i},\mleft(e^{tK}-1\mright)x_{j}\right\rangle \right| & =\left|\left\langle x_{i},\mleft(\mleft(\cosh\mleft(-tK\mright)-1\mright)-\sinh\mleft(-tK\mright)\mright)x_{j}\right\rangle \right|\\
 & \leq\max\left\{ \left\langle x_{i},\mleft(\cosh\mleft(-tK\mright)-1\mright)x_{j}\right\rangle ,\left\langle x_{i},\sinh\mleft(-tK\mright)x_{j}\right\rangle \right\} \leq\frac{\left\langle x_{i},v\right\rangle \left\langle v,x_{j}\right\rangle }{\lambda_{i}+\lambda_{j}}.\nonumber 
\end{align}
$\hfill\square$

\subsection{Matrix Element Estimates for $A\mleft(t\mright)$ and $B\mleft(t\mright)$}

We now consider operators $A\mleft(t\mright),B\mleft(t\mright):V\rightarrow V$
defined by
\begin{align}
A\mleft(t\mright) & =\frac{1}{2}\mleft(e^{tK}\mleft(h+2P_{v}\mright)e^{tK}+e^{-tK}he^{-tK}\mright)-h\\
B\mleft(t\mright) & =\frac{1}{2}\mleft(e^{tK}\mleft(h+2P_{v}\mright)e^{tK}-e^{-tK}he^{-tK}\mright)\nonumber 
\end{align}
for $t\in\left[0,1\right]$. We decompose these as
\begin{align}
A\mleft(t\mright) & =A_{h}\mleft(t\mright)+e^{tK}P_{v}e^{tK}\\
B\mleft(t\mright) & =\mleft(1-t\mright)P_{v}+B_{h}\mleft(t\mright)+e^{tK}P_{v}e^{tK}-P_{v}\nonumber 
\end{align}
where, with
\begin{equation}
C_{K}\mleft(t\mright)=\cosh\mleft(-tK\mright)-1\quad\text{and}\quad S_{K}\mleft(t\mright)=\sinh\mleft(-tK\mright),
\end{equation}
the operators $A_{h}\mleft(t\mright)$ and $B_{h}\mleft(t\mright)$ are
given by
\begin{align}
A_{h}\mleft(t\mright) & =\cosh\mleft(-tK\mright)\,h\cosh\mleft(-tK\mright)+\sinh\mleft(-tK\mright)\,h\sinh\mleft(-tK\mright)-h\\
 & =\left\{ h,C_{K}\mleft(t\mright)\right\} +S_{K}\mleft(t\mright)\,h\,S_{K}\mleft(t\mright)+C_{K}\mleft(t\mright)\,h\,C_{K}\mleft(t\mright)\nonumber 
\end{align}
and
\begin{align}
B_{h}\mleft(t\mright) & =-\sinh\mleft(-tK\mright)\,h\cosh\mleft(-tK\mright)-\cosh\mleft(-tK\mright)\,h\sinh\mleft(-tK\mright)+tP_{v}\\
 & =tP_{v}-\left\{ h,S_{K}\mleft(t\mright)\right\} -S_{K}\mleft(t\mright)\,h\,C_{K}\mleft(t\mright)-C_{K}\mleft(t\mright)\,h\,S_{K}\mleft(t\mright).\nonumber 
\end{align}
We begin by estimating the $e^{tK}P_{v}e^{tK}$ terms:
\begin{prop}
\label{prop:etKPetK-PEstimate}For all $1\leq i,j\leq n$ and $t\in\left[0,1\right]$
it holds that
\[
\left|\left\langle x_{i},\mleft(e^{tK}P_{v}e^{tK}-P_{v}\mright)x_{j}\right\rangle \right|\leq\mleft(2+\left\langle v,h^{-1}v\right\rangle \mright)\left\langle v,h^{-1}v\right\rangle \left\langle x_{i},v\right\rangle \left\langle v,x_{j}\right\rangle .
\]
\end{prop}

\textbf{Proof:} Writing
\begin{equation}
e^{tK}P_{v}e^{tK}-P_{v}=\left\{ P_{v},e^{tK}-1\right\} +\mleft(e^{tK}-1\mright)P_{v}\mleft(e^{tK}-1\mright)
\end{equation}
we see that
\begin{align}
\left\langle x_{i},\mleft(e^{tK}P_{v}e^{tK}-P_{v}\mright)x_{j}\right\rangle  & =\left\langle x_{i},v\right\rangle \left\langle \mleft(e^{tK}-1\mright)v,x_{j}\right\rangle +\left\langle x_{i},\mleft(e^{tK}-1\mright)v\right\rangle \left\langle v,x_{j}\right\rangle \\
 & +\left\langle x_{i},\mleft(e^{tK}-1\mright)v\right\rangle \left\langle \mleft(e^{tK}-1\mright)v,x_{j}\right\rangle .\nonumber 
\end{align}
Now, by Proposition \ref{prop:tDependentElementEstimates} we can
for any $1\leq i\leq n$ estimate
\begin{align}
\left|\left\langle x_{i},\mleft(e^{tK}-1\mright)v\right\rangle \right| & =\left|\sum_{j=1}^{n}\left\langle x_{i},\mleft(e^{tK}-1\mright)x_{j}\right\rangle \left\langle x_{j},v\right\rangle \right|\leq\sum_{j=1}^{n}\frac{\left\langle x_{i},v\right\rangle \left\langle v,x_{j}\right\rangle }{\lambda_{i}+\lambda_{j}}\left\langle x_{j},v\right\rangle \\
 & \leq\left\langle x_{i},v\right\rangle \sum_{j=1}^{n}\frac{\left|\left\langle x_{j},v\right\rangle \right|^{2}}{\lambda_{j}}=\left\langle x_{i},v\right\rangle \left\langle v,h^{-1}v\right\rangle \nonumber 
\end{align}
whence the claim follows.

$\hfill\square$

Note that for $\left\langle x_{i},e^{tK}P_{v}e^{tK}x_{j}\right\rangle $
this in particular implies the bound
\begin{equation}
\left|\left\langle x_{i},e^{tK}P_{v}e^{tK}x_{j}\right\rangle \right|\leq\mleft(1+\left\langle v,h^{-1}v\right\rangle \mright)^{2}\left\langle x_{i},v\right\rangle \left\langle v,x_{j}\right\rangle .\label{eq:etKPetKEstimate}
\end{equation}
We now consider $A_{h}\mleft(t\mright)$ and $B_{h}\mleft(t\mright)$:
\begin{prop}
\label{prop:AhBhEstimate}For all $1\leq i,j\leq n$ and $t\in\left[0,1\right]$
it holds that
\[
\left|\left\langle x_{i},A_{h}\mleft(t\mright)x_{j}\right\rangle \right|,\left|\left\langle x_{i},B_{h}\mleft(t\mright)x_{j}\right\rangle \right|\leq4\left\langle v,h^{-1}v\right\rangle \left\langle x_{i},v\right\rangle \left\langle v,x_{j}\right\rangle .
\]
\end{prop}

\textbf{Proof:} The estimates of Proposition \ref{prop:tDependentElementEstimates}
imply that
\begin{align}
\left|\left\langle x_{i},\left\{ h,C_{K}\mleft(t\mright)\right\} x_{j}\right\rangle \right| & =\mleft(\lambda_{i}+\lambda_{j}\mright)\left|\left\langle x_{i},C_{K}\mleft(t\mright)x_{j}\right\rangle \right|\leq\mleft(\lambda_{i}+\lambda_{j}\mright)\frac{\left\langle v,h^{-1}v\right\rangle }{1+2\left\langle v,h^{-1}v\right\rangle }\frac{\left\langle x_{i},v\right\rangle \left\langle v,x_{j}\right\rangle }{\lambda_{i}+\lambda_{j}}\nonumber \\
 & \leq\left\langle v,h^{-1}v\right\rangle \left\langle x_{i},v\right\rangle \left\langle v,x_{j}\right\rangle 
\end{align}
and
\begin{align}
\left|\left\langle x_{i},S_{K}\mleft(t\mright)\,h\,S_{K}\mleft(t\mright)x_{j}\right\rangle \right| & =\left|\sum_{k=1}^{n}\lambda_{k}\left\langle x_{i},S_{K}\mleft(t\mright)x_{k}\right\rangle \left\langle x_{k},S_{K}\mleft(t\mright)x_{j}\right\rangle \right|\leq\sum_{k=1}^{n}\lambda_{k}\frac{\left\langle x_{i},v\right\rangle \left\langle v,x_{k}\right\rangle }{\lambda_{i}+\lambda_{k}}\frac{\left\langle x_{k},v\right\rangle \left\langle v,x_{j}\right\rangle }{\lambda_{k}+\lambda_{j}}\nonumber \\
 & \leq\left\langle x_{i},v\right\rangle \left\langle v,x_{j}\right\rangle \sum_{k=1}^{n}\frac{\left|\left\langle x_{k},v\right\rangle \right|^{2}}{\lambda_{k}}=\left\langle v,h^{-1}v\right\rangle \left\langle x_{i},v\right\rangle \left\langle v,x_{j}\right\rangle .
\end{align}
The latter estimate only relied on the inequality
\begin{equation}
\left|\left\langle x_{i},S_{K}\mleft(t\mright)x_{j}\right\rangle \right|\leq\frac{\left\langle x_{i},v\right\rangle \left\langle v,x_{j}\right\rangle }{\lambda_{i}+\lambda_{j}},
\end{equation}
which is also true for $C_{K}\mleft(t\mright)$, so the terms
\begin{equation}
C_{K}\mleft(t\mright)\,h\,C_{K}\mleft(t\mright),\quad C_{K}\mleft(t\mright)\,h\,S_{K}\mleft(t\mright)\quad\text{and}\quad S_{K}\mleft(t\mright)\,h\,C_{K}\mleft(t\mright)
\end{equation}
also obey this estimate. It thus only remains to bound $tP_{v}-\left\{ h,S_{K}\mleft(t\mright)\right\} $.
From Proposition \ref{prop:tDependentElementEstimates} we see that
\begin{equation}
\frac{\left\langle x_{i},v\right\rangle \left\langle v,x_{j}\right\rangle }{1+2\left\langle v,h^{-1}v\right\rangle }t\leq\left\langle x_{i},\left\{ h,S_{K}\mleft(t\mright)\right\} x_{j}\right\rangle \leq\left\langle x_{i},v\right\rangle \left\langle v,x_{j}\right\rangle t
\end{equation}
whence
\begin{align}
 & \quad\;\,\left|\left\langle x_{i},\mleft(tP_{v}-\left\{ h,S_{K}\mleft(t\mright)\right\} \mright)x_{j}\right\rangle \right|=\left\langle x_{i},P_{v}x_{j}\right\rangle t-\left\langle x_{i},\left\{ h,S_{K}\mleft(t\mright)\right\} x_{j}\right\rangle \nonumber \\
 & \leq\mleft(1-\frac{1}{1+2\left\langle v,h^{-1}v\right\rangle }\mright)\left\langle x_{i},v\right\rangle \left\langle v,x_{j}\right\rangle t=\frac{2\left\langle v,h^{-1}v\right\rangle }{1+2\left\langle v,h^{-1}v\right\rangle }\left\langle x_{i},v\right\rangle \left\langle v,x_{j}\right\rangle t\\
 & \leq2\left\langle v,h^{-1}v\right\rangle \left\langle x_{i},v\right\rangle \left\langle v,x_{j}\right\rangle .\nonumber 
\end{align}
$\hfill\square$

Combining equation (\ref{eq:etKPetKEstimate}) and Proposition \ref{prop:AhBhEstimate}
we conclude the following:
\begin{prop}
\label{prop:AtBtEstimates}For all $1\leq i,j\leq n$ and $t\in\left[0,1\right]$
it holds that
\[
\left|\left\langle x_{i},A\mleft(t\mright)x_{j}\right\rangle \right|,\,\left|\left\langle x_{i},B\mleft(t\mright)x_{j}\right\rangle \right|\leq3\mleft(1+\left\langle v,h^{-1}v\right\rangle \mright)^{2}\left\langle x_{i},v\right\rangle \left\langle v,x_{j}\right\rangle .
\]
\end{prop}

\subsubsection*{Analysis of $\left\{ K,B\mleft(t\mright)\right\} $ and $\int_{0}^{1}B\mleft(t\mright)\,dt$}

We end by estimating $\left\{ K,B\mleft(t\mright)\right\} $ and $\int_{0}^{1}B\mleft(t\mright)\,dt$,
the latter of which will be needed for the analysis of the exchange
contribution in the next section.

First is $\left\{ K,B\mleft(t\mright)\right\} $:
\begin{prop}
\label{prop:KBtEstimate}For all $1\leq i,j\leq n$ and $t\in\left[0,1\right]$
it holds that
\[
\left|\left\langle x_{i},\left\{ K,B\mleft(t\mright)\right\} x_{j}\right\rangle \right|\leq6\mleft(1+\left\langle v,h^{-1}v\right\rangle \mright)^{2}\left\langle v,h^{-1}v\right\rangle \left\langle x_{i},v\right\rangle \left\langle v,x_{j}\right\rangle .
\]
\end{prop}

\textbf{Proof:} Using the Propositions \ref{prop:KElementEstimates}
and \ref{prop:AtBtEstimates} we see that
\begin{align}
\left|\left\langle x_{i},KB\mleft(t\mright)x_{j}\right\rangle \right| & =\left|\sum_{k=1}^{n}\left\langle x_{i},Kx_{k}\right\rangle \left\langle x_{k},B\mleft(t\mright)x_{j}\right\rangle \right|\leq3\mleft(1+\left\langle v,h^{-1}v\right\rangle \mright)^{2}\sum_{k=1}^{n}\frac{\left\langle x_{i},v\right\rangle \left\langle v,x_{k}\right\rangle }{\lambda_{i}+\lambda_{k}}\left\langle x_{k},v\right\rangle \left\langle v,x_{j}\right\rangle \nonumber \\
 & \leq3\mleft(1+\left\langle v,h^{-1}v\right\rangle \mright)^{2}\sum_{k=1}^{n}\frac{\left|\left\langle x_{k},v\right\rangle \right|^{2}}{\lambda_{k}}\left\langle x_{i},v\right\rangle \left\langle v,x_{j}\right\rangle \\
 & =3\mleft(1+\left\langle v,h^{-1}v\right\rangle \mright)^{2}\left\langle v,h^{-1}v\right\rangle \left\langle x_{i},v\right\rangle \left\langle v,x_{j}\right\rangle .\nonumber 
\end{align}
This estimate is also valid for $\left|\left\langle x_{i},B\mleft(t\mright)Kx_{j}\right\rangle \right|$
so the claim follows.

$\hfill\square$

Finally is $\int_{0}^{1}B\mleft(t\mright)\,dt$:
\begin{prop}
\label{prop:OneBodyExchangeContribution}For all $1\leq i,j\leq n$
it holds that
\[
\left|\left\langle x_{i},\mleft(\int_{0}^{1}B\mleft(t\mright)\,dt\mright)x_{j}\right\rangle -\frac{1}{2}\left\langle x_{i},v\right\rangle \left\langle v,x_{j}\right\rangle \right|\leq\mleft(6+\left\langle v,h^{-1}v\right\rangle \mright)\left\langle v,h^{-1}v\right\rangle \left\langle x_{i},v\right\rangle \left\langle v,x_{j}\right\rangle .
\]
\end{prop}

\textbf{Proof:} Noting that $\frac{1}{2}\left\langle x_{i},v\right\rangle \left\langle v,x_{j}\right\rangle =\frac{1}{2}\left\langle x_{i},P_{v}x_{j}\right\rangle $
and that
\begin{align}
\int_{0}^{1}B\mleft(t\mright)\,dt-\frac{1}{2}P_{v} & =\int_{0}^{1}\mleft(\mleft(1-t\mright)P_{v}+B_{h}\mleft(t\mright)+e^{tK}P_{v}e^{tK}-P_{v}\mright)\,dt-\frac{1}{2}P_{v}\\
 & =\int_{0}^{1}\mleft(B_{h}\mleft(t\mright)+e^{tK}P_{v}e^{tK}-P_{v}\mright)\,dt\nonumber 
\end{align}
we can estimate using the Propositions \ref{prop:etKPetK-PEstimate}
and \ref{prop:AhBhEstimate} that
\begin{align}
\left|\left\langle x_{i},\mleft(\int_{0}^{1}B\mleft(t\mright)\,dt-\frac{1}{2}P_{v}\mright)x_{j}\right\rangle \right| & \leq\int_{0}^{1}\left|\left\langle x_{i},B_{h}\mleft(t\mright)x_{j}\right\rangle \right|dt+\int_{0}^{1}\left|\left\langle x_{i},\mleft(e^{tK}P_{v}e^{tK}-P_{v}\mright)x_{j}\right\rangle \right|dt\nonumber \\
 & \leq\mleft(6+\left\langle v,h^{-1}v\right\rangle \mright)\left\langle v,h^{-1}v\right\rangle \left\langle x_{i},v\right\rangle \left\langle v,x_{j}\right\rangle .
\end{align}
$\hfill\square$

\subsubsection*{Insertion of the Particular Operators $h_{k}$ and $P_{k}$}

Recall that the particular operators we must consider are $h_{k},P_{k}:\ell^{2}\mleft(L_{k}\mright)\rightarrow\ell^{2}\mleft(L_{k}\mright)$
defined by
\begin{equation}
\begin{array}{ccccccc}
h_{k}e_{p} & = & \lambda_{k,p}e_{p} &  & \lambda_{k,p} & = & \frac{1}{2}\mleft(\left|p\right|^{2}-\left|p-k\right|^{2}\mright)\\
P_{k}\mleft(\cdot\mright) & = & \left\langle v_{k},\cdot\right\rangle v_{k} &  & v_{k} & = & \sqrt{\frac{s\hat{V}_{k}k_{F}^{-1}}{2\,\mleft(2\pi\mright)^{3}}}\sum_{p\in L_{k}}e_{p}.
\end{array}
\end{equation}
For these we have that
\begin{equation}
\left\langle v_{k},h_{k}^{-1}v_{k}\right\rangle =\frac{s\hat{V}_{k}k_{F}^{-1}}{2\,\mleft(2\pi\mright)^{3}}\sum_{p\in L_{k}}\lambda_{k,p}^{-1}.
\end{equation}
In appendix section \ref{sec:RiemannSumEstimates} we obtain the following
estimates for sums of the form $\sum_{p\in L_{k}}\lambda_{k,p}^{\beta}$:
\begin{prop}
\label{prop:RiemannSumEstimates}For any $k\in\mathbb{Z}_{\ast}^{3}$
and $\beta\in\left[-1,0\right]$ it holds that
\[
\sum_{p\in L_{k}}\lambda_{k,p}^{\beta}\leq C_{\beta}\begin{cases}
k_{F}^{2+\beta}\left|k\right|^{1+\beta} & \left|k\right|\leq2k_{F}\\
k_{F}^{3}\left|k\right|^{2\beta} & \left|k\right|>2k_{F}
\end{cases}
\]
for a constant $C_{\beta}>0$ independent of $k$ and $k_{F}$.
\end{prop}

In particular, it holds that $\sum_{p\in L_{k}}\lambda_{k,p}^{-1}\leq Ck_{F}\min\,\{1,k_{F}^{2}\left|k\right|^{-2}\}$,
so
\begin{equation}
\left\langle v_{k},h_{k}^{-1}v_{k}\right\rangle \leq C\hat{V}_{k}
\end{equation}
for a constant $C>0$ depending only on $s$. Additionally, independently
of $p$ and $q$ it holds that
\begin{equation}
\left\langle e_{p},v_{k}\right\rangle \left\langle v_{k},e_{q}\right\rangle =\frac{s\hat{V}_{k}k_{F}^{-1}}{2\,\mleft(2\pi\mright)^{3}}
\end{equation}
and for any $t\geq0$
\begin{equation}
\left\langle v,h\mleft(h^{2}+t^{2}\mright)^{-1}v\right\rangle =\frac{s\hat{V}_{k}k_{F}^{-1}}{2\,\mleft(2\pi\mright)^{3}}\sum_{p\in L_{k}}\frac{\lambda_{k,p}}{\lambda_{k,p}^{2}+t^{2}}.
\end{equation}
Inserting these quantities into the statements of the Propositions
\ref{prop:GeneralBosonizableContribution}, \ref{prop:KElementEstimates},
\ref{prop:AtBtEstimates} and \ref{prop:KBtEstimate} yields Theorem
\ref{them:OneBodyEstimates}, noting also that by Proposition \ref{prop:KElementEstimates}
\begin{align}
\left\Vert K_{k}\right\Vert _{\mathrm{HS}} & =\sqrt{\sum_{p,q\in L_{k}}\left|\left\langle e_{p},K_{k}e_{q}\right\rangle \right|^{2}}\leq\frac{s\hat{V}_{k}k_{F}^{-1}}{2\,\mleft(2\pi\mright)^{3}}\sqrt{\sum_{p,q\in L_{k}}\frac{1}{\mleft(\lambda_{k,p}+\lambda_{k,q}\mright)^{2}}}\leq\frac{s\hat{V}_{k}k_{F}^{-1}}{2\,\mleft(2\pi\mright)^{3}}\sum_{p\in L_{k}}\lambda_{k,p}^{-1}\\
 & \leq C\hat{V}_{k}\min\,\{1,k_{F}^{2}\left|k\right|^{-2}\}.\nonumber 
\end{align}

\section{\label{sec:AnalysisofExchangeTerms}Analysis of Exchange Terms}

In this section we analyze the \textit{exchange terms}, by which we
mean the quantities of the expression
\begin{equation}
\sum_{k\in\mathbb{Z}_{\ast}^{3}}\int_{0}^{1}e^{\mleft(1-t\mright)\mathcal{K}}\mleft(\varepsilon_{k}\mleft(\left\{ K_{k},B_{k}\mleft(t\mright)\right\} \mright)+2\,\mathrm{Re}\mleft(\mathcal{E}_{k}^{1}\mleft(A_{k}\mleft(t\mright)\mright)\mright)+2\,\mathrm{Re}\mleft(\mathcal{E}_{k}^{2}\mleft(B_{k}\mleft(t\mright)\mright)\mright)\mright)e^{-\mleft(1-t\mright)\mathcal{K}}dt
\end{equation}
which appears in Theorem \ref{thm:DiagonalizationoftheBosonizableTerms}
- the name is apt, as these enter our calculations due to the presence
of the exchange correction $\varepsilon_{k,l}\mleft(p;q\mright)$ of
the quasi-bosonic commutation relations.

To be more precise, what we consider in this section are the operators
$\varepsilon_{k}\mleft(\left\{ K_{k},B_{k}\mleft(t\mright)\right\} \mright)$,
$\mathcal{E}_{k}^{1}\mleft(A_{k}\mleft(t\mright)\mright)$ and $\mathcal{E}_{k}^{2}\mleft(B_{k}\mleft(t\mright)\mright)$
- the effect of the integration will be handled by Gronwall estimates
in the next section.

The exchange terms are primarily to be regarded as error terms, and
the main result of this section is the following estimates for them:
\begin{thm}
\label{them:ExchangeTermsEstimates}For any $\Psi\in\mathcal{H}_{N}$
and $t\in\left[0,1\right]$ it holds that
\begin{align*}
\left|\sum_{k\in\mathbb{Z}_{\ast}^{3}}\left\langle \Psi,\varepsilon_{k}\mleft(\left\{ K_{k},B_{k}\mleft(t\mright)\right\} \mright)\Psi\right\rangle \right| & \leq Ck_{F}^{-1}\left\langle \Psi,\mathcal{N}_{E}\Psi\right\rangle \\
\sum_{k\in\mathbb{Z}_{\ast}^{3}}\left|\left\langle \Psi,\mathcal{E}_{k}^{1}\mleft(A_{k}\mleft(t\mright)\mright)\Psi\right\rangle \right| & \leq C\sqrt{\sum_{k\in\mathbb{Z}_{\ast}^{3}}\hat{V}_{k}^{2}\min\left\{ \left|k\right|,k_{F}\right\} }\left\langle \Psi,\mleft(\mathcal{N}_{E}^{3}+1\mright)\Psi\right\rangle \\
\sum_{k\in\mathbb{Z}_{\ast}^{3}}\left|\left\langle \Psi,\mleft(\mathcal{E}_{k}^{2}\mleft(B_{k}\mleft(t\mright)\mright)-\left\langle \psi_{F},\mathcal{E}_{k}^{2}\mleft(B_{k}\mleft(t\mright)\mright)\psi_{F}\right\rangle \mright)\Psi\right\rangle \right| & \leq C\sqrt{\sum_{k\in\mathbb{Z}_{\ast}^{3}}\hat{V}_{k}^{2}\min\left\{ \left|k\right|,k_{F}\right\} }\left\langle \Psi,\mathcal{N}_{E}^{3}\Psi\right\rangle 
\end{align*}
for a constant $C>0$ depending only on $\sum_{k\in\mathbb{Z}_{\ast}^{3}}\hat{V}_{k}^{2}$
and $s$.
\end{thm}

Note the presence of the constant terms $\left\langle \psi_{F},\mathcal{E}_{k}^{2}\mleft(B_{k}\mleft(t\mright)\mright)\psi_{F}\right\rangle $
in the final estimate of the theorem. By adding and subtracting these,
we see that
\begin{equation}
\mathrm{Exchange\,Terms}=\sum_{k\in\mathbb{Z}_{\ast}^{3}}\int_{0}^{1}\left\langle \psi_{F},2\,\mathrm{Re}\mleft(\mathcal{E}_{k}^{2}\mleft(B_{k}\mleft(t\mright)\mright)\mright)\psi_{F}\right\rangle dt+\mathrm{Error\,Terms}.
\end{equation}
The quantity $\sum_{k\in\mathbb{Z}_{\ast}^{3}}\int_{0}^{1}\left\langle \psi_{F},2\,\mathrm{Re}\mleft(\mathcal{E}_{k}^{2}\mleft(B_{k}\mleft(t\mright)\mright)\mright)\psi_{F}\right\rangle dt$
is the \textit{exchange contribution} (to the correlation energy),
which is not generally negligible for singular potentials $V$. We
end the section by determining the leading behavior of these:
\begin{prop}
\label{prop:LeadingExchangeContribution}It holds that
\[
\left|\sum_{k\in\mathbb{Z}_{\ast}^{3}}\int_{0}^{1}\left\langle \psi_{F},2\,\mathrm{Re}\mleft(\mathcal{E}_{k}^{2}\mleft(B_{k}\mleft(t\mright)\mright)\mright)\psi_{F}\right\rangle dt-E_{\mathrm{corr},\mathrm{ex}}\right|\leq C\sqrt{\sum_{k\in\mathbb{Z}_{\ast}^{3}}\hat{V}_{k}^{2}\min\left\{ \left|k\right|,k_{F}\right\} }
\]
for a constant $C>0$ depending only on $\sum_{k\in\mathbb{Z}_{\ast}^{3}}\hat{V}_{k}^{2}$
and $s$, where
\[
E_{\mathrm{corr},\mathrm{ex}}=\frac{sk_{F}^{-2}}{4\,\mleft(2\pi\mright)^{6}}\sum_{k,l\in\mathbb{Z}_{\ast}^{3}}\hat{V}_{k}\hat{V}_{l}\sum_{p,q\in L_{k}\cap L_{l}}\frac{\delta_{p+q,k+l}}{\lambda_{k,p}+\lambda_{k,q}}.
\]
\end{prop}

\subsubsection*{Analysis of $\varepsilon_{k}$ Terms}

Let us first consider terms of the form $\sum_{k\in\mathbb{Z}_{\ast}^{3}}\varepsilon_{k}\mleft(A_{k}\mright)$,
where we recall that $\varepsilon_{k}\mleft(A_{k}\mright)$ is given
by
\begin{equation}
\varepsilon_{k}\mleft(A_{k}\mright)=-\frac{1}{s}\sum_{p\in L_{k}}^{\sigma}\left\langle e_{p},A_{k}e_{p}\right\rangle \mleft(c_{p,\sigma}^{\ast}c_{p,\sigma}+c_{p-k,\sigma}c_{p-k,\sigma}^{\ast}\mright).
\end{equation}
When summing over $k\in\mathbb{Z}_{\ast}^{3}$, we can split the sum
into two parts and interchange the summations as follows:
\begin{align}
-\sum_{k\in\mathbb{Z}_{\ast}^{3}}\varepsilon_{k}\mleft(A_{k}\mright) & =\frac{1}{s}\sum_{k\in\mathbb{Z}_{\ast}^{3}}\sum_{p\in L_{k}}^{\sigma}\left\langle e_{p},A_{k}e_{p}\right\rangle c_{p,\sigma}^{\ast}c_{p,\sigma}+\frac{1}{s}\sum_{k\in\mathbb{Z}_{\ast}^{3}}\sum_{q\in\mleft(L_{k}-k\mright)}^{\sigma}\left\langle e_{q+k},A_{k}e_{q+k}\right\rangle c_{q,\sigma}c_{q,\sigma}^{\ast}\\
 & =\frac{1}{s}\sum_{p\in B_{F}^{c}}^{\sigma}\mleft(\sum_{k\in\mathbb{Z}_{\ast}^{3}}1_{L_{k}}\mleft(p\mright)\left\langle e_{p},A_{k}e_{p}\right\rangle \mright)c_{p,\sigma}^{\ast}c_{p,\sigma}+\frac{1}{s}\sum_{q\in B_{F}}^{\sigma}\mleft(\sum_{k\in\mathbb{Z}_{\ast}^{3}}1_{L_{k}}\mleft(q+k\mright)\left\langle e_{q+k},A_{k}e_{q+k}\right\rangle \mright)c_{q,\sigma}c_{q,\sigma}^{\ast}.\nonumber 
\end{align}
Recalling that the excitation number operator is given by
\begin{equation}
\mathcal{N}_{E}=\sum_{p\in B_{F}^{c}}^{\sigma}c_{p,\sigma}^{\ast}c_{p,\sigma}=\sum_{q\in B_{F}}^{\sigma}c_{q,\sigma}c_{q,\sigma}^{\ast}
\end{equation}
on $\mathcal{H}_{N}$, we can then immediately conclude that
\begin{align}
\pm\sum_{k\in\mathbb{Z}_{\ast}^{3}}\varepsilon_{k}\mleft(A_{k}\mright) & \leq\frac{1}{s}\mleft(\sup_{p\in B_{F}^{c}}\sum_{k\in\mathbb{Z}_{\ast}^{3}}1_{L_{k}}\mleft(p\mright)\left|\left\langle e_{p},A_{k}e_{p}\right\rangle \right|+\sup_{q\in B_{F}}\sum_{k\in\mathbb{Z}_{\ast}^{3}}1_{L_{k}}\mleft(q+k\mright)\left|\left\langle e_{q+k},A_{k}e_{q+k}\right\rangle \right|\mright)\mathcal{N}_{E}\nonumber \\
 & \leq\frac{2}{s}\mleft(\sum_{k\in\mathbb{Z}_{\ast}^{3}}\sup_{p\in L_{k}}\left|\left\langle e_{p},A_{k}e_{p}\right\rangle \right|\mright)\mathcal{N}_{E}.
\end{align}
By the estimates of the previous section we thus obtain the first
estimate of Theorem \ref{them:ExchangeTermsEstimates}:
\begin{prop}
For any $\Psi\in\mathcal{H}_{N}$ and $t\in\left[0,1\right]$ it holds
that
\[
\left|\sum_{k\in\mathbb{Z}_{\ast}^{3}}\left\langle \Psi,\varepsilon_{k}\mleft(\left\{ K_{k},B_{k}\mleft(t\mright)\right\} \mright)\Psi\right\rangle \right|\leq Ck_{F}^{-1}\left\langle \Psi,\mathcal{N}_{E}\Psi\right\rangle 
\]
for a constant $C>0$ depending only on $\sum_{k\in\mathbb{Z}_{\ast}^{3}}\hat{V}_{k}^{2}$
and $s$.
\end{prop}

\textbf{Proof:} By Theorem \ref{them:OneBodyEstimates} we have that
\begin{equation}
\left|\left\langle e_{p},\left\{ K_{k},B_{k}\mleft(t\mright)\right\} e_{q}\right\rangle \right|\leq C\mleft(1+\hat{V}_{k}^{2}\mright)\hat{V}_{k}^{2}k_{F}^{-1},\quad k\in\mathbb{Z}_{\ast}^{3},\,p,q\in L_{k},
\end{equation}
for a constant $C>0$ depending only on $s$, so
\begin{align}
 & \quad\;\left|\sum_{k\in\mathbb{Z}_{\ast}^{3}}\left\langle \Psi,\varepsilon_{k}\mleft(\left\{ K_{k},B_{k}\mleft(t\mright)\right\} \mright)\Psi\right\rangle \right|\leq\frac{2}{s}\mleft(\sum_{k\in\mathbb{Z}_{\ast}^{3}}\sup_{p\in L_{k}}\left|\left\langle e_{p},\left\{ K_{k},B_{k}\mleft(t\mright)\right\} e_{p}\right\rangle \right|\mright)\left\langle \Psi,\mathcal{N}_{E}\Psi\right\rangle \\
 & \leq Ck_{F}^{-1}\sum_{k\in\mathbb{Z}_{\ast}^{3}}\mleft(1+\hat{V}_{k}^{2}\mright)\hat{V}_{k}^{2}\left\langle \Psi,\mathcal{N}_{E}\Psi\right\rangle \leq Ck_{F}^{-1}\mleft(1+\Vert\hat{V}\Vert_{\infty}^{2}\mright)\sum_{k\in\mathbb{Z}_{\ast}^{3}}\hat{V}_{k}^{2}\left\langle \Psi,\mathcal{N}_{E}\Psi\right\rangle .\nonumber 
\end{align}
As $\Vert\hat{V}\Vert_{\infty}^{2}\leq\Vert\hat{V}\Vert_{2}^{2}=\sum_{k\in\mathbb{Z}_{\ast}^{3}}\hat{V}_{k}^{2}$
the claim follows.

$\hfill\square$

\subsection{Analysis of $\mathcal{E}_{k}^{1}$ Terms}

We consider terms of the form
\begin{equation}
\mathcal{E}_{k}^{1}\mleft(A_{k}\mright)=\sum_{l\in\mathbb{Z}_{\ast}^{3}}\sum_{p\in L_{k}}\sum_{q\in L_{l}}b_{k}^{\ast}\mleft(A_{k}e_{p}\mright)\left\{ \varepsilon_{k,l}\mleft(e_{p};e_{q}\mright),b_{-l}^{\ast}\mleft(K_{-l}e_{-q}\mright)\right\} .
\end{equation}
Recalling that $\varepsilon_{k,l}\mleft(e_{p};e_{q}\mright)$ is given
by
\begin{equation}
\varepsilon_{k,l}\mleft(e_{p};e_{q}\mright)=-\frac{1}{s}\sum_{\sigma=1}^{s}\mleft(\delta_{p,q}c_{q-l,\sigma}c_{p-k,\sigma}^{\ast}+\delta_{p-k,q-l}c_{q,\sigma}^{\ast}c_{p,\sigma}\mright)
\end{equation}
we see that $\mathcal{E}_{k}^{1}\mleft(A_{k}\mright)$ splits into two
sums as
\begin{align}
-s\,\mathcal{E}_{k}^{1}\mleft(A_{k}\mright) & =\sum_{l\in\mathbb{Z}_{\ast}^{3}}\sum_{p\in L_{k}}^{\sigma}\sum_{q\in L_{l}}b_{k}^{\ast}\mleft(A_{k}e_{p}\mright)\left\{ \delta_{p,q}c_{q-l,\sigma}c_{p-k,\sigma}^{\ast},b_{-l}^{\ast}\mleft(K_{-l}e_{-q}\mright)\right\} \nonumber \\
 & +\sum_{l\in\mathbb{Z}_{\ast}^{3}}\sum_{p\in\mleft(L_{k}-k\mright)}^{\sigma}\sum_{q\in\mleft(L_{l}-l\mright)}b_{k}^{\ast}\mleft(A_{k}e_{p+k}\mright)\left\{ \delta_{p,q}c_{q+l,\sigma}^{\ast}c_{p+k,\sigma},b_{-l}^{\ast}\mleft(K_{-l}e_{-q-l}\mright)\right\} \\
 & =\sum_{l\in\mathbb{Z}_{\ast}^{3}}\sum_{p\in L_{k}\cap L_{l}}^{\sigma}b_{k}^{\ast}\mleft(A_{k}e_{p}\mright)\left\{ c_{p-l,\sigma}c_{p-k,\sigma}^{\ast},b_{-l}^{\ast}\mleft(K_{-l}e_{-p}\mright)\right\} \nonumber \\
 & +\sum_{l\in\mathbb{Z}_{\ast}^{3}}\sum_{p\in\mleft(L_{k}-k\mright)\cap\mleft(L_{l}-l\mright)}^{\sigma}b_{k}^{\ast}\mleft(A_{k}e_{p+k}\mright)\left\{ c_{p+l,\sigma}^{\ast}c_{p+k,\sigma},b_{-l}^{\ast}\mleft(K_{-l}e_{-p-l}\mright)\right\} .\nonumber 
\end{align}
The two sums on the right-hand side have the same ``schematic form'':
They can both be written as
\begin{equation}
\sum_{l\in\mathbb{Z}_{\ast}^{3}}\sum_{p\in S_{k}\cap S_{l}}^{\sigma}b_{k}^{\ast}\mleft(A_{k}e_{p_{1}}\mright)\left\{ \tilde{c}_{p_{2},\sigma}^{\ast}\tilde{c}_{p_{3},\sigma},b_{-l}^{\ast}\mleft(K_{-l}e_{p_{4}}\mright)\right\} ,\quad\tilde{c}_{p}=\begin{cases}
c_{p} & p\in B_{F}^{c}\\
c_{p}^{\ast} & p\in B_{F}
\end{cases},\label{eq:calE1FirstSchematicForm}
\end{equation}
where the index set is either the lune $S_{k}=L_{k}$ or the set of
corresponding hole states $S_{k}=L_{k}-k,$ and depending on this
index set the variables $p_{1},p_{2},p_{3},p_{4}$ are given by
\begin{equation}
\mleft(p_{1},p_{2},p_{3},p_{4}\mright)=\begin{cases}
\mleft(p,p-l,p-k,-p\mright) & S_{k}=L_{k}\\
\mleft(p+k,p+l,p+k,-p-l\mright) & S_{k}=L_{k}-k
\end{cases}.
\end{equation}
Note that in either case $p_{1}$, $p_{3}$ depend only on $p$ and
$k$, while $p_{2}$, $p_{4}$ depend only on $p$ and $l$. Additionally,
$p_{1}$ is always an element of $L_{k}$ and $p_{4}$ is always an
element of $L_{-l}$.

Since $b_{k,p}=s^{-\frac{1}{2}}\sum_{\sigma=1}^{s}c_{p-k,\sigma}^{\ast}c_{p,\sigma}=s^{-\frac{1}{2}}\sum_{\sigma=1}^{s}\tilde{c}_{p-k,\sigma}\tilde{c}_{p,\sigma}$
it is easily seen that $\left[b,\tilde{c}\right]=0$, so in normal-ordering
(with respect to $\psi_{F}$) the summand of equation (\ref{eq:calE1FirstSchematicForm})
we find
\begin{align}
 & \quad\;b_{k}^{\ast}\mleft(A_{k}e_{p_{1}}\mright)\left\{ \tilde{c}_{p_{2},\sigma}^{\ast}\tilde{c}_{p_{3},\sigma},b_{-l}^{\ast}\mleft(K_{-l}e_{p_{4}}\mright)\right\} \nonumber \\
 & =b_{k}^{\ast}\mleft(A_{k}e_{p_{1}}\mright)\tilde{c}_{p_{2},\sigma}^{\ast}\tilde{c}_{p_{3},\sigma}b_{-l}^{\ast}\mleft(K_{-l}e_{p_{4}}\mright)+b_{k}^{\ast}\mleft(A_{k}e_{p_{1}}\mright)b_{-l}^{\ast}\mleft(K_{-l}e_{p_{4}}\mright)\tilde{c}_{p_{2},\sigma}^{\ast}\tilde{c}_{p_{3},\sigma}\\
 & =2\,\tilde{c}_{p_{2},\sigma}^{\ast}b_{k}^{\ast}\mleft(A_{k}e_{p_{1}}\mright)b_{-l}^{\ast}\mleft(K_{-l}e_{p_{4}}\mright)\tilde{c}_{p_{3},\sigma}+\tilde{c}_{p_{2},\sigma}^{\ast}b_{k}^{\ast}\mleft(A_{k}e_{p_{1}}\mright)\left[\tilde{c}_{p_{3},\sigma},b_{-l}^{\ast}\mleft(K_{-l}e_{p_{4}}\mright)\right].\nonumber 
\end{align}
To bound a sum of the form $\sum_{k\in\mathbb{Z}_{\ast}^{3}}\mathcal{E}_{1}^{k}\mleft(A_{k}\mright)$
it thus suffices to estimate the two schematic forms
\begin{align}
 & \sum_{k,l\in\mathbb{Z}_{\ast}^{3}}\sum_{p\in S_{k}\cap S_{l}}^{\sigma}\tilde{c}_{p_{2},\sigma}^{\ast}b_{k}^{\ast}\mleft(A_{k}e_{p_{1}}\mright)b_{-l}^{\ast}\mleft(K_{-l}e_{p_{4}}\mright)\tilde{c}_{p_{3},\sigma}\label{eq:calE1SchematicForms}\\
 & \sum_{k,l\in\mathbb{Z}_{\ast}^{3}}\sum_{p\in S_{k}\cap S_{l}}^{\sigma}\tilde{c}_{p_{2},\sigma}^{\ast}b_{k}^{\ast}\mleft(A_{k}e_{p_{1}}\mright)\left[b_{-l}\mleft(K_{-l}e_{p_{4}}\mright),\tilde{c}_{p_{3},\sigma}^{\ast}\right]^{\ast}.\nonumber 
\end{align}

\subsubsection*{Preliminary Estimates}

We prepare for the estimation of these schematic forms by deriving
some auxilliary bounds for the operators involved.

Recall that for any $k\in\mathbb{Z}_{\ast}^{3}$ and $\varphi\in\ell^{2}\mleft(L_{k}\mright)$,
the excitation operator $b_{k}\mleft(\varphi\mright)$ is given by
\begin{equation}
b_{k}\mleft(\varphi\mright)=\sum_{p\in L_{k}}\left\langle \varphi,e_{p}\right\rangle b_{k,p}=\frac{1}{\sqrt{s}}\sum_{p\in L_{k}}^{\sigma}\left\langle \varphi,e_{p}\right\rangle c_{p-k,\sigma}^{\ast}c_{p,\sigma}.
\end{equation}
We observe that the exchange correction $\varepsilon_{k,k}\mleft(\varphi;\varphi\mright)$
arising from the commutator $\left[b_{k}\mleft(\varphi\mright),b_{k}^{\ast}\mleft(\varphi\mright)\right]$
is non-positive: Indeed, this is given by
\begin{align}
\varepsilon_{k,k}\mleft(\varphi;\varphi\mright) & =-\frac{1}{s}\sum_{p,q\in L_{k}}^{\sigma}\left\langle \varphi,e_{p}\right\rangle \left\langle e_{q},\varphi\right\rangle \mleft(\delta_{p,q}c_{q-k,\sigma}c_{p-k,\sigma}^{\ast}+\delta_{p-k,q-k}c_{q,\sigma}^{\ast}c_{p,\sigma}\mright)\\
 & =-\frac{1}{s}\sum_{p\in L_{k}}^{\sigma}\left|\left\langle e_{p},\varphi\right\rangle \right|^{2}\mleft(c_{p-k,\sigma}c_{p-k,\sigma}^{\ast}+c_{p,\sigma}^{\ast}c_{p,\sigma}\mright)\leq0.\nonumber 
\end{align}
Using this we can bound both $b_{k}\mleft(\varphi\mright)$ and $b_{k}^{\ast}\mleft(\varphi\mright)$
as follows:
\begin{prop}
\label{prop:bbastEstimates}For any $k\in\mathbb{Z}_{\ast}^{3}$,
$\varphi\in\ell^{2}\mleft(L_{k}\mright)$ and $\Psi\in\mathcal{H}_{N}$
it holds that
\[
\left\Vert b_{k}\mleft(\varphi\mright)\Psi\right\Vert \leq\left\Vert \varphi\right\Vert \Vert\mathcal{N}_{k}^{\frac{1}{2}}\Psi\Vert,\quad\left\Vert b_{k}^{\ast}\mleft(\varphi\mright)\Psi\right\Vert \leq\left\Vert \varphi\right\Vert \Vert\mleft(\mathcal{N}_{k}+1\mright)^{\frac{1}{2}}\Psi\Vert,
\]
where $\mathcal{N}_{k}=\sum_{p\in L_{k}}b_{k,p}^{\ast}b_{k,p}$.
\end{prop}

\textbf{Proof:} By the triangle and Cauchy-Schwarz inequalities we
immediately obtain
\begin{equation}
\left\Vert b_{k}\mleft(\varphi\mright)\Psi\right\Vert \leq\sum_{p\in L_{k}}\left|\left\langle \varphi,e_{p}\right\rangle \right|\left\Vert b_{k,p}\Psi\right\Vert \leq\left\Vert \varphi\right\Vert \sqrt{\sum_{p\in L_{k}}\left\Vert b_{k,p}\Psi\right\Vert ^{2}}=\left\Vert \varphi\right\Vert \Vert\mathcal{N}_{k}^{\frac{1}{2}}\Psi\Vert
\end{equation}
and the bound for $\left\Vert b_{k}^{\ast}\mleft(\varphi\mright)\Psi\right\Vert $
now follows from this, since the above observation implies that
\begin{equation}
b_{k}\mleft(\varphi\mright)b_{k}^{\ast}\mleft(\varphi\mright)=b_{k}^{\ast}\mleft(\varphi\mright)b_{k}\mleft(\varphi\mright)+\left\langle \varphi,\varphi\right\rangle +\varepsilon_{k,k}\mleft(\varphi;\varphi\mright)\leq\left\Vert \varphi\right\Vert ^{2}\mleft(\mathcal{N}_{k}+1\mright).
\end{equation}
$\hfill\square$

Note that the operator
\begin{equation}
\mathcal{N}_{k}=\sum_{p\in L_{k}}b_{k,p}^{\ast}b_{k,p}=\frac{1}{s}\sum_{p\in L_{k}}^{\sigma,\tau}c_{p,\sigma}^{\ast}c_{p-k,\sigma}c_{p-k,\tau}^{\ast}c_{p,\tau}
\end{equation}
can be estimated directly in terms of $\mathcal{N}_{E}$ as $\mathcal{N}_{k}\leq\mathcal{N}_{E}$,
since for any $\Psi\in\mathcal{H}_{N}$
\begin{align}
\left\langle \Psi,\mathcal{N}_{k}\Psi\right\rangle  & =\sum_{p\in L_{k}}\left\Vert b_{k,p}\Psi\right\Vert ^{2}=\sum_{p\in L_{k}}\left\Vert \frac{1}{\sqrt{s}}\sum_{\sigma=1}^{s}c_{p-k,\sigma}^{\ast}c_{p,\sigma}\Psi\right\Vert ^{2}\leq\sum_{p\in L_{k}}\mleft(\frac{1}{\sqrt{s}}\sum_{\sigma=1}^{s}\left\Vert c_{p-k,\sigma}^{\ast}c_{p,\sigma}\Psi\right\Vert \mright)^{2}\label{eq:calNkSimpleBound}\\
 & \leq\sum_{p\in L_{k}}^{\sigma}\left\Vert c_{p-k,\sigma}^{\ast}c_{p,\sigma}\Psi\right\Vert ^{2}\leq\sum_{p\in L_{k}}^{\sigma}\left\Vert c_{p,\sigma}\Psi\right\Vert ^{2}\leq\left\langle \Psi,\mathcal{N}_{E}\Psi\right\rangle \nonumber 
\end{align}
by the usual fermionic estimate. Below we will generally only use
this cruder estimate, but $\mathcal{N}_{k}$ is useful for some bounds
since it can be summed over $k\in\mathbb{Z}_{\ast}^{3}$: By rearranging
the summations one concludes that
\begin{align}
 & \sum_{k\in\mathbb{Z}_{\ast}^{3}}\left\langle \Psi,\mathcal{N}_{k}\Psi\right\rangle \leq\sum_{k\in\mathbb{Z}_{\ast}^{3}}\sum_{p\in L_{k}}^{\sigma}\left\Vert c_{p-k,\sigma}^{\ast}c_{p,\sigma}\Psi\right\Vert ^{2}=\sum_{k\in\mathbb{Z}_{\ast}^{3}}\sum_{p\in L_{k}}^{\sigma}\left\langle \Psi,c_{p,\sigma}^{\ast}c_{p,\sigma}c_{p-k,\sigma}c_{p-k,\sigma}^{\ast}\Psi\right\rangle \nonumber \\
 & \;=\left\langle \Psi,\sum_{p\in B_{F}^{c}}^{\sigma}c_{p,\sigma}^{\ast}c_{p,\sigma}\sum_{k\in\mathbb{Z}_{\ast}^{3}}1_{L_{k}}\mleft(p\mright)c_{p-k,\sigma}c_{p-k,\sigma}^{\ast}\Psi\right\rangle \\
 & \;=\left\langle \Psi,\sum_{p\in B_{F}^{c}}^{\sigma}c_{p,\sigma}^{\ast}c_{p,\sigma}\sum_{k\in\mleft(B_{F}+p\mright)}c_{p-k,\sigma}c_{p-k,\sigma}^{\ast}\Psi\right\rangle =\left\langle \Psi,\sum_{p\in B_{F}^{c}}^{\sigma}c_{p,\sigma}^{\ast}c_{p,\sigma}\sum_{q\in B_{F}}c_{q,\sigma}c_{q,\sigma}^{\ast}\Psi\right\rangle ,\nonumber 
\end{align}
so noting that $\sum_{q\in B_{F}}c_{q,\sigma}c_{q,\sigma}^{\ast}=\mathcal{N}_{E}-\sum_{q\in B_{F}}^{\tau\neq\sigma}c_{q,\tau}c_{q,\tau}^{\ast}$
we can estimate
\begin{align}
\sum_{k\in\mathbb{Z}_{\ast}^{3}}\left\langle \Psi,\mathcal{N}_{k}\Psi\right\rangle  & \leq\left\langle \Psi,\sum_{p\in B_{F}^{c}}^{\sigma}c_{p,\sigma}^{\ast}c_{p,\sigma}\mathcal{N}_{E}\Psi\right\rangle -\left\langle \Psi,\sum_{p\in B_{F}^{c}}^{\sigma}c_{p,\sigma}^{\ast}c_{p,\sigma}\sum_{q\in B_{F}}^{\tau\neq\sigma}c_{q,\tau}c_{q,\tau}^{\ast}\Psi\right\rangle \\
 & =\left\langle \Psi,\mathcal{N}_{E}^{2}\Psi\right\rangle -\sum_{p\in B_{F}^{c}}^{\sigma}\left\langle \Psi,c_{p,\sigma}^{\ast}\mleft(\sum_{q\in B_{F}}^{\tau\neq\sigma}c_{q,\tau}c_{q,\tau}^{\ast}\mright)c_{p,\sigma}\Psi\right\rangle \leq\left\langle \Psi,\mathcal{N}_{E}^{2}\Psi\right\rangle \nonumber 
\end{align}
i.e. $\sum_{k\in\mathbb{Z}_{\ast}^{3}}\mathcal{N}_{k}\leq\mathcal{N}_{E}^{2}$.
(Equality even holds for $s=1$.)

We also note that for any $\Psi\in\mathcal{H}_{N}$ and $p\in\mathbb{Z}^{3}$
\begin{align}
\sum_{\sigma=1}^{s}\Vert\mathcal{N}_{k}^{\frac{1}{2}}\tilde{c}_{p,\sigma}\Psi\Vert^{2} & \leq\sum_{\sigma=1}^{s}\Vert\tilde{c}_{p,\sigma}\mathcal{N}_{k}^{\frac{1}{2}}\Psi\Vert^{2}\leq\sum_{\sigma=1}^{s}\Vert\tilde{c}_{p,\sigma}\mathcal{N}_{E}^{\frac{1}{2}}\Psi\Vert^{2}\\
\sum_{\sigma=1}^{s}\Vert\mleft(\mathcal{N}_{k}+1\mright)^{\frac{1}{2}}\tilde{c}_{p,\sigma}\Psi\Vert^{2} & \leq\sum_{\sigma=1}^{s}\Vert\tilde{c}_{p,\sigma}\mleft(\mathcal{N}_{k}+1\mright)^{\frac{1}{2}}\Psi\Vert^{2}\leq\sum_{\sigma=1}^{s}\Vert\tilde{c}_{p,\sigma}\mleft(\mathcal{N}_{E}+1\mright)^{\frac{1}{2}}\Psi\Vert^{2},\nonumber 
\end{align}
as follows by the inequality (considering $p\in B_{F}^{c}$ for definiteness)
\begin{align}
\sum_{\sigma=1}^{s}\tilde{c}_{p,\sigma}^{\ast}\mathcal{N}_{k}\tilde{c}_{p,\sigma} & =\frac{1}{s}\sum_{q\in L_{k}}^{\sigma,\tau,\rho}c_{p,\sigma}^{\ast}c_{q,\tau}^{\ast}c_{q-k,\tau}c_{q-k,\rho}^{\ast}c_{q,\rho}c_{p,\sigma}=\frac{1}{s}\sum_{q\in L_{k}}^{\sigma,\tau,\rho}c_{q,\tau}^{\ast}c_{q-k,\tau}c_{q-k,\rho}^{\ast}\mleft(c_{q,\rho}c_{p,\sigma}^{\ast}-\delta_{p,q}\delta_{\sigma,\tau}\mright)c_{p,\sigma}\nonumber \\
 & =\mathcal{N}_{k}\sum_{\sigma=1}^{s}c_{p,\sigma}^{\ast}c_{p,\sigma}-\frac{1}{s}\sum_{\sigma,\tau=1}^{s}1_{L_{k}}\mleft(p\mright)c_{p,\tau}^{\ast}c_{p-k,\tau}c_{p-k,\sigma}^{\ast}c_{p,\sigma}\\
 & =\mathcal{N}_{k}\sum_{\sigma=1}^{s}c_{p,\sigma}^{\ast}c_{p,\sigma}-1_{L_{k}}\mleft(p\mright)b_{k,p}^{\ast}b_{k,p}\leq\mathcal{N}_{k}\sum_{\sigma=1}^{s}c_{p,\sigma}^{\ast}c_{p,\sigma}\nonumber 
\end{align}
and the fact that $\sum_{\sigma=1}^{s}\left[\tilde{c}_{p,\sigma}^{\ast}\tilde{c}_{p,\sigma},\mathcal{N}_{k}\right]=0=\sum_{\sigma=1}^{s}\left[\tilde{c}_{p,\sigma}^{\ast}\tilde{c}_{p,\sigma},\mathcal{N}_{E}\right]$.
Similarly\footnote{There is a slight ambiguity here: $\mathcal{N}_{E}=\sum_{p\in B_{F}^{c}}^{\sigma}c_{p,\sigma}^{\ast}c_{p,\sigma}=\sum_{q\in B_{F}}^{\sigma}c_{q,\sigma}c_{q,\sigma}^{\ast}$
holds on $\mathcal{H}_{N}$, but an element such as $\tilde{c}_{p,\sigma}\Psi$
belongs to $\mathcal{H}_{N\pm1}$. This is of no importance, however,
since these inequalities hold no matter if $\mathcal{N}_{E}$ is understood
as $\sum_{p\in B_{F}^{c}}^{\sigma}c_{p,\sigma}^{\ast}c_{p,\sigma}$
or $\sum_{q\in B_{F}}^{\sigma}c_{q,\sigma}c_{q,\sigma}^{\ast}$. On
the same note, the estimate of equation (\ref{eq:calNkSimpleBound})
is valid for either case even if $\Psi\in\mathcal{H}_{N\pm1}$.}
\begin{equation}
\sum_{\sigma=1}^{s}\Vert\mathcal{N}_{E}^{\frac{1}{2}}\tilde{c}_{p,\sigma}\Psi\Vert^{2}\leq\sum_{\sigma=1}^{s}\Vert\tilde{c}_{p,\sigma}\mathcal{N}_{E}^{\frac{1}{2}}\Psi\Vert^{2},\quad\sum_{\sigma=1}^{s}\Vert\mleft(\mathcal{N}_{E}+1\mright)^{\frac{1}{2}}\tilde{c}_{p,\sigma}\Psi\Vert^{2}\leq\sum_{\sigma=1}^{s}\Vert\tilde{c}_{p,\sigma}\mleft(\mathcal{N}_{E}+1\mright)^{\frac{1}{2}}\Psi\Vert^{2}.\label{eq:NumberOperatorRearrangement}
\end{equation}
To analyze the commutator term $\left[b_{-l}\mleft(K_{-l}e_{p_{4}}\mright),\tilde{c}_{p_{3},\sigma}^{\ast}\right]$
we calculate a general identity: For any $l\in\mathbb{Z}_{\ast}^{3}$,
$\psi\in\ell^{2}\mleft(L_{l}\mright)$ and $p\in\mathbb{Z}^{3}$
\begin{align}
\left[b_{l}\mleft(\psi\mright),\tilde{c}_{p,\sigma}^{\ast}\right] & =\frac{1}{\sqrt{s}}\sum_{q\in L_{l}}^{\tau}\left\langle \psi,e_{q}\right\rangle \begin{cases}
\left[c_{q-l,\tau}^{\ast}c_{q,\tau},c_{p,\sigma}\right] & p\in B_{F}\\
\left[c_{q-l,\tau}^{\ast}c_{q,\tau},c_{p,\sigma}^{\ast}\right] & p\in B_{F}^{c}
\end{cases}=\frac{1}{\sqrt{s}}\sum_{q\in L_{l}}^{\tau}\left\langle \psi,e_{q}\right\rangle \begin{cases}
-c_{q,\tau}\left\{ c_{q-l,\tau}^{\ast},c_{p,\sigma}\right\}  & p\in B_{F}\\
c_{q-l,\tau}^{\ast}\left\{ c_{q,\tau},c_{p,\sigma}^{\ast}\right\}  & p\in B_{F}^{c}
\end{cases}\nonumber \\
 & =\begin{cases}
-1_{L_{l}}\mleft(p+l\mright)s^{-\frac{1}{2}}\left\langle \psi,e_{p+l}\right\rangle \tilde{c}_{p+l,\sigma} & p\in B_{F}\\
1_{L_{l}}\mleft(p\mright)s^{-\frac{1}{2}}\left\langle \psi,e_{p}\right\rangle \tilde{c}_{p-l,\sigma} & p\in B_{F}^{c}
\end{cases},\label{eq:bcastCommutator}
\end{align}
so for our particular commutator we obtain
\begin{equation}
\left[b_{-l}\mleft(K_{-l}e_{p_{4}}\mright),\tilde{c}_{p_{3},\sigma}^{\ast}\right]=\begin{cases}
-1_{L_{-l}}\mleft(p_{3}-l\mright)s^{-\frac{1}{2}}\left\langle K_{-l}e_{p_{4}},e_{p_{3}-l}\right\rangle \tilde{c}_{p_{3}-l,\sigma} & S_{k}=L_{k}\\
1_{L_{-l}}\mleft(p_{3}\mright)s^{-\frac{1}{2}}\left\langle K_{-l}e_{p_{4}},e_{p_{3}}\right\rangle \tilde{c}_{p_{3}+l,\sigma} & S_{k}=L_{k}-k
\end{cases}.\label{eq:calE1CommutatorTermIdentity}
\end{equation}
It will be crucial to our estimates that the prefactors obey the following:
\begin{prop}
\label{prop:calE1KSplitEstimate}For any $k,l\in\mathbb{Z}_{\ast}^{3}$
and $p\in S_{k}\cap S_{l}$ it holds that
\[
\left|1_{L_{-l}}\mleft(p_{3}-l\mright)s^{-\frac{1}{2}}\left\langle K_{-l}e_{p_{4}},e_{p_{3}-l}\right\rangle \right|\leq C\hat{V}_{-l}k_{F}^{-1}\frac{1_{L_{-k}}\mleft(p_{2}-k\mright)1_{L_{-l}}\mleft(p_{3}-l\mright)}{\sqrt{\lambda_{k,p_{1}}+\lambda_{-k,p_{2}-k}}\sqrt{\lambda_{-l,p_{3}-l}+\lambda_{-l,p_{4}}}},\quad S_{k}=L_{k},
\]
and
\[
\left|1_{L_{-l}}\mleft(p_{3}\mright)s^{-\frac{1}{2}}\left\langle K_{-l}e_{p_{4}},e_{p_{3}}\right\rangle \right|\leq C\hat{V}_{-l}k_{F}^{-1}\frac{1_{L_{-k}}\mleft(p_{2}\mright)1_{L_{-l}}\mleft(p_{3}\mright)}{\sqrt{\lambda_{k,p_{1}}+\lambda_{-k,p_{2}}}\sqrt{\lambda_{-l,p_{3}}+\lambda_{-l,p_{4}}}},\quad S_{k}=L_{k}-k,
\]
for a constant $C>0$ depending only on $s$.
\end{prop}

\textbf{Proof:} Recall that $p_{1},p_{2},p_{3},p_{4}$ are given by
\begin{equation}
\mleft(p_{1},p_{2},p_{3},p_{4}\mright)=\begin{cases}
\mleft(p,p-l,p-k,-p\mright) & S_{k}=L_{k}\\
\mleft(p+k,p+l,p+k,-p-l\mright) & S_{k}=L_{k}-k
\end{cases}.
\end{equation}
From this we see that for any $p\in S_{k}\cap S_{l}$
\begin{align}
\begin{cases}
1_{L_{-l}}\mleft(p_{3}-l\mright) & S_{k}=L_{k}\\
1_{L_{-l}}\mleft(p_{3}\mright) & S_{k}=L_{k}-k
\end{cases} & =\begin{cases}
1_{B_{F}^{c}}\mleft(p-k-l\mright)1_{B_{F}}\mleft(p-k\mright) & S_{k}=L_{k}\\
1_{B_{F}^{c}}\mleft(p+k\mright)1_{B_{F}}\mleft(p+k+l\mright) & S_{k}=L_{k}-k
\end{cases}\nonumber \\
 & =\begin{cases}
1_{B_{F}^{c}}\mleft(p-l-k\mright)1_{B_{F}}\mleft(p-l\mright) & S_{k}=L_{k}\\
1_{B_{F}^{c}}\mleft(p+l\mright)1_{B_{F}}\mleft(p+l+k\mright) & S_{k}=L_{k}-k
\end{cases}\\
 & =\begin{cases}
1_{L_{-k}}\mleft(p_{2}-k\mright) & S_{k}=L_{k}\\
1_{L_{-k}}\mleft(p_{2}\mright) & S_{k}=L_{k}-k
\end{cases}\nonumber 
\end{align}
where the assumption that $p\in S_{k}\cap S_{l}$ enters to ensure
that $1_{B_{F}}\mleft(p-k\mright)=1=1_{B_{F}}\mleft(p-l\mright)$ or $1_{B_{F}^{c}}\mleft(p+k\mright)=1=1_{B_{F}^{c}}\mleft(p+l\mright)$,
respectively. Importantly this also implies that, when combined with
such an indicator function, we also have the identity
\begin{align}
 & \quad\,\begin{cases}
\lambda_{-l,p_{3}-l}+\lambda_{-l,p_{4}} & S_{k}=L_{k}\\
\lambda_{-l,p_{3}}+\lambda_{-l,p_{4}} & S_{k}=L_{k}-k
\end{cases}\nonumber \\
 & =\frac{1}{2}\begin{cases}
\left|p-k-l\right|^{2}-\left|p-k-l+l\right|^{2}+\left|-p\right|^{2}-\left|-p+l\right|^{2} & S_{k}=L_{k}\\
\left|p+k\right|^{2}-\left|p+k+l\right|^{2}+\left|-p-l\right|^{2}-\left|-p-l+l\right|^{2} & S_{k}=L_{k}-k
\end{cases}\\
 & =\frac{1}{2}\begin{cases}
\left|p\right|^{2}-\left|p-k\right|^{2}+\left|p-l-k\right|^{2}-\left|p-l-k+k\right|^{2} & S_{k}=L_{k}\\
\left|p+k\right|^{2}-\left|p+k-k\right|^{2}+\left|p+l\right|^{2}-\left|p+l+k\right|^{2} & S_{k}=L_{k}-k
\end{cases}\nonumber \\
 & =\begin{cases}
\lambda_{k,p_{1}}+\lambda_{-k,p_{2}-k} & S_{k}=L_{k}\\
\lambda_{k,p_{1}}+\lambda_{-k,p_{2}} & S_{k}=L_{k}-k
\end{cases}.\nonumber 
\end{align}
The claim now follows by applying these identities to the estimates
\begin{align}
\left|1_{L_{-l}}\mleft(p_{3}-l\mright)s^{-\frac{1}{2}}\left\langle K_{-l}e_{p_{4}},e_{p_{3}-l}\right\rangle \right| & \leq C\frac{1_{L_{-l}}\mleft(p_{3}-l\mright)\hat{V}_{-l}k_{F}^{-1}}{\lambda_{-l,p_{3}-l}+\lambda_{-l,p_{4}}},\qquad\;\;S_{k}=L_{k},\\
\left|1_{L_{-l}}\mleft(p_{3}\mright)s^{-\frac{1}{2}}\left\langle K_{-l}e_{p_{4}},e_{p_{3}}\right\rangle \right| & \leq C\frac{1_{L_{-l}}\mleft(p_{3}\mright)\hat{V}_{-l}k_{F}^{-1}}{\lambda_{-l,p_{3}}+\lambda_{-l,p_{4}}},\qquad\qquad S_{k}=L_{k}-k,\nonumber 
\end{align}
which are given by Theorem \ref{them:OneBodyEstimates}.

$\hfill\square$

Below we will only use the simpler bound
\begin{equation}
\begin{cases}
\left|1_{L_{-l}}\mleft(p_{3}-l\mright)s^{-\frac{1}{2}}\left\langle K_{-l}e_{p_{4}},e_{p_{3}-l}\right\rangle \right| & S_{k}=L_{k}\\
\left|1_{L_{-l}}\mleft(p_{3}\mright)s^{-\frac{1}{2}}\left\langle K_{-l}e_{p_{4}},e_{p_{3}}\right\rangle \right| & S_{k}=L_{k}-k
\end{cases}\leq C\frac{\hat{V}_{-l}k_{F}^{-1}}{\sqrt{\lambda_{k,p_{1}}\lambda_{-l,p_{4}}}}\label{eq:calE1SimpleSplitInequality}
\end{equation}
but for the $\mathcal{E}_{k}^{2}$ terms the more general ones will
be needed.

\subsubsection*{Estimation of $\sum_{k\in\mathbb{Z}_{\ast}^{3}}\mathcal{E}_{k}^{1}\mleft(A_{k}\mleft(t\mright)\mright)$}

Now the main estimate of this subsection:
\begin{prop}
\label{prop:calE1Estimates}For any collection of symmetric operators
$\mleft(A_{k}\mright)$ and $\Psi\in\mathcal{H}_{N}$ it holds that
\begin{align*}
\sum_{k,l\in\mathbb{Z}_{\ast}^{3}}\sum_{p\in S_{k}\cap S_{l}}^{\sigma}\left|\left\langle \Psi,\tilde{c}_{p_{2},\sigma}^{\ast}b_{k}^{\ast}\mleft(A_{k}e_{p_{1}}\mright)b_{-l}^{\ast}\mleft(K_{-l}e_{p_{4}}\mright)\tilde{c}_{p_{3},\sigma}\Psi\right\rangle \right| & \leq C\sqrt{\sum_{k\in\mathbb{Z}_{\ast}^{3}}\max_{p\in L_{k}}\left\Vert A_{k}e_{p}\right\Vert ^{2}}\Vert\mleft(\mathcal{N}_{E}+1\mright)^{\frac{3}{2}}\Psi\Vert^{2}\\
\sum_{k,l\in\mathbb{Z}_{\ast}^{3}}\sum_{p\in S_{k}\cap S_{l}}^{\sigma}\left|\left\langle \Psi,\tilde{c}_{p_{2},\sigma}^{\ast}b_{k}^{\ast}\mleft(A_{k}e_{p_{1}}\mright)\left[b_{-l}\mleft(K_{-l}e_{p_{4}}\mright),\tilde{c}_{p_{3},\sigma}^{\ast}\right]^{\ast}\Psi\right\rangle \right| & \leq Ck_{F}^{-\frac{1}{2}}\sqrt{\sum_{k\in\mathbb{Z}_{\ast}^{3}}\Vert A_{k}h_{k}^{-\frac{1}{2}}\Vert_{\mathrm{HS}}^{2}}\left\Vert \mleft(\mathcal{N}_{E}+1\mright)\Psi\right\Vert ^{2}
\end{align*}
for a constant $C>0$ depending only on $\sum_{k\in\mathbb{Z}_{\ast}^{3}}\hat{V}_{k}^{2}$
and $s$.
\end{prop}

\textbf{Proof:} Using the triangle and Cauchy-Schwarz inequalities
and Proposition \ref{prop:bbastEstimates} we estimate
\begin{align}
 & \quad\,\sum_{k,l\in\mathbb{Z}_{\ast}^{3}}\sum_{p\in S_{k}\cap S_{l}}^{\sigma}\left|\left\langle \Psi,\tilde{c}_{p_{2},\sigma}^{\ast}b_{k}^{\ast}\mleft(A_{k}e_{p_{1}}\mright)b_{-l}^{\ast}\mleft(K_{-l}e_{p_{4}}\mright)\tilde{c}_{p_{3},\sigma}\Psi\right\rangle \right|\nonumber \\
 & \leq\sum_{k,l\in\mathbb{Z}_{\ast}^{3}}\sum_{p\in S_{k}\cap S_{l}}^{\sigma}\left\Vert b_{k}\mleft(A_{k}e_{p_{1}}\mright)\tilde{c}_{p_{2},\sigma}\Psi\right\Vert \left\Vert b_{-l}^{\ast}\mleft(K_{-l}e_{p_{4}}\mright)\tilde{c}_{p_{3},\sigma}\Psi\right\Vert \nonumber \\
 & \leq\sum_{k\in\mathbb{Z}_{\ast}^{3}}\sum_{p\in S_{k}}^{\sigma}\sum_{l\in\mathbb{Z}_{\ast}^{3}}1_{S_{l}}\mleft(p\mright)\left\Vert A_{k}e_{p_{1}}\right\Vert \left\Vert K_{-l}e_{p_{4}}\right\Vert \Vert\mathcal{N}_{k}^{\frac{1}{2}}\tilde{c}_{p_{2},\sigma}\Psi\Vert\Vert\mleft(\mathcal{N}_{-l}+1\mright)^{\frac{1}{2}}\tilde{c}_{p_{3},\sigma}\Psi\Vert\\
 & \leq\sum_{k\in\mathbb{Z}_{\ast}^{3}}\mleft(\max_{p\in L_{k}}\left\Vert A_{k}e_{p}\right\Vert \mright)\sum_{p\in S_{k}}\sqrt{\sum_{\sigma=1}^{s}\Vert\tilde{c}_{p_{3},\sigma}\mleft(\mathcal{N}_{E}+1\mright)^{\frac{1}{2}}\Psi\Vert^{2}}\sqrt{\sum_{l\in\mathbb{Z}_{\ast}^{3}}1_{S_{l}}\mleft(p\mright)\left\Vert K_{-l}e_{p_{4}}\right\Vert ^{2}}\sqrt{\sum_{l\in\mathbb{Z}_{\ast}^{3}}^{\sigma}1_{S_{l}}\mleft(p\mright)\Vert\tilde{c}_{p_{2},\sigma}\mathcal{N}_{k}^{\frac{1}{2}}\Psi\Vert^{2}}\nonumber \\
 & \leq\sum_{k\in\mathbb{Z}_{\ast}^{3}}\mleft(\max_{p\in L_{k}}\left\Vert A_{k}e_{p}\right\Vert \mright)\Vert\mathcal{N}_{E}^{\frac{1}{2}}\mathcal{N}_{k}^{\frac{1}{2}}\Psi\Vert\sqrt{\sum_{p\in S_{k}}^{\sigma}\Vert\tilde{c}_{p_{3},\sigma}\mleft(\mathcal{N}_{E}+1\mright)^{\frac{1}{2}}\Psi\Vert^{2}}\sqrt{\sum_{p\in S_{k}}\sum_{l\in\mathbb{Z}_{\ast}^{3}}1_{S_{l}}\mleft(p\mright)\left\Vert K_{-l}e_{p_{4}}\right\Vert ^{2}}\nonumber \\
 & \leq\sqrt{\sum_{k\in\mathbb{Z}_{\ast}^{3}}\max_{p\in L_{k}}\left\Vert A_{k}e_{p}\right\Vert ^{2}}\sqrt{\sum_{l\in\mathbb{Z}_{\ast}^{3}}\left\Vert K_{l}\right\Vert _{\mathrm{HS}}^{2}}\left\Vert \mleft(\mathcal{N}_{E}+1\mright)\Psi\right\Vert \sqrt{\sum_{k\in\mathbb{Z}_{\ast}^{3}}\Vert\mathcal{N}_{E}^{\frac{1}{2}}\mathcal{N}_{k}^{\frac{1}{2}}\Psi\Vert^{2}}\nonumber \\
 & =\sqrt{\sum_{k\in\mathbb{Z}_{\ast}^{3}}\max_{p\in L_{k}}\left\Vert A_{k}e_{p}\right\Vert ^{2}}\sqrt{\sum_{l\in\mathbb{Z}_{\ast}^{3}}\left\Vert K_{l}\right\Vert _{\mathrm{HS}}^{2}}\left\Vert \mleft(\mathcal{N}_{E}+1\mright)\Psi\right\Vert \Vert\mathcal{N}_{E}^{\frac{3}{2}}\Psi\Vert\nonumber 
\end{align}
and the first bound now follows by recalling that $\left\Vert K_{l}\right\Vert _{\mathrm{HS}}^{2}\leq C\hat{V}_{l}$.
For the second we have by the equations (\ref{eq:calE1CommutatorTermIdentity})
and (\ref{eq:calE1SimpleSplitInequality}) that
\begin{align}
 & \quad\,\sum_{k,l\in\mathbb{Z}_{\ast}^{3}}\sum_{p\in S_{k}\cap S_{l}}^{\sigma}\left|\left\langle \Psi,\tilde{c}_{p_{2},\sigma}^{\ast}b_{k}^{\ast}\mleft(A_{k}e_{p_{1}}\mright)\left[b_{-l}\mleft(K_{-l}e_{p_{4}}\mright),\tilde{c}_{p_{3},\sigma}^{\ast}\right]^{\ast}\Psi\right\rangle \right|\nonumber \\
 & \leq\sum_{k,l\in\mathbb{Z}_{\ast}^{3}}\sum_{p\in S_{k}\cap S_{l}}^{\sigma}\left\Vert \left[b_{-l}\mleft(K_{-l}e_{p_{4}}\mright),\tilde{c}_{p_{3},\sigma}^{\ast}\right]\tilde{c}_{p_{2},\sigma}\Psi\right\Vert \left\Vert b_{k}^{\ast}\mleft(A_{k}e_{p_{1}}\mright)\Psi\right\Vert \nonumber \\
 & \leq C\sum_{l\in\mathbb{Z}_{\ast}^{3}}\sum_{p\in S_{l}}^{\sigma}\sum_{k\in\mathbb{Z}_{\ast}^{3}}1_{S_{k}}\mleft(p\mright)\left\Vert A_{k}e_{p_{1}}\right\Vert \frac{\hat{V}_{-l}k_{F}^{-1}}{\sqrt{\lambda_{k,p_{1}}\lambda_{-l,p_{4}}}}\left\Vert \tilde{c}_{p_{3}\mp l,\sigma}\tilde{c}_{p_{2},\sigma}\Psi\right\Vert \Vert\mleft(\mathcal{N}_{k}+1\mright)^{\frac{1}{2}}\Psi\Vert\\
 & \leq Ck_{F}^{-1}\Vert\mleft(\mathcal{N}_{E}+1\mright)^{\frac{1}{2}}\Psi\Vert\sum_{p}\sum_{l\in\mathbb{Z}_{\ast}^{3}}\frac{1_{S_{l}}\mleft(p\mright)\hat{V}_{-l}}{\sqrt{\lambda_{-l,p_{4}}}}\sqrt{\sum_{k\in\mathbb{Z}_{\ast}^{3}}1_{S_{k}}\mleft(p\mright)\Vert A_{k}h_{k}^{-\frac{1}{2}}e_{p_{1}}\Vert^{2}}\sqrt{\sum_{k\in\mathbb{Z}_{\ast}^{3}}1_{S_{k}}\mleft(p\mright)\mleft(\sum_{\sigma=1}^{s}\left\Vert \tilde{c}_{p_{3}\mp l,\sigma}\tilde{c}_{p_{2},\sigma}\Psi\right\Vert \mright)^{2}}\nonumber \\
 & \leq Ck_{F}^{-1}\Vert\mleft(\mathcal{N}_{E}+1\mright)^{\frac{1}{2}}\Psi\Vert\sum_{p}\sqrt{\sum_{k\in\mathbb{Z}_{\ast}^{3}}1_{S_{k}}\mleft(p\mright)\Vert A_{k}h_{k}^{-\frac{1}{2}}e_{p_{1}}\Vert^{2}}\sqrt{\sum_{l\in\mathbb{Z}_{\ast}^{3}}1_{S_{l}}\mleft(p\mright)\frac{\hat{V}_{-l}^{2}}{\lambda_{-l,p_{4}}}}\sqrt{\sum_{l\in\mathbb{Z}_{\ast}^{3}}^{\sigma}1_{S_{l}}\mleft(p\mright)\Vert\tilde{c}_{p_{2},\sigma}\mathcal{N}_{E}^{\frac{1}{2}}\Psi\Vert^{2}}\nonumber \\
 & \leq Ck_{F}^{-1}\Vert\mleft(\mathcal{N}_{E}+1\mright)^{\frac{1}{2}}\Psi\Vert\left\Vert \mathcal{N}_{E}\Psi\right\Vert \sqrt{\sum_{k\in\mathbb{Z}_{\ast}^{3}}\sum_{p\in S_{k}}\Vert A_{k}h_{k}^{-\frac{1}{2}}e_{p_{1}}\Vert^{2}}\sqrt{\sum_{l\in\mathbb{Z}_{\ast}^{3}}\hat{V}_{-l}^{2}\sum_{p\in S_{l}}\frac{1}{\lambda_{-l,p_{4}}}}\nonumber \\
 & \leq Ck_{F}^{-1}\sqrt{\sum_{k\in\mathbb{Z}_{\ast}^{3}}\Vert A_{k}h_{k}^{-\frac{1}{2}}\Vert_{\mathrm{HS}}^{2}}\sqrt{\sum_{l\in\mathbb{Z}_{\ast}^{3}}\hat{V}_{l}^{2}\sum_{p\in L_{l}}\frac{1}{\lambda_{l,p}}}\Vert\mleft(\mathcal{N}_{E}+1\mright)^{\frac{1}{2}}\Psi\Vert\left\Vert \mathcal{N}_{E}\Psi\right\Vert \nonumber 
\end{align}
where we noted that $\left\Vert A_{k}e_{p_{1}}\right\Vert \lambda_{k,p_{1}}^{-\frac{1}{2}}=\Vert A_{k}h_{k}^{-\frac{1}{2}}e_{p_{1}}\Vert$
and also estimated
\begin{align}
\sum_{k\in\mathbb{Z}_{\ast}^{3}}1_{S_{k}}\mleft(p\mright)\mleft(\sum_{\sigma=1}^{s}\left\Vert \tilde{c}_{p_{3}\mp l,\sigma}\tilde{c}_{p_{2},\sigma}\Psi\right\Vert \mright)^{2} & \leq s\sum_{k\in\mathbb{Z}_{\ast}^{3}}^{\sigma}1_{S_{k}}\mleft(p\mright)\left\Vert \tilde{c}_{p_{3}\mp l,\sigma}\tilde{c}_{p_{2},\sigma}\Psi\right\Vert ^{2}\leq C\sum_{k\in\mathbb{Z}_{\ast}^{3}}^{\sigma,\tau}1_{S_{k}}\mleft(p\mright)\left\Vert \tilde{c}_{p_{3}\mp l,\tau}\tilde{c}_{p_{2},\sigma}\Psi\right\Vert ^{2}\\
 & \leq C\sum_{\sigma=1}^{s}\Vert\mathcal{N}_{E}^{\frac{1}{2}}\tilde{c}_{p_{2},\sigma}\Psi\Vert^{2}.\nonumber 
\end{align}
The claim follows as $\sum_{p\in L_{l}}\lambda_{l,p}^{-1}\leq Ck_{F}$.

$\hfill\square$

The bound on $\sum_{k\in\mathbb{Z}_{\ast}^{3}}\mathcal{E}_{k}^{1}\mleft(A_{k}\mleft(t\mright)\mright)$
of Theorem \ref{them:ExchangeTermsEstimates} now follows by our matrix
element estimates:
\begin{prop}
\label{prop:ParticularcalE1Estimates}For any $\Psi\in\mathcal{H}_{N}$
and $t\in\left[0,1\right]$ it holds that
\[
\sum_{k\in\mathbb{Z}_{\ast}^{3}}\left|\left\langle \Psi,\mathcal{E}_{k}^{1}\mleft(A_{k}\mleft(t\mright)\mright)\Psi\right\rangle \right|\leq C\sqrt{\sum_{k\in\mathbb{Z}_{\ast}^{3}}\hat{V}_{k}^{2}\min\left\{ \left|k\right|,k_{F}\right\} }\left\langle \Psi,\mleft(\mathcal{N}_{E}^{3}+1\mright)\Psi\right\rangle 
\]
for a constant $C>0$ depending only on $\sum_{k\in\mathbb{Z}_{\ast}^{3}}\hat{V}_{k}^{2}$
and $s$.
\end{prop}

\textbf{Proof:} By Theorem \ref{them:OneBodyEstimates} we have
\begin{equation}
\left|\left\langle e_{p},A_{k}\mleft(t\mright)e_{q}\right\rangle \right|\leq C\mleft(1+\hat{V}_{k}^{2}\mright)\hat{V}_{k}k_{F}^{-1},\quad k\in\mathbb{Z}_{\ast}^{3},\,p,q\in L_{k},
\end{equation}
so
\begin{align}
\sum_{k\in\mathbb{Z}_{\ast}^{3}}\max_{p\in L_{k}}\left\Vert A_{k}\mleft(t\mright)e_{p}\right\Vert ^{2} & =\sum_{k\in\mathbb{Z}_{\ast}^{3}}\max_{p\in L_{k}}\sum_{q\in L_{k}}\left|\left\langle e_{q},A_{k}\mleft(t\mright)e_{p}\right\rangle \right|^{2}\leq Ck_{F}^{-2}\sum_{k\in\mathbb{Z}_{\ast}^{3}}\mleft(1+\hat{V}_{k}^{2}\mright)^{2}\hat{V}_{k}^{2}\left|L_{k}\right|\\
 & \leq Ck_{F}^{-2}\sum_{k\in\mathbb{Z}_{\ast}^{3}}\mleft(\hat{V}_{k}^{2}+\hat{V}_{k}^{6}\mright)\min\left\{ k_{F}^{2}\left|k\right|,k_{F}^{3}\right\} \leq C\mleft(1+\Vert\hat{V}\Vert_{\infty}^{4}\mright)\sum_{k\in\mathbb{Z}_{\ast}^{3}}\hat{V}_{k}^{2}\min\left\{ \left|k\right|,k_{F}\right\} \nonumber 
\end{align}
where we used that $\left|L_{k}\right|\leq C\min\left\{ k_{F}^{2}\left|k\right|,k_{F}^{3}\right\} $.
Likewise
\begin{align}
\sum_{k\in\mathbb{Z}_{\ast}^{3}}\Vert A_{k}\mleft(t\mright)h_{k}^{-\frac{1}{2}}\Vert_{\mathrm{HS}}^{2} & =\sum_{k\in\mathbb{Z}_{\ast}^{3}}\sum_{p,q\in L_{k}}\left|\left\langle e_{p},A_{k}\mleft(t\mright)h_{k}^{-\frac{1}{2}}e_{q}\right\rangle \right|^{2}\leq Ck_{F}^{-2}\sum_{k\in\mathbb{Z}_{\ast}^{3}}\mleft(1+\hat{V}_{k}^{2}\mright)^{2}\hat{V}_{k}^{2}\left|L_{k}\right|\sum_{q\in L_{k}}\frac{1}{\lambda_{k,q}}\nonumber \\
 & \leq Ck_{F}\mleft(1+\Vert\hat{V}\Vert_{\infty}^{4}\mright)\sum_{k\in\mathbb{Z}_{\ast}^{3}}\hat{V}_{k}^{2}\min\left\{ \left|k\right|,k_{F}\right\} 
\end{align}
since $\sum_{q\in L_{k}}\lambda_{k,q}^{-1}\leq Ck_{F}$. Inserting
these estimates into Proposition \ref{prop:calE1Estimates} yields
the claim.

$\hfill\square$

\subsection{Analysis of $\mathcal{E}_{k}^{2}$ Terms}

Now we come to the terms
\begin{equation}
\mathcal{E}_{k}^{2}\mleft(B_{k}\mright)=\frac{1}{2}\sum_{l\in\mathbb{Z}_{\ast}^{3}}\sum_{p\in L_{k}}\sum_{q\in L_{l}}\left\{ b_{k}\mleft(B_{k}e_{p}\mright),\left\{ \varepsilon_{-k,-l}\mleft(e_{-p};e_{-q}\mright),b_{l}^{\ast}\mleft(K_{l}e_{q}\mright)\right\} \right\} .
\end{equation}
We will analyze these similarly to the $\mathcal{E}_{k}^{1}\mleft(A_{k}\mright)$
terms. Noting that
\begin{equation}
\varepsilon_{-k,-l}\mleft(e_{-p};e_{-q}\mright)=-\frac{1}{s}\sum_{\sigma=1}^{s}\mleft(\delta_{p,q}c_{-q+l,\sigma}c_{-p+k,\sigma}^{\ast}+\delta_{p-k,q-l}c_{-q,\sigma}^{\ast}c_{-p,\sigma}\mright)
\end{equation}
we find that $\mathcal{E}_{k}^{2}\mleft(B_{k}\mright)$ splits into
two sums as
\begin{align}
-2s\,\mathcal{E}_{k}^{2}\mleft(B_{k}\mright) & =\sum_{l\in\mathbb{Z}_{\ast}^{3}}\sum_{p\in L_{k}}^{\sigma}\sum_{q\in L_{l}}\left\{ b_{k}\mleft(B_{k}e_{p}\mright),\left\{ \delta_{p,q}c_{-q+l,\sigma}c_{-p+k,\sigma}^{\ast},b_{l}^{\ast}\mleft(K_{l}e_{q}\mright)\right\} \right\} \nonumber \\
 & +\sum_{l\in\mathbb{Z}_{\ast}^{3}}\sum_{p\in\mleft(L_{k}-k\mright)}^{\sigma}\sum_{q\in\mleft(L_{l}-l\mright)}\left\{ b_{k}\mleft(B_{k}e_{p+k}\mright),\left\{ \delta_{p,q}c_{-q-l,\sigma}^{\ast}c_{-p-k,\sigma},b_{l}^{\ast}\mleft(K_{l}e_{q+l}\mright)\right\} \right\} \\
 & =\sum_{l\in\mathbb{Z}_{\ast}^{3}}\sum_{p\in L_{k}\cap L_{l}}^{\sigma}\left\{ b_{k}\mleft(B_{k}e_{p}\mright),\left\{ c_{-p+l,\sigma}c_{-p+k,\sigma}^{\ast},b_{l}^{\ast}\mleft(K_{l}e_{p}\mright)\right\} \right\} \nonumber \\
 & +\sum_{l\in\mathbb{Z}_{\ast}^{3}}\sum_{p\in\mleft(L_{k}-k\mright)\cap\mleft(L_{l}-l\mright)}^{\sigma}\left\{ b_{k}\mleft(B_{k}e_{p+k}\mright),\left\{ c_{-p-l,\sigma}^{\ast}c_{-p-k,\sigma},b_{l}^{\ast}\mleft(K_{l}e_{p+l}\mright)\right\} \right\} \nonumber 
\end{align}
and again these share a common schematic form, namely
\begin{equation}
\sum_{l\in\mathbb{Z}_{\ast}^{3}}\sum_{p\in S_{k}\cap S_{l}}^{\sigma}\left\{ b_{k}\mleft(B_{k}e_{p_{1}}\mright),\left\{ \tilde{c}_{p_{2},\sigma}^{\ast}\tilde{c}_{p_{3},\sigma},b_{l}^{\ast}\mleft(K_{l}e_{p_{4}}\mright)\right\} \right\} 
\end{equation}
where the momenta are now
\begin{equation}
\mleft(p_{1},p_{2},p_{3},p_{4}\mright)=\begin{cases}
\mleft(p,-p+l,-p+k,p\mright) & S_{k}=L_{k}\\
\mleft(p+k,-p-l,-p-k,p+l\mright) & S_{k}=L_{k}-k
\end{cases}.
\end{equation}
Again $p_{1}$, $p_{3}$ only depend on $p$ and $k$ while $p_{2}$,
$p_{4}$ only depend on $p$ and $l$.

We normal order the summand: As
\begin{align}
 & \quad\;b_{k}\mleft(B_{k}e_{p_{1}}\mright)\left\{ \tilde{c}_{p_{2},\sigma}^{\ast}\tilde{c}_{p_{3},\sigma},b_{l}^{\ast}\mleft(K_{l}e_{p_{4}}\mright)\right\} \nonumber \\
 & =\tilde{c}_{p_{2},\sigma}^{\ast}b_{k}\mleft(B_{k}e_{p_{1}}\mright)\left\{ \tilde{c}_{p_{3},\sigma},b_{l}^{\ast}\mleft(K_{l}e_{p_{4}}\mright)\right\} +\left[b_{k}\mleft(B_{k}e_{p_{1}}\mright),\tilde{c}_{p_{2},\sigma}^{\ast}\right]\left\{ \tilde{c}_{p_{3},\sigma},b_{l}^{\ast}\mleft(K_{l}e_{p_{4}}\mright)\right\} \nonumber \\
 & =2\,\tilde{c}_{p_{2},\sigma}^{\ast}b_{k}\mleft(B_{k}e_{p_{1}}\mright)b_{l}^{\ast}\mleft(K_{l}e_{p_{4}}\mright)\tilde{c}_{p_{3},\sigma}+\tilde{c}_{p_{2},\sigma}^{\ast}b_{k}\mleft(B_{k}e_{p_{1}}\mright)\left[b_{l}\mleft(K_{l}e_{p_{4}}\mright),\tilde{c}_{p_{3},\sigma}^{\ast}\right]^{\ast}\nonumber \\
 & +2\left[b_{k}\mleft(B_{k}e_{p_{1}}\mright),\tilde{c}_{p_{2},\sigma}^{\ast}\right]b_{l}^{\ast}\mleft(K_{l}e_{p_{4}}\mright)\tilde{c}_{p_{3},\sigma}+\left[b_{k}\mleft(B_{k}e_{p_{1}}\mright),\tilde{c}_{p_{2},\sigma}^{\ast}\right]\left[b_{l}\mleft(K_{l}e_{p_{4}}\mright),\tilde{c}_{p_{3},\sigma}^{\ast}\right]^{\ast}\\
 & =2\,\tilde{c}_{p_{2},\sigma}^{\ast}b_{l}^{\ast}\mleft(K_{l}e_{p_{4}}\mright)b_{k}\mleft(B_{k}e_{p_{1}}\mright)\tilde{c}_{p_{3},\sigma}+2\,\tilde{c}_{p_{2},\sigma}^{\ast}\left[b_{k}\mleft(B_{k}e_{p_{1}}\mright),b_{l}^{\ast}\mleft(K_{l}e_{p_{4}}\mright)\right]\tilde{c}_{p_{3},\sigma}\nonumber \\
 & +\tilde{c}_{p_{2},\sigma}^{\ast}\left[b_{l}\mleft(K_{l}e_{p_{4}}\mright),\tilde{c}_{p_{3},\sigma}^{\ast}\right]^{\ast}b_{k}\mleft(B_{k}e_{p_{1}}\mright)+\tilde{c}_{p_{2},\sigma}^{\ast}\left[b_{k}\mleft(B_{k}e_{p_{1}}\mright),\left[b_{l}\mleft(K_{l}e_{p_{4}}\mright),\tilde{c}_{p_{3},\sigma}^{\ast}\right]^{\ast}\right]\nonumber \\
 & +2\,b_{l}^{\ast}\mleft(K_{l}e_{p_{4}}\mright)\left[b_{k}\mleft(B_{k}e_{p_{1}}\mright),\tilde{c}_{p_{2},\sigma}^{\ast}\right]\tilde{c}_{p_{3},\sigma}+2\left[b_{l}\mleft(K_{l}e_{p_{4}}\mright),\left[b_{k}\mleft(B_{k}e_{p_{1}}\mright),\tilde{c}_{p_{2},\sigma}^{\ast}\right]^{\ast}\right]^{\ast}\tilde{c}_{p_{3},\sigma}\nonumber \\
 & -\left[b_{l}\mleft(K_{l}e_{p_{4}}\mright),\tilde{c}_{p_{3},\sigma}^{\ast}\right]^{\ast}\left[b_{k}\mleft(B_{k}e_{p_{1}}\mright),\tilde{c}_{p_{2},\sigma}^{\ast}\right]+\left\{ \left[b_{k}\mleft(B_{k}e_{p_{1}}\mright),\tilde{c}_{p_{2},\sigma}^{\ast}\right],\left[b_{l}\mleft(K_{l}e_{p_{4}}\mright),\tilde{c}_{p_{3},\sigma}^{\ast}\right]^{\ast}\right\} \nonumber 
\end{align}
and simply
\begin{align}
 & \quad\;\left\{ \tilde{c}_{p_{2},\sigma}^{\ast}\tilde{c}_{p_{3},\sigma},b_{l}^{\ast}\mleft(K_{l}e_{p_{4}}\mright)\right\} b_{k}\mleft(B_{k}e_{p_{1}}\mright)=\tilde{c}_{p_{2},\sigma}^{\ast}\left\{ \tilde{c}_{p_{3},\sigma},b_{l}^{\ast}\mleft(K_{l}e_{p_{4}}\mright)\right\} b_{k}\mleft(B_{k}e_{p_{1}}\mright)\\
 & =2\,\tilde{c}_{p_{2},\sigma}^{\ast}b_{l}^{\ast}\mleft(K_{l}e_{p_{4}}\mright)b_{k}\mleft(B_{k}e_{p_{1}}\mright)\tilde{c}_{p_{3},\sigma}+\tilde{c}_{p_{2},\sigma}^{\ast}\left[b_{l}\mleft(K_{l}e_{p_{4}}\mright),\tilde{c}_{p_{3},\sigma}^{\ast}\right]^{\ast}b_{k}\mleft(B_{k}e_{p_{1}}\mright)\nonumber 
\end{align}
the summand decomposes into 8 schematic forms as
\begin{align}
 & \quad\left\{ b_{k}\mleft(B_{k}e_{p_{1}}\mright),\left\{ \tilde{c}_{p_{2},\sigma}^{\ast}\tilde{c}_{p_{3},\sigma},b_{l}^{\ast}\mleft(K_{l}e_{p_{4}}\mright)\right\} \right\} \nonumber \\
 & =4\,\tilde{c}_{p_{2},\sigma}^{\ast}b_{l}^{\ast}\mleft(K_{l}e_{p_{4}}\mright)b_{k}\mleft(B_{k}e_{p_{1}}\mright)\tilde{c}_{p_{3},\sigma}+2\,\tilde{c}_{p_{2},\sigma}^{\ast}\left[b_{k}\mleft(B_{k}e_{p_{1}}\mright),b_{l}^{\ast}\mleft(K_{l}e_{p_{4}}\mright)\right]\tilde{c}_{p_{3},\sigma}\nonumber \\
 & +2\,\tilde{c}_{p_{2},\sigma}^{\ast}\left[b_{l}\mleft(K_{l}e_{p_{4}}\mright),\tilde{c}_{p_{3},\sigma}^{\ast}\right]^{\ast}b_{k}\mleft(B_{k}e_{p_{1}}\mright)+2\,b_{l}^{\ast}\mleft(K_{l}e_{p_{4}}\mright)\left[b_{k}\mleft(B_{k}e_{p_{1}}\mright),\tilde{c}_{p_{2},\sigma}^{\ast}\right]\tilde{c}_{p_{3},\sigma}\\
 & +\tilde{c}_{p_{2},\sigma}^{\ast}\left[b_{k}\mleft(B_{k}e_{p_{1}}\mright),\left[b_{l}\mleft(K_{l}e_{p_{4}}\mright),\tilde{c}_{p_{3},\sigma}^{\ast}\right]^{\ast}\right]+2\left[b_{l}\mleft(K_{l}e_{p_{4}}\mright),\left[b_{k}\mleft(B_{k}e_{p_{1}}\mright),\tilde{c}_{p_{2},\sigma}^{\ast}\right]^{\ast}\right]^{\ast}\tilde{c}_{p_{3},\sigma}\nonumber \\
 & -\left[b_{l}\mleft(K_{l}e_{p_{4}}\mright),\tilde{c}_{p_{3},\sigma}^{\ast}\right]^{\ast}\left[b_{k}\mleft(B_{k}e_{p_{1}}\mright),\tilde{c}_{p_{2},\sigma}^{\ast}\right]+\left\{ \left[b_{k}\mleft(B_{k}e_{p_{1}}\mright),\tilde{c}_{p_{2},\sigma}^{\ast}\right],\left[b_{l}\mleft(K_{l}e_{p_{4}}\mright),\tilde{c}_{p_{3},\sigma}^{\ast}\right]^{\ast}\right\} .\nonumber 
\end{align}
Of these it should be noted that only the last one is proportional
to a constant (i.e. does not contain any creation or annihilation
operators). As the rest annihilate $\psi_{F}$, it follows that (when
summed) the constant term yields precisely $\left\langle \psi_{F},\mathcal{E}_{k}^{2}\mleft(B_{k}\mright)\psi_{F}\right\rangle $,
whence bounding the other terms amounts to estimating the operator
\begin{equation}
\mathcal{E}_{k}^{2}\mleft(B_{k}\mright)-\left\langle \psi_{F},\mathcal{E}_{k}^{2}\mleft(B_{k}\mright)\psi_{F}\right\rangle 
\end{equation}
as in the statement of Theorem \ref{them:ExchangeTermsEstimates}.

\subsubsection*{Estimation of the Top Terms}

We begin by bounding the ``top'' terms
\[
\sum_{k,l\in\mathbb{Z}_{\ast}^{3}}\sum_{p\in S_{k}\cap S_{l}}^{\sigma}\tilde{c}_{p_{2},\sigma}^{\ast}b_{l}^{\ast}\mleft(K_{l}e_{p_{4}}\mright)b_{k}\mleft(B_{k}e_{p_{1}}\mright)\tilde{c}_{p_{3},\sigma}\quad\text{and}\quad\sum_{k,l\in\mathbb{Z}_{\ast}^{3}}\sum_{p\in S_{k}\cap S_{l}}^{\sigma}\tilde{c}_{p_{2},\sigma}^{\ast}\left[b_{k}\mleft(B_{k}e_{p_{1}}\mright),b_{l}^{\ast}\mleft(K_{l}e_{p_{4}}\mright)\right]\tilde{c}_{p_{3},\sigma}.
\]
By the quasi-bosonic commutation relations, the commutator term reduces
to
\begin{align}
 & \quad\sum_{k,l\in\mathbb{Z}_{\ast}^{3}}\sum_{p\in S_{k}\cap S_{l}}^{\sigma}\tilde{c}_{p_{2},\sigma}^{\ast}\left[b_{k}\mleft(B_{k}e_{p_{1}}\mright),b_{l}^{\ast}\mleft(K_{l}e_{p_{4}}\mright)\right]\tilde{c}_{p_{3},\sigma}\\
 & =\sum_{k\in\mathbb{Z}_{\ast}^{3}}\sum_{p\in S_{k}}^{\sigma}\left\langle B_{k}e_{p_{1}},K_{k}e_{p_{1}}\right\rangle \tilde{c}_{p_{3},\sigma}^{\ast}\tilde{c}_{p_{3},\sigma}+\sum_{k,l\in\mathbb{Z}_{\ast}^{3}}\sum_{p\in S_{k}\cap S_{l}}^{\sigma}\tilde{c}_{p_{2},\sigma}^{\ast}\varepsilon_{k,l}\mleft(B_{k}e_{p_{1}};K_{l}e_{p_{4}}\mright)\tilde{c}_{p_{3},\sigma}\nonumber 
\end{align}
where we used that $p_{1}=p_{4}$ and $p_{2}=p_{3}$ when $k=l$.
Now, the exchange correction of the second sum splits as
\begin{align}
-s\,\varepsilon_{k,l}\mleft(B_{k}e_{p_{1}};K_{l}e_{p_{4}}\mright) & =\sum_{q\in L_{k}}^{\tau}\sum_{q'\in L_{l}}\left\langle B_{k}e_{p_{1}},e_{q}\right\rangle \left\langle e_{q'},K_{l}e_{p_{4}}\right\rangle \mleft(\delta_{q,q'}c_{q'-l,\tau}c_{q-k,\tau}^{\ast}+\delta_{q-k,q'-l}c_{q',\tau}^{\ast}c_{q,\tau}\mright)\nonumber \\
 & =\sum_{q\in L_{k}\cap L_{l}}^{\tau}\left\langle B_{k}e_{p_{1}},e_{q}\right\rangle \left\langle e_{q},K_{l}e_{p_{4}}\right\rangle \tilde{c}_{q-l,\tau}^{\ast}\tilde{c}_{q-k,\tau}\\
 & +\sum_{q\in\mleft(L_{k}-k\mright)\cap\mleft(L_{l}-l\mright)}^{\tau}\left\langle B_{k}e_{p_{1}},e_{q+k}\right\rangle \left\langle e_{q+l},K_{l}e_{p_{4}}\right\rangle \tilde{c}_{q+l,\tau}^{\ast}\tilde{c}_{q+k,\tau}\nonumber 
\end{align}
which are both of the schematic form $\sum_{q\in S_{k}^{\prime}\cap S_{l}^{\prime}}^{\tau}\left\langle B_{k}e_{p_{1}},e_{q_{1}}\right\rangle \left\langle e_{q_{4}},K_{l}e_{p_{4}}\right\rangle \tilde{c}_{q_{2},\tau}^{\ast}\tilde{c}_{q_{3},\tau}$.

To estimate $\sum_{k,l\in\mathbb{Z}_{\ast}^{3}}\sum_{p\in S_{k}\cap S_{l}}^{\sigma}\tilde{c}_{p_{2},\sigma}^{\ast}\varepsilon_{k,l}\mleft(B_{k}e_{p_{1}};K_{l}e_{p_{4}}\mright)\tilde{c}_{p_{3},\sigma}$
it thus suffices to consider
\begin{equation}
\sum_{k,l\in\mathbb{Z}_{\ast}^{3}}\sum_{p\in S_{k}\cap S_{l}}^{\sigma}\sum_{q\in S_{k}^{\prime}\cap S_{l}^{\prime}}^{\tau}\left\langle B_{k}e_{p_{1}},e_{q_{1}}\right\rangle \left\langle e_{q_{4}},K_{l}e_{p_{4}}\right\rangle \tilde{c}_{p_{2},\sigma}^{\ast}\tilde{c}_{q_{2},\tau}^{\ast}\tilde{c}_{q_{3},\tau}\tilde{c}_{p_{3},\sigma}.
\end{equation}
The estimates for the top terms are as follows:
\begin{prop}
\label{prop:calE2TopEstimates}For any collection of symmetric operators
$\mleft(B_{k}\mright)$ and $\Psi\in\mathcal{H}_{N}$ it holds that
\begin{align*}
\sum_{k,l\in\mathbb{Z}_{\ast}^{3}}\sum_{p\in S_{k}\cap S_{l}}^{\sigma}\left|\left\langle \Psi,\tilde{c}_{p_{2},\sigma}^{\ast}b_{l}^{\ast}\mleft(K_{l}e_{p_{4}}\mright)b_{k}\mleft(B_{k}e_{p_{1}}\mright)\tilde{c}_{p_{3},\sigma}\Psi\right\rangle \right| & \leq C\sqrt{\sum_{k\in\mathbb{Z}_{\ast}^{3}}\max_{p\in L_{k}}\left\Vert B_{k}e_{p}\right\Vert ^{2}}\Vert\mathcal{N}_{E}^{\frac{3}{2}}\Psi\Vert^{2}\\
\sum_{k,l\in\mathbb{Z}_{\ast}^{3}}\sum_{p\in S_{k}\cap S_{l}}^{\sigma}\left|\left\langle \Psi,\tilde{c}_{p_{2},\sigma}^{\ast}\left[b_{k}\mleft(B_{k}e_{p_{1}}\mright),b_{l}^{\ast}\mleft(K_{l}e_{p_{4}}\mright)\right]\tilde{c}_{p_{3},\sigma}\Psi\right\rangle \right| & \leq C\sqrt{\sum_{k\in\mathbb{Z}_{\ast}^{3}}\sum_{p\in L_{k}}\max_{q\in L_{k}}\left|\left\langle e_{p},B_{k}e_{q}\right\rangle \right|^{2}}\left\Vert \mathcal{N}_{E}\Psi\right\Vert ^{2}
\end{align*}
for a constant $C>0$ depending only on $\sum_{k\in\mathbb{Z}_{\ast}^{3}}\hat{V}_{k}^{2}$
and $s$.
\end{prop}

\textbf{Proof:} The first term we can estimate as in Proposition \ref{prop:calE1Estimates}
by
\begin{align}
 & \quad\,\sum_{k,l\in\mathbb{Z}_{\ast}^{3}}\sum_{p\in S_{k}\cap S_{l}}^{\sigma}\left|\left\langle \Psi,\tilde{c}_{p_{2},\sigma}^{\ast}b_{l}^{\ast}\mleft(K_{l}e_{p_{4}}\mright)b_{k}\mleft(B_{k}e_{p_{1}}\mright)\tilde{c}_{p_{3},\sigma}\Psi\right\rangle \right|\nonumber \\
 & \leq\sum_{k,l\in\mathbb{Z}_{\ast}^{3}}\sum_{p\in S_{k}\cap S_{l}}^{\sigma}\left\Vert b_{l}\mleft(K_{l}e_{p_{4}}\mright)\tilde{c}_{p_{2},\sigma}\Psi\right\Vert \left\Vert b_{k}\mleft(B_{k}e_{p_{1}}\mright)\tilde{c}_{p_{3},\sigma}\Psi\right\Vert \nonumber \\
 & \leq\sum_{k\in\mathbb{Z}_{\ast}^{3}}\sum_{p\in S_{k}}^{\sigma}\sum_{l\in\mathbb{Z}_{\ast}^{3}}1_{S_{l}}\mleft(p\mright)\left\Vert B_{k}e_{p_{1}}\right\Vert \left\Vert K_{l}e_{p_{4}}\right\Vert \Vert\mathcal{N}_{l}^{\frac{1}{2}}\tilde{c}_{p_{2},\sigma}\Psi\Vert\Vert\mathcal{N}_{k}^{\frac{1}{2}}\tilde{c}_{p_{3},\sigma}\Psi\Vert\\
 & \leq\sum_{k\in\mathbb{Z}_{\ast}^{3}}\mleft(\max_{p\in L_{k}}\left\Vert B_{k}e_{p}\right\Vert \mright)\sum_{p\in S_{k}}\sqrt{\sum_{\sigma=1}^{s}\Vert\tilde{c}_{p_{3},\sigma}\mathcal{N}_{k}^{\frac{1}{2}}\Psi\Vert^{2}}\sqrt{\sum_{l\in\mathbb{Z}_{\ast}^{3}}1_{S_{l}}\mleft(p\mright)\left\Vert K_{l}e_{p_{4}}\right\Vert ^{2}}\sqrt{\sum_{l\in\mathbb{Z}_{\ast}^{3}}^{\sigma}1_{S_{l}}\mleft(p\mright)\Vert\tilde{c}_{p_{2},\sigma}\mathcal{N}_{E}^{\frac{1}{2}}\Psi\Vert^{2}}\nonumber \\
 & \leq\left\Vert \mathcal{N}_{E}\Psi\right\Vert \sum_{k\in\mathbb{Z}_{\ast}^{3}}\mleft(\max_{p\in L_{k}}\left\Vert B_{k}e_{p}\right\Vert \mright)\sqrt{\sum_{p\in S_{k}}^{\sigma}\Vert\tilde{c}_{p_{3},\sigma}\mathcal{N}_{k}^{\frac{1}{2}}\Psi\Vert^{2}}\sqrt{\sum_{p\in S_{k}}\sum_{l\in\mathbb{Z}_{\ast}^{3}}1_{S_{l}}\mleft(p\mright)\left\Vert K_{l}e_{p_{4}}\right\Vert ^{2}}\nonumber \\
 & \leq\sqrt{\sum_{l\in\mathbb{Z}_{\ast}^{3}}\left\Vert K_{l}\right\Vert _{\mathrm{HS}}^{2}}\left\Vert \mathcal{N}_{E}\Psi\right\Vert \sum_{k\in\mathbb{Z}_{\ast}^{3}}\mleft(\max_{p\in L_{k}}\left\Vert B_{k}e_{p}\right\Vert \mright)\Vert\mathcal{N}_{E}^{\frac{1}{2}}\mathcal{N}_{k}^{\frac{1}{2}}\Psi\Vert\nonumber \\
 & \leq\sqrt{\sum_{k\in\mathbb{Z}_{\ast}^{3}}\max_{p\in L_{k}}\left\Vert B_{k}e_{p}\right\Vert ^{2}}\sqrt{\sum_{l\in\mathbb{Z}_{\ast}^{3}}\left\Vert K_{l}\right\Vert _{\mathrm{HS}}^{2}}\left\Vert \mathcal{N}_{E}\Psi\right\Vert \Vert\mathcal{N}_{E}^{\frac{3}{2}}\Psi\Vert.\nonumber 
\end{align}
For the commutator term we first consider $\sum_{k\in\mathbb{Z}_{\ast}^{3}}\sum_{p\in S_{k}}^{\sigma}\left\langle B_{k}e_{p_{1}},K_{k}e_{p_{1}}\right\rangle \tilde{c}_{p_{3},\sigma}^{\ast}\tilde{c}_{p_{3},\sigma}$:
This is trivially bounded by
\begin{align}
\sum_{k\in\mathbb{Z}_{\ast}^{3}}\sum_{p\in S_{k}}^{\sigma}\left|\left\langle B_{k}e_{p_{1}},K_{k}e_{p_{1}}\right\rangle \left\langle \Psi,\tilde{c}_{p_{3},\sigma}^{\ast}\tilde{c}_{p_{3},\sigma}\Psi\right\rangle \right| & \leq\sum_{k\in\mathbb{Z}_{\ast}^{3}}\max_{p\in L_{k}}\left|\left\langle B_{k}e_{p},K_{k}e_{p}\right\rangle \right|\sum_{p\in S_{k}}^{\sigma}\left\langle \Psi,\tilde{c}_{p_{3},\sigma}^{\ast}\tilde{c}_{p_{3},\sigma}\Psi\right\rangle \\
 & \leq\sum_{k\in\mathbb{Z}_{\ast}^{3}}\max_{p\in L_{k}}\left|\left\langle e_{p},B_{k}K_{k}e_{p}\right\rangle \right|\left\langle \Psi,\mathcal{N}_{E}\Psi\right\rangle \nonumber 
\end{align}
and by the matrix element estimate for $K_{k}$ of Theorem \ref{them:OneBodyEstimates}
we have for any $p\in L_{k}$ that
\begin{align}
\left|\left\langle B_{k}e_{p},K_{k}e_{p}\right\rangle \right| & \leq\sum_{q\in L_{k}}\left|\left\langle B_{k}e_{p},e_{q}\right\rangle \right|\left|\left\langle e_{q},K_{k}e_{p}\right\rangle \right|\leq C\sum_{q\in L_{k}}\left|\left\langle e_{p},B_{k}e_{q}\right\rangle \right|\frac{\hat{V}_{k}k_{F}^{-1}}{\lambda_{k,q}+\lambda_{k,p}}\\
 & \leq C\hat{V}_{k}k_{F}^{-1}\mleft(\max_{q\in L_{k}}\left|\left\langle e_{p},B_{k}e_{q}\right\rangle \right|\mright)\sum_{q\in L_{k}}\frac{1}{\lambda_{k,q}}\leq C\hat{V}_{k}\max_{q\in L_{k}}\left|\left\langle e_{p},B_{k}e_{q}\right\rangle \right|\nonumber 
\end{align}
since $\sum_{q\in L_{k}}\lambda_{k,q}^{-1}\leq Ck_{F}$. Consequently
\begin{align}
\sum_{k\in\mathbb{Z}_{\ast}^{3}}\sum_{p\in S_{k}}^{\sigma}\left|\left\langle B_{k}e_{p_{1}},K_{k}e_{p_{1}}\right\rangle \left\langle \Psi,\tilde{c}_{p_{3},\sigma}^{\ast}\tilde{c}_{p_{3},\sigma}\Psi\right\rangle \right| & \leq C\sum_{k\in\mathbb{Z}_{\ast}^{3}}\hat{V}_{k}\mleft(\max_{p,q\in L_{k}}\left|\left\langle e_{p},B_{k}e_{q}\right\rangle \right|\mright)\left\langle \Psi,\mathcal{N}_{E}\Psi\right\rangle \\
 & \leq C\sqrt{\sum_{k\in\mathbb{Z}_{\ast}^{3}}\hat{V}_{k}^{2}}\sqrt{\sum_{k\in\mathbb{Z}_{\ast}^{3}}\max_{p,q\in L_{k}}\left|\left\langle e_{p},B_{k}e_{q}\right\rangle \right|^{2}}\left\langle \Psi,\mathcal{N}_{E}\Psi\right\rangle \nonumber 
\end{align}
and clearly $\max_{p,q\in L_{k}}\left|\left\langle e_{p},B_{k}e_{q}\right\rangle \right|^{2}\leq\sum_{p\in L_{k}}\max_{q\in L_{k}}\left|\left\langle e_{p},B_{k}e_{q}\right\rangle \right|^{2}$.
Finally
\begin{align}
 & \quad\,\sum_{k,l\in\mathbb{Z}_{\ast}^{3}}\sum_{p\in S_{k}\cap S_{l}}^{\sigma}\sum_{q\in S_{k}^{\prime}\cap S_{l}^{\prime}}^{\tau}\left|\left\langle B_{k}e_{p_{1}},e_{q_{1}}\right\rangle \left\langle e_{q_{4}},K_{l}e_{p_{4}}\right\rangle \left\langle \Psi,\tilde{c}_{p_{2},\sigma}^{\ast}\tilde{c}_{q_{2},\tau}^{\ast}\tilde{c}_{q_{3},\tau}\tilde{c}_{p_{3},\sigma}\Psi\right\rangle \right|\nonumber \\
 & \leq\sum_{k,l\in\mathbb{Z}_{\ast}^{3}}\sum_{p\in S_{k}\cap S_{l}}^{\sigma}\sum_{q\in S_{k}^{\prime}\cap S_{l}^{\prime}}^{\tau}\left|\left\langle B_{k}e_{p_{1}},e_{q_{1}}\right\rangle \right|\left|\left\langle e_{q_{4}},K_{l}e_{p_{4}}\right\rangle \right|\left\Vert \tilde{c}_{q_{2},\tau}\tilde{c}_{p_{2},\sigma}\Psi\right\Vert \left\Vert \tilde{c}_{q_{3},\tau}\tilde{c}_{p_{3},\sigma}\Psi\right\Vert \\
 & \leq\sqrt{\sum_{k,l\in\mathbb{Z}_{\ast}^{3}}\sum_{p\in S_{k}\cap S_{l}}^{\sigma}\sum_{q\in S_{k}^{\prime}\cap S_{l}^{\prime}}^{\tau}\left|\left\langle B_{k}e_{p_{1}},e_{q_{1}}\right\rangle \right|^{2}\left\Vert \tilde{c}_{q_{2},\tau}\tilde{c}_{p_{2},\sigma}\Psi\right\Vert ^{2}}\sqrt{\sum_{k,l\in\mathbb{Z}_{\ast}^{3}}\sum_{p\in S_{k}\cap S_{l}}^{\sigma}\sum_{q\in S_{k}^{\prime}\cap S_{l}^{\prime}}^{\tau}\left|\left\langle e_{q_{4}},K_{l}e_{p_{4}}\right\rangle \right|^{2}\left\Vert \tilde{c}_{q_{3},\tau}\tilde{c}_{p_{3},\sigma}\Psi\right\Vert ^{2}}\nonumber \\
 & \leq\sqrt{\sum_{k\in\mathbb{Z}_{\ast}^{3}}\sum_{p\in S_{k}}\max_{q\in L_{k}}\left|\left\langle e_{p_{1}},B_{k}e_{q}\right\rangle \right|^{2}\sum_{l\in\mathbb{Z}_{\ast}^{3}}^{\sigma}1_{S_{l}}\mleft(p\mright)\Vert\tilde{c}_{p_{2},\sigma}\mathcal{N}_{E}^{\frac{1}{2}}\Psi\Vert^{2}}\sqrt{\sum_{l\in\mathbb{Z}_{\ast}^{3}}\sum_{p\in S_{l}}\left\Vert K_{l}e_{p_{4}}\right\Vert ^{2}\sum_{k\in\mathbb{Z}_{\ast}^{3}}^{\sigma}1_{S_{k}}\mleft(p\mright)\left\Vert \tilde{c}_{p_{3},\sigma}\Psi\right\Vert ^{2}}\nonumber \\
 & \leq\sqrt{\sum_{k\in\mathbb{Z}_{\ast}^{3}}\sum_{p\in L_{k}}\max_{q\in L_{k}}\left|\left\langle e_{p},B_{k}e_{q}\right\rangle \right|^{2}}\sqrt{\sum_{l\in\mathbb{Z}_{\ast}^{3}}\left\Vert K_{l}\right\Vert _{\mathrm{HS}}^{2}}\Vert\mathcal{N}_{E}^{\frac{1}{2}}\Psi\Vert\left\Vert \mathcal{N}_{E}\Psi\right\Vert \nonumber 
\end{align}
whence the claim follows as $\left\Vert K_{l}\right\Vert _{\mathrm{HS}}\leq C\hat{V}_{l}$.

$\hfill\square$

\subsubsection*{Estimation of the Single Commutator Terms}

For the single commutator terms
\[
\sum_{k,l\in\mathbb{Z}_{\ast}^{3}}\sum_{p\in S_{k}\cap S_{l}}^{\sigma}\tilde{c}_{p_{2},\sigma}^{\ast}\left[b_{l}\mleft(K_{l}e_{p_{4}}\mright),\tilde{c}_{p_{3},\sigma}^{\ast}\right]^{\ast}b_{k}\mleft(B_{k}e_{p_{1}}\mright)\quad\text{and}\quad\sum_{k,l\in\mathbb{Z}_{\ast}^{3}}\sum_{p\in S_{k}\cap S_{l}}^{\sigma}b_{l}^{\ast}\mleft(K_{l}e_{p_{4}}\mright)\left[b_{k}\mleft(B_{k}e_{p_{1}}\mright),\tilde{c}_{p_{2},\sigma}^{\ast}\right]\tilde{c}_{p_{3},\sigma}
\]
we note that by equation (\ref{eq:bcastCommutator}), the commutator
$\left[b_{l}\mleft(K_{l}e_{p_{4}}\mright),\tilde{c}_{p_{3},\sigma}^{\ast}\right]$
is given by
\begin{equation}
\left[b_{l}\mleft(K_{l}e_{p_{4}}\mright),\tilde{c}_{p_{3},\sigma}^{\ast}\right]=\begin{cases}
-1_{L_{l}}\mleft(p_{3}+l\mright)s^{-\frac{1}{2}}\left\langle K_{l}e_{p_{4}},e_{p_{3}+l}\right\rangle \tilde{c}_{p_{3}+l,\sigma} & S_{k}=L_{k}\\
1_{L_{l}}\mleft(p_{3}\mright)s^{-\frac{1}{2}}\left\langle K_{l}e_{p_{4}},e_{p_{3}}\right\rangle \tilde{c}_{p_{3}-l,\sigma} & S_{k}=L_{k}-k
\end{cases}.\label{eq:Klep4cp3Commutator}
\end{equation}
The prefactors again obey an estimate as in Proposition \ref{prop:calE1KSplitEstimate}:
\begin{prop}
\label{prop:calE2KSplitEstimate}For any $k,l\in\mathbb{Z}_{\ast}^{3}$
and $p\in S_{k}\cap S_{l}$ it holds that
\[
\left|1_{L_{l}}\mleft(p_{3}+l\mright)s^{-\frac{1}{2}}\left\langle K_{l}e_{p_{4}},e_{p_{3}+l}\right\rangle \right|\leq C\hat{V}_{l}k_{F}^{-1}\frac{1_{L_{k}}\mleft(p_{2}+k\mright)1_{L_{l}}\mleft(p_{3}+l\mright)}{\sqrt{\lambda_{k,p_{1}}+\lambda_{k,p_{2}+k}}\sqrt{\lambda_{l,p_{3}+l}+\lambda_{l,p_{4}}}},\quad S_{k}=L_{k},
\]
and
\[
\left|1_{L_{l}}\mleft(p_{3}\mright)s^{-\frac{1}{2}}\left\langle K_{l}e_{p_{4}},e_{p_{3}}\right\rangle \right|\leq C\hat{V}_{l}k_{F}^{-1}\frac{1_{L_{k}}\mleft(p_{2}\mright)1_{L_{l}}\mleft(p_{3}\mright)}{\sqrt{\lambda_{k,p_{1}}+\lambda_{k,p_{2}}}\sqrt{\lambda_{l,p_{3}}+\lambda_{l,p_{4}}}},\quad S_{k}=L_{k}-k,
\]
for a constant $C>0$ depending only on $s$.
\end{prop}

The proof is essentially the same as that of Proposition \ref{prop:calE1KSplitEstimate}
(indeed, this proposition can be obtained directly from the former
by appropriate substition, but some care must be used since the $p_{i}$'s
differ in their definition).

For the single commutator terms we again only need the simpler bound
\begin{equation}
\begin{cases}
\left|1_{L_{l}}\mleft(p_{3}+l\mright)s^{-\frac{1}{2}}\left\langle K_{l}e_{p_{4}},e_{p_{3}+l}\right\rangle \right| & S_{k}=L_{k}\\
\left|1_{L_{l}}\mleft(p_{3}\mright)s^{-\frac{1}{2}}\left\langle K_{l}e_{p_{4}},e_{p_{3}}\right\rangle \right| & S_{k}=L_{k}-k
\end{cases}\leq C\frac{\hat{V}_{l}k_{F}^{-1}}{\sqrt{\lambda_{k,p_{1}}\lambda_{l,p_{4}}}}
\end{equation}
but the full one will be needed for the double commutator terms below.
Now the estimate:
\begin{prop}
\label{prop:calE2SingleCommutatorEstimates}For any collection of
symmetric operators $\mleft(B_{k}\mright)$ and $\Psi\in\mathcal{H}_{N}$
it holds that
\begin{align*}
\sum_{k,l\in\mathbb{Z}_{\ast}^{3}}\sum_{p\in S_{k}\cap S_{l}}^{\sigma}\left|\left\langle \Psi,\tilde{c}_{p_{2},\sigma}^{\ast}\left[b_{l}\mleft(K_{l}e_{p_{4}}\mright),\tilde{c}_{p_{3},\sigma}^{\ast}\right]^{\ast}b_{k}\mleft(B_{k}e_{p_{1}}\mright)\Psi\right\rangle \right| & \leq Ck_{F}^{-\frac{1}{2}}\sqrt{\sum_{k\in\mathbb{Z}_{\ast}^{3}}\Vert B_{k}h_{k}^{-\frac{1}{2}}\Vert_{\mathrm{HS}}^{2}}\left\Vert \mathcal{N}_{E}\Psi\right\Vert ^{2}\\
\sum_{k,l\in\mathbb{Z}_{\ast}^{3}}\sum_{p\in S_{k}\cap S_{l}}^{\sigma}\left|\left\langle \Psi,b_{l}^{\ast}\mleft(K_{l}e_{p_{4}}\mright)\left[b_{k}\mleft(B_{k}e_{p_{1}}\mright),\tilde{c}_{p_{2},\sigma}^{\ast}\right]\tilde{c}_{p_{3},\sigma}\Psi\right\rangle \right| & \leq C\sqrt{\sum_{k\in\mathbb{Z}_{\ast}^{3}}\sum_{p\in L_{k}}\max_{q\in L_{k}}\left|\left\langle e_{p},B_{k}e_{q}\right\rangle \right|^{2}}\left\Vert \mathcal{N}_{E}\Psi\right\Vert ^{2}
\end{align*}
for a constant $C>0$ depending only on $\sum_{k\in\mathbb{Z}_{\ast}^{3}}\hat{V}_{k}^{2}$
and $s$.
\end{prop}

\textbf{Proof:} As in the second estimate of Proposition \ref{prop:calE1Estimates}
we have
\begin{align}
 & \quad\,\sum_{k,l\in\mathbb{Z}_{\ast}^{3}}\sum_{p\in S_{k}\cap S_{l}}^{\sigma}\left|\left\langle \Psi,\tilde{c}_{p_{2},\sigma}^{\ast}\left[b_{l}\mleft(K_{l}e_{p_{4}}\mright),\tilde{c}_{p_{3},\sigma}^{\ast}\right]^{\ast}b_{k}\mleft(B_{k}e_{p_{1}}\mright)\Psi\right\rangle \right|\nonumber \\
 & \leq\sum_{k,l\in\mathbb{Z}_{\ast}^{3}}\sum_{p\in S_{k}\cap S_{l}}^{\sigma}\left\Vert \left[b_{l}\mleft(K_{l}e_{p_{4}}\mright),\tilde{c}_{p_{3},\sigma}^{\ast}\right]\tilde{c}_{p_{2},\sigma}\Psi\right\Vert \left\Vert b_{k}\mleft(B_{k}e_{p_{1}}\mright)\Psi\right\Vert \nonumber \\
 & \leq C\sum_{l\in\mathbb{Z}_{\ast}^{3}}\sum_{p\in S_{l}}^{\sigma}\sum_{k\in\mathbb{Z}_{\ast}^{3}}1_{S_{k}}\mleft(p\mright)\left\Vert B_{k}e_{p_{1}}\right\Vert \frac{\hat{V}_{l}k_{F}^{-1}}{\sqrt{\lambda_{k,p_{1}}\lambda_{l,p_{4}}}}\left\Vert \tilde{c}_{p_{3}\pm l,\sigma}\tilde{c}_{p_{2},\sigma}\Psi\right\Vert \Vert\mathcal{N}_{k}^{\frac{1}{2}}\Psi\Vert\\
 & \leq Ck_{F}^{-1}\Vert\mathcal{N}_{E}^{\frac{1}{2}}\Psi\Vert\sum_{p}\sum_{l\in\mathbb{Z}_{\ast}^{3}}\frac{1_{S_{l}}\mleft(p\mright)\hat{V}_{l}}{\sqrt{\lambda_{l,p_{4}}}}\sqrt{\sum_{k\in\mathbb{Z}_{\ast}^{3}}1_{S_{k}}\mleft(p\mright)\Vert B_{k}h_{k}^{-\frac{1}{2}}e_{p_{1}}\Vert^{2}}\sqrt{\sum_{k\in\mathbb{Z}_{\ast}^{3}}^{\sigma,\tau}1_{S_{k}}\mleft(p\mright)\left\Vert \tilde{c}_{p_{3}\pm l,\tau}\tilde{c}_{p_{2},\sigma}\Psi\right\Vert ^{2}}\nonumber \\
 & \leq Ck_{F}^{-1}\Vert\mathcal{N}_{E}^{\frac{1}{2}}\Psi\Vert\sum_{p}\sqrt{\sum_{k\in\mathbb{Z}_{\ast}^{3}}1_{S_{k}}\mleft(p\mright)\Vert B_{k}h_{k}^{-\frac{1}{2}}e_{p_{1}}\Vert^{2}}\sqrt{\sum_{l\in\mathbb{Z}_{\ast}^{3}}1_{S_{l}}\mleft(p\mright)\frac{\hat{V}_{l}^{2}}{\lambda_{l,p_{4}}}}\sqrt{\sum_{l\in\mathbb{Z}_{\ast}^{3}}^{\sigma}1_{S_{l}}\mleft(p\mright)\Vert\tilde{c}_{p_{2},\sigma}\mathcal{N}_{E}^{\frac{1}{2}}\Psi\Vert^{2}}\nonumber \\
 & \leq Ck_{F}^{-1}\Vert\mathcal{N}_{E}^{\frac{1}{2}}\Psi\Vert\left\Vert \mathcal{N}_{E}\Psi\right\Vert \sqrt{\sum_{k\in\mathbb{Z}_{\ast}^{3}}\sum_{p\in S_{k}}\Vert B_{k}h_{k}^{-\frac{1}{2}}e_{p_{1}}\Vert^{2}}\sqrt{\sum_{l\in\mathbb{Z}_{\ast}^{3}}\hat{V}_{l}^{2}\sum_{p\in S_{l}}\frac{1}{\lambda_{l,p_{4}}}}\nonumber \\
 & \leq Ck_{F}^{-\frac{1}{2}}\sqrt{\sum_{k\in\mathbb{Z}_{\ast}^{3}}\Vert B_{k}h_{k}^{-\frac{1}{2}}\Vert_{\mathrm{HS}}^{2}}\sqrt{\sum_{l\in\mathbb{Z}_{\ast}^{3}}\hat{V}_{l}^{2}}\Vert\mathcal{N}_{E}^{\frac{1}{2}}\Psi\Vert\left\Vert \mathcal{N}_{E}\Psi\right\Vert .\nonumber 
\end{align}
By equation (\ref{eq:bcastCommutator}) it holds that
\begin{equation}
\left[b_{k}\mleft(B_{k}e_{p_{1}}\mright),\tilde{c}_{p_{2},\sigma}^{\ast}\right]=\begin{cases}
-1_{L_{k}}\mleft(p_{2}+k\mright)s^{-\frac{1}{2}}\left\langle B_{k}e_{p_{1}},e_{p_{2}+k}\right\rangle \tilde{c}_{p_{2}+k,\sigma} & p\in B_{F}\\
1_{L_{k}}\mleft(p_{2}\mright)s^{-\frac{1}{2}}\left\langle B_{k}e_{p_{1}},e_{p_{2}}\right\rangle \tilde{c}_{p_{2}-k,\sigma} & p\in B_{F}^{c}
\end{cases}\label{eq:Bkep1cp2Commutator}
\end{equation}
so the second term can be bounded as
\begin{align}
 & \quad\,\sum_{k,l\in\mathbb{Z}_{\ast}^{3}}\sum_{p\in S_{k}\cap S_{l}}^{\sigma}\left|\left\langle \Psi,b_{l}^{\ast}\mleft(K_{l}e_{p_{4}}\mright)\left[b_{k}\mleft(B_{k}e_{p_{1}}\mright),\tilde{c}_{p_{2},\sigma}^{\ast}\right]\tilde{c}_{p_{3},\sigma}\Psi\right\rangle \right|\nonumber \\
 & \leq\sum_{k,l\in\mathbb{Z}_{\ast}^{3}}\sum_{p\in S_{k}\cap S_{l}}^{\sigma}\left\Vert b_{l}\mleft(K_{l}e_{p_{4}}\mright)\Psi\right\Vert \left\Vert \left[b_{k}\mleft(B_{k}e_{p_{1}}\mright),\tilde{c}_{p_{2},\sigma}^{\ast}\right]\tilde{c}_{p_{3},\sigma}\Psi\right\Vert \nonumber \\
 & \leq\sum_{k\in\mathbb{Z}_{\ast}^{3}}\sum_{p\in S_{k}}^{\sigma}\sum_{l\in\mathbb{Z}_{\ast}^{3}}1_{S_{l}}\mleft(p\mright)\mleft(\max_{q\in L_{k}}\left|\left\langle e_{p_{1}},B_{k}e_{q}\right\rangle \right|\mright)\left\Vert K_{l}e_{p_{4}}\right\Vert \Vert\mathcal{N}_{l}^{\frac{1}{2}}\Psi\Vert\left\Vert \tilde{c}_{p_{2}\pm k,\sigma}\tilde{c}_{p_{3},\sigma}\Psi\right\Vert \\
 & \leq\Vert\mathcal{N}_{E}^{\frac{1}{2}}\Psi\Vert\sum_{p}\sum_{k\in\mathbb{Z}_{\ast}^{3}}1_{S_{k}}\mleft(p\mright)\mleft(\max_{q\in L_{k}}\left|\left\langle e_{p_{1}},B_{k}e_{q}\right\rangle \right|\mright)\sqrt{\sum_{l\in\mathbb{Z}_{\ast}^{3}}1_{S_{l}}\mleft(p\mright)\left\Vert K_{l}e_{p_{4}}\right\Vert ^{2}}\sqrt{\sum_{l\in\mathbb{Z}_{\ast}^{3}}^{\sigma,\tau}1_{S_{l}}\mleft(p\mright)\left\Vert \tilde{c}_{p_{2}\pm k,\tau}\tilde{c}_{p_{3},\sigma}\Psi\right\Vert ^{2}}\nonumber \\
 & \leq\Vert\mathcal{N}_{E}^{\frac{1}{2}}\Psi\Vert\sum_{p}\sqrt{\sum_{l\in\mathbb{Z}_{\ast}^{3}}1_{S_{l}}\mleft(p\mright)\left\Vert K_{l}e_{p_{4}}\right\Vert ^{2}}\sqrt{\sum_{k\in\mathbb{Z}_{\ast}^{3}}1_{S_{k}}\mleft(p\mright)\mleft(\max_{q\in L_{k}}\left|\left\langle e_{p_{1}},B_{k}e_{q}\right\rangle \right|^{2}\mright)}\sqrt{\sum_{k\in\mathbb{Z}_{\ast}^{3}}^{\sigma}1_{S_{k}}\mleft(p\mright)\Vert\tilde{c}_{p_{3},\sigma}\mathcal{N}_{E}^{\frac{1}{2}}\Psi\Vert^{2}}\nonumber \\
 & \leq\Vert\mathcal{N}_{E}^{\frac{1}{2}}\Psi\Vert\left\Vert \mathcal{N}_{E}\Psi\right\Vert \sqrt{\sum_{l\in\mathbb{Z}_{\ast}^{3}}\sum_{p\in S_{l}}\left\Vert K_{l}e_{p_{4}}\right\Vert ^{2}}\sqrt{\sum_{k\in\mathbb{Z}_{\ast}^{3}}\sum_{p\in S_{k}}\max_{q\in L_{k}}\left|\left\langle e_{p_{1}},B_{k}e_{q}\right\rangle \right|^{2}}\nonumber \\
 & \leq\sqrt{\sum_{k\in\mathbb{Z}_{\ast}^{3}}\sum_{p\in L_{k}}\max_{q\in L_{k}}\left|\left\langle e_{p},B_{k}e_{q}\right\rangle \right|^{2}}\sqrt{\sum_{l\in\mathbb{Z}_{\ast}^{3}}\left\Vert K_{l}\right\Vert _{\mathrm{HS}}^{2}}\,\Vert\mathcal{N}_{E}^{\frac{1}{2}}\Psi\Vert\left\Vert \mathcal{N}_{E}\Psi\right\Vert .\nonumber 
\end{align}
$\hfill\square$

\subsubsection*{Estimation of the Double Commutator Terms}

Finally we have the double commutator terms
\[
\sum_{k,l\in\mathbb{Z}_{\ast}^{3}}\sum_{p\in S_{k}\cap S_{l}}^{\sigma}\tilde{c}_{p_{2},\sigma}^{\ast}\left[b_{k}\mleft(B_{k}e_{p_{1}}\mright),\left[b_{l}\mleft(K_{l}e_{p_{4}}\mright),\tilde{c}_{p_{3},\sigma}^{\ast}\right]^{\ast}\right],\quad\sum_{k,l\in\mathbb{Z}_{\ast}^{3}}\sum_{p\in S_{k}\cap S_{l}}^{\sigma}\left[b_{l}\mleft(K_{l}e_{p_{4}}\mright),\left[b_{k}\mleft(B_{k}e_{p_{1}}\mright),\tilde{c}_{p_{2},\sigma}^{\ast}\right]^{\ast}\right]^{\ast}\tilde{c}_{p_{3},\sigma}
\]
and
\[
\sum_{k,l\in\mathbb{Z}_{\ast}^{3}}\sum_{p\in S_{k}\cap S_{l}}^{\sigma}\left[b_{l}\mleft(K_{l}e_{p_{4}}\mright),\tilde{c}_{p_{3},\sigma}^{\ast}\right]^{\ast}\left[b_{k}\mleft(B_{k}e_{p_{1}}\mright),\tilde{c}_{p_{2},\sigma}^{\ast}\right].
\]
An identity for the iterated commutators is obtained by applying the
identity of equation (\ref{eq:bcastCommutator}) to itself: For any
$k,l\in\mathbb{Z}_{\ast}^{3}$, $\varphi\in\ell^{2}\mleft(L_{k}\mright)$,
$\psi\in\ell^{2}\mleft(L_{l}\mright)$ and $p\in\mathbb{Z}_{\ast}^{3}$
\begin{align}
\left[b_{k}\mleft(\varphi\mright),\left[b_{l}\mleft(\psi\mright),\tilde{c}_{p,\sigma}^{\ast}\right]^{\ast}\right] & =\begin{cases}
-1_{L_{l}}\mleft(p+l\mright)s^{-\frac{1}{2}}\left\langle e_{p+l},\psi\right\rangle \left[b_{k}\mleft(\varphi\mright),\tilde{c}_{p+l,\sigma}^{\ast}\right] & p\in B_{F}\\
1_{L_{l}}\mleft(p\mright)s^{-\frac{1}{2}}\left\langle e_{p},\psi\right\rangle \left[b_{k}\mleft(\varphi\mright),\tilde{c}_{p-l,\sigma}^{\ast}\right] & p\in B_{F}^{c}
\end{cases}\label{eq:bbcastastIdentity}\\
 & =\begin{cases}
-1_{L_{k}}\mleft(p+l\mright)1_{L_{l}}\mleft(p+l\mright)s^{-1}\left\langle \varphi,e_{p+l}\right\rangle \left\langle e_{p+l},\psi\right\rangle \tilde{c}_{p+l-k,\sigma} & p\in B_{F}\\
-1_{L_{k}}\mleft(p-l+k\mright)1_{L_{l}}\mleft(p\mright)s^{-1}\left\langle \varphi,e_{p-l+k}\right\rangle \left\langle e_{p},\psi\right\rangle \tilde{c}_{p-l+k,\sigma} & p\in B_{F}^{c}
\end{cases}.\nonumber 
\end{align}
The estimates are the following:
\begin{prop}
\label{prop:calE2DoubleCommutatorEstimates}For any collection of
symmetric operators $\mleft(B_{k}\mright)$ and $\Psi\in\mathcal{H}_{N}$
it holds that
\begin{align*}
\sum_{k,l\in\mathbb{Z}_{\ast}^{3}}\sum_{p\in S_{k}\cap S_{l}}^{\sigma}\left|\left\langle \Psi,\tilde{c}_{p_{2},\sigma}^{\ast}\left[b_{k}\mleft(B_{k}e_{p_{1}}\mright),\left[b_{l}\mleft(K_{l}e_{p_{4}}\mright),\tilde{c}_{p_{3},\sigma}^{\ast}\right]^{\ast}\right]\Psi\right\rangle \right|,\\
\sum_{k,l\in\mathbb{Z}_{\ast}^{3}}\sum_{p\in S_{k}\cap S_{l}}^{\sigma}\left|\left\langle \Psi,\left[b_{l}\mleft(K_{l}e_{p_{4}}\mright),\left[b_{k}\mleft(B_{k}e_{p_{1}}\mright),\tilde{c}_{p_{2},\sigma}^{\ast}\right]^{\ast}\right]^{\ast}\tilde{c}_{p_{3},\sigma}\Psi\right\rangle \right|,\\
\sum_{k,l\in\mathbb{Z}_{\ast}^{3}}\sum_{p\in S_{k}\cap S_{l}}^{\sigma}\left|\left\langle \Psi,\left[b_{l}\mleft(K_{l}e_{p_{4}}\mright),\tilde{c}_{p_{3},\sigma}^{\ast}\right]^{\ast}\left[b_{k}\mleft(B_{k}e_{p_{1}}\mright),\tilde{c}_{p_{2},\sigma}^{\ast}\right]\Psi\right\rangle \right|,
\end{align*}
are all bounded by $Ck_{F}^{-\frac{1}{2}}\sqrt{\sum_{k\in\mathbb{Z}_{\ast}^{3}}\max_{p\in L_{k}}\Vert h_{k}^{-\frac{1}{2}}B_{k}e_{p}\Vert^{2}}\left\langle \Psi,\mathcal{N}_{E}\Psi\right\rangle $
for a constant $C>0$ depending only on $\sum_{k\in\mathbb{Z}_{\ast}^{3}}\hat{V}_{k}^{2}$
and $s$.
\end{prop}

\textbf{Proof:} For these estimates we consider only the case $S_{k}=L_{k}$
for the sake of clarity, i.e. we let
\begin{equation}
\mleft(p_{1},p_{2},p_{3},p_{4}\mright)=\mleft(p,-p+l,-p+k,p\mright);
\end{equation}
the case $S_{k}=L_{k}-k$ can be handled by similar manipulations.

Using the identity of equation (\ref{eq:bbcastastIdentity}) we start
by estimating (by the bound of Proposition \ref{prop:calE2KSplitEstimate})
\begin{align}
 & \quad\,\sum_{k,l\in\mathbb{Z}_{\ast}^{3}}\sum_{p\in L_{k}\cap L_{l}}^{\sigma}\left|\left\langle \Psi,\tilde{c}_{p_{2},\sigma}^{\ast}\left[b_{k}\mleft(B_{k}e_{p_{1}}\mright),\left[b_{l}\mleft(K_{l}e_{p_{4}}\mright),\tilde{c}_{p_{3},\sigma}^{\ast}\right]^{\ast}\right]\Psi\right\rangle \right|\nonumber \\
 & =\sum_{k,l\in\mathbb{Z}_{\ast}^{3}}\sum_{p\in L_{k}\cap L_{l}}^{\sigma}\left|1_{L_{k}}\mleft(p_{3}+l\mright)1_{L_{l}}\mleft(p_{3}+l\mright)s^{-1}\left\langle B_{k}e_{p_{1}},e_{p_{3}+l}\right\rangle \left\langle e_{p_{3}+l},K_{l}e_{p_{4}}\right\rangle \left\langle \Psi,\tilde{c}_{p_{2},\sigma}^{\ast}\tilde{c}_{p_{3}+l-k,\sigma}\Psi\right\rangle \right|\nonumber \\
 & \leq C\sum_{k,l\in\mathbb{Z}_{\ast}^{3}}\sum_{p\in L_{k}\cap L_{l}}^{\sigma}1_{L_{k}}\mleft(p_{3}+l\mright)\left|\left\langle B_{k}e_{p_{1}},e_{p_{3}+l}\right\rangle \right|\frac{\hat{V}_{l}k_{F}^{-1}1_{L_{k}}\mleft(p_{2}+k\mright)1_{L_{l}}\mleft(p_{3}+l\mright)}{\sqrt{\lambda_{k,p_{1}}+\lambda_{k,p_{2}+k}}\sqrt{\lambda_{l,p_{3}+l}+\lambda_{l,p_{4}}}}\left\langle \Psi,\tilde{c}_{p_{2},\sigma}^{\ast}\tilde{c}_{p_{2},\sigma}\Psi\right\rangle \nonumber \\
 & \leq Ck_{F}^{-1}\sum_{l\in\mathbb{Z}_{\ast}^{3}}\hat{V}_{l}\sum_{p\in L_{l}}^{\sigma}\sqrt{\sum_{k\in\mathbb{Z}_{\ast}^{3}}1_{L_{k}}\mleft(p\mright)1_{L_{k}}\mleft(p_{3}+l\mright)\left|\left\langle e_{p},h_{k}^{-\frac{1}{2}}B_{k}e_{p_{3}+l}\right\rangle \right|^{2}}\sqrt{\sum_{k\in\mathbb{Z}_{\ast}^{3}}\frac{1_{L_{l}}\mleft(p_{3}+l\mright)}{\lambda_{l,p_{3}+l}}}\left\langle \Psi,\tilde{c}_{-p+l,\sigma}^{\ast}\tilde{c}_{-p+l,\sigma}\Psi\right\rangle \nonumber \\
 & \leq Ck_{F}^{-\frac{1}{2}}\sum_{l\in\mathbb{Z}_{\ast}^{3}}\hat{V}_{l}\sum_{p\in\mleft(L_{l}-l\mright)}^{\sigma}\sqrt{\sum_{k\in\mathbb{Z}_{\ast}^{3}}1_{L_{k}}\mleft(p+l\mright)1_{L_{k}}\mleft(p_{3}\mright)\left|\left\langle e_{p+l},h_{k}^{-\frac{1}{2}}B_{k}e_{p_{3}}\right\rangle \right|^{2}}\left\langle \Psi,\tilde{c}_{-p,\sigma}^{\ast}\tilde{c}_{-p,\sigma}\Psi\right\rangle \\
 & \leq Ck_{F}^{-\frac{1}{2}}\sum_{p\in B_{F}}^{\sigma}\sqrt{\sum_{l\in\mathbb{Z}_{\ast}^{3}}\hat{V}_{l}^{2}}\sqrt{\sum_{k,l\in\mathbb{Z}_{\ast}^{3}}1_{L_{k}}\mleft(p+l\mright)1_{L_{k}}\mleft(p_{3}\mright)\left|\left\langle e_{p+l},h_{k}^{-\frac{1}{2}}B_{k}e_{p_{3}}\right\rangle \right|^{2}}\left\langle \Psi,\tilde{c}_{-p,\sigma}^{\ast}\tilde{c}_{-p,\sigma}\Psi\right\rangle \nonumber \\
 & \leq Ck_{F}^{-\frac{1}{2}}\sqrt{\sum_{k\in\mathbb{Z}_{\ast}^{3}}\max_{p\in L_{k}}\Vert h_{k}^{-\frac{1}{2}}B_{k}e_{p}\Vert^{2}}\sqrt{\sum_{l\in\mathbb{Z}_{\ast}^{3}}\hat{V}_{l}^{2}}\left\langle \Psi,\mathcal{N}_{E}\Psi\right\rangle \nonumber 
\end{align}
where we used that $\sum_{k\in\mathbb{Z}_{\ast}^{3}}1_{L_{l}}\mleft(p_{3}+l\mright)\lambda_{l,p_{3}+l}^{-1}\leq\sum_{q\in L_{l}}\lambda_{l,q}^{-1}\leq Ck_{F}$.

From equation (\ref{eq:bbcastastIdentity}) we have
\begin{align*}
\left[b_{l}\mleft(K_{l}e_{p_{4}}\mright),\left[b_{k}\mleft(B_{k}e_{p_{1}}\mright),\tilde{c}_{p,\sigma}^{\ast}\right]^{\ast}\right] & =-1_{L_{l}}\mleft(p_{2}+k\mright)1_{L_{k}}\mleft(p_{2}+k\mright)s^{-1}\left\langle K_{l}e_{p_{4}},e_{p_{2}+k}\right\rangle \left\langle e_{p_{2}+k},B_{k}e_{p_{1}}\right\rangle \tilde{c}_{p_{2}+k-l,\sigma}\\
 & =-1_{L_{k}}\mleft(p_{2}+k\mright)1_{L_{l}}\mleft(p_{3}+l\mright)s^{-1}\left\langle K_{l}e_{p_{4}},e_{p_{3}+l}\right\rangle \left\langle e_{p_{2}+k},B_{k}e_{p_{1}}\right\rangle \tilde{c}_{p_{3},\sigma}
\end{align*}
so the second term can be similarly estimated as
\begin{align}
 & \qquad\,\sum_{k,l\in\mathbb{Z}_{\ast}^{3}}\sum_{p\in S_{k}\cap S_{l}}^{\sigma}\left|\left\langle \Psi,\left[b_{l}\mleft(K_{l}e_{p_{4}}\mright),\left[b_{k}\mleft(B_{k}e_{p_{1}}\mright),\tilde{c}_{p_{2},\sigma}^{\ast}\right]^{\ast}\right]^{\ast}\tilde{c}_{p_{3},\sigma}\Psi\right\rangle \right|\nonumber \\
 & \leq C\sum_{k,l\in\mathbb{Z}_{\ast}^{3}}\sum_{p\in L_{k}\cap L_{l}}^{\sigma}\frac{\hat{V}_{l}k_{F}^{-1}1_{L_{k}}\mleft(p_{2}+k\mright)1_{L_{l}}\mleft(p_{3}+l\mright)}{\sqrt{\lambda_{k,p_{1}}+\lambda_{k,p_{2}+k}}\sqrt{\lambda_{l,p_{3}+l}+\lambda_{l,p_{4}}}}\left|\left\langle e_{p_{2}+k},B_{k}e_{p_{1}}\right\rangle \right|\left\langle \Psi,\tilde{c}_{p_{3},\sigma}^{\ast}\tilde{c}_{p_{3},\sigma}\Psi\right\rangle \nonumber \\
 & \leq Ck_{F}^{-1}\sum_{k\in\mathbb{Z}_{\ast}^{3}}\sum_{p\in L_{k}}^{\sigma}\sqrt{\sum_{l\in\mathbb{Z}_{\ast}^{3}}1_{L_{l}}\mleft(p\mright)\frac{\hat{V}_{l}^{2}}{\lambda_{l,p_{4}}}}\sqrt{\sum_{l\in\mathbb{Z}_{\ast}^{3}}1_{L_{k}}\mleft(p_{2}+k\mright)\left|\left\langle e_{p_{2}+k},h_{k}^{-\frac{1}{2}}B_{k}e_{p_{1}}\right\rangle \right|^{2}}\left\langle \Psi,\tilde{c}_{-p+k,\sigma}^{\ast}\tilde{c}_{-p+k,\sigma}\Psi\right\rangle \nonumber \\
 & \leq Ck_{F}^{-1}\sum_{p\in B_{F}}^{\sigma}\sum_{k\in\mathbb{Z}_{\ast}^{3}}1_{L_{k}-k}\mleft(p\mright)\sqrt{\sum_{l\in\mathbb{Z}_{\ast}^{3}}\hat{V}_{l}^{2}\frac{1_{L_{l}}\mleft(p+k\mright)}{\lambda_{l,p+k}}}\Vert h_{k}^{-\frac{1}{2}}B_{k}e_{p+k}\Vert\left\langle \Psi,\tilde{c}_{-p,\sigma}^{\ast}\tilde{c}_{-p,\sigma}\Psi\right\rangle \\
 & \leq Ck_{F}^{-1}\sum_{p\in B_{F}}^{\sigma}\sqrt{\sum_{l\in\mathbb{Z}_{\ast}^{3}}\hat{V}_{l}^{2}\sum_{k\in\mathbb{Z}_{\ast}^{3}}\frac{1_{L_{l}}\mleft(p+k\mright)}{\lambda_{l,p+k}}}\sqrt{\sum_{k\in\mathbb{Z}_{\ast}^{3}}\Vert h_{k}^{-\frac{1}{2}}B_{k}e_{p+k}\Vert^{2}}\left\langle \Psi,\tilde{c}_{-p,\sigma}^{\ast}\tilde{c}_{-p,\sigma}\Psi\right\rangle \nonumber \\
 & \leq Ck_{F}^{-\frac{1}{2}}\sqrt{\sum_{k\in\mathbb{Z}_{\ast}^{3}}\max_{p\in L_{k}}\Vert h_{k}^{-\frac{1}{2}}B_{k}e_{p}\Vert^{2}}\sqrt{\sum_{l\in\mathbb{Z}_{\ast}^{3}}\hat{V}_{l}^{2}}\left\langle \Psi,\mathcal{N}_{E}\Psi\right\rangle .\nonumber 
\end{align}
Finally, from the equations (\ref{eq:Klep4cp3Commutator}) and (\ref{eq:Bkep1cp2Commutator})
we see that
\[
\left[b_{l}\mleft(K_{l}e_{p_{4}}\mright),\tilde{c}_{p_{3},\sigma}^{\ast}\right]^{\ast}\left[b_{k}\mleft(B_{k}e_{p_{1}}\mright),\tilde{c}_{p_{2},\sigma}^{\ast}\right]=1_{L_{k}}\mleft(p_{2}+k\mright)1_{L_{l}}\mleft(p_{3}+l\mright)s^{-1}\left\langle B_{k}e_{p_{1}},e_{p_{2}+k}\right\rangle \left\langle e_{p_{3}+l},K_{l}e_{p_{4}}\right\rangle \tilde{c}_{p_{3}+l,\sigma}^{\ast}\tilde{c}_{p_{2}+k,\sigma}
\]
so we estimate
\begin{align}
 & \qquad\,\sum_{k,l\in\mathbb{Z}_{\ast}^{3}}\sum_{p\in S_{k}\cap S_{l}}^{\sigma}\left|\left\langle \Psi,\left[b_{l}\mleft(K_{l}e_{p_{4}}\mright),\tilde{c}_{p_{3},\sigma}^{\ast}\right]^{\ast}\left[b_{k}\mleft(B_{k}e_{p_{1}}\mright),\tilde{c}_{p_{2},\sigma}^{\ast}\right]\Psi\right\rangle \right|\nonumber \\
 & \leq C\sum_{k,l\in\mathbb{Z}_{\ast}^{3}}\sum_{p\in L_{k}\cap L_{l}}^{\sigma}\frac{\hat{V}_{l}k_{F}^{-1}1_{L_{k}}\mleft(p_{2}+k\mright)1_{L_{l}}\mleft(p_{3}+l\mright)}{\sqrt{\lambda_{k,p_{1}}+\lambda_{k,p_{2}+k}}\sqrt{\lambda_{l,p_{3}+l}+\lambda_{l,p_{4}}}}\left|\left\langle B_{k}e_{p_{1}},e_{p_{2}+k}\right\rangle \right|\left\langle \Psi,\tilde{c}_{p_{3}+l,\sigma}^{\ast}\tilde{c}_{p_{2}+k,\sigma}\Psi\right\rangle \\
 & \leq Ck_{F}^{-1}\sum_{p\in B_{F}^{c}}^{\sigma}\sum_{k,l\in\mathbb{Z}_{\ast}^{3}}1_{L_{k}\cap L_{l}}\mleft(p\mright)1_{L_{k}\cap L_{l}}\mleft(-p+k+l\mright)\frac{\hat{V}_{l}}{\sqrt{\lambda_{l,p}}}\left|\left\langle e_{p},h_{k}^{-\frac{1}{2}}B_{k}e_{-p+k+l}\right\rangle \right|\left\langle \Psi,\tilde{c}_{-p+k+l,\sigma}^{\ast}\tilde{c}_{-p+k+l,\sigma}\Psi\right\rangle \nonumber \\
 & =Ck_{F}^{-1}\sum_{p\in B_{F}^{c}}^{\sigma}\sum_{k,l\in\mathbb{Z}_{\ast}^{3}}1_{L_{k}\cap L_{l}}\mleft(p+k+l\mright)1_{L_{k}\cap L_{l}}\mleft(-p\mright)\frac{\hat{V}_{l}}{\sqrt{\lambda_{l,p+k+l}}}\left|\left\langle e_{p+k+l},h_{k}^{-\frac{1}{2}}B_{k}e_{-p}\right\rangle \right|\left\langle \Psi,\tilde{c}_{-p,\sigma}^{\ast}\tilde{c}_{-p,\sigma}\Psi\right\rangle \nonumber \\
 & \leq Ck_{F}^{-1}\sum_{p\in B_{F}^{c}}^{\sigma}\sqrt{\sum_{k,l\in\mathbb{Z}_{\ast}^{3}}1_{L_{k}}\mleft(p+k+l\mright)1_{L_{k}}\mleft(-p\mright)\left|\left\langle e_{p+k+l},h_{k}^{-\frac{1}{2}}B_{k}e_{-p}\right\rangle \right|^{2}}\nonumber \\
 & \qquad\cdot\sqrt{\sum_{k,l\in\mathbb{Z}_{\ast}^{3}}\hat{V}_{l}^{2}\frac{1_{L_{l}}\mleft(p+k+l\mright)}{\lambda_{l,p+k+l}}}\left\langle \Psi,\tilde{c}_{-p,\sigma}^{\ast}\tilde{c}_{-p,\sigma}\Psi\right\rangle \leq Ck_{F}^{-\frac{1}{2}}\sqrt{\sum_{k\in\mathbb{Z}_{\ast}^{3}}\max_{p\in L_{k}}\Vert h_{k}^{-\frac{1}{2}}B_{k}e_{p}\Vert^{2}}\sqrt{\sum_{l\in\mathbb{Z}_{\ast}^{3}}\hat{V}_{l}^{2}}\left\langle \Psi,\mathcal{N}_{E}\Psi\right\rangle .\nonumber 
\end{align}
$\hfill\square$

The $\mathcal{E}_{k}^{2}$ bound of Theorem \ref{them:ExchangeTermsEstimates}
now follows:
\begin{prop}
For any $\Psi\in\mathcal{H}_{N}$ and $t\in\left[0,1\right]$ it holds
that
\[
\sum_{k\in\mathbb{Z}_{\ast}^{3}}\left|\left\langle \Psi,\mleft(\mathcal{E}_{k}^{2}\mleft(B_{k}\mleft(t\mright)\mright)-\left\langle \psi_{F},\mathcal{E}_{k}^{2}\mleft(B_{k}\mleft(t\mright)\mright)\psi_{F}\right\rangle \mright)\Psi\right\rangle \right|\leq C\sqrt{\sum_{k\in\mathbb{Z}_{\ast}^{3}}\hat{V}_{k}^{2}\min\left\{ \left|k\right|,k_{F}\right\} }\left\langle \Psi,\mathcal{N}_{E}^{3}\Psi\right\rangle 
\]
for a constant $C>0$ depending only on $\sum_{k\in\mathbb{Z}_{\ast}^{3}}\hat{V}_{k}^{2}$
and $s$.
\end{prop}

\textbf{Proof:} Clearly
\begin{equation}
\max_{p\in L_{k}}\left\Vert B_{k}e_{p}\right\Vert ^{2}\leq\sum_{p\in L_{k}}\max_{q\in L_{k}}\left|\left\langle e_{p},B_{k}e_{q}\right\rangle \right|^{2},\quad\max_{p\in L_{k}}\Vert h_{k}^{-\frac{1}{2}}B_{k}e_{p}\Vert^{2}\leq\Vert B_{k}h_{k}^{-\frac{1}{2}}\Vert_{\mathrm{HS}}^{2},
\end{equation}
for any $B_{k}$, and as our estimate for $B_{k}\mleft(t\mright)$ in
Theorem \ref{them:OneBodyEstimates} is the same as that for $A_{k}\mleft(t\mright)$,
the bounds
\begin{equation}
\sum_{k\in\mathbb{Z}_{\ast}^{3}}\sum_{p\in L_{k}}\max_{q\in L_{k}}\left|\left\langle e_{p},B_{k}e_{q}\right\rangle \right|^{2},\,k_{F}^{-1}\sum_{k\in\mathbb{Z}_{\ast}^{3}}\Vert B_{k}h_{k}^{-\frac{1}{2}}\Vert_{\mathrm{HS}}^{2}\leq C\mleft(1+\Vert\hat{V}\Vert_{\infty}^{4}\mright)\sum_{k\in\mathbb{Z}_{\ast}^{3}}\hat{V}_{k}^{2}\min\left\{ \left|k\right|,k_{F}\right\} 
\end{equation}
follow exactly as those of Proposition \ref{prop:ParticularcalE1Estimates}.
Insertion into the Propositions \ref{prop:calE2TopEstimates}, \ref{prop:calE2SingleCommutatorEstimates}
and \ref{prop:calE2DoubleCommutatorEstimates} yields the claim.

$\hfill\square$

\subsection{Analysis of the Exchange Contribution}

Finally we determine the leading order of the exchange contribution.
To begin we derive a general formula for a quantity of the form $\left\langle \psi_{F},\mathcal{E}_{k}^{2}\mleft(B_{k}\mright)\psi_{F}\right\rangle $:
We can write
\begin{align}
-2\left\langle \psi_{F},\mathcal{E}_{k}^{2}\mleft(B_{k}\mright)\psi_{F}\right\rangle  & =-\sum_{l\in\mathbb{Z}_{\ast}^{3}}\sum_{p\in L_{k}}\sum_{q\in L_{l}}\left\langle \psi_{F},b_{k}\mleft(B_{k}e_{p}\mright)\varepsilon_{-k,-l}\mleft(e_{-p};e_{-q}\mright)b_{l}^{\ast}\mleft(K_{l}e_{q}\mright)\psi_{F}\right\rangle \nonumber \\
 & =\frac{1}{s}\sum_{l\in\mathbb{Z}_{\ast}^{3}}\sum_{p\in L_{k}\cap L_{l}}^{\sigma}\left\langle \psi_{F},b_{k}\mleft(B_{k}e_{p}\mright)\tilde{c}_{-p+l,\sigma}^{\ast}\tilde{c}_{-p+k,\sigma}b_{l}^{\ast}\mleft(K_{l}e_{p}\mright)\psi_{F}\right\rangle \\
 & +\frac{1}{s}\sum_{l\in\mathbb{Z}_{\ast}^{3}}\sum_{p\in\mleft(L_{k}-k\mright)\cap\mleft(L_{l}-l\mright)}^{\sigma}\left\langle \psi_{F},b_{k}\mleft(B_{k}e_{p+k}\mright)\tilde{c}_{-p-l,\sigma}^{\ast}\tilde{c}_{-p-k,\sigma}b_{l}^{\ast}\mleft(K_{l}e_{p+l}\mright)\psi_{F}\right\rangle \nonumber \\
 & =:A+B\nonumber 
\end{align}
where, using equation (\ref{eq:bcastCommutator}) in the form
\begin{equation}
\left[b_{l}\mleft(\psi\mright),\tilde{c}_{p,\sigma}^{\ast}\right]=\begin{cases}
-s^{-\frac{1}{2}}\sum_{q\in L_{l}}\delta_{p,q-l}\left\langle \psi,e_{q}\right\rangle \tilde{c}_{q,\sigma} & p\in B_{F}\\
s^{-\frac{1}{2}}\sum_{q\in\mleft(L_{l}-l\mright)}\delta_{p,q+l}\left\langle \psi,e_{q+l}\right\rangle \tilde{c}_{q,\sigma} & p\in B_{F}^{c}
\end{cases},
\end{equation}
the terms $A$ and $B$ are given by
\begin{align}
A & =\frac{1}{s}\sum_{l\in\mathbb{Z}_{\ast}^{3}}\sum_{p\in L_{k}\cap L_{l}}^{\sigma}\left\langle \psi_{F},\left[b_{k}\mleft(B_{k}e_{p}\mright),\tilde{c}_{-p+l,\sigma}^{\ast}\right]\left[b_{l}\mleft(K_{l}e_{p}\mright),\tilde{c}_{-p+k,\sigma}^{\ast}\right]^{\ast}\psi_{F}\right\rangle \\
 & =\frac{1}{s^{2}}\sum_{l\in\mathbb{Z}_{\ast}^{3}}\sum_{p\in L_{k}\cap L_{l}}^{\sigma}\left\langle \psi_{F},\mleft(\sum_{q\in L_{k}}\delta_{-p+l,q-k}\left\langle B_{k}e_{p},e_{q}\right\rangle \tilde{c}_{q,\sigma}\mright)\mleft(\sum_{q'\in L_{l}}\delta_{-p+k,q'-l}\left\langle e_{q'},K_{l}e_{p}\right\rangle \tilde{c}_{q',\sigma}^{\ast}\mright)\psi_{F}\right\rangle \nonumber \\
 & =\frac{1}{s}\sum_{l\in\mathbb{Z}_{\ast}^{3}}\sum_{p,q\in L_{k}\cap L_{l}}\delta_{p+q,k+l}\left\langle e_{p},B_{k}e_{q}\right\rangle \left\langle e_{q},K_{l}e_{p}\right\rangle \nonumber 
\end{align}
and
\begin{align}
B & =\frac{1}{s}\sum_{l\in\mathbb{Z}_{\ast}^{3}}\sum_{p\in\mleft(L_{k}-k\mright)\cap\mleft(L_{l}-l\mright)}^{\sigma}\left\langle \psi_{F},\left[b_{k}\mleft(B_{k}e_{p+k}\mright),\tilde{c}_{-p-l,\sigma}^{\ast}\right]\left[b_{l}\mleft(K_{l}e_{p+l}\mright),\tilde{c}_{-p-k,\sigma}^{\ast}\right]^{\ast}\psi_{F}\right\rangle \\
 & =\frac{1}{s}\sum_{l\in\mathbb{Z}_{\ast}^{3}}\sum_{p,q\in\mleft(L_{k}-k\mright)\cap\mleft(L_{l}-l\mright)}\delta_{-p-q,k+l}\left\langle e_{p+k},B_{k}e_{q+k}\right\rangle \left\langle e_{q+l},K_{l}e_{p+l}\right\rangle .\nonumber 
\end{align}
Although it is not obvious, there holds the identity $A=B$. To see
this we rewrite both terms: First, for $A$, note that the presence
of the $\delta_{p+q,k+l}$ makes the $L_{l}$ of the summation $p,q\in L_{k}\cap L_{l}$
redundant: For any $p,q\in B_{F}^{c}$ there holds the equivalence
\begin{equation}
p,q\in L_{p+q-k}\Leftrightarrow p,q\in L_{k}
\end{equation}
by the trivial identities
\begin{equation}
\left|p-k\right|=\left|q-\mleft(p+q-k\mright)\right|,\quad\left|q-k\right|=\left|p-\mleft(p+q-k\mright)\right|,
\end{equation}
so $A$ can be written as
\begin{equation}
A=\frac{1}{s}\sum_{p,q\in L_{k}}\sum_{l\in\mathbb{Z}_{\ast}^{3}}\delta_{p+q,k+l}\left\langle e_{p},B_{k}e_{q}\right\rangle \left\langle e_{q},K_{l}e_{p}\right\rangle =\frac{1}{s}\sum_{p,q\in L_{k}}\left\langle e_{p},B_{k}e_{q}\right\rangle \left\langle e_{q},K_{p+q-k}e_{p}\right\rangle .
\end{equation}
A similar observation applies to $B$: For any $p,q\in B_{F}$ we
likewise have
\begin{equation}
p,q\in\mleft(L_{-p-q-k}+p+q+k\mright)\Leftrightarrow p+k,q+k\in L_{p+q+k}\Leftrightarrow p,q\in\mleft(L_{k}-k\mright)
\end{equation}
so
\begin{align}
B & =\frac{1}{s}\sum_{p,q\in\mleft(L_{k}-k\mright)}\sum_{l\in\mathbb{Z}_{\ast}^{3}}\delta_{-p-q,k+l}\left\langle e_{p+k},B_{k}e_{q+k}\right\rangle \left\langle e_{q+l},K_{l}e_{p+l}\right\rangle \\
 & =\frac{1}{s}\sum_{p,q\in\mleft(L_{k}-k\mright)}\left\langle e_{p+k},B_{k}e_{q+k}\right\rangle \left\langle e_{-p-k},K_{-p-q-k}e_{-q-k}\right\rangle =\frac{1}{s}\sum_{p,q\in L_{k}}\left\langle e_{p},B_{k}e_{q}\right\rangle \left\langle e_{q},K_{p+q-k}e_{p}\right\rangle \nonumber 
\end{align}
where we lastly used that the kernels $K_{k}$ obey
\begin{equation}
\left\langle e_{-p},K_{-k}e_{-q}\right\rangle =\left\langle e_{p},K_{k}e_{q}\right\rangle =\left\langle e_{q},K_{k}e_{p}\right\rangle ,\quad k\in\mathbb{Z}_{\ast}^{3},\,p,q\in L_{k}.
\end{equation}
In all we thus have the identity
\begin{align}
\left\langle \psi_{F},\mathcal{E}_{k}^{2}\mleft(B_{k}\mright)\psi_{F}\right\rangle  & =-\frac{1}{s}\sum_{l\in\mathbb{Z}_{\ast}^{3}}\sum_{p,q\in L_{k}\cap L_{l}}\delta_{p+q,k+l}\left\langle e_{p},B_{k}e_{q}\right\rangle \left\langle e_{q},K_{l}e_{p}\right\rangle \\
 & =-\frac{1}{s}\sum_{p,q\in L_{k}}\left\langle e_{p},B_{k}e_{q}\right\rangle \left\langle e_{q},K_{p+q-k}e_{p}\right\rangle .\nonumber 
\end{align}
Our matrix element estimates of the last section now yield the following:
\begin{prop*}[\ref{prop:LeadingExchangeContribution}]
It holds that
\[
\left|\sum_{k\in\mathbb{Z}_{\ast}^{3}}\int_{0}^{1}\left\langle \psi_{F},2\,\mathrm{Re}\mleft(\mathcal{E}_{k}^{2}\mleft(B_{k}\mleft(t\mright)\mright)\mright)\psi_{F}\right\rangle dt-E_{\mathrm{corr},\mathrm{ex}}\right|\leq C\sqrt{\sum_{k\in\mathbb{Z}_{\ast}^{3}}\hat{V}_{k}^{2}\min\left\{ \left|k\right|,k_{F}\right\} }
\]
for a constant $C>0$ depending only on $\sum_{k\in\mathbb{Z}_{\ast}^{3}}\hat{V}_{k}^{2}$
and $s$, where
\[
E_{\mathrm{corr},\mathrm{ex}}=\frac{sk_{F}^{-2}}{4\,\mleft(2\pi\mright)^{6}}\sum_{k,l\in\mathbb{Z}_{\ast}^{3}}\hat{V}_{k}\hat{V}_{l}\sum_{p,q\in L_{k}\cap L_{l}}\frac{\delta_{p+q,k+l}}{\lambda_{k,p}+\lambda_{k,q}}.
\]
\end{prop*}
\textbf{Proof:} Since all the one-body operators are real-valued we
can drop the $\mathrm{Re}\mleft(\cdot\mright)$ and apply the above
identity for
\begin{align}
 & \qquad\;\sum_{k\in\mathbb{Z}_{\ast}^{3}}\int_{0}^{1}\left\langle \psi_{F},2\,\mathrm{Re}\mleft(\mathcal{E}_{k}^{2}\mleft(B_{k}\mleft(t\mright)\mright)\mright)\psi_{F}\right\rangle dt=\sum_{k\in\mathbb{Z}_{\ast}^{3}}2\left\langle \psi_{F},\mathcal{E}_{k}^{2}\mleft(\int_{0}^{1}B_{k}\mleft(t\mright)\,dt\mright)\psi_{F}\right\rangle \\
 & =\frac{2}{s}\sum_{k,l\in\mathbb{Z}_{\ast}^{3}}\sum_{p,q\in L_{k}\cap L_{l}}\delta_{p+q,k+l}\left\langle e_{p},\mleft(\int_{0}^{1}B_{k}\mleft(t\mright)\,dt\mright)e_{q}\right\rangle \left\langle e_{q},\mleft(-K_{l}\mright)e_{p}\right\rangle .\nonumber 
\end{align}
Now, note that $E_{\mathrm{corr},\mathrm{ex}}$ can be written as
\begin{equation}
E_{\mathrm{corr},\mathrm{ex}}=\frac{1}{s}\sum_{k,l\in\mathbb{Z}_{\ast}^{3}}\sum_{p,q\in L_{k}\cap L_{l}}\delta_{p+q,k+l}\frac{s\hat{V}_{k}k_{F}^{-1}}{2\,\mleft(2\pi\mright)^{3}}\frac{s\hat{V}_{l}k_{F}^{-1}}{2\,\mleft(2\pi\mright)^{3}}\frac{1}{\lambda_{l,p}+\lambda_{l,q}}
\end{equation}
since, much as in Proposition \ref{prop:calE1KSplitEstimate}, the
$\delta_{p+q,k+l}$ implies the following identity for the denominators:
\begin{align}
\lambda_{l,p}+\lambda_{l,q} & =\frac{1}{2}\mleft(\left|p\right|^{2}-\left|p-l\right|^{2}\mright)+\frac{1}{2}\mleft(\left|q\right|^{2}-\left|q-l\right|^{2}\mright)\\
 & =\frac{1}{2}\mleft(\left|p\right|^{2}-\left|q-k\right|^{2}\mright)+\frac{1}{2}\mleft(\left|q\right|^{2}-\left|p-k\right|^{2}\mright)=\lambda_{k,p}+\lambda_{k,q}.\nonumber 
\end{align}
We thus see that
\begin{align}
 & \qquad\;\sum_{k\in\mathbb{Z}_{\ast}^{3}}\int_{0}^{1}\left\langle \psi_{F},2\,\mathrm{Re}\mleft(\mathcal{E}_{k}^{2}\mleft(B_{k}\mleft(t\mright)\mright)\mright)\psi_{F}\right\rangle dt-E_{\mathrm{corr},\mathrm{ex}}\nonumber \\
 & =\frac{2}{s}\sum_{k,l\in\mathbb{Z}_{\ast}^{3}}\sum_{p,q\in L_{k}\cap L_{l}}\delta_{p+q,k+l}\mleft(\left\langle e_{p},\mleft(\int_{0}^{1}B_{k}\mleft(t\mright)\,dt\mright)e_{q}\right\rangle -\frac{s\hat{V}_{k}k_{F}^{-1}}{4\,\mleft(2\pi\mright)^{3}}\mright)\left\langle e_{q},\mleft(-K_{l}\mright)e_{p}\right\rangle \\
 & +\frac{1}{s}\sum_{k,l\in\mathbb{Z}_{\ast}^{3}}\sum_{p,q\in L_{k}\cap L_{l}}\delta_{p+q,k+l}\frac{s\hat{V}_{k}k_{F}^{-1}}{2\,\mleft(2\pi\mright)^{3}}\mleft(\left\langle e_{q},\mleft(-K_{l}\mright)e_{p}\right\rangle -\frac{s\hat{V}_{l}k_{F}^{-1}}{2\,\mleft(2\pi\mright)^{3}}\frac{1}{\lambda_{l,p}+\lambda_{l,q}}\mright)=:A+B.\nonumber 
\end{align}
We estimate $A$ and $B$. By the matrix element estimates of Theorem
\ref{them:OneBodyEstimates} we have that (using our freedom to replace
$\lambda_{l,p}+\lambda_{l,q}$ by $\lambda_{k,p}+\lambda_{k,q}$)
\begin{align}
\left|A\right| & \leq C\sum_{k,l\in\mathbb{Z}_{\ast}^{3}}\sum_{p,q\in L_{k}\cap L_{l}}\delta_{p+q,k+l}\mleft(1+\hat{V}_{k}\mright)\hat{V}_{k}^{2}k_{F}^{-1}\frac{\hat{V}_{l}k_{F}^{-1}}{\lambda_{l,p}+\lambda_{l,q}}\nonumber \\
 & \leq Ck_{F}^{-2}\mleft(1+\Vert\hat{V}\Vert_{\infty}\mright)\sum_{k\in\mathbb{Z}_{\ast}^{3}}\hat{V}_{k}^{2}\sum_{p\in L_{k}}\frac{1}{\sqrt{\lambda_{k,p}}}\sum_{q\in L_{k}}\frac{\hat{V}_{p+q-k}}{\sqrt{\lambda_{k,q}}}\\
 & \leq Ck_{F}^{-\frac{3}{2}}\mleft(1+\Vert\hat{V}\Vert_{\infty}\mright)\sqrt{\sum_{l\in\mathbb{Z}_{\ast}^{3}}\hat{V}_{l}^{2}}\sum_{k\in\mathbb{Z}_{\ast}^{3}}\hat{V}_{k}^{2}\sum_{p\in L_{k}}\frac{1}{\sqrt{\lambda_{k,p}}}\nonumber \\
 & \leq C\mleft(1+\Vert\hat{V}\Vert_{\infty}\mright)\sqrt{\sum_{l\in\mathbb{Z}_{\ast}^{3}}\hat{V}_{l}^{2}}\sum_{k\in\mathbb{Z}_{\ast}^{3}}\hat{V}_{k}^{2}\left|k\right|^{\frac{1}{2}}\min\,\{1,k_{F}^{\frac{3}{2}}\left|k\right|^{-\frac{3}{2}}\}\nonumber 
\end{align}
where we applied the inequality $\sum_{q\in L_{k}}\lambda_{k,q}^{-1}\leq Ck_{F}$
and also used that Proposition \ref{prop:RiemannSumEstimates} implies
that
\begin{equation}
\sum_{p\in L_{k}}\frac{1}{\sqrt{\lambda_{k,p}}}\leq Ck_{F}^{\frac{3}{2}}\left|k\right|^{\frac{1}{2}}\min\,\{1,k_{F}^{\frac{3}{2}}\left|k\right|^{-\frac{3}{2}}\}
\end{equation}
for a $C>0$ independent of all quantities. By Cauchy-Schwarz we can
further estimate
\begin{align}
\sum_{k\in\mathbb{Z}_{\ast}^{3}}\hat{V}_{k}^{2}\left|k\right|^{\frac{1}{2}}\min\,\{1,k_{F}^{\frac{3}{2}}\left|k\right|^{-\frac{3}{2}}\} & \leq\sqrt{\sum_{k\in\mathbb{Z}_{\ast}^{3}}\hat{V}_{k}^{2}}\sqrt{\sum_{k\in\mathbb{Z}_{\ast}^{3}}\hat{V}_{k}^{2}\left|k\right|\min\,\{1,k_{F}^{3}\left|k\right|^{-3}\}}\\
 & \leq\sqrt{\sum_{k\in\mathbb{Z}_{\ast}^{3}}\hat{V}_{k}^{2}}\sqrt{\sum_{k\in\mathbb{Z}_{\ast}^{3}}\hat{V}_{k}^{2}\min\left\{ \left|k\right|,k_{F}\right\} }\nonumber 
\end{align}
for the bound of the statement. By similar estimation also
\begin{equation}
\left|B\right|\leq C\sum_{k,l\in\mathbb{Z}_{\ast}^{3}}\sum_{p,q\in L_{k}\cap L_{l}}\delta_{p+q,k+l}\hat{V}_{k}k_{F}^{-1}\frac{\hat{V}_{l}^{2}k_{F}^{-1}}{\lambda_{l,p}+\lambda_{l,q}}\leq C\sqrt{\sum_{k\in\mathbb{Z}_{\ast}^{3}}\hat{V}_{k}^{2}}\sum_{l\in\mathbb{Z}_{\ast}^{3}}\hat{V}_{l}^{2}\left|l\right|^{\frac{1}{2}}\min\,\{1,k_{F}^{\frac{3}{2}}\left|l\right|^{-\frac{3}{2}}\}
\end{equation}
and the claim follows.

$\hfill\square$

\section{\label{sec:EstimationoftheNon-BosonizableTermsandGronwallEstimates}Estimation
of the Non-Bosonizable Terms and Gronwall Estimates}

In this section we perform the final work which will allow us to conclude
Theorem \ref{them:MainTheorem}.

The main content of this section lies in the estimation of the non-bosonizable
terms, which we recall are the cubic and quartic terms
\begin{align}
\mathcal{C} & =\frac{k_{F}^{-1}}{\mleft(2\pi\mright)^{3}}\sum_{k\in\mathbb{Z}_{\ast}^{3}}\hat{V}_{k}\,\mathrm{Re}\mleft(\mleft(B_{k}+B_{-k}^{\ast}\mright)^{\ast}D_{k}\mright)\\
\mathcal{Q} & =\frac{k_{F}^{-1}}{2\,\mleft(2\pi\mright)^{3}}\sum_{k\in\mathbb{Z}_{\ast}^{3}}\hat{V}_{k}\mleft(D_{k}^{\ast}D_{k}-\sum_{p\in L_{k}}^{\sigma}\mleft(c_{p,\sigma}^{\ast}c_{p,\sigma}+c_{p-k,\sigma}c_{p-k,\sigma}^{\ast}\mright)\mright).\nonumber 
\end{align}
The cubic terms $\mathcal{C}$ will not present a big obstacle to
us: As was first noted in \cite{BenNamPorSchSei-20} (in their formulation),
the expectation value of these in fact vanish identically with respect
to the type of trial state we will consider. The bulk of the work
will thus be to estimate the quartic terms. We prove the following
bounds:
\begin{thm}
\label{thm:NBTermEstimates}It holds that $\mathcal{Q}=G+\mathcal{Q}_{\mathrm{LR}}+\mathcal{Q}_{\mathrm{SR}}$
where for any $\Psi\in\mathcal{H}_{N}$
\begin{align*}
\left|\left\langle \Psi,G\Psi\right\rangle \right| & \leq C\sqrt{\sum_{k\in\mathbb{Z}_{\ast}^{3}}\hat{V}_{k}^{2}\min\left\{ \left|k\right|,k_{F}\right\} }\left\langle \Psi,\mathcal{N}_{E}\Psi\right\rangle \\
\left|\left\langle \Psi,\mathcal{Q}_{\mathrm{LR}}\Psi\right\rangle \right| & \leq C\sqrt{\sum_{k\in\mathbb{Z}_{\ast}^{3}}\hat{V}_{k}^{2}\min\left\{ \left|k\right|,k_{F}\right\} }\left\langle \Psi,\mathcal{N}_{E}^{2}\Psi\right\rangle 
\end{align*}
and $e^{\mathcal{K}}\mathcal{Q}_{\mathrm{SR}}e^{-\mathcal{K}}=\mathcal{Q}_{\mathrm{SR}}+\int_{0}^{1}e^{t\mathcal{K}}\mleft(2\,\mathrm{Re}\mleft(\mathcal{G}\mright)\mright)e^{-t\mathcal{K}}dt$
for an operator $\mathcal{G}$ obeying
\[
\left|\left\langle \Psi,\mathcal{G}\Psi\right\rangle \right|\leq C\sqrt{\sum_{k\in\mathbb{Z}_{\ast}^{3}}\hat{V}_{k}^{2}\min\left\{ \left|k\right|,k_{F}\right\} }\left\langle \Psi,\mleft(\mathcal{N}_{E}^{3}+1\mright)\Psi\right\rangle ,
\]
$C>0$ being a constant depending only on $\sum_{k\in\mathbb{Z}_{\ast}^{3}}\hat{V}_{k}^{2}$.
\end{thm}

(Again there are some technical questions which arise due to the unboundedness
of $\mathcal{Q}_{\mathrm{SR}}$. We consider these in appendix section
\ref{subsec:TransformationofcalQSR}.)

With these all the general bounds are established. As all our error
estimates are with respect to $\mathcal{N}_{E}$ and powers thereof,
it then only remains to control the effect which the transformation
$e^{\mathcal{K}}$ has on these. By a standard Gronwall-type argument
this control will follow from the estimate of Proposition \ref{prop:calKTildeBound},
and we then end the section by concluding Theorem \ref{them:MainTheorem}.

\subsubsection*{Analysis of the Cubic Terms}

Expanding the $\mathrm{Re}\mleft(\cdot\mright)$, the cubic terms are
\begin{equation}
\mathcal{C}=\frac{k_{F}^{-1}}{2\,\mleft(2\pi\mright)^{3}}\sum_{k\in\mathbb{Z}_{\ast}^{3}}\hat{V}_{k}\mleft(\mleft(B_{k}^{\ast}+B_{-k}\mright)D_{k}+D_{k}^{\ast}\mleft(B_{k}+B_{-k}^{\ast}\mright)\mright).
\end{equation}
The operators $B_{k}$ can be written simply as $B_{k}=s\sum_{p\in L_{k}}b_{k,p}$
in terms of the excitation operators $b_{k,p}=s^{-\frac{1}{2}}\sum_{\sigma=1}^{s}c_{p-k,\sigma}^{\ast}c_{p,\sigma}$,
whence it is easily seen that
\begin{equation}
\left[\mathcal{N}_{E},B_{k}\right]=-B_{k},\quad\left[\mathcal{N}_{E},B_{k}^{\ast}\right]=B_{k}^{\ast}.
\end{equation}
As a consequence, $B_{k}$ maps the eigenspace $\left\{ \mathcal{N}_{E}=M\right\} $
into $\left\{ \mathcal{N}_{E}=M-1\right\} $ and $B_{k}^{\ast}$ maps
$\left\{ \mathcal{N}_{E}=M\right\} $ into $\left\{ \mathcal{N}_{E}=M+1\right\} $.
Meanwhile, the operators $D_{k}$ preserve the eigenspaces: Writing
$D_{k}=D_{1,k}+D_{2,k}$ for
\begin{align}
D_{1,k} & =\mathrm{d\Gamma}\mleft(P_{B_{F}}e^{-ik\cdot x}P_{B_{F}}\mright)=\sum_{p,q\in B_{F}}^{\sigma}\delta_{p,q-k}c_{p,\sigma}^{\ast}c_{q,\sigma}=-\sum_{q\in B_{F}\cap\mleft(B_{F}+k\mright)}^{\sigma}\tilde{c}_{q,\sigma}^{\ast}\tilde{c}_{q-k,\sigma}\\
D_{2,k} & =\mathrm{d\Gamma}\mleft(P_{B_{F}^{c}}e^{-ik\cdot x}P_{B_{F}^{c}}\mright)=\sum_{p,q\in B_{F}^{c}}^{\sigma}\delta_{p,q-k}c_{p,\sigma}^{\ast}c_{q,\sigma}=\sum_{p\in B_{F}^{c}\cap\mleft(B_{F}^{c}-k\mright)}^{\sigma}\tilde{c}_{p,\sigma}^{\ast}\tilde{c}_{p+k,\sigma}\nonumber 
\end{align}
these annihilate and create one hole or excitation, respectively,
whence $\left[\mathcal{N}_{E},D_{k}\right]=0=\left[\mathcal{N}_{E},D_{k}^{\ast}\right]$.

It follows that $\mathcal{C}$ maps the eigenspace $\left\{ \mathcal{N}_{E}=M\right\} $
into $\left\{ \mathcal{N}_{E}=M-1\right\} \oplus\left\{ \mathcal{N}_{E}=M+1\right\} $.
Decomposing $\mathcal{H}_{N}$ orthogonally as $\mathcal{H}_{N}=\mathcal{H}_{N}^{\mathrm{even}}\oplus\mathcal{H}_{N}^{\mathrm{odd}}$
for
\begin{equation}
\mathcal{H}_{N}^{\mathrm{even}}=\bigoplus_{m=0}^{\infty}\left\{ \mathcal{N}_{E}=2m\right\} ,\quad\mathcal{H}_{N}^{\mathrm{odd}}=\bigoplus_{m=0}^{\infty}\left\{ \mathcal{N}_{E}=2m+1\right\} ,
\end{equation}
we thus see that $\mathcal{C}$ maps each subspace into the other.
On the other hand, since our transformation kernel $\mathcal{K}$
is of the form
\begin{equation}
\mathcal{K}=\frac{1}{2}\sum_{l\in\mathbb{Z}_{\ast}^{3}}\sum_{p,q\in L_{l}}\left\langle e_{p},K_{l}e_{q}\right\rangle \mleft(b_{l,p}b_{-l,-q}-b_{-l,-q}^{\ast}b_{l,p}^{\ast}\mright)
\end{equation}
we note that $\mathcal{K}$ maps each $\left\{ \mathcal{N}_{E}=M\right\} $
into $\left\{ \mathcal{N}_{E}=M-2\right\} \oplus\left\{ \mathcal{N}_{E}=M+2\right\} $,
hence $\mathcal{K}$ preserves $\mathcal{H}_{N}^{\mathrm{even}}$
and $\mathcal{H}_{N}^{\mathrm{odd}}$, and so too does the transformation
$e^{-\mathcal{K}}$. As any eigenstate $\Psi\in\mathcal{H}_{N}$ of
$\mathcal{N}_{E}$ is contained in either $\mathcal{H}_{N}^{\mathrm{even}}$
or $\mathcal{H}_{N}^{\mathrm{odd}}$, and these are orthogonal, we
conclude the following:
\begin{prop}
\label{prop:VanishingofCubicTerms}For any eigenstate $\Psi$ of $\mathcal{N}_{E}$
it holds that
\[
\left\langle e^{-\mathcal{K}}\Psi,\mathcal{C}e^{-\mathcal{K}}\Psi\right\rangle =0.
\]
\end{prop}

\subsection{Analysis of the Quartic Terms}

Now we consider the quartic terms
\begin{equation}
\mathcal{Q}=\frac{k_{F}^{-1}}{2\,\mleft(2\pi\mright)^{3}}\sum_{k\in\mathbb{Z}_{\ast}^{3}}\hat{V}_{k}\mleft(D_{k}^{\ast}D_{k}-\sum_{p\in L_{k}}^{\sigma}\mleft(c_{p,\sigma}^{\ast}c_{p,\sigma}+c_{p-k,\sigma}c_{p-k,\sigma}^{\ast}\mright)\mright).
\end{equation}
We begin by rewriting these: Recalling the decomposition $D_{k}=D_{1,k}+D_{2,k}$
above, we calculate
\begin{align}
D_{1,k}^{\ast}D_{1,k} & =\sum_{p,q\in B_{F}\cap\mleft(B_{F}+k\mright)}^{\sigma,\tau}\tilde{c}_{p-k,\sigma}^{\ast}\tilde{c}_{p,\sigma}\tilde{c}_{q,\tau}^{\ast}\tilde{c}_{q-k,\tau}=\sum_{p,q\in B_{F}\cap\mleft(B_{F}+k\mright)}^{\sigma,\tau}\tilde{c}_{p-k,\sigma}^{\ast}\tilde{c}_{q,\tau}^{\ast}\tilde{c}_{q-k,\tau}\tilde{c}_{p,\sigma}+\sum_{q\in B_{F}\cap\mleft(B_{F}+k\mright)}^{\sigma}\tilde{c}_{q-k,\sigma}^{\ast}\tilde{c}_{q-k,\sigma}\nonumber \\
 & =\sum_{p,q\in B_{F}\cap\mleft(B_{F}+k\mright)}^{\sigma,\tau}\tilde{c}_{p-k,\sigma}^{\ast}\tilde{c}_{q,\tau}^{\ast}\tilde{c}_{q-k,\tau}\tilde{c}_{p,\sigma}+\sum_{q\in B_{F}}^{\sigma}1_{B_{F}}\mleft(q+k\mright)\tilde{c}_{q,\sigma}^{\ast}\tilde{c}_{q,\sigma}\label{eq:D1kastD1k}
\end{align}
and similarly
\begin{align}
D_{2,k}^{\ast}D_{2,k} & =\sum_{p,q\in B_{F}^{c}\cap\mleft(B_{F}^{c}-k\mright)}^{\sigma,\tau}\tilde{c}_{p+k,\sigma}^{\ast}\tilde{c}_{p,\sigma}\tilde{c}_{q,\tau}^{\ast}\tilde{c}_{q+k,\tau}=\sum_{p,q\in B_{F}^{c}\cap\mleft(B_{F}^{c}-k\mright)}^{\sigma,\tau}\tilde{c}_{p+k,\sigma}^{\ast}\tilde{c}_{q,\tau}^{\ast}\tilde{c}_{q+k,\tau}\tilde{c}_{p,\sigma}+\sum_{p\in B_{F}^{c}}^{\sigma}1_{B_{F}^{c}}\mleft(p-k\mright)\tilde{c}_{p,\sigma}^{\ast}\tilde{c}_{p,\sigma}\nonumber \\
 & =\sum_{p,q\in B_{F}^{c}\cap\mleft(B_{F}^{c}-k\mright)}^{\sigma,\tau}\tilde{c}_{p+k,\sigma}^{\ast}\tilde{c}_{q,\tau}^{\ast}\tilde{c}_{q+k,\tau}\tilde{c}_{p,\sigma}+\mathcal{N}_{E}-\sum_{p\in B_{F}^{c}}^{\sigma}1_{B_{F}}\mleft(p-k\mright)\tilde{c}_{p,\sigma}^{\ast}\tilde{c}_{p,\sigma}.
\end{align}
For any $k\in\mathbb{Z}_{\ast}^{3}$ we can likewise write $\sum_{p\in L_{k}}^{\sigma}\mleft(c_{p,\sigma}^{\ast}c_{p,\sigma}+c_{p-k,\sigma}c_{p-k,\sigma}^{\ast}\mright)$
in the form
\begin{align}
\sum_{p\in L_{k}}^{\sigma}\mleft(c_{p,\sigma}^{\ast}c_{p,\sigma}+c_{p-k,\sigma}c_{p-k,\sigma}^{\ast}\mright) & =\sum_{p\in B_{F}^{c}}^{\sigma}1_{B_{F}}\mleft(p-k\mright)\tilde{c}_{p,\sigma}^{\ast}\tilde{c}_{p,\sigma}+\sum_{q\in B_{F}}^{\sigma}1_{B_{F}^{c}}\mleft(q+k\mright)\tilde{c}_{q,\sigma}^{\ast}\tilde{c}_{q,\sigma}\\
 & =\sum_{p\in B_{F}^{c}}^{\sigma}1_{B_{F}}\mleft(p-k\mright)\tilde{c}_{p,\sigma}^{\ast}\tilde{c}_{p,\sigma}+\mathcal{N}_{E}-\sum_{q\in B_{F}}^{\sigma}1_{B_{F}}\mleft(q+k\mright)\tilde{c}_{q,\sigma}^{\ast}\tilde{c}_{q,\sigma}.\nonumber 
\end{align}
Noting that $D_{1,k}=0$ for $\left|k\right|>2k_{F}$, as then $B_{F}\cap\mleft(B_{F}+k\mright)=\emptyset$,
we thus obtain the decomposition
\begin{equation}
\mathcal{Q}=G+\mathcal{Q}_{\mathrm{LR}}+\mathcal{Q}_{\mathrm{SR}}
\end{equation}
where $G$ is
\begin{equation}
G=\frac{k_{F}^{-1}}{\mleft(2\pi\mright)^{3}}\sum_{k\in\mathbb{Z}_{\ast}^{3}}\hat{V}_{k}\mleft(\sum_{q\in B_{F}}^{\sigma}1_{B_{F}}\mleft(q+k\mright)\tilde{c}_{q,\sigma}^{\ast}\tilde{c}_{q,\sigma}-\sum_{p\in B_{F}^{c}}^{\sigma}1_{B_{F}}\mleft(p-k\mright)\tilde{c}_{p,\sigma}^{\ast}\tilde{c}_{p,\sigma}\mright),
\end{equation}
the \textit{long-range terms} $\mathcal{Q}_{\mathrm{LR}}$ are given
by
\begin{equation}
\mathcal{Q}_{\mathrm{LR}}=\frac{k_{F}^{-1}}{2\,\mleft(2\pi\mright)^{3}}\sum_{k\in\overline{B}\mleft(0,2k_{F}\mright)\cap\mathbb{Z}_{\ast}^{3}}\hat{V}_{k}\mleft(\sum_{p,q\in B_{F}\cap\mleft(B_{F}+k\mright)}^{\sigma,\tau}\tilde{c}_{p-k,\sigma}^{\ast}\tilde{c}_{q,\tau}^{\ast}\tilde{c}_{q-k,\tau}\tilde{c}_{p,\sigma}+D_{1,k}^{\ast}D_{2,k}+D_{2,k}^{\ast}D_{1,k}\mright)\label{eq:calQLRDefinition}
\end{equation}
and the \textit{short-range terms} $\mathcal{Q}_{\mathrm{SR}}$ are
\begin{equation}
\mathcal{Q}_{\mathrm{SR}}=\frac{k_{F}^{-1}}{2\,\mleft(2\pi\mright)^{3}}\sum_{k\in\mathbb{Z}_{\ast}^{3}}\hat{V}_{k}\sum_{p,q\in B_{F}^{c}\cap\mleft(B_{F}^{c}-k\mright)}^{\sigma,\tau}\tilde{c}_{p+k,\sigma}^{\ast}\tilde{c}_{q,\tau}^{\ast}\tilde{c}_{q+k,\tau}\tilde{c}_{p,\sigma}.
\end{equation}

\subsubsection*{Estimation of $G$ and $\mathcal{Q}_{\mathrm{LR}}$}

$G$ and the long-range terms are easily controlled: First, interchanging
the summations we can write $G$ as
\begin{equation}
G=\frac{k_{F}^{-1}}{\mleft(2\pi\mright)^{3}}\sum_{q\in B_{F}}^{\sigma}\mleft(\sum_{k\in\mleft(B_{F}-q\mright)\cap\mathbb{Z}_{\ast}^{3}}\hat{V}_{k}\mright)\tilde{c}_{q,\sigma}^{\ast}\tilde{c}_{q,\sigma}-\frac{k_{F}^{-1}}{\mleft(2\pi\mright)^{3}}\sum_{p\in B_{F}^{c}}^{\sigma}\mleft(\sum_{k\in\mleft(B_{F}+p\mright)\cap\mathbb{Z}_{\ast}^{3}}\hat{V}_{k}\mright)\tilde{c}_{p,\sigma}^{\ast}\tilde{c}_{p,\sigma}
\end{equation}
from which it is obvious that $G$ obeys
\begin{equation}
\pm G\leq\max_{p\in\mathbb{Z}_{\ast}^{3}}\mleft(\frac{k_{F}^{-1}}{\mleft(2\pi\mright)^{3}}\sum_{k\in\mleft(B_{F}+p\mright)\cap\mathbb{Z}_{\ast}^{3}}\hat{V}_{k}\mright)\mathcal{N}_{E}.
\end{equation}
This implies the following:
\begin{prop}
For any $\Psi\in\mathcal{H}_{N}$ it holds that
\[
\left|\left\langle \Psi,G\Psi\right\rangle \right|\leq C\sqrt{\sum_{k\in\mathbb{Z}_{\ast}^{3}}\hat{V}_{k}^{2}\min\left\{ \left|k\right|,k_{F}\right\} }\left\langle \Psi,\mathcal{N}_{E}\Psi\right\rangle 
\]
for a constant $C>0$ independent of all quantities.
\end{prop}

\textbf{Proof:} For any $p\in\mathbb{Z}^{3}$ we estimate by Cauchy-Schwarz
\begin{align}
\sum_{k\in\mleft(B_{F}+p\mright)\cap\mathbb{Z}_{\ast}^{3}}\hat{V}_{k} & \leq\sqrt{\sum_{k\in\mleft(B_{F}+p\mright)\cap\mathbb{Z}_{\ast}^{3}}\hat{V}_{k}^{2}\min\left\{ \left|k\right|,k_{F}\right\} }\sqrt{\sum_{k\in\mleft(B_{F}+p\mright)\cap\mathbb{Z}_{\ast}^{3}}\min\left\{ \left|k\right|,k_{F}\right\} ^{-1}}\label{eq:VkSumBound}\\
 & \leq\sqrt{\sum_{k\in\mathbb{Z}_{\ast}^{3}}\hat{V}_{k}^{2}\min\left\{ \left|k\right|,k_{F}\right\} }\sqrt{\sum_{k\in B_{F}\backslash\left\{ 0\right\} }\left|k\right|^{-1}+k_{F}^{-1}}\nonumber 
\end{align}
where we lastly used that $k\mapsto\min\left\{ \left|k\right|,k_{F}\right\} ^{-1}$
is radially non-increasing and that $\mleft(B_{F}+p\mright)\cap\mathbb{Z}_{\ast}^{3}$
contains at most $\left|B_{F}\right|$ points. As it is well-known
that $\sum_{k\in\overline{B}\mleft(0,R\mright)\backslash\left\{ 0\right\} }\left|k\right|^{-1}\leq CR^{2}$
as $R\rightarrow\infty$ the bound follows.

$\hfill\square$

$\mathcal{Q}_{\mathrm{LR}}$ can be handled in a similar manner:
\begin{prop}
For any $\Psi\in\mathcal{H}_{N}$ it holds that
\[
\left|\left\langle \Psi,\mathcal{Q}_{\mathrm{LR}}\Psi\right\rangle \right|\leq C\sqrt{\sum_{k\in\mathbb{Z}_{\ast}^{3}}\hat{V}_{k}^{2}\min\left\{ \left|k\right|,k_{F}\right\} }\left\langle \Psi,\mathcal{N}_{E}^{2}\Psi\right\rangle 
\]
for a constant $C>0$ independent of all quantities.
\end{prop}

\textbf{Proof:} Consider the first term in the parenthesis of equation
(\ref{eq:calQLRDefinition}): For any $k\in\mathbb{Z}_{\ast}^{3}$
we can estimate
\begin{align}
 & \qquad\,\sum_{p,q\in B_{F}\cap\mleft(B_{F}+k\mright)}^{\sigma,\tau}\left|\left\langle \Psi,\tilde{c}_{p-k,\sigma}^{\ast}\tilde{c}_{q,\tau}^{\ast}\tilde{c}_{q-k,\tau}\tilde{c}_{p,\sigma}\Psi\right\rangle \right|\leq\sum_{p,q\in B_{F}\cap\mleft(B_{F}+k\mright)}^{\sigma,\tau}\left\Vert \tilde{c}_{q,\tau}\tilde{c}_{p-k,\sigma}\Psi\right\Vert \left\Vert \tilde{c}_{q-k,\tau}\tilde{c}_{p,\sigma}\Psi\right\Vert \label{eq:QLRMainEstimate}\\
 & \leq\sqrt{\sum_{p,q\in B_{F}\cap\mleft(B_{F}+k\mright)}^{\sigma,\tau}\left\Vert \tilde{c}_{q,\tau}\tilde{c}_{p-k,\sigma}\Psi\right\Vert ^{2}}\sqrt{\sum_{p,q\in B_{F}\cap\mleft(B_{F}+k\mright)}^{\sigma,\tau}\left\Vert \tilde{c}_{q-k,\tau}\tilde{c}_{p,\sigma}\Psi\right\Vert ^{2}}\leq\left\langle \Psi,\mathcal{N}_{E}^{2}\Psi\right\rangle .\nonumber 
\end{align}
As e.g.
\begin{align}
D_{1,k}^{\ast}D_{2,k} & =\sum_{p\in B_{F}^{c}\cap\mleft(B_{F}^{c}-k\mright)}^{\sigma}\sum_{q\in B_{F}\cap\mleft(B_{F}+k\mright)}^{\tau}\tilde{c}_{q-k,\tau}^{\ast}\tilde{c}_{q,\tau}\tilde{c}_{p,\sigma}^{\ast}\tilde{c}_{p+k,\sigma}\\
 & =\sum_{p\in B_{F}^{c}\cap\mleft(B_{F}^{c}-k\mright)}^{\sigma}\sum_{q\in B_{F}\cap\mleft(B_{F}+k\mright)}^{\tau}\tilde{c}_{p,\sigma}^{\ast}\tilde{c}_{q-k,\tau}^{\ast}\tilde{c}_{q,\tau}\tilde{c}_{p+k,\sigma}\nonumber 
\end{align}
since $B_{F}$ and $B_{F}^{c}$ are disjoint, the terms $D_{1,k}^{\ast}D_{2,k}$
and $D_{2,k}^{\ast}D_{1,k}$ can be handled similarly, whence
\begin{equation}
\left|\left\langle \Psi,\mathcal{Q}_{\mathrm{LR}}\Psi\right\rangle \right|\leq\frac{3k_{F}^{-1}}{2\,\mleft(2\pi\mright)^{3}}\mleft(\sum_{k\in\overline{B}\mleft(0,2k_{F}\mright)\cap\mathbb{Z}_{\ast}^{3}}\hat{V}_{k}\mright)\left\langle \Psi,\mathcal{N}_{E}^{2}\Psi\right\rangle \leq C\sqrt{\sum_{k\in\mathbb{Z}_{\ast}^{3}}\hat{V}_{k}^{2}\min\left\{ \left|k\right|,k_{F}\right\} }\left\langle \Psi,\mathcal{N}_{E}^{2}\Psi\right\rangle 
\end{equation}
where $\sum_{k\in\overline{B}\mleft(0,2k_{F}\mright)\cap\mathbb{Z}_{\ast}^{3}}\hat{V}_{k}$
was bounded as in equation (\ref{eq:VkSumBound}).

$\hfill\square$

\subsubsection*{Analysis of $\mathcal{Q}_{\mathrm{SR}}$}

Lastly we come to
\begin{equation}
\mathcal{Q}_{\mathrm{SR}}=\frac{k_{F}^{-1}}{2\,\mleft(2\pi\mright)^{3}}\sum_{k\in\mathbb{Z}_{\ast}^{3}}\hat{V}_{k}\sum_{p,q\in B_{F}^{c}\cap\mleft(B_{F}^{c}-k\mright)}^{\sigma,\tau}\tilde{c}_{p+k,\sigma}^{\ast}\tilde{c}_{q,\tau}^{\ast}\tilde{c}_{q+k,\tau}\tilde{c}_{p,\sigma}.
\end{equation}
Recall that the transformation kernel $\mathcal{K}$ can be written
as $\mathcal{K}=\tilde{\mathcal{K}}-\tilde{\mathcal{K}}^{\ast}$ for
\begin{equation}
\tilde{\mathcal{K}}=\frac{1}{2}\sum_{l\in\mathbb{Z}_{\ast}^{3}}\sum_{p,q\in L_{l}}\left\langle e_{p},K_{l}e_{q}\right\rangle b_{l,p}b_{-l,-q}=\frac{1}{2}\sum_{l\in\mathbb{Z}_{\ast}^{3}}\sum_{q\in L_{l}}b_{l}\mleft(K_{l}e_{q}\mright)b_{-l,-q}.
\end{equation}
To determine $e^{\mathcal{K}}\mathcal{Q}_{\mathrm{SR}}e^{-\mathcal{K}}$
we will need the commutator $\left[\mathcal{K},\mathcal{Q}_{\mathrm{SR}}\right]=2\,\mathrm{Re}\mleft([\tilde{\mathcal{K}},\mathcal{Q}_{\mathrm{SR}}]\mright)$.
Noting that for any $p\in B_{F}^{c}$ and $l\in\mathbb{Z}_{\ast}^{3}$,
$q\in L_{l}$, we have
\begin{equation}
\left[b_{l,q},\tilde{c}_{p,\sigma}^{\ast}\right]=\frac{1}{\sqrt{s}}\sum_{\tau=1}^{s}\left[c_{q-l,\tau}^{\ast}c_{q,\tau},c_{p,\sigma}^{\ast}\right]=\frac{1}{\sqrt{s}}\sum_{\tau=1}^{s}\delta_{p,q}\delta_{\sigma,\tau}c_{q-l,\tau}^{\ast}=\frac{1}{\sqrt{s}}\delta_{p,q}\tilde{c}_{q-l,\sigma},
\end{equation}
we deduce (with the help of Lemma \ref{lemma:TraceFormLemma}) that
\begin{align}
[\tilde{\mathcal{K}},\tilde{c}_{p,\sigma}^{\ast}] & =\frac{1}{2}\sum_{l\in\mathbb{Z}_{\ast}^{3}}\sum_{q\in L_{l}}\mleft(b_{l}\mleft(K_{l}e_{q}\mright)\left[b_{-l,-q},\tilde{c}_{p,\sigma}^{\ast}\right]+\left[b_{l}\mleft(K_{l}e_{q}\mright),\tilde{c}_{p,\sigma}^{\ast}\right]b_{-l,-q}\mright)\nonumber \\
 & =\frac{1}{2}\sum_{l\in\mathbb{Z}_{\ast}^{3}}\sum_{q\in L_{l}}\mleft(b_{l}\mleft(K_{l}e_{q}\mright)\left[b_{-l,-q},\tilde{c}_{p,\sigma}^{\ast}\right]+\left[b_{l,q},\tilde{c}_{p,\sigma}^{\ast}\right]b_{-l}\mleft(K_{-l}e_{-q}\mright)\mright)\label{eq:calKtildecpastCommutator}\\
 & =\frac{1}{2\sqrt{s}}\sum_{l\in\mathbb{Z}_{\ast}^{3}}\sum_{q\in L_{l}}\mleft(b_{l}\mleft(K_{l}e_{q}\mright)\delta_{p,-q}\tilde{c}_{-q+l,\sigma}+\delta_{p,q}\tilde{c}_{q-l,\sigma}b_{-l}\mleft(K_{-l}e_{-q}\mright)\mright)\nonumber \\
 & =\frac{1}{\sqrt{s}}\sum_{l\in\mathbb{Z}_{\ast}^{3}}\sum_{q\in L_{l}}\delta_{p,-q}b_{l}\mleft(K_{l}e_{q}\mright)\tilde{c}_{-q+l,\sigma}=\frac{1}{\sqrt{s}}\sum_{l\in\mathbb{Z}_{\ast}^{3}}1_{L_{l}}\mleft(-p\mright)b_{l}\mleft(K_{l}e_{-p}\mright)\tilde{c}_{p+l,\sigma}.\nonumber 
\end{align}
Using this we conclude the following:
\begin{prop}
It holds that $e^{\mathcal{K}}\mathcal{Q}_{\mathrm{SR}}e^{-\mathcal{K}}=\mathcal{Q}_{\mathrm{SR}}+\int_{0}^{1}e^{t\mathcal{K}}\mleft(2\,\mathrm{Re}\mleft(\mathcal{G}\mright)\mright)e^{-t\mathcal{K}}dt$
for
\begin{align*}
\mathcal{G} & =\frac{s^{-\frac{1}{2}}k_{F}^{-1}}{\mleft(2\pi\mright)^{3}}\sum_{k,l\in\mathbb{Z}_{\ast}^{3}}\hat{V}_{k}\sum_{p,q\in B_{F}^{c}\cap\mleft(B_{F}^{c}+k\mright)}^{\sigma,\tau}1_{L_{l}}\mleft(q\mright)\tilde{c}_{p,\sigma}^{\ast}b_{l}\mleft(K_{l}e_{q}\mright)\tilde{c}_{-q+l,\tau}\tilde{c}_{-q+k,\tau}\tilde{c}_{p-k,\sigma}\\
 & +\frac{s^{-1}k_{F}^{-1}}{2\,\mleft(2\pi\mright)^{3}}\sum_{k,l\in\mathbb{Z}_{\ast}^{3}}\hat{V}_{k}\sum_{p,q\in B_{F}^{c}\cap\mleft(B_{F}^{c}+k\mright)}^{\sigma,\tau}1_{L_{l}}\mleft(p\mright)1_{L_{l}}\mleft(q\mright)\left\langle K_{l}e_{q},e_{p}\right\rangle \tilde{c}_{p-l,\sigma}\tilde{c}_{-q+l,\tau}\tilde{c}_{-q+k,\tau}\tilde{c}_{p-k,\sigma}.
\end{align*}
\end{prop}

\textbf{Proof:} By the fundamental theorem of calculus
\begin{equation}
e^{\mathcal{K}}\mathcal{Q}_{\mathrm{SR}}e^{-\mathcal{K}}=\mathcal{Q}_{\mathrm{SR}}+\int_{0}^{1}e^{t\mathcal{K}}\left[\mathcal{K},\mathcal{Q}_{\mathrm{SR}}\right]e^{-t\mathcal{K}}dt
\end{equation}
and as noted $\left[\mathcal{K},\mathcal{Q}_{\mathrm{SR}}\right]=2\,\mathrm{Re}\mleft([\tilde{\mathcal{K}},\mathcal{Q}_{\mathrm{SR}}]\mright)$.
Using equation (\ref{eq:calKtildecpastCommutator}) we compute that
$\mathcal{G}:=[\tilde{\mathcal{K}},\mathcal{Q}_{\mathrm{SR}}]$ is
given by
\begin{align}
\mathcal{G} & =\frac{k_{F}^{-1}}{2\,\mleft(2\pi\mright)^{3}}\sum_{k\in\mathbb{Z}_{\ast}^{3}}\hat{V}_{k}\sum_{p\in B_{F}^{c}\cap\mleft(B_{F}^{c}+k\mright)}^{\sigma}\sum_{q\in B_{F}^{c}\cap\mleft(B_{F}^{c}-k\mright)}^{\tau}\mleft(\tilde{c}_{p,\sigma}^{\ast}[\tilde{\mathcal{K}},\tilde{c}_{q,\tau}^{\ast}]+[\tilde{\mathcal{K}},\tilde{c}_{p,\sigma}^{\ast}]\tilde{c}_{q,\tau}^{\ast}\mright)\tilde{c}_{q+k,\tau}\tilde{c}_{p-k,\sigma}\nonumber \\
 & =\frac{s^{-\frac{1}{2}}k_{F}^{-1}}{2\,\mleft(2\pi\mright)^{3}}\sum_{k,l\in\mathbb{Z}_{\ast}^{3}}\hat{V}_{k}\sum_{p\in B_{F}^{c}\cap\mleft(B_{F}^{c}+k\mright)}^{\sigma}\sum_{q\in B_{F}^{c}\cap\mleft(B_{F}^{c}-k\mright)}^{\tau}1_{L_{l}}\mleft(-q\mright)\tilde{c}_{p,\sigma}^{\ast}b_{l}\mleft(K_{l}e_{-q}\mright)\tilde{c}_{q+l,\tau}\tilde{c}_{q+k,\tau}\tilde{c}_{p-k,\sigma}\nonumber \\
 & +\frac{s^{-\frac{1}{2}}k_{F}^{-1}}{2\,\mleft(2\pi\mright)^{3}}\sum_{k,l\in\mathbb{Z}_{\ast}^{3}}\hat{V}_{k}\sum_{p\in B_{F}^{c}\cap\mleft(B_{F}^{c}+k\mright)}^{\sigma}\sum_{q\in B_{F}^{c}\cap\mleft(B_{F}^{c}-k\mright)}^{\tau}1_{L_{l}}\mleft(-p\mright)b_{l}\mleft(K_{l}e_{-p}\mright)\tilde{c}_{p+l,\sigma}\tilde{c}_{q,\tau}^{\ast}\tilde{c}_{q+k,\tau}\tilde{c}_{p-k,\sigma}\\
 & =\frac{s^{-\frac{1}{2}}k_{F}^{-1}}{2\,\mleft(2\pi\mright)^{3}}\sum_{k,l\in\mathbb{Z}_{\ast}^{3}}\hat{V}_{k}\sum_{p\in B_{F}^{c}\cap\mleft(B_{F}^{c}+k\mright)}^{\sigma}\sum_{q\in B_{F}^{c}\cap\mleft(B_{F}^{c}-k\mright)}^{\tau}1_{L_{l}}\mleft(-q\mright)\left\{ b_{l}\mleft(K_{l}e_{-q}\mright),\tilde{c}_{p,\sigma}^{\ast}\right\} \tilde{c}_{q+l,\tau}\tilde{c}_{q+k,\tau}\tilde{c}_{p-k,\sigma}\nonumber \\
 & =\frac{s^{-\frac{1}{2}}k_{F}^{-1}}{2\,\mleft(2\pi\mright)^{3}}\sum_{k,l\in\mathbb{Z}_{\ast}^{3}}\hat{V}_{k}\sum_{p,q\in B_{F}^{c}\cap\mleft(B_{F}^{c}+k\mright)}^{\sigma,\tau}1_{L_{l}}\mleft(q\mright)\left\{ b_{l}\mleft(K_{l}e_{q}\mright),\tilde{c}_{p,\sigma}^{\ast}\right\} \tilde{c}_{-q+l,\tau}\tilde{c}_{-q+k,\tau}\tilde{c}_{p-k,\sigma},\nonumber 
\end{align}
where we for the third inequality substituted $\mleft(p,\sigma\mright)\leftrightarrow\mleft(q,\tau\mright)$
and $k\rightarrow-k$ in the second sum, noting that then
\begin{equation}
1_{L_{l}}\mleft(-q\mright)\tilde{c}_{q+l,\tau}\tilde{c}_{p,\sigma}^{\ast}\tilde{c}_{p-k,\sigma}\tilde{c}_{q+k,\tau}=1_{L_{l}}\mleft(-q\mright)\tilde{c}_{p,\sigma}^{\ast}\tilde{c}_{q+l,\tau}\tilde{c}_{q+k,\tau}\tilde{c}_{p-k,\sigma}
\end{equation}
as the indicator function (and summation range) ensures that $q+l\neq p$.

By the identity of equation (\ref{eq:bcastCommutator}) the anti-commutator
is given by
\begin{equation}
\left\{ b_{l}\mleft(K_{l}e_{q}\mright),\tilde{c}_{p,\sigma}^{\ast}\right\} =2\,\tilde{c}_{p,\sigma}^{\ast}b_{l}\mleft(K_{l}e_{q}\mright)+1_{L_{l}}\mleft(p\mright)s^{-\frac{1}{2}}\left\langle K_{l}e_{q},e_{p}\right\rangle \tilde{c}_{p-l,\sigma}
\end{equation}
which is inserted into the previous equation for the claim.

$\hfill\square$

We bound the $\mathcal{G}$ operator as follows:
\begin{prop}
\label{prop:calGEstimate}For any $\Psi\in\mathcal{H}_{N}$ it holds
that
\[
\left|\left\langle \Psi,\mathcal{G}\Psi\right\rangle \right|\leq C\sqrt{\sum_{k\in\mathbb{Z}_{\ast}^{3}}\hat{V}_{k}^{2}\min\left\{ \left|k\right|,k_{F}\right\} }\left\langle \Psi,\mleft(\mathcal{N}_{E}^{3}+1\mright)\Psi\right\rangle 
\]
for a constant $C>0$ depending only on $\sum_{k\in\mathbb{Z}_{\ast}^{3}}\hat{V}_{k}^{2}$.
\end{prop}

\textbf{Proof:} Using Proposition \ref{prop:bbastEstimates} we estimate
the sum of the first term of $\mathcal{G}$ as
\begin{align}
 & \quad\,\sum_{k,l\in\mathbb{Z}_{\ast}^{3}}\hat{V}_{k}\sum_{p,q\in B_{F}^{c}\cap\mleft(B_{F}^{c}+k\mright)}^{\sigma,\tau}1_{L_{l}}\mleft(q\mright)\left|\left\langle \Psi,\tilde{c}_{p,\sigma}^{\ast}b_{l}\mleft(K_{l}e_{q}\mright)\tilde{c}_{-q+l,\tau}\tilde{c}_{-q+k,\tau}\tilde{c}_{p-k,\sigma}\Psi\right\rangle \right|\nonumber \\
 & \leq\sum_{k,l\in\mathbb{Z}_{\ast}^{3}}\hat{V}_{k}\sum_{p,q\in B_{F}^{c}\cap\mleft(B_{F}^{c}+k\mright)}^{\sigma,\tau}1_{L_{l}}\mleft(q\mright)\left\Vert b_{l}^{\ast}\mleft(K_{l}e_{q}\mright)\tilde{c}_{p,\sigma}\Psi\right\Vert \left\Vert \tilde{c}_{-q+l,\tau}\tilde{c}_{-q+k,\tau}\tilde{c}_{p-k,\sigma}\Psi\right\Vert \nonumber \\
 & \leq\sum_{k,l\in\mathbb{Z}_{\ast}^{3}}\hat{V}_{k}\sum_{p,q\in B_{F}^{c}\cap\mleft(B_{F}^{c}+k\mright)}^{\sigma,\tau}1_{L_{l}}\mleft(q\mright)\left\Vert K_{l}e_{q}\right\Vert \Vert\tilde{c}_{p,\sigma}\mleft(\mathcal{N}_{E}+1\mright)^{\frac{1}{2}}\Psi\Vert\left\Vert \tilde{c}_{p-k,\sigma}\tilde{c}_{-q+l,\tau}\tilde{c}_{-q+k,\tau}\Psi\right\Vert \nonumber \\
 & \leq\left\Vert \mleft(\mathcal{N}_{E}+1\mright)\Psi\right\Vert \sum_{l\in\mathbb{Z}_{\ast}^{3}}\sum_{q\in L_{l}}^{\tau}\left\Vert K_{l}e_{q}\right\Vert \sum_{k\in\mathbb{Z}_{\ast}^{3}}1_{B_{F}^{c}+k}\mleft(q\mright)\hat{V}_{k}\Vert\tilde{c}_{-q+k,\tau}\tilde{c}_{-q+l,\tau}\mathcal{N}_{E}^{\frac{1}{2}}\Psi\Vert\\
 & \leq\sqrt{\sum_{k\in\mathbb{Z}_{\ast}^{3}}\hat{V}_{k}^{2}}\left\Vert \mleft(\mathcal{N}_{E}+1\mright)\Psi\right\Vert \sum_{l\in\mathbb{Z}_{\ast}^{3}}\sum_{q\in L_{l}}^{\tau}\left\Vert K_{l}e_{q}\right\Vert \left\Vert \tilde{c}_{-q+l,\tau}\mathcal{N}_{E}\Psi\right\Vert \nonumber \\
 & \leq\sqrt{s}\sqrt{\sum_{k\in\mathbb{Z}_{\ast}^{3}}\hat{V}_{k}^{2}}\mleft(\sum_{l\in\mathbb{Z}_{\ast}^{3}}\left\Vert K_{l}\right\Vert _{\mathrm{HS}}\mright)\left\Vert \mleft(\mathcal{N}_{E}+1\mright)\Psi\right\Vert \Vert\mathcal{N}_{E}^{\frac{3}{2}}\Psi\Vert.\nonumber 
\end{align}
Now, the $\left\Vert K_{k}\right\Vert _{\mathrm{HS}}$ estimate of
Theorem \ref{them:OneBodyEstimates} and Cauchy-Schwarz lets us estimate
\[
\sum_{k\in\mathbb{Z}_{\ast}^{3}}\left\Vert K_{k}\right\Vert _{\mathrm{HS}}\leq C\sum_{k\in\mathbb{Z}_{\ast}^{3}}\hat{V}_{k}\min\,\{1,k_{F}^{2}\left|k\right|^{-2}\}\leq C\sqrt{\sum_{k\in\mathbb{Z}_{\ast}^{3}}\frac{\min\,\{1,k_{F}^{4}\left|k\right|^{-4}\}}{\min\left\{ \left|k\right|,k_{F}\right\} }}\sqrt{\sum_{k\in\mathbb{Z}_{\ast}^{3}}\hat{V}_{k}^{2}\min\left\{ \left|k\right|,k_{F}\right\} },
\]
and
\begin{equation}
\sum_{k\in\mathbb{Z}_{\ast}^{3}}\frac{\min\,\{1,k_{F}^{4}\left|k\right|^{-4}\}}{\min\left\{ \left|k\right|,k_{F}\right\} }=\sum_{k\in B_{F}\backslash\left\{ 0\right\} }\frac{1}{\left|k\right|}+k_{F}^{3}\sum_{k\in\mathbb{Z}_{\ast}^{3}\backslash B_{F}}\frac{1}{\left|k\right|^{4}}\leq Ck_{F}^{2}
\end{equation}
for a constant $C>0$ independent of all quantities, so in all the
first term of $\mathcal{G}$ obeys
\begin{align}
 & \;\frac{s^{-\frac{1}{2}}k_{F}^{-1}}{2\,\mleft(2\pi\mright)^{3}}\sum_{k,l\in\mathbb{Z}_{\ast}^{3}}\hat{V}_{k}\sum_{p,q\in B_{F}^{c}\cap\mleft(B_{F}^{c}+k\mright)}1_{L_{l}}\mleft(q\mright)\left|\left\langle \Psi,\tilde{c}_{p}^{\ast}b_{l}\mleft(K_{l}e_{q}\mright)\tilde{c}_{-q+l}\tilde{c}_{-q+k}\tilde{c}_{p-k}\Psi\right\rangle \right|\\
 & \leq C\sqrt{\sum_{k\in\mathbb{Z}_{\ast}^{3}}\hat{V}_{k}^{2}}\sqrt{\sum_{k\in\mathbb{Z}_{\ast}^{3}}\hat{V}_{k}^{2}\min\left\{ \left|k\right|,k_{F}\right\} }\left\Vert \mleft(\mathcal{N}_{E}+1\mright)\Psi\right\Vert \Vert\mathcal{N}_{E}^{\frac{3}{2}}\Psi\Vert.\nonumber 
\end{align}
Similarly, for the second term (using simply that $\left\Vert \tilde{c}_{p-l,\sigma}\right\Vert _{\mathrm{Op}}=1$
at the beginning)
\begin{align}
 & \quad\sum_{k,l\in\mathbb{Z}_{\ast}^{3}}\hat{V}_{k}\sum_{p,q\in B_{F}^{c}\cap\mleft(B_{F}^{c}+k\mright)}^{\sigma,\tau}1_{L_{l}}\mleft(p\mright)1_{L_{l}}\mleft(q\mright)\left|\left\langle K_{l}e_{q},e_{p}\right\rangle \left\langle \Psi,\tilde{c}_{p-l,\sigma}\tilde{c}_{-q+l,\tau}\tilde{c}_{-q+k,\tau}\tilde{c}_{p-k,\sigma}\Psi\right\rangle \right|\nonumber \\
 & \leq\left\Vert \Psi\right\Vert \sum_{k,l\in\mathbb{Z}_{\ast}^{3}}\hat{V}_{k}\sum_{p,q\in B_{F}^{c}\cap\mleft(B_{F}^{c}+k\mright)}^{\sigma,\tau}1_{L_{l}}\mleft(p\mright)1_{L_{l}}\mleft(q\mright)\left|\left\langle K_{l}e_{q},e_{p}\right\rangle \right|\left\Vert \tilde{c}_{p-k,\sigma}\tilde{c}_{-q+l,\tau}\tilde{c}_{-q+k,\tau}\Psi\right\Vert \\
 & \leq\sqrt{s}\left\Vert \Psi\right\Vert \sum_{l\in\mathbb{Z}_{\ast}^{3}}\sum_{q\in L_{l}}^{\tau}\left\Vert K_{l}e_{q}\right\Vert \sum_{k\in\mathbb{Z}_{\ast}^{3}}1_{B_{F}^{c}+k}\mleft(q\mright)\hat{V}_{k}\Vert\tilde{c}_{-q+k,\tau}\tilde{c}_{-q+l,\tau}\mathcal{N}_{E}^{\frac{1}{2}}\Psi\Vert\nonumber \\
 & \leq s\sqrt{\sum_{k\in\mathbb{Z}_{\ast}^{3}}\hat{V}_{k}^{2}}\mleft(\sum_{l\in\mathbb{Z}_{\ast}^{3}}\left\Vert K_{l}\right\Vert _{\mathrm{HS}}\mright)\left\Vert \Psi\right\Vert \Vert\mathcal{N}_{E}^{\frac{3}{2}}\Psi\Vert.\nonumber 
\end{align}
$\hfill\square$

\subsection{Gronwall Estimates}

We now establish control over the operators $e^{\mathcal{K}}\mathcal{N}_{E}^{m}e^{-\mathcal{K}}$
for $m=1,2,3$. Consider first the mapping $t\mapsto e^{t\mathcal{K}}\mathcal{N}_{E}e^{-t\mathcal{K}}$:
Noting that for any $\Psi\in\mathcal{H}_{N}$
\begin{equation}
\frac{d}{dt}\left\langle \Psi,e^{t\mathcal{K}}\mleft(\mathcal{N}_{E}+1\mright)e^{-t\mathcal{K}}\Psi\right\rangle =\left\langle \Psi,e^{-t\mathcal{K}}\left[\mathcal{K},\mathcal{N}_{E}\right]e^{-t\mathcal{K}}\Psi\right\rangle ,
\end{equation}
Gronwall's lemma implies that to bound $e^{t\mathcal{K}}\mleft(\mathcal{N}_{E}+1\mright)e^{-t\mathcal{K}}$
it suffices to control $\left[\mathcal{K},\mathcal{N}_{E}\right]$
with respect to $\mathcal{N}_{E}+1$ itself. We determine the commutator:
As $\mathcal{K}=\mathcal{\tilde{K}}-\mathcal{\tilde{K}}^{\ast}$ for
\begin{equation}
\tilde{\mathcal{K}}=\frac{1}{2}\sum_{l\in\mathbb{Z}_{\ast}^{3}}\sum_{p,q\in L_{l}}\left\langle e_{p},K_{l}e_{q}\right\rangle b_{l,p}b_{-l,-q}
\end{equation}
and $\left[b_{l,p},\mathcal{N}_{E}\right]=b_{l,p}$ it holds that
$[\tilde{\mathcal{K}},\mathcal{N}_{E}]=2\,\tilde{\mathcal{K}}$, whence
\begin{equation}
\left[\mathcal{K},\mathcal{N}_{E}\right]=2\,\mathrm{Re}\mleft([\tilde{\mathcal{K}},\mathcal{N}_{E}]\mright)=2\,\tilde{\mathcal{K}}+2\,\tilde{\mathcal{K}}^{\ast}.
\end{equation}
The estimate of Proposition \ref{prop:calKTildeBound} immediately
yields that
\begin{equation}
\pm\left[\mathcal{K},\mathcal{N}_{E}\right]\leq C\mleft(\mathcal{N}_{E}+1\mright)\label{eq:calKNeCommutatorBound}
\end{equation}
for a constant $C>0$ depending only on $\sum_{k\in\mathbb{Z}_{\ast}^{3}}\hat{V}_{k}^{2}$
and $s$, whence by Gronwall's lemma
\begin{equation}
\left\langle \Psi,e^{t\mathcal{K}}\mleft(\mathcal{N}_{E}+1\mright)e^{-t\mathcal{K}}\Psi\right\rangle \leq e^{C\left|t\right|}\left\langle \Psi,\mleft(\mathcal{N}_{E}+1\mright)\Psi\right\rangle \leq C'\left\langle \Psi,\mleft(\mathcal{N}_{E}+1\mright)\Psi\right\rangle ,\quad\left|t\right|\leq1.
\end{equation}
This proves the bound for $\mathcal{N}_{E}$; for $\mathcal{N}_{E}^{2}$
we will as in \cite{ChrHaiNam-21} apply the following lemma:
\begin{lem}
\label{lemma:CommutatorRootEstimate}Let $A,B,Z$ be given with $A>0$,
$Z\geq0$ and $\left[A,Z\right]=0$. Then if $\pm\left[A,\left[A,B\right]\right]\leq Z$
it holds that
\[
\pm[A^{\frac{1}{2}},[A^{\frac{1}{2}},B]]\leq\frac{1}{4}A^{-1}Z.
\]
\end{lem}

We include the proof in appendix section \ref{subsec:ASquareRootEstimationResult}.

The estimates are as follows:
\begin{prop}
\label{prop:GronwallEstimates}For any $\Psi\in\mathcal{H}_{N}$ and
$\left|t\right|\leq1$ it holds that
\[
\left\langle e^{-t\mathcal{K}}\Psi,\mleft(\mathcal{N}_{E}^{m}+1\mright)e^{-t\mathcal{K}}\Psi\right\rangle \leq C\left\langle \Psi,\mleft(\mathcal{N}_{E}^{m}+1\mright)\Psi\right\rangle ,\quad m=1,2,3,
\]
for a constant $C>0$ depending only on $\sum_{k\in\mathbb{Z}_{\ast}^{3}}\hat{V}_{k}^{2}$
and $s$.
\end{prop}

\textbf{Proof:} The case of $m=1$ was proved above. For $m=2$ it
suffices to control $\left[\mathcal{K},\mathcal{N}_{E}^{2}\right]$
in terms of $\mathcal{N}_{E}^{2}+1$; by the identity $\left\{ A,B\right\} =A^{\frac{1}{2}}BA^{\frac{1}{2}}+[A^{\frac{1}{2}},[A^{\frac{1}{2}},B]]$
we can write
\begin{align}
\left[\mathcal{K},\mathcal{N}_{E}^{2}\right] & =\mathcal{N}_{E}\left[\mathcal{K},\mathcal{N}_{E}\right]+\left[\mathcal{K},\mathcal{N}_{E}\right]\mathcal{N}_{E}=\left\{ \mathcal{N}_{E},\left[\mathcal{K},\mathcal{N}_{E}\right]\right\} =\left\{ \mathcal{N}_{E}+1,\left[\mathcal{K},\mathcal{N}_{E}\right]\right\} -2\left[\mathcal{K},\mathcal{N}_{E}\right]\\
 & =\mleft(\mathcal{N}_{E}+1\mright)^{\frac{1}{2}}\left[\mathcal{K},\mathcal{N}_{E}\right]\mleft(\mathcal{N}_{E}+1\mright)^{\frac{1}{2}}+[\mleft(\mathcal{N}_{E}+1\mright)^{\frac{1}{2}},[\mleft(\mathcal{N}_{E}+1\mright)^{\frac{1}{2}},\left[\mathcal{K},\mathcal{N}_{E}\right]]]-2\left[\mathcal{K},\mathcal{N}_{E}\right]\nonumber 
\end{align}
and note that the commutator $[\tilde{\mathcal{K}},\mathcal{N}_{E}]=2\,\tilde{\mathcal{K}}$
also implies that
\begin{equation}
\left[\mathcal{N}_{E},\left[\mathcal{N}_{E},\left[\mathcal{K},\mathcal{N}_{E}\right]\right]\right]=4\left[\mathcal{K},\mathcal{N}_{E}\right],
\end{equation}
so by Lemma \ref{lemma:CommutatorRootEstimate} and equation (\ref{eq:calKNeCommutatorBound})
\begin{equation}
\pm\left[\mathcal{K},\mathcal{N}_{E}^{2}\right]\leq C\mleft(\mleft(\mathcal{N}_{E}+1\mright)^{2}+1+\mleft(\mathcal{N}_{E}+1\mright)\mright)\leq C'\mleft(\mathcal{N}_{E}^{2}+1\mright).
\end{equation}
Similarly, for $\mathcal{N}_{E}^{3}$,
\begin{align}
\left[\mathcal{K},\mathcal{N}_{E}^{3}\right] & =\mathcal{N}_{E}^{2}\left[\mathcal{K},\mathcal{N}_{E}\right]+\mathcal{N}_{E}\left[\mathcal{K},\mathcal{N}_{E}\right]\mathcal{N}_{E}+\left[\mathcal{K},\mathcal{N}_{E}\right]\mathcal{N}_{E}^{2}\nonumber \\
 & =3\,\mathcal{N}_{E}\left[\mathcal{K},\mathcal{N}_{E}\right]\mathcal{N}_{E}+\mathcal{N}_{E}\left[\mathcal{N}_{E},\left[\mathcal{K},\mathcal{N}_{E}\right]\right]+\left[\left[\mathcal{K},\mathcal{N}_{E}\right],\mathcal{N}_{E}\right]\mathcal{N}_{E}\\
 & =3\,\mathcal{N}_{E}\left[\mathcal{K},\mathcal{N}_{E}\right]\mathcal{N}_{E}+\left[\mathcal{N}_{E},\left[\mathcal{N}_{E},\left[\mathcal{K},\mathcal{N}_{E}\right]\right]\right]=3\,\mathcal{N}_{E}\left[\mathcal{K},\mathcal{N}_{E}\right]\mathcal{N}_{E}+4\left[\mathcal{K},\mathcal{N}_{E}\right]\nonumber 
\end{align}
implies that
\begin{equation}
\pm\left[\mathcal{K},\mathcal{N}_{E}^{3}\right]\leq C\mleft(\mathcal{N}_{E}\mleft(\mathcal{N}_{E}+1\mright)\mathcal{N}_{E}+\mleft(\mathcal{N}_{E}+1\mright)\mright)\leq C'\mleft(\mathcal{N}_{E}^{3}+1\mright)
\end{equation}
hence the $m=3$ bound.

$\hfill\square$

\subsubsection*{Conclusion of Theorem \ref{them:MainTheorem}}

We can now conclude:
\begin{thm*}[\ref{them:MainTheorem}]
It holds that
\[
\inf\sigma\mleft(H_{N}\mright)\leq E_{F}+E_{\mathrm{corr},\mathrm{bos}}+E_{\mathrm{corr},\mathrm{ex}}+C\sqrt{\sum_{k\in\mathbb{Z}_{\ast}^{3}}\hat{V}_{k}^{2}\min\left\{ \left|k\right|,k_{F}\right\} },\quad k_{F}\rightarrow\infty,
\]
for a constant $C>0$ depending only on $\sum_{k\in\mathbb{Z}_{\ast}^{3}}\hat{V}_{k}^{2}$
and $s$.
\end{thm*}
\textbf{Proof:} By the variational principle applied to the trial
state $e^{-\mathcal{K}}\psi_{F}$ we have by Proposition \ref{prop:LocalizedHamiltonian}
and the Theorems \ref{thm:DiagonalizationoftheBosonizableTerms},
\ref{them:OneBodyEstimates} and \ref{thm:NBTermEstimates} that
\begin{align*}
 & \inf\sigma\mleft(H_{N}\mright)\leq E_{F}+\left\langle \psi_{F},e^{\mathcal{K}}\mleft(H_{\mathrm{kin}}^{\prime}+\sum_{k\in\mathbb{Z}_{\ast}^{3}}\frac{\hat{V}_{k}k_{F}^{-1}}{2\,\mleft(2\pi\mright)^{3}}\mleft(2B_{k}^{\ast}B_{k}+B_{k}B_{-k}+B_{-k}^{\ast}B_{k}^{\ast}\mright)\mright)e^{-\mathcal{K}}\psi_{F}\right\rangle \\
 & \qquad\qquad\qquad\qquad\qquad\qquad\qquad\qquad\qquad\qquad\qquad\qquad+\left\langle \psi_{F},e^{\mathcal{K}}\mathcal{C}e^{-\mathcal{K}}\psi_{F}\right\rangle +\left\langle \psi_{F},e^{\mathcal{K}}\mathcal{Q}e^{-\mathcal{K}}\psi_{F}\right\rangle \\
 & =E_{F}+E_{\mathrm{corr},\mathrm{bos}}+\left\langle \psi_{F},H_{\mathrm{kin}}^{\prime}\psi_{F}\right\rangle +2\sum_{k\in\mathbb{Z}_{\ast}^{3}}\left\langle \psi_{F},Q_{1}^{k}\mleft(e^{-K_{k}}h_{k}e^{-K_{k}}-h_{k}\mright)\psi_{F}\right\rangle \\
 & +\sum_{k\in\mathbb{Z}_{\ast}^{3}}\int_{0}^{1}\left\langle e^{-\mleft(1-t\mright)\mathcal{K}}\psi_{F},\mleft(\varepsilon_{k}\mleft(\left\{ K_{k},B_{k}\mleft(t\mright)\right\} \mright)+2\,\mathrm{Re}\mleft(\mathcal{E}_{k}^{1}\mleft(A_{k}\mleft(t\mright)\mright)\mright)+2\,\mathrm{Re}\mleft(\mathcal{E}_{k}^{2}\mleft(B_{k}\mleft(t\mright)\mright)\mright)\mright)e^{-\mleft(1-t\mright)\mathcal{K}}\psi_{F}\right\rangle dt\\
 & +\left\langle e^{\mathcal{K}}\psi_{F},\mleft(G+\mathcal{Q}_{\mathrm{LR}}\mright)e^{-\mathcal{K}}\psi_{F}\right\rangle +\left\langle \psi_{F},\mathcal{Q}_{\mathrm{SR}}\psi_{F}\right\rangle +\int_{0}^{1}\left\langle e^{-t\mathcal{K}}\psi_{F},\mleft(2\,\mathrm{Re}\mleft(\mathcal{G}\mright)\mright)e^{-t\mathcal{K}}\psi_{F}\right\rangle dt\\
 & =E_{F}+E_{\mathrm{corr},\mathrm{bos}}+E_{\mathrm{corr},\mathrm{ex}}+\epsilon_{1}+\epsilon_{2}+\epsilon_{3},
\end{align*}
where we also used that
\begin{equation}
H_{\mathrm{kin}}^{\prime}\psi_{F}=Q_{1}^{k}\mleft(A\mright)\psi_{F}=\mathcal{Q}_{\mathrm{SR}}\psi_{F}=0
\end{equation}
and that $\left\langle \psi_{F},e^{\mathcal{K}}\mathcal{C}e^{-\mathcal{K}}\psi_{F}\right\rangle =0$
by Proposition \ref{prop:VanishingofCubicTerms}. The errors $\epsilon_{1}$,
$\epsilon_{2}$ and $\epsilon_{3}$ obey
\begin{equation}
\epsilon_{1}=\sum_{k\in\mathbb{Z}_{\ast}^{3}}\int_{0}^{1}\left\langle \psi_{F},2\,\mathrm{Re}\mleft(\mathcal{E}_{k}^{2}\mleft(B_{k}\mleft(t\mright)\mright)\mright)\psi_{F}\right\rangle dt-E_{\mathrm{corr},\mathrm{ex}}\leq C\sum_{k\in\mathbb{Z}_{\ast}^{3}}\sqrt{\sum_{k\in\mathbb{Z}_{\ast}^{3}}\hat{V}_{k}^{2}\min\left\{ \left|k\right|,k_{F}\right\} }
\end{equation}
by Proposition \ref{prop:LeadingExchangeContribution},
\begin{align}
\epsilon_{2} & =\sum_{k\in\mathbb{Z}_{\ast}^{3}}\int_{0}^{1}\left\langle e^{-\mleft(1-t\mright)\mathcal{K}}\psi_{F},\mleft(\varepsilon_{k}\mleft(\left\{ K_{k},B_{k}\mleft(t\mright)\right\} \mright)+2\,\mathrm{Re}\mleft(\mathcal{E}_{k}^{1}\mleft(A_{k}\mleft(t\mright)\mright)\mright)\mright)e^{-\mleft(1-t\mright)\mathcal{K}}\psi_{F}\right\rangle dt\nonumber \\
 & +\sum_{k\in\mathbb{Z}_{\ast}^{3}}\int_{0}^{1}\left\langle e^{-\mleft(1-t\mright)\mathcal{K}}\psi_{F},\mleft(2\,\mathrm{Re}\mleft(\mathcal{E}_{k}^{2}\mleft(B_{k}\mleft(t\mright)\mright)-\left\langle \psi_{F},\mathcal{E}_{k}^{2}\mleft(B_{k}\mleft(t\mright)\mright)\psi_{F}\right\rangle \mright)\mright)e^{-\mleft(1-t\mright)\mathcal{K}}\psi_{F}\right\rangle dt\\
 & \leq Ck_{F}^{-1}+C\sqrt{\sum_{k\in\mathbb{Z}_{\ast}^{3}}\hat{V}_{k}^{2}\min\left\{ \left|k\right|,k_{F}\right\} }\leq C\sqrt{\sum_{k\in\mathbb{Z}_{\ast}^{3}}\hat{V}_{k}^{2}\min\left\{ \left|k\right|,k_{F}\right\} }\nonumber 
\end{align}
by Theorem \ref{them:ExchangeTermsEstimates}, and
\begin{align}
\epsilon_{3} & =\left\langle e^{-\mathcal{K}}\psi_{F},\mleft(G+\mathcal{Q}_{\mathrm{LR}}\mright)e^{-\mathcal{K}}\psi_{F}\right\rangle +\int_{0}^{1}\left\langle e^{-t\mathcal{K}}\psi_{F},\mleft(2\,\mathrm{Re}\mleft(\mathcal{G}\mright)\mright)e^{-t\mathcal{K}}\psi_{F}\right\rangle dt\\
 & \leq C\sqrt{\sum_{k\in\mathbb{Z}_{\ast}^{3}}\hat{V}_{k}^{2}\min\left\{ \left|k\right|,k_{F}\right\} }\nonumber 
\end{align}
by Theorem \ref{thm:NBTermEstimates}, where we for the last error
terms also used that
\begin{equation}
\left\langle e^{-t\mathcal{K}}\psi_{F},\mleft(\mathcal{N}_{E}^{m}+1\mright)e^{-t\mathcal{K}}\psi_{F}\right\rangle \leq C,\quad\left|t\right|\leq1,\,m=1,2,3,
\end{equation}
as follows by Proposition \ref{prop:GronwallEstimates}.

\section{\label{sec:ExtensiontoAttractivePotentials}Extension to Attractive
Potentials}

We now make the observation that the result of Theorem \ref{them:MainTheorem}
generalizes to weakly attractive potentials.

To determine under what conditions we can do this, let us consider
where we applied the assumption $\hat{V}_{k}\geq0$. This condition
did not enter anywhere in Section \ref{prop:LocalizedHamiltonian},
so the conclusion of that section, i.e. the representation
\begin{equation}
H_{N}=E_{F}+H_{\mathrm{kin}}^{\prime}+\sum_{k\in\mathbb{Z}_{\ast}^{3}}\frac{\hat{V}_{k}k_{F}^{-1}}{2\,\mleft(2\pi\mright)^{3}}\mleft(2B_{k}^{\ast}B_{k}+B_{k}B_{-k}+B_{-k}^{\ast}B_{k}^{\ast}\mright)+\mathcal{C}+\mathcal{Q},
\end{equation}
continues to hold. The first time we applied the condition was in
Section \ref{sec:OverviewofBosonicBogolubovTransformations}, when
we wrote the bosonizable interaction terms in the form
\begin{equation}
\sum_{k\in\mathbb{Z}_{\ast}^{3}}\frac{\hat{V}_{k}k_{F}^{-1}}{2\,\mleft(2\pi\mright)^{3}}\mleft(2B_{k}^{\ast}B_{k}+B_{k}B_{-k}+B_{-k}^{\ast}B_{k}^{\ast}\mright)=\sum_{k\in\mathbb{Z}_{\ast}^{3}}\mleft(2\,Q_{1}^{k}\mleft(P_{k}\mright)+Q_{2}^{k}\mleft(P_{k}\mright)\mright),
\end{equation}
since we defined $P_{k}:\ell^{2}\mleft(L_{k}\mright)\rightarrow\ell^{2}\mleft(L_{k}\mright)$
to act as $P_{k}\mleft(\cdot\mright)=\left\langle v_{k},\cdot\right\rangle v_{k}$
for $v_{k}=\sqrt{\frac{s\hat{V}_{k}k_{F}^{-1}}{2\,\mleft(2\pi\mright)^{3}}}\sum_{p\in L_{k}}e_{p}$.
This definition was made to ensure that
\begin{equation}
\left\langle e_{p},P_{k}e_{q}\right\rangle =\left\langle e_{p},v_{k}\right\rangle \left\langle v_{k},e_{q}\right\rangle =\frac{s\hat{V}_{k}k_{F}^{-1}}{2\,\mleft(2\pi\mright)^{3}},\quad p,q\in L_{k},
\end{equation}
but it is clear that this can still be enforced by a slight modification:
If we more generally define $P_{k}$ and $v_{k}$ by
\begin{equation}
P_{k}\mleft(\cdot\mright)=\mathrm{sgn}(\hat{V}_{k})\left\langle v_{k},\cdot\right\rangle v_{k},\quad v_{k}=\sqrt{\frac{s|\hat{V}_{k}|k_{F}^{-1}}{2\,\mleft(2\pi\mright)^{3}}}\sum_{p\in L_{k}}e_{p},
\end{equation}
then we recover the previous definition for $\hat{V}_{k}\geq0$, but
now also have that $\left\langle e_{p},P_{k}e_{q}\right\rangle =\frac{s\hat{V}_{k}k_{F}^{-1}}{2\,\mleft(2\pi\mright)^{3}}$
even if $\hat{V}_{k}<0$.

As the calculations of Section \ref{sec:OverviewofBosonicBogolubovTransformations}
were purely algebraic, we see that the conclusion, i.e. the existence
of a unitary transformation $e^{\mathcal{K}}$ such that
\begin{align}
 & \;\,e^{\mathcal{K}}\mleft(H_{\mathrm{kin}}^{\prime}+\sum_{k\in\mathbb{Z}_{\ast}^{3}}\frac{\hat{V}_{k}k_{F}^{-1}}{2\,\mleft(2\pi\mright)^{3}}\mleft(2B_{k}^{\ast}B_{k}+B_{k}B_{-k}+B_{-k}^{\ast}B_{k}^{\ast}\mright)\mright)e^{-\mathcal{K}}\nonumber \\
 & =\sum_{k\in\mathbb{Z}_{\ast}^{3}}\mathrm{tr}\mleft(e^{-K_{k}}h_{k}e^{-K_{k}}-h_{k}-P_{k}\mright)+H_{\mathrm{kin}}^{\prime}+2\sum_{k\in\mathbb{Z}_{\ast}^{3}}Q_{1}^{k}\mleft(e^{-K_{k}}h_{k}e^{-K_{k}}-h_{k}\mright)\\
 & +\sum_{k\in\mathbb{Z}_{\ast}^{3}}\int_{0}^{1}e^{\mleft(1-t\mright)\mathcal{K}}\mleft(\varepsilon_{k}\mleft(\left\{ K_{k},B_{k}\mleft(t\mright)\right\} \mright)+2\,\mathrm{Re}\mleft(\mathcal{E}_{k}^{1}\mleft(A_{k}\mleft(t\mright)\mright)\mright)+2\,\mathrm{Re}\mleft(\mathcal{E}_{k}^{2}\mleft(B_{k}\mleft(t\mright)\mright)\mright)\mright)e^{-\mleft(1-t\mright)\mathcal{K}}dt,\nonumber 
\end{align}
continues to hold (keeping the new definition of $P_{k}$ in mind),
\textit{provided} the diagonalizing kernels
\begin{equation}
K_{k}=-\frac{1}{2}\log\mleft(h_{k}^{-\frac{1}{2}}\mleft(h_{k}^{\frac{1}{2}}\mleft(h_{k}+2P_{k}\mright)h_{k}^{\frac{1}{2}}\mright)^{\frac{1}{2}}h_{k}^{-\frac{1}{2}}\mright)
\end{equation}
are still well-defined when $\hat{V}_{k}<0$.

This is the condition that $h_{k}+2P_{k}=h_{k}-2P_{v_{k}}>0$. By
the Sherman-Morrison formula (Lemma \ref{lemma:ShermanMorrison} -
as well as motonotony of $t\mapsto h_{k}+tP_{k}$) this is the case
if and only if
\begin{equation}
1-2\left\langle v_{k},h_{k}^{-1}v_{k}\right\rangle >0
\end{equation}
which can be expanded and rearranged to
\begin{equation}
\hat{V}_{k}>-\frac{\mleft(2\pi\mright)^{3}}{sk_{F}^{-1}\sum_{p\in L_{k}}\lambda_{k,p}^{-1}},\quad k\in\mathbb{Z}_{\ast}^{3}.\label{eq:PositivityCondition}
\end{equation}
In appendix section \ref{sec:RiemannSumEstimates} we prove the following
asymptotic behaviour of the Riemann sum $\sum_{p\in L_{k}}\lambda_{k,p}^{-1}$:
\begin{prop}
\label{prop:betaeq-1Asymptotics}For any $\gamma\in\mleft(0,\frac{1}{11}\mright)$
and $k\in\overline{B}\mleft(0,k_{F}^{\gamma}\mright)$ it holds that
\[
\sum_{p\in L_{k}}\lambda_{k,p}^{-1}=2\pi k_{F}+O\mleft(\log\mleft(k_{F}\mright)^{\frac{5}{3}}k_{F}^{\frac{1}{3}\mleft(2+11\gamma\mright)}\mright),\quad k_{F}\rightarrow\infty.
\]
\end{prop}

The condition of equation (\ref{eq:PositivityCondition}) thus asymptotically
amounts to
\begin{equation}
\hat{V}_{k}>-\frac{4\pi^{2}}{s},\quad k\in\mathbb{Z}_{\ast}^{3},
\end{equation}
but as in the statement of Theorem \ref{them:AttractiveGeneralization}
we will for the purposes of analysis make the slightly stronger assumption
that
\begin{equation}
\hat{V}_{k}\geq-\mleft(1-\epsilon\mright)\frac{4\pi^{2}}{s},\quad k\in\mathbb{Z}_{\ast}^{3},
\end{equation}
for some $\epsilon>0$. With this we can uniformly bound $1-2\left\langle v_{k},h_{k}^{-1}v_{k}\right\rangle $
away from $0$:
\begin{lem}
\label{lemma:AttractiveLowerBound}Let $\sum_{k\in\mathbb{Z}_{\ast}^{3}}\hat{V}_{k}^{2}<\infty$
and $\hat{V}_{k}\geq-\mleft(1-\epsilon\mright)\frac{4\pi^{2}}{s}$ for
all $k\in\mathbb{Z}_{\ast}^{3}$. Then
\[
\inf_{\left\{ k\in\mathbb{Z}_{\ast}^{3}\mid\hat{V}_{k}<0\right\} }\mleft(1-2\left\langle v_{k},h_{k}^{-1}v_{k}\right\rangle \mright)\geq C,\quad k_{F}\rightarrow\infty,
\]
for a constant $C>0$ depending only on $\epsilon$.
\end{lem}

\textbf{Proof:} Expanding the definitions and applying Proposition
\ref{prop:betaeq-1Asymptotics}, we have for all $k\in\overline{B}(0,k_{F}^{\frac{1}{20}})$
(say) with $\hat{V}_{k}<0$ that
\begin{align}
1-2\left\langle v_{k},h_{k}^{-1}v_{k}\right\rangle  & =1-\frac{s|\hat{V}_{k}|k_{F}^{-1}}{\mleft(2\pi\mright)^{3}}\sum_{p\in L_{k}}\lambda_{k,p}^{-1}\geq1-\mleft(1-\epsilon\mright)\frac{4\pi^{2}}{s}\frac{sk_{F}^{-1}}{\mleft(2\pi\mright)^{3}}\sum_{p\in L_{k}}\lambda_{k,p}^{-1}\nonumber \\
 & =\epsilon+\mleft(1-\epsilon\mright)\frac{k_{F}^{-1}}{2\pi}\mleft(2\pi k_{F}-\sum_{p\in L_{k}}\lambda_{k,p}^{-1}\mright)\geq\epsilon-C\log\mleft(k_{F}\mright)^{\frac{5}{3}}k_{F}^{-\frac{1}{3}\mleft(1-\frac{11}{20}\mright)}\\
 & \geq C'\nonumber 
\end{align}
as $k_{F}\rightarrow\infty$ for some $C'>0$ depending only on $\epsilon$.
If instead $k\in\mathbb{Z}_{\ast}^{3}\backslash\overline{B}(0,k_{F}^{\frac{1}{20}})$
we may note that by the general bound $\sum_{p\in L_{k}}\lambda_{k,p}^{-1}\leq Ck_{F}$,
we can always estimate
\begin{equation}
1-2\left\langle v_{k},h_{k}^{-1}v_{k}\right\rangle \geq1-Cs|\hat{V}_{k}|,
\end{equation}
so noting that
\begin{equation}
\sup_{k\in\mathbb{Z}_{\ast}^{3}\backslash\overline{B}(0,k_{F}^{\frac{1}{20}})}|\hat{V}_{k}|\leq\sqrt{\sum_{k\in\mathbb{Z}_{\ast}^{3}\backslash\overline{B}(0,k_{F}^{\frac{1}{20}})}\hat{V}_{k}^{2}}\rightarrow0,\quad k_{F}\rightarrow\infty,
\end{equation}
since $\sum_{k\in\mathbb{Z}_{\ast}^{3}}\hat{V}_{k}^{2}<\infty$ we
see that we can for $k_{F}$ sufficiently large assume that $\sup_{k\in\mathbb{Z}_{\ast}^{3}\backslash\overline{B}(0,k_{F}^{\frac{1}{20}})}\mleft(1-2\left\langle v_{k},h_{k}^{-1}v_{k}\right\rangle \mright)\geq\frac{1}{2}$
(say), so either way the claim holds.

$\hfill\square$

We remark that a similar argument shows that our condition on $\hat{V}_{k}$
is nearly optimal, in the sense that if for some $k\in\mathbb{Z}_{\ast}^{3}$
it holds that $\hat{V}_{k}<-\frac{4\pi^{2}}{s}$, then the asymptotic
result of Proposition \ref{prop:betaeq-1Asymptotics} in fact implies
that
\begin{equation}
1-2\left\langle v_{k},h_{k}^{-1}v_{k}\right\rangle <0
\end{equation}
for all sufficiently large $k_{F}$, in which case the corresponding
term of $E_{\mathrm{corr},\mathrm{bos}}$ is not even well-defined
as the integrand involves
\begin{equation}
\log\mleft(1+\frac{s\hat{V}_{k}k_{F}^{-1}}{\mleft(2\pi\mright)^{3}}\sum_{p\in L_{k}}\frac{\lambda_{k,p}}{\lambda_{k,p}^{2}+t^{2}}\mright)=\log\mleft(1-2\left\langle v_{k},h_{k}\mleft(h_{k}^{2}+t^{2}\mright)^{-1}v_{k}\right\rangle \mright).
\end{equation}
The condition $\hat{V}_{k}\geq-\mleft(1-\epsilon\mright)\frac{4\pi^{2}}{s}$
thus ensures that our diagonalization procedure (and $E_{\mathrm{corr},\mathrm{bos}}$)
remains well-defined, but it is not immediately clear how the one-body
estimates of Section \ref{sec:AnalysisofOne-BodyOperators} are to
be modified for the attractive case.

This is the main information that is needed for the generalization
to attractive potentials, but it turns out that Theorem \ref{them:OneBodyEstimates}
continues to hold almost exactly as stated before, the only difference
being an $\epsilon$-dependence and the substitution $\hat{V}_{k}\rightarrow|\hat{V}_{k}|$
in the error terms:
\begin{prop}
\label{prop:AttractiveOneBodyEstimates}It holds for any $k\in\mathbb{Z}_{\ast}^{3}$
that
\[
\mathrm{tr}\mleft(e^{-K_{k}}h_{k}e^{-K_{k}}-h_{k}-P_{k}\mright)=\frac{1}{\pi}\int_{0}^{\infty}F\mleft(\frac{s\hat{V}_{k}k_{F}^{-1}}{\mleft(2\pi\mright)^{3}}\sum_{p\in L_{k}}\frac{\lambda_{k,p}}{\lambda_{k,p}^{2}+t^{2}}\mright)dt,
\]
where $F\mleft(x\mright)=\log\mleft(1+x\mright)-x$. Furthermore, as $k_{F}\rightarrow\infty$,
\[
\left\Vert K_{k}\right\Vert _{\mathrm{HS}}\leq C|\hat{V}_{k}|\min\left\{ 1,k_{F}^{2}\left|k\right|^{-2}\right\} 
\]
and for all $p,q\in L_{k}$ and $t\in\left[0,1\right]$
\begin{align*}
\left|\left\langle e_{p},K_{k}e_{q}\right\rangle \right| & \leq C\frac{|\hat{V}_{k}|k_{F}^{-1}}{\lambda_{k,p}+\lambda_{k,q}}\\
\left|\left\langle e_{p},\mleft(-K_{k}\mright)e_{q}\right\rangle -\frac{s\hat{V}_{k}k_{F}^{-1}}{2\,\mleft(2\pi\mright)^{3}}\frac{1}{\lambda_{k,p}+\lambda_{k,q}}\right| & \leq C\frac{\hat{V}_{k}^{2}k_{F}^{-1}}{\lambda_{k,p}+\lambda_{k,q}}\\
\left|\left\langle e_{p},A_{k}\mleft(t\mright)e_{q}\right\rangle \right|,\left|\left\langle e_{p},B_{k}\mleft(t\mright)e_{q}\right\rangle \right| & \leq C\mleft(1+\hat{V}_{k}^{2}\mright)|\hat{V}_{k}|k_{F}^{-1}\\
\left|\left\langle e_{p},\mleft(\int_{0}^{1}B_{k}\mleft(t\mright)\,dt\mright)e_{q}\right\rangle -\frac{s\hat{V}_{k}k_{F}^{-1}}{4\,\mleft(2\pi\mright)^{3}}\right| & \leq C\mleft(1+|\hat{V}_{k}|\mright)\hat{V}_{k}^{2}k_{F}^{-1}\\
\left|\left\langle e_{p},\left\{ K_{k},B_{k}\mleft(t\mright)\right\} e_{q}\right\rangle \right| & \leq C\mleft(1+\hat{V}_{k}^{2}\mright)\hat{V}_{k}^{2}k_{F}^{-1}
\end{align*}
for a constant $C>0$ depending only on $s$ and $\epsilon$.
\end{prop}

We momentarily postpone the proof to subsection \ref{subsec:One-BodyEstimatesforAttractiveModes}
below.

With these estimates we are essentially done, since the computations
of the Sections \ref{sec:AnalysisofExchangeTerms} and \ref{sec:EstimationoftheNon-BosonizableTermsandGronwallEstimates}
only relied on these, as well as the triangle and Cauchy-Schwarz inequalities.
Whenever the triangle inequality was applied, the only difference
that is required for attractive potentials is that $\hat{V}_{k}$
is substituted with $|\hat{V}_{k}|$, but since we generally apply
the Cauchy-Schwarz inequality to estimate in terms of $\hat{V}_{k}^{2}$
this makes no difference in the end.

The only modification to Theorem \ref{them:MainTheorem} that is necessary
to generalize to the condition $\hat{V}_{k}>-\mleft(1-\epsilon\mright)\frac{4\pi^{2}}{s}$
is therefore that the constant in the error term is $\epsilon$-dependent,
which is Theorem \ref{them:AttractiveGeneralization}.

\subsection{\label{subsec:One-BodyEstimatesforAttractiveModes}One-Body Estimates
for Attractive Modes}

To prove Proposition \ref{prop:AttractiveOneBodyEstimates} we return
to the general setting of Section \ref{sec:AnalysisofOne-BodyOperators},
i.e. we consider an $n$-dimensional Hilbert space $\mleft(V,\left\langle \cdot,\cdot\right\rangle \mright)$,
a positive self-adjoint operator $h:V\rightarrow V$ with eigenbasis
$\mleft(x_{i}\mright)_{i=1}^{n}$ and a vector $v\in V$ such that $\left\langle x_{i},v\right\rangle \geq0$,
$1\leq i\leq n$.

The calculations of this subsection are very reminiscent of those
of Section \ref{sec:AnalysisofOne-BodyOperators}, and for that reason
we will adopt a brisk pacing, mainly pointing out the necessary modifications
- these will mainly be various sign reversals.

We let $K:V\rightarrow V$ be given by
\begin{equation}
K=-\frac{1}{2}\log\mleft(h^{-\frac{1}{2}}\mleft(h^{\frac{1}{2}}\mleft(h-2P_{v}\mright)h^{\frac{1}{2}}\mright)^{\frac{1}{2}}h^{-\frac{1}{2}}\mright)=-\frac{1}{2}\log\mleft(h^{-\frac{1}{2}}\mleft(h^{2}-2P_{h^{\frac{1}{2}}v}\mright)^{\frac{1}{2}}h^{-\frac{1}{2}}\mright);
\end{equation}
we assume that $1-2\left\langle v,h^{-1}v\right\rangle >0$ so that
$K$ is well-defined. In this case we have that $e^{-2K}$ and $e^{2K}$
are given by
\begin{align}
e^{-2K} & =h^{-\frac{1}{2}}\mleft(h^{2}-2P_{h^{\frac{1}{2}}v}\mright)^{\frac{1}{2}}h^{-\frac{1}{2}}\\
e^{2K} & =h^{\frac{1}{2}}\mleft(h^{-2}+\frac{2}{1-2\left\langle v,h^{-1}v\right\rangle }P_{h^{-\frac{3}{2}}v}\mright)^{\frac{1}{2}}h^{\frac{1}{2}}\nonumber 
\end{align}
and it follows from Proposition \ref{prop:SquareRootofAOneDimensionalPerturbation}
that $\mathrm{tr}\mleft(e^{-K}he^{-K}-h+P_{v}\mright)$ is given by
\begin{equation}
\mathrm{tr}\mleft(e^{-K}he^{-K}-h+P_{v}\mright)=\frac{1}{\pi}\int_{0}^{\infty}F\mleft(-2\left\langle v,h\mleft(h^{2}+t^{2}\mright)^{-1}v\right\rangle \mright)dt,\quad F\mleft(x\mright)=\log\mleft(1+x\mright)-x.
\end{equation}
The operators $e^{-2K}$ and $e^{2K}$ obey the following matrix element
estimates:
\begin{prop}
\label{prop:Attractivee2KBounds}For all $1\leq i,j\leq n$ it holds
that
\[
2\frac{\left\langle x_{i},v\right\rangle \left\langle v,x_{j}\right\rangle }{\lambda_{i}+\lambda_{j}}\leq\left\langle x_{i},\mleft(1-e^{-2K}\mright)x_{j}\right\rangle ,\left\langle x_{i},\mleft(e^{2K}-1\mright)x_{j}\right\rangle \leq\frac{2}{1-2\left\langle v,h^{-1}v\right\rangle }\frac{\left\langle x_{i},v\right\rangle \left\langle v,x_{j}\right\rangle }{\lambda_{i}+\lambda_{j}}.
\]
\end{prop}

\textbf{Proof:} By Proposition \ref{prop:SquareRootofAOneDimensionalPerturbation}
we have that
\begin{align}
1-e^{-2K} & =1-h^{-\frac{1}{2}}\mleft(h-\frac{4}{\pi}\int_{0}^{\infty}\frac{t^{2}}{1-2\left\langle h^{\frac{1}{2}}v,\mleft(h^{2}+t^{2}\mright)^{-1}h^{\frac{1}{2}}v\right\rangle }P_{\mleft(h^{2}+t^{2}\mright)^{-1}h^{\frac{1}{2}}v}dt\mright)h^{-\frac{1}{2}}\\
 & =\frac{4}{\pi}\int_{0}^{\infty}\frac{t^{2}}{1-2\left\langle v,h\mleft(h^{2}+t^{2}\mright)^{-1}v\right\rangle }P_{\mleft(h^{2}+t^{2}\mright)^{-1}v}dt\nonumber 
\end{align}
and now it holds that
\begin{equation}
1\leq\frac{1}{1-2\left\langle v,h\mleft(h^{2}+t^{2}\mright)^{-1}v\right\rangle }\leq\frac{1}{1-2\left\langle v,h^{-1}v\right\rangle },\quad t\geq0,
\end{equation}
whence the element estimate follows as in Proposition \ref{prop:e-2Ke2KElementEstimates}.
Similarly, for $e^{2K}$,
\begin{align}
e^{2K}-1 & =h^{\frac{1}{2}}\mleft(h^{-1}+\frac{4}{\pi}\int_{0}^{\infty}\frac{t^{2}}{1-2\left\langle v,h^{-1}v\right\rangle +2\left\langle h^{-\frac{3}{2}}v,\mleft(h^{-2}+t^{2}\mright)^{-1}h^{-\frac{3}{2}}v\right\rangle }P_{\mleft(h^{-2}+t^{2}\mright)^{-1}h^{-\frac{3}{2}}v}dt\mright)h^{\frac{1}{2}}-1\\
 & =\frac{4}{\pi}\int_{0}^{\infty}\frac{t^{2}}{1-2\left\langle v,h^{-1}\mleft(h^{-2}+t^{2}\mright)^{-1}v\right\rangle t^{2}}P_{\mleft(h^{-2}+t^{2}\mright)^{-1}h^{-1}v}dt\nonumber 
\end{align}
so the claim follows as
\begin{equation}
1\leq\frac{1}{1-2\left\langle v,h^{-1}\mleft(h^{-2}+t^{2}\mright)^{-1}v\right\rangle t^{2}}\leq\frac{1}{1-2\left\langle v,h^{-1}v\right\rangle },\quad t\geq0.
\end{equation}
$\hfill\square$

As in Corollary \ref{coro:HyperbolicBounds} we can then conclude
the bounds
\begin{align}
\left\langle x_{i},\sinh\mleft(2K\mright)x_{j}\right\rangle  & \leq\frac{2}{1-2\left\langle v,h^{-1}v\right\rangle }\frac{\left\langle x_{i},v\right\rangle \left\langle v,x_{j}\right\rangle }{\lambda_{i}+\lambda_{j}}\\
\left\langle x_{i},\mleft(\cosh\mleft(2K\mright)-1\mright)x_{j}\right\rangle  & \leq\frac{2\left\langle v,h^{-1}v\right\rangle }{1-2\left\langle v,h^{-1}v\right\rangle }\frac{\left\langle x_{i},v\right\rangle \left\langle v,x_{j}\right\rangle }{\lambda_{i}+\lambda_{j}},\nonumber 
\end{align}
and we note that $\cosh\mleft(2K\mright)$ also obeys
\begin{align}
\left\langle x_{i},\mleft(\cosh\mleft(2K\mright)-1\mright)x_{j}\right\rangle  & =\frac{1}{2}\mleft(\left\langle x_{i},\mleft(e^{2K}-1\mright)x_{j}\right\rangle -\left\langle x_{i},\mleft(1-e^{-2K}\mright)x_{j}\right\rangle \mright)\\
 & \leq\frac{1}{2}\left\langle x_{i},\mleft(e^{2K}-1\mright)x_{j}\right\rangle \leq\frac{1}{1-2\left\langle v,h^{-1}v\right\rangle }\frac{\left\langle x_{i},v\right\rangle \left\langle v,x_{j}\right\rangle }{\lambda_{i}+\lambda_{j}}\nonumber 
\end{align}
so in fact
\begin{equation}
\left\langle x_{i},\mleft(\cosh\mleft(2K\mright)-1\mright)x_{j}\right\rangle \leq\frac{\min\left\{ 1,2\left\langle v,h^{-1}v\right\rangle \right\} }{1-2\left\langle v,h^{-1}v\right\rangle }\frac{\left\langle x_{i},v\right\rangle \left\langle v,x_{j}\right\rangle }{\lambda_{i}+\lambda_{j}}.
\end{equation}
By the same arguments used in Proposition \ref{prop:KElementEstimates},
it follows from Proposition \ref{prop:Attractivee2KBounds} that $K$
obeys the following elementwise bounds:
\begin{prop}
\label{prop:AttractiveKBounds}For any $1\leq i,j\leq n$ it holds
that
\[
\frac{\left\langle x_{i},v\right\rangle \left\langle v,x_{j}\right\rangle }{\lambda_{i}+\lambda_{j}}\leq\left\langle x_{i},Kx_{j}\right\rangle \leq\frac{1}{1-2\left\langle v,h^{-1}v\right\rangle }\frac{\left\langle x_{i},v\right\rangle \left\langle v,x_{j}\right\rangle }{\lambda_{i}+\lambda_{j}}.
\]
\end{prop}

As this in particular implies that $\left\langle x_{i},Kx_{j}\right\rangle \geq0$
for all $1\leq i,j\leq n$, it follows that the functions
\begin{equation}
t\mapsto\left\langle x_{i},\mleft(e^{tK}-1\mright)x_{j}\right\rangle ,\,\left\langle x_{i},\sinh\mleft(tK\mright)x_{j}\right\rangle ,\,\left\langle x_{i},\mleft(\sinh\mleft(tK\mright)-tK\mright)x_{j}\right\rangle ,\,\left\langle x_{i},\mleft(\cosh\mleft(tK\mright)-1\mright)x_{j}\right\rangle ,
\end{equation}
are non-negative and convex, whence we obtain the following analogue
of Proposition \ref{prop:tDependentElementEstimates}:
\begin{prop}
\label{prop:AttractivetDependentElementEstimates}For all $1\leq i,j\leq n$
and $t\in\left[0,1\right]$ it holds that
\begin{align*}
\frac{\left\langle x_{i},v\right\rangle \left\langle v,x_{j}\right\rangle }{\lambda_{i}+\lambda_{j}}t\leq\left\langle x_{i},\sinh\mleft(tK\mright)x_{j}\right\rangle  & \leq\frac{1}{1-2\left\langle v,h^{-1}v\right\rangle }\frac{\left\langle x_{i},v\right\rangle \left\langle v,x_{j}\right\rangle }{\lambda_{i}+\lambda_{j}}t\\
0\leq\left\langle x_{i},\mleft(\cosh\mleft(tK\mright)-1\mright)x_{j}\right\rangle  & \leq\frac{\min\left\{ 1,\left\langle v,h^{-1}v\right\rangle \right\} }{1-2\left\langle v,h^{-1}v\right\rangle }\frac{\left\langle x_{i},v\right\rangle \left\langle v,x_{j}\right\rangle }{\lambda_{i}+\lambda_{j}}\\
0\leq\left\langle x_{i},\mleft(e^{tK}-1\mright)x_{j}\right\rangle  & \leq\frac{1}{1-2\left\langle v,h^{-1}v\right\rangle }\frac{\left\langle x_{i},v\right\rangle \left\langle v,x_{j}\right\rangle }{\lambda_{i}+\lambda_{j}}.
\end{align*}
\end{prop}

\subsubsection*{Estimation of $A\mleft(t\mright)$ and $B\mleft(t\mright)$}

We thus come to the estimation of $A\mleft(t\mright)$ and $B\mleft(t\mright)$,
which are now given by
\begin{align}
A\mleft(t\mright) & =\frac{1}{2}\mleft(e^{tK}\mleft(h-2P_{v}\mright)e^{tK}+e^{-tK}he^{-tK}\mright)-h\\
B\mleft(t\mright) & =\frac{1}{2}\mleft(e^{tK}\mleft(h-2P_{v}\mright)e^{tK}-e^{-tK}he^{-tK}\mright).\nonumber 
\end{align}
As in Section \ref{sec:AnalysisofOne-BodyOperators} we decompose
these as
\begin{align}
A\mleft(t\mright) & =A_{h}\mleft(t\mright)-e^{tK}P_{v}e^{tK}\\
B\mleft(t\mright) & =-\mleft(1-t\mright)P_{v}+B_{h}\mleft(t\mright)-e^{tK}P_{v}e^{tK}+P_{v}\nonumber 
\end{align}
for
\begin{align}
A_{h}\mleft(t\mright) & =\cosh\mleft(tK\mright)\,h\cosh\mleft(tK\mright)+\sinh\mleft(tK\mright)\,h\sinh\mleft(tK\mright)-h\\
 & =\left\{ h,C_{K}\mleft(t\mright)\right\} +S_{K}\mleft(t\mright)\,h\,S_{K}\mleft(t\mright)+C_{K}\mleft(t\mright)\,h\,C_{K}\mleft(t\mright)\nonumber 
\end{align}
and
\begin{align}
B_{h}\mleft(t\mright) & =\sinh\mleft(tK\mright)\,h\cosh\mleft(tK\mright)+\cosh\mleft(tK\mright)\,h\sinh\mleft(tK\mright)-tP_{v}\\
 & =\left\{ h,S_{K}\mleft(t\mright)\right\} -tP_{v}+S_{K}\mleft(t\mright)\,h\,C_{K}\mleft(t\mright)+C_{K}\mleft(t\mright)\,h\,S_{K}\mleft(t\mright),\nonumber 
\end{align}
where $S_{K}\mleft(t\mright)$ and $C_{K}\mleft(t\mright)$ are now given
by
\begin{equation}
C_{K}\mleft(t\mright)=\cosh\mleft(tK\mright)-1\quad\text{and}\quad S_{K}\mleft(t\mright)=\sinh\mleft(tK\mright).
\end{equation}
Since the only effective difference between the statement of Proposition
\ref{prop:AttractivetDependentElementEstimates} and that of Proposition
\ref{prop:tDependentElementEstimates} is a factor of $\mleft(1-2\left\langle v,h^{-1}v\right\rangle \mright)^{-1}$,
the bound of Proposition \ref{prop:etKPetK-PEstimate} generalizes
as (using also the trivial estimate $1\leq\mleft(1-2\left\langle v,h^{-1}v\right\rangle \mright)^{-1}$)
\begin{equation}
\left|\left\langle x_{i},\mleft(e^{tK}P_{v}e^{tK}-P_{v}\mright)x_{j}\right\rangle \right|\leq\frac{\mleft(2+\left\langle v,h^{-1}v\right\rangle \mright)\left\langle v,h^{-1}v\right\rangle }{\mleft(1-2\left\langle v,h^{-1}v\right\rangle \mright)^{2}}\left\langle x_{i},v\right\rangle \left\langle v,x_{j}\right\rangle .
\end{equation}
Consequently also
\begin{equation}
\left|\left\langle x_{i},e^{tK}P_{v}e^{tK}x_{j}\right\rangle \right|\leq\mleft(\frac{1+\left\langle v,h^{-1}v\right\rangle }{1-2\left\langle v,h^{-1}v\right\rangle }\mright)^{2}\left\langle x_{i},v\right\rangle \left\langle v,x_{j}\right\rangle 
\end{equation}
and by the same argument
\begin{equation}
\left|\left\langle x_{i},\left\{ h,C_{K}\mleft(t\mright)\right\} x_{j}\right\rangle \right|\leq\frac{\left\langle v,h^{-1}v\right\rangle }{1-2\left\langle v,h^{-1}v\right\rangle }\left\langle x_{i},v\right\rangle \left\langle v,x_{j}\right\rangle 
\end{equation}
and
\begin{equation}
\left|\left\langle x_{i},S_{K}\mleft(t\mright)\,h\,S_{K}\mleft(t\mright)x_{j}\right\rangle \right|\leq\frac{\left\langle v,h^{-1}v\right\rangle }{\mleft(1-2\left\langle v,h^{-1}v\right\rangle \mright)^{2}}\left\langle x_{i},v\right\rangle \left\langle v,x_{j}\right\rangle ,
\end{equation}
the latter extending also to the operators $C_{K}\mleft(t\mright)\,h\,C_{K}\mleft(t\mright)$,
$S_{K}\mleft(t\mright)\,h\,C_{K}\mleft(t\mright)$ and $C_{K}\mleft(t\mright)\,h\,S_{K}\mleft(t\mright)$.

Proposition \ref{prop:AttractivetDependentElementEstimates} finally
implies that
\begin{align}
\left|\left\langle x_{i},\mleft(\left\{ h,S_{K}\mleft(t\mright)\right\} -tP_{v}\mright)x_{j}\right\rangle \right| & =\left\langle x_{i},\left\{ h,S_{K}\mleft(t\mright)\right\} x_{j}\right\rangle -\left\langle x_{i},P_{v}x_{j}\right\rangle t\nonumber \\
 & \leq\mleft(\frac{1}{1-2\left\langle v,h^{-1}v\right\rangle }-1\mright)\left\langle x_{i},v\right\rangle \left\langle v,x_{j}\right\rangle t\\
 & =\frac{2\left\langle v,h^{-1}v\right\rangle }{1-2\left\langle v,h^{-1}v\right\rangle }\left\langle x_{i},v\right\rangle \left\langle v,x_{j}\right\rangle t,\nonumber 
\end{align}
so combining all the estimates we conclude the following analogue
of the Propositions \ref{prop:AhBhEstimate} and \ref{prop:AtBtEstimates}:
\begin{prop}
For all $1\leq i,j\leq n$ it holds that
\begin{align*}
\left|\left\langle x_{i},A_{h}\mleft(t\mright)x_{j}\right\rangle \right|,\left|\left\langle x_{i},B_{h}\mleft(t\mright)x_{j}\right\rangle \right| & \leq\frac{4\left\langle v,h^{-1}v\right\rangle }{\mleft(1-2\left\langle v,h^{-1}v\right\rangle \mright)^{2}}\left\langle x_{i},v\right\rangle \left\langle v,x_{j}\right\rangle \\
\left|\left\langle x_{i},A\mleft(t\mright)x_{j}\right\rangle \right|,\left|\left\langle x_{i},B\mleft(t\mright)x_{j}\right\rangle \right| & \leq3\mleft(\frac{1+\left\langle v,h^{-1}v\right\rangle }{1-2\left\langle v,h^{-1}v\right\rangle }\mright)^{2}\left\langle x_{i},v\right\rangle \left\langle v,x_{j}\right\rangle .
\end{align*}
\end{prop}

These estimates again only differ from those of Section \ref{sec:AnalysisofOne-BodyOperators}
by a factor of $\mleft(1-2\left\langle v,h^{-1}v\right\rangle \mright)^{-2}$,
so the statements of the Propositions \ref{prop:KBtEstimate} and
\ref{prop:OneBodyExchangeContribution} likewise generalize as
\begin{equation}
\left|\left\langle x_{i},\left\{ K,B\mleft(t\mright)\right\} x_{j}\right\rangle \right|\leq\frac{\mleft(6+\left\langle v,h^{-1}v\right\rangle \mright)\left\langle v,h^{-1}v\right\rangle }{\mleft(1-2\left\langle v,h^{-1}v\right\rangle \mright)^{3}}\left\langle x_{i},v\right\rangle \left\langle v,x_{j}\right\rangle 
\end{equation}
and
\begin{equation}
\left|\left\langle x_{i},\mleft(\int_{0}^{1}B\mleft(t\mright)\,dt\mright)x_{j}\right\rangle +\frac{1}{2}\left\langle x_{i},v\right\rangle \left\langle v,x_{j}\right\rangle \right|\leq\frac{\mleft(6+\left\langle v,h^{-1}v\right\rangle \mright)\left\langle v,h^{-1}v\right\rangle }{\mleft(1-2\left\langle v,h^{-1}v\right\rangle \mright)^{2}}\left\langle x_{i},v\right\rangle \left\langle v,x_{j}\right\rangle ,
\end{equation}
respectively.

\subsubsection*{Conclusion of Proposition \ref{prop:AttractiveOneBodyEstimates}}

We have now obtained estimates similar to those of Section \ref{sec:AnalysisofOne-BodyOperators},
with only two differences: First, the left-hand sides differ by a
sign whenever $v$ (or rather $P_{v}$) appears. This only serves
to negative the absolute value of $|\hat{V}_{k}|$ in our new definition
of $v_{k}$ and $P_{k}$, however, which is the reason that $|\hat{V}_{k}|$
only appears on the right-hand sides of Proposition \ref{prop:AttractiveOneBodyEstimates}.

The second difference (apart from the absolute value) is various factors
of $\mleft(1-2\left\langle v,h^{-1}v\right\rangle \mright)^{-1}$. By
Lemma \ref{lemma:AttractiveLowerBound} we can however estimate
\begin{equation}
1-2\left\langle v_{k},h_{k}^{-1}v_{k}\right\rangle \geq C
\end{equation}
uniformly in $k$ for a $C>0$ depending only on $\epsilon$, whence
also $\mleft(1-2\left\langle v_{k},h_{k}^{-1}v_{k}\right\rangle \mright)^{-1}\leq C'$
depending only on $\epsilon$. Absorbing this dependence into the
overall constant yields Proposition \ref{prop:AttractiveOneBodyEstimates}.

\section{\label{sec:OverviewoftheOperatorResult}Overview of the Operator
Result}

In this section we review the main points which lead to the conclusion
of Theorem \ref{them:OperatorStatement}.

We first present a general outline of the approach, and then consider
the main points in greater detail in the rest of the section. As in
\cite{ChrHaiNam-21} we will focus on the case $s=1$ for simplicity,
and assume as in the theorem that $\sum_{k\in\mathbb{Z}_{\ast}^{3}}\hat{V}_{k}\left|k\right|<\infty$.

First we should note that the statement in \cite{ChrHaiNam-21} is
slightly more general than that of Theorem \ref{them:OperatorStatement},
in that with respect to the decomposition
\begin{equation}
\mathcal{U}H_{N}\mathcal{U}^{\ast}=E_{F}+E_{\mathrm{corr},\mathrm{bos}}+H_{\mathrm{eff}}+\mathcal{E}
\end{equation}
the error operator $\mathcal{E}$ is shown to generally obey
\begin{equation}
\pm\mathcal{E}\leq Ck_{F}^{-\frac{1}{94}+\varepsilon}\mleft(k_{F}+H_{\mathrm{kin}}^{\prime}+k_{F}^{-1}\mathcal{N}_{E}H_{\mathrm{kin}}^{\prime}\mright)
\end{equation}
with respect to $D\mleft(H_{\mathrm{kin}}^{\prime}\mright)$, and not
just the low-lying eigenstates. The particular statement of Theorem
\ref{them:OperatorStatement} then follows by \textit{a priori} bounds
on such states: Define a normalized state $\Psi\in D\mleft(H_{\mathrm{kin}}^{\prime}\mright)$
to be low-lying (with respect to $H_{N}$) if
\begin{equation}
\left\langle \Psi,H_{N}\Psi\right\rangle \leq E_{F}+\kappa k_{F}
\end{equation}
for some fixed $\kappa>0$. Then the following holds:
\begin{prop}
\label{prop:APrioriEigenstateBounds}For any low-lying eigenstate
$\Psi\in D\mleft(H_{\mathrm{kin}}^{\prime}\mright)$ it holds that
\[
\left\langle \Psi,\mathcal{N}_{E}\Psi\right\rangle \leq\left\langle \Psi,H_{\mathrm{kin}}^{\prime}\Psi\right\rangle \leq Ck_{F},\quad\left\langle \Psi,\mathcal{N}_{E}H_{\mathrm{kin}}^{\prime}\Psi\right\rangle \leq Ck_{F}^{2},
\]
for a constant $C>0$ depending only on $\sum_{k\in\mathbb{Z}_{\ast}^{3}}\hat{V}_{k}\left|k\right|$
and $\kappa$.
\end{prop}

Let us comment on the quality of these estimates: That $\left\langle \Psi,H_{\mathrm{kin}}^{\prime}\Psi\right\rangle \leq O\mleft(k_{F}\mright)$
is presumably optimal, since $H_{\mathrm{kin}}^{\prime}$ enters directly
in $H_{N}-E_{F}$ and we already know that $\inf\mleft(\sigma\mleft(H_{N}\mright)\mright)\sim E_{F}+O\mleft(k_{F}\mright)$.
The bound $\left\langle \Psi,\mathcal{N}_{E}\Psi\right\rangle \leq O\mleft(k_{F}\mright)$
is likely far from optimal, however, since the trial state we applied
for the upper bound had only $\left\langle \Psi,\mathcal{N}_{E}\Psi\right\rangle \leq O\mleft(1\mright)$.
(It can also be shown that for this state, $\left\langle \Psi,\mathcal{N}_{E}H_{\mathrm{kin}}^{\prime}\Psi\right\rangle \leq O\mleft(k_{F}\mright)$.)

This point is important for the estimation of error terms later on,
since it means that in order to bound these well, they must be bounded
in terms of $H_{\mathrm{kin}}^{\prime}$ to the greatest extent possible,
rather than just $\mathcal{N}_{E}$ and its powers (as we have done
for the upper bound).

\subsubsection*{Decomposition of the Hamiltonian}

With these \textit{a priori} bounds at our disposal we can turn to
the Hamiltonian proper. Here we must at the outset make a slight modification
compared to the decomposition of Theorem \ref{prop:LocalizedHamiltonian}:
We now write
\begin{equation}
H_{N}^{\prime}=H_{\mathrm{kin}}^{\prime}+\sum_{k\in\overline{B}\mleft(0,k_{F}^{\gamma}\mright)\cap\mathbb{Z}_{\ast}^{3}}\mleft(2\,Q_{1}^{k}\mleft(P_{k}\mright)+Q_{2}^{k}\mleft(P_{k}\mright)\mright)+\mathrm{ND}+\mathcal{C}+\mathcal{Q}
\end{equation}
for some $\gamma>0$ to be optimized at the end, where the non-diagonalized
terms $\mathrm{ND}$ are the tail of the interaction terms,
\begin{equation}
\mathrm{ND}=\frac{k_{F}^{-1}}{2\,\mleft(2\pi\mright)^{3}}\sum_{k\in\mathbb{Z}_{\ast}^{3}\backslash\overline{B}\mleft(0,k_{F}^{\gamma}\mright)}\hat{V}_{k}\mleft(2B_{k}^{\ast}B_{k}+B_{k}B_{-k}+B_{-k}^{\ast}B_{k}^{\ast}\mright).
\end{equation}
We do this as we will later on need to estimate Riemann sums which
are more singular than $\sum_{p\in L_{k}}\lambda_{k,p}^{-1}$, and
these we can only establish for $\left|k\right|$ sufficiently small
compared to $k_{F}$. This necessitates a cut-off in the transformation,
hence in the number of terms we can diagonalize for a given $k_{F}$.
As $\overline{B}\mleft(0,k_{F}^{\gamma}\mright)\cap\mathbb{Z}_{\ast}^{3}$
exhausts $\mathbb{Z}_{\ast}^{3}$ when $k_{F}\rightarrow\infty$,
all terms are ``eventually'' diagonalized, but the tail terms of
$\mathrm{ND}$ must be treated as errors rather than included in the
transformation.

The non-bosonizable terms $\mathcal{C}$ and $\mathcal{Q}$ are likewise
bounded prior to the transformation. This is a difficult task since,
as mentioned above, these are to be bounded in terms of the kinetic
operator. Nonetheless we obtain the following:
\begin{prop}
\label{prop:APrioriErrorBounds}It holds that
\begin{align*}
\pm\mathrm{ND} & \leq Ck_{F}^{-\frac{\gamma}{2}}\mleft(k_{F}+H_{\mathrm{kin}}^{\prime}\mright)\\
\pm\mleft(\mathcal{C}+\mathcal{Q}\mright) & \leq C\log\mleft(k_{F}\mright)^{\frac{1}{9}}k_{F}^{-\frac{1}{18}}\mleft(H_{\mathrm{kin}}^{\prime}+k_{F}^{-1}\mathcal{N}_{E}H_{\mathrm{kin}}^{\prime}\mright)
\end{align*}
as $k_{F}\rightarrow\infty$ for a constant $C>0$ depending only
on $\sum_{k\in\mathbb{Z}_{\ast}^{3}}\hat{V}_{k}\left|k\right|$.
\end{prop}

We remark that in the end it will be $\mathrm{ND}$ which is the dominant
error term of $\mathcal{E}$ - the Riemann sum estimates impose the
condition $\gamma<\frac{1}{47}$, whence $\pm\mathrm{ND}\leq Ck_{F}^{-\frac{1}{94}+\varepsilon}\mleft(k_{F}+H_{\mathrm{kin}}^{\prime}\mright)$.
This is not surprising since the non-diagonalizable terms \textit{do}
contribute to the correlation energy, we simply lack singular Riemann
sum estimates which are sufficiently uniform in $k$ to meaningfully
extract this.

\subsubsection*{Analysis of Bosonizable Terms}

With these bounds the remaining analysis reduces entirely to the (now
cut-off) bosonizable terms. For these, Theorem \ref{thm:DiagonalizationoftheBosonizableTerms}
continues to hold in the form
\begin{align}
 & \;\,e^{\mathcal{K}}\mleft(H_{\mathrm{kin}}^{\prime}+\sum_{k\in\overline{B}\mleft(0,k_{F}^{\gamma}\mright)\cap\mathbb{Z}_{\ast}^{3}}\frac{\hat{V}_{k}k_{F}^{-1}}{2\,\mleft(2\pi\mright)^{3}}\mleft(2B_{k}^{\ast}B_{k}+B_{k}B_{-k}+B_{-k}^{\ast}B_{k}^{\ast}\mright)\mright)e^{-\mathcal{K}}\nonumber \\
 & =\sum_{k\in\overline{B}\mleft(0,k_{F}^{\gamma}\mright)\cap\mathbb{Z}_{\ast}^{3}}\mathrm{tr}\mleft(E_{k}-h_{k}-P_{k}\mright)+H_{\mathrm{kin}}^{\prime}+2\sum_{k\in\overline{B}\mleft(0,k_{F}^{\gamma}\mright)\cap\mathbb{Z}_{\ast}^{3}}Q_{1}^{k}\mleft(E_{k}-h_{k}\mright)\\
 & +\sum_{k\in\overline{B}\mleft(0,k_{F}^{\gamma}\mright)\cap\mathbb{Z}_{\ast}^{3}}\int_{0}^{1}e^{\mleft(1-t\mright)\mathcal{K}}\mleft(\varepsilon_{k}\mleft(\left\{ K_{k},B_{k}\mleft(t\mright)\right\} \mright)+2\,\mathrm{Re}\mleft(\mathcal{E}_{k}^{1}\mleft(A_{k}\mleft(t\mright)\mright)\mright)+2\,\mathrm{Re}\mleft(\mathcal{E}_{k}^{2}\mleft(B_{k}\mleft(t\mright)\mright)\mright)\mright)e^{-\mleft(1-t\mright)\mathcal{K}}dt\nonumber 
\end{align}
where $E_{k}=e^{-K_{k}}h_{k}e^{-K_{k}}$. The cut-off means that we
only recover part of $E_{\mathrm{corr},\mathrm{bos}}$, but the remainder
is of lower order as $k_{F}\rightarrow\infty$. Additionally, the
following kinetic estimates of the exchange terms, and Gronwall estimates
for the kinetic operators, can be derived:
\begin{prop}
\label{prop:OperatorDiagonalizationControl}It holds that
\begin{align*}
\sum_{k\in\overline{B}\mleft(0,k_{F}^{\gamma}\mright)\cap\mathbb{Z}_{\ast}^{3}}\mathrm{tr}\mleft(E_{k}-h_{k}-P_{k}\mright) & =E_{\mathrm{corr},\mathrm{bos}}+O\mleft(k_{F}^{1-\gamma}\mright)\\
\pm\,\mathrm{Exchange\,Terms} & \leq C\log\mleft(k_{F}\mright)^{\frac{2}{3}}k_{F}^{\frac{8}{3}\gamma-\frac{1}{3}}\mleft(k_{F}+H_{\mathrm{kin}}^{\prime}+k_{F}^{-1}\mathcal{N}_{E}H_{\mathrm{kin}}^{\prime}\mright)
\end{align*}
and for any $t\in\left[-1,1\right]$
\begin{align*}
e^{t\mathcal{K}}H_{\mathrm{kin}}^{\prime}e^{-t\mathcal{K}} & \leq C\mleft(H_{\mathrm{kin}}^{\prime}+k_{F}\mright)\\
e^{t\mathcal{K}}\mathcal{N}_{E}H_{\mathrm{kin}}^{\prime}e^{-t\mathcal{K}} & \leq C\mleft(\mathcal{N}_{E}H_{\mathrm{kin}}^{\prime}+k_{F}H_{\mathrm{kin}}^{\prime}+k_{F}\mright)
\end{align*}
for a constant $C>0$ depending only on $\sum_{k\in\mathbb{Z}_{\ast}^{3}}\hat{V}_{k}\left|k\right|$.
\end{prop}

This leaves only $H_{\mathrm{kin}}^{\prime}+2\sum_{k\in\overline{B}\mleft(0,k_{F}^{\gamma}\mright)\cap\mathbb{Z}_{\ast}^{3}}Q_{1}^{k}\mleft(E_{k}-h_{k}\mright)$.
Now, if we were only considering a lower bound, it would be tempting
to think that we are done, since $E_{k}=e^{-K_{k}}h_{k}e^{-K_{k}}=e^{-K_{k}}h_{k}^{\frac{1}{2}}h_{k}^{\frac{1}{2}}e^{-K_{k}}$
is isospectral to
\begin{equation}
\tilde{E}_{k}=h_{k}^{\frac{1}{2}}e^{-K_{k}}e^{-K_{k}}h_{k}^{\frac{1}{2}}=h_{k}^{\frac{1}{2}}e^{-2K_{k}}h_{k}^{\frac{1}{2}}=\mleft(h_{k}^{2}+2P_{h_{k}^{\frac{1}{2}}v_{k}}\mright)^{\frac{1}{2}}
\end{equation}
and $\tilde{E}_{k}\geq h_{k}$, so one might suspect that $E_{k}\geq h_{k}$
which would imply that $Q_{1}^{k}\mleft(E_{k}-h_{k}\mright)\geq0$.
This is not so, however - $E_{k}-h_{k}$ is not non-negative.

To get around this issue we consider a second transformation $e^{\mathcal{J}}:\mathcal{H}_{N}\rightarrow\mathcal{H}_{N}$
for $\mathcal{J}$ of the form
\begin{equation}
\mathcal{J}=\sum_{k\in\overline{B}\mleft(0,k_{F}^{\gamma}\mright)\cap\mathbb{Z}_{\ast}^{3}}\sum_{p,q\in L_{k}}\left\langle e_{p},J_{k}e_{q}\right\rangle b_{k,p}^{\ast}b_{k,q}=\sum_{k\in\overline{B}\mleft(0,k_{F}^{\gamma}\mright)\cap\mathbb{Z}_{\ast}^{3}}\sum_{p\in L_{k}}b_{k}^{\ast}\mleft(J_{k}e_{p}\mright)b_{k,p}\label{eq:calJDefinition}
\end{equation}
where we take $J_{k}:\ell^{2}\mleft(L_{k}\mright)\rightarrow\ell^{2}\mleft(L_{k}\mright)$,
$k\in\mathbb{Z}_{\ast}^{3}$, to be a collection of \textit{skew}-symmetric
operators. It follows that $\mathcal{J}$ is also skew-symmetric,
as
\begin{equation}
\mathcal{J}^{\ast}=\sum_{k\in\overline{B}\mleft(0,k_{F}^{\gamma}\mright)\cap\mathbb{Z}_{\ast}^{3}}\sum_{p\in L_{k}}b_{k}^{\ast}\mleft(e_{p}\mright)b_{k}\mleft(J_{k}e_{p}\mright)=\sum_{k\in\overline{B}\mleft(0,k_{F}^{\gamma}\mright)\cap\mathbb{Z}_{\ast}^{3}}\sum_{p\in L_{k}}b_{k}^{\ast}\mleft(J_{k}^{\ast}e_{p}\mright)b_{k}\mleft(e_{p}\mright)=-\mathcal{J},
\end{equation}
so $e^{\mathcal{J}}$ is a unitary transformation.

In the exact bosonic case, a transformation of such a form obeys $e^{\mathcal{J}}\mathrm{d}\Gamma\mleft(A\mright)e^{-\mathcal{J}}=\mathrm{d}\Gamma\mleft(e^{J}Ae^{-J}\mright)$.
We thus take the operators $J_{k}$ to be the principal logarithms
of the operators $U_{k}$, given by
\begin{equation}
U_{k}=\mleft(h_{k}^{\frac{1}{2}}e^{-2K_{k}}h_{k}^{\frac{1}{2}}\mright)^{\frac{1}{2}}h_{k}^{-\frac{1}{2}}e^{K_{k}},
\end{equation}
which precisely act by taking $E_{k}$ to $\tilde{E}_{k}$:
\begin{align}
U_{k}E_{k}U_{k}^{\ast} & =\mleft(h_{k}^{\frac{1}{2}}e^{-2K_{k}}h_{k}^{\frac{1}{2}}\mright)^{\frac{1}{2}}h_{k}^{-\frac{1}{2}}e^{K_{k}}e^{-K_{k}}h_{k}e^{-K_{k}}e^{K_{k}}h_{k}^{-\frac{1}{2}}\mleft(h_{k}^{\frac{1}{2}}e^{-2K_{k}}h_{k}^{\frac{1}{2}}\mright)^{\frac{1}{2}}\\
 & =\mleft(h_{k}^{\frac{1}{2}}e^{-2K_{k}}h_{k}^{\frac{1}{2}}\mright)^{\frac{1}{2}}\mleft(h_{k}^{\frac{1}{2}}e^{-2K_{k}}h_{k}^{\frac{1}{2}}\mright)^{\frac{1}{2}}=h_{k}^{\frac{1}{2}}e^{-2K_{k}}h_{k}^{\frac{1}{2}}=\tilde{E}_{k}.\nonumber 
\end{align}
It can then be shown to hold that
\begin{align}
 & \;\,e^{\mathcal{J}}\mleft(H_{\mathrm{kin}}^{\prime}+2\sum_{k\in\overline{B}\mleft(0,k_{F}^{\gamma}\mright)\cap\mathbb{Z}_{\ast}^{3}}Q_{1}^{k}\mleft(E_{k}-h_{k}\mright)\mright)e^{-\mathcal{J}}\label{eq:SecondTransformationAction}\\
 & =H_{\mathrm{kin}}^{\prime}+2\sum_{k\in\overline{B}\mleft(0,k_{F}^{\gamma}\mright)\cap\mathbb{Z}_{\ast}^{3}}Q_{1}^{k}\mleft(\tilde{E}_{k}-h_{k}\mright)+2\sum_{k\in\overline{B}\mleft(0,k_{F}^{\gamma}\mright)\cap\mathbb{Z}_{\ast}^{3}}\int_{0}^{1}e^{\mleft(1-t\mright)\mathcal{J}}\mathcal{E}_{k}^{3}\mleft(E_{k}\mleft(t\mright)\mright)e^{-\mleft(1-t\mright)\mathcal{J}}dt\nonumber 
\end{align}
where $\mathcal{E}_{k}^{3}\mleft(\cdot\mright)$ is of a similar form
to $\mathcal{E}_{k}^{1}\mleft(\cdot\mright)$ and $\mathcal{E}_{k}^{2}\mleft(\cdot\mright)$
of the first transformation, while $E_{k}\mleft(t\mright):\ell^{2}\mleft(L_{k}\mright)\rightarrow\ell^{2}\mleft(L_{k}\mright)$
is given by
\begin{equation}
E_{k}\mleft(t\mright)=e^{tJ_{k}}e^{-K_{k}}h_{k}e^{-K_{k}}e^{-tJ_{k}}-h_{k}.
\end{equation}
The following estimate for the error term, and Gronwall estimates
for the kinetic operators with respect to the second transformation,
can then be obtained:
\begin{prop}
\label{prop:SecondTransformationExchangeTermEstimate}It holds for
all $0<\gamma<\frac{1}{47}$ that
\[
\pm\sum_{k\in\overline{B}\mleft(0,k_{F}^{\gamma}\mright)\cap\mathbb{Z}_{\ast}^{3}}\mathcal{E}_{k}^{3}\mleft(E_{k}\mleft(t\mright)\mright)\leq C\log\mleft(k_{F}\mright)^{\frac{5}{3}}k_{F}^{\mleft(5+\frac{2}{3}\mright)\gamma-\frac{1}{3}}\mleft(H_{\mathrm{kin}}^{\prime}+k_{F}^{-1}\mathcal{N}_{E}H_{\mathrm{kin}}^{\prime}\mright)
\]
and for $t\in\left[-1,1\right]$
\begin{align*}
e^{t\mathcal{J}}H_{\mathrm{kin}}^{\prime}e^{-t\mathcal{J}} & \leq CH_{\mathrm{kin}}^{\prime}\\
e^{t\mathcal{J}}\mathcal{N}_{E}H_{\mathrm{kin}}^{\prime}e^{-t\mathcal{J}} & \leq C\mathcal{N}_{E}H_{\mathrm{kin}}^{\prime}
\end{align*}
for a constant $C>0$ depending only on $\sum_{k\in\mathbb{Z}_{\ast}^{3}}\hat{V}_{k}\left|k\right|$.
\end{prop}

As mentioned, the condition $\gamma<\frac{1}{47}$ enters in the estimation
of one-body estimates for $E_{k}\mleft(t\mright)$ and the Gronwall
argument - the Gronwall argument is particularly sensitive to this,
as the exponential prefactor diverges as $k_{F}\rightarrow\infty$
if these are not estimated optimally.

Theorem \ref{them:OperatorStatement} now follows by taking $\mathcal{U}=e^{\mathcal{J}}e^{\mathcal{K}}$,
for which
\begin{align}
\mathcal{U}H_{N}\mathcal{U}^{\ast} & =E_{F}+E_{\mathrm{corr},\mathrm{bos}}+H_{\mathrm{kin}}^{\prime}+2\sum_{k\in\overline{B}\mleft(0,k_{F}^{\gamma}\mright)\cap\mathbb{Z}_{\ast}^{3}}Q_{1}^{k}\mleft(\tilde{E}_{k}-h_{k}\mright)\nonumber \\
 & +\sum_{k\in\overline{B}\mleft(0,k_{F}^{\gamma}\mright)\cap\mathbb{Z}_{\ast}^{3}}\mathrm{tr}\mleft(E_{k}-h_{k}-P_{k}\mright)-E_{\mathrm{corr},\mathrm{bos}}+\mathcal{U}\mleft(\mathrm{ND}+\mathcal{C}+\mathcal{Q}\mright)\mathcal{U}^{\ast}\\
 & +e^{\mathcal{J}}\mleft(\mathrm{Exchange\,Terms}\mright)e^{-\mathcal{J}}+2\sum_{k\in\overline{B}\mleft(0,k_{F}^{\gamma}\mright)\cap\mathbb{Z}_{\ast}^{3}}\int_{0}^{1}e^{\mleft(1-t\mright)\mathcal{J}}\mathcal{E}_{k}^{3}\mleft(E_{k}\mleft(t\mright)\mright)e^{-\mleft(1-t\mright)\mathcal{J}}dt;\nonumber 
\end{align}
by the estimates obtained, the terms on the second and third lines
are bounded by
\begin{equation}
C\mleft(k_{F}^{-\frac{\gamma}{2}}+\log\mleft(k_{F}\mright)^{\frac{5}{3}}k_{F}^{\mleft(5+\frac{2}{3}\mright)\gamma-\frac{1}{3}}+\log\mleft(k_{F}\mright)^{\frac{1}{9}}k_{F}^{-\frac{1}{18}}\mright)\mleft(k_{F}+H_{\mathrm{kin}}^{\prime}+k_{F}^{-1}\mathcal{N}_{E}H_{\mathrm{kin}}^{\prime}\mright)
\end{equation}
which is optimized as $\gamma\rightarrow\frac{1}{47}$ for the prefactor
$k_{F}^{-\frac{1}{94}+\varepsilon}$, $\varepsilon>0$\@. It then
only remains to estimate the tail of $\sum_{k\in\mathbb{Z}_{\ast}^{3}}Q_{1}^{k}\mleft(\tilde{E}_{k}-h_{k}\mright)$,
but it is not too difficult to show that these obey
\begin{equation}
\pm\sum_{k\in\mathbb{Z}_{\ast}^{3}\backslash\overline{B}\mleft(0,k_{F}^{\gamma}\mright)}Q_{1}^{k}\mleft(\tilde{E}_{k}-h_{k}\mright)\leq Ck_{F}^{-\gamma}H_{\mathrm{kin}}^{\prime}
\end{equation}
and so are likewise negligible.

\subsection{\textit{A Priori} Bounds}

In this subsection we prove Proposition \ref{prop:APrioriEigenstateBounds}.
For the sake of brevity we will write $H_{N}^{\prime}=H_{N}-E_{F}$,
so the definition of a low-lying state is simply that
\begin{equation}
\left\langle \Psi,H_{N}^{\prime}\Psi\right\rangle \leq\kappa k_{F}.
\end{equation}
First we obtain an \textit{a priori} bound for $H_{N}^{\prime}$ itself.
Recall that we in Section \ref{sec:LocalizationoftheHamiltonian}
found that
\begin{equation}
H_{\mathrm{kin}}=\left\langle \psi_{F},H_{\mathrm{kin}}\psi_{F}\right\rangle +H_{\mathrm{kin}}^{\prime}
\end{equation}
and note that it follows from the equations (\ref{eq:HintFactorizedForm})
and (\ref{eq:FSInteractionEnergy}) that
\begin{equation}
H_{\mathrm{int}}=\left\langle \psi_{F},H_{\mathrm{int}}\psi_{F}\right\rangle +\frac{1}{2\,\mleft(2\pi\mright)^{3}}\sum_{k\in\mathbb{Z}_{\ast}^{3}}\hat{V}_{k}\mleft(\mathrm{d}\Gamma\mleft(e^{-ik\cdot x}\mright)^{\ast}\mathrm{d}\Gamma\mleft(e^{-ik\cdot x}\mright)-\left|L_{k}\right|\mright)
\end{equation}
so
\begin{equation}
H_{N}^{\prime}=H_{\mathrm{kin}}^{\prime}+\frac{k_{F}^{-1}}{2\,\mleft(2\pi\mright)^{3}}\sum_{k\in\mathbb{Z}_{\ast}^{3}}\hat{V}_{k}\mleft(\mathrm{d}\Gamma\mleft(e^{-ik\cdot x}\mright)^{\ast}\mathrm{d}\Gamma\mleft(e^{-ik\cdot x}\mright)-\left|L_{k}\right|\mright).
\end{equation}
As trivially $\mathrm{d}\Gamma\mleft(e^{-ik\cdot x}\mright)^{\ast}\mathrm{d}\Gamma\mleft(e^{-ik\cdot x}\mright)\geq0$
we can thus apply the bound $\left|L_{k}\right|\leq Ck_{F}^{2}\left|k\right|$
to conclude that
\begin{equation}
H_{N}^{\prime}\ge H_{\mathrm{kin}}^{\prime}-\frac{k_{F}^{-1}}{2\,\mleft(2\pi\mright)^{3}}\sum_{k\in\mathbb{Z}_{\ast}^{3}}\hat{V}_{k}\left|L_{k}\right|\geq H_{\mathrm{kin}}^{\prime}-C'k_{F}\sum_{k\in\mathbb{Z}_{\ast}^{3}}\hat{V}_{k}\left|k\right|=H_{\mathrm{kin}}^{\prime}-Ck_{F}\label{eq:HNPrimeLowerBound}
\end{equation}
for a constant $C>0$ depending only on $\sum_{k\in\mathbb{Z}_{\ast}^{3}}\hat{V}_{k}\left|k\right|$.

This immediately implies that the correlation energy is (at most)
of order $k_{F}$ in this case, but more crucial is the implied bound
on $H_{\mathrm{kin}}^{\prime}$: Equation (\ref{eq:HNPrimeLowerBound})
implies that any low-lying state must obey
\begin{equation}
\left\langle \Psi,H_{\mathrm{kin}}^{\prime}\Psi\right\rangle \leq\left\langle \Psi,H_{N}^{\prime}\Psi\right\rangle +Ck_{F}\leq\mleft(C+\kappa\mright)k_{F},
\end{equation}
which is our first \textit{a priori} bound.

This in turn yields an \textit{a priori} bound for $\left\langle \Psi,\mathcal{N}_{E}\Psi\right\rangle $
as well, for recall that we found that the particle-hole symmetry
allowed us to express
\begin{equation}
H_{\mathrm{kin}}^{\prime}=\sum_{p\in B_{F}^{c}}\left|p\right|^{2}c_{p}^{\ast}c_{p}-\sum_{p\in B_{F}}\left|p\right|^{2}c_{p}c_{p}^{\ast}
\end{equation}
in the manifestly positive form
\begin{equation}
H_{\mathrm{kin}}^{\prime}=\sum_{p\in B_{F}^{c}}\mleft(\left|p\right|^{2}-k_{F}^{2}\mright)c_{p}^{\ast}c_{p}+\sum_{p\in B_{F}}\mleft(k_{F}^{2}-\left|p\right|^{2}\mright)c_{p}c_{p}^{\ast}.\label{eq:ManifestlyPositiveHkin}
\end{equation}
This particular form is not useful, as the prefactors in the sums
can be arbitrarily small. The only condition we used to obtain this
was however that $\left|p\right|\geq k_{F}\geq\left|q\right|$, so
the same argument shows that for any $\zeta\in\left[\sup_{q\in B_{F}}\left|q\right|^{2},\inf_{p\in B_{F}^{c}}\left|p\right|^{2}\right]$
it holds that
\begin{equation}
H_{\mathrm{kin}}^{\prime}=\sum_{p\in B_{F}^{c}}\mleft(\left|p\right|^{2}-\zeta\mright)c_{p}^{\ast}c_{p}+\sum_{p\in B_{F}}\mleft(\zeta-\left|p\right|^{2}\mright)c_{p}c_{p}^{\ast},\label{eq:GeneralPositiveHkinPrimeDecomposition}
\end{equation}
and choosing $\zeta=\frac{1}{2}\mleft(\inf_{p\in B_{F}^{c}}\left|p\right|^{2}+\sup_{q\in B_{F}}\left|q\right|^{2}\mright)$
we have
\begin{equation}
\inf_{p\in\mathbb{Z}^{3}}|\left|p\right|^{2}-\zeta|\geq\frac{1}{2}\label{eq:ZetaDefiningProperty}
\end{equation}
since $\inf_{p\in B_{F}^{c}}\left|p\right|^{2}-\sup_{q\in B_{F}}\left|q\right|^{2}\geq1$
as $\left|p\right|^{2},\left|q\right|^{2}\in\mathbb{Z}$ but $\left|q\right|^{2}<\left|p\right|^{2}$
for any $p\in B_{F}^{c}$ and $q\in B_{F}$.

We thus conclude the general operator inequality (first noted in \cite{BenNamPorSchSei-21})
\begin{equation}
H_{\mathrm{kin}}^{\prime}\geq\frac{1}{2}\sum_{p\in B_{F}^{c}}c_{p}^{\ast}c_{p}+\frac{1}{2}\sum_{p\in B_{F}}c_{p}c_{p}^{\ast}=\mathcal{N}_{E}
\end{equation}
and conclude the following:
\begin{prop}
For any low-lying state $\Psi\in D\mleft(H_{\mathrm{kin}}^{\prime}\mright)$
it holds that
\[
\left\langle \Psi,\mathcal{N}_{E}\Psi\right\rangle \leq\left\langle \Psi,H_{\mathrm{kin}}^{\prime}\Psi\right\rangle \leq\mleft(C+\kappa\mright)k_{F}
\]
for a $C>0$ depending only on $\sum_{k\in\mathbb{Z}_{\ast}^{3}}\hat{V}_{k}\left|k\right|$.
\end{prop}

\subsubsection*{Bootstrapped Bounds for Eigenstates}

In the particular case that $\Psi$ is additionally an eigenstate
we can also obtain an \textit{a priori} bound on $\left\langle \Psi,\mathcal{N}_{E}H_{\mathrm{kin}}^{\prime}\Psi\right\rangle $
by employing a bootstrapping argument (similar to an idea of \cite{Seiringer-11}).
It turns out to be easier to bound $\left\langle \Psi,\mathcal{N}_{E}^{2}H_{\mathrm{kin}}^{\prime}\Psi\right\rangle $
and then obtain $\left\langle \Psi,\mathcal{N}_{E}H_{\mathrm{kin}}^{\prime}\Psi\right\rangle $
as a corollary, so let us consider this: First, by equation (\ref{eq:HNPrimeLowerBound}),
we have the operator inequality
\begin{align}
\mathcal{N}_{E}^{2}H_{\mathrm{kin}}^{\prime} & =\mathcal{N}_{E}H_{\mathrm{kin}}^{\prime}\mathcal{N}_{E}\leq\mathcal{N}_{E}H_{N}^{\prime}\mathcal{N}_{E}+Ck_{F}\mathcal{N}_{E}^{2}\\
 & =\frac{1}{2}\mleft(\mathcal{N}_{E}^{2}H_{N}^{\prime}+H_{N}^{\prime}\mathcal{N}_{E}^{2}-\left[\mathcal{N}_{E},\left[\mathcal{N}_{E},H_{N}^{\prime}\right]\right]\mright)+Ck_{F}\mathcal{N}_{E}^{2},\nonumber 
\end{align}
so if $\Psi\in D\mleft(H_{\mathrm{kin}}^{\prime}\mright)$ is an eigenstate
of $H_{N}$ such that $H_{N}^{\prime}\Psi=E\Psi$, it holds that
\begin{align}
\left\langle \Psi,\mathcal{N}_{E}^{2}H_{\mathrm{kin}}^{\prime}\Psi\right\rangle  & \leq\mleft(E+Ck_{F}\mright)\left\langle \Psi,\mathcal{N}_{E}^{2}\Psi\right\rangle -\frac{1}{2}\left\langle \Psi,\left[\mathcal{N}_{E},\left[\mathcal{N}_{E},H_{N}^{\prime}\right]\right]\Psi\right\rangle \label{eq:NE2HKinonEigenstate}\\
 & \leq\mleft(E+Ck_{F}\mright)\left\langle \Psi,\mathcal{N}_{E}H_{\mathrm{kin}}^{\prime}\Psi\right\rangle -\frac{1}{2}\left\langle \Psi,\left[\mathcal{N}_{E},\left[\mathcal{N}_{E},H_{N}^{\prime}\right]\right]\Psi\right\rangle \nonumber 
\end{align}
where we also used that $\mathcal{N}_{E}\leq H_{\mathrm{kin}}^{\prime}$
for the first term.

We must therefore consider $\left[\mathcal{N}_{E},\left[\mathcal{N}_{E},H_{N}^{\prime}\right]\right]$.
Note that by the decomposition of Proposition \ref{prop:LocalizedHamiltonian},
we can write
\begin{equation}
H_{N}^{\prime}=H_{\Delta}+\frac{k_{F}^{-1}}{2\,\mleft(2\pi\mright)^{3}}\sum_{k\in\mathbb{Z}_{\ast}^{3}}\hat{V}_{k}\mleft(B_{k}B_{-k}+B_{-k}^{\ast}B_{k}^{\ast}\mright)+\mathcal{C}
\end{equation}
for
\begin{equation}
H_{\Delta}=H_{\mathrm{kin}}^{\prime}+\frac{k_{F}^{-1}}{\mleft(2\pi\mright)^{3}}\sum_{k\in\mathbb{Z}_{\ast}^{3}}\hat{V}_{k}B_{k}^{\ast}B_{k}+\mathcal{Q},
\end{equation}
where we recall that the cubic terms $\mathcal{C}$ are given by
\begin{equation}
\mathcal{C}=\frac{k_{F}^{-1}}{2\,\mleft(2\pi\mright)^{3}}\sum_{k\in\mathbb{Z}_{\ast}^{3}}\hat{V}_{k}\mleft(\mleft(B_{k}^{\ast}+B_{-k}\mright)D_{k}+D_{k}^{\ast}\mleft(B_{k}+B_{-k}^{\ast}\mright)\mright).
\end{equation}
As remarked at the start of Section \ref{sec:EstimationoftheNon-BosonizableTermsandGronwallEstimates}
there holds the commutators
\begin{equation}
\left[\mathcal{N}_{E},B_{k}\right]=-B_{k},\quad\left[\mathcal{N}_{E},B_{k}^{\ast}\right]=B_{k}^{\ast},\quad\left[\mathcal{N}_{E},D_{k}\right]=0=\left[\mathcal{N}_{E},D_{k}^{\ast}\right],
\end{equation}
which imply that $\left[\mathcal{N}_{E},H_{\Delta}\right]=0$ and
thus
\begin{align}
\left[\mathcal{N}_{E},\left[\mathcal{N}_{E},H_{N}^{\prime}\right]\right] & =\frac{k_{F}^{-1}}{2\,\mleft(2\pi\mright)^{3}}\sum_{k\in\mathbb{Z}_{\ast}^{3}}\hat{V}_{k}\mleft(4B_{k}B_{-k}+4B_{-k}^{\ast}B_{k}^{\ast}+\mleft(B_{k}^{\ast}+B_{-k}\mright)D_{k}+D_{k}^{\ast}\mleft(B_{k}+B_{-k}^{\ast}\mright)\mright)\\
 & =\frac{k_{F}^{-1}}{\mleft(2\pi\mright)^{3}}\sum_{k\in\mathbb{Z}_{\ast}^{3}}\hat{V}_{k}\,\mathrm{Re}\mleft(4B_{k}B_{-k}+\mleft(B_{k}^{\ast}+B_{-k}\mright)D_{k}\mright).\nonumber 
\end{align}
We note the following estimates for the $B_{k}$ and $D_{k}$ operators
(the \textit{kinetic bound} on $B_{k}$ was first obtained in \cite{HaiPorRex-20}):
\begin{prop}
\label{prop:KineticBkEstimates}For any $k\in\mathbb{Z}_{\ast}^{3}$
and $\Psi\in D\mleft(H_{\mathrm{kin}}^{\prime}\mright)$ it holds that
\begin{align*}
\left\Vert B_{k}\Psi\right\Vert ^{2} & \leq Ck_{F}\left\langle \Psi,H_{\mathrm{kin}}^{\prime}\Psi\right\rangle \\
\left\Vert B_{k}^{\ast}\Psi\right\Vert ^{2} & \leq C\mleft(k_{F}\left\langle \Psi,H_{\mathrm{kin}}^{\prime}\Psi\right\rangle +k_{F}^{2}\left|k\right|\left\Vert \Psi\right\Vert ^{2}\mright)\\
\left\Vert D_{k}\Psi\right\Vert ^{2} & \leq8\left\langle \Psi,\mathcal{N}_{E}^{2}\Psi\right\rangle 
\end{align*}
for a constant $C>0$ independent of all quantities.
\end{prop}

\textbf{Proof:} For $B_{k}$ we can by Cauchy-Schwarz estimate
\begin{equation}
\left\Vert B_{k}\Psi\right\Vert =\left\Vert \sum_{p\in L_{k}}b_{k,p}\Psi\right\Vert \leq\sum_{p\in L_{k}}\left\Vert b_{k,p}\Psi\right\Vert \leq\sqrt{\sum_{p\in L_{k}}\lambda_{k,p}^{-1}}\sqrt{\sum_{p\in L_{k}}\lambda_{k,p}\left\Vert b_{k,p}\Psi\right\Vert ^{2}}\leq Ck_{F}^{\frac{1}{2}}\sqrt{\sum_{p\in L_{k}}\lambda_{k,p}\left\Vert b_{k,p}\Psi\right\Vert ^{2}}
\end{equation}
where we also used that $\sum_{p\in L_{k}}\lambda_{k,p}^{-1}\leq Ck_{F}$.
For the remaining factor we expand and bound as
\begin{align}
\sum_{p\in L_{k}}\lambda_{k,p}\left\Vert b_{k,p}\Psi\right\Vert ^{2} & =\frac{1}{2}\sum_{p\in L_{k}}\mleft(\left|p\right|^{2}-\left|p-k\right|^{2}\mright)\left\Vert c_{p-k}^{\ast}c_{p}\Psi\right\Vert ^{2}\nonumber \\
 & =\frac{1}{2}\sum_{p\in L_{k}}\mleft(\left|p\right|^{2}-k_{F}^{2}\mright)\left\Vert c_{p-k}^{\ast}c_{p}\Psi\right\Vert ^{2}+\frac{1}{2}\sum_{p\in L_{k}}\mleft(k_{F}^{2}-\left|p-k\right|^{2}\mright)\left\Vert c_{p-k}^{\ast}c_{p}\Psi\right\Vert ^{2}\\
 & \leq\frac{1}{2}\sum_{p\in L_{k}}\mleft(\left|p\right|^{2}-k_{F}^{2}\mright)\left\Vert c_{p}\Psi\right\Vert ^{2}+\frac{1}{2}\sum_{p\in L_{k}}\mleft(k_{F}^{2}-\left|p-k\right|^{2}\mright)\left\Vert c_{p-k}^{\ast}\Psi\right\Vert ^{2}\nonumber \\
 & =\frac{1}{2}\left\langle \Psi,H_{\mathrm{kin}}^{\prime}\Psi\right\rangle \nonumber 
\end{align}
where we applied the representation of $H_{\mathrm{kin}}^{\prime}$
given by equation (\ref{eq:ManifestlyPositiveHkin}). This implies
the first bound. The second then follows as the commutator of equation
(\ref{eq:BkBkastCommutator}) shows that
\begin{align}
\left\Vert B_{k}^{\ast}\Psi\right\Vert ^{2} & =\left\langle \Psi,B_{k}B_{k}^{\ast}\Psi\right\rangle =\left\langle \Psi,B_{k}^{\ast}B_{k}\Psi\right\rangle +\left\langle \Psi,\left[B_{k},B_{k}^{\ast}\right]\Psi\right\rangle \\
 & \leq\left\Vert B_{k}\Psi\right\Vert ^{2}+\left|L_{k}\right|\left\Vert \Psi\right\Vert ^{2}\leq C\mleft(k_{F}\left\langle \Psi,H_{\mathrm{kin}}^{\prime}\Psi\right\rangle +k_{F}^{2}\left|k\right|\left\Vert \Psi\right\Vert ^{2}\mright).\nonumber 
\end{align}
For $D_{k}$, recall the decomposition $D_{k}=D_{1,k}+D_{2,k}$ we
used in Section \ref{sec:EstimationoftheNon-BosonizableTermsandGronwallEstimates}.
As $\left\Vert D_{k}\Psi\right\Vert ^{2}\leq2\left\Vert D_{1,k}\Psi\right\Vert ^{2}+2\left\Vert D_{2,k}\Psi\right\Vert ^{2}$
it suffices to bound $D_{1,k}$ and $D_{2,k}$. Equation (\ref{eq:D1kastD1k})
says that (with $s=1$)
\begin{equation}
D_{1,k}^{\ast}D_{1,k}=\sum_{p,q\in B_{F}\cap\mleft(B_{F}+k\mright)}c_{p-k}c_{q}c_{q-k}^{\ast}c_{p}^{\ast}+\sum_{q\in B_{F}}1_{B_{F}}\mleft(q+k\mright)c_{q}c_{q}^{\ast}
\end{equation}
and the first term we bounded in equation (\ref{eq:QLRMainEstimate})
as
\begin{equation}
\sum_{p,q\in B_{F}\cap\mleft(B_{F}+k\mright)}\left\langle \Psi,c_{p-k}c_{q}c_{q-k}^{\ast}c_{p}^{\ast}\Psi\right\rangle \leq\left\langle \Psi,\mathcal{N}_{E}^{2}\Psi\right\rangle 
\end{equation}
while the second term trivially obeys
\begin{equation}
\sum_{q\in B_{F}}1_{B_{F}}\mleft(q+k\mright)c_{q}c_{q}^{\ast}\leq\mathcal{N}_{E}\leq\mathcal{N}_{E}^{2},
\end{equation}
so $\left\Vert D_{1,k}\Psi\right\Vert ^{2}\leq2\left\langle \Psi,\mathcal{N}_{E}^{2}\Psi\right\rangle =2\left\Vert \mathcal{N}_{E}\Psi\right\Vert ^{2}$.
$\left\Vert D_{2,k}\Psi\right\Vert ^{2}$ can be bounded similarly
for the claim.

$\hfill\square$

A bound on $\left[\mathcal{N}_{E},\left[\mathcal{N}_{E},H_{N}^{\prime}\right]\right]$
immediately follows:
\begin{prop}
It holds that
\[
\pm\left[\mathcal{N}_{E},\left[\mathcal{N}_{E},H_{N}^{\prime}\right]\right]\leq C\mleft(k_{F}+H_{\mathrm{kin}}^{\prime}+k_{F}^{-1}\mathcal{N}_{E}^{2}\mright)
\]
for a constant $C>0$ depending only on $\sum_{k\in\mathbb{Z}_{\ast}^{3}}\hat{V}_{k}\left|k\right|$.
\end{prop}

The main eigenstate bound can then be obtained:
\begin{prop}
For any normalized eigenstate $\Psi$ of $H_{N}$ with $H_{N}^{\prime}\Psi=E\Psi$
it holds that
\[
\left\langle \Psi,\mathcal{N}_{E}^{2}H_{\mathrm{kin}}^{\prime}\Psi\right\rangle \leq C\max\left\{ E,k_{F}\right\} ^{3}
\]
for a constant $C>0$ depending only on $\sum_{k\in\mathbb{Z}_{\ast}^{3}}\hat{V}_{k}\left|k\right|$.
\end{prop}

\textbf{Proof:} Inserting the previous estimate into equation (\ref{eq:NE2HKinonEigenstate}),
we obtain
\begin{align}
\left\langle \Psi,\mathcal{N}_{E}^{2}H_{\mathrm{kin}}^{\prime}\Psi\right\rangle  & \leq\mleft(E+Ck_{F}\mright)\left\langle \Psi,\mathcal{N}_{E}H_{\mathrm{kin}}^{\prime}\Psi\right\rangle +C\left\langle \Psi,\mleft(k_{F}+H_{\mathrm{kin}}^{\prime}+k_{F}^{-1}\mathcal{N}_{E}^{2}\mright)\Psi\right\rangle \label{eq:NEsqHKinEstimate2}\\
 & \leq Ck_{F}+C\max\left\{ E,k_{F}\right\} \left\langle \Psi,\mathcal{N}_{E}H_{\mathrm{kin}}^{\prime}\Psi\right\rangle \nonumber 
\end{align}
where we also used that $H_{\mathrm{kin}}^{\prime},\,\mathcal{N}_{E}^{2}\leq\mathcal{N}_{E}H_{\mathrm{kin}}^{\prime}$
to simplify the expression. Now, by the Cauchy-Schwarz inequality
for $H_{\mathrm{kin}}$ we can estimate
\begin{equation}
\left\langle \Psi,\mathcal{N}_{E}H_{\mathrm{kin}}^{\prime}\Psi\right\rangle \leq\sqrt{\left\langle \Psi,H_{\mathrm{kin}}^{\prime}\Psi\right\rangle \left\langle \Psi,\mathcal{N}_{E}H_{\mathrm{kin}}^{\prime}\mathcal{N}_{E}\Psi\right\rangle }\leq\sqrt{C\max\left\{ E,k_{F}\right\} }\sqrt{\left\langle \Psi,\mathcal{N}_{E}^{2}H_{\mathrm{kin}}^{\prime}\Psi\right\rangle }\label{eq:CauchySchwarzforHKin}
\end{equation}
where we also applied the inequality $H_{\mathrm{kin}}^{\prime}\leq H_{N}^{\prime}+Ck_{F}$.
It follows by the Cauchy inequality that
\begin{align}
\max\left\{ E,k_{F}\right\} \left\langle \Psi,\mathcal{N}_{E}H_{\mathrm{kin}}^{\prime}\Psi\right\rangle  & \leq C\mleft(\max\left\{ E,k_{F}\right\} \mright)^{\frac{3}{2}}\sqrt{\left\langle \Psi,\mathcal{N}_{E}^{2}H_{\mathrm{kin}}^{\prime}\Psi\right\rangle }\\
 & \leq C\max\left\{ E,k_{F}\right\} ^{3}+\frac{1}{2}\left\langle \Psi,\mathcal{N}_{E}^{2}H_{\mathrm{kin}}^{\prime}\Psi\right\rangle \nonumber 
\end{align}
which upon insertion into equation (\ref{eq:NEsqHKinEstimate2}) upon
rearrangement yields
\begin{equation}
\left\langle \Psi,\mathcal{N}_{E}^{2}H_{\mathrm{kin}}^{\prime}\Psi\right\rangle \leq2\mleft(Ck_{F}+C\max\left\{ E,k_{F}\right\} ^{3}\mright)\le C\max\left\{ E,k_{F}\right\} ^{3}.
\end{equation}
$\hfill\square$

We can now conclude the desired estimate:
\begin{cor}
For any low-lying eigenstate $\Psi\in D\mleft(H_{\mathrm{kin}}^{\prime}\mright)$
it holds that
\[
\left\langle \Psi,\mathcal{N}_{E}H_{\mathrm{kin}}^{\prime}\Psi\right\rangle \leq Ck_{F}^{2}
\]
for a constant $C>0$ depending only on $\sum_{k\in\mathbb{Z}_{\ast}^{3}}\hat{V}_{k}\left|k\right|$
and $\kappa$.
\end{cor}

\textbf{Proof:} Estimating as in equation (\ref{eq:CauchySchwarzforHKin})
we have by the proposition that
\begin{equation}
\left\langle \Psi,\mathcal{N}_{E}H_{\mathrm{kin}}^{\prime}\Psi\right\rangle \leq\sqrt{\left\langle \Psi,H_{\mathrm{kin}}^{\prime}\Psi\right\rangle \left\langle \Psi,\mathcal{N}_{E}^{2}H_{\mathrm{kin}}^{\prime}\Psi\right\rangle }\leq C\sqrt{\max\left\{ \kappa k_{F},k_{F}\right\} ^{4}}\leq Ck_{F}^{2}.
\end{equation}
$\hfill\square$

\subsection{Bounding the Non-Diagonalized and Non-Bosonizable Terms}

We consider the bounds of Proposition \ref{prop:APrioriErrorBounds}.
The non-diagonalized terms
\begin{equation}
\mathrm{ND}=\frac{k_{F}^{-1}}{2\,\mleft(2\pi\mright)^{3}}\sum_{k\in\mathbb{Z}_{\ast}^{3}\backslash\overline{B}\mleft(0,k_{F}^{\gamma}\mright)}\hat{V}_{k}\mleft(2B_{k}^{\ast}B_{k}+B_{k}B_{-k}+B_{-k}^{\ast}B_{k}^{\ast}\mright)
\end{equation}
can be immediately estimated by Proposition \ref{prop:KineticBkEstimates}
as
\begin{align}
\left|\left\langle \Psi,\mathrm{ND}\Psi\right\rangle \right| & \leq\frac{k_{F}^{-1}}{\mleft(2\pi\mright)^{3}}\sum_{k\in\mathbb{Z}_{\ast}^{3}\backslash\overline{B}\mleft(0,k_{F}^{\gamma}\mright)}\hat{V}_{k}\mleft(\left\Vert B_{k}\Psi\right\Vert ^{2}+\left\Vert B_{k}^{\ast}\Psi\right\Vert \left\Vert B_{-k}\Psi\right\Vert \mright)\nonumber \\
 & \leq Ck_{F}^{-1}\sum_{k\in\mathbb{Z}_{\ast}^{3}\backslash\overline{B}\mleft(0,k_{F}^{\gamma}\mright)}\hat{V}_{k}\sqrt{k_{F}\left\langle \Psi,H_{\mathrm{kin}}^{\prime}\Psi\right\rangle \mleft(k_{F}\left\langle \Psi,H_{\mathrm{kin}}^{\prime}\Psi\right\rangle +k_{F}^{2}\left|k\right|\left\Vert \Psi\right\Vert ^{2}\mright)}\\
 & \leq C\mleft(\sum_{k\in\mathbb{Z}_{\ast}^{3}\backslash\overline{B}\mleft(0,k_{F}^{\gamma}\mright)}\hat{V}_{k}\left|k\right|^{\frac{1}{2}}\mright)\mleft(\left\langle \Psi,H_{\mathrm{kin}}^{\prime}\Psi\right\rangle +k_{F}\left\Vert \Psi\right\Vert ^{2}\mright)\nonumber \\
 & \leq Ck_{F}^{-\frac{\gamma}{2}}\mleft(\sum_{k\in\mathbb{Z}_{\ast}^{3}}\hat{V}_{k}\left|k\right|\mright)\mleft(\left\langle \Psi,H_{\mathrm{kin}}^{\prime}\Psi\right\rangle +k_{F}\left\Vert \Psi\right\Vert ^{2}\mright)\nonumber 
\end{align}
for any $\Psi\in D\mleft(H_{\mathrm{kin}}^{\prime}\mright)$, i.e.
\begin{equation}
\pm\mathrm{ND}\leq Ck_{F}^{-\frac{\gamma}{2}}\mleft(k_{F}+H_{\mathrm{kin}}^{\prime}\mright).
\end{equation}
We again recall the non-bosonizable terms (for $s=1$):
\begin{align}
\mathcal{C} & =\frac{k_{F}^{-1}}{\mleft(2\pi\mright)^{3}}\sum_{k\in\mathbb{Z}_{\ast}^{3}}\hat{V}_{k}\,\mathrm{Re}\mleft(\mleft(B_{k}+B_{-k}^{\ast}\mright)^{\ast}D_{k}\mright)\\
\mathcal{Q} & =\frac{k_{F}^{-1}}{2\,\mleft(2\pi\mright)^{3}}\sum_{k\in\mathbb{Z}_{\ast}^{3}}\hat{V}_{k}\mleft(D_{k}^{\ast}D_{k}-\sum_{p\in L_{k}}\mleft(c_{p}^{\ast}c_{p}+c_{p-k}c_{p-k}^{\ast}\mright)\mright).\nonumber 
\end{align}
When $\sum_{k\in\mathbb{Z}_{\ast}^{3}}\hat{V}_{k}<\infty$ the second
terms of $\mathcal{Q}$ are entirely negligible, as
\begin{equation}
0\leq\frac{k_{F}^{-1}}{2\,\mleft(2\pi\mright)^{3}}\sum_{k\in\mathbb{Z}_{\ast}^{3}}\hat{V}_{k}\sum_{p\in L_{k}}\mleft(c_{p}^{\ast}c_{p}+c_{p-k}c_{p-k}^{\ast}\mright)\leq\frac{k_{F}^{-1}}{\mleft(2\pi\mright)^{3}}\mleft(\sum_{k\in\mathbb{Z}_{\ast}^{3}}\hat{V}_{k}\mright)\mathcal{N}_{E}\leq Ck_{F}^{-1}\mathcal{N}_{E}
\end{equation}
so we may disregard these. For the remaining terms we rewrite $\mathcal{C}$:
Straightforward computation shows that $\left[B_{-k},D_{k}\right]=0=\left[B_{-k}^{\ast},D_{k}^{\ast}\right]$
for any $k\in\mathbb{Z}_{\ast}^{3}$, and as furthermore $D_{k}^{\ast}=D_{-k}$
we can write $\mathcal{C}$ as
\begin{align}
\mathcal{C} & =\frac{k_{F}^{-1}}{2\,\mleft(2\pi\mright)^{3}}\sum_{k\in\mathbb{Z}_{\ast}^{3}}\hat{V}_{k}\mleft(\mleft(B_{k}^{\ast}+B_{-k}\mright)D_{k}+D_{k}^{\ast}\mleft(B_{k}+B_{-k}^{\ast}\mright)\mright)\nonumber \\
 & =\frac{k_{F}^{-1}}{2\,\mleft(2\pi\mright)^{3}}\sum_{k\in\mathbb{Z}_{\ast}^{3}}\hat{V}_{k}\mleft(B_{k}^{\ast}D_{k}+D_{k}B_{-k}+D_{k}^{\ast}B_{k}+B_{-k}^{\ast}D_{k}^{\ast}\mright)\\
 & =\frac{k_{F}^{-1}}{\mleft(2\pi\mright)^{3}}\sum_{k\in\mathbb{Z}_{\ast}^{3}}\hat{V}_{k}\mleft(B_{k}^{\ast}D_{k}+D_{k}^{\ast}B_{k}\mright)\nonumber 
\end{align}
so the terms that we need to control are
\begin{equation}
\mathrm{NB}=\frac{k_{F}^{-1}}{\mleft(2\pi\mright)^{3}}\sum_{k\in\mathbb{Z}_{\ast}^{3}}\hat{V}_{k}\mleft(B_{k}^{\ast}D_{k}+D_{k}^{\ast}B_{k}+\frac{1}{2}D_{k}^{\ast}D_{k}\mright).
\end{equation}
Dividing the summation range into $k\in\overline{B}\mleft(0,k_{F}^{\delta}\mright)\cap\mathbb{Z}_{\ast}^{3}$
and $k\in\mathbb{Z}_{\ast}^{3}\backslash\overline{B}\mleft(0,k_{F}^{\delta}\mright)$
for some $\delta>0$, we write $\mathrm{NB}=\mathrm{NB}_{1}+\mathrm{NB}_{2}$
and estimate $\mathrm{NB}_{2}$ using Proposition \ref{prop:KineticBkEstimates}
as
\begin{align}
\pm\mathrm{NB}_{2} & \leq Ck_{F}^{-1}\sum_{k\in\mathbb{Z}_{\ast}^{3}\backslash\overline{B}\mleft(0,k_{F}^{\delta}\mright)}\hat{V}_{k}\mleft(B_{k}^{\ast}B_{k}+D_{k}^{\ast}D_{k}\mright)\leq Ck_{F}^{-1}\sum_{k\in\mathbb{Z}_{\ast}^{3}\backslash\overline{B}\mleft(0,k_{F}^{\delta}\mright)}\hat{V}_{k}\mleft(k_{F}H_{\mathrm{kin}}^{\prime}+\mathcal{N}_{E}^{2}\mright)\\
 & \leq C\mleft(\sum_{k\in\mathbb{Z}_{\ast}^{3}\backslash\overline{B}\mleft(0,k_{F}^{\delta}\mright)}\hat{V}_{k}\mright)\mleft(H_{\mathrm{kin}}^{\prime}+k_{F}^{-1}\mathcal{N}_{E}H_{\mathrm{kin}}^{\prime}\mright)\leq Ck_{F}^{-\delta}\mleft(H_{\mathrm{kin}}^{\prime}+k_{F}^{-1}\mathcal{N}_{E}H_{\mathrm{kin}}^{\prime}\mright).\nonumber 
\end{align}
For $\mathrm{NB}_{1}$ we note that by Cauchy-Schwarz and the $B_{k}$
estimate of Proposition \ref{prop:KineticBkEstimates},
\begin{equation}
\left|\left\langle \Psi,\mleft(B_{k}^{\ast}D_{k}+D_{k}^{\ast}B_{k}+\frac{1}{2}D_{k}^{\ast}D_{k}\mright)\Psi\right\rangle \right|\leq C\mleft(k_{F}^{\frac{1}{2}}\sqrt{\left\langle \Psi,H_{\mathrm{kin}}^{\prime}\Psi\right\rangle }+\left\Vert D_{k}\Psi\right\Vert \mright)\left\Vert D_{k}\Psi\right\Vert ,\label{eq:DkTermEstimate}
\end{equation}
so it suffices to obtain an improved $D_{k}$ estimate for small $k$.

\subsubsection*{Detailed Analysis of $D_{k}$}

We begin by noting the following:
\begin{prop}
\label{prop:PreciseDkEstimate}For all $k\in\mathbb{Z}_{\ast}^{3}$
and any $\lambda>0$ it holds that
\[
D_{k}^{\ast}D_{k}\leq C\mleft(1+\left|S_{k,\lambda}^{1}\right|+\left|S_{k,\lambda}^{2}\right|\mright)\mathcal{N}_{E}+C\lambda^{-\frac{1}{2}}\mathcal{N}_{E}H_{\mathrm{kin}}^{\prime}
\]
for a constant $C>0$ independent of $k$ and $k_{F}$, where
\begin{align*}
S_{k,\lambda}^{1} & =\left\{ p\in B_{F}\cap\mleft(B_{F}+k\mright)\mid\max\left\{ |\left|p\right|^{2}-\zeta|,|\left|p-k\right|^{2}-\zeta|\right\} <\lambda\right\} \\
S_{k,\lambda}^{2} & =\left\{ p\in B_{F}^{c}\cap\mleft(B_{F}^{c}+k\mright)\mid\max\left\{ |\left|p\right|^{2}-\zeta|,|\left|p-k\right|^{2}-\zeta|\right\} <\lambda\right\} .
\end{align*}
\end{prop}

\textbf{Proof:} It suffices to consider $D_{1,k}$ and $D_{2,k}$;
we focus on $D_{1,k}$. Recall again that
\begin{equation}
D_{1,k}^{\ast}D_{1,k}=\sum_{p,q\in B_{F}\cap\mleft(B_{F}+k\mright)}c_{p-k}c_{q}c_{q-k}^{\ast}c_{p}^{\ast}+\sum_{q\in B_{F}}1_{B_{F}}\mleft(q+k\mright)c_{q}c_{q}^{\ast}
\end{equation}
so for any $\Psi\in D\mleft(H_{\mathrm{kin}}^{\prime}\mright)$
\begin{align}
\left\Vert D_{1,k}\Psi\right\Vert ^{2} & \leq\sum_{p,q\in B_{F}\cap\mleft(B_{F}+k\mright)}\left\Vert c_{q}^{\ast}c_{p-k}^{\ast}\Psi\right\Vert \left\Vert c_{q-k}^{\ast}c_{p}^{\ast}\Psi\right\Vert +\left\langle \Psi,\mathcal{N}_{E}\Psi\right\rangle \\
 & \leq\sum_{p\in B_{F}\cap\mleft(B_{F}+k\mright)}\Vert c_{p-k}^{\ast}\mathcal{N}_{E}^{\frac{1}{2}}\Psi\Vert\Vert c_{p}^{\ast}\mathcal{N}_{E}^{\frac{1}{2}}\Psi\Vert+\left\langle \Psi,\mathcal{N}_{E}\Psi\right\rangle .\nonumber 
\end{align}
To estimate the sum, we decompose $B_{F}\cap\mleft(B_{F}+k\mright)=S_{k,\lambda}^{1}\cup S_{\geq\lambda}^{1}$
where $S_{k,\lambda}^{1}$ is as in the statement of the theorem,
and
\begin{equation}
S_{\geq\lambda}^{1}=\left\{ p\in B_{F}\cap\mleft(B_{F}+k\mright)\mid\max\left\{ |\left|p\right|^{2}-\zeta|,|\left|p-k\right|^{2}-\zeta|\right\} \geq\lambda\right\} .
\end{equation}
By this definition and equation (\ref{eq:ZetaDefiningProperty}) it
holds for all $p\in S_{\geq\lambda}^{1}$ that
\begin{equation}
\sqrt{|\left|p\right|^{2}-\zeta|}\sqrt{|\left|p-k\right|^{2}-\zeta|}\geq2^{-\frac{1}{2}}\lambda^{\frac{1}{2}}
\end{equation}
so we can estimate
\begin{align}
 & \sum_{p\in B_{F}\cap\mleft(B_{F}+k\mright)}\Vert c_{p-k}^{\ast}\mathcal{N}_{E}^{\frac{1}{2}}\Psi\Vert\Vert c_{p}^{\ast}\mathcal{N}_{E}^{\frac{1}{2}}\Psi\Vert\nonumber \\
 & \leq\left|S_{k,\lambda}^{1}\right|\Vert\mathcal{N}_{E}^{\frac{1}{2}}\Psi\Vert^{2}+\sqrt{2}\lambda^{-\frac{1}{2}}\sum_{p\in S_{\geq\lambda}^{1}}\sqrt{|\left|p\right|^{2}-\zeta|}\sqrt{|\left|p-k\right|^{2}-\zeta|}\Vert c_{p-k}^{\ast}\mathcal{N}_{E}^{\frac{1}{2}}\Psi\Vert\Vert c_{p}^{\ast}\mathcal{N}_{E}^{\frac{1}{2}}\Psi\Vert\\
 & \leq\left|S_{k,\lambda}^{1}\right|\left\langle \Psi,\mathcal{N}_{E}\Psi\right\rangle +\sqrt{2}\lambda^{-\frac{1}{2}}\sqrt{\sum_{p\in S_{\geq\lambda}^{1}}|\left|p\right|^{2}-\zeta|\Vert c_{p}^{\ast}\mathcal{N}_{E}^{\frac{1}{2}}\Psi\Vert^{2}}\sqrt{\sum_{p\in S_{\geq\lambda}^{1}}|\left|p-k\right|^{2}-\zeta|\Vert c_{p-k}^{\ast}\mathcal{N}_{E}^{\frac{1}{2}}\Psi\Vert^{2}}\nonumber \\
 & \leq\left|S_{k,\lambda}^{1}\right|\left\langle \Psi,\mathcal{N}_{E}\Psi\right\rangle +\sqrt{2}\lambda^{-\frac{1}{2}}\left\langle \Psi,\mathcal{N}_{E}H_{\mathrm{kin}}^{\prime}\Psi\right\rangle \nonumber 
\end{align}
by equation (\ref{eq:GeneralPositiveHkinPrimeDecomposition}), whence
the claim follows.

$\hfill\square$

By employing precise lattice point counting techniques of the same
kind used in appendix section \ref{subsec:PreciseEstimates}, the
following was obtained in \cite{ChrHaiNam-21}:
\begin{prop}
For all $k\in\overline{B}\mleft(0,k_{F}\mright)\cap\mathbb{Z}_{\ast}^{3}$
and $0<\lambda\leq\frac{1}{6}k_{F}^{2}$ (depending on $k$ and $k_{F}$)
it holds that
\[
\left|S_{k,\lambda}^{1}\right|+\left|S_{k,\lambda}^{2}\right|\leq C\mleft(\left|k\right|^{-1}\lambda+\left|k\right|^{3+\frac{2}{3}}\log\mleft(k_{F}\mright)^{\frac{2}{3}}k_{F}^{\frac{2}{3}}\mright)\mleft(\lambda+\left|k\right|\mright),\quad k_{F}\rightarrow\infty,
\]
for a constant $C>0$ independent of all quantities.
\end{prop}

From this and Proposition \ref{prop:PreciseDkEstimate} one can then
conclude a stronger $D_{k}$ bound:
\begin{prop}
For all $k\in\overline{B}\mleft(0,k_{F}^{\delta}\mright)\cap\mathbb{Z}_{\ast}^{3}$,
$0<\delta<\frac{2}{31}$, it holds that
\[
D_{k}^{\ast}D_{k}\leq C\left|k\right|^{\frac{11}{9}}\log\mleft(k_{F}\mright)^{\frac{2}{9}}k_{F}^{\frac{8}{9}}\mleft(H_{\mathrm{kin}}^{\prime}+k_{F}^{-1}\mathcal{N}_{E}H_{\mathrm{kin}}^{\prime}\mright),\quad k_{F}\rightarrow\infty,
\]
for a constant $C>0$ independent of all quantities.
\end{prop}

It now follows from equation (\ref{eq:DkTermEstimate}) that
\begin{equation}
\left|\left\langle \Psi,\mleft(B_{k}^{\ast}D_{k}+D_{k}^{\ast}B_{k}+\frac{1}{2}D_{k}^{\ast}D_{k}\mright)\Psi\right\rangle \right|\leq C\left|k\right|\log\mleft(k_{F}\mright)^{\frac{1}{9}}k_{F}^{\frac{17}{18}}\left\langle \Psi,\mleft(H_{\mathrm{kin}}^{\prime}+k_{F}^{-1}\mathcal{N}_{E}H_{\mathrm{kin}}^{\prime}\mright)\Psi\right\rangle 
\end{equation}
for $\left|k\right|\leq k_{F}^{\delta}$, $\delta<\frac{2}{31}$,
whence
\begin{equation}
\pm\mathrm{NB}_{1}\leq C\log\mleft(k_{F}\mright)^{\frac{1}{9}}k_{F}^{-\frac{1}{18}}\mleft(H_{\mathrm{kin}}^{\prime}+k_{F}^{-1}\mathcal{N}_{E}H_{\mathrm{kin}}^{\prime}\mright).
\end{equation}
As $\frac{2}{31}>\frac{1}{18}$, $\delta$ can be chosen such that
the $\mathrm{NB}_{2}$ bound matches this one, yielding Proposition
\ref{prop:APrioriErrorBounds}.

\subsection{Controlling the Diagonalization}

We begin by considering the tail estimate for $E_{\mathrm{corr},\mathrm{bos}}$.
Recall that by Theorem \ref{them:OneBodyEstimates} (with $s=1$)
\begin{equation}
\mathrm{tr}\mleft(E_{k}-h_{k}-P_{k}\mright)=\frac{1}{\pi}\int_{0}^{\infty}F\mleft(\frac{\hat{V}_{k}k_{F}^{-1}}{\mleft(2\pi\mright)^{3}}\sum_{p\in L_{k}}\frac{\lambda_{k,p}}{\lambda_{k,p}^{2}+t^{2}}\mright)dt,\quad F\mleft(x\mright)=\log\mleft(1+x\mright)-x.
\end{equation}
As $F$ obeys $\left|F\mleft(x\mright)\right|\leq\frac{1}{2}x^{2}$
for $x\geq0$, we may estimate
\begin{align}
\left|\mathrm{tr}\mleft(E_{k}-h_{k}-P_{k}\mright)\right| & \leq\frac{1}{2\pi}\int_{0}^{\infty}\mleft(\frac{\hat{V}_{k}k_{F}^{-1}}{\mleft(2\pi\mright)^{3}}\sum_{p\in L_{k}}\frac{\lambda_{k,p}}{\lambda_{k,p}^{2}+t^{2}}\mright)^{2}dt=\frac{\hat{V}_{k}^{2}k_{F}^{-2}}{\mleft(2\pi\mright)^{7}}\sum_{p,q\in L_{k}}\int_{0}^{\infty}\frac{\lambda_{k,p}}{\lambda_{k,p}^{2}+t^{2}}\frac{\lambda_{k,q}}{\lambda_{k,q}^{2}+t^{2}}dt\nonumber \\
 & =\frac{\hat{V}_{k}^{2}k_{F}^{-2}}{4\,\mleft(2\pi\mright)^{6}}\sum_{p,q\in L_{k}}\frac{1}{\lambda_{k,p}+\lambda_{k,q}}\leq\frac{\hat{V}_{k}^{2}k_{F}^{-2}}{4\,\mleft(2\pi\mright)^{6}}\mleft(\sum_{p\in L_{k}}\frac{1}{\sqrt{\lambda_{k,p}}}\mright)^{2}\\
 & \leq\frac{\hat{V}_{k}^{2}k_{F}^{-2}}{4\,\mleft(2\pi\mright)^{6}}\mleft(Ck_{F}^{\frac{3}{2}}\left|k\right|^{\frac{1}{2}}\mright)^{2}\leq Ck_{F}\hat{V}_{k}^{2}\left|k\right|\nonumber 
\end{align}
where we used the integral identity
\begin{equation}
\int_{0}^{\infty}\frac{a}{a^{2}+t^{2}}\frac{b}{b^{2}+t^{2}}dt=\frac{\pi}{2}\frac{1}{a+b},\quad a,b>0,
\end{equation}
and the estimate $\sum_{p\in L_{k}}\lambda_{k,p}^{-\frac{1}{2}}\leq Ck_{F}^{\frac{3}{2}}\left|k\right|^{\frac{1}{2}}$.
Consequently $\sum_{k\in\overline{B}\mleft(0,k_{F}^{\gamma}\mright)\cap\mathbb{Z}_{\ast}^{3}}\mathrm{tr}\mleft(E_{k}-h_{k}-P_{k}\mright)-E_{\mathrm{corr},\mathrm{bos}}$
is bounded by
\begin{align}
\sum_{k\in\mathbb{Z}_{\ast}^{3}\backslash\overline{B}\mleft(0,k_{F}^{\gamma}\mright)}\left|\mathrm{tr}\mleft(E_{k}-h_{k}-P_{k}\mright)\right| & \leq Ck_{F}\sum_{k\in\mathbb{Z}_{\ast}^{3}\backslash\overline{B}\mleft(0,k_{F}^{\gamma}\mright)}\hat{V}_{k}^{2}\left|k\right|\leq Ck_{F}^{1-\gamma}\sum_{k\in\mathbb{Z}_{\ast}^{3}\backslash\overline{B}\mleft(0,k_{F}^{\gamma}\mright)}\hat{V}_{k}^{2}\left|k\right|^{2}\\
 & \leq Ck_{F}^{1-\gamma}\mleft(\sum_{k\in\mathbb{Z}_{\ast}^{3}}\hat{V}_{k}\left|k\right|\mright)^{2}\leq Ck_{F}^{1-\gamma}\nonumber 
\end{align}
as claimed in Proposition \ref{prop:OperatorDiagonalizationControl},
where we used that $\sum_{k}\left|a_{k}\right|^{2}\leq\mleft(\sum_{k}\left|a_{k}\right|\mright)^{2}$.

We will not prove the exchange term bounds of the proposition here,
but let us mention the idea behind kinetic estimation: The thing to
note is that the idea of the kinetic estimate of Proposition \ref{prop:KineticBkEstimates}
immediately generalizes as
\begin{align}
\left\Vert b_{k}\mleft(\varphi\mright)\Psi\right\Vert  & \leq\sum_{p\in L_{k}}\left|\left\langle \varphi,e_{p}\right\rangle \right|\left\Vert c_{p-k}^{\ast}c_{p}\Psi\right\Vert \leq\sqrt{\sum_{p\in L_{k}}\lambda_{k,p}^{-1}\left|\left\langle \varphi,e_{p}\right\rangle \right|^{2}}\sqrt{\sum_{p\in L_{k}}\lambda_{k,p}\left\Vert c_{p-k}^{\ast}c_{p}\Psi\right\Vert ^{2}}\\
 & \leq2^{-\frac{1}{2}}\sqrt{\left\langle \varphi,h_{k}^{-1}\varphi\right\rangle \left\langle \Psi,H_{\mathrm{kin}}^{\prime}\Psi\right\rangle }\nonumber 
\end{align}
and so, as $\varepsilon_{k,k}\mleft(\varphi;\varphi\mright)\leq0$,
also
\begin{equation}
\left\Vert b_{k}^{\ast}\mleft(\varphi\mright)\Psi\right\Vert ^{2}\leq\left\Vert b_{k}\mleft(\varphi\mright)\Psi\right\Vert ^{2}+\left\Vert \varphi\right\Vert ^{2}\left\Vert \Psi\right\Vert ^{2}\leq\frac{1}{2}\left\langle \varphi,h_{k}^{-1}\varphi\right\rangle \left\langle \Psi,H_{\mathrm{kin}}^{\prime}\Psi\right\rangle +\left\Vert \varphi\right\Vert ^{2}\left\Vert \Psi\right\Vert ^{2},
\end{equation}
so for any $\Psi\in D\mleft(H_{\mathrm{kin}}^{\prime}\mright)$
\begin{equation}
\left\Vert b_{k}\mleft(\varphi\mright)\Psi\right\Vert \leq\Vert h_{k}^{-\frac{1}{2}}\varphi\Vert\sqrt{\left\langle \Psi,H_{\mathrm{kin}}^{\prime}\Psi\right\rangle },\quad\left\Vert b_{k}^{\ast}\mleft(\varphi\mright)\Psi\right\Vert \leq\Vert h_{k}^{-\frac{1}{2}}\varphi\Vert\sqrt{\left\langle \Psi,H_{\mathrm{kin}}^{\prime}\Psi\right\rangle }+\left\Vert \varphi\right\Vert \left\Vert \Psi\right\Vert .\label{eq:bkbkastKineticEstimates}
\end{equation}
These inequalities allow us to arbitrage between the one-body and
many-body kinetic operators. As we have good control on both the one-body
quantities and the many-body kinetic energy, this is a significant
improvement over pure $\mathcal{N}_{E}$ estimates given our poor
control of this quantity.

To illustrate the application of these bounds, let us derive the Gronwall
estimate for $e^{t\mathcal{K}}H_{\mathrm{kin}}^{\prime}e^{-t\mathcal{K}}$;
this amounts to controlling
\begin{equation}
\left[\mathcal{K},H_{\mathrm{kin}}^{\prime}\right]=\sum_{k\in\overline{B}\mleft(0,k_{F}^{\gamma}\mright)\cap\mathbb{Z}_{\ast}^{3}}Q_{2}^{k}\mleft(\left\{ K_{k},h_{k}\right\} \mright)
\end{equation}
in terms of $H_{\mathrm{kin}}^{\prime}+k_{F}$. We derive a general
kinetic bound for a $Q_{2}^{k}\mleft(B\mright)$ operator: By the kinetic
estimate
\begin{align}
\left|\left\langle \Psi,Q_{2}^{k}\mleft(B\mright)\Psi\right\rangle \right| & \leq2\sum_{p\in L_{k}}\left|\left\langle \Psi,b_{k}\mleft(Be_{p}\mright)b_{-k,-p}\Psi\right\rangle \right|\leq2\sum_{p\in L_{k}}\left\Vert b_{k}^{\ast}\mleft(Be_{p}\mright)\Psi\right\Vert \left\Vert b_{-k,-p}\Psi\right\Vert \\
 & \leq2\sum_{p\in L_{k}}\mleft(\Vert h_{k}^{-\frac{1}{2}}Be_{p}\Vert\sqrt{\left\langle \Psi,H_{\mathrm{kin}}^{\prime}\Psi\right\rangle }+\left\Vert Be_{p}\right\Vert \left\Vert \Psi\right\Vert \mright)\left\Vert b_{-k,-p}\Psi\right\Vert ,\nonumber 
\end{align}
and by Cauchy-Schwarz we have that
\begin{align}
\sum_{p\in L_{k}}\Vert h_{k}^{-\frac{1}{2}}Be_{p}\Vert\left\Vert b_{-k,-p}\Psi\right\Vert  & \leq\sqrt{\sum_{p\in L_{k}}\lambda_{k,p}^{-1}\Vert h_{k}^{-\frac{1}{2}}Be_{p}\Vert^{2}}\sqrt{\sum_{p\in L_{k}}\lambda_{-k,-p}\left\Vert b_{-k,-p}\Psi\right\Vert ^{2}}\\
 & \leq\sqrt{\sum_{p\in L_{k}}\Vert h_{k}^{-\frac{1}{2}}Bh_{k}^{-\frac{1}{2}}e_{p}\Vert^{2}}\sqrt{\left\langle \Psi,H_{\mathrm{kin}}^{\prime}\Psi\right\rangle }=\Vert h_{k}^{-\frac{1}{2}}Bh_{k}^{-\frac{1}{2}}\Vert_{\mathrm{HS}}\sqrt{\left\langle \Psi,H_{\mathrm{kin}}^{\prime}\Psi\right\rangle }\nonumber 
\end{align}
and similarly
\begin{align}
\sum_{p\in L_{k}}\left\Vert Be_{p}\right\Vert \left\Vert b_{-k,-p}\Psi\right\Vert  & \leq\sqrt{\sum_{p\in L_{k}}\lambda_{k,p}^{-\frac{1}{2}}\left\Vert Be_{p}\right\Vert ^{2}}\sqrt{\sum_{p\in L_{k}}\lambda_{-k,-p}\left\Vert b_{-k,-p}\Psi\right\Vert ^{2}}\\
 & \leq\Vert Bh_{k}^{-\frac{1}{2}}\Vert_{\mathrm{HS}}\sqrt{\left\langle \Psi,H_{\mathrm{kin}}^{\prime}\Psi\right\rangle },\nonumber 
\end{align}
whence
\begin{equation}
\left|\left\langle \Psi,Q_{2}^{k}\mleft(B\mright)\Psi\right\rangle \right|\leq\Vert h_{k}^{-\frac{1}{2}}Bh_{k}^{-\frac{1}{2}}\Vert_{\mathrm{HS}}\left\langle \Psi,H_{\mathrm{kin}}^{\prime}\Psi\right\rangle +\Vert Bh_{k}^{-\frac{1}{2}}\Vert_{\mathrm{HS}}\left\Vert \Psi\right\Vert \sqrt{\left\langle \Psi,H_{\mathrm{kin}}^{\prime}\Psi\right\rangle }.
\end{equation}
For $B=\left\{ K_{k},h_{k}\right\} $, it follows from our one-body
operator estimates that
\begin{equation}
\Vert h_{k}^{-\frac{1}{2}}\left\{ K_{k},h_{k}\right\} h_{k}^{-\frac{1}{2}}\Vert_{\mathrm{HS}}\leq C\hat{V}_{k},\quad\Vert\left\{ K_{k},h_{k}\right\} h_{k}^{-\frac{1}{2}}\Vert_{\mathrm{HS}}\leq Ck_{F}^{\frac{1}{2}}\hat{V}_{k}\left|k\right|^{\frac{1}{2}},
\end{equation}
so
\begin{align}
\left|\left\langle \Psi,Q_{2}^{k}\mleft(\left\{ K_{k},h_{k}\right\} \mright)\Psi\right\rangle \right| & \leq C\hat{V}_{k}\left\langle \Psi,H_{\mathrm{kin}}^{\prime}\Psi\right\rangle +Ck_{F}^{\frac{1}{2}}\hat{V}_{k}\left|k\right|^{\frac{1}{2}}\left\Vert \Psi\right\Vert \sqrt{\left\langle \Psi,H_{\mathrm{kin}}^{\prime}\Psi\right\rangle }\\
 & \leq C\hat{V}_{k}\left|k\right|^{\frac{1}{2}}\left\langle \Psi,\mleft(H_{\mathrm{kin}}^{\prime}+k_{F}\mright)\Psi\right\rangle \nonumber 
\end{align}
i.e. $\pm\,Q_{2}^{k}\mleft(\left\{ K_{k},h_{k}\right\} \mright)\leq C\hat{V}_{k}\left|k\right|^{\frac{1}{2}}\mleft(H_{\mathrm{kin}}^{\prime}+k_{F}\mright)$,
whence
\begin{equation}
\pm\left[\mathcal{K},H_{\mathrm{kin}}^{\prime}\right]\leq C\mleft(\sum_{k\in\overline{B}\mleft(0,k_{F}^{\gamma}\mright)\cap\mathbb{Z}_{\ast}^{3}}\hat{V}_{k}\left|k\right|^{\frac{1}{2}}\mright)\mleft(H_{\mathrm{kin}}^{\prime}+k_{F}\mright)\leq C\mleft(H_{\mathrm{kin}}^{\prime}+k_{F}\mright)
\end{equation}
as desired.

\subsection{The Second Transformation}

In this last subsection we consider the one-body operator estimates
needed to control the second transformation. First note that for $\mathcal{J}$
as defined by equation (\ref{eq:calJDefinition}), computation using
the quasi-bosonic commutation relations as in Section \ref{sec:DiagonalizationoftheBosonizableTerms}
establishes that $\mathcal{J}$ obeys
\begin{equation}
\left[\mathcal{J},b_{k}\mleft(\varphi\mright)\right]=b_{k}\mleft(J_{k}\varphi\mright)+\sum_{l\in\overline{B}\mleft(0,k_{F}^{\gamma}\mright)\cap\mathbb{Z}_{\ast}^{3}}\sum_{q\in L_{l}}\varepsilon_{k,l}\mleft(\varphi;e_{q}\mright)b_{l}\mleft(J_{l}e_{q}\mright)
\end{equation}
hence
\begin{equation}
\left[\mathcal{J},Q_{1}^{k}\mleft(A\mright)\right]=Q_{1}^{k}\mleft(\left[J_{k},A\right]\mright)+\mathcal{E}_{3}^{k}\mleft(A\mright)
\end{equation}
for symmetric $A:\ell^{2}\mleft(L_{k}\mright)\rightarrow\ell^{2}\mleft(L_{k}\mright)$,
where $\mathcal{E}_{3}^{k}\mleft(A\mright)$ is given by
\begin{equation}
\mathcal{E}_{3}^{k}\mleft(A\mright)=2\sum_{l\in\overline{B}\mleft(0,k_{F}^{\gamma}\mright)\cap\mathbb{Z}_{\ast}^{3}}\sum_{p\in L_{k}}\sum_{q\in L_{l}}\mathrm{Re}\mleft(b_{k}^{\ast}\mleft(Ae_{p}\mright)\varepsilon_{k,l}\mleft(e_{p};e_{q}\mright)b_{l}\mleft(J_{l}e_{q}\mright)\mright).
\end{equation}
We estimate a generic term of $\mathcal{E}_{3}^{k}\mleft(A\mright)$
using the kinetic bound of equation (\ref{eq:bkbkastKineticEstimates})
in the manner of \cite{ChrHaiNam-21}: We have
\begin{align}
 & \quad\;\sum_{l\in\overline{B}\mleft(0,k_{F}^{\gamma}\mright)\cap\mathbb{Z}_{\ast}^{3}}\sum_{p\in L_{k}\cap L_{l}}\left|\left\langle \Psi,b_{k}^{\ast}\mleft(Ae_{p}\mright)c_{p-l}c_{p-k}^{\ast}b_{l}\mleft(J_{l}e_{p}\mright)\Psi\right\rangle \right|\nonumber \\
 & \leq\sum_{l\in\overline{B}\mleft(0,k_{F}^{\gamma}\mright)\cap\mathbb{Z}_{\ast}^{3}}\sum_{p\in L_{k}\cap L_{l}}\left\Vert b_{k}\mleft(Ae_{p}\mright)c_{p-l}^{\ast}\Psi\right\Vert \left\Vert b_{l}\mleft(J_{l}e_{p}\mright)c_{p-k}^{\ast}\Psi\right\Vert \nonumber \\
 & \leq\sum_{l\in\overline{B}\mleft(0,k_{F}^{\gamma}\mright)\cap\mathbb{Z}_{\ast}^{3}}\sum_{p\in L_{k}\cap L_{l}}\Vert h_{k}^{-\frac{1}{2}}Ae_{p}\Vert\Vert h_{l}^{-\frac{1}{2}}J_{l}e_{p}\Vert\sqrt{\left\langle \Psi,c_{p-l}H_{\mathrm{kin}}^{\prime}c_{p-l}^{\ast}\Psi\right\rangle }\sqrt{\left\langle \Psi,c_{p-k}H_{\mathrm{kin}}^{\prime}c_{p-k}^{\ast}\Psi\right\rangle }\\
 & \leq\mleft(\max_{p\in L_{k}}\Vert h_{k}^{-\frac{1}{2}}Ae_{p}\Vert\mright)\sqrt{\left\langle \Psi,H_{\mathrm{kin}}^{\prime}\Psi\right\rangle }\sum_{l\in\overline{B}\mleft(0,k_{F}^{\gamma}\mright)\cap\mathbb{Z}_{\ast}^{3}}\sum_{p\in L_{k}\cap L_{l}}\Vert h_{l}^{-\frac{1}{2}}J_{l}e_{p}\Vert\sqrt{\left\langle \Psi,c_{p-l}H_{\mathrm{kin}}^{\prime}c_{p-l}^{\ast}\Psi\right\rangle }\nonumber \\
 & \leq\mleft(\max_{p\in L_{k}}\Vert h_{k}^{-\frac{1}{2}}Ae_{p}\Vert\mright)\mleft(\sum_{l\in\overline{B}\mleft(0,k_{F}^{\gamma}\mright)\cap\mathbb{Z}_{\ast}^{3}}\Vert h_{l}^{-\frac{1}{2}}J_{l}\Vert_{\mathrm{HS}}\mright)\sqrt{\left\langle \Psi,H_{\mathrm{kin}}^{\prime}\Psi\right\rangle \left\langle \Psi,\mathcal{N}_{E}H_{\mathrm{kin}}^{\prime}\Psi\right\rangle }.\nonumber 
\end{align}
Controlling the error term of the transformation of equation (\ref{eq:SecondTransformationAction})
thus requires us to estimate one-body quantities of the form $\max_{p\in L_{k}}\Vert h_{k}^{-\frac{1}{2}}E_{k}\mleft(t\mright)e_{p}\Vert$,
where $E_{k}\mleft(t\mright)$ is given by
\begin{equation}
E_{k}\mleft(t\mright)=e^{tJ_{k}}e^{-K_{k}}h_{k}e^{-K_{k}}e^{-tJ_{k}}-h_{k}.
\end{equation}
We consider this in the abstract one-body setting of Section \ref{sec:AnalysisofOne-BodyOperators}.
In this case, the unitary transformation $U$ is given by
\begin{equation}
U=\mleft(h^{2}+2P_{h^{\frac{1}{2}}v}\mright)^{\frac{1}{4}}h^{-\frac{1}{2}}e^{K},
\end{equation}
and by using the integral identity
\begin{equation}
a^{\frac{1}{4}}=\frac{2\sqrt{2}}{\pi}\int_{0}^{\infty}\mleft(1-\frac{t^{4}}{a+t^{4}}\mright)\,dt,\quad a>0,
\end{equation}
one can derive a representation formula for an operator of the form
$\mleft(A+gP_{w}\mright)^{\frac{1}{4}}$ similar to that of Proposition
\ref{prop:SquareRootofAOneDimensionalPerturbation} with the following
consequence:
\begin{prop}
For all $1\leq i,j\leq n$ it holds that
\[
\left|\left\langle x_{i},\mleft(\mleft(h^{2}+2P_{h^{\frac{1}{2}}v}\mright)^{\frac{1}{4}}-h^{\frac{1}{2}}\mright)x_{j}\right\rangle \right|\leq2\frac{\sqrt{\lambda_{i}\lambda_{j}}}{\sqrt{\lambda_{i}}+\sqrt{\lambda_{j}}}\frac{\left\langle x_{i},v\right\rangle \left\langle v,x_{j}\right\rangle }{\lambda_{i}+\lambda_{j}}.
\]
\end{prop}

This implies the following elementwise bounds for $U$:
\begin{prop}
\label{prop:UElementEstimates}For all $1\leq i,j\leq n$ it holds
that
\[
\left|\left\langle x_{i},\mleft(U-1\mright)x_{j}\right\rangle \right|,\,\left|\left\langle x_{i},\mleft(U^{\ast}-1\mright)x_{j}\right\rangle \right|\leq3\mleft(1+\left\langle v,h^{-1}v\right\rangle \mright)\frac{\left\langle x_{i},v\right\rangle \left\langle v,x_{j}\right\rangle }{\lambda_{i}+\lambda_{j}}.
\]
\end{prop}

\textbf{Proof:} It suffices to consider $U-1$. Writing
\begin{align}
U-1 & =\mleft(\mleft(h^{2}+2P_{h^{\frac{1}{2}}v}\mright)^{\frac{1}{4}}-h^{\frac{1}{2}}\mright)h^{-\frac{1}{2}}e^{K}+h^{\frac{1}{2}}h^{-\frac{1}{2}}e^{K}-1\\
 & =\mleft(e^{K}-1\mright)+\mleft(\mleft(h^{2}+2P_{h^{\frac{1}{2}}v}\mright)^{\frac{1}{4}}-h^{\frac{1}{2}}\mright)h^{-\frac{1}{2}}+\mleft(\mleft(h^{2}+2P_{h^{\frac{1}{2}}v}\mright)^{\frac{1}{4}}-h^{\frac{1}{2}}\mright)h^{-\frac{1}{2}}\mleft(e^{K}-1\mright)\nonumber 
\end{align}
we estimate each part in turn. Firstly, we already know that
\begin{equation}
\left|\left\langle x_{i},\mleft(e^{K}-1\mright)x_{j}\right\rangle \right|\leq\frac{\left\langle x_{i},v\right\rangle \left\langle v,x_{j}\right\rangle }{\lambda_{i}+\lambda_{j}}
\end{equation}
by Proposition \ref{prop:tDependentElementEstimates}. Meanwhile,
by the previous proposition
\begin{equation}
\left|\left\langle x_{i},\mleft(\mleft(h^{2}+2P_{h^{\frac{1}{2}}v}\mright)^{\frac{1}{4}}-h^{\frac{1}{2}}\mright)h^{-\frac{1}{2}}x_{j}\right\rangle \right|\leq\frac{1}{\sqrt{\lambda_{j}}}\frac{2\sqrt{\lambda_{i}\lambda_{j}}}{\sqrt{\lambda_{i}}+\sqrt{\lambda_{j}}}\frac{\left\langle x_{i},v\right\rangle \left\langle v,x_{j}\right\rangle }{\lambda_{i}+\lambda_{j}}\leq2\frac{\left\langle x_{i},v\right\rangle \left\langle v,x_{j}\right\rangle }{\lambda_{i}+\lambda_{j}}
\end{equation}
and using both of these estimates we also find that
\begin{align}
 & \;\left|\left\langle x_{i},\mleft(\mleft(h^{2}+2P_{h^{\frac{1}{2}}v}\mright)^{\frac{1}{4}}-h^{\frac{1}{2}}\mright)h^{-\frac{1}{2}}\mleft(e^{K}-1\mright)x_{j}\right\rangle \right|\nonumber \\
 & \leq\sum_{k=1}^{n}\left|\left\langle x_{i},\mleft(\mleft(h^{2}+2P_{h^{\frac{1}{2}}v}\mright)^{\frac{1}{4}}-h^{\frac{1}{2}}\mright)h^{-\frac{1}{2}}x_{k}\right\rangle \right|\left|\left\langle x_{k},\mleft(e^{K}-1\mright)x_{j}\right\rangle \right|\\
 & \leq2\sum_{k=1}^{n}\frac{\left\langle x_{i},v\right\rangle \left\langle v,x_{k}\right\rangle }{\lambda_{i}+\lambda_{k}}\frac{\left\langle x_{k},v\right\rangle \left\langle v,x_{j}\right\rangle }{\lambda_{k}+\lambda_{j}}\leq2\mleft(\sum_{k=1}^{n}\frac{\left|\left\langle x_{k},v\right\rangle \right|^{2}}{\lambda_{k}}\mright)\frac{\left\langle x_{i},v\right\rangle \left\langle v,x_{j}\right\rangle }{\lambda_{i}+\lambda_{j}}\nonumber \\
 & =2\left\langle v,h^{-1}v\right\rangle \frac{\left\langle x_{i},v\right\rangle \left\langle v,x_{j}\right\rangle }{\lambda_{i}+\lambda_{j}}\nonumber 
\end{align}
where we used that $\mleft(a+c\mright)^{-1}\mleft(b+c\mright)^{-1}\leq c^{-1}\mleft(a+b\mright)^{-1}$
for $a,b,c>0$ (as $c\mleft(a+b\mright)\leq\mleft(a+c\mright)\mleft(b+c\mright)$
follows by expansion). Combining the estimates yields the claim.

$\hfill\square$

Recall that for the particular operators $h_{k}$ and $P_{v_{k}}$
it holds that $\left\langle v_{k},h_{k}^{-1}v_{k}\right\rangle \leq C\hat{V}_{k}$,
so for the purposes of estimation this matrix element estimate for
$U=e^{J}$ is almost as good as that for $e^{K}$ of Proposition \ref{prop:tDependentElementEstimates}.
Unlike that proposition, however, we can not extend this to $e^{tJ}$
for general $t\in\left[0,1\right]$, as we now lack the required monotonicity.

We can work around this by finding a way to reduce estimates involving
$e^{tJ}$ to ones involving $U$. To provide a concrete example, let
us consider a term we would need to control $\mathcal{E}_{3}^{k}\mleft(E_{k}\mleft(t\mright)\mright)$:
In the general setting, we consider $E\mleft(t\mright)$ defined by
\begin{equation}
E\mleft(t\mright)=e^{tJ}e^{-K}he^{-K}e^{-tJ}-h=\mleft(e^{tJ}he^{-tJ}-h\mright)+e^{tJ}\mleft(e^{-K}he^{-K}-h\mright)e^{-tJ}=:E_{1}\mleft(t\mright)+E_{2}\mleft(t\mright)
\end{equation}
and decompose $E_{1}\mleft(t\mright)$ further as
\begin{equation}
E_{1}\mleft(t\mright)=\mleft(e^{tJ}-1\mright)h+h\mleft(e^{-tJ}-1\mright)+\mleft(e^{tJ}-1\mright)h\mleft(e^{-tJ}-1\mright).
\end{equation}
We consider the first term, and so need to estimate $\max_{1\leq i\leq n}\Vert h^{-\frac{1}{2}}\mleft(e^{tJ}-1\mright)hx_{i}\Vert$.
As mentioned we are to find a way to replace $e^{tJ}-1$ by $U-1$
(and possibly $U^{\ast}-1$). Now, $J$ is the principal logarithm
of $U$, and as $U$ is unitary, hence normal, and we are working
on a finite-dimensional space (which we now consider as a complex
vector space), there exists an orthonormal basis $\mleft(w_{j}\mright)_{j=1}^{n}$
and real numbers $\mleft(\theta_{j}\mright)_{j=1}^{n}\subset\mleft[-\pi,\pi\mright)$
such that
\begin{equation}
e^{\pm tJ}w_{j}=e^{\pm it\theta_{j}}w_{j},\quad1\leq j\leq n.
\end{equation}
With respect to this basis, our task thus amounts to estimating $e^{it\theta}-1$
in terms of $e^{i\theta}-1$ and $e^{-i\theta}-1$. To that end we
note the following: There exists a $C>0$ such that for all $t\in\left[-1,1\right]$
and $\theta\in\left[-\pi,\pi\right]$
\begin{equation}
\left|\mleft(e^{it\theta}-1\mright)-t\mleft(e^{i\theta}-1\mright)+\frac{t\mleft(1-t\mright)}{2}\mleft(e^{i\theta}+e^{-i\theta}-2\mright)\right|\leq C\left|e^{i\theta}-1\right|^{3}.\label{eq:CubicExponentialBound}
\end{equation}
(There is a particular reason for why we want a cubic error bound
- we will explain this at the end.)

This bound follows by considering the series expansion for $e^{x}$
and compactness of $\left[-1,1\right]\times\left[-\pi,\pi\right]$.
Motivated by this, we define the operator $F_{t}$ for $t\in\left[0,1\right]$
by
\begin{equation}
F_{t}=t\mleft(U-1\mright)-\frac{t\mleft(1-t\mright)}{2}\mleft(U+U^{\ast}-2\mright).
\end{equation}
We then have the following:
\begin{prop}
For any $T:V\rightarrow V$, $x\in V$, $m\in\left\{ 1,2\right\} $
and $t\in\left[0,1\right]$ it holds that
\[
\left\Vert T\mleft(e^{tJ}-1-F_{t}\mright)x\right\Vert ,\,\left\Vert T\mleft(e^{-tJ}-1-F_{t}^{\ast}\mright)x\right\Vert \leq C\left\Vert T\mleft(U-1\mright)^{m}\right\Vert _{\mathrm{HS}}\left\Vert \mleft(U-1\mright)^{3-m}x\right\Vert 
\]
and for all $1\leq i,j\leq n$
\[
\left|\left\langle x_{i},F_{t}x_{j}\right\rangle \right|,\,\left|\left\langle x_{i},F_{t}^{\ast}x_{j}\right\rangle \right|\leq C\mleft(1+\left\langle v,h^{-1}v\right\rangle \mright)\frac{\left\langle x_{i},v\right\rangle \left\langle v,x_{j}\right\rangle }{\lambda_{i}+\lambda_{j}}
\]
for a constant $C>0$ independent of all quantities.
\end{prop}

\textbf{Proof:} It suffices to consider $\left\Vert T\mleft(e^{tJ}-1-F_{t}\mright)x\right\Vert $.
By orthonormal expansion using the basis $\mleft(w_{j}\mright)_{j=1}^{n}$,
it holds by equation (\ref{eq:CubicExponentialBound}) and Cauchy-Schwarz
that
\begin{align}
 & \;\left\Vert T\mleft(e^{tJ}-1-F_{t}\mright)x\right\Vert ^{2}=\sum_{j=1}^{n}\left|\left\langle w_{j},T\mleft(e^{tJ}-1-F_{t}\mright)x\right\rangle \right|^{2}=\sum_{j=1}^{n}\left|\sum_{k=1}^{n}\left\langle w_{j},T\mleft(e^{tJ}-1-F_{t}\mright)w_{k}\right\rangle \left\langle w_{k},x\right\rangle \right|^{2}\nonumber \\
 & =\sum_{j=1}^{n}\left|\sum_{k=1}^{n}\mleft(\mleft(e^{it\theta_{k}}-1\mright)-t\mleft(e^{i\theta_{k}}-1\mright)+\frac{t\mleft(1-t\mright)}{2}\mleft(e^{i\theta_{k}}+e^{-i\theta_{k}}-2\mright)\mright)\left\langle w_{j},Tw_{k}\right\rangle \left\langle w_{k},x\right\rangle \right|^{2}\nonumber \\
 & \leq C\sum_{j=1}^{n}\mleft(\sum_{k=1}^{n}\left|e^{i\theta_{k}}-1\right|^{3}\left|\left\langle w_{j},Tw_{k}\right\rangle \right|\left|\left\langle w_{k},x\right\rangle \right|\mright)^{2}\\
 & \leq C\sum_{j=1}^{n}\mleft(\sum_{k=1}^{n}\left|e^{i\theta_{k}}-1\right|^{2m}\left|\left\langle w_{j},Tw_{k}\right\rangle \right|^{2}\mright)\mleft(\sum_{k=1}^{n}\left|e^{i\theta_{k}}-1\right|^{2\mleft(3-m\mright)}\left|\left\langle w_{k},x\right\rangle \right|^{2}\mright)\nonumber \\
 & =C\mleft(\sum_{j,k=1}^{n}\left|\left\langle w_{j},T\mleft(U-1\mright)^{m}w_{k}\right\rangle \right|^{2}\mright)\mleft(\sum_{k=1}^{n}\left|\left\langle w_{k},\mleft(U-1\mright)^{3-m}x\right\rangle \right|^{2}\mright)\nonumber \\
 & =C\left\Vert T\mleft(U-1\mright)^{m}\right\Vert _{\mathrm{HS}}^{2}\left\Vert \mleft(U-1\mright)^{3-m}x\right\Vert ^{2}\nonumber 
\end{align}
which implies the first claim. The elementwise estimates for $F_{t}$
and $F_{t}^{\ast}$ follow immediately from Proposition \ref{prop:UElementEstimates}.

$\hfill\square$

By the proposition we then have that
\begin{align}
\Vert h^{-\frac{1}{2}}\mleft(e^{tJ}-1\mright)hx_{i}\Vert & \leq\Vert h^{-\frac{1}{2}}F_{t}hx_{i}\Vert+\Vert h^{-\frac{1}{2}}\mleft(e^{tJ}-1-F_{t}\mright)hx_{i}\Vert\\
 & \leq\Vert h^{-\frac{1}{2}}F_{t}hx_{i}\Vert+C\Vert h^{-\frac{1}{2}}\mleft(U-1\mright)^{2}\Vert_{\mathrm{HS}}\left\Vert \mleft(U-1\mright)hx_{i}\right\Vert \nonumber 
\end{align}
and so have reduced the estimation to operators which we have good
control over. We can estimate that
\begin{align}
 & \quad\;\,\Vert h^{-\frac{1}{2}}F_{t}hx_{i}\Vert^{2}=\sum_{j=1}^{n}\left|\left\langle x_{j},h^{-\frac{1}{2}}F_{t}hx_{i}\right\rangle \right|^{2}=\sum_{j=1}^{n}\frac{\lambda_{i}^{2}}{\lambda_{j}}\left|\left\langle x_{j},F_{t}x_{i}\right\rangle \right|^{2}\\
 & \leq C\mleft(1+\left\langle v,h^{-1}v\right\rangle \mright)^{2}\sum_{j=1}^{n}\frac{\lambda_{i}^{2}}{\lambda_{j}}\left|\frac{\left\langle x_{i},v\right\rangle \left\langle v,x_{j}\right\rangle }{\lambda_{i}+\lambda_{j}}\right|^{2}\leq C\mleft(1+\left\langle v,h^{-1}v\right\rangle \mright)^{2}\left\langle v,h^{-1}v\right\rangle \left|\left\langle x_{i},v\right\rangle \right|^{2}\nonumber 
\end{align}
and likewise
\begin{align}
\left\Vert \mleft(U-1\mright)hx_{i}\right\Vert ^{2} & =\sum_{j=1}^{n}\lambda_{i}^{2}\left|\left\langle x_{j},\mleft(U-1\mright)x_{i}\right\rangle \right|^{2}\leq C\mleft(1+\left\langle v,h^{-1}v\right\rangle \mright)^{2}\sum_{j=1}^{n}\lambda_{i}^{2}\left|\frac{\left\langle x_{i},v\right\rangle \left\langle v,x_{j}\right\rangle }{\lambda_{i}+\lambda_{j}}\right|^{2}\\
 & \leq C\mleft(1+\left\langle v,h^{-1}v\right\rangle \mright)^{2}\left\Vert v\right\Vert ^{2}\left|\left\langle x_{i},v\right\rangle \right|^{2},\nonumber 
\end{align}
while
\begin{align}
\Vert h^{-\frac{1}{2}}\mleft(U-1\mright)^{2}\Vert_{\mathrm{HS}}^{2} & =\sum_{i,j=1}^{n}\left|\left\langle x_{i},h^{-\frac{1}{2}}\mleft(U-1\mright)^{2}x_{j}\right\rangle \right|^{2}=\sum_{i,j=1}^{n}\frac{1}{\lambda_{i}}\left|\sum_{k=1}^{n}\left\langle x_{i},\mleft(U-1\mright)x_{k}\right\rangle \left\langle x_{k},\mleft(U-1\mright)x_{j}\right\rangle \right|^{2}\nonumber \\
 & \leq C\mleft(1+\left\langle v,h^{-1}v\right\rangle \mright)^{4}\sum_{i,j=1}^{n}\frac{1}{\lambda_{i}}\left|\sum_{k=1}^{n}\frac{\left\langle x_{i},v\right\rangle \left\langle v,x_{k}\right\rangle }{\lambda_{i}+\lambda_{k}}\frac{\left\langle x_{k},v\right\rangle \left\langle v,x_{j}\right\rangle }{\lambda_{k}+\lambda_{j}}\right|^{2}\\
 & \leq C\mleft(1+\left\langle v,h^{-1}v\right\rangle \mright)^{4}\sum_{i,j=1}^{n}\frac{\left|\left\langle x_{i},v\right\rangle \right|^{2}}{\lambda_{i}^{\frac{5}{4}}}\frac{\left|\left\langle v,x_{j}\right\rangle \right|^{2}}{\lambda_{j}^{\frac{5}{4}}}\mleft(\sum_{k=1}^{n}\frac{\left|\left\langle x_{k},v\right\rangle \right|^{2}}{\lambda_{k}^{\frac{7}{8}}\lambda_{k}^{\frac{3}{8}}}\mright)^{2}\nonumber \\
 & =C\mleft(1+\left\langle v,h^{-1}v\right\rangle \mright)^{4}\left\langle v,h^{-\frac{5}{4}}v\right\rangle ^{4},\nonumber 
\end{align}
so in all
\begin{equation}
\Vert h^{-\frac{1}{2}}\mleft(e^{tJ}-1\mright)hx_{i}\Vert\leq C\mleft(1+\left\langle v,h^{-1}v\right\rangle \mright)^{3}\mleft(\sqrt{\left\langle v,h^{-1}v\right\rangle }+\left\Vert v_{k}\right\Vert \left\langle v,h^{-\frac{5}{4}}v\right\rangle ^{2}\mright)\left|\left\langle x_{i},v\right\rangle \right|.
\end{equation}
For the particular operators $h_{k}$ and $P_{v_{k}}$, this implies
that
\begin{equation}
\max_{p\in L_{k}}\Vert h_{k}^{-\frac{1}{2}}\mleft(e^{tJ_{k}}-1\mright)h_{k}e_{p}\Vert\leq Ck_{F}^{-\frac{1}{2}}\mleft(1+\hat{V}_{k}\mright)^{3}\hat{V}_{k}\mleft(1+k_{F}^{\frac{1}{2}}\left|k\right|^{\frac{1}{2}}\left\langle v_{k},h_{k}^{-\frac{5}{4}}v_{k}\right\rangle ^{2}\mright).
\end{equation}
The inner product is
\begin{equation}
\left\langle v_{k},h_{k}^{-\frac{5}{4}}v_{k}\right\rangle =\frac{\hat{V}_{k}k_{F}^{-1}}{2\,\mleft(2\pi\mright)^{3}}\sum_{p\in L_{k}}\lambda_{k,p}^{-\frac{5}{4}}
\end{equation}
and this Riemann sum is more singular than what we consider in appendix
section \ref{sec:RiemannSumEstimates}. Nonetheless, the methods used
therein - in particular, the summation formula of Proposition \ref{prop:RiemannSummationFormula}
- implies the following:
\begin{prop}
For all $k\in\overline{B}\mleft(0,k_{F}^{\gamma}\mright)\cap\mathbb{Z}_{\ast}^{3}$,
$0<\gamma<\frac{1}{47}$, it holds that
\[
\sum_{p\in L_{k}}\lambda_{k,p}^{-\frac{5}{4}}\leq Ck_{F}^{\frac{3}{4}}\left|k\right|^{-\frac{1}{4}}
\]
for a constant $C>0$ depending only on $\gamma$.
\end{prop}

With this we arrive at
\begin{equation}
\max_{p\in L_{k}}\Vert h_{k}^{-\frac{1}{2}}\mleft(e^{tJ_{k}}-1\mright)h_{k}e_{p}\Vert\leq Ck_{F}^{-\frac{1}{2}}\mleft(1+\hat{V}_{k}\mright)^{5}\hat{V}_{k}
\end{equation}
provided $\gamma<\frac{1}{47}$, which is sufficient for the purposes
of Proposition \ref{prop:SecondTransformationExchangeTermEstimate}.

Finally, regarding the bound of (\ref{eq:CubicExponentialBound}),
it likewise holds that
\begin{equation}
\left|\mleft(e^{it\theta}-1\mright)-t\mleft(e^{i\theta}-1\mright)\right|\leq C\left|e^{i\theta}-1\right|^{2}
\end{equation}
and so, considering simply $F_{t}^{\prime}=t\mleft(U-1\mright)$, that
e.g.
\begin{equation}
\Vert h^{-\frac{1}{2}}\mleft(e^{tJ}-1-F_{t}^{\prime}\mright)hx_{i}\Vert\leq C\Vert h^{-\frac{1}{2}}\mleft(U-1\mright)\Vert_{\mathrm{HS}}\left\Vert \mleft(U-1\mright)hx_{i}\right\Vert .
\end{equation}
The issue with this lies in the fact that we now have to deal with
$\Vert h^{-\frac{1}{2}}\mleft(U-1\mright)\Vert_{\mathrm{HS}}$ instead
of $\Vert h^{-\frac{1}{2}}\mleft(U-1\mright)^{2}\Vert_{\mathrm{HS}}$;
this can be estimated similarly, but with the result
\begin{equation}
\Vert h^{-\frac{1}{2}}\mleft(U-1\mright)\Vert_{\mathrm{HS}}\leq C\mleft(1+\left\langle v,h^{-1}v\right\rangle \mright)^{2}\left\langle v,h^{-\frac{3}{2}}v\right\rangle .
\end{equation}
Formally - i.e. if one replaces the Riemann sums with integrals -
it is true that $\left\langle v_{k},h_{k}^{-\frac{3}{2}}v_{k}\right\rangle \sim\left\langle v_{k},h_{k}^{-\frac{5}{4}}v_{k}\right\rangle ^{2}$
with respect to $k_{F}$, and so there should not be a difference.
The result of appendix section \ref{sec:RiemannSumEstimates} however
only extends (optimally) to Riemann sums of the form $\sum_{p\in L_{k}}\lambda_{k,p}^{\beta}$
for $\beta>-\frac{4}{3}$, and so $\left\langle v_{k},h_{k}^{-\frac{3}{2}}v_{k}\right\rangle $
it outside the range which we are able to control, even with a cut-off
in $k$.

\section{\label{sec:PlasmonModesoftheEffectiveHamiltonian}Plasmon Modes of
the Effective Hamiltonian}

In this final section we consider the effective operator
\begin{equation}
H_{\mathrm{eff}}=H_{\mathrm{kin}}^{\prime}+2\sum_{k\in\mathbb{Z}_{\ast}^{3}}Q_{k}^{1}\mleft(\tilde{E}_{k}-h_{k}\mright)=H_{\mathrm{kin}}^{\prime}+2\sum_{k\in\mathbb{Z}_{\ast}^{3}}\sum_{p\in L_{k}}b_{k}^{\ast}\mleft(\mleft(\tilde{E}_{k}-h_{k}\mright)e_{p}\mright)b_{k,p},
\end{equation}
where $\tilde{E}_{k}=(h_{k}^{2}+2P_{h_{k}^{\frac{1}{2}}v_{k}})^{\frac{1}{2}}$,
in detail. As we will consider $H_{\mathrm{eff}}$ in isolation from
the proper Hamiltonian, we will now omit the mean-field scaling factor
$k_{F}^{-1}$ - concretely this means that $v_{k}\in\ell^{2}\mleft(L_{k}\mright)$
is now given by
\begin{equation}
v_{k}=\frac{s\hat{V}_{k}}{2\,\mleft(2\pi\mright)^{3}}\sum_{p\in L_{k}}e_{p}.
\end{equation}
For this section we will fix a $k\in B_{F}$, let $\phi\in\ell^{2}\mleft(L_{k}\mright)$
denote the normalized eigenvector of $2\tilde{E}_{k}$ corresponding
to the greatest eigenvalue $\epsilon_{k}$, and define $\Psi_{M}\in\left\{ \mathcal{N}_{E}=M\right\} $
by
\begin{equation}
\Psi_{M}=b_{k}^{\ast}\mleft(\phi\mright)^{M}\psi_{F},\quad M\in\mathbb{N}_{0}.
\end{equation}

(For the statements of certain propositions below we will understand
$\Psi_{-1},\Psi_{-2}=0$.)

The main result of this section is the following bound for $\hat{\Psi}_{M}$:
\begin{thm}
\label{them:GeneralPlasmonStates}There exists constants $c,C>0$
such that if $\hat{V}_{k}>ck_{F}^{-1}$ it holds for all $M\leq Ck_{F}^{2}\left|k\right|$
that $\hat{\Psi}_{M}=\left\Vert \Psi_{M}\right\Vert ^{-1}\Psi_{M}$
obeys
\[
\Vert\mleft(H_{\mathrm{eff}}-M\epsilon_{k}\mright)\hat{\Psi}_{M}\Vert\leq C'\sqrt{\sum_{l\in\mathbb{Z}_{\ast}^{3}}\min\left\{ 1,k_{F}\hat{V}_{l},k_{F}^{3}\hat{V}_{l}\left|l\right|^{-2}\right\} \hat{V}_{l}\left|l\right|^{2}}\frac{M^{\frac{5}{2}}}{\sqrt{k_{F}}\left|k\right|},\quad k_{F}\rightarrow\infty,
\]
where $\epsilon_{k}$ denotes the greatest eigenvalue of $2\tilde{E}_{k}$,
which obeys $\epsilon_{k}\geq c's^{\frac{1}{2}}k_{F}^{\frac{3}{2}}\left|k\right|\hat{V}_{k}^{\frac{1}{2}}$
and
\[
0\leq\epsilon_{k}-2\sqrt{2\left\langle v_{k},h_{k}v_{k}\right\rangle +\frac{\left\langle v_{k},h_{k}^{3}v_{k}\right\rangle }{\left\langle v_{k},h_{k}v_{k}\right\rangle }}\leq C'k_{F}^{-\frac{1}{2}}\left|k\right|\hat{V}_{k}^{-\frac{3}{2}},\quad k_{F}\rightarrow\infty,
\]
for constants $c',C'>0$. The constants $c,c',C,C'$ are independent
of all quantities.
\end{thm}

Note that Theorem \ref{them:PlasmonStates} is an immediate consequence
of this result: For $\hat{V}_{k}=g\left|k\right|^{-2}$, the condition
$\hat{V}_{k}>ck_{F}^{-1}$ becomes
\begin{equation}
\left|k\right|<\sqrt{\frac{g}{c}k_{F}}
\end{equation}
which is ensured for all $k\in\overline{B}\mleft(0,k_{F}^{\delta}\mright)\cap\mathbb{Z}_{\ast}^{3}$
for $k_{F}$ suffiently large provided $\delta\in\mleft(0,\frac{1}{2}\mright)$.
That $M\leq k_{F}^{\varepsilon}$ for $\varepsilon\in\mleft(0,2\mright)$
similarly ensures that $M\leq Ck_{F}^{2}\left|k\right|$ for $k_{F}$
sufficiently large, so the conditions of the theorem hold, and the
sum can be estimated as
\begin{align}
 & \;\;\;\sum_{l\in\mathbb{Z}_{\ast}^{3}}\min\left\{ 1,k_{F}\hat{V}_{l},k_{F}^{3}\hat{V}_{l}\left|l\right|^{-2}\right\} \hat{V}_{l}\left|l\right|^{2}\leq\max\left\{ 1,g\right\} \sum_{l\in\mathbb{Z}_{\ast}^{3}}\min\left\{ 1,k_{F}\left|l\right|^{-2},k_{F}^{3}\left|l\right|^{-4}\right\} \nonumber \\
 & \leq C\mleft(\sum_{l\in\overline{B}\mleft(0,\sqrt{k_{F}}\mright)\cap\mathbb{Z}_{\ast}^{3}}1+k_{F}\sum_{l\in B_{F}\backslash\overline{B}\mleft(0,\sqrt{k_{F}}\mright)}\left|l\right|^{-2}+k_{F}^{3}\sum_{l\in\mathbb{Z}_{\ast}^{3}\backslash B_{F}}\left|l\right|^{-4}\mright)\\
 & \leq C\mleft(\mleft(\sqrt{k_{F}}\mright)^{3}+k_{F}^{2}+k_{F}^{2}\mright)\leq Ck_{F}^{2}.\nonumber 
\end{align}
The statement regarding $\epsilon_{k}$ follows by expanding the inner
products $\left\langle v_{k},h_{k}^{\beta}v_{k}\right\rangle $ and
inserting $\hat{V}_{k}=g\left|k\right|^{-2}$.

\subsection{Properties of the Plasmon State $\Psi_{M}$}

Owing to the inequality (which in the exact bosonic case would be
an equality)
\begin{equation}
\left\Vert \Psi_{M}\right\Vert ^{2}=\left\Vert b_{k}^{\ast}\mleft(\phi\mright)\Psi_{M-1}\right\Vert ^{2}\leq\Vert\mleft(\mathcal{N}_{E}+1\mright)^{\frac{1}{2}}\Psi_{M-1}\Vert^{2}=M\left\Vert \Psi_{M-1}\right\Vert ^{2}
\end{equation}
we can control the ratio $\left\Vert \Psi_{M}\right\Vert ^{-1}\left\Vert \Psi_{M-1}\right\Vert $
well from below, but for the purposes of Theorem \ref{them:GeneralPlasmonStates}
it is an upper bound which will be needed. To that end we begin by
noting the following:
\begin{lem}
\label{lemma:PsiMComputation}For any $p\in B_{F}^{c}$, $q\in B_{F}$,
$1\leq\sigma\leq s$ and $M\in\mathbb{N}$ it holds that
\begin{align*}
c_{p,\sigma}\Psi_{M} & =1_{L_{k}}\mleft(p\mright)Ms^{-\frac{1}{2}}\left\langle e_{p},\phi\right\rangle c_{p-k,\sigma}\Psi_{M-1}\\
c_{q,\sigma}^{\ast}\Psi_{M} & =-1_{L_{k}}\mleft(q+k\mright)Ms^{-\frac{1}{2}}\left\langle e_{q+k},\phi\right\rangle c_{q+k,\sigma}^{\ast}\Psi_{M-1}.
\end{align*}
As a consequence it holds for any $l\in\mathbb{Z}_{\ast}^{3}$ and
$p\in L_{l}$ that
\[
b_{l,p}\Psi_{M}=\delta_{k,l}M\left\langle e_{p},\phi\right\rangle \Psi_{M-1}+1_{L_{k}}\mleft(p\mright)\frac{M\mleft(M-1\mright)}{s^{\frac{3}{2}}}\sum_{q\in L_{k}}^{\sigma}\delta_{p-l,q-k}\left\langle e_{p},\phi\right\rangle \left\langle e_{q},\phi\right\rangle c_{q,\sigma}^{\ast}c_{p-k,\sigma}\Psi_{M-2}.
\]
\end{lem}

\textbf{Proof:} By equation (\ref{eq:bcastCommutator}) we have
\begin{align}
\left[c_{p,\sigma},b_{k}^{\ast}\mleft(\phi\mright)\right] & =1_{L_{k}}\mleft(p\mright)s^{-\frac{1}{2}}\left\langle e_{p},\phi\right\rangle c_{p-k,\sigma}\\
\left[c_{q,\sigma}^{\ast},b_{k}^{\ast}\mleft(\phi\mright)\right] & =-1_{L_{k}}\mleft(q+k\mright)s^{-\frac{1}{2}}\left\langle e_{q+k},\phi\right\rangle c_{q+k,\sigma}^{\ast},\nonumber 
\end{align}
so
\begin{align}
c_{p,\sigma}\Psi_{M} & =c_{p,\sigma}b_{k}^{\ast}\mleft(\phi\mright)^{M}\psi_{F}=b_{k}^{\ast}\mleft(\phi\mright)^{M}c_{p,\sigma}\psi_{F}+\sum_{j=0}^{M-1}b_{k}^{\ast}\mleft(\phi\mright)^{j}\left[c_{p,\sigma},b_{k}^{\ast}\mleft(\phi\mright)\right]b_{k}^{\ast}\mleft(\phi\mright)^{M-j-1}\psi_{F}\nonumber \\
 & =\sum_{j=0}^{M-1}b_{k}^{\ast}\mleft(\phi\mright)^{j}\mleft(1_{L_{k}}\mleft(p\mright)s^{-\frac{1}{2}}\left\langle e_{p},\phi\right\rangle c_{p-k,\sigma}\mright)b_{k}^{\ast}\mleft(\phi\mright)^{M-j-1}\psi_{F}\\
 & =1_{L_{k}}\mleft(p\mright)Ms^{-\frac{1}{2}}\left\langle e_{p},\phi\right\rangle c_{p-k,\sigma}b_{k}^{\ast}\mleft(\phi\mright)^{M-1}\psi_{F}=1_{L_{k}}\mleft(p\mright)Ms^{-\frac{1}{2}}\left\langle e_{p},\phi\right\rangle c_{p-k,\sigma}\Psi_{M-1}\nonumber 
\end{align}
and likewise for $c_{q,\sigma}^{\ast}\Psi_{M}$, $q\in B_{F}$. The
expression for $b_{l,p}\Psi_{M}$ then follows as
\begin{align}
b_{l,p}\Psi_{M} & =\frac{1}{\sqrt{s}}\sum_{\sigma=1}^{s}c_{p-l,\sigma}^{\ast}c_{p,\sigma}\Psi_{M}=\frac{M}{s}\sum_{\sigma=1}^{s}1_{L_{k}}\mleft(p\mright)\left\langle e_{p},\phi\right\rangle c_{p-l,\sigma}^{\ast}c_{p-k,\sigma}\Psi_{M-1}\nonumber \\
 & =\delta_{k,l}1_{L_{k}}\mleft(p\mright)\frac{M}{s}\sum_{\sigma=1}^{s}\left\langle e_{p},\phi\right\rangle \Psi_{M-1}-1_{L_{k}}\mleft(p\mright)\frac{M}{s}\sum_{\sigma=1}^{s}\left\langle e_{p},\phi\right\rangle c_{p-k,\sigma}c_{p-l,\sigma}^{\ast}\Psi_{M-1}\\
 & =\delta_{k,l}M\left\langle e_{p},\phi\right\rangle \Psi_{M-1}-1_{L_{k}}\mleft(p\mright)\frac{M\mleft(M-1\mright)}{s^{\frac{3}{2}}}\sum_{\sigma=1}^{s}1_{L_{k}}\mleft(p-l+k\mright)\left\langle e_{p},\phi\right\rangle \left\langle e_{p-l+k},\phi\right\rangle c_{p-k,\sigma}c_{p-l+k,\sigma}^{\ast}\Psi_{M-2}\nonumber \\
 & =\delta_{k,l}M\left\langle e_{p},\phi\right\rangle \Psi_{M-1}+1_{L_{k}}\mleft(p\mright)\frac{M\mleft(M-1\mright)}{s^{\frac{3}{2}}}\sum_{q\in L_{k}}^{\sigma}\delta_{p-l,q-k}\left\langle e_{p},\phi\right\rangle \left\langle e_{q},\phi\right\rangle c_{q,\sigma}^{\ast}c_{p-k,\sigma}\Psi_{M-2}\nonumber 
\end{align}
where we used the identity $1_{L_{k}}\mleft(p-l+k\mright)f\mleft(p-l+k\mright)=\sum_{q\in L_{k}}\delta_{p-l,q-k}f\mleft(q\mright)$
to rewrite the second term.

$\hfill\square$

This implies the following bound:
\begin{cor}
\label{coro:RelativeNormLowerBound}For any $M\in\mathbb{N}$ it holds
that
\[
\left\Vert \Psi_{M}\right\Vert ^{2}\geq M\mleft(1-\frac{M-1}{s}\left\Vert \phi\right\Vert _{\infty}^{2}\mright)\left\Vert \Psi_{M-1}\right\Vert ^{2}
\]
where $\left\Vert \phi\right\Vert _{\infty}=\sup_{p\in L_{k}}\left|\left\langle e_{p},\phi\right\rangle \right|$.
\end{cor}

\textbf{Proof:} We estimate
\begin{align}
\left\Vert \Psi_{M}\right\Vert ^{2} & =\left\langle \Psi_{M-1},b\mleft(\phi\mright)\Psi_{M}\right\rangle =\frac{1}{\sqrt{s}}\sum_{p\in L_{k}}^{\sigma}\left\langle \phi,e_{p}\right\rangle \left\langle \Psi_{M-1},c_{p-k,\sigma}^{\ast}c_{p,\sigma}\Psi_{M}\right\rangle \nonumber \\
 & =\frac{M}{s}\sum_{p\in L_{k}}^{\sigma}\left|\left\langle e_{p},\phi\right\rangle \right|^{2}\left\langle \Psi_{M-1},c_{p-k,\sigma}^{\ast}c_{p-k,\sigma}\Psi_{M-1}\right\rangle \\
 & =\frac{M}{s}\sum_{p\in L_{k}}^{\sigma}\left|\left\langle e_{p},\phi\right\rangle \right|^{2}\left\Vert \Psi_{M-1}\right\Vert ^{2}-\frac{M}{s}\sum_{p\in L_{k}}^{\sigma}\left|\left\langle e_{p},\phi\right\rangle \right|^{2}\left\langle \Psi_{M-1},c_{p-k,\sigma}c_{p-k,\sigma}^{\ast}\Psi_{M-1}\right\rangle \nonumber \\
 & \geq M\left\Vert \Psi_{M-1}\right\Vert ^{2}-\frac{M}{s}\left\Vert \phi\right\Vert _{\infty}^{2}\left\langle \Psi_{M-1},\mathcal{N}_{E}\Psi_{M-1}\right\rangle =M\mleft(1-\frac{M-1}{s}\left\Vert \phi\right\Vert _{\infty}^{2}\mright)\left\Vert \Psi_{M-1}\right\Vert ^{2}\nonumber 
\end{align}
where we used that $\mathcal{N}_{E}\Psi_{M-1}=\mleft(M-1\mright)\Psi_{M-1}$.

$\hfill\square$

Note that this bound actually applies to all (normalized) $\varphi\in\ell^{2}\mleft(L_{k}\mright)$
in the form
\begin{equation}
\Vert b_{k}^{\ast}\mleft(\varphi\mright)^{M}\psi_{F}\Vert^{2}\geq M\mleft(1-\frac{M-1}{s}\left\Vert \varphi\right\Vert _{\infty}^{2}\mright)\Vert b_{k}^{\ast}\mleft(\varphi\mright)^{M-1}\psi_{F}\Vert^{2}
\end{equation}
- this is even optimal, with equality holding for all $\varphi$ which
are uniformly supported on some $S\subset L_{k}$ in the sense that
\begin{equation}
\left|\left\langle e_{p},\varphi\right\rangle \right|=\begin{cases}
\left|S\right|^{-\frac{1}{2}} & p\in S\\
0 & p\in L_{k}\backslash S
\end{cases}.
\end{equation}
Although $\phi$ is not uniformly supported, we will see below that
it is ``almost completely delocalized'' as
\begin{equation}
\left\Vert \phi\right\Vert _{\infty}\leq C\left|L_{k}\right|^{-\frac{1}{2}},
\end{equation}
so the corollary and the inequality $\left\Vert \Psi_{M}\right\Vert ^{2}\leq M\left\Vert \Psi_{M-1}\right\Vert ^{2}$
implies that
\begin{equation}
1\leq\frac{M\left\Vert \Psi_{M-1}\right\Vert ^{2}}{\left\Vert \Psi_{M}\right\Vert ^{2}}\leq\frac{1}{1-C\frac{M}{s\left|L_{k}\right|}}\leq1+C'\frac{M}{\left|L_{k}\right|},\quad M\ll\left|L_{k}\right|,
\end{equation}
i.e. $M\left\Vert \Psi_{M}\right\Vert ^{-2}\left\Vert \Psi_{M-1}\right\Vert ^{2}\sim1$
for all $M\ll\left|L_{k}\right|\sim O\mleft(k_{F}^{2}\left|k\right|\mright)$.

\subsubsection*{The Action of $H_{\mathrm{eff}}$ on $\Psi_{M}$}

Having established control on the state $\Psi_{M}$ itself we now
turn to the action of $H_{\mathrm{eff}}$ upon it:
\begin{prop}
\label{prop:AlmostEigenstateIdentity}For all $M\in\mathbb{N}$ it
holds that
\[
\left\Vert \mleft(H_{\mathrm{eff}}-M\epsilon_{k}\mright)\Psi_{M}\right\Vert =\frac{2M\mleft(M-1\mright)}{s^{\frac{3}{2}}}\left\Vert \mathcal{E}\Psi_{M-2}\right\Vert 
\]
where $\mathcal{E}:\mathcal{H}_{N}\rightarrow\mathcal{H}_{N}$ is
given by
\[
\mathcal{E}=\sum_{p,q\in L_{k}}^{\sigma}\left\langle e_{p},\phi\right\rangle \left\langle e_{q},\phi\right\rangle \mleft(\sum_{l\in\mathbb{Z}_{\ast}^{3}}\delta_{p-l,q-k}1_{L_{l}}\mleft(p\mright)b_{l}^{\ast}\mleft(\mleft(\tilde{E}_{l}-h_{l}\mright)e_{p}\mright)\mright)c_{q,\sigma}^{\ast}c_{p-k,\sigma}.
\]
\end{prop}

\textbf{Proof:} By the commutation relation $\left[H_{\mathrm{kin}}^{\prime},b_{k}^{\ast}\mleft(\phi\mright)\right]=2b_{k}^{\ast}\mleft(h_{k}\phi\mright)$
it follows as $H_{\mathrm{kin}}^{\prime}\psi_{F}=0$ that
\begin{equation}
H_{\mathrm{kin}}^{\prime}\Psi_{M}=Mb_{k}^{\ast}\mleft(2h_{k}\phi\mright)\Psi_{M-1},
\end{equation}
so applying Lemma \ref{lemma:PsiMComputation} we find
\begin{align}
H_{\mathrm{eff}}\Psi_{M} & =H_{\mathrm{kin}}^{\prime}\Psi_{M}+2\sum_{l\in\mathbb{Z}_{\ast}^{3}}\sum_{p\in L_{l}}b_{l}^{\ast}\mleft(\mleft(\tilde{E}_{l}-h_{l}\mright)e_{p}\mright)b_{l,p}\Psi_{M}\nonumber \\
 & =Mb_{k}^{\ast}\mleft(2h_{k}\phi\mright)\Psi_{M-1}+2M\sum_{l\in\mathbb{Z}_{\ast}^{3}}\sum_{p\in L_{l}}\delta_{k,l}b_{l}^{\ast}\mleft(\mleft(\tilde{E}_{l}-h_{l}\mright)e_{p}\mright)\left\langle e_{p},\phi\right\rangle \Psi_{M-1}\nonumber \\
 & +\frac{2M\mleft(M-1\mright)}{s^{\frac{3}{2}}}\sum_{l\in\mathbb{Z}_{\ast}^{3}}\sum_{p\in L_{k}\cap L_{l}}\sum_{q\in L_{k}}^{\sigma}\delta_{p-l,q-k}\left\langle e_{p},\phi\right\rangle \left\langle e_{q},\phi\right\rangle b_{l}^{\ast}\mleft(\mleft(\tilde{E}_{l}-h_{l}\mright)e_{p}\mright)c_{q,\sigma}^{\ast}c_{p-k,\sigma}\Psi_{M-2}\\
 & =Mb_{k}^{\ast}\mleft(2h_{k}\phi\mright)\Psi_{M-1}+Mb_{k}^{\ast}\mleft(2\mleft(\tilde{E}_{k}-h_{k}\mright)\phi\mright)\Psi_{M-1}\nonumber \\
 & +\frac{2M\mleft(M-1\mright)}{s^{\frac{3}{2}}}\sum_{p,q\in L_{k}}^{\sigma}\left\langle e_{p},\phi\right\rangle \left\langle e_{q},\phi\right\rangle \mleft(\sum_{l\in\mathbb{Z}_{\ast}^{3}}\delta_{p-l,q-k}1_{L_{l}}\mleft(p\mright)b_{l}^{\ast}\mleft(\mleft(\tilde{E}_{l}-h_{l}\mright)e_{p}\mright)\mright)c_{q,\sigma}^{\ast}c_{p-k,\sigma}\Psi_{M-2}\nonumber \\
 & =Mb_{k}^{\ast}\mleft(2\tilde{E}_{k}\phi\mright)\Psi_{M-1}+\frac{2M\mleft(M-1\mright)}{s^{\frac{3}{2}}}\mathcal{E}\Psi_{M-2}.\nonumber 
\end{align}
By our choice of $\phi$ the claim now follows as
\begin{equation}
Mb_{k}^{\ast}\mleft(2\tilde{E}_{k}\phi\mright)\Psi_{M-1}=M\epsilon_{k}b_{k}^{\ast}\mleft(\phi\mright)\Psi_{M-1}=M\epsilon_{k}\Psi_{M}.
\end{equation}
$\hfill\square$

To bound $\left\Vert \mathcal{E}\Psi_{M-2}\right\Vert $ we note the
following generalization of Proposition \ref{prop:bbastEstimates}:
\begin{prop}
For any collection of vectors $\varphi_{k}\in\ell^{2}\mleft(L_{k}\mright)$,
$k\in\mathbb{Z}_{\ast}^{3}$, with $\sum_{k\in\mathbb{Z}_{\ast}^{3}}\left\Vert \varphi_{k}\right\Vert ^{2}<\infty$
it holds for all $\Psi\in\mathcal{H}_{N}$ that
\[
\left\Vert \sum_{k\in\mathbb{Z}_{\ast}^{3}}b_{k}\mleft(\varphi_{k}\mright)\Psi\right\Vert \leq\sqrt{\sum_{k\in\mathbb{Z}_{\ast}^{3}}\left\Vert \varphi_{k}\right\Vert ^{2}}\Vert\mathcal{N}_{E}^{\frac{1}{2}}\Psi\Vert,\quad\left\Vert \sum_{k\in\mathbb{Z}_{\ast}^{3}}b_{k}^{\ast}\mleft(\varphi_{k}\mright)\Psi\right\Vert \leq\sqrt{\sum_{k\in\mathbb{Z}_{\ast}^{3}}\left\Vert \varphi_{k}\right\Vert ^{2}}\Vert\mleft(\mathcal{N}_{E}+1\mright)^{\frac{1}{2}}\Psi\Vert.
\]
\end{prop}

\textbf{Proof:} By the triangle and Cauchy-Schwarz inequalities and
the usual fermionic estimate we can bound
\begin{align}
\left\Vert \sum_{k\in\mathbb{Z}_{\ast}^{3}}b_{k}\mleft(\varphi_{k}\mright)\Psi\right\Vert  & =\frac{1}{\sqrt{s}}\left\Vert \sum_{k\in\mathbb{Z}_{\ast}^{3}}\sum_{p\in L_{k}}^{\sigma}\left\langle \varphi,e_{p}\right\rangle c_{p-k,\sigma}^{\ast}c_{p,\sigma}\Psi\right\Vert =\frac{1}{\sqrt{s}}\left\Vert \sum_{p\in B_{F}^{c}}^{\sigma}\mleft(\sum_{k\in\mathbb{Z}_{\ast}^{3}}1_{L_{k}}\mleft(p\mright)\left\langle \varphi,e_{p}\right\rangle c_{p-k,\sigma}^{\ast}\mright)c_{p,\sigma}\Psi\right\Vert \\
 & \leq\frac{1}{\sqrt{s}}\sum_{p\in B_{F}^{c}}^{\sigma}\left\Vert \mleft(\sum_{k\in\mathbb{Z}_{\ast}^{3}}1_{L_{k}}\mleft(p\mright)\left\langle \varphi,e_{p}\right\rangle c_{p-k,\sigma}^{\ast}\mright)c_{p,\sigma}\Psi\right\Vert \leq\frac{1}{\sqrt{s}}\sum_{p\in B_{F}^{c}}^{\sigma}\sqrt{\sum_{k\in\mathbb{Z}_{\ast}^{3}}1_{L_{k}}\mleft(p\mright)\left|\left\langle \varphi,e_{p}\right\rangle \right|^{2}}\left\Vert c_{p,\sigma}\Psi\right\Vert \nonumber \\
 & \leq\sqrt{\sum_{p\in B_{F}^{c}}\sum_{k\in\mathbb{Z}_{\ast}^{3}}1_{L_{k}}\mleft(p\mright)\left|\left\langle \varphi,e_{p}\right\rangle \right|^{2}}\sqrt{\sum_{p\in B_{F}^{c}}^{\sigma}\left\Vert c_{p,\sigma}\Psi\right\Vert ^{2}}=\sqrt{\sum_{k\in\mathbb{Z}_{\ast}^{3}}\left\Vert \varphi_{k}\right\Vert ^{2}}\Vert\mathcal{N}_{E}^{\frac{1}{2}}\Psi\Vert\nonumber 
\end{align}
for the first claim. The second follows from this, since
\begin{align}
\mleft(\sum_{k\in\mathbb{Z}_{\ast}^{3}}b_{k}\mleft(\varphi_{k}\mright)\mright)\mleft(\sum_{k\in\mathbb{Z}_{\ast}^{3}}b_{k}\mleft(\varphi_{k}\mright)\mright)^{\ast} & =\mleft(\sum_{k\in\mathbb{Z}_{\ast}^{3}}b_{k}\mleft(\varphi_{k}\mright)\mright)^{\ast}\mleft(\sum_{k\in\mathbb{Z}_{\ast}^{3}}b_{k}\mleft(\varphi_{k}\mright)\mright)+\sum_{k\in\mathbb{Z}_{\ast}^{3}}\left\Vert \varphi_{k}\right\Vert ^{2}+\sum_{k,l\in\mathbb{Z}_{\ast}^{3}}\varepsilon_{k,l}\mleft(\varphi_{k};\varphi_{l}\mright)\nonumber \\
 & \leq\mleft(\sum_{k\in\mathbb{Z}_{\ast}^{3}}\left\Vert \varphi_{k}\right\Vert ^{2}\mright)\mleft(\mathcal{N}_{E}+1\mright)+\sum_{k,l\in\mathbb{Z}_{\ast}^{3}}\varepsilon_{k,l}\mleft(\varphi_{k};\varphi_{l}\mright)
\end{align}
and we claim that $\sum_{k,l\in\mathbb{Z}_{\ast}^{3}}\varepsilon_{k,l}\mleft(\varphi_{k};\varphi_{l}\mright)\leq0$.
Indeed, as
\begin{equation}
\varepsilon_{k,l}\mleft(\varphi_{k};\varphi_{l}\mright)=-\frac{1}{s}\sum_{p\in L_{k}}^{\sigma}\sum_{q\in L_{l}}\left\langle \varphi_{k},e_{p}\right\rangle \left\langle e_{q},\varphi_{l}\right\rangle \mleft(\delta_{p,q}c_{q-l,\sigma}c_{p-k,\sigma}^{\ast}+\delta_{p-k,q-l}c_{q,\sigma}^{\ast}c_{p,\sigma}\mright)
\end{equation}
we see that for the sum corresponding to the $\delta_{p,q}c_{q-l,\sigma}c_{p-k,\sigma}^{\ast}$
terms,
\begin{align}
 & \quad\,\sum_{k,l\in\mathbb{Z}_{\ast}^{3}}\sum_{p\in L_{k}}^{\sigma}\sum_{q\in L_{l}}\left\langle \varphi_{k},e_{p}\right\rangle \left\langle e_{q},\varphi_{l}\right\rangle \delta_{p,q}c_{q-l,\sigma}c_{p-k,\sigma}^{\ast}=\sum_{k,l\in\mathbb{Z}_{\ast}^{3}}\sum_{p\in L_{k}\cap L_{l}}^{\sigma}\left\langle \varphi_{k},e_{p}\right\rangle \left\langle e_{p},\varphi_{l}\right\rangle c_{p-l,\sigma}c_{p-k,\sigma}^{\ast}\\
 & =\sum_{p\in B_{F}^{c}}^{\sigma}\mleft(\sum_{l\in\mathbb{Z}_{\ast}^{3}}1_{L_{l}}\mleft(p\mright)\left\langle \varphi_{l},e_{p}\right\rangle c_{p-l,\sigma}^{\ast}\mright)^{\ast}\mleft(\sum_{k\in\mathbb{Z}_{\ast}^{3}}1_{L_{k}}\mleft(p\mright)\left\langle \varphi_{k},e_{p}\right\rangle c_{p-k,\sigma}^{\ast}\mright)\geq0,\nonumber 
\end{align}
and a similar observation applies to the $\delta_{p-k,q-l}c_{q,\sigma}^{\ast}c_{p,\sigma}$
terms.

$\hfill\square$

We can then bound $\left\Vert \mathcal{E}\Psi_{M-2}\right\Vert $
in the following form:
\begin{prop}
For all $M\in\mathbb{N}$ it holds that
\[
\left\Vert \mathcal{E}\Psi_{M-2}\right\Vert \leq M\sqrt{M-1}s^{\frac{1}{2}}\left\Vert \phi\right\Vert _{\infty}^{2}\sqrt{\sum_{l\in\mathbb{Z}_{\ast}^{3}}\Vert\tilde{E}_{l}-h_{l}\Vert_{\mathrm{HS}}^{2}}\left\Vert \Psi_{M-2}\right\Vert .
\]
\end{prop}

\textbf{Proof:} Write $B_{p,q}=\sum_{l\in\mathbb{Z}_{\ast}^{3}}\delta_{p-l,q-k}1_{L_{l}}\mleft(p\mright)b_{l}\mleft(\mleft(\tilde{E}_{l}-h_{l}\mright)e_{p}\mright)$
for brevity, so that
\begin{equation}
\mathcal{E}=\sum_{p,q\in L_{k}}^{\sigma}\left\langle e_{p},\phi\right\rangle \left\langle e_{q},\phi\right\rangle B_{p,q}^{\ast}c_{q,\sigma}^{\ast}c_{p-k,\sigma},
\end{equation}
and note that by the previous proposition, the operators $B_{p,q}^{\ast}$
obey
\begin{align}
\sum_{p,q\in L_{k}}\left\Vert B_{p,q}^{\ast}\Psi_{M-2}\right\Vert ^{2} & \leq\sum_{p,q\in L_{k}}\sum_{l\in\mathbb{Z}_{\ast}^{3}}\delta_{p-l,q-k}1_{L_{l}}\mleft(p\mright)\left\Vert \mleft(\tilde{E}_{l}-h_{l}\mright)e_{p}\right\Vert ^{2}\Vert\mleft(\mathcal{N}_{E}+1\mright)^{\frac{1}{2}}\Psi_{M-2}\Vert^{2}\nonumber \\
 & =\mleft(M-1\mright)\sum_{l\in\mathbb{Z}_{\ast}^{3}}\sum_{p\in L_{k}\cap L_{l}}\left\Vert \mleft(\tilde{E}_{l}-h_{l}\mright)e_{p}\right\Vert ^{2}\left\Vert \Psi_{M-2}\right\Vert ^{2}\\
 & \leq\mleft(M-1\mright)\sum_{l\in\mathbb{Z}_{\ast}^{3}}\Vert\tilde{E}_{l}-h_{l}\Vert_{\mathrm{HS}}^{2}\left\Vert \Psi_{M-2}\right\Vert ^{2}.\nonumber 
\end{align}
Due to the identity
\begin{equation}
\left[c_{p-k,\sigma}^{\ast}c_{q,\sigma},c_{q',\tau}^{\ast}c_{p'-k,\tau}\right]=\delta_{p,p'}^{\sigma,\tau}\delta_{q,q'}^{\sigma,\tau}-\delta_{p,p'}^{\sigma,\tau}c_{q',\tau}^{\ast}c_{q,\sigma}-\delta_{q,q'}^{\sigma,\tau}c_{p'-k,\tau}c_{p-k,\sigma}^{\ast}
\end{equation}
it follows by a partial normal-ordering of $\mathcal{E}^{\ast}\mathcal{E}$
that
\begin{align}
\left\Vert \mathcal{E}\Psi_{M-2}\right\Vert ^{2} & =\sum_{p,p',q,q'\in L_{k}}^{\sigma,\tau}\left\langle \phi,e_{p}\right\rangle \left\langle \phi,e_{q}\right\rangle \left\langle e_{p'},\phi\right\rangle \left\langle e_{q'},\phi\right\rangle \left\langle \Psi_{M-2},c_{p-k,\sigma}^{\ast}c_{q,\sigma}B_{p,q}B_{p',q'}^{\ast}c_{q',\tau}^{\ast}c_{p'-k,\tau}\Psi_{M-2}\right\rangle \nonumber \\
 & =\sum_{p,p',q,q'\in L_{k}}^{\sigma,\tau}\left\langle \phi,e_{p}\right\rangle \left\langle \phi,e_{q}\right\rangle \left\langle e_{p'},\phi\right\rangle \left\langle e_{q'},\phi\right\rangle \left\langle c_{p'-k,\tau}^{\ast}c_{q',\tau}B_{p,q}^{\ast}\Psi_{M-2},c_{p-k,\sigma}^{\ast}c_{q,\sigma}B_{p',q'}^{\ast}\Psi_{M-2}\right\rangle \nonumber \\
 & -\sum_{p,q,q'\in L_{k}}^{\sigma}\left|\left\langle e_{p},\phi\right\rangle \right|^{2}\left\langle \phi,e_{q}\right\rangle \left\langle e_{q'},\phi\right\rangle \left\langle c_{q',\sigma}B_{p,q}^{\ast}\Psi_{M-2},c_{q,\sigma}B_{p,q'}^{\ast}\Psi_{M-2}\right\rangle \\
 & -\sum_{p,p',q\in L_{k}}^{\sigma}\left|\left\langle e_{q},\phi\right\rangle \right|^{2}\left\langle \phi,e_{p}\right\rangle \left\langle e_{p'},\phi\right\rangle \left\langle c_{p'-k,\sigma}^{\ast}B_{p,q}^{\ast}\Psi_{M-2},c_{p-k,\sigma}^{\ast}B_{p',q}^{\ast}\Psi_{M-2}\right\rangle \nonumber \\
 & +\sum_{p,q\in L_{k}}^{\sigma}\left|\left\langle e_{p},\phi\right\rangle \right|^{2}\left|\left\langle e_{q},\phi\right\rangle \right|^{2}\left\Vert B_{p,q}^{\ast}\Psi_{M-2}\right\Vert ^{2}\nonumber \\
 & =:T_{1}+T_{2}+T_{3}+T_{4}.\nonumber 
\end{align}
We bound these terms individually. For $T_{1}$ we have by Cauchy-Schwarz
that
\begin{align}
\left|T_{1}\right| & \leq\left\Vert \phi\right\Vert _{\infty}^{4}\sum_{p,p',q,q'\in L_{k}}^{\sigma,\tau}\left\Vert c_{p'-k,\tau}^{\ast}c_{q',\tau}B_{p,q}^{\ast}\Psi_{M-2}\right\Vert \left\Vert c_{p-k,\sigma}^{\ast}c_{q,\sigma}B_{p',q'}^{\ast}\Psi_{M-2}\right\Vert \nonumber \\
 & \leq s\left\Vert \phi\right\Vert _{\infty}^{4}\sum_{p,p',q,q'\in L_{k}}^{\sigma}\left\Vert c_{p-k,\sigma}^{\ast}c_{q,\sigma}B_{p',q'}^{\ast}\Psi_{M-2}\right\Vert ^{2}\leq s\left\Vert \phi\right\Vert _{\infty}^{4}\sum_{p',q'\in L_{k}}\sum_{p,q\in L_{k}}^{\sigma,\tau}\left\Vert c_{p-k,\sigma}^{\ast}c_{q,\tau}B_{p',q'}^{\ast}\Psi_{M-2}\right\Vert ^{2}\\
 & =s\left\Vert \phi\right\Vert _{\infty}^{4}\sum_{p',q'\in L_{k}}\left\Vert \mathcal{N}_{E}B_{p',q'}^{\ast}\Psi_{M-2}\right\Vert ^{2}=\mleft(M-1\mright)^{2}s\left\Vert \phi\right\Vert _{\infty}^{4}\sum_{p',q'\in L_{k}}\left\Vert B_{p',q'}^{\ast}\Psi_{M-2}\right\Vert ^{2}\nonumber \\
 & \leq\mleft(M-1\mright)^{3}s\left\Vert \phi\right\Vert _{\infty}^{4}\sum_{l\in\mathbb{Z}_{\ast}^{3}}\Vert\tilde{E}_{l}-h_{l}\Vert_{\mathrm{HS}}^{2}\left\Vert \Psi_{M-2}\right\Vert ^{2}\nonumber 
\end{align}
and similarly for $T_{2}$
\begin{align}
\left|T_{2}\right| & \leq\left\Vert \phi\right\Vert _{\infty}^{4}\sum_{p,q,q'\in L_{k}}^{\sigma}\left\Vert c_{q',\sigma}B_{p,q}^{\ast}\Psi_{M-2}\right\Vert \left\Vert c_{q,\sigma}B_{p,q'}^{\ast}\Psi_{M-2}\right\Vert \leq\left\Vert \phi\right\Vert _{\infty}^{4}\sum_{p,q,q'\in L_{k}}^{\sigma}\left\Vert c_{q',\sigma}B_{p,q}^{\ast}\Psi_{M-2}\right\Vert ^{2}\nonumber \\
 & \leq\left\Vert \phi\right\Vert _{\infty}^{4}\sum_{p,q\in L_{k}}\Vert\mathcal{N}_{E}^{\frac{1}{2}}B_{p,q}^{\ast}\Psi_{M-2}\Vert^{2}=\left\Vert \phi\right\Vert _{\infty}^{4}\mleft(M-1\mright)\sum_{p,q\in L_{k}}\Vert B_{p,q}^{\ast}\Psi_{M-2}\Vert^{2}\\
 & \leq\mleft(M-1\mright)^{2}\left\Vert \phi\right\Vert _{\infty}^{4}\sum_{l\in\mathbb{Z}_{\ast}^{3}}\Vert\tilde{E}_{l}-h_{l}\Vert_{\mathrm{HS}}^{2}\left\Vert \Psi_{M-2}\right\Vert ^{2}.\nonumber 
\end{align}
$T_{3}$ obeys the same bound and obviously $\left|T_{4}\right|\leq\mleft(M-1\mright)s\left\Vert \phi\right\Vert _{\infty}^{4}\sum_{l\in\mathbb{Z}_{\ast}^{3}}\Vert\tilde{E}_{l}-h_{l}\Vert_{\mathrm{HS}}^{2}\left\Vert \Psi_{M-2}\right\Vert ^{2}$.
The claim follows by combining these estimates.

$\hfill\square$

We summarize this subsection in the following:
\begin{prop}
\label{prop:FirstPlasmonStateBound}It holds for all $M\in\mathbb{N}$
with $M<\left\Vert \phi\right\Vert _{\infty}^{-2}$ that $\hat{\Psi}_{M}=\left\Vert \Psi_{M}\right\Vert ^{-1}\Psi_{M}$
obeys
\[
\Vert\mleft(H_{\mathrm{eff}}-M\epsilon_{k}\mright)\hat{\Psi}_{M}\Vert\leq\frac{2}{\left\Vert \phi\right\Vert _{\infty}^{-2}-M}\sqrt{s^{-2}\sum_{l\in\mathbb{Z}_{\ast}^{3}}\Vert\tilde{E}_{l}-h_{l}\Vert_{\mathrm{HS}}^{2}}M^{\frac{5}{2}}.
\]
\end{prop}

\textbf{Proof:} By inserting the previous estimate into the statement
of Proposition \ref{prop:AlmostEigenstateIdentity} we obtain
\begin{equation}
\Vert\mleft(H_{\mathrm{eff}}-M\epsilon_{k}\mright)\hat{\Psi}_{M}\Vert\leq2M^{2}\mleft(M-1\mright)^{\frac{3}{2}}\left\Vert \phi\right\Vert _{\infty}^{2}\sqrt{s^{-2}\sum_{l\in\mathbb{Z}_{\ast}^{3}}\Vert\tilde{E}_{l}-h_{l}\Vert_{\mathrm{HS}}^{2}}\frac{\left\Vert \Psi_{M-2}\right\Vert }{\left\Vert \Psi_{M}\right\Vert },
\end{equation}
and by the lower bound of Corollary \ref{coro:RelativeNormLowerBound}
it holds that
\begin{equation}
\frac{\left\Vert \Psi_{M-2}\right\Vert }{\left\Vert \Psi_{M}\right\Vert }=\frac{\left\Vert \Psi_{M-2}\right\Vert }{\left\Vert \Psi_{M-1}\right\Vert }\frac{\left\Vert \Psi_{M-1}\right\Vert }{\left\Vert \Psi_{M}\right\Vert }\leq\frac{1}{\sqrt{M\mleft(M-1\mright)}}\frac{1}{1-M\left\Vert \phi\right\Vert _{\infty}^{2}}
\end{equation}
for $M<\left\Vert \phi\right\Vert _{\infty}^{-2}$.

$\hfill\square$

\subsection{Estimates of One-Body Quantities}

To conclude Theorem \ref{them:GeneralPlasmonStates} it only remains
to control the one-body quantities $\left\Vert \phi\right\Vert _{\infty}^{2}$,
$\Vert\tilde{E}_{l}-h_{l}\Vert_{\mathrm{HS}}^{2}$ and $\epsilon_{k}$.
To this end we return a final time to the setting of Section \ref{sec:AnalysisofOne-BodyOperators}
and consider $\tilde{E}:V\rightarrow V$ given by
\begin{equation}
\tilde{E}=\mleft(h^{2}+2P_{h^{\frac{1}{2}}v}\mright)^{\frac{1}{2}}
\end{equation}
with normalized eigenvector $\phi\in V$ (chosen such that $\left\langle h^{\frac{1}{2}}v,\phi\right\rangle \geq0$)
corresponding to the greatest eigenvalue $\epsilon$ of $2\tilde{E}$.
Below it will be more convenient to work in terms of the greatest
eigenvalue $\varepsilon$ of $\tilde{E}^{2}$; the eigenvalues are
simply related by $\epsilon=2\sqrt{\epsilon}$.

The eigenvalue equation for $\varepsilon$ is
\begin{equation}
\varepsilon\phi=\tilde{E}^{2}\phi=\mleft(h^{2}+2P_{h^{\frac{1}{2}}v}\mright)\phi=h^{2}\phi+2\left\langle h^{\frac{1}{2}}v,\phi\right\rangle h^{\frac{1}{2}}v
\end{equation}
and assuming that $\varepsilon>\max_{1\leq i\leq n}\lambda_{i}^{2}=:\lambda_{\max}^{2}$
this can be rearranged to
\begin{equation}
\phi=2\left\langle h^{\frac{1}{2}}v,\phi\right\rangle \mleft(\varepsilon-h^{2}\mright)^{-1}h^{\frac{1}{2}}v.
\end{equation}
As $\phi$ is by assumption normalized and $\left\langle h^{\frac{1}{2}}v,\phi\right\rangle \geq0$,
this implies that $\phi$ is determined with $\varepsilon$ as the
only unknown quantity by the formula
\begin{equation}
\phi=\Vert\mleft(\varepsilon-h^{2}\mright)^{-1}h^{\frac{1}{2}}v\Vert^{-1}\mleft(\varepsilon-h^{2}\mright)^{-1}h^{\frac{1}{2}}v.
\end{equation}
In particular, the components of $\phi$ with respect to the eigenvectors
$\mleft(x_{i}\mright)_{i=1}^{n}$ of $h$ obey
\begin{equation}
\left\langle x_{i},\phi\right\rangle =\frac{1}{\Vert\mleft(\varepsilon-h^{2}\mright)^{-1}h^{\frac{1}{2}}v\Vert}\frac{\sqrt{\lambda_{i}}}{\varepsilon-\lambda_{i}^{2}}\left\langle x_{i},v\right\rangle ,\quad1\leq i\leq n.\label{eq:phiComponentIdentity}
\end{equation}
To ensure that $\varepsilon>\lambda_{\max}^{2}$, note that by the
variational principle there holds the inequality
\begin{equation}
\varepsilon\geq\frac{\left\langle h^{\frac{1}{2}}v,\mleft(h^{2}+2P_{h^{\frac{1}{2}}v}\mright)h^{\frac{1}{2}}v\right\rangle }{\left\langle h^{\frac{1}{2}}v,h^{\frac{1}{2}}v\right\rangle }=2\left\langle v,hv\right\rangle +\frac{\left\langle v,h^{3}v\right\rangle }{\left\langle v,hv\right\rangle }
\end{equation}
so $\varepsilon>\lambda_{\max}^{2}$ is assured if $2\left\langle v,hv\right\rangle >\lambda_{\max}^{2}$.
Under this condition we then have the following bound:
\begin{cor}
\label{coro:phiinfNormBound}Provided $2\left\langle v,hv\right\rangle >\lambda_{\max}^{2}$
it holds that
\[
\left\Vert \phi\right\Vert _{\infty}\leq\frac{2\sqrt{\left\langle v,hv\right\rangle \lambda_{\max}}}{2\left\langle v,hv\right\rangle -\lambda_{\max}^{2}}\left\Vert v\right\Vert _{\infty}.
\]
\end{cor}

\textbf{Proof:} As $t\mapsto t\mleft(t-\lambda_{\max}^{2}\mright)^{-1}$
is decreasing for $t\in\mleft(\lambda_{\max}^{2},\infty\mright)$, equation
(\ref{eq:phiComponentIdentity}) shows that for any $1\leq i\leq n$
\begin{align}
\left|\left\langle x_{i},\phi\right\rangle \right| & \leq\frac{1}{\frac{1}{\varepsilon}\Vert h^{\frac{1}{2}}v\Vert}\frac{\sqrt{\lambda_{\max}}}{\varepsilon-\lambda_{\max}^{2}}\left|\left\langle x_{i},v\right\rangle \right|=\frac{\varepsilon}{\varepsilon-\lambda_{\max}^{2}}\frac{\sqrt{\lambda_{\max}}}{\sqrt{\left\langle v,hv\right\rangle }}\left|\left\langle x_{i},v\right\rangle \right|\\
 & \leq\frac{2\left\langle v,hv\right\rangle }{2\left\langle v,hv\right\rangle -\lambda_{\max}^{2}}\frac{\sqrt{\lambda_{\max}}}{\sqrt{\left\langle v,hv\right\rangle }}\left|\left\langle x_{i},v\right\rangle \right|=\frac{2\sqrt{\left\langle v,hv\right\rangle \lambda_{\max}}}{2\left\langle v,hv\right\rangle -\lambda_{\max}^{2}}\left|\left\langle x_{i},v\right\rangle \right|.\nonumber 
\end{align}
$\hfill\square$

Under the same assumption we can also control $\varepsilon$ well:
\begin{prop}
\label{prop:varepsilonBounds}Provided $2\left\langle v,hv\right\rangle >\lambda_{\max}^{2}$
it holds that
\[
2\left\langle v,hv\right\rangle +\frac{\left\langle v,h^{3}v\right\rangle }{\left\langle v,hv\right\rangle }\leq\varepsilon\leq2\left\langle v,hv\right\rangle +\frac{\left\langle v,h^{3}v\right\rangle }{\left\langle v,hv\right\rangle }+\frac{4\left\langle v,h^{3}v\right\rangle \lambda_{\max}^{2}}{\mleft(2\left\langle v,hv\right\rangle -\lambda_{\max}^{2}\mright)^{2}}.
\]
\end{prop}

\textbf{Proof:} We noted the lower bound above. For the upper bound
we estimate
\begin{align}
\left\langle \phi,h^{2}\phi\right\rangle  & =\frac{\left\langle v,h^{3}\mleft(\varepsilon-h^{2}\mright)^{-2}v\right\rangle }{\left\langle v,h\mleft(\varepsilon-h^{2}\mright)^{-2}v\right\rangle }\leq\mleft(\frac{\varepsilon}{\varepsilon-\lambda_{\max}^{2}}\mright)^{2}\frac{\left\langle v,h^{3}v\right\rangle }{\left\langle v,hv\right\rangle }\leq\mleft(\frac{2\left\langle v,hv\right\rangle }{2\left\langle v,hv\right\rangle -\lambda_{\max}^{2}}\mright)^{2}\frac{\left\langle v,h^{3}v\right\rangle }{\left\langle v,hv\right\rangle }\nonumber \\
 & =\frac{4\left\langle v,hv\right\rangle ^{2}}{\mleft(2\left\langle v,hv\right\rangle -\lambda_{\max}^{2}\mright)^{2}}\frac{\left\langle v,h^{3}v\right\rangle }{\left\langle v,hv\right\rangle }=\frac{\left\langle v,h^{3}v\right\rangle }{\left\langle v,hv\right\rangle }+\frac{4\left\langle v,hv\right\rangle ^{2}-\mleft(2\left\langle v,hv\right\rangle -\lambda_{\max}^{2}\mright)^{2}}{\mleft(2\left\langle v,hv\right\rangle -\lambda_{\max}^{2}\mright)^{2}}\frac{\left\langle v,h^{3}v\right\rangle }{\left\langle v,hv\right\rangle }\\
 & \leq\frac{\left\langle v,h^{3}v\right\rangle }{\left\langle v,hv\right\rangle }+\frac{4\left\langle v,hv\right\rangle \lambda_{\max}^{2}}{\mleft(2\left\langle v,hv\right\rangle -\lambda_{\max}^{2}\mright)^{2}}\frac{\left\langle v,h^{3}v\right\rangle }{\left\langle v,hv\right\rangle }=\frac{\left\langle v,h^{3}v\right\rangle }{\left\langle v,hv\right\rangle }+\frac{4\left\langle v,h^{3}v\right\rangle \lambda_{\max}^{2}}{\mleft(2\left\langle v,hv\right\rangle -\lambda_{\max}^{2}\mright)^{2}}\nonumber 
\end{align}
and see that by the eigevalue equation for $\varepsilon$ and the
Cauchy-Schwarz inequality in the form $\left|\left\langle h^{\frac{1}{2}}v,\phi\right\rangle \right|^{2}\leq\left\langle v,hv\right\rangle $,
\begin{equation}
\varepsilon=\left\langle \phi,h^{2}\phi\right\rangle +2\left|\left\langle h^{\frac{1}{2}}v,\phi\right\rangle \right|^{2}\leq2\left\langle v,hv\right\rangle +\frac{\left\langle v,h^{3}v\right\rangle }{\left\langle v,hv\right\rangle }+\frac{4\left\langle v,h^{3}v\right\rangle \lambda_{\max}^{2}}{\mleft(2\left\langle v,hv\right\rangle -\lambda_{\max}^{2}\mright)^{2}}.
\end{equation}
$\hfill\square$

Lastly we bound $\Vert\tilde{E}-h\Vert_{\mathrm{HS}}^{2}$:
\begin{prop}
\label{prop:Etilde-hHSEstimate}It holds that
\[
\Vert\tilde{E}-h\Vert_{\mathrm{HS}}^{2}\leq\min\left\{ 2\left\langle v,hv\right\rangle ,\left\Vert v\right\Vert ^{4}\right\} .
\]
\end{prop}

\textbf{Proof:} The first bound is easily obtained as
\begin{align}
\Vert\tilde{E}-h\Vert_{\mathrm{HS}}^{2} & =\mathrm{tr}\mleft(\mleft(\tilde{E}-h\mright)^{2}\mright)=\mathrm{tr}\mleft(\tilde{E}^{2}-2\{\tilde{E},h\}+h^{2}\mright)=2\,\mathrm{tr}\mleft(h^{2}+P_{h^{\frac{1}{2}}v}-h^{\frac{1}{2}}\tilde{E}h^{\frac{1}{2}}\mright)\\
 & \leq2\,\mathrm{tr}\mleft(P_{h^{\frac{1}{2}}v}\mright)=2\left\langle v,hv\right\rangle \nonumber 
\end{align}
since $h\leq\tilde{E}$ implies that $h^{2}-h^{\frac{1}{2}}\tilde{E}h^{\frac{1}{2}}\leq0$.
For the second, note that $\tilde{E}=h^{\frac{1}{2}}e^{-2K}h^{\frac{1}{2}}$
whence Proposition \ref{prop:e-2Ke2KElementEstimates} affords us
the elementwise estimate
\begin{equation}
\left|\left\langle x_{i},\mleft(\tilde{E}-h\mright)x_{j}\right\rangle \right|=\sqrt{\lambda_{i}\lambda_{j}}\left|\left\langle x_{i},\mleft(e^{-2K}-1\mright)x_{j}\right\rangle \right|\leq\frac{2\sqrt{\lambda_{i}\lambda_{j}}}{\lambda_{i}+\lambda_{j}}\left\langle x_{i},v\right\rangle \left\langle v,x_{j}\right\rangle \leq\left\langle x_{i},v\right\rangle \left\langle v,x_{j}\right\rangle 
\end{equation}
for $1\leq i,j\leq n$, so
\begin{equation}
\Vert\tilde{E}-h\Vert_{\mathrm{HS}}^{2}=\sum_{i,j=1}^{n}\left|\left\langle x_{i},\mleft(\tilde{E}-h\mright)x_{j}\right\rangle \right|^{2}\leq\sum_{i,j=1}^{n}\left|\left\langle x_{i},v\right\rangle \left\langle v,x_{j}\right\rangle \right|^{2}=\left\Vert v\right\Vert ^{4}.
\end{equation}
$\hfill\square$

\subsection{Final Details}

We now insert the particular operators $h_{k}$ and $P_{k}$. For
the quantity $2\left\langle v_{k},h_{k}v_{k}\right\rangle -\lambda_{k,\max}^{2}$,
we note that the inequalities defining $L_{k}$ imply that
\begin{equation}
\lambda_{k,p}=k\cdot p-\frac{1}{2}\left|k\right|^{2}=k\cdot\mleft(p-k\mright)+\frac{1}{2}\left|k\right|^{2}\leq\left|k\right|\mleft(k_{F}+\frac{1}{2}\left|k\right|\mright),\quad p\in L_{k},
\end{equation}
so
\begin{equation}
\lambda_{k,\max}^{2}\leq Ck_{F}^{2}\left|k\right|^{2}
\end{equation}
as we assumed that $k\in B_{F}$. The quantity $2\left\langle v_{k},h_{k}v_{k}\right\rangle $
is
\begin{equation}
2\left\langle v_{k},h_{k}v_{k}\right\rangle =\frac{s\hat{V}_{k}}{\mleft(2\pi\mright)^{3}}\sum_{p\in L_{k}}\lambda_{k,p}
\end{equation}
and for a lower bound we prove the following in appendix section \ref{subsec:RiemannSumLowerBounds}:
\begin{prop}
\label{prop:RiemannSumLowerBound}For all $k\in B_{F}$ and $\beta\in\left\{ 0\right\} \cup\mleft[1,\infty\mright)$
it holds that
\[
\sum_{p\in L_{k}}\lambda_{k,p}^{\beta}\geq ck_{F}^{2+\beta}\left|k\right|^{1+\beta},\quad k_{F}\rightarrow\infty,
\]
for a $c>0$ depending only on $\beta$.
\end{prop}

It follows that
\begin{equation}
2\left\langle v_{k},h_{k}v_{k}\right\rangle -\lambda_{k,\max}^{2}\geq c_{1}sk_{F}^{3}\left|k\right|^{2}\mleft(\hat{V}_{k}-c_{1}^{-1}k_{F}^{-1}\mright),
\end{equation}
so if $\hat{V}_{k}>ck_{F}$ for $c=2c_{1}^{-1}$, say, it holds that
\begin{equation}
2\left\langle v_{k},h_{k}v_{k}\right\rangle -\lambda_{k,\max}^{2}\geq c'sk_{F}^{3}\left|k\right|^{2}\hat{V}_{k}
\end{equation}
for some $c'>0$ independent of all quantities. For $\beta\in\mleft[0,\infty\mright)$
we also have that
\begin{equation}
\left\langle v_{k},h_{k}^{\beta}v_{k}\right\rangle =\frac{s\hat{V}_{k}}{\mleft(2\pi\mright)^{3}}\sum_{p\in L_{k}}\lambda_{k,p}^{\beta}\leq\frac{s\hat{V}_{k}}{\mleft(2\pi\mright)^{3}}\left|L_{k}\right|\lambda_{k,\max}^{\beta}\leq C'sk_{F}^{2+\beta}\left|k\right|^{1+\beta}\hat{V}_{k}
\end{equation}
so Corollary \ref{coro:phiinfNormBound} allows us to bound $\left\Vert \phi\right\Vert _{\infty}$
as
\begin{equation}
\left\Vert \phi\right\Vert _{\infty}\leq\frac{2\sqrt{\left\langle v_{k},h_{k}v_{k}\right\rangle \lambda_{k,\max}}}{2\left\langle v_{k},h_{k}v_{k}\right\rangle -\lambda_{k,\max}^{2}}\left\Vert v_{k}\right\Vert _{\infty}\leq C'\frac{\sqrt{\mleft(sk_{F}^{3}\left|k\right|^{2}\hat{V}_{k}\mright)\mleft(k_{F}\left|k\right|\mright)}}{sk_{F}^{3}\left|k\right|^{2}\hat{V}_{k}}\mleft(s\hat{V}_{k}\mright)^{\frac{1}{2}}=\frac{C'}{k_{F}\left|k\right|^{\frac{1}{2}}}.
\end{equation}
Note that since $\left|L_{k}\right|\sim O\mleft(k_{F}^{2}\left|k\right|\mright)$,
$\phi$ is indeed almost completely delocalized, and we can estimate
that
\begin{equation}
\left\Vert \phi\right\Vert _{\infty}^{-2}-M\geq C'k_{F}^{2}\left|k\right|
\end{equation}
for all $M\in\mathbb{N}$ such that $M\leq Ck_{F}^{2}\left|k\right|$
for some $C$ also independent of all quantities.

Finally, by Proposition \ref{prop:Etilde-hHSEstimate},
\begin{align}
\Vert\tilde{E}_{l}-h_{l}\Vert_{\mathrm{HS}}^{2} & \leq\min\left\{ 2\left\langle v_{l},h_{l}v_{l}\right\rangle ,\left\Vert v_{l}\right\Vert ^{4}\right\} \leq C\min\left\{ sk_{F}^{3}\hat{V}_{l}\left|l\right|^{2},s^{2}\hat{V}_{l}^{2}\left|L_{l}\right|^{2}\right\} \nonumber \\
 & \leq Cs^{2}\min\left\{ k_{F}^{3}\hat{V}_{l}\left|l\right|^{2},k_{F}^{4}\hat{V}_{l}^{2}\left|l\right|^{2},k_{F}^{6}\hat{V}_{l}^{2}\right\} \\
 & =Cs^{2}\min\left\{ 1,k_{F}\hat{V}_{l},k_{F}^{3}\hat{V}_{l}\left|l\right|^{-2}\right\} k_{F}^{3}\hat{V}_{l}\left|l\right|^{2}\nonumber 
\end{align}
for any $l\in\mathbb{Z}_{\ast}^{3}$. Inserting these bounds into
Proposition \ref{prop:FirstPlasmonStateBound} yields the first claim
of Theorem \ref{them:GeneralPlasmonStates}:
\begin{prop}
There exists constants $c,C>0$ such that if $\hat{V}_{k}>ck_{F}^{-1}$
it holds for all $M\leq Ck_{F}^{2}\left|k\right|$ that $\hat{\Psi}_{M}=\left\Vert \Psi_{M}\right\Vert ^{-1}\Psi_{M}$
obeys
\[
\Vert\mleft(H_{\mathrm{eff}}-M\epsilon_{k}\mright)\hat{\Psi}_{M}\Vert\leq C'\sqrt{\sum_{l\in\mathbb{Z}_{\ast}^{3}}\min\left\{ 1,k_{F}\hat{V}_{l},k_{F}^{3}\hat{V}_{l}\left|l\right|^{-2}\right\} \hat{V}_{l}\left|l\right|^{2}}\frac{M^{\frac{5}{2}}}{\sqrt{k_{F}}\left|k\right|}
\]
for a constant $C'>0$. $c,C,C'$ are independent of all quantities.
\end{prop}

\subsubsection*{The Eigenvalue $\epsilon_{k}$}

For $\epsilon_{k}$ we have by Proposition \ref{prop:varepsilonBounds}
that (recalling the relation $\epsilon=2\sqrt{\varepsilon}$)
\begin{equation}
2\sqrt{2\left\langle v_{k},h_{k}v_{k}\right\rangle +\frac{\left\langle v_{k},h_{k}^{3}v_{k}\right\rangle }{\left\langle v_{k},h_{k}v_{k}\right\rangle }}\leq\epsilon_{k}\leq2\sqrt{2\left\langle v_{k},h_{k}v_{k}\right\rangle +\frac{\left\langle v_{k},h_{k}^{3}v_{k}\right\rangle }{\left\langle v_{k},h_{k}v_{k}\right\rangle }+\frac{4\left\langle v_{k},h_{k}^{3}v_{k}\right\rangle \lambda_{k,\max}^{2}}{\mleft(2\left\langle v_{k},h_{k}v_{k}\right\rangle -\lambda_{k,\max}^{2}\mright)^{2}}}.
\end{equation}
The lower bound given in Theorem \ref{them:GeneralPlasmonStates}
is then immediate since
\begin{equation}
\epsilon_{k}\geq2\sqrt{2\left\langle v_{k},h_{k}v_{k}\right\rangle }\geq c's^{\frac{1}{2}}k_{F}^{\frac{3}{2}}\left|k\right|\hat{V}_{k}^{\frac{1}{2}}
\end{equation}
as above, while the inequality $\sqrt{a+b}-\sqrt{a}\leq\frac{b}{2\sqrt{a}}$
yields the upper bound
\begin{align}
\epsilon_{k}-2\sqrt{2\left\langle v_{k},h_{k}v_{k}\right\rangle +\frac{\left\langle v_{k},h_{k}^{3}v_{k}\right\rangle }{\left\langle v_{k},hv_{k}\right\rangle }} & \leq\frac{1}{\sqrt{2\left\langle v_{k},h_{k}v_{k}\right\rangle +\frac{\left\langle v_{k},h_{k}^{3}v_{k}\right\rangle }{\left\langle v_{k},hv_{k}\right\rangle }}}\frac{4\left\langle v_{k},h_{k}^{3}v_{k}\right\rangle \lambda_{k,\max}^{2}}{\mleft(2\left\langle v_{k},h_{k}v_{k}\right\rangle -\lambda_{k,\max}^{2}\mright)^{2}}\nonumber \\
 & \leq C\frac{1}{\sqrt{sk_{F}^{3}\left|k\right|^{2}\hat{V}_{k}}}\frac{\mleft(sk_{F}^{5}\left|k\right|^{4}\hat{V}_{k}\mright)\mleft(k_{F}\left|k\right|\mright)^{2}}{\mleft(sk_{F}^{3}\left|k\right|^{2}\hat{V}_{k}\mright)^{2}}\\
 & =Ck_{F}^{-\frac{1}{2}}\left|k\right|\hat{V}_{k}^{-\frac{3}{2}}.\nonumber 
\end{align}
For the form given in Theorem \ref{them:PlasmonStates} for $\hat{V}_{k}=g\left|k\right|^{-2}$,
note that expanding the inner products gives
\begin{align}
\epsilon & \sim2\sqrt{2\left\langle v_{k},h_{k}v_{k}\right\rangle +\frac{\left\langle v_{k},h_{k}^{3}v_{k}\right\rangle }{\left\langle v_{k},hv_{k}\right\rangle }}=2\sqrt{2\frac{s\hat{V}_{k}}{2\,\mleft(2\pi\mright)^{3}}\sum_{p\in L_{k}}\lambda_{k,p}+\frac{\frac{s\hat{V}_{k}}{2\,\mleft(2\pi\mright)^{3}}\sum_{p\in L_{k}}\lambda_{k,p}^{3}}{\frac{s\hat{V}_{k}}{2\,\mleft(2\pi\mright)^{3}}\sum_{p\in L_{k}}\lambda_{k,p}}}\\
 & =2\sqrt{\frac{s}{\mleft(2\pi\mright)^{3}}\frac{g}{\left|k\right|^{2}}\sum_{p\in L_{k}}\lambda_{k,p}+\frac{\sum_{p\in L_{k}}\lambda_{k,p}^{3}}{\sum_{p\in L_{k}}\lambda_{k,p}}}\nonumber 
\end{align}
and formally replacing the Riemann sums by integrals according to
equation (\ref{eq:ContinuumLuneIntegrals}) shows that
\begin{equation}
\sum_{p\in L_{k}}\lambda_{k,p}\sim\frac{2\pi}{3}k_{F}^{3}\left|k\right|^{2},\quad\sum_{p\in L_{k}}\lambda_{k,p}^{3}\sim\frac{2\pi}{5}k_{F}^{5}\left|k\right|^{4}+\frac{\pi}{6}k_{F}^{3}\left|k\right|^{6}\approx\frac{2\pi}{5}k_{F}^{5}\left|k\right|^{4},
\end{equation}
whence
\begin{align}
\epsilon_{k} & \sim2\sqrt{\frac{s}{\mleft(2\pi\mright)^{3}}\frac{g}{\left|k\right|^{2}}\frac{2\pi}{3}k_{F}^{3}\left|k\right|^{2}+\frac{\frac{2\pi}{5}k_{F}^{5}\left|k\right|^{4}}{\frac{2\pi}{3}k_{F}^{3}\left|k\right|^{2}}}=\sqrt{2g\mleft(\frac{1}{\mleft(2\pi\mright)^{3}}\frac{4\pi s}{3}k_{F}^{3}\mright)+\frac{12}{5}k_{F}^{2}\left|k\right|^{2}}\\
 & \sim\sqrt{2gn+\frac{12}{5}k_{F}^{2}\left|k\right|^{2}}\nonumber 
\end{align}
for $n=\frac{N}{\mleft(2\pi\mright)^{3}}=\frac{s\left|B_{F}\right|}{\mleft(2\pi\mright)^{3}}\sim\frac{1}{\mleft(2\pi\mright)^{3}}\frac{4\pi s}{3}k_{F}^{3}$.

\appendix

\section{Some Functional Analysis Results}

\subsection{\label{subsec:TheSquareRootofaRankOnePerturbtation}The Square Root
of a Rank One Perturbation}

Let $\left\langle V,\left\langle \cdot,\cdot\right\rangle \right\rangle $
be an $n$-dimensional Hilbert space. With the notation
\begin{equation}
P_{w}\mleft(\cdot\mright)=\left\langle w,\cdot\right\rangle w,\quad w\in V,
\end{equation}
we recall the Sherman-Morrison formula:
\begin{lem}
Let $A:V\rightarrow V$ be an invertible operator. Then for any $w\in V$
and $g\in\mathbb{C}$, the operator $A+gP_{w}$ is invertible if and
only if $\left\langle w,A^{-1}w\right\rangle \neq g^{-1}$, in which
case the inverse is given by
\[
\mleft(A+gP_{w}\mright)^{-1}=A^{-1}-\frac{g}{1+g\left\langle w,A^{-1}w\right\rangle }P_{A^{-1}w}.
\]
\end{lem}

By applying this we conclude the following representation (first presented
in \cite{BenNamPorSchSei-20}):
\begin{prop*}[\ref{prop:SquareRootofAOneDimensionalPerturbation}]
Let $A:V\rightarrow V$ be a positive self-adjoint operator. Then
for any $w\in V$ and $g\in\mathbb{R}$ such that $A+gP_{w}>0$ it
holds that
\[
\mleft(A+gP_{w}\mright)^{\frac{1}{2}}=A^{\frac{1}{2}}+\frac{2g}{\pi}\int_{0}^{\infty}\frac{t^{2}}{1+g\left\langle w,\mleft(A+t^{2}\mright)^{-1}w\right\rangle }P_{\mleft(A+t^{2}\mright)^{-1}w}dt
\]
and
\[
\mathrm{tr}\mleft(\mleft(A+gP_{w}\mright)^{\frac{1}{2}}\mright)=\mathrm{tr}\mleft(A^{\frac{1}{2}}\mright)+\frac{1}{\pi}\int_{0}^{\infty}\log\mleft(1+g\left\langle w,\mleft(A+t^{2}\mright)^{-1}w\right\rangle \mright)dt.
\]
\end{prop*}
\textbf{Proof:} For any $a>0$ there holds the integral identity
\begin{equation}
\sqrt{a}=\frac{2}{\pi}\int_{0}^{\infty}\frac{a}{a+t^{2}}dt=\frac{2}{\pi}\int_{0}^{\infty}\mleft(1-\frac{t^{2}}{a+t^{2}}\mright)dt
\end{equation}
so by the spectral theorem the same is true for a positive operator
$A$, provided the fraction is understood as a resolvent. As the Sherman-Morrison
formula lets us write
\begin{equation}
\mleft(A+gP_{w}+t^{2}\mright)^{-1}=\mleft(A+t^{2}\mright)^{-1}-\frac{g}{1+g\left\langle w,\mleft(A+t^{2}\mright)^{-1}w\right\rangle }P_{\mleft(A+t^{2}\mright)^{-1}w}
\end{equation}
for any $t\geq0$, we thus conclude that
\begin{align}
\mleft(A+gP_{w}\mright)^{\frac{1}{2}} & =\frac{2}{\pi}\int_{0}^{\infty}\mleft(1-t^{2}\mleft(A+gP_{w}+t^{2}\mright)^{-1}\mright)dt\nonumber \\
 & =\frac{2}{\pi}\int_{0}^{\infty}\mleft(1-t^{2}\mleft(\mleft(A+t^{2}\mright)^{-1}-\frac{g}{1+g\left\langle w,\mleft(A+t^{2}\mright)^{-1}w\right\rangle }P_{\mleft(A+t^{2}\mright)^{-1}w}\mright)\mright)dt\\
 & =\frac{2}{\pi}\int_{0}^{\infty}\mleft(1-t^{2}\mleft(A+t^{2}\mright)^{-1}\mright)\,dt+\frac{2g}{\pi}\int_{0}^{\infty}\frac{t^{2}}{1+g\left\langle w,\mleft(A+t^{2}\mright)^{-1}w\right\rangle }P_{\mleft(A+t^{2}\mright)^{-1}w}dt\nonumber \\
 & =A^{\frac{1}{2}}+\frac{2g}{\pi}\int_{0}^{\infty}\frac{t^{2}}{1+g\left\langle w,\mleft(A+t^{2}\mright)^{-1}w\right\rangle }P_{\mleft(A+t^{2}\mright)^{-1}w}dt.\nonumber 
\end{align}
The trace formula now follows by partial integration as
\begin{align}
\mathrm{tr}\mleft(\mleft(A+gP_{w}\mright)^{\frac{1}{2}}-A^{\frac{1}{2}}\mright) & =\frac{2g}{\pi}\int_{0}^{\infty}\frac{t^{2}}{1+g\left\langle w,\mleft(A+t^{2}\mright)^{-1}w\right\rangle }\Vert\mleft(A+t^{2}\mright)^{-1}w\Vert^{2}dt=\frac{1}{\pi}\int_{0}^{\infty}t\frac{2gt\left\langle w,\mleft(A+t^{2}\mright)^{-2}w\right\rangle }{1+g\left\langle w,\mleft(A+t^{2}\mright)^{-1}w\right\rangle }dt\nonumber \\
 & =-\frac{1}{\pi}\left[t\log\mleft(1+g\left\langle w,\mleft(A+t^{2}\mright)^{-1}w\right\rangle \mright)\right]_{0}^{\infty}+\frac{1}{\pi}\int_{0}^{\infty}\log\mleft(1+g\left\langle w,\mleft(A+t^{2}\mright)^{-1}w\right\rangle \mright)dt\nonumber \\
 & =\frac{1}{\pi}\int_{0}^{\infty}\log\mleft(1+g\left\langle w,\mleft(A+t^{2}\mright)^{-1}w\right\rangle \mright)dt
\end{align}
since $\left|\log\mleft(1+g\left\langle w,\mleft(A+t^{2}\mright)^{-1}w\right\rangle \mright)\right|\leq\left|g\left\langle w,\mleft(A+t^{2}\mright)^{-1}w\right\rangle \right|\leq Ct^{-2}$
for $t\rightarrow\infty$.

$\hfill\square$

\subsection{A Square Root Estimation Result\label{subsec:ASquareRootEstimationResult}}
\begin{lem*}[\ref{lemma:CommutatorRootEstimate}]
Let $A,B,Z$ be given with $A>0$, $Z\geq0$ and $\left[A,Z\right]=0$.
Then if $\pm\left[A,\left[A,B\right]\right]\leq Z$ it holds that
\[
\pm[A^{\frac{1}{2}},[A^{\frac{1}{2}},B]]\leq\frac{1}{4}A^{-1}Z.
\]
\end{lem*}
\textbf{Proof:} Applying the identity $A^{\frac{1}{2}}=\frac{2}{\pi}\int_{0}^{\infty}\mleft(1-t^{2}\mleft(A+t^{2}\mright)^{-1}\mright)\,dt$
as above, we find that
\begin{align}
[A^{\frac{1}{2}},B] & =\frac{2}{\pi}\int_{0}^{\infty}\left[1-t^{2}\mleft(A+t^{2}\mright)^{-1},B\right]dt=-\frac{2}{\pi}\int_{0}^{\infty}\left[\mleft(A+t^{2}\mright)^{-1},B\right]t^{2}dt\nonumber \\
 & =\frac{2}{\pi}\int_{0}^{\infty}\mleft(A+t^{2}\mright)^{-1}\left[A+t^{2},B\right]\mleft(A+t^{2}\mright)^{-1}t^{2}dt\\
 & =\frac{2}{\pi}\int_{0}^{\infty}\mleft(A+t^{2}\mright)^{-1}\left[A,B\right]\mleft(A+t^{2}\mright)^{-1}t^{2}dt\nonumber 
\end{align}
where we also used the general identity $\left[A^{-1},B\right]=-A^{-1}\left[A,B\right]A^{-1}$.
Iterating this formula we conclude that
\begin{align}
[A^{\frac{1}{2}},[A^{\frac{1}{2}},B]] & =\frac{2}{\pi}\int_{0}^{\infty}\mleft(A+t^{2}\mright)^{-1}[A,[A^{\frac{1}{2}},B]]\mleft(A+t^{2}\mright)^{-1}t^{2}dt\\
 & =\mleft(\frac{2}{\pi}\mright)^{2}\int_{0}^{\infty}\mleft(A+t^{2}\mright)^{-1}\mleft(A+s^{2}\mright)^{-1}\left[A,\left[A,B\right]\right]\mleft(A+s^{2}\mright)^{-1}\mleft(A+t^{2}\mright)^{-1}s^{2}t^{2}dt\nonumber 
\end{align}
whence the asumptions imply that
\begin{align}
\pm[A^{\frac{1}{2}},[A^{\frac{1}{2}},B]] & \leq\mleft(\frac{2}{\pi}\mright)^{2}\int_{0}^{\infty}\mleft(A+t^{2}\mright)^{-1}\mleft(A+s^{2}\mright)^{-1}Z\mleft(A+s^{2}\mright)^{-1}\mleft(A+t^{2}\mright)^{-1}s^{2}t^{2}dt\\
 & =\mleft(\frac{2}{\pi}\int_{0}^{\infty}\mleft(A+t^{2}\mright)^{-2}t^{2}dt\mright)^{2}Z=\mleft(\frac{1}{2}A^{-\frac{1}{2}}\mright)^{2}Z=\frac{1}{4}A^{-1}Z\nonumber 
\end{align}
as the identity $\int_{0}^{\infty}\frac{t^{2}}{\mleft(a+t^{2}\mright)^{2}}dt=\frac{\pi}{4}a^{-\frac{1}{2}}$,
$a>0$, similarly yields that $\int_{0}^{\infty}\mleft(A+t^{2}\mright)^{-2}t^{2}dt=\frac{\pi}{4}A^{-\frac{1}{2}}$.

$\hfill\square$

\subsection{\label{subsec:TransformationofUnboundedOperators}Operators of the
Form $e^{zK}Ae^{-zK}$ for Unbounded $A$}

We prove the following:
\begin{prop}
\label{prop:ezKAe-zKProperties}Let $X$ be a Banach space, $A:D\mleft(A\mright)\rightarrow X$
be a closed operator and let $K:X\rightarrow X$ be a bounded operator
which preserves $D\mleft(A\mright)$. Suppose that $AK:D\mleft(A\mright)\rightarrow X$
is $A$-bounded.

Then for every $z\in\mathbb{C}$ the operator $e^{zK}:X\rightarrow X$
likewise preserves $D\mleft(A\mright)$ and $e^{zK}Ae^{-zK}:D\mleft(A\mright)\rightarrow X$
is closed. If additionally $X$ is a Hilbert space, $A$ is self-adjoint
and $K$ is skew-symmetric then $e^{tK}Ae^{-tK}$ is self-adjoint
for all $t\in\mathbb{R}$.

Furthermore, for every $x\in D\mleft(A\mright)$ the mapping $z\mapsto e^{zK}Ae^{-zK}x$
is complex differentiable and $C^{1}$ with
\[
\frac{d}{dz}e^{zK}Ae^{-zK}x=e^{zK}\left[K,A\right]e^{-zK}x.
\]
\end{prop}

For the remainder of this section we impose the following assumptions:
$A:D\mleft(A\mright)\rightarrow X$ is a closed operator on a Banach
space $X$ and $K:X\rightarrow X$ is a bounded operator on $X$,
which preserves $D\mleft(A\mright)$ such that $AK:D\mleft(A\mright)\rightarrow X$
is $A$-bounded according to
\begin{equation}
\left\Vert AKx\right\Vert \leq a\left\Vert Ax\right\Vert +b\left\Vert x\right\Vert ,\quad x\in D\mleft(A\mright),
\end{equation}
for some $a,b\geq0$.

\subsubsection*{Well-Definedness of $e^{zK}Ae^{-zK}$}

We begin with a lemma:
\begin{lem}
Under the assumptions on $A$ and $K$, the operator $AK^{m}:D\mleft(A\mright)\rightarrow X$
is $A$-bounded for any $m\in\mathbb{N}_{0}$ with
\[
\left\Vert AK^{m}x\right\Vert \leq a^{m}\left\Vert Ax\right\Vert +mc^{m-1}b\left\Vert x\right\Vert ,\quad x\in D\mleft(A\mright),
\]
for $c=\max\left\{ a,\left\Vert K\right\Vert _{\mathrm{Op}}\right\} $.
\end{lem}

\textbf{Proof:} The claim is clearly true for $m=0,1$ (by assumption).
We prove the general claim by induction: Suppose that case $m-1$
holds. Then we obtain case $m$ by estimating
\begin{align}
\left\Vert AK^{m}x\right\Vert  & =\left\Vert AK\mleft(K^{m-1}x\mright)\right\Vert \leq a\left\Vert AK^{m-1}x\right\Vert +b\left\Vert K^{m-1}x\right\Vert \nonumber \\
 & \leq a\mleft(a^{m-1}\left\Vert Ax\right\Vert +\mleft(m-1\mright)c^{m-2}b\left\Vert x\right\Vert \mright)+b\left\Vert K^{m-1}\right\Vert _{\mathrm{Op}}\left\Vert x\right\Vert \\
 & \leq a^{m}\left\Vert Ax\right\Vert +\mleft(\mleft(m-1\mright)ac^{m-2}+\left\Vert K\right\Vert _{\mathrm{Op}}^{m-1}\mright)b\left\Vert x\right\Vert \nonumber \\
 & \leq a^{m}\left\Vert Ax\right\Vert +mc^{m-1}b\left\Vert x\right\Vert .\nonumber 
\end{align}
$\hfill\square$

We can now conclude the first part of Proposition \ref{prop:ezKAe-zKProperties},
namely that $e^{zK}$ preserves $D\mleft(A\mright)$ for any $z\in\mathbb{C}$,
so that $e^{zK}Ae^{-zK}:D\mleft(A\mright)\rightarrow X$ is well-defined.
For use below we prove the following more general statement:
\begin{prop}
\label{prop:EntireFunctionBound}Under the assumptions on $A$ and
$K$, it holds for any entire function $f\mleft(z\mright)=\sum_{m=0}^{\infty}d_{m}z^{m}$
with $d_{m}\geq0$, $m\in\mathbb{N}_{0}$, that $f\mleft(zK\mright):X\rightarrow X$
also preserves $D\mleft(A\mright)$ for any $z\in\mathbb{C}$, and that
$Af\mleft(zK\mright):D\mleft(A\mright)\rightarrow X$ is $A$-bounded
as
\[
\left\Vert Af\mleft(zK\mright)x\right\Vert \leq f\mleft(a\left|z\right|\mright)\left\Vert Ax\right\Vert +b\left|z\right|f'\mleft(c\left|z\right|\mright)\left\Vert x\right\Vert ,\quad x\in D\mleft(A\mright),
\]
for $c=\max\left\{ a,\left\Vert K\right\Vert _{\mathrm{Op}}\right\} $.
\end{prop}

\textbf{Proof:} By definition of $f\mleft(zK\mright)=\sum_{m=0}^{\infty}d_{m}\mleft(zK\mright)^{m}$
we can for any $x\in D\mleft(A\mright)$ express $f\mleft(zK\mright)x$
as the limit
\begin{equation}
f\mleft(zK\mright)x=\sum_{m=0}^{\infty}d_{m}\mleft(zK\mright)^{m}x=\lim_{k\rightarrow\infty}\sum_{m=0}^{k}d_{m}z^{m}K^{m}x=\lim_{k\rightarrow\infty}y_{k}
\end{equation}
where $y_{k}=\sum_{m=0}^{k}d_{m}z^{m}K^{m}x$, $k\in\mathbb{N}$.

Since $K$ preserves $D\mleft(A\mright)$, so too does $K^{m}$ for
any $m\in\mathbb{N}_{0}$, whence $y_{k}\in D\mleft(A\mright)$ for
every $k\in\mathbb{N}$. In order to prove that $f\mleft(zK\mright)x$
is an element of $D\mleft(A\mright)$ it thus suffices to prove that
the sequence
\begin{equation}
Ay_{k}=\sum_{m=0}^{k}d_{m}z^{m}AK^{m}x,\quad k\in\mathbb{N},
\end{equation}
converges. As $X$ is a Banach space this is ensured if $\sum_{m=0}^{\infty}\left\Vert d_{m}z^{m}AK^{m}x\right\Vert <\infty$.
By the lemma this is indeed the case, as we may estimate
\begin{align}
\sum_{m=0}^{\infty}\left\Vert d_{m}z^{m}AK^{m}x\right\Vert  & =\sum_{m=0}^{\infty}d_{m}\left|z\right|^{m}\left\Vert AK^{m}x\right\Vert \leq\sum_{m=0}^{\infty}d_{m}\left|z\right|^{m}a^{m}\left\Vert Ax\right\Vert +\sum_{m=0}^{\infty}md_{m}\left|z\right|^{m}c^{m-1}b\left\Vert x\right\Vert \nonumber \\
 & =\sum_{m=0}^{\infty}d_{m}\mleft(a\left|z\right|\mright)^{m}\left\Vert Ax\right\Vert +b\left|z\right|\sum_{m=0}^{\infty}md_{m}\mleft(c\left|z\right|\mright)^{m-1}\left\Vert x\right\Vert \\
 & =f\mleft(a\left|z\right|\mright)\left\Vert Ax\right\Vert +b\left|z\right|f'\mleft(c\left|z\right|\mright)\left\Vert x\right\Vert .\nonumber 
\end{align}
We can then similarly conclude the $A$-boundedness as
\begin{equation}
\left\Vert Af\mleft(zK\mright)x\right\Vert =\lim_{n\rightarrow\infty}\left\Vert Ay_{n}\right\Vert \leq\sum_{m=0}^{\infty}\left\Vert d_{m}z^{m}AK^{m}x\right\Vert \leq f\mleft(a\left|z\right|\mright)\left\Vert Ax\right\Vert +b\left|z\right|f'\mleft(c\left|z\right|\mright)\left\Vert x\right\Vert .
\end{equation}
$\hfill\square$

\subsubsection*{Qualitative Properties of $e^{zK}Ae^{-zK}$}

Having ensured that $e^{zK}Ae^{-zK}$ is well-defined, we now show
the second part of Proposition \ref{prop:ezKAe-zKProperties}, i.e.
that $e^{zK}Ae^{-zK}$ also inherits the properties of $A$:
\begin{prop}
Under the assumptions on $A$ and $K$, the operator $e^{zK}Ae^{-zK}:D\mleft(A\mright)\rightarrow X$
is closed for any $z\in\mathbb{C}$.
\end{prop}

\textbf{Proof:} Let $\mleft(x_{k}\mright)_{k=1}^{\infty}\subset D\mleft(A\mright)$
be a sequence such that $x_{k}\rightarrow x$ and $e^{zK}Ae^{-zK}x_{k}\rightarrow y$
for some $x,y\in X$. We must show that $x\in D\mleft(A\mright)$ and
$y=e^{zK}Ae^{-zK}x$.

By boundedness of $K$, hence of $e^{-zK}$, it holds that also $e^{-zK}x_{k}\rightarrow e^{-zK}x$,
and similarly
\begin{equation}
Ae^{-zK}x_{k}=e^{-zK}\mleft(e^{zK}Ae^{-zK}x\mright)\rightarrow e^{-zK}y,
\end{equation}
so by closedness of $A$, $e^{-zK}x\in D\mleft(A\mright)$ and $Ae^{-zK}x=e^{-zK}y$.
Since $e^{zK}$ preserves $D\mleft(A\mright)$, it follows that also
$x=e^{zK}\mleft(e^{-zK}x\mright)\in D\mleft(A\mright)$, and furthermore
\begin{equation}
e^{zK}Ae^{-zK}x=e^{zK}\mleft(Ae^{-zK}x\mright)=e^{zK}\mleft(e^{-zK}y\mright)=y
\end{equation}
as was to be shown.

$\hfill\square$

If $A$ is a self-adjoint operator on a Hilbert space, self-adjointness
is also inherited (for appropriate $tK$):
\begin{prop}
Suppose that $X$ is a Hilbert space, that $A$ is self-adjoint and
that $K$ is skew-symmetric. Then under the assumptions on $A$ and
$K$, the operator $e^{tK}Ae^{-tK}:D\mleft(A\mright)\rightarrow X$
is self-adjoint for any $t\in\mathbb{R}$.
\end{prop}

\textbf{Proof:} The assumptions clearly imply that $e^{tK}Ae^{-tK}$
is at least symmetric. Letting $x\in D\mleft(\mleft(e^{tK}Ae^{-tK}\mright)^{\ast}\mright)$
be arbitrary, we must thus show that $x\in D\mleft(e^{tK}Ae^{-tK}\mright)=D\mleft(A\mright)$.

The assumption is that there exists a $z\in X$ such that
\begin{equation}
\left\langle x,e^{tK}Ae^{-tK}y\right\rangle =\left\langle z,y\right\rangle ,\quad y\in D\mleft(A\mright).
\end{equation}
Rearranging this, we have
\begin{equation}
\left\langle e^{-tK}x,A\mleft(e^{-tK}y\mright)\right\rangle =\left\langle x,e^{tK}Ae^{-tK}y\right\rangle =\left\langle z,y\right\rangle =\left\langle e^{-tK}z,\mleft(e^{-tK}y\mright)\right\rangle ,\quad y\in D\mleft(A\mright),
\end{equation}
which implies that $e^{-tK}x\in D\mleft(A^{\ast}\mright)=D\mleft(A\mright)$
by self-adjointness of $A$, hence $x\in D\mleft(A\mright)$ as in the
previous proposition.

$\hfill\square$

\subsubsection*{Differentiability of $z\protect\mapsto e^{zK}Ae^{-zK}x$}

Finally we come to the last part of Proposition \ref{prop:ezKAe-zKProperties},
which is the statement regarding the mapping $z\mapsto e^{zK}Ae^{-zK}x$
for $x\in D\mleft(A\mright)$. We begin by observing that this is indeed
differentiable:
\begin{prop}
Under the assumptions on $A$ and $K$, it holds for every $x\in D\mleft(A\mright)$
that the mapping $z\mapsto e^{zK}Ae^{-zK}$, $z\in\mathbb{C}$, is
complex differentiable with derivative
\[
\frac{d}{dz}e^{zK}Ae^{-zK}x=e^{zK}\left[K,A\right]e^{-zK}x.
\]
\end{prop}

\textbf{Proof:} The claim is that for any $z_{0}\in\mathbb{C}$
\begin{equation}
\left\Vert \frac{e^{zK}Ae^{-zK}x-e^{z_{0}K}Ae^{-z_{0}K}x}{z-z_{0}}-e^{z_{0}K}\left[K,A\right]e^{-z_{0}K}x\right\Vert \rightarrow0,\quad z\rightarrow z_{0}.
\end{equation}
By the identity
\begin{align}
e^{zK}Ae^{-zK}x-e^{z_{0}K}Ae^{-z_{0}K}x & =\mleft(e^{zK}-e^{z_{0}K}\mright)Ae^{-z_{0}K}x+e^{z_{0}K}A\mleft(e^{-zK}-e^{-z_{0}K}\mright)x+\mleft(e^{zK}-e^{z_{0}K}\mright)A\mleft(e^{-zK}-e^{-z_{0}K}\mright)x\nonumber \\
 & =e^{z_{0}K}\mleft(e^{\mleft(z-z_{0}\mright)K}-1\mright)Ae^{-z_{0}K}x+e^{z_{0}K}A\mleft(e^{-\mleft(z-z_{0}\mright)K}-1\mright)e^{-z_{0}K}x\\
 & +\mleft(e^{zK}-e^{z_{0}K}\mright)A\mleft(e^{-\mleft(z-z_{0}\mright)K}-1\mright)e^{-z_{0}K}x\nonumber 
\end{align}
we see that we can write the argument of $\left\Vert \cdot\right\Vert $
of the previous equation as a sum of three terms:
\begin{align}
 & \quad\;\frac{e^{zK}Ae^{-zK}x-e^{z_{0}K}Ae^{-z_{0}K}x}{z-z_{0}}-e^{z_{0}K}\left[K,A\right]e^{-z_{0}K}x\nonumber \\
 & =e^{z_{0}K}\mleft(\frac{e^{\mleft(z-z_{0}\mright)K}-1}{z-z_{0}}-K\mright)Ae^{-z_{0}K}x+e^{z_{0}K}A\mleft(\frac{e^{-\mleft(z-z_{0}\mright)K}-1}{z-z_{0}}+K\mright)e^{-z_{0}K}x\\
 & +\mleft(e^{zK}-e^{z_{0}K}\mright)A\mleft(\frac{e^{-\mleft(z-z_{0}\mright)K}-1}{z-z_{0}}\mright)e^{-z_{0}K}x.\nonumber 
\end{align}
We show that each term converges to $0$ separately as $z\rightarrow z_{0}$.
First we have
\begin{equation}
\left\Vert e^{z_{0}K}\mleft(\frac{e^{\mleft(z-z_{0}\mright)K}-1}{z-z_{0}}-K\mright)Ae^{-z_{0}K}x\right\Vert \leq\left\Vert e^{z_{0}K}\right\Vert _{\mathrm{Op}}\left\Vert \frac{e^{\mleft(z-z_{0}\mright)K}-1}{z-z_{0}}-K\right\Vert _{\mathrm{Op}}\left\Vert Ae^{-z_{0}K}x\right\Vert 
\end{equation}
which vanishes as $\left.\frac{d}{dz}\right\vert _{z=0}e^{zK}=K$
in operator norm by boundedness of $K$. For the second we estimate
using Proposition \ref{prop:EntireFunctionBound} with $f\mleft(z'\mright)=e^{z'}-1-z'$
and $z'=-\mleft(z-z_{0}\mright)$ that
\begin{align}
 & \quad\;\left\Vert e^{z_{0}K}A\mleft(\frac{e^{-\mleft(z-z_{0}\mright)K}-1}{z-z_{0}}+K\mright)e^{-z_{0}K}x\right\Vert \leq\frac{\left\Vert e^{z_{0}K}\right\Vert _{\mathrm{Op}}}{\left|z-z_{0}\right|}\left\Vert A\mleft(e^{-\mleft(z-z_{0}\mright)K}-1+K\mleft(z-z_{0}\mright)\mright)e^{-z_{0}K}x\right\Vert \nonumber \\
 & \leq\frac{\left\Vert e^{z_{0}K}\right\Vert _{\mathrm{Op}}}{\left|z-z_{0}\right|}\mleft(\mleft(e^{a\left|z-z_{0}\right|}-1-a\left|z-z_{0}\right|\mright)\left\Vert Ae^{-z_{0}K}x\right\Vert +b\left|z-z_{0}\right|\mleft(e^{c\left|z-z_{0}\right|}-1\mright)\left\Vert e^{-z_{0}K}x\right\Vert \mright)\\
 & =\left\Vert e^{z_{0}K}\right\Vert _{\mathrm{Op}}\mleft(\mleft(\frac{e^{a\left|z-z_{0}\right|}-1}{\left|z-z_{0}\right|}-1\mright)\left\Vert Ae^{-z_{0}K}x\right\Vert +b\mleft(e^{c\left|z-z_{0}\right|}-1\mright)\left\Vert e^{-z_{0}K}x\right\Vert \mright)\nonumber 
\end{align}
which likewise vanishes since $z\mapsto e^{z}$ is continuous and
$\left.\frac{d}{dz}\right\vert _{z=0}e^{z}=1$. Similarly, for the
last term we can apply Proposition \ref{prop:EntireFunctionBound}
with $f\mleft(z'\mright)=e^{z'}-1$ to bound
\begin{align}
 & \quad\;\left\Vert \mleft(e^{zK}-e^{z_{0}K}\mright)A\mleft(\frac{e^{-\mleft(z-z_{0}\mright)K}-1}{z-z_{0}}\mright)e^{-z_{0}K}x\right\Vert \leq\frac{\left\Vert e^{zK}-e^{z_{0}K}\right\Vert _{\mathrm{Op}}}{\left|z-z_{0}\right|}\left\Vert A\mleft(e^{-\mleft(z-z_{0}\mright)K}-1\mright)e^{-z_{0}K}x\right\Vert \nonumber \\
 & \leq\frac{\left\Vert e^{zK}-e^{z_{0}K}\right\Vert _{\mathrm{Op}}}{\left|z-z_{0}\right|}\mleft(\mleft(e^{a\left|z-z_{0}\right|}-1\mright)\left\Vert Ae^{-z_{0}K}x\right\Vert +b\left|z-z_{0}\right|e^{c\left|z-z_{0}\right|}\left\Vert e^{-z_{0}K}x\right\Vert \mright)\label{eq:AezK-1Bound}\\
 & =\left\Vert e^{zK}-e^{z_{0}K}\right\Vert _{\mathrm{Op}}\mleft(\frac{e^{a\left|z-z_{0}\right|}-1}{\left|z-z_{0}\right|}\left\Vert Ae^{-z_{0}K}x\right\Vert +be^{c\left|z-z_{0}\right|}\left\Vert e^{-z_{0}K}x\right\Vert \mright)\nonumber 
\end{align}
which vanishes since the term in parenthesis is uniformly bounded
for $z$ near $z_{0}$ by differentiability of $z\mapsto e^{z}$ while
$e^{zK}\rightarrow e^{z_{0}K}$ as $z\rightarrow z_{0}$ by boundedness
of $K$.

$\hfill\square$

A similar argument now shows that the derivative is even continuous:
\begin{prop}
Under the assumptions on $A$ and $K$, it holds for every $x\in D\mleft(A\mright)$
that the mapping $z\mapsto e^{zK}Ae^{-zK}$, $z\in\mathbb{C}$, is
$C^{1}$.
\end{prop}

\textbf{Proof:} We must show that for any $z_{0}\in\mathbb{C}$
\begin{equation}
\left\Vert e^{zK}\left[K,A\right]e^{-zK}x-e^{z_{0}K}\left[K,A\right]e^{-z_{0}K}x\right\Vert \rightarrow0,\quad z\rightarrow z_{0}.
\end{equation}
As in the previous proposition we can write the argument of $\left\Vert \cdot\right\Vert $
as a sum of three terms:
\begin{align}
e^{zK}\left[K,A\right]e^{-zK}x-e^{z_{0}K}\left[K,A\right]e^{-z_{0}K}x & =e^{z_{0}K}\mleft(e^{\mleft(z-z_{0}\mright)K}-1\mright)\left[K,A\right]e^{-z_{0}K}x+e^{z_{0}K}\left[K,A\right]\mleft(e^{-\mleft(z-z_{0}\mright)K}-1\mright)e^{-z_{0}K}x\nonumber \\
 & +\mleft(e^{zK}-e^{z_{0}K}\mright)\left[K,A\right]\mleft(e^{-\mleft(z-z_{0}\mright)K}-1\mright)e^{-z_{0}K}x.
\end{align}
The first term vanishes as
\begin{equation}
\left\Vert e^{z_{0}K}\mleft(e^{\mleft(z-z_{0}\mright)K}-1\mright)\left[K,A\right]e^{-z_{0}K}x\right\Vert \leq\left\Vert e^{z_{0}K}\right\Vert _{\mathrm{Op}}\left\Vert \mleft(e^{\mleft(z-z_{0}\mright)K}-1\mright)\left[K,A\right]e^{-z_{0}K}x\right\Vert 
\end{equation}
and $e^{\mleft(z-z_{0}\mright)K}\rightarrow1$ as $z\rightarrow z_{0}$
while $\left[K,A\right]e^{-z_{0}K}x$ is a fixed vector. For the other
two terms we note that since
\begin{align}
\left\Vert e^{z_{0}K}\left[K,A\right]\mleft(e^{-\mleft(z-z_{0}\mright)K}-1\mright)e^{-z_{0}K}x\right\Vert  & \leq\left\Vert e^{z_{0}K}\right\Vert _{\mathrm{Op}}\left\Vert \left[K,A\right]\mleft(e^{-\mleft(z-z_{0}\mright)K}-1\mright)e^{-z_{0}K}x\right\Vert \\
\left\Vert \mleft(e^{zK}-e^{z_{0}K}\mright)\left[K,A\right]\mleft(e^{-\mleft(z-z_{0}\mright)K}-1\mright)e^{-z_{0}K}x\right\Vert  & \leq\left\Vert \mleft(e^{zK}-e^{z_{0}K}\mright)\right\Vert _{\mathrm{Op}}\left\Vert \left[K,A\right]\mleft(e^{-\mleft(z-z_{0}\mright)K}-1\mright)e^{-z_{0}K}x\right\Vert \nonumber 
\end{align}
it suffices to prove that $\left\Vert \left[K,A\right]\mleft(e^{-\mleft(z-z_{0}\mright)K}-1\mright)e^{-z_{0}K}x\right\Vert \rightarrow0$.
By boundedness of $K$, the assumed $A$-boundedness of $AK$ implies
that $\left[K,A\right]=KA-AK$ is also $A$-bounded, since
\begin{equation}
\left\Vert \left[K,A\right]x\right\Vert \leq\left\Vert KAx\right\Vert +\left\Vert AKx\right\Vert \leq\mleft(\left\Vert K\right\Vert _{\mathrm{Op}}+a\mright)\left\Vert Ax\right\Vert +b\left\Vert x\right\Vert ,\quad x\in D\mleft(A\mright),
\end{equation}
so
\begin{align}
\left\Vert \left[K,A\right]\mleft(e^{-\mleft(z-z_{0}\mright)K}-1\mright)e^{-z_{0}K}x\right\Vert  & \leq\mleft(\left\Vert K\right\Vert _{\mathrm{Op}}+a\mright)\left\Vert A\mleft(e^{-\mleft(z-z_{0}\mright)K}-1\mright)e^{-z_{0}K}x\right\Vert \\
 & +b\left\Vert \mleft(e^{-\mleft(z-z_{0}\mright)K}-1\mright)e^{-z_{0}K}x\right\Vert \nonumber 
\end{align}
and again $\left\Vert \mleft(e^{-\mleft(z-z_{0}\mright)K}-1\mright)e^{-z_{0}K}x\right\Vert \rightarrow0$
while ~$\left\Vert A\mleft(e^{-\mleft(z-z_{0}\mright)K}-1\mright)e^{-z_{0}K}x\right\Vert $
is seen to vanish when $z\rightarrow z_{0}$ as in equation (\ref{eq:AezK-1Bound}).

$\hfill\square$

\section{\label{sec:RiemannSumEstimates}Riemann Sum Estimates}

In this section we establish three results. The first is the following
general bound on sums of the form $\sum_{p\in L_{k}}\lambda_{k,p}^{\beta}$:
\begin{prop}[\ref{prop:RiemannSumEstimates}]
\label{prop:RiemannSumEstimatesAppendix}For any $k\in\mathbb{Z}_{\ast}^{3}$
and $\beta\in\left[-1,0\right]$ it holds that
\[
\sum_{p\in L_{k}}\lambda_{k,p}^{\beta}\leq C\begin{cases}
k_{F}^{2+\beta}\left|k\right|^{1+\beta} & \left|k\right|<2k_{F}\\
k_{F}^{3}\left|k\right|^{2\beta} & \left|k\right|\geq2k_{F}
\end{cases},\quad k_{F}\rightarrow\infty,
\]
for a constant $C>0$ depending only on $\beta$.
\end{prop}

The second result is the precise asymptotic behaviour of $\sum_{p\in L_{k}}\lambda_{k,p}^{-1}$
for small $k$:
\begin{prop}[\ref{prop:betaeq-1Asymptotics}]
\label{prop:betaeq-1AsymptoticsAppendix}For any $\gamma\in\mleft(0,\frac{1}{11}\mright)$
and $k\in\overline{B}\mleft(0,k_{F}^{\gamma}\mright)$ it holds that
\[
\sum_{p\in L_{k}}\lambda_{k,p}^{-1}=2\pi k_{F}+O\mleft(\log\mleft(k_{F}\mright)^{\frac{5}{3}}k_{F}^{\frac{1}{3}\mleft(2+11\gamma\mright)}\mright),\quad k_{F}\rightarrow\infty.
\]
\end{prop}

Finally we prove the following lower bounds for the sums $\sum_{p\in L_{k}}\lambda_{k,p}^{\beta}$:
\begin{prop}[\ref{prop:RiemannSumLowerBound}]
\label{prop:RiemannSumLowerBoundsAppendix}For all $k\in B_{F}$
and $\beta\in\left\{ 0\right\} \cup\mleft[1,\infty\mright)$ it holds
that
\[
\sum_{p\in L_{k}}\lambda_{k,p}^{\beta}\geq ck_{F}^{2+\beta}\left|k\right|^{1+\beta},\quad k_{F}\rightarrow\infty,
\]
for a $c>0$ depending only on $\beta$.
\end{prop}

\subsubsection*{Some General Riemann Sum Estimation Results}

To prove these propositions we first note some general Riemann sum
estimation results.

Let $S\subset\mathbb{R}^{n}$, $n\in\mathbb{N}$, be given, define
for $k\in\mathbb{Z}^{n}$ the translated unit cube $\mathcal{C}_{k}$
by
\begin{equation}
\mathcal{C}_{k}=\left[-2^{-1},2^{-1}\right]+k
\end{equation}
and let $\mathcal{C}_{S}=\bigcup_{k\in S\cap\mathbb{Z}^{n}}\mathcal{C}_{k}$
denote the union of the cubes centered at the lattice points contained
in $S$. We then note the following:
\begin{lem}
Let $f\in C\mleft(\mathcal{C}_{S}\mright)$ be a function which is convex
on $\mathcal{C}_{k}$ for all $k\in S\cap\mathbb{Z}^{n}$. Then
\[
\sum_{k\in S\cap\mathbb{Z}^{n}}f\mleft(k\mright)\leq\int_{\mathcal{C}_{S}}f\mleft(p\mright)\,dp.
\]
\end{lem}

\textbf{Proof:} As a convex function admits a supporting hyperplane
at every interior point of its domain, there exists for every $k\in S\cap\mathbb{Z}^{n}$
a $c\in\mathbb{R}^{n}$ such that
\begin{equation}
f\mleft(p\mright)\geq f\mleft(k\mright)+c\cdot\mleft(p-k\mright),\quad p\in\mathcal{C}_{k},
\end{equation}
and so integration yields
\begin{equation}
\int_{\mathcal{C}_{k}}f\mleft(p\mright)\,dp\ge\int_{\mathcal{C}_{k}}f\mleft(k\mright)\,dp+\int_{\mathcal{C}_{k}}c\cdot\mleft(p-k\mright)\,dp=f\mleft(k\mright)
\end{equation}
as $\int_{\mathcal{C}_{k}}c\cdot\mleft(p-k\mright)\,dp=0$ by antisymmetry.
Consequently
\begin{equation}
\sum_{k\in S\cap\mathbb{Z}^{n}}f\mleft(k\mright)\leq\sum_{k\in S\cap\mathbb{Z}^{n}}\int_{\mathcal{C}_{k}}f\mleft(p\mright)\,dp=\int_{\mathcal{C}_{S}}f\mleft(k\mright)\,dp.
\end{equation}
$\hfill\square$

This lemma lets us replace a sum by an integral, but over an integration
domain $\mathcal{C}_{S}$ which will generally be complicated. An
exception is the $n=1$ case which we record in the following (generalizing
also the statement to any lattice spacing $l$):
\begin{cor}
\label{coro:1DRiemannSumLemma}Let for $a,b\in\mathbb{Z}$ and $l>0$
a convex function $f\in C\mleft(\left[la-\frac{1}{2}l,lb+\frac{1}{2}l\right]\mright)$
be given. Then
\[
\sum_{m=a}^{b}f\mleft(lm\mright)l\leq\int_{la-\frac{1}{2}l}^{lb+\frac{1}{2}l}f\mleft(x\mright)\,dx.
\]
\end{cor}

For $n\neq1$ we instead require an additional step that lets us replace
$\mathcal{C}_{S}$ by a simpler integration domain. Define $S_{+}\subset\mathbb{R}^{n}$
by
\begin{equation}
S_{+}=\left\{ p\in\mathbb{R}^{n}\mid\inf_{q\in S}\left|p-q\right|\leq\frac{\sqrt{n}}{2}\right\} .
\end{equation}
Observe that $\mathcal{C}_{S}\subset S_{+}$: Indeed, for any $p\in\mathcal{C}_{S}$
there exists by assumption a $k\in S\cap\mathbb{Z}^{n}$ such that
$p\in\mathcal{C}_{S}$; consequently
\begin{equation}
\inf_{q\in S}\left|p-q\right|\leq\left|p-k\right|\leq\frac{\sqrt{n}}{2}
\end{equation}
since every point of a unit cube is a distance at most $\frac{\sqrt{n}}{2}$
from its center. The containment $\mathcal{C}_{S}\subset S_{+}$ and
the lemma now easily imply the following:
\begin{cor}
\label{coro:nDRiemannSumLemma}Let $f\in C\mleft(S_{+}\mright)$ be
a positive function which is convex on $\mathcal{C}_{k}$ for all
$k\in S\cap\mathbb{Z}^{n}$. Then
\[
\sum_{k\in S\cap\mathbb{Z}^{n}}f\mleft(k\mright)\leq\int_{S_{+}}f\mleft(p\mright)\,dp.
\]
\end{cor}

Note that in the particular case that $f$ is identically $1$ this
yields a bound on the lattice points contained in $S$:
\begin{equation}
\left|S\cap\mathbb{Z}^{n}\right|\leq\mathrm{Vol}\mleft(S_{+}\mright).
\end{equation}

\subsection{Simple Upper Bounds for $\beta\in\left[-1,0\right]$}

We now consider the sums $\sum_{p\in L_{k}}\lambda_{k,p}^{\beta}$.
In this subsection we prove the $\beta\in\mleft(-1,0\mright]$ statement
of Proposition \ref{prop:RiemannSumEstimatesAppendix}, i.e. that
\begin{equation}
\sum_{p\in L_{k}}\lambda_{k,p}^{\beta}\leq C\begin{cases}
k_{F}^{2+\beta}\left|k\right|^{1+\beta} & \left|k\right|<2k_{F}\\
k_{F}^{3}\left|k\right|^{2\beta} & \left|k\right|\geq2k_{F}
\end{cases},\quad\beta\in\mleft(-1,0\mright],
\end{equation}
as well as the partial statement for $\beta=-1$ that
\begin{equation}
\sum_{p\in L_{k}}\lambda_{k,p}^{-1}\leq C\begin{cases}
\mleft(1+\left|k\right|^{-1}\log\mleft(k_{F}\mright)\mright)k_{F} & \left|k\right|<2k_{F}\\
k_{F}^{3}\left|k\right|^{-2} & \left|k\right|\geq2k_{F}
\end{cases}.
\end{equation}
The improvement of the latter estimate to $\sum_{p\in L_{k}}\lambda_{k,p}^{-1}\leq Ck_{F}$
for $\left|k\right|<2k_{F}$ will be handled by more precise estimates
later in the section.

Recall that the lunes $L_{k}$ are given by
\begin{equation}
L_{k}=\left\{ p\in\mathbb{Z}^{3}\mid\left|p-k\right|\leq k_{F}<\left|p\right|\right\} =S\cap\mathbb{Z}^{3}
\end{equation}
where $S=\overline{B}\mleft(k,k_{F}\mright)\backslash\overline{B}\mleft(0,k_{F}\mright)$.
The relevant integrand for our Riemann sums,
\begin{equation}
p\mapsto\lambda_{k,p}^{\beta}=\mleft(\frac{1}{2}\mleft(\left|p\right|^{2}-\left|p-k\right|^{2}\mright)\mright)^{\beta}=\left|k\right|^{\beta}\mleft(\hat{k}\cdotp p-\frac{1}{2}\left|k\right|\mright)^{\beta}
\end{equation}
is convex on $\left\{ p\in\mathbb{R}^{3}\mid\hat{k}\cdot p>\frac{1}{2}\left|k\right|\right\} $
but singular when $\hat{k}\cdot p=\frac{1}{2}\left|k\right|$. We
must therefore introduce a cut-off to the Riemann sum $\sum_{p\in L_{k}}\lambda_{k,p}^{\beta}$:
We write $L_{k}=L_{k}^{1}\cup L_{k}^{2}$ for
\begin{align}
L_{k}^{1} & =\left\{ p\in L_{k}\mid\hat{k}\cdot p\leq\frac{1}{2}\left|k\right|+1+\frac{\sqrt{3}}{2}\right\} \\
L_{k}^{2} & =\left\{ p\in L_{k}\mid\hat{k}\cdot p>\frac{1}{2}\left|k\right|+1+\frac{\sqrt{3}}{2}\right\} .\nonumber 
\end{align}
Then also $L_{k}^{i}=S^{i}\cap\mathbb{Z}^{3}$, $i=1,2$, for
\begin{align}
S^{1} & =\left\{ p\in S\mid\hat{k}\cdot p\leq\frac{1}{2}\left|k\right|+1+\frac{\sqrt{3}}{2}\right\} \\
S^{2} & =\left\{ p\in S\mid\hat{k}\cdot p>\frac{1}{2}\left|k\right|+1+\frac{\sqrt{3}}{2}\right\} \nonumber 
\end{align}
so we can by Corollary \ref{coro:nDRiemannSumLemma} estimate that
\begin{align}
\sum_{p\in L_{k}}\lambda_{k,p}^{\beta} & =\sum_{p\in L_{k}^{1}}\lambda_{k,p}^{\beta}+\sum_{p\in L_{k}^{2}}\lambda_{k,p}^{\beta}\leq\mleft(\sup_{p\in L_{k}^{1}}\lambda_{k,p}^{\beta}\mright)\left|L_{k}^{1}\right|+\int_{S_{+}^{2}}\left|k\right|^{\beta}\mleft(\hat{k}\cdotp p-\frac{1}{2}\left|k\right|\mright)^{\beta}dp\label{eq:FirstRiemannSumSimplification}\\
 & \leq2^{-\beta}\,\mathrm{Vol}\mleft(S_{+}^{1}\mright)+\left|k\right|^{\beta}\int_{S_{+}^{2}}\mleft(\hat{k}\cdotp p-\frac{1}{2}\left|k\right|\mright)^{\beta}dp.\nonumber 
\end{align}
Here we also used the observation that
\begin{equation}
\lambda_{k,p}=\frac{1}{2}\mleft(\left|p\right|^{2}-\left|p-k\right|^{2}\mright)\geq\frac{1}{2}
\end{equation}
for all $k\in\mathbb{Z}_{\ast}^{3}$ and $p\in L_{k}$, as $\left|p\right|^{2}$
and $\left|p-k\right|^{2}$ are then non-equal integers.

To estimate $\mathrm{Vol}\mleft(S_{+}^{1}\mright)$ and the integral
over $S_{+}^{2}$ we will replace these by simpler sets once more:
Let $\mathcal{S}\subset\mathbb{R}^{3}$ be given by
\begin{equation}
\mathcal{S}=\overline{B}\mleft(k,k_{F}+\frac{\sqrt{3}}{2}\mright)\backslash B\mleft(0,k_{F}-\frac{\sqrt{3}}{2}\mright)
\end{equation}
and define the subsets $\mathcal{S}^{1},\mathcal{S}^{2}\subset\mathcal{S}$
by
\begin{align}
\mathcal{S}^{1} & =\left\{ p\in\mathcal{S}\mid-\frac{\sqrt{3}}{2}\leq\hat{k}\cdot p-\frac{1}{2}\left|k\right|\leq1+\sqrt{3}\right\} \\
\mathcal{S}^{2} & =\left\{ p\in\mathcal{S}\mid1\leq\hat{k}\cdot p-\frac{1}{2}\left|k\right|\right\} .\nonumber 
\end{align}
Then we have the following:
\begin{prop}
It holds that
\[
S_{+}^{1}\subset\mathcal{S}^{1}\quad\text{and}\quad S_{+}^{2}\subset\mathcal{S}^{2}.
\]
\end{prop}

\textbf{Proof:} We first show that $S_{+}\subset\mathcal{S}$: Let
$p\in S_{+}=\left\{ p'\in\mathbb{R}^{3}\mid\inf_{q\in S}\left|p'-q\right|\leq\frac{\sqrt{3}}{2}\right\} $
be arbitrary. Then we can for any $q\in S$ estimate that
\begin{align}
\left|p\right| & \geq\left|q\right|-\left|p-q\right|>k_{F}-\left|p-q\right|\\
\left|p-k\right| & \leq\left|q-k\right|+\left|p-q\right|\leq k_{F}+\left|p-q\right|\nonumber 
\end{align}
whence taking the supremum and infimum over $q\in S$ yields
\begin{equation}
\left|p\right|\geq k_{F}-\frac{\sqrt{3}}{2},\quad\left|p-k\right|\leq k_{F}+\frac{\sqrt{3}}{2},
\end{equation}
which is to say that $p\in\mathcal{S}$ as claimed. Supposing then
that $p\in S_{+}^{1}$ we furthermore note that for any $q\in S^{1}$,
Cauchy-Schwarz implies that
\begin{equation}
\hat{k}\cdot p-\frac{1}{2}\left|k\right|=\hat{k}\cdot q-\frac{1}{2}\left|k\right|+\hat{k}\cdot\mleft(p-q\mright)\leq1+\frac{\sqrt{3}}{2}+\left|p-q\right|
\end{equation}
and similarly
\begin{equation}
\hat{k}\cdot p-\frac{1}{2}\left|k\right|=\hat{k}\cdot q-\frac{1}{2}\left|k\right|+\hat{k}\cdot\mleft(p-q\mright)\geq-\left|p-q\right|
\end{equation}
so taking the supremum and infimum over $q\in S^{1}$ again yields
\begin{equation}
-\frac{\sqrt{3}}{2}\leq\hat{k}\cdot p-\frac{1}{2}\left|k\right|\leq1+\sqrt{3}
\end{equation}
i.e. $p\in\mathcal{S}^{1}$. That $S_{+}^{2}\subset\mathcal{S}^{2}$
follows similarly.

$\hfill\square$

By this proposition it now follows from equation (\ref{eq:FirstRiemannSumSimplification})
that
\begin{equation}
\sum_{p\in L_{k}}\lambda_{k,p}^{\beta}\leq2^{-\beta}\,\mathrm{Vol}\mleft(\mathcal{S}^{1}\mright)+\left|k\right|^{\beta}\int_{\mathcal{S}^{2}}\mleft(\hat{k}\cdotp p-\frac{1}{2}\left|k\right|\mright)^{\beta}dp.
\end{equation}
To compute $\mathrm{Vol}\mleft(\mathcal{S}^{1}\mright)$ and the integral
over $\mathcal{S}^{2}$ we will integrate along the $\hat{k}$-axis,
so we must now consider the behaviour of the ``slices''
\begin{equation}
\mathcal{S}_{t}=\mathcal{S}\cap\left\{ p\in\mathbb{R}^{3}\mid\hat{k}\cdot p=t\right\} .
\end{equation}

\subsubsection*{The Case $\left|k\right|<2k_{F}$}

Suppose first that $\left|k\right|<2k_{F}$. Then when moving along
the $\hat{k}$-axis, it holds that
\begin{equation}
\inf\mleft(\left\{ t\mid\mathcal{S}_{t}\neq\emptyset\right\} \mright)=\begin{cases}
-\mleft(k_{F}+\frac{\sqrt{3}}{2}\mright)+\left|k\right| & \left|k\right|\leq\sqrt{3}\\
\frac{1}{2}\left|k\right|-\left|k\right|^{-1}\sqrt{3}\,k_{F} & \left|k\right|>\sqrt{3}
\end{cases}
\end{equation}
where the first case corresponds to the case that $B\mleft(0,k_{F}-\frac{\sqrt{3}}{2}\mright)$
is entirely contained in $\overline{B}\mleft(0,k_{F}+\frac{\sqrt{3}}{2}\mright)$.

As the lower end of $\mathcal{S}^{1}$ is at $t=\frac{1}{2}\left|k\right|-\frac{\sqrt{3}}{2}$,
we need not consider this case, since $\left\{ \hat{k}\cdotp p=\frac{1}{2}\left|k\right|-\frac{\sqrt{3}}{2}\right\} $
will intersect both $\overline{B}\mleft(k,k_{F}+\frac{\sqrt{3}}{2}\mright)$
and $B\mleft(0,k_{F}-\frac{\sqrt{3}}{2}\mright)$ anyway. In this case
the slice $\mathcal{S}_{t}$ forms an annulus, and elementary trigonometry
shows that
\begin{align}
\mathrm{Area}\mleft(\mathcal{S}_{t}\mright) & =\pi\mleft(\mleft(k_{F}+\frac{\sqrt{3}}{2}\mright)^{2}-\mleft(t-\left|k\right|\mright)^{2}\mright)-\pi\mleft(\mleft(k_{F}-\frac{\sqrt{3}}{2}\mright)^{2}-t^{2}\mright)\\
 & =\pi\mleft(2\sqrt{3}\,k_{F}-\mleft(\left|k\right|^{2}-2\left|k\right|t\mright)\mright)=2\pi\mleft(\left|k\right|\mleft(t-\frac{1}{2}\left|k\right|\mright)+\sqrt{3}\,k_{F}\mright)\nonumber 
\end{align}
for $\frac{1}{2}\left|k\right|-\frac{\sqrt{3}}{2}\leq t\leq k_{F}-\frac{\sqrt{3}}{2}$,
with $t=k_{F}-\frac{\sqrt{3}}{2}$ corresponding to the ``upper end''
of $B\mleft(0,k_{F}-\frac{\sqrt{3}}{2}\mright)$. Thereafter the planes
intersect only $\overline{B}\mleft(k,k_{F}+\frac{\sqrt{3}}{2}\mright)$,
whence
\begin{align}
\mathrm{Area}\mleft(\mathcal{S}_{t}\mright) & =\pi\mleft(\mleft(k_{F}+\frac{\sqrt{3}}{2}\mright)^{2}-\mleft(t-\left|k\right|\mright)^{2}\mright)=\pi\mleft(\mleft(k_{F}+\frac{\sqrt{3}}{2}\mright)^{2}-t^{2}+2\left|k\right|\mleft(t-\frac{1}{2}\left|k\right|\mright)\mright)\nonumber \\
 & =2\pi\mleft(\left|k\right|\mleft(t-\frac{1}{2}\left|k\right|\mright)+\sqrt{3}\,k_{F}\mright)+\pi\mleft(\mleft(k_{F}-\frac{\sqrt{3}}{2}\mright)^{2}-t^{2}\mright)\\
 & \leq2\pi\mleft(\left|k\right|\mleft(t-\frac{1}{2}\left|k\right|\mright)+\sqrt{3}\,k_{F}\mright)\nonumber 
\end{align}
for $k_{F}-\frac{\sqrt{3}}{2}\leq t\leq k_{F}+\frac{\sqrt{3}}{2}+\left|k\right|$.

With this we can now prove the $\left|k\right|<2k_{F}$ bounds:
\begin{prop}
For all $k\in B\mleft(0,2k_{F}\mright)\cap\mathbb{Z}_{\ast}^{3}$ and
$\beta\in\left[-1,0\right]$ it holds that
\[
\sum_{p\in L_{k}}\lambda_{k,p}^{\beta}\leq C\begin{cases}
k_{F}^{2+\beta}\left|k\right|^{1+\beta} & \beta\in\mleft(-1,0\mright]\\
\mleft(1+\left|k\right|^{-1}\log\mleft(k_{F}\mright)\mright)k_{F} & \beta=-1
\end{cases},\quad k_{F}\rightarrow\infty,
\]
for a constant $C>0$ depending only on $\beta$.
\end{prop}

\textbf{Proof:} Recall that
\begin{equation}
\sum_{p\in L_{k}}\lambda_{k,p}^{\beta}\leq2^{-\beta}\,\mathrm{Vol}\mleft(\mathcal{S}^{1}\mright)+\left|k\right|^{\beta}\int_{\mathcal{S}^{2}}\mleft(\hat{k}\cdotp p-\frac{1}{2}\left|k\right|\mright)^{\beta}dp.
\end{equation}
The volume of $\mathcal{S}^{1}$ obeys
\begin{align}
\mathrm{Vol}\mleft(\mathcal{S}^{1}\mright) & =\int_{\frac{1}{2}\left|k\right|-\frac{\sqrt{3}}{2}}^{\frac{1}{2}\left|k\right|+1+\sqrt{3}}\mathrm{Area}\mleft(\mathcal{S}_{t}\mright)\,dt\leq2\pi\int_{\frac{1}{2}\left|k\right|-\frac{\sqrt{3}}{2}}^{\frac{1}{2}\left|k\right|+1+\sqrt{3}}\mleft(\left|k\right|\mleft(t-\frac{1}{2}\left|k\right|\mright)+\sqrt{3}\,k_{F}\mright)\,dt\\
 & =2\pi\int_{-\frac{\sqrt{3}}{2}}^{1+\sqrt{3}}\mleft(\left|k\right|t+\sqrt{3}\,k_{F}\mright)\,dt\leq C\mleft(\left|k\right|+k_{F}\mright)\leq Ck_{F},\quad k_{F}\rightarrow\infty,\nonumber 
\end{align}
which is $O\mleft(k_{F}^{2+\beta}\left|k\right|^{1+\beta}\mright)$
for all $\beta\in\left[-1,0\right]$. For $\beta\in\mleft(-1,0\mright]$
the integral is
\begin{align}
\int_{\mathcal{S}^{2}}\mleft(\hat{k}\cdotp p-\frac{1}{2}\left|k\right|\mright)^{\beta}dp & =\int_{\frac{1}{2}\left|k\right|+1}^{k_{F}+\frac{\sqrt{3}}{2}+\left|k\right|}\mleft(t-\frac{1}{2}\left|k\right|\mright)^{\beta}\mathrm{Area}\mleft(\mathcal{S}_{t}\mright)\,dt\nonumber \\
 & \leq2\pi\int_{\frac{1}{2}\left|k\right|+1}^{k_{F}+\frac{\sqrt{3}}{2}+\left|k\right|}\mleft(t-\frac{1}{2}\left|k\right|\mright)^{\beta}\mleft(\left|k\right|\mleft(t-\frac{1}{2}\left|k\right|\mright)+\sqrt{3}\,k_{F}\mright)\,dt\nonumber \\
 & =2\pi\mleft(\left|k\right|\int_{1}^{k_{F}+\frac{\sqrt{3}}{2}+\frac{1}{2}\left|k\right|}t^{1+\beta}dt+\sqrt{3}\,k_{F}\int_{1}^{k_{F}+\frac{\sqrt{3}}{2}+\frac{1}{2}\left|k\right|}t^{\beta}dt\mright)\\
 & \leq2\pi\mleft(\frac{\left|k\right|}{2+\beta}\mleft(k_{F}+\frac{\sqrt{3}}{2}+\frac{1}{2}\left|k\right|\mright)^{2+\beta}+\frac{\sqrt{3}\,k_{F}}{1+\beta}\mleft(k_{F}+\frac{\sqrt{3}}{2}+\frac{1}{2}\left|k\right|\mright)^{1+\beta}\mright)\nonumber \\
 & \leq2\pi\mleft(\frac{1}{2+\beta}k_{F}^{2+\beta}\left|k\right|+\frac{\sqrt{3}}{1+\beta}k_{F}^{2+\beta}\mright)\leq Ck_{F}^{2+\beta}\left|k\right|,\quad k_{F}\rightarrow\infty,\nonumber 
\end{align}
while the $\beta=-1$ case is
\begin{align}
\int_{\mathcal{S}^{2}}\mleft(\hat{k}\cdotp p-\frac{1}{2}\left|k\right|\mright)^{-1}dp & \leq2\pi\mleft(\left|k\right|\int_{1}^{k_{F}+\frac{\sqrt{3}}{2}+\frac{1}{2}\left|k\right|}1\,dt+\sqrt{3}\,k_{F}\int_{1}^{k_{F}+\frac{\sqrt{3}}{2}+\frac{1}{2}\left|k\right|}t^{-1}dt\mright)\nonumber \\
 & \leq2\pi\mleft(\left|k\right|\mleft(k_{F}+\frac{\sqrt{3}}{2}+\frac{1}{2}\left|k\right|\mright)+\sqrt{3}\,k_{F}\log\mleft(k_{F}+\frac{\sqrt{3}}{2}+\frac{1}{2}\left|k\right|\mright)\mright)\\
 & \leq C\left|k\right|\mleft(1+\left|k\right|^{-1}\log\mleft(k_{F}\mright)\mright)k_{F}.\nonumber 
\end{align}
Combining the estimates yields the claim.

$\hfill\square$

\subsubsection*{The Case $\left|k\right|\protect\geq2k_{F}$}

Now suppose instead that $\left|k\right|\geq2k_{F}$. In this case
the lune $S=\overline{B}\mleft(k,k_{F}\mright)\backslash\overline{B}\mleft(0,k_{F}\mright)$
degenerates into a ball, and so we simply have that
\begin{equation}
S_{+}=\mathcal{S}=\overline{B}\mleft(k,k_{F}+\frac{\sqrt{3}}{2}\mright).
\end{equation}
Now, if $\frac{1}{2}\left|k\right|\geq k_{F}+1+\frac{\sqrt{3}}{2}$
then every $p\in\mathcal{S}$ satisfies $\hat{k}\cdot p-\frac{1}{2}\left|k\right|\geq1$
and the cut-off set $\mathcal{S}^{1}$ is unnecessary. If this is
not the case then it still holds that
\begin{equation}
\sum_{p\in L_{k}}\lambda_{k,p}^{\beta}\leq2^{-\beta}\,\mathrm{Vol}\mleft(\mathcal{S}^{1}\mright)+\left|k\right|^{\beta}\int_{\mathcal{S}^{2}}\mleft(\hat{k}\cdotp p-\frac{1}{2}\left|k\right|\mright)^{\beta}dp
\end{equation}
for
\begin{align}
\mathcal{S}^{1} & =\left\{ p\in\mathcal{S}\mid\hat{k}\cdot p-\frac{1}{2}\left|k\right|\leq1+\sqrt{3}\right\} \\
\mathcal{S}^{2} & =\left\{ p\in\mathcal{S}\mid1\leq\hat{k}\cdot p-\frac{1}{2}\left|k\right|\right\} ,\nonumber 
\end{align}
and we may easily estimate $\mathrm{Vol}\mleft(\mathcal{S}^{1}\mright)$
as $\mathcal{S}^{1}$ is now seen to be a spherical cap of radius
$k_{F}+\frac{\sqrt{3}}{2}$ and height
\begin{equation}
\mleft(\frac{1}{2}\left|k\right|+1+\sqrt{3}\mright)-\mleft(\left|k\right|-k_{F}-\frac{\sqrt{3}}{2}\mright)=k_{F}-\frac{1}{2}\left|k\right|+1+\frac{3\sqrt{3}}{2}\leq1+\frac{3\sqrt{3}}{2}
\end{equation}
whence
\begin{equation}
\mathrm{Vol}\mleft(\mathcal{S}^{1}\mright)\leq\frac{\pi}{3}\mleft(1+\frac{3\sqrt{3}}{2}\mright)\mleft(3\mleft(k_{F}+\frac{\sqrt{3}}{2}\mright)-1-\frac{3\sqrt{3}}{2}\mright)\leq Ck_{F}
\end{equation}
which is again $O\mleft(k_{F}^{2+\beta}\left|k\right|^{1+\beta}\mright)$.
We thus only need to estimate the integral for the $\left|k\right|\geq2k_{F}$
bounds:
\begin{prop}
For all $k\in\mathbb{Z}_{\ast}^{3}\backslash B\mleft(0,2k_{F}\mright)$
and $\beta\in\left[-1,0\right]$ it holds that
\[
\sum_{p\in L_{k}}\lambda_{k,p}^{\beta}\leq Ck_{F}^{3}\left|k\right|^{2\beta},\quad k_{F}\rightarrow\infty,
\]
for a constant $C>0$ depending only on $\beta$.
\end{prop}

\textbf{Proof:} We again note that
\begin{equation}
\mathrm{Area}\mleft(\mathcal{S}_{t}\mright)=\pi\mleft(\mleft(k_{F}+\frac{\sqrt{3}}{2}\mright)^{2}-\mleft(t-\left|k\right|\mright)^{2}\mright),
\end{equation}
now for $\left|k\right|-k_{F}-\frac{\sqrt{3}}{2}\leq t\leq\left|k\right|+k_{F}+\frac{\sqrt{3}}{2}$.
If $\frac{1}{2}\left|k\right|\leq k_{F}+1+\frac{\sqrt{3}}{2}$ we
just saw that the contribution coming from the cut-off set $\mathcal{S}^{1}$
is negligible, while the integral term is
\begin{align}
\left|k\right|^{\beta}\int_{\mathcal{S}^{2}}\mleft(\hat{k}\cdotp p-\frac{1}{2}\left|k\right|\mright)^{\beta}dp & =\left|k\right|^{\beta}\int_{\frac{1}{2}\left|k\right|+1}^{k_{F}+\frac{\sqrt{3}}{2}+\left|k\right|}\mleft(t-\frac{1}{2}\left|k\right|\mright)^{\beta}\mathrm{Area}\mleft(\mathcal{S}_{t}\mright)\,dt\\
 & \leq Ck_{F}^{2+\beta}\left|k\right|^{1+\beta},\quad k_{F}\rightarrow\infty,\nonumber 
\end{align}
as calculated in the previous proposition, which is $O\mleft(k_{F}^{3}\left|k\right|^{2\beta}\mright)$
since $\frac{1}{2}\left|k\right|\leq k_{F}+1+\frac{\sqrt{3}}{2}$.
(Here we also used that for $\beta=-1$, the term $\left|k\right|^{-1}\log\mleft(k_{F}\mright)$
can be disregarded when $\left|k\right|\geq2k_{F}$.)

If $\frac{1}{2}\left|k\right|>k_{F}+1+\frac{\sqrt{3}}{2}$ then we
simply have
\begin{equation}
\sum_{p\in L_{k}}\lambda_{k,p}^{\beta}\leq\left|k\right|^{\beta}\int_{\mathcal{S}}\mleft(\hat{k}\cdot p-\frac{1}{2}\left|k\right|\mright)^{\beta}dp=\left|k\right|^{\beta}\int_{\left|k\right|-k_{F}-\frac{\sqrt{3}}{2}}^{\left|k\right|+k_{F}+\frac{\sqrt{3}}{2}}\mleft(t-\frac{1}{2}\left|k\right|\mright)^{\beta}\mathrm{Area}\mleft(\mathcal{S}_{t}\mright)\,dt,
\end{equation}
and noting that
\begin{equation}
\mleft(t-\left|k\right|\mright)^{2}=\mleft(t-\frac{1}{2}\left|k\right|\mright)^{2}-\left|k\right|\mleft(t-\frac{1}{2}\left|k\right|\mright)+\frac{1}{4}\left|k\right|^{2}
\end{equation}
we can now estimate $\mathrm{Area}\mleft(\mathcal{S}_{t}\mright)$ as
\begin{align}
\mathrm{Area}\mleft(\mathcal{S}_{t}\mright) & =\pi\mleft(\mleft(k_{F}+\frac{\sqrt{3}}{2}\mright)^{2}-\mleft(t-\left|k\right|\mright)^{2}\mright)\nonumber \\
 & =\pi\mleft(\mleft(k_{F}+\frac{\sqrt{3}}{2}\mright)^{2}-\mleft(t-\frac{1}{2}\left|k\right|\mright)^{2}+\left|k\right|\mleft(t-\frac{1}{2}\left|k\right|\mright)-\frac{1}{4}\left|k\right|^{2}\mright)\\
 & =\pi\mleft(\left|k\right|\mleft(t-\frac{1}{2}\left|k\right|^{2}\mright)-\mleft(\frac{1}{4}\left|k\right|^{2}-\mleft(k_{F}+\frac{\sqrt{3}}{2}\mright)^{2}\mright)-\mleft(t-\frac{1}{2}\left|k\right|\mright)^{2}\mright)\nonumber \\
 & \leq\pi\left|k\right|\mleft(t-\frac{1}{2}\left|k\right|\mright).\nonumber 
\end{align}
Consequently
\begin{align}
\sum_{p\in L_{k}}\lambda_{k,p}^{\beta} & \leq\pi\left|k\right|^{1+\beta}\int_{\left|k\right|-k_{F}-\frac{\sqrt{3}}{2}}^{\left|k\right|+k_{F}+\frac{\sqrt{3}}{2}}\mleft(t-\frac{1}{2}\left|k\right|\mright)^{1+\beta}dt=\pi\left|k\right|^{1+\beta}\int_{\frac{1}{2}\left|k\right|-k_{F}-\frac{\sqrt{3}}{2}}^{\frac{1}{2}\left|k\right|+k_{F}+\frac{\sqrt{3}}{2}}t^{1+\beta}dt\nonumber \\
 & =\frac{\pi}{2+\beta}\left|k\right|^{1+\beta}\mleft(\mleft(\frac{1}{2}\left|k\right|+k_{F}+\frac{\sqrt{3}}{2}\mright)^{2+\beta}-\mleft(\frac{1}{2}\left|k\right|-k_{F}-\frac{\sqrt{3}}{2}\mright)^{2+\beta}\mright)\\
 & \leq C\left|k\right|^{3+2\beta}.\nonumber 
\end{align}
If additionally $\left|k\right|\leq3\,k_{F}$ (say) then this is $O\mleft(k_{F}^{3}\left|k\right|^{2\beta}\mright)$,
and if not then we can nonetheless trivially estimate
\begin{align}
\sum_{p\in L_{k}}\lambda_{k,p}^{\beta} & \leq\left|k\right|^{\beta}\int_{\mathcal{S}}\mleft(\hat{k}\cdot p-\frac{1}{2}\left|k\right|\mright)^{\beta}dp\leq\left|k\right|^{\beta}\mleft(\inf_{p\in\mathcal{S}}\mleft(\hat{k}\cdot p-\frac{1}{2}\left|k\right|\mright)\mright)^{\beta}\int_{\mathcal{S}}1\,dp\nonumber \\
 & \leq\left|k\right|^{\beta}\mleft(\left|k\right|-k_{F}-\frac{\sqrt{3}}{2}-\frac{1}{2}\left|k\right|\mright)^{\beta}\mathrm{Vol}\mleft(\overline{B}\mleft(0,k_{F}+\frac{\sqrt{3}}{2}\mright)\mright)\\
 & \leq Ck_{F}^{3}\left|k\right|^{\beta}\mleft(\frac{1}{2}\left|k\right|-\frac{1}{3}\left|k\right|-\frac{\sqrt{3}}{2}\mright)^{\beta}\leq Ck_{F}^{3}\left|k\right|^{2\beta},\quad k_{F}\rightarrow\infty,\nonumber 
\end{align}
for the claim.

$\hfill\square$

\subsection{Some Lattice Concepts}

To improve upon our bound on $\sum_{p\in L_{k}}\lambda_{k,p}^{-1}$
(and in particular to establish its asymptotic behaviour) we will
need some results regarding lattices, which we now review.

A lattice $\Lambda$ in a real $n$-dimensional vector space $V$
is defined to be a subset of $V$ with the following property: There
exists a basis $\mleft(v_{i}\mright)_{i=1}^{n}$ of $V$ such that $\Lambda$
equals the integral span of $\mleft(v_{i}\mright)_{i=1}^{n}$, i.e.
\begin{equation}
\Lambda=\left\{ \sum_{i=1}^{n}m_{i}v_{i}\mid m_{1},\ldots,m_{n}\in\mathbb{Z}\right\} .
\end{equation}
Given a basis $\mleft(v_{i}\mright)_{i=1}^{n}$, the right-hand side
of this equation always defines a lattice, called the lattice generated
by $\mleft(v_{i}\mright)_{i=1}^{n}$, and denoted by $\left\langle v_{1},\ldots,v_{n}\right\rangle $.
Two different bases $\mleft(v_{i}\mright)_{i=1}^{n}$ and $\mleft(w_{i}\mright)_{i=1}^{n}$
may generate the same lattice, in which case the following is well-known:
\begin{prop}
Let $\mleft(v_{i}\mright)_{i=1}^{n}$ and $\mleft(w_{i}\mright)_{i=1}^{n}$
be bases of $V$. Then $\left\langle v_{1},\ldots,v_{n}\right\rangle =\left\langle w_{1},\ldots,w_{n}\right\rangle $
if and only if the transition matrix $\mleft(T_{i,j}\mright)_{i,j=1}^{n}$,
defined by the relation
\[
w_{i}=\sum_{j=1}^{n}T_{i,j}v_{j},\quad1\leq i\leq n,
\]
has integer entries and determinant $\pm1$.
\end{prop}

This result has an important consequence when $V$ is endowed with
an inner product: Then one can define the hypervolume of the parallelepiped
spanned by $\mleft(v_{i}\mright)_{i=1}^{n}$ by
\begin{equation}
\left|\det\mleft(\begin{array}{ccc}
\left\langle e_{1},v_{1}\right\rangle  & \cdots & \left\langle e_{n},v_{1}\right\rangle \\
\vdots & \ddots & \vdots\\
\left\langle e_{1},v_{n}\right\rangle  & \cdots & \left\langle e_{n},v_{n}\right\rangle 
\end{array}\mright)\right|=\sqrt{\det\mleft(\begin{array}{ccc}
\left\langle v_{1},v_{1}\right\rangle  & \cdots & \left\langle v_{n},v_{1}\right\rangle \\
\vdots & \ddots & \vdots\\
\left\langle v_{1},v_{n}\right\rangle  & \cdots & \left\langle v_{n},v_{n}\right\rangle 
\end{array}\mright)}
\end{equation}
for any orthonormal basis $\mleft(e_{i}\mright)_{i=1}^{n}$ (the expression
on the right-hand side follows by orthonormal expansion). It is however
a general fact that if two bases $\mleft(v_{i}\mright)_{i=1}^{n}$ and
$\mleft(w_{i}\mright)_{i=1}^{n}$ are related by a transition matrix
$T$, then
\begin{equation}
\det\mleft(\begin{array}{ccc}
\left\langle e_{1},w_{1}\right\rangle  & \cdots & \left\langle e_{n},w_{1}\right\rangle \\
\vdots & \ddots & \vdots\\
\left\langle e_{1},w_{n}\right\rangle  & \cdots & \left\langle e_{n},w_{n}\right\rangle 
\end{array}\mright)=\det\mleft(T\mright)\det\mleft(\begin{array}{ccc}
\left\langle e_{1},v_{1}\right\rangle  & \cdots & \left\langle e_{n},v_{1}\right\rangle \\
\vdots & \ddots & \vdots\\
\left\langle e_{1},v_{n}\right\rangle  & \cdots & \left\langle e_{n},v_{n}\right\rangle 
\end{array}\mright)
\end{equation}
whence one concludes the following:
\begin{prop}
Let $\Lambda$ be a lattice in $\mleft(V,\left\langle \cdot,\cdot\right\rangle \mright)$
and let $\mleft(v_{i}\mright)_{i=1}^{n}$ generate $\Lambda$. Then
the quantity
\[
d\mleft(\Lambda\mright)=\left|\det\mleft(\begin{array}{ccc}
\left\langle e_{1},v_{1}\right\rangle  & \cdots & \left\langle e_{n},v_{1}\right\rangle \\
\vdots & \ddots & \vdots\\
\left\langle e_{1},v_{n}\right\rangle  & \cdots & \left\langle e_{n},v_{n}\right\rangle 
\end{array}\mright)\right|=\sqrt{\det\mleft(\begin{array}{ccc}
\left\langle v_{1},v_{1}\right\rangle  & \cdots & \left\langle v_{n},v_{1}\right\rangle \\
\vdots & \ddots & \vdots\\
\left\langle v_{1},v_{n}\right\rangle  & \cdots & \left\langle v_{n},v_{n}\right\rangle 
\end{array}\mright)}
\]
is an invariant of $\Lambda$, independent of the choice of generators
$\mleft(v_{i}\mright)_{i=1}^{n}$.
\end{prop}

The quantity $d\mleft(\Lambda\mright)$ is referred to as the covolume
of $\Lambda$.

For a lattice $\Lambda$ in an inner product space $V$, one defines
the succesive minima (relative to $\overline{B}\mleft(0,1\mright)$),
$\mleft(\lambda_{i}\mright)_{i=1}^{n}$, by
\begin{equation}
\lambda_{i}=\inf\mleft(\left\{ \lambda\mid\overline{B}\mleft(0,\lambda\mright)\cap\Lambda\text{ contains \ensuremath{i} linearly independent vectors}\right\} \mright),\quad1\leq i\leq n.
\end{equation}
A well-known theorem due to Minkowski relates succesive minima and
covolumes:
\begin{thm}[Minkowski's Second Theorem]
Let $\Lambda$ be a lattice in an $n$-dimensional inner product
space $V$. Then it holds that
\[
\frac{2^{n}d\mleft(\Lambda\mright)}{n!\mathrm{Vol}\mleft(\overline{B}\mleft(0,1\mright)\mright)}\leq\lambda_{1}\cdots\lambda_{n}\leq\frac{2^{n}d\mleft(\Lambda\mright)}{\mathrm{Vol}\mleft(\overline{B}\mleft(0,1\mright)\mright)}.
\]
\end{thm}

Note that although the quantity $\lambda_{n}$ is such that $\overline{B}\mleft(0,\lambda_{n}\mright)\cap\Lambda$
contains $n$ linearly independent vectors, it is not ensured that
these can be chosen to generate $\Lambda$. For $n=2$ this is nonetheless
the case:
\begin{prop}
\label{prop:2DGeneratorChoice}Let $\Lambda$ be a lattice in a $2$-dimensional
inner product space $V$. Then there exists vectors $v_{1},v_{2}\in\Lambda$
which generate $\Lambda$ such that
\[
\left\Vert v_{1}\right\Vert \left\Vert v_{2}\right\Vert \leq\frac{4}{\pi}d\mleft(\lambda\mright).
\]
\end{prop}

\textbf{Proof:} By definition there exists linearly independent vectors
$v_{1},v_{2}\in\Lambda$ such that $\left\Vert v_{1}\right\Vert \leq\lambda_{1}$,
$\left\Vert v_{2}\right\Vert \leq\lambda_{2}$, and by Minkowski's
second theorem $\left\Vert v_{1}\right\Vert \left\Vert v_{2}\right\Vert \leq\frac{4}{\pi}d\mleft(\lambda\mright)$.
We argue that $v_{1}$ and $v_{2}$ must generate $\Lambda$.

Suppose otherwise, i.e. let $v\in\Lambda$ be such that $v\neq m_{1}v_{1}+m_{2}v_{2}$
for $m_{1},m_{2}\in\mathbb{Z}$. As $v_{1}$ and $v_{2}$ are linearly
independent and $\dim\mleft(V\mright)=2$ they span $V$, so we can
nonetheless write $v=c_{1}v_{1}+c_{2}v_{2}$ for some $c_{1},c_{2}\in\mathbb{R}$.
By subtracting integer multiplies of $v_{1}$ and $v_{2}$ we may
further assume that $\left|c_{1}\right|,\left|c_{2}\right|\leq\frac{1}{2}$.

As $\left\langle v_{1},v_{2}\right\rangle <\left\Vert v_{1}\right\Vert \left\Vert v_{2}\right\Vert $
by Cauchy-Schwarz (the inequality being strict due to linear independence)
we can then estimate that
\begin{align}
\left\Vert v\right\Vert ^{2} & =\left|c_{1}\right|^{2}\left\Vert v_{1}\right\Vert ^{2}+\left|c_{2}\right|^{2}\left\Vert v_{2}\right\Vert ^{2}+2c_{1}c_{2}\left\langle v_{1},v_{2}\right\rangle <\left|c_{1}\right|^{2}\left\Vert v_{1}\right\Vert ^{2}+\left|c_{2}\right|^{2}\left\Vert v_{2}\right\Vert ^{2}+2\left|c_{1}\right|\left|c_{2}\right|\left\Vert v_{1}\right\Vert \left\Vert v_{2}\right\Vert \nonumber \\
 & =\mleft(\left|c_{1}\right|\left\Vert v_{1}\right\Vert +\left|c_{2}\right|\left\Vert v_{2}\right\Vert \mright)^{2}\leq\mleft(\frac{1}{2}\lambda_{2}+\frac{1}{2}\lambda_{2}\mright)^{2}=\lambda_{2}^{2},
\end{align}
i.e. $\left\Vert v\right\Vert <\lambda_{2}$. But this contradicts
the minimality of $\lambda_{2}$, so such a $v$ can not exist.

$\hfill\square$

\subsubsection*{The Sublattice Orthogonal to a Vector $k\in\mathbb{Z}^{3}$}

Consider $\mathbb{Z}^{3}$ as a lattice in $\mathbb{R}^{3}$, endowed
with the usual dot product. Let $k=\mleft(k_{1},k_{2},k_{3}\mright)\in\mathbb{Z}_{\ast}^{3}$
be arbitrary, and write $\hat{k}=\left|k\right|^{-1}k$. We now characterize
sets of the form
\begin{equation}
\{p\in\mathbb{Z}^{3}\mid\hat{k}\cdotp p=t\},\quad t\in\mathbb{R}.
\end{equation}
For this we note the following well-known result on linear Diophantine
equations:
\begin{thm}
Let $\mleft(k_{1},k_{2},k_{3}\mright)\in\mathbb{Z}_{\ast}^{3}$ and
$c\in\mathbb{Z}$ be given. Then the linear Diophantine equation
\[
k_{1}m_{1}+k_{2}m_{2}+k_{3}m_{3}=c
\]
is solvable with $\mleft(m_{1},m_{2},m_{3}\mright)\in\mathbb{Z}^{3}$
if and only if $c$ is a multiple of $\gcd\mleft(k_{1},k_{2},k_{3}\mright)$.

If this is the case then there exists linearly independent vectors
$v_{1},v_{2}\in\mathbb{Z}^{3}$, which are independent of $c$, such
that if $\mleft(m_{1}^{\ast},m_{2}^{\ast},m_{3}^{\ast}\mright)$ is
any particular solution of the equation then all solutions are given
by
\[
\left\{ \mleft(m_{1},m_{2},m_{3}\mright)\in\mathbb{Z}^{3}\mid k_{1}m_{1}+k_{2}m_{2}+k_{3}m_{3}=c\right\} =\mleft(m_{1}^{\ast},m_{2}^{\ast},m_{3}^{\ast}\mright)+\left\{ a_{1}v_{1}+a_{2}v_{2}\mid a_{1},a_{2}\in\mathbb{Z}\right\} .
\]
\end{thm}

This theorem implies the following:
\begin{prop}
\label{prop:PlaneDecompositionofZ3}Let $k=\mleft(k_{1},k_{2},k_{3}\mright)\in\mathbb{Z}_{\ast}^{3}$
and define $l=\left|k\right|^{-1}\gcd\mleft(k_{1},k_{2},k_{3}\mright)$.
Then there holds the disjoint union of non-empty sets
\[
\mathbb{Z}^{3}=\bigcup_{m\in\mathbb{Z}}\{p\in\mathbb{Z}^{3}\mid\hat{k}\cdotp p=lm\}
\]
and $\{p\in\mathbb{Z}^{3}\mid\hat{k}\cdotp p=0\}$ is a lattice in
$k^{\perp}=\{p\in\mathbb{R}^{3}\mid\hat{k}\cdotp p=0\}$.
\end{prop}

\textbf{Proof:} Clearly $\mathbb{Z}^{3}=\bigcup_{t\in\mathbb{R}}\{p\in\mathbb{Z}^{3}\mid\hat{k}\cdotp p=t\}$
so we must determine for which values of $t$ the set $\{p\in\mathbb{Z}^{3}\mid\hat{k}\cdotp p=t\}$
is non-empty. For an arbitrary $p=\mleft(p_{1},p_{2},p_{3}\mright)\in\mathbb{Z}^{3}$
the equation $\hat{k}\cdot p=t$ is equivalent with
\begin{equation}
k_{1}p_{1}+k_{2}p_{2}+k_{3}p_{3}=\left|k\right|t
\end{equation}
and as the left-hand side is the sum of products of integers, the
right-hand side must likewise be an integer, i.e. $t=\left|k\right|^{-1}c$
for some $c\in\mathbb{Z}$. By the theorem it must then hold that
$c=\gcd\mleft(k_{1},k_{2},k_{3}\mright)\cdot m$ for some $m\in\mathbb{Z}$,
i.e.
\begin{equation}
t=\left|k\right|^{-1}\gcd\mleft(k_{1},k_{2},k_{3}\mright)\cdot m=lm.
\end{equation}
As $p$ was arbitrary we see that $\mathbb{Z}^{3}=\bigcup_{m\in\mathbb{Z}}\{p\in\mathbb{Z}^{3}\mid\hat{k}\cdotp p=lm\}$
as claimed. That all sets $\{p\in\mathbb{Z}^{3}\mid\hat{k}\cdotp p=lm\}$
are non-empty likewise follows from the theorem, as does the existence
of linearly independent $v_{1},v_{2}\in\mathbb{Z}^{3}$ such that
\begin{equation}
\{p\in\mathbb{Z}^{3}\mid\hat{k}\cdotp p=lm\}=q+\left\{ a_{1}v_{1}+a_{2}v_{2}\mid a_{1},a_{2}\in\mathbb{Z}\right\} 
\end{equation}
for any particular $q\in\{p\in\mathbb{Z}^{3}\mid\hat{k}\cdotp p=lm\}$.
Taking $q=0$ as a particular solution, we see that
\begin{equation}
\{p\in\mathbb{Z}^{3}\mid\hat{k}\cdotp p=0\}=\left\{ a_{1}v_{1}+a_{2}v_{2}\mid a_{1},a_{2}\in\mathbb{Z}\right\} 
\end{equation}
which is precisely the statement that $\{p\in\mathbb{Z}^{3}\mid\hat{k}\cdotp p=0\}$
is a lattice (in $k^{\perp}$).

$\hfill\square$

The covolume $d\mleft(\{p\in\mathbb{Z}^{3}\mid\hat{k}\cdotp p=0\}\mright)=\sqrt{\left\Vert v_{1}\right\Vert ^{2}\left\Vert v_{2}\right\Vert ^{2}-\mleft(v_{1}\cdot v_{2}\mright)^{2}}$
is given by the following:
\begin{prop}
\label{prop:CovolumeofPerpLattice}For any generators $v_{1},v_{2}\in\mathbb{Z}^{3}$
of $\{p\in\mathbb{Z}^{3}\mid\hat{k}\cdotp p=0\}$ it holds that
\[
d\mleft(\{p\in\mathbb{Z}^{3}\mid\hat{k}\cdotp p=0\}\mright)=\sqrt{\left\Vert v_{1}\right\Vert ^{2}\left\Vert v_{2}\right\Vert ^{2}-\mleft(v_{1}\cdot v_{2}\mright)^{2}}=l^{-1}.
\]
\end{prop}

\textbf{Proof:} Let $w\in\{p\in\mathbb{Z}^{3}\mid\hat{k}\cdotp p=l\}$
be arbitrary. Then by linearity
\begin{equation}
\{p\in\mathbb{Z}^{3}\mid\hat{k}\cdotp p=lm\}=mw+\{p\in\mathbb{Z}^{3}\mid\hat{k}\cdotp p=0\}
\end{equation}
for any $m\in\mathbb{Z}$, so by the previous proposition
\begin{equation}
\mathbb{Z}^{3}=\bigcup_{m\in\mathbb{Z}}\mleft(mw+\{p\in\mathbb{Z}^{3}\mid\hat{k}\cdotp p=0\}\mright)=\left\{ m_{1}v_{1}+m_{2}v_{2}+m_{3}w\mid m_{1},m_{2},m_{3}\in\mathbb{Z}\right\} ,
\end{equation}
i.e. $\mleft(v_{1},v_{2},w\mright)$ is a set of generators for $\mathbb{Z}^{3}$.
Let $\mleft(e_{1},e_{2}\mright)$ be an orthonormal basis for $k^{\perp}$
so that $(e_{1},e_{2},\hat{k})$ forms an orthonormal basis for $\mathbb{R}^{3}$.
Then
\begin{align}
d\mleft(\mathbb{Z}^{3}\mright) & =\left|\det\mleft(\begin{array}{ccc}
\mleft(e_{1}\cdot v_{1}\mright) & \mleft(e_{2}\cdot v_{1}\mright) & (\hat{k}\cdot v_{1})\\
\mleft(e_{1}\cdot v_{2}\mright) & \mleft(e_{2}\cdot v_{2}\mright) & (\hat{k}\cdot v_{2})\\
\mleft(e_{1}\cdot w\mright) & \mleft(e_{w}\cdot w\mright) & (\hat{k}\cdot w)
\end{array}\mright)\right|=\left|\det\mleft(\begin{array}{ccc}
\mleft(e_{1}\cdot v_{1}\mright) & \mleft(e_{2}\cdot v_{1}\mright) & 0\\
\mleft(e_{1}\cdot v_{2}\mright) & \mleft(e_{2}\cdot v_{2}\mright) & 0\\
\mleft(e_{1}\cdot w\mright) & \mleft(e_{w}\cdot w\mright) & l
\end{array}\mright)\right|\\
 & =l\left|\det\mleft(\begin{array}{cc}
\mleft(e_{1}\cdot v_{1}\mright) & \mleft(e_{2}\cdot v_{1}\mright)\\
\mleft(e_{1}\cdot v_{2}\mright) & \mleft(e_{2}\cdot v_{2}\mright)
\end{array}\mright)\right|=l\cdot d\mleft(\{p\in\mathbb{Z}^{3}\mid\hat{k}\cdotp p=0\}\mright)\nonumber 
\end{align}
and as it is clear that $d\mleft(\mathbb{Z}^{3}\mright)=1$ the result
follows.

$\hfill\square$

Finally we note that Proposition \ref{prop:2DGeneratorChoice} implies
a bound on the norms of a generating set of $\{p\in\mathbb{Z}^{3}\mid\hat{k}\cdotp p=0\}$:
\begin{cor}
\label{coro:SmallGenerators}There exists a constant $C>0$ independent
of $k$ such that $\{p\in\mathbb{Z}^{3}\mid\hat{k}\cdotp p=0\}$ admits
generators $v_{1}$ and $v_{2}$ obeying
\[
\left\Vert v_{1}\right\Vert ^{2}+\left\Vert v_{2}\right\Vert ^{2}\leq Cl^{-2}.
\]
\end{cor}

\textbf{Proof:} By the proposition there exists generators $v_{1},v_{2}$
such that
\begin{equation}
\left\Vert v_{1}\right\Vert \left\Vert v_{2}\right\Vert \leq\frac{4}{\pi}d\mleft(\lambda\mright)=\frac{4}{\pi}l^{-1},
\end{equation}
and as every $v\in\mathbb{Z}_{\ast}^{3}$ obeys $\left\Vert v\right\Vert \geq1$
this implies that $\left\Vert v_{1}\right\Vert ,\left\Vert v_{2}\right\Vert \leq\frac{4}{\pi}l^{-1}$.
Consequently
\begin{equation}
\left\Vert v_{1}\right\Vert ^{2}+\left\Vert v_{2}\right\Vert ^{2}\leq\frac{32}{\pi^{2}}l^{-2}=Cl^{-2}.
\end{equation}
$\hfill\square$

\subsection{\label{subsec:PreciseEstimates}Precise Estimates}

Throughout this section we let $k=\mleft(k_{1},k_{2},k_{3}\mright)\in B\mleft(0,2k_{F}\mright)\cap\mathbb{Z}_{\ast}^{3}$
be fixed and write $\hat{k}=\left|k\right|^{-1}k$ for brevity.

We now decompose the lune
\begin{equation}
L_{k}=\left\{ p\in\mathbb{Z}^{3}\mid\left|p-k\right|\leq k_{F}<\left|p\right|\right\} 
\end{equation}
along the $\{\hat{k}\cdotp p=t\}$ planes. Note that for any $p\in L_{k}$
it holds that
\begin{equation}
k\cdotp p=\frac{1}{2}\mleft(\left|p\right|^{2}-\left|p-k\right|^{2}+\left|k\right|^{2}\mright)>\frac{1}{2}\left|k\right|^{2}
\end{equation}
and that
\begin{equation}
k\cdotp p=k\cdotp\mleft(p-k\mright)+\left|k\right|^{2}\leq\left|k\right|\mleft(k_{F}+\left|k\right|\mright)
\end{equation}
so
\begin{equation}
\frac{1}{2}\left|k\right|<\hat{k}\cdot p\leq k_{F}+\left|k\right|.
\end{equation}
Let $l=\left|k\right|^{-1}\gcd\mleft(k_{1},k_{2},k_{3}\mright)$ as
in Proposition \ref{prop:PlaneDecompositionofZ3}, and let $m^{\ast}$
be the least integer and $M^{\ast}$ the greatest integer such that
\begin{equation}
\frac{1}{2}\left|k\right|<lm^{\ast},\quad lM^{\ast}\leq k_{F}+\left|k\right|.
\end{equation}
It then follows by the decomposition of Proposition \ref{prop:PlaneDecompositionofZ3}
that $L_{k}$ can be expressed as the disjoint union
\begin{equation}
L_{k}=\bigcup_{m=m^{\ast}}^{M^{\ast}}L_{k}^{m}
\end{equation}
where the subsets $L_{k}^{m}$ are given by
\begin{equation}
L_{k}^{m}=\{p\in L_{k}\mid\hat{k}\cdot p=lm\},\quad m^{\ast}\leq m\leq M^{\ast}.
\end{equation}
Consequently, a Riemann sum of the form $\sum_{p\in L_{k}}f\mleft(\lambda_{k,p}\mright)$
can be written as
\begin{equation}
\sum_{p\in L_{k}}f\mleft(\lambda_{k,p}\mright)=\sum_{m=m^{\ast}}^{M^{\ast}}\sum_{p\in L_{k}^{m}}f\mleft(\left|k\right|\mleft(\hat{k}\cdot p-\frac{1}{2}\left|k\right|\mright)\mright)=\sum_{m=m^{\ast}}^{M^{\ast}}f\mleft(\left|k\right|\mleft(lm-\frac{1}{2}\left|k\right|\mright)\mright)\left|L_{k}^{m}\right|.\label{eq:GeneralRiemannSumExpression}
\end{equation}
To proceed we must analyze $\left|L_{k}^{m}\right|$, the number of
points contained in $L_{k}^{m}$. For this, note that by expanding
and rearranging the inequalities defining $L_{k}$, we may equivalently
express it as
\begin{equation}
L_{k}=\left\{ p\in\mathbb{Z}^{3}\mid k_{F}^{2}<\left|p\right|^{2}\leq k_{F}^{2}-\left|k\right|^{2}+2k\cdot p\right\} .
\end{equation}
Letting $P_{\perp}:\mathbb{R}^{3}\rightarrow k^{\perp}$ denote the
orthogonal projection onto $k^{\perp}$, it holds that $\left|p\right|^{2}=\left|P_{\perp}p\right|^{2}+(\hat{k}\cdot p)^{2}$,
whence
\begin{align}
L_{k} & =\{p\in\mathbb{Z}^{3}\mid k_{F}^{2}-(\hat{k}\cdot p)^{2}<\left|P_{\perp}p\right|^{2}\leq k_{F}^{2}-\left|k\right|^{2}+2k\cdot p-(\hat{k}\cdot p)^{2}\}\\
 & =\{p\in\mathbb{Z}^{3}\mid k_{F}^{2}-(\hat{k}\cdot p)^{2}<\left|P_{\perp}p\right|^{2}\leq k_{F}^{2}-(\hat{k}\cdotp p-\left|k\right|)^{2}\}\nonumber 
\end{align}
so the sets $L_{k}^{m}=L_{k}\cap\{p\in\mathbb{Z}^{3}\mid\hat{k}\cdot p=lm\}$
can be written as
\begin{align}
L_{k}^{m} & =\left\{ p\in\mathbb{Z}^{3}\mid\hat{k}\cdotp p=lm,\,k_{F}^{2}-\mleft(lm\mright)^{2}<\left|P_{\perp}p\right|^{2}\leq k_{F}^{2}-\mleft(lm-\left|k\right|\mright)^{2}\right\} \\
 & =\left\{ p\in\mathbb{Z}^{3}\mid\hat{k}\cdotp p=lm,\,\mleft(R_{1}^{m}\mright)^{2}<\left|P_{\perp}p\right|^{2}\leq\mleft(R_{2}^{m}\mright)^{2}\right\} \nonumber 
\end{align}
where the real numbers $R_{1}^{m}$ and $R_{2}^{m}$ are given by
\begin{equation}
R_{1}^{m}=\sqrt{k_{F}^{2}-\mleft(lm\mright)^{2}},\quad R_{2}^{m}=\sqrt{k_{F}^{2}-\mleft(lm-\left|k\right|\mright)^{2}},\quad m^{\ast}\leq m\leq M^{\ast}.
\end{equation}
Let $v_{1},v_{2}\in\mathbb{Z}^{3}$ generate $\{p\in\mathbb{Z}^{3}\mid\hat{k}\cdot p=0\}$.
For a fixed $m$, let $q\in\{p\in\mathbb{Z}^{3}\mid\hat{k}\cdot p=lm\}$
be arbitrary. Then Proposition \ref{prop:PlaneDecompositionofZ3}
asserts that $p\in\mathbb{Z}^{3}$ is an element of $\{p\in\mathbb{Z}^{3}\mid\hat{k}\cdot p=lm\}$
if and only if it can be written as
\begin{equation}
p=a_{1}v_{1}+a_{2}v_{2}+q,\quad a_{1},a_{2}\in\mathbb{Z}.
\end{equation}
As $v_{1}$ and $v_{2}$ span $k^{\perp}$ it must hold that $P_{\perp}q=b_{1}v_{1}+b_{2}v_{2}$
for some $b_{1},b_{2}\in\mathbb{R}$, whence we see that $P_{\perp}p$
is of the form
\begin{equation}
P_{\perp}p=a_{1}v_{1}+a_{2}v_{2}+P_{\perp}q=\mleft(a_{1}+b_{1}\mright)v_{1}+\mleft(a_{2}+b_{2}\mright)v_{2},
\end{equation}
and so we can express $\left|L_{k}^{m}\right|$ as
\begin{align}
\left|L_{k}^{m}\right| & =\left|\left\{ \mleft(a_{1},a_{2}\mright)\in\mathbb{Z}^{2}\mid\mleft(R_{1}^{m}\mright)^{2}<\mleft(a_{1}+b_{1}\mright)^{2}\left\Vert v_{1}\right\Vert ^{2}+\mleft(a_{2}+b_{2}\mright)^{2}\left\Vert v_{2}\right\Vert ^{2}+2\mleft(a_{1}+b_{1}\mright)\mleft(a_{2}+b_{2}\mright)\mleft(v_{1}\cdot v_{2}\mright)\leq\mleft(R_{2}^{m}\mright)^{2}\right\} \right|\nonumber \\
 & =\left|\left\{ \mleft(x,y\mright)\in\mleft(\mathbb{R}^{2}+\mleft(b_{1},b_{2}\mright)\mright)\mid\mleft(R_{1}^{m}\mright)^{2}<\left\Vert v_{1}\right\Vert ^{2}x^{2}+\left\Vert v_{2}\right\Vert ^{2}y^{2}+2\mleft(v_{1}\cdot v_{2}\mright)xy\leq\mleft(R_{2}^{m}\mright)^{2}\right\} \cap\mathbb{Z}^{2}\right|\\
 & =\left|\mleft(E_{2}^{m}\backslash E_{1}^{m}+\mleft(b_{1},b_{2}\mright)\mright)\cap\mathbb{Z}^{2}\right|\nonumber 
\end{align}
where the sets $E_{1}^{m}$ and $E_{2}^{m}$ are given by
\begin{equation}
E_{i}^{m}=\left\{ \mleft(x,y\mright)\in\mathbb{R}^{2}\mid\left\Vert v_{1}\right\Vert ^{2}x^{2}+\left\Vert v_{2}\right\Vert ^{2}y^{2}+2\mleft(v_{1}\cdot v_{2}\mright)xy\leq\mleft(R_{i}^{m}\mright)^{2}\right\} ,\quad i=1,2.\label{eq:EimDefinition}
\end{equation}

\subsubsection*{Lattice Point Estimation}

The sets $E_{i}^{m}$ are seen to be (closed interiors of) ellipses,
and analyzing $\left|L_{k}^{m}\right|$ amounts to estimating the
lattice points enclosed by these. To do this we will apply the following
general result:
\begin{thm}[\cite{GelFondLinnik-66}]
Let $K\subset\mathbb{R}^{2}$ be a compact, strictly convex set with
$C^{2}$ boundary and let $\partial K$ have minimal and maximal radii
of curvature $0<r_{1}\leq r_{2}$. If $r_{2}\geq1$ then
\[
\left|\left|K\cap\mathbb{Z}^{2}\right|-\mathrm{Area}\mleft(K\mright)\right|\leq C\frac{r_{2}}{r_{1}}r_{2}^{\frac{2}{3}}\log\mleft(1+2\sqrt{2}r_{2}^{\frac{1}{2}}\mright)^{\frac{2}{3}}
\]
for a constant $C>0$ independent of all quantities.
\end{thm}

This result follows from the techniques of Chapter 8 of \cite{GelFondLinnik-66},
but is not explicitly stated in this fashion. Giving a proof of this
result is out of the scope of this thesis, but a detailed derivation
is available upon request.

In our present case we note that this implies that for any ellipse
$E\subset\mathbb{R}^{2}$, it holds that
\begin{equation}
\left|\left|E\cap\mathbb{Z}^{2}\right|-\mathrm{Area}\mleft(E\mright)\right|\leq C\mleft(1+\frac{r_{2}}{r_{1}}r_{2}^{\frac{2}{3}}\log\mleft(1+2\sqrt{2}r_{2}^{\frac{1}{2}}\mright)^{\frac{2}{3}}\mright),
\end{equation}
the $r_{2}\leq1$ case being accounted for by the constant term. It
follows that $\left|L_{k}^{m}\right|$ obeys
\begin{equation}
\left|L_{k}^{m}\right|=\mathrm{Area}\mleft(E_{2}^{m}\backslash E_{1}^{m}\mright)+O\mleft(1+\frac{r_{2}}{r_{1}}r_{2}^{\frac{2}{3}}\log\mleft(1+2\sqrt{2}r_{2}^{\frac{1}{2}}\mright)^{\frac{2}{3}}+\frac{r_{2}^{\prime}}{r_{1}^{\prime}}\mleft(r_{2}^{\prime}\mright)^{\frac{2}{3}}\log\mleft(1+2\sqrt{2}\mleft(r_{2}^{\prime}\mright)^{\frac{1}{2}}\mright)^{\frac{2}{3}}\mright)\label{eq:LkmPointsEstimate}
\end{equation}
where $r_{1},r_{1}^{\prime}$ and $r_{2},r_{2}^{\prime}$ are the
minimal and maximal radii of curvature of $E_{1}^{m}$ and $E_{2}^{m}$,
respectively.

We thus need to obtain some information on the geometry of the ellipses
$E_{i}^{m}$. Consulting a reference on conic sections, one finds
that the semi axes $a_{i}\geq b_{i}>0$ of $E_{i}^{m}$, as defined
by equation (\ref{eq:EimDefinition}), are given by
\begin{align}
a_{i} & =\sqrt{2}R_{i}^{m}\mleft(\left\Vert v_{1}\right\Vert ^{2}+\left\Vert v_{2}\right\Vert ^{2}-\sqrt{\mleft(\left\Vert v_{1}\right\Vert ^{2}-\left\Vert v_{2}\right\Vert ^{2}\mright)^{2}+4\mleft(v_{1}\cdot v_{2}\mright)^{2}}\mright)^{-\frac{1}{2}}\\
b_{i} & =\sqrt{2}R_{i}^{m}\mleft(\left\Vert v_{1}\right\Vert ^{2}+\left\Vert v_{2}\right\Vert ^{2}+\sqrt{\mleft(\left\Vert v_{1}\right\Vert ^{2}-\left\Vert v_{2}\right\Vert ^{2}\mright)^{2}+4\mleft(v_{1}\cdot v_{2}\mright)^{2}}\mright)^{-\frac{1}{2}}.\nonumber 
\end{align}
We can then describe the geometry of the ellipses in terms of $k$
and $m$:
\begin{prop}
It holds that
\[
\mathrm{Area}\mleft(E_{2}^{m}\backslash E_{1}^{m}\mright)=\begin{cases}
2\pi\left|k\right|\mleft(lm-\frac{1}{2}\left|k\right|\mright)l & lm^{\ast}\leq lm\leq k_{F}\\
\pi\mleft(k_{F}^{2}-\mleft(lm-\left|k\right|\mright)^{2}\mright)l & k_{F}<lm\leq lM^{\ast}
\end{cases}
\]
and the minimal and maximal radii of curvature $0<r_{1}\leq r_{2}$
of either of $E_{1}^{m}$, $E_{2}^{m}$ can be assumed to obey the
estimates
\[
\frac{r_{2}}{r_{1}}\leq Cl^{-3},\quad r_{2}\leq Cl^{-1}k_{F},
\]
for a constant $C>0$ independent of all quantities.
\end{prop}

\textbf{Proof:} The area enclosed by an ellipse with semi-axes $a$
and $b$ is $\pi ab$, so for $lm^{\ast}\leq lm\leq k_{F}$, when
$\emptyset\neq E_{1}^{m}\subset E_{2}^{m}$,
\begin{align}
\mathrm{Area}\mleft(E_{2}^{m}\backslash E_{1}^{m}\mright) & =\frac{2\pi\mleft(\mleft(R_{2}^{m}\mright)^{2}-\mleft(R_{1}^{m}\mright)^{2}\mright)}{\sqrt{\mleft(\left\Vert v_{1}\right\Vert ^{2}+\left\Vert v_{2}\right\Vert ^{2}\mright)^{2}-\mleft(\mleft(\left\Vert v_{1}\right\Vert ^{2}-\left\Vert v_{2}\right\Vert ^{2}\mright)^{2}+4\mleft(v_{1}\cdot v_{2}\mright)^{2}\mright)}}\nonumber \\
 & =\frac{2\pi\mleft(k_{F}^{2}-\mleft(lm-\left|k\right|\mright)^{2}-\mleft(k_{F}^{2}-\mleft(lm\mright)^{2}\mright)\mright)}{\sqrt{4\left\Vert v_{1}\right\Vert ^{2}\left\Vert v_{2}\right\Vert ^{2}-4\mleft(v_{1}\cdot v_{2}\mright)^{2}}}\\
 & =\frac{2\pi\mleft(\left|k\right|\mleft(2lm-\left|k\right|\mright)\mright)}{2l^{-1}}=2\pi\left|k\right|\mleft(lm-\frac{1}{2}\left|k\right|\mright)l,\nonumber 
\end{align}
where we used that $\sqrt{\left\Vert v_{1}\right\Vert ^{2}\left\Vert v_{2}\right\Vert ^{2}-\mleft(v_{1}\cdot v_{2}\mright)^{2}}=l^{-1}$
by Proposition \ref{prop:CovolumeofPerpLattice}, while for $k_{F}<lm\leq lM^{\ast}$,
when $E_{1}^{m}=\emptyset$,
\begin{equation}
\mathrm{Area}\mleft(E_{2}^{m}\backslash E_{1}^{m}\mright)=\mathrm{Area}\mleft(E_{2}^{m}\mright)=\frac{2\pi\mleft(R_{2}^{m}\mright)^{2}}{2l^{-1}}=\pi\mleft(k_{F}^{2}-\mleft(lm-\left|k\right|\mright)^{2}\mright)l.
\end{equation}
For the radii of curvature we note that for an ellipse with semi axes
$a\geq b>0$ these are given by $r_{1}=a^{-1}b^{2}$ and $r_{2}=b^{-1}a^{2}$,
respectively, so for either of $E_{i}^{m}$ we can estimate that
\begin{align}
\frac{r_{2}}{r_{1}} & =\mleft(\frac{a_{i}}{b_{i}}\mright)^{3}=\mleft(\frac{\left\Vert v_{1}\right\Vert ^{2}+\left\Vert v_{2}\right\Vert ^{2}+\sqrt{\mleft(\left\Vert v_{1}\right\Vert ^{2}-\left\Vert v_{2}\right\Vert ^{2}\mright)^{2}+4\mleft(v_{1}\cdot v_{2}\mright)^{2}}}{\left\Vert v_{1}\right\Vert ^{2}+\left\Vert v_{2}\right\Vert ^{2}-\sqrt{\mleft(\left\Vert v_{1}\right\Vert ^{2}-\left\Vert v_{2}\right\Vert ^{2}\mright)^{2}+4\mleft(v_{1}\cdot v_{2}\mright)^{2}}}\mright)^{\frac{3}{2}}\nonumber \\
 & =\mleft(\frac{\mleft(\left\Vert v_{1}\right\Vert ^{2}+\left\Vert v_{2}\right\Vert ^{2}+\sqrt{\mleft(\left\Vert v_{1}\right\Vert ^{2}-\left\Vert v_{2}\right\Vert ^{2}\mright)^{2}+4\mleft(v_{1}\cdot v_{2}\mright)^{2}}\mright)^{2}}{\left\Vert v_{1}\right\Vert ^{2}+\left\Vert v_{2}\right\Vert ^{2}-\mleft(\mleft(\left\Vert v_{1}\right\Vert ^{2}-\left\Vert v_{2}\right\Vert ^{2}\mright)^{2}+4\mleft(v_{1}\cdot v_{2}\mright)^{2}\mright)}\mright)^{\frac{3}{2}}\\
 & =\mleft(\frac{1}{4}\frac{\mleft(\left\Vert v_{1}\right\Vert ^{2}+\left\Vert v_{2}\right\Vert ^{2}+\sqrt{\mleft(\left\Vert v_{1}\right\Vert ^{2}+\left\Vert v_{2}\right\Vert ^{2}\mright)^{2}-4\mleft(\left\Vert v_{1}\right\Vert ^{2}\left\Vert v_{2}\right\Vert ^{2}-\mleft(v_{1}\cdot v_{2}\mright)^{2}\mright)}\mright)^{2}}{\left\Vert v_{1}\right\Vert ^{2}\left\Vert v_{2}\right\Vert ^{2}-\mleft(v_{1}\cdot v_{2}\mright)^{2}}\mright)^{\frac{3}{2}}\nonumber \\
 & \leq\mleft(\frac{l^{2}}{4}\mleft(2\mleft(\left\Vert v_{1}\right\Vert ^{2}+\left\Vert v_{2}\right\Vert ^{2}\mright)\mright)^{2}\mright)^{\frac{3}{2}}=\mleft(\left\Vert v_{1}\right\Vert ^{2}+\left\Vert v_{2}\right\Vert ^{2}\mright)^{3}l^{3}\nonumber 
\end{align}
and that
\begin{align}
r_{2} & =\frac{a_{i}^{2}}{b_{i}}=\sqrt{2}R_{i}^{m}\frac{\sqrt{\left\Vert v_{1}\right\Vert ^{2}+\left\Vert v_{2}\right\Vert ^{2}+\sqrt{\mleft(\left\Vert v_{1}\right\Vert ^{2}-\left\Vert v_{2}\right\Vert ^{2}\mright)^{2}+4\mleft(v_{1}\cdot v_{2}\mright)^{2}}}}{\left\Vert v_{1}\right\Vert ^{2}+\left\Vert v_{2}\right\Vert ^{2}-\sqrt{\mleft(\left\Vert v_{1}\right\Vert ^{2}-\left\Vert v_{2}\right\Vert ^{2}\mright)^{2}+4\mleft(v_{1}\cdot v_{2}\mright)^{2}}}\nonumber \\
 & =\sqrt{2}R_{i}^{m}\frac{\mleft(\left\Vert v_{1}\right\Vert ^{2}+\left\Vert v_{2}\right\Vert ^{2}+\sqrt{\mleft(\left\Vert v_{1}\right\Vert ^{2}-\left\Vert v_{2}\right\Vert ^{2}\mright)^{2}+4\mleft(v_{1}\cdot v_{2}\mright)^{2}}\mright)^{\frac{3}{2}}}{\left\Vert v_{1}\right\Vert ^{2}+\left\Vert v_{2}\right\Vert ^{2}-\mleft(\mleft(\left\Vert v_{1}\right\Vert ^{2}-\left\Vert v_{2}\right\Vert ^{2}\mright)^{2}+4\mleft(v_{1}\cdot v_{2}\mright)^{2}\mright)}\\
 & \leq\sqrt{2}R_{i}^{m}\frac{\mleft(2\mleft(\left\Vert v_{1}\right\Vert ^{2}+\left\Vert v_{2}\right\Vert ^{2}\mright)\mright)^{\frac{3}{2}}}{4l^{-2}}=\mleft(\left\Vert v_{1}\right\Vert ^{2}+\left\Vert v_{2}\right\Vert ^{2}\mright)^{\frac{3}{2}}l^{2}R_{i}^{m}.\nonumber 
\end{align}
Corollary \ref{coro:SmallGenerators} asserts that $v_{1}$ and $v_{2}$
can be chosen to obey $\left\Vert v_{1}\right\Vert ^{2}+\left\Vert v_{2}\right\Vert ^{2}\leq Cl^{-2}$,
in which case these estimates become
\begin{equation}
\frac{r_{2}}{r_{1}}\leq\mleft(Cl^{-2}\mright)^{3}l^{3}\leq Cl^{-3},\quad r_{2}\leq\mleft(Cl^{-2}\mright)^{\frac{3}{2}}l^{2}R_{i}^{m}\leq Cl^{-1}k_{F},
\end{equation}
as claimed (using also that $R_{i}^{m}\leq k_{F}$ for all $m^{\ast}\leq m\leq M^{\ast}$).

$\hfill\square$

Noting that $l$ obeys
\begin{equation}
l^{-1}=\frac{\left|k\right|}{\gcd\mleft(k_{1},k_{2},k_{3}\mright)}\leq\left|k\right|
\end{equation}
we can by equation (\ref{eq:LkmPointsEstimate}) and the proposition
estimate that
\begin{align}
\left|\left|L_{k}^{m}\right|-\mathrm{Area}\mleft(E_{2}^{m}\backslash E_{1}^{m}\mright)\right| & \leq C\mleft(1+l^{-3}\mleft(l^{-1}k_{F}\mright)^{\frac{2}{3}}\log\mleft(1+\mleft(l^{-1}k_{F}\mright)^{\frac{1}{2}}\mright)^{\frac{2}{3}}\mright)\nonumber \\
 & \leq C\mleft(1+\left|k\right|^{3+\frac{2}{3}}k_{F}^{\frac{2}{3}}\log\mleft(1+\sqrt{\left|k\right|k_{F}}\mright)^{\frac{2}{3}}\mright)\\
 & \leq C\left|k\right|^{3+\frac{2}{3}}\log\mleft(k_{F}\mright)^{\frac{2}{3}}k_{F}^{\frac{2}{3}},\quad k_{F}\rightarrow\infty,\nonumber 
\end{align}
for a constant $C>0$ independent of all quantities, which is to say
\begin{equation}
\left|L_{k}^{m}\right|=\begin{cases}
2\pi\left|k\right|\mleft(lm-\frac{1}{2}\left|k\right|\mright)l & lm^{\ast}\leq lm\leq k_{F}\\
\pi\mleft(k_{F}^{2}-\mleft(lm-\left|k\right|\mright)^{2}\mright)l & k_{F}<lm\leq lM^{\ast}
\end{cases}+O\mleft(\left|k\right|^{3+\frac{2}{3}}\log\mleft(k_{F}\mright)^{\frac{2}{3}}k_{F}^{\frac{2}{3}}\mright).
\end{equation}

\subsubsection*{The Summation Formula}

From equation (\ref{eq:GeneralRiemannSumExpression}) we can now conclude
a general summation formula:
\begin{prop}
\label{prop:RiemannSummationFormula}For all $k=\mleft(k_{1},k_{2},k_{3}\mright)\in\mathbb{Z}_{\ast}^{3}$
with $\left|k\right|<2k_{F}$ and $f:\mleft(0,\infty\mright)\rightarrow\mathbb{R}$
it holds that
\begin{align*}
\sum_{p\in L_{k}}f\mleft(\lambda_{k,p}\mright) & =2\pi\left|k\right|\sum_{m=m^{\ast}}^{M}f\mleft(\left|k\right|\mleft(lm-\frac{1}{2}\left|k\right|\mright)\mright)\mleft(lm-\frac{1}{2}\left|k\right|\mright)l\\
 & +\pi\sum_{m=M+1}^{M^{\ast}}f\mleft(\left|k\right|\mleft(lm-\frac{1}{2}\left|k\right|\mright)\mright)\mleft(k_{F}^{2}-\mleft(lm-\left|k\right|\mright)^{2}\mright)l\\
 & +O\mleft(\left|k\right|^{3+\frac{2}{3}}\log\mleft(k_{F}\mright)^{\frac{2}{3}}k_{F}^{\frac{2}{3}}\sum_{m=m^{\ast}}^{M^{\ast}}\left|f\mleft(\left|k\right|\mleft(lm-\frac{1}{2}\left|k\right|\mright)\mright)\right|\mright)
\end{align*}
as $k_{F}\rightarrow\infty$, where $l=\left|k\right|^{-1}\gcd\mleft(k_{1},k_{2},k_{3}\mright)$
and $m^{\ast}$ is the least integer and $M$, $M^{\ast}$ the greatest
integers for which
\[
\frac{1}{2}\left|k\right|<lm^{\ast},\quad lM\leq k_{F},\quad lM^{\ast}\leq k_{F}+\left|k\right|.
\]
\end{prop}

Note that the two first terms are exactly what one would expect from
the continuum case, since
\begin{align}
\int_{\overline{B}\mleft(k,k_{F}\mright)\backslash\overline{B}\mleft(0,k_{F}\mright)}f\mleft(k\cdot p-\frac{1}{2}\left|k\right|^{2}\mright)dp & =2\pi\left|k\right|\int_{\frac{1}{2}\left|k\right|}^{k_{F}}f\mleft(\left|k\right|\mleft(t-\frac{1}{2}\left|k\right|\mright)\mright)\mleft(t-\frac{1}{2}\left|k\right|\mright)dt\label{eq:ContinuumLuneIntegrals}\\
 & +\pi\int_{k_{F}}^{k_{F}+\left|k\right|}f\mleft(\left|k\right|\mleft(t-\frac{1}{2}\left|k\right|\mright)\mright)\mleft(k_{F}^{2}-\mleft(t-\left|k\right|\mright)^{2}\mright)dt.\nonumber 
\end{align}
The summation formula thus allows us to convert the $3$-dimensional
Riemann sum $\sum_{p\in L_{k}}f\mleft(\lambda_{k,p}\mright)$ into the
$1$-dimensional Riemann sums corresponding to the integrals above,
up to an additional error term.

We can then finally conclude the precise estimate of Proposition \ref{prop:betaeq-1AsymptoticsAppendix}:
\begin{prop}
For all $k\in B\mleft(0,2k_{F}\mright)$ it holds that
\[
\left|\sum_{p\in L_{k}}\lambda_{k,p}^{-1}-2\pi k_{F}\right|\leq C\left|k\right|^{3+\frac{2}{3}}\log\mleft(k_{F}\mright)^{\frac{5}{3}}k_{F}^{\frac{2}{3}},\quad k_{F}\rightarrow\infty,
\]
for a constant $C>0$ independent of all quantities.
\end{prop}

\textbf{Proof:} By the summation formula we have that
\begin{align}
\sum_{p\in L_{k}}\lambda_{k,p}^{-1} & =2\pi\left|k\right|\sum_{m=m^{\ast}}^{M}\frac{lm-\frac{1}{2}\left|k\right|}{\left|k\right|\mleft(lm-\frac{1}{2}\left|k\right|\mright)}l+\pi\sum_{m=M+1}^{M^{\ast}}\frac{k_{F}^{2}-\mleft(lm-\left|k\right|\mright)^{2}}{\left|k\right|\mleft(lm-\frac{1}{2}\left|k\right|\mright)}l\\
 & +O\mleft(\left|k\right|^{2+\frac{2}{3}}\log\mleft(k_{F}\mright)^{\frac{2}{3}}k_{F}^{\frac{2}{3}}\sum_{m=m^{\ast}}^{M^{\ast}}\frac{1}{lm-\frac{1}{2}\left|k\right|}\mright).\nonumber 
\end{align}
The first sum is what contributes the term $2\pi k_{F}$, as we can
estimate
\begin{align}
\left|2\pi\left|k\right|\sum_{m=m^{\ast}}^{M}\frac{lm-\frac{1}{2}\left|k\right|}{\left|k\right|\mleft(lm-\frac{1}{2}\left|k\right|\mright)}l-2\pi k_{F}\right| & =2\pi\left|\sum_{m=m^{\ast}}^{M}l-k_{F}\right|=2\pi\left|\mleft(lM-lm^{\ast}+l\mright)-k_{F}\right|\nonumber \\
 & \leq2\pi\mleft(l\mleft(m^{\ast}-1\mright)+\left|lM-k_{F}\right|+2l\mright)\\
 & \leq2\pi\mleft(\frac{1}{2}\left|k\right|+3\mright)\leq C\left|k\right|\nonumber 
\end{align}
which is $O\mleft(\left|k\right|^{3+\frac{2}{3}}\log\mleft(k_{F}\mright)^{\frac{5}{3}}k_{F}^{\frac{2}{3}}\mright)$
as $k_{F}\rightarrow\infty$ (above we also used that $l\leq1$).
Noting that
\begin{equation}
k_{F}^{2}-\mleft(lm-\left|k\right|\mright)^{2}=k_{F}^{2}-\mleft(lm\mright)^{2}+2\left|k\right|\mleft(lm-\frac{1}{2}\left|k\right|\mright)\leq2\left|k\right|\mleft(lm-\frac{1}{2}\left|k\right|\mright)
\end{equation}
for $m\geq M+1$, we can similarly estimate the second sum as
\begin{align}
0 & \leq\pi\sum_{m=M+1}^{M^{\ast}}\frac{k_{F}^{2}-\mleft(lm-\left|k\right|\mright)^{2}}{\left|k\right|\mleft(lm-\frac{1}{2}\left|k\right|\mright)}l=2\pi\sum_{m=M+1}^{M^{\ast}}l=2\pi\mleft(lM^{\ast}-lM+l\mright)\\
 & =2\pi\mleft(lM^{\ast}-l\mleft(M+1\mright)+2l\mright)\leq2\pi\mleft(k_{F}+\left|k\right|-k_{F}+2\mright)\leq C\left|k\right|.\nonumber 
\end{align}
For the main error term we first note that $lm^{\ast}-\frac{1}{2}\left|k\right|\geq\frac{1}{2}\left|k\right|^{-1}$,
as the definition of $m^{\ast}$ implies that
\begin{equation}
2\gcd\mleft(k_{1},k_{2},k_{3}\mright)m^{\ast}>\left|k\right|^{2}
\end{equation}
so as both sides are integers
\begin{equation}
2\gcd\mleft(k_{1},k_{2},k_{3}\mright)m^{\ast}\geq\left|k\right|^{2}+1\Leftrightarrow lm^{\ast}\geq\frac{1}{2}\left|k\right|+\frac{1}{2}\left|k\right|^{-1}.
\end{equation}
We can thus apply Corollary \ref{coro:1DRiemannSumLemma} to estimate
\begin{align}
\sum_{m=m^{\ast}}^{M^{\ast}}\frac{1}{lm-\frac{1}{2}\left|k\right|} & =\frac{1}{lm^{\ast}-\frac{1}{2}\left|k\right|}+l^{-1}\sum_{m=m^{\ast}+1}^{M^{\ast}}\frac{1}{lm-\frac{1}{2}\left|k\right|}l\leq2\left|k\right|+l^{-1}\int_{lm^{\ast}+\frac{1}{2}l}^{lM^{\ast}+\frac{1}{2}l}\frac{1}{x-\frac{1}{2}\left|k\right|}dx\nonumber \\
 & \leq C\left|k\right|\mleft(1+\log\mleft(\frac{lM^{\ast}+\frac{1}{2}l-\frac{1}{2}\left|k\right|}{lm^{\ast}+\frac{1}{2}l-\frac{1}{2}\left|k\right|}\mright)\mright)\leq C\left|k\right|\mleft(1+\log\mleft(\frac{k_{F}+\left|k\right|+\frac{1}{2}l-\frac{1}{2}\left|k\right|}{\frac{1}{2}l}\mright)\mright)\\
 & \leq C\left|k\right|\mleft(1+\log\mleft(\left|k\right|k_{F}\mright)\mright)\leq C\left|k\right|\log\mleft(k_{F}\mright),\quad k_{F}\rightarrow\infty,\nonumber 
\end{align}
where we also used that $l^{-1}\leq\left|k\right|$. In all the last
error term thus obeys
\begin{equation}
\left|k\right|^{2+\frac{2}{3}}\log\mleft(k_{F}\mright)^{\frac{2}{3}}k_{F}^{\frac{2}{3}}\sum_{m=m^{\ast}}^{M^{\ast}}\frac{1}{lm-\frac{1}{2}\left|k\right|}\leq C\left|k\right|^{3+\frac{2}{3}}\log\mleft(k_{F}\mright)^{\frac{5}{3}}k_{F}^{\frac{2}{3}}
\end{equation}
and the claim follows by combining the estimates.

$\hfill\square$

Note that the condition $\left|k\right|\leq k_{F}^{\gamma}$, $\gamma\in\mleft(0,\frac{1}{11}\mright)$,
of the statement of Proposition \ref{prop:betaeq-1AsymptoticsAppendix}
arises to ensure that the error term is always $o\mleft(k_{F}\mright)$.
Although we must require this condition to control the precise asymptotics,
we can however still conclude the bound
\begin{equation}
\sum_{p\in L_{k}}\lambda_{k,p}^{-1}\leq Ck_{F},\quad\left|k\right|<2k_{F},
\end{equation}
of Proposition \ref{prop:RiemannSumEstimatesAppendix}, since it at
least shows that $\sum_{p\in L_{k}}\lambda_{k,p}^{-1}$ is $O\mleft(k_{F}\mright)$
for $\left|k\right|<k_{F}^{\frac{1}{20}}$ (say), and we previously
established the bound
\begin{equation}
\sum_{p\in L_{k}}\lambda_{k,p}^{-1}\leq C\mleft(1+\left|k\right|^{-1}\log\mleft(k_{F}\mright)\mright)k_{F},\quad\left|k\right|<2k_{F},
\end{equation}
of which the right-hand side is also $O\mleft(k_{F}\mright)$ if $\left|k\right|\geq k_{F}^{\frac{1}{20}}$,
so either way the claimed estimate holds.

\subsection{\label{subsec:RiemannSumLowerBounds}Lower Bounds for $\beta\in\left\{ 0\right\} \cup\mleft[1,\infty\mright)$}

For the lower bound of Proposition \ref{prop:RiemannSumLowerBoundsAppendix}
we must similarly divide our analysis into a ``small $k$'' and
a ``large $k$'' part. The result of Proposition \ref{prop:RiemannSummationFormula}
is sufficiently precise that we can obtain the small $k$ estimate
almost immediately by the following lower bound for $1$-dimensional
Riemann sums of convex functions:
\begin{lem}
\label{lemma:1DRiemannSumLowerBound}Let for $a,b\in\mathbb{Z}$ and
$l>0$ a convex function $f\in C\mleft(\left[la,lb\right]\mright)$
be given. Then
\[
\sum_{m=a}^{b}f\mleft(lm\mright)l\geq\int_{la}^{lb}f\mleft(x\mright)\,dx+\frac{l}{2}\mleft(f\mleft(la\mright)+f\mleft(lb\mright)\mright).
\]
\end{lem}

\textbf{Proof:} Convexity implies that for every $m\in\left\{ a,a+1,\ldots,b-1\right\} $,
\begin{equation}
f\mleft(x\mright)\leq\mleft(1-\mleft(l^{-1}x-m\mright)\mright)f\mleft(lm\mright)+\mleft(l^{-1}x-m\mright)f\mleft(l\mleft(m+1\mright)\mright),\quad x\in\left[lm,l\mleft(m+1\mright)\right],
\end{equation}
so
\begin{align}
\int_{lm}^{l\mleft(m+1\mright)}f\mleft(x\mright)\,dx & \leq\mleft(\int_{lm}^{l\mleft(m+1\mright)}\mleft(1-\mleft(l^{-1}x-m\mright)\mright)\,dx\mright)f\mleft(lm\mright)+\mleft(\int_{lm}^{l\mleft(m+1\mright)}\mleft(l^{-1}x-m\mright)\,dx\mright)f\mleft(l\mleft(m+1\mright)\mright)\nonumber \\
 & =f\mleft(lm\mright)l\int_{0}^{1}\mleft(1-x\mright)\,dx+f\mleft(l\mleft(m+1\mright)\mright)l\int_{0}^{1}x\,dx\\
 & =\frac{1}{2}\mleft(f\mleft(lm\mright)l+f\mleft(l\mleft(m+1\mright)\mright)l\mright)\nonumber 
\end{align}
whence
\begin{align}
\sum_{m=a}^{b}f\mleft(lm\mright)l & =\frac{l}{2}\mleft(f\mleft(la\mright)+f\mleft(lb\mright)\mright)+\sum_{m=a}^{b-1}\frac{1}{2}\mleft(f\mleft(lm\mright)l+f\mleft(l\mleft(m+1\mright)\mright)l\mright)\\
 & \geq\frac{l}{2}\mleft(f\mleft(la\mright)+f\mleft(lb\mright)\mright)+\int_{la}^{lb}f\mleft(x\mright)\,dx.\nonumber 
\end{align}
$\hfill\square$

By applying this we obtain the following:
\begin{prop}
For all $k\in B\mleft(0,2k_{F}\mright)$ and $\beta\in\left\{ 0\right\} \cup\mleft[1,\infty\mright)$
it holds that
\[
\sum_{p\in L_{k}}\lambda_{k,p}^{\beta}\geq c\mleft(\mleft(1-\frac{1}{2}k_{F}^{-1}\left|k\right|\mright)^{2+\beta}-C\left|k\right|^{3+\frac{2}{3}}\log\mleft(k_{F}\mright)^{\frac{2}{3}}k_{F}^{-\frac{1}{3}}\mright)k_{F}^{2+\beta}\left|k\right|^{1+\beta},\quad k_{F}\rightarrow\infty,
\]
for constants $c,C>0$ depending only on $\beta$.
\end{prop}

\textbf{Proof:} By Proposition \ref{prop:RiemannSummationFormula}
it holds that
\begin{equation}
\sum_{p\in L_{k}}\lambda_{k,p}^{\beta}\geq2\pi\left|k\right|^{1+\beta}\sum_{m=m^{\ast}}^{M}\mleft(lm-\frac{1}{2}\left|k\right|\mright)^{1+\beta}l-C\left|k\right|^{\beta+3+\frac{2}{3}}\log\mleft(k_{F}\mright)^{\frac{2}{3}}k_{F}^{\frac{2}{3}}\sum_{m=m^{\ast}}^{M^{\ast}}\mleft(lm-\frac{1}{2}\left|k\right|\mright)^{\beta}
\end{equation}
where we discarded the second sum as every term of this is non-negative.
By the previous lemma we can bound
\begin{align}
\sum_{m=m^{\ast}}^{M}\mleft(lm-\frac{1}{2}\left|k\right|\mright)^{1+\beta}l & \geq\int_{lm^{\ast}}^{lM}\mleft(x-\frac{1}{2}\left|k\right|\mright)^{1+\beta}dx+\frac{l}{2}\mleft(\mleft(lM-\frac{1}{2}\left|k\right|\mright)^{1+\beta}+\mleft(lm^{\ast}-\frac{1}{2}\left|k\right|\mright)^{1+\beta}\mright)\nonumber \\
 & \geq\frac{1}{2+\beta}\mleft(\mleft(lM-\frac{1}{2}\left|k\right|\mright)^{2+\beta}-\mleft(lm^{\ast}-\frac{1}{2}\left|k\right|\mright)^{2+\beta}\mright)\\
 & \geq\frac{1}{2+\beta}\mleft(\mleft(k_{F}-\frac{1}{2}\left|k\right|-l\mright)^{2+\beta}-l^{2+\beta}\mright)\geq c\mleft(1-\frac{1}{2}k_{F}^{-1}\left|k\right|\mright)^{2+\beta}k_{F}^{2+\beta}\nonumber 
\end{align}
as $k_{F}\text{\ensuremath{\rightarrow\infty}}$, where we used that
$l\leq1$ and that by the definition of $m^{\ast}$ and $M$,
\begin{equation}
l\mleft(m^{\ast}-1\mright)\leq\frac{1}{2}\left|k\right|,\quad k_{F}<l\mleft(M+1\mright).
\end{equation}
Meanwhile, Corollary \ref{coro:1DRiemannSumLemma} lets us bound the
sum of the error term as
\begin{align}
\sum_{m=m^{\ast}}^{M^{\ast}}\mleft(lm-\frac{1}{2}\left|k\right|\mright)^{\beta} & \leq l^{-1}\int_{lm^{\ast}-\frac{1}{2}l}^{lM^{\ast}+\frac{1}{2}l}\mleft(x-\frac{1}{2}\left|k\right|\mright)^{\beta}dx=\frac{l^{-1}}{1+\beta}\mleft(\mleft(lM^{\ast}-\frac{1}{2}\left|k\right|\mright)^{1+\beta}-\mleft(lm^{\ast}-\frac{1}{2}\left|k\right|\mright)^{1+\beta}\mright)\nonumber \\
 & \leq\frac{\left|k\right|}{1+\beta}\mleft(k_{F}+\frac{1}{2}\left|k\right|\mright)^{1+\beta}\leq C\left|k\right|k_{F}^{1+\beta},\quad k_{F}\rightarrow\infty,
\end{align}
and combining the estimates yields the claim.

$\hfill\square$

As was the case for our precise bound on $\sum_{p\in L_{k}}\lambda_{k,p}^{-1}$,
this implies that
\begin{equation}
\sum_{p\in L_{k}}\lambda_{k,p}^{\beta}\geq ck_{F}^{2+\beta}\left|k\right|^{1+\beta},\quad k_{F}\rightarrow\infty,
\end{equation}
uniformly for $\left|k\right|\leq k_{F}^{\gamma}$, $\gamma\in\mleft(0,\frac{1}{11}\mright)$,
but to extend this to all $k\in B_{F}$ we must also establish some
simpler bounds for larger $k$.

\subsubsection*{Large $k$ Estimates}

We begin by observing that
\begin{equation}
\sum_{p\in L_{k}}\lambda_{k,p}^{\beta}\geq\left|k\right|^{\beta}\int_{\bigcup_{q\in L_{k}}\mathcal{C}_{q}}\max\left\{ \mleft(\hat{k}\cdot p-\frac{1}{2}\left|k\right|-\frac{\sqrt{3}}{2}\mright)^{\beta},0\right\} dp\label{eq:SimpleLowerBoundAppendixEstimate1}
\end{equation}
where we recall that $\mathcal{C}_{q}=\left[-2^{-1},2^{-1}\right]+q$.
Indeed, for any $p\in\mathcal{C}_{q}$ it holds that
\begin{align}
\lambda_{k,q} & =\frac{1}{2}\mleft(\left|q\right|^{2}-\left|q-k\right|^{2}\mright)=\left|k\right|\mleft(\hat{k}\cdot q-\frac{1}{2}\left|k\right|\mright)\\
 & =\left|k\right|\mleft(\hat{k}\cdot p-\frac{1}{2}\left|k\right|-\hat{k}\cdot\mleft(p-q\mright)\mright)\geq\left|k\right|\mleft(\hat{k}\cdot p-\frac{1}{2}\left|k\right|-\frac{\sqrt{3}}{2}\mright)\nonumber 
\end{align}
by Cauchy-Schwarz, as $p\in\mathcal{C}_{q}$ implies that $\left|p-q\right|\leq\frac{\sqrt{3}}{2}$
as also used earlier. We then note the following inclusion:
\begin{prop}
For any $\epsilon>0$ it holds that
\[
\mathcal{S}_{\epsilon}=\overline{B}\mleft(k,k_{F}-\frac{\sqrt{3}}{2}-\epsilon\mright)\backslash\overline{B}\mleft(0,k_{F}+\frac{\sqrt{3}}{2}+\epsilon\mright)\subset\bigcup_{q\in L_{k}}\mathcal{C}_{q}.
\]
\end{prop}

\textbf{Proof:} We first show that $S_{-}\subset\bigcup_{q\in L_{k}}\mathcal{C}_{q}$
where $S_{-}$ is given by
\begin{equation}
S_{-}=\left\{ p\in\mathbb{R}^{3}\mid\inf_{q\in\mathbb{R}^{3}\backslash\mleft(\overline{B}\mleft(k,k_{F}\mright)\backslash\overline{B}\mleft(0,k_{F}\mright)\mright)}\left|p-q\right|>\frac{\sqrt{3}}{2}\right\} .
\end{equation}
Indeed, for any $p\in\mathbb{R}^{3}$ we have that $\mathcal{C}_{p}\cap\mathbb{Z}^{3}\neq\emptyset$,
so if additionally $p\in S_{-}$ then it holds for $q'\in\mathcal{C}_{p}\cap\mathbb{Z}^{3}$
that
\begin{equation}
\inf_{q\in\mathbb{R}^{3}\backslash\mleft(\overline{B}\mleft(k,k_{F}\mright)\backslash\overline{B}\mleft(0,k_{F}\mright)\mright)}\left|q'-q\right|\geq\inf_{q\in\mathbb{R}^{3}\backslash\mleft(\overline{B}\mleft(k,k_{F}\mright)\backslash\overline{B}\mleft(0,k_{F}\mright)\mright)}\left|p-q\right|-\left|q'-p\right|>0
\end{equation}
hence $q'\in\mathbb{Z}^{3}\cap\mleft(\overline{B}\mleft(k,k_{F}\mright)\backslash\overline{B}\mleft(0,k_{F}\mright)\mright)=L_{k}$.
As $q'\in\mathcal{C}_{p}\Leftrightarrow p\in\mathcal{C}_{q'}$ by
symmetry of the cube, this shows that $p\in\bigcup_{q\in L_{k}}\mathcal{C}_{q}$.

Now it holds that $\mathcal{S}_{\epsilon}\subset S_{-}$, as $p\in\mathcal{S}_{\epsilon}$
implies that if $q\in\mathbb{R}^{3}\backslash\mleft(\overline{B}\mleft(k,k_{F}\mright)\backslash\overline{B}\mleft(0,k_{F}\mright)\mright)=\mleft(\mathbb{R}^{3}\backslash\overline{B}\mleft(k,k_{F}\mright)\mright)\cup\overline{B}\mleft(0,k_{F}\mright)$
then at least one of the inequalities
\begin{align}
\left|p-q\right| & \geq\left|\left|p-k\right|-\left|q-k\right|\right|=\left|q-k\right|-\left|p-k\right|>k_{F}-k_{F}+\frac{\sqrt{3}}{2}+\epsilon=\frac{\sqrt{3}}{2}+\epsilon\\
\left|p-q\right| & \geq\left|\left|p\right|-\left|q\right|\right|=\left|p\right|-\left|q\right|>k_{F}+\frac{\sqrt{3}}{2}+\epsilon-k_{F}=\frac{\sqrt{3}}{2}+\epsilon\nonumber 
\end{align}
are valid, according to whether $q\in\mathbb{R}^{3}\backslash\overline{B}\mleft(k,k_{F}\mright)$
or $q\in\overline{B}\mleft(0,k_{F}\mright)$, hence
\begin{equation}
\inf_{q\in\mathbb{R}^{3}\backslash\mleft(\overline{B}\mleft(k,k_{F}\mright)\backslash\overline{B}\mleft(0,k_{F}\mright)\mright)}\left|p-q\right|\geq\frac{\sqrt{3}}{2}+\epsilon>\frac{\sqrt{3}}{2}
\end{equation}
i.e. $p\in S_{-}\subset\bigcup_{q\in L_{k}}\mathcal{C}_{q}$.

$\hfill\square$

From equation (\ref{eq:SimpleLowerBoundAppendixEstimate1}) we can
now obtain
\begin{align}
\sum_{p\in L_{k}}\lambda_{k,p}^{\beta} & \geq\limsup_{\epsilon\rightarrow0^{+}}\left|k\right|^{\beta}\int_{\mathcal{S}_{\epsilon}}\max\left\{ \mleft(\hat{k}\cdot p-\frac{1}{2}\left|k\right|-\frac{\sqrt{3}}{2}\mright)^{\beta},0\right\} dp\\
 & =\left|k\right|^{\beta}\int_{\mathcal{S}}\mleft(\hat{k}\cdot p-\frac{1}{2}\left|k\right|-\frac{\sqrt{3}}{2}\mright)^{\beta}dp\nonumber 
\end{align}
for $\mathcal{S}=\overline{B}\mleft(k,k_{F}-\frac{\sqrt{3}}{2}\mright)\backslash\overline{B}\mleft(0,k_{F}+\frac{\sqrt{3}}{2}\mright)$,
where we also used that $\hat{k}\cdot p\geq\frac{1}{2}\left|k\right|+\frac{\sqrt{3}}{2}$
for $p\in\mathcal{S}$. Note that $\mathcal{S}=\emptyset$ unless
$\left|k\right|>\sqrt{3}$.

Similar to what we did for the simple upper bounds, we consider the
slices $\mathcal{S}_{t}=\mathcal{S}\cap\left\{ \hat{k}\cdot p=t\right\} $:
The area of $\mathcal{S}_{t}$ is
\begin{align}
\mathrm{Area}\mleft(\mathcal{S}_{t}\mright) & =\pi\mleft(\mleft(k_{F}-\frac{\sqrt{3}}{2}\mright)^{2}-\mleft(t-\left|k\right|\mright)\mright)-\pi\mleft(\mleft(k_{F}+\frac{\sqrt{3}}{2}\mright)^{2}-t^{2}\mright)\\
 & =2\pi\mleft(\left|k\right|\mleft(t-\frac{1}{2}\left|k\right|\mright)-\sqrt{3}k_{F}\mright)\nonumber 
\end{align}
for $\frac{1}{2}\left|k\right|+\sqrt{3}\left|k\right|^{-1}k_{F}\leq t\leq k_{F}+\frac{\sqrt{3}}{2}$;
the area for $t\geq k_{F}+\frac{\sqrt{3}}{2}$ is unnecessary since
the integrand under consideration is non-negative and we are looking
for a lower bound. We can then estimate as follows:
\begin{prop}
For all $k\in B\mleft(0,2k_{F}\mright)\backslash\overline{B}\mleft(0,\sqrt{3}\mright)$
and $\beta\in\left\{ 0\right\} \cup\mleft[1,\infty\mright)$ it holds
that
\[
\sum_{p\in L_{k}}\lambda_{k,p}^{\beta}\geq c\mleft(\mleft(1-\frac{1}{2}k_{F}^{-1}\left|k\right|\mright)^{2+\beta}-C\mleft(\left|k\right|^{-1}\mleft(1-\frac{1}{2}k_{F}^{-1}\left|k\right|\mright)^{1+\beta}+\left|k\right|^{-\mleft(2+\beta\mright)}\mright)\mright)k_{F}^{2+\beta}\left|k\right|^{1+\beta}
\]
as $k_{F}\rightarrow\infty$ for constants $c,C>0$ depending only
on $\beta$.
\end{prop}

\textbf{Proof:} By the considerations above
\begin{align}
\sum_{p\in L_{k}}\lambda_{k,p}^{\beta} & \geq2\pi\left|k\right|^{\beta}\int_{\frac{1}{2}\left|k\right|+\sqrt{3}\left|k\right|^{-1}k_{F}}^{k_{F}+\frac{\sqrt{3}}{2}}\mleft(t-\frac{1}{2}\left|k\right|-\frac{\sqrt{3}}{2}\mright)^{\beta}\mleft(\left|k\right|\mleft(t-\frac{1}{2}\left|k\right|\mright)-\sqrt{3}k_{F}\mright)dt\nonumber \\
 & \geq2\pi\left|k\right|^{\beta}\mleft(\left|k\right|\int_{\sqrt{3}\left|k\right|^{-1}k_{F}-\frac{\sqrt{3}}{2}}^{k_{F}-\frac{1}{2}\left|k\right|}t^{1+\beta}dt-\sqrt{3}k_{F}\int_{\sqrt{3}\left|k\right|^{-1}k_{F}-\frac{\sqrt{3}}{2}}^{k_{F}-\frac{1}{2}\left|k\right|}t^{\beta}dt\mright)\\
 & \geq2\pi\left|k\right|^{\beta}\mleft(\frac{\left|k\right|}{2+\beta}\mleft(\mleft(k_{F}-\frac{1}{2}\left|k\right|\mright)^{2+\beta}-\mleft(\sqrt{3}\left|k\right|^{-1}k_{F}\mright)^{2+\beta}\mright)-\frac{\sqrt{3}k_{F}}{1+\beta}\mleft(k_{F}-\frac{1}{2}\left|k\right|\mright)^{1+\beta}\mright)\nonumber \\
 & \geq c\mleft(\mleft(1-\frac{1}{2}k_{F}^{-1}\left|k\right|\mright)^{2+\beta}-C\mleft(\left|k\right|^{-1}\mleft(1-\frac{1}{2}k_{F}^{-1}\left|k\right|\mright)^{1+\beta}+\left|k\right|^{-\mleft(2+\beta\mright)}\mright)\mright)k_{F}^{2+\beta}\left|k\right|^{1+\beta}.\nonumber 
\end{align}
$\hfill\square$

This implies that
\begin{equation}
\sum_{p\in L_{k}}\lambda_{k,p}^{\beta}\geq ck_{F}^{2+\beta}\left|k\right|^{1+\beta},\quad k_{F}\rightarrow\infty,
\end{equation}
uniformly for $k_{F}^{\gamma}<\left|k\right|<k_{F}$, $\gamma>0$,
which combined with the small $k$ result yields Proposition \ref{prop:RiemannSumLowerBoundsAppendix}.

\section{\label{sec:CarefulJustificationoftheTransformationFormulas}Careful
Justification of the Transformation Formulas}

In this section we give a more detailed justification of the transformation
identities which we derived in the sections \ref{sec:DiagonalizationoftheBosonizableTerms}
and \ref{sec:EstimationoftheNon-BosonizableTermsandGronwallEstimates}
for the operator $\mathcal{K}$. Although we proved in Section \ref{sec:ControllingtheTransformationKernel}
that
\begin{equation}
\mathcal{K}=\frac{1}{2}\sum_{l\in\mathbb{Z}_{\ast}^{3}}\sum_{p,q\in L_{l}}\left\langle e_{p},K_{l}e_{q}\right\rangle \mleft(b_{l,p}b_{-l,-q}-b_{-l,-q}^{\ast}b_{l,p}^{\ast}\mright)\label{eq:AppendixcalKDefinition}
\end{equation}
defines a bounded operator whenever $\sum_{l\in\mathbb{Z}_{\ast}^{3}}\left\Vert K_{l}\right\Vert _{\mathrm{HS}}^{2}<\infty$,
and so most of the subleties involving unbounded operators can be
avoided, the fact that the operators we apply the transformation to
are themselves unbounded still raises some technical questions.

The first transformation rules we consider are those for the bosonizable
terms
\begin{equation}
H_{\mathrm{B}}=H_{\mathrm{kin}}^{\prime}+\sum_{k\in\mathbb{Z}_{\ast}^{3}}\mleft(2Q_{1}^{k}\mleft(P_{k}\mright)+Q_{2}^{k}\mleft(P_{k}\mright)\mright).
\end{equation}
In this section we prove the following precise statement for these:
\begin{prop}
The transformation $e^{-\mathcal{K}}$ preserves $D\mleft(H_{\mathrm{kin}}^{\prime}\mright)$,
$e^{\mathcal{K}}H_{\mathrm{B}}e^{-\mathcal{K}}:D\mleft(H_{\mathrm{kin}}^{\prime}\mright)\rightarrow\mathcal{H}_{N}$
is self-adjoint and both $H_{\mathrm{B}}-H_{\mathrm{kin}}^{\prime}$
and $e^{\mathcal{K}}H_{\mathrm{B}}e^{-\mathcal{K}}-H_{\mathrm{kin}}^{\prime}$
extend to bounded operators on all of $\mathcal{H}_{N}$.
\end{prop}

In words, the transformation of the bosonizable terms does indeed
make rigorous sense, and the tranformation does not generate any ``new''
unboundedness, in so far as $H_{\mathrm{kin}}^{\prime}$ is the only
unbounded part of $H_{\mathrm{B}}$ both before and after the transformation.

The second transformation formula we consider is the one concerning
$\mathcal{Q}_{\mathrm{SR}}$. Here we will prove the following:
\begin{prop}
$\mathcal{Q_{\mathrm{SR}}}$and $e^{\mathcal{K}}\mathcal{Q}_{\mathrm{SR}}e^{-\mathcal{K}}$
are well-defined in quadratic form sense on $D\mleft(H_{\mathrm{kin}}^{\prime}\mright)$
and $e^{\mathcal{K}}\mathcal{Q}_{\mathrm{SR}}e^{-\mathcal{K}}-Q_{\mathrm{SR}}$
extends to a bounded operator on all of $\mathcal{H}_{N}$.
\end{prop}

Due to a technical point we will not verify whether the transformation
identity is valid on an operator level, but it is valid in the quadratic
form sense (which is all we apply in the main text) and again the
transformation does not generate any new unboundedness.

As we are chiefly concerned with qualitative properties of operators
in this section, we will generally estimate rather roughly and not
keep track of $k_{F}$ and $s$ dependencies. In this case the bound
of Proposition \ref{prop:calKNumberBound} can simply be summarized
as
\begin{equation}
\left\Vert \mathcal{K}\right\Vert _{\mathrm{Op}}\leq C\sqrt{\sum_{l\in\mathbb{Z}_{\ast}^{3}}\left\Vert K_{l}\right\Vert _{\mathrm{HS}}^{2}}\label{eq:AppendixcalKBound}
\end{equation}
for any operator of the form of equation (\ref{eq:AppendixcalKDefinition}),
since (as also remarked in Section \ref{sec:ControllingtheTransformationKernel})
$\mathcal{N}_{E}\leq N\leq Csk_{F}^{3}\leq C'$.

\subsubsection*{Elaboration on the Well-Definedness of $\mathcal{K}$ }

On the same note, let us also elaborate on \textit{how} this bound
implies that $\mathcal{K}$ is well-defined - since this is a sum
of infinitely many terms, this is not immediately clear, and so the
bound of equation (\ref{eq:AppendixcalKBound}) might only constitute
a formal calculation.

The reason this is not so is that Proposition \ref{prop:calKNumberBound}
applies to \textit{any} operator of the form of equation (\ref{eq:AppendixcalKDefinition}),
and so if we for $R\in\mathbb{N}$ define $\mathcal{K}_{R}$ by
\begin{equation}
\mathcal{K}_{R}=\frac{1}{2}\sum_{l\in\overline{B}\mleft(0,R\mright)\cap\mathbb{Z}_{\ast}^{3}}\sum_{p,q\in L_{l}}\left\langle e_{p},K_{l}e_{q}\right\rangle \mleft(b_{l,p}b_{-l,-q}-b_{-l,-q}^{\ast}b_{l,p}^{\ast}\mright),
\end{equation}
i.e. let $\mathcal{K}_{R}$ be a cut-off version of $\mathcal{K}$,
then this is \textit{a priori} well-defined, as the summation is now
only over finitely many terms. The bound then certainly applies in
this case to show that
\begin{equation}
\left\Vert \mathcal{K}_{R}\right\Vert _{\mathrm{Op}}\leq C\sqrt{\sum_{l\in\overline{B}\mleft(0,R\mright)\cap\mathbb{Z}_{\ast}^{3}}\left\Vert K_{l}\right\Vert _{\mathrm{HS}}^{2}}.
\end{equation}
This implies that if the limit $\mathcal{K}=\lim_{R\rightarrow\infty}\mathcal{K}_{R}$
exists then it obeys the claimed bound. Existence is however automatically
guaranteed by the same argument, as $\mleft(\mathcal{K}_{R}\mright)_{R=1}^{\infty}$
is in fact Cauchy: For any $r,R\in\mathbb{N}$, the difference $\mathcal{K}_{R}-\mathcal{K}_{r}$
is also of the form of equation (\ref{eq:AppendixcalKDefinition}),
whence (assuming that $r\leq R$ for definiteness)
\begin{equation}
\left\Vert \mathcal{K}_{R}-\mathcal{K}_{r}\right\Vert _{\mathrm{Op}}\leq C\sqrt{\sum_{l\in\overline{B}\mleft(0,R\mright)\backslash\overline{B}\mleft(0,r\mright)\cap\mathbb{Z}_{\ast}^{3}}\left\Vert K_{l}\right\Vert _{\mathrm{HS}}^{2}}\leq C\sqrt{\sum_{l\in\mathbb{Z}_{\ast}^{3}\backslash\overline{B}\mleft(0,r\mright)}\left\Vert K_{l}\right\Vert _{\mathrm{HS}}^{2}}
\end{equation}
which implies the Cauchy property.

For our argument we considered the particular cut-off sets $\overline{B}\mleft(0,R\mright)\cap\mathbb{Z}_{\ast}^{3}$,
but an argument similar to this last one shows that the limit exists
for, and is independent of, any particular exhaustion of $\mathbb{Z}_{\ast}^{3}$,
so $\mathcal{K}$ is indeed unambigously defined.

\subsection{Transformation of Quadratic Operators}

We begin by considering the transformation law for quadratic operators.
This is greatly simplified by the fact that these are in fact bounded
- not only are
\begin{align}
Q_{k}^{1}\mleft(A\mright) & =\sum_{p,q\in L_{k}}\left\langle e_{p},Ae_{q}\right\rangle b_{k,p}^{\ast}b_{k,q}\\
Q_{k}^{2}\mleft(B\mright) & =\sum_{p,q\in L_{k}}\left\langle e_{p},Be_{q}\right\rangle \mleft(b_{k,p}b_{-k,-q}+b_{-k,-q}^{\ast}b_{k,p}^{\ast}\mright)\nonumber 
\end{align}
bounded for any $k\in\mathbb{Z}_{\ast}^{3}$ and $A,B:\ell^{2}\mleft(L_{k}\mright)\rightarrow\ell^{2}\mleft(L_{k}\mright)$
simply by virtue of being sums of finitely many terms of bounded operators,
the infinite sums $\sum_{k\in\mathbb{Z}_{\ast}^{3}}Q_{1}^{k}\mleft(A_{k}\mright)$
and $\sum_{k\in\mathbb{Z}_{\ast}^{3}}Q_{2}^{k}\mleft(B_{k}\mright)$
also define bounded operators, as we claim the following holds:
\begin{prop}
\label{prop:QuadraticSumBounds}For any collections of symmetric operators
$\mleft(A_{k}\mright)$, $\mleft(B_{k}\mright)$ and $\Psi\in\mathcal{H}_{N}$
it holds that
\begin{align*}
\left|\sum_{k\in\mathbb{Z}_{\ast}^{3}}\left\langle \Psi,Q_{1}^{k}\mleft(A_{k}\mright)\Psi\right\rangle \right| & \leq\sqrt{3}\sqrt{\sum_{k\in\mathbb{Z}_{\ast}^{3}}\left\Vert A_{k}\right\Vert _{\mathrm{HS}}^{2}}\left\langle \Psi,\mathcal{N}_{E}\Psi\right\rangle \\
\left|\sum_{k\in\mathbb{Z}_{\ast}^{3}}\left\langle \Psi,Q_{2}^{k}\mleft(B_{k}\mright)\Psi\right\rangle \right| & \leq2\sqrt{5}\sqrt{\sum_{k\in\mathbb{Z}_{\ast}^{3}}\left\Vert B_{k}\right\Vert _{\mathrm{HS}}^{2}}\left\langle \Psi,\mleft(\mathcal{N}_{E}+1\mright)\Psi\right\rangle .
\end{align*}
\end{prop}

Qualitatively this implies that
\begin{equation}
\left\Vert \sum_{k\in\mathbb{Z}_{\ast}^{3}}Q_{1}^{k}\mleft(A_{k}\mright)\right\Vert _{\mathrm{Op}}\leq C\sqrt{\sum_{k\in\mathbb{Z}_{\ast}^{3}}\left\Vert A_{k}\right\Vert _{\mathrm{HS}}^{2}},\quad\left\Vert \sum_{k\in\mathbb{Z}_{\ast}^{3}}Q_{2}^{k}\mleft(B_{k}\mright)\right\Vert _{\mathrm{Op}}\leq C\sqrt{\sum_{k\in\mathbb{Z}_{\ast}^{3}}\left\Vert B_{k}\right\Vert _{\mathrm{HS}}^{2}}.
\end{equation}
(Here we also use the assumed symmetry of $\mleft(A_{k}\mright)$ and
$\mleft(B_{k}\mright)$, though this isn't necessary.)

The same argument we just illustrated with $\mathcal{K}$ thus implies
that these sums are well-defined bounded operators provided the right-hand
sides are finite.

Before we turn to the transformation law, let us prove this proposition.
First we note that we have effectively already proven the $Q_{2}^{k}$
bound, since we can write
\begin{equation}
Q_{2}^{k}\mleft(B\mright)=\sum_{p,q\in L_{k}}\left\langle e_{p},Be_{q}\right\rangle \mleft(b_{k,p}b_{-k,-q}+b_{-k,-q}^{\ast}b_{k,p}^{\ast}\mright)=2\,\mathrm{Re}\mleft(\tilde{Q}_{2}^{k}\mleft(B\mright)\mright)
\end{equation}
for
\begin{equation}
\tilde{Q}_{2}^{k}\mleft(B\mright)=\sum_{p,q\in L_{k}}\left\langle e_{p},Be_{q}\right\rangle b_{k,p}b_{-k,-q},
\end{equation}
and $\sum_{k\in\mathbb{Z}_{\ast}^{3}}\tilde{Q}_{2}^{k}\mleft(B_{k}\mright)$
is (up to a factor of $2$) of the same form as $\tilde{\mathcal{K}}$
in Proposition \ref{prop:calKTildeBound}, whence
\begin{equation}
\left|\sum_{k\in\mathbb{Z}_{\ast}^{3}}\left\langle \Psi,Q_{2}^{k}\mleft(B_{k}\mright)\Psi\right\rangle \right|\leq2\left|\sum_{k\in\mathbb{Z}_{\ast}^{3}}\left\langle \Psi,\tilde{Q}_{2}^{k}\mleft(B_{k}\mright)\Psi\right\rangle \right|\leq2\sqrt{5}\sqrt{\sum_{k\in\mathbb{Z}_{\ast}^{3}}\left\Vert B_{k}\right\Vert _{\mathrm{HS}}^{2}}\left\langle \Psi,\mleft(\mathcal{N}_{E}+1\mright)\Psi\right\rangle .
\end{equation}
The $Q_{1}^{k}$ bound follows similarly to how we obtained Proposition
\ref{prop:calKTildeBound} (although simpler, as there is less computation
necessary): Writing
\begin{align}
\sum_{k\in\mathbb{Z}_{\ast}^{3}}Q_{1}^{k}\mleft(A_{k}\mright) & =\sum_{k\in\mathbb{Z}_{\ast}^{3}}\sum_{p,q\in L_{k}}\left\langle e_{p},A_{k}e_{q}\right\rangle b_{k,p}^{\ast}b_{k,q}=\frac{1}{\sqrt{s}}\sum_{k\in\mathbb{Z}_{\ast}^{3}}\sum_{p,q\in L_{k}}^{\sigma}\left\langle e_{p},A_{k}e_{q}\right\rangle b_{k,p}^{\ast}c_{q-k,\sigma}^{\ast}c_{q,\sigma}\\
 & =\frac{1}{\sqrt{s}}\sum_{q\in B_{F}^{c}}^{\sigma}\mleft(\sum_{k\in\mathbb{Z}_{\ast}^{3}}\sum_{p\in L_{k}}1_{L_{k}}\mleft(q\mright)\left\langle e_{p},A_{k}e_{q}\right\rangle b_{k,p}^{\ast}c_{q-k,\sigma}^{\ast}\mright)c_{q,\sigma}\nonumber 
\end{align}
we can bound
\begin{align}
\sum_{k\in\mathbb{Z}_{\ast}^{3}}\left\langle \Psi,Q_{1}^{k}\mleft(A_{k}\mright)\Psi\right\rangle  & =\frac{1}{\sqrt{s}}\sum_{q\in B_{F}^{c}}^{\sigma}\left\langle \mleft(\sum_{k\in\mathbb{Z}_{\ast}^{3}}\sum_{p\in L_{k}}1_{L_{k}}\mleft(q\mright)\left\langle A_{k}e_{q},e_{p}\right\rangle c_{q-k,\sigma}b_{k,p}\mright)\Psi,c_{q,\sigma}\Psi\right\rangle \nonumber \\
 & \leq\frac{1}{\sqrt{s}}\sqrt{\sum_{q\in B_{F}^{c}}^{\sigma}\left\Vert \sum_{k\in\mathbb{Z}_{\ast}^{3}}\sum_{p\in L_{k}}1_{L_{k}}\mleft(q\mright)\left\langle A_{k}e_{q},e_{p}\right\rangle c_{q-k,\sigma}b_{k,p}\Psi\right\Vert ^{2}}\sqrt{\sum_{q\in B_{F}^{c}}^{\sigma}\left\Vert c_{q,\sigma}\Psi\right\Vert ^{2}}\label{eq:SumofQ1kAkEstimate}\\
 & =\frac{1}{\sqrt{s}}\sqrt{\sum_{q\in B_{F}^{c}}^{\sigma}\left\Vert \sum_{k\in\mathbb{Z}_{\ast}^{3}}\sum_{p\in L_{k}}1_{L_{k}}\mleft(q\mright)\left\langle A_{k}e_{q},e_{p}\right\rangle c_{q-k,\sigma}b_{k,p}\Psi\right\Vert ^{2}}\sqrt{\left\langle \Psi,\mathcal{N}_{E}\Psi\right\rangle }\nonumber 
\end{align}
and note that
\begin{align}
 & \quad\sum_{k\in\mathbb{Z}_{\ast}^{3}}\sum_{p\in L_{k}}1_{L_{k}}\mleft(q\mright)\left\langle A_{k}e_{q},e_{p}\right\rangle c_{q-k,\sigma}b_{k,p}=\frac{1}{\sqrt{s}}\sum_{k\in\mathbb{Z}_{\ast}^{3}}\sum_{p\in L_{k}}^{\tau}1_{L_{k}}\mleft(q\mright)\left\langle A_{k}e_{q},e_{p}\right\rangle c_{q-k,\sigma}c_{p-k,\tau}^{\ast}c_{p,\tau}\\
 & =\frac{1}{\sqrt{s}}\sum_{p'\in B_{F}^{c}}^{\tau}\sum_{q',r'\in B_{F}}\mleft(\sum_{k\in\mathbb{Z}_{\ast}^{3}}\sum_{p\in L_{k}}\delta_{p,p'}\delta_{p-k,q'}\delta_{q-k,r'}1_{L_{k}}\mleft(q\mright)\left\langle A_{k}e_{q},e_{p}\right\rangle \mright)c_{r',\sigma}c_{q',\tau}^{\ast}c_{p',\tau}\nonumber 
\end{align}
so that it suffices to consider expressions of the form
\begin{equation}
\frac{1}{\sqrt{s}}\sum_{p\in B_{F}^{c}}^{\tau}\sum_{q,r\in B_{F}}A_{p,q,r}c_{r,\sigma}c_{q,\tau}^{\ast}c_{p,\tau}.
\end{equation}
We calculate the following commutator:
\begin{lem}
For any $p,p'\in B_{F}^{c}$, $q,q',r',r'\in B_{F}$ and $1\le\sigma,\tau,\tau'\leq s$
it holds that
\begin{align*}
\left\{ \mleft(c_{r,\sigma}c_{q,\tau}^{\ast}c_{p,\tau}\mright)^{\ast},c_{r',\sigma}c_{q',\tau'}^{\ast}c_{p',\tau'}\right\}  & =\delta_{p,p'}^{\tau,\tau'}c_{r',\sigma}c_{q',\tau'}^{\ast}c_{q,\tau}c_{r,\sigma}^{\ast}+\delta_{q,q'}^{\tau,\tau'}c_{r',\sigma}c_{p',\tau'}c_{p,\tau}^{\ast}c_{r,\sigma}^{\ast}\\
 & +\delta_{r,r'}c_{p,\tau}^{\ast}c_{q,\tau}c_{q',\tau'}^{\ast}c_{p',\tau'}-\delta_{p,p'}^{\tau,\tau}\delta_{q,q'}^{\tau,\tau'}c_{r',\sigma}c_{r,\sigma}^{\ast}.
\end{align*}
\end{lem}

\textbf{Proof:} Repeatedly applying the CAR we find
\begin{align}
 & \quad\;\;\mleft(c_{r,\sigma}c_{q,\tau}^{\ast}c_{p,\tau}\mright)^{\ast}c_{r',\sigma}c_{q',\tau'}^{\ast}c_{p',\tau'}=c_{p,\tau}^{\ast}c_{q,\tau}c_{r,\sigma}^{\ast}c_{r',\sigma}c_{q',\tau'}^{\ast}c_{p',\tau'}\nonumber \\
 & =-c_{p,\tau}^{\ast}c_{q,\tau}c_{r',\sigma}c_{r,\sigma}^{\ast}c_{q',\tau'}^{\ast}c_{p',\tau'}+\delta_{r,r'}c_{p,\tau}^{\ast}c_{q,\tau}c_{q',\tau'}^{\ast}c_{p',\tau'}\nonumber \\
 & =-c_{r',\sigma}c_{p,\tau}^{\ast}c_{q,\tau}c_{q',\tau'}^{\ast}c_{p',\tau'}c_{r,\sigma}^{\ast}+\delta_{r,r'}c_{p,\tau}^{\ast}c_{q,\tau}c_{q',\tau'}^{\ast}c_{p',\tau'}\\
 & =c_{r',\sigma}c_{p,\tau}^{\ast}c_{q',\tau'}^{\ast}c_{q,\tau}c_{p',\tau'}c_{r,\sigma}^{\ast}-\delta_{q,q'}^{\tau,\tau'}c_{r',\sigma}c_{p,\tau}^{\ast}c_{p',\tau'}c_{r,\sigma}^{\ast}+\delta_{r,r'}c_{p,\tau}^{\ast}c_{q,\tau}c_{q',\tau'}^{\ast}c_{p',\tau'}\nonumber \\
 & =c_{r',\sigma}c_{q',\tau'}^{\ast}c_{p,\tau}^{\ast}c_{p',\tau'}c_{q,\tau}c_{r,\sigma}^{\ast}+\delta_{q,q'}^{\tau,\tau'}c_{r',\sigma}c_{p',\tau'}c_{p,\tau}^{\ast}c_{r,\sigma}^{\ast}+\delta_{r,r'}c_{p,\tau}^{\ast}c_{q,\tau}c_{q',\tau'}^{\ast}c_{p',\tau'}-\delta_{p,p'}^{\tau,\tau'}\delta_{q,q'}^{\tau,\tau'}c_{r',\sigma}c_{r,\sigma}^{\ast}\nonumber \\
 & =-c_{r',\sigma}c_{q',\tau'}^{\ast}c_{p',\tau'}c_{p,\tau}^{\ast}c_{q,\tau}c_{r,\sigma}^{\ast}+\delta_{p,p'}^{\tau,\tau'}c_{r',\sigma}c_{q',\tau'}^{\ast}c_{q,\tau}c_{r,\sigma}^{\ast}+\delta_{q,q'}^{\tau,\tau'}c_{r',\sigma}c_{p',\tau'}c_{p,\tau}^{\ast}c_{r,\sigma}^{\ast}+\delta_{r,r'}c_{p,\tau}^{\ast}c_{q,\tau}c_{q',\tau'}^{\ast}c_{p',\tau'}\nonumber \\
 & -\delta_{p,p'}^{\tau,\tau}\delta_{q,q'}^{\tau,\tau'}c_{r',\sigma}c_{r,\sigma}^{\ast}.\nonumber 
\end{align}
$\hfill\square$

The bound on $\sum_{p\in B_{F}^{c}}^{\tau}\sum_{q,r\in B_{F}}A_{p,q,r}c_{r,\sigma}c_{q,\tau}^{\ast}c_{p,\tau}$
now follows:
\begin{prop}
Let $A_{p,q,r}\in\mathbb{C}$ for $p\in B_{F}^{c}$ and $q,r\in B_{F}$
with $\sum_{p\in B_{F}}\sum_{q,r\in B_{F}}\left|A_{p,q,r}\right|^{2}<\infty$
be given. Then for any $\Psi\in\mathcal{H}_{N}$
\[
\frac{1}{s}\sum_{\sigma=1}^{s}\left\Vert \sum_{p\in B_{F}^{c}}^{\tau}\sum_{q,r\in B_{F}}A_{p,q,r}c_{r,\sigma}c_{q,\tau}^{\ast}c_{p,\tau}\Psi\right\Vert ^{2}\leq3s\sum_{p\in B_{F}^{c}}\sum_{q,r\in B_{F}}\left|A_{p,q,r}\right|^{2}\left\langle \Psi,\mathcal{N}_{E}\Psi\right\rangle .
\]
\end{prop}

\textbf{Proof:} Arguing as in Proposition \ref{prop:CubicSumBound}
and applying the lemma, we estimate
\begin{align}
 & \quad\;\frac{1}{s}\sum_{\sigma=1}^{s}\left\Vert \sum_{p\in B_{F}^{c}}^{\tau}\sum_{q,r\in B_{F}}A_{p,q,r}c_{r,\sigma}c_{q,\tau}^{\ast}c_{p,\tau}\Psi\right\Vert ^{2}\nonumber \\
 & \leq\frac{1}{s}\sum_{p,p'\in B_{F}^{c}}^{\sigma,\tau,\tau'}\sum_{q,q',r,r'\in B_{F}}\overline{A_{p,q,r}}A_{p',q',r'}\left\langle \Psi,\left\{ \mleft(c_{r,\sigma}c_{q,\tau}^{\ast}c_{p,\tau}\mright)^{\ast},c_{r',\sigma}c_{q',\tau'}^{\ast}c_{p',\tau'}\right\} \Psi\right\rangle \nonumber \\
 & =\frac{1}{s}\sum_{p\in B_{F}^{c}}^{\sigma,\tau}\left\Vert \sum_{q,r\in B_{F}}\overline{A_{p,q,r}}c_{q,\tau}c_{r,\sigma}^{\ast}\Psi\right\Vert ^{2}+\frac{1}{s}\sum_{q\in B_{F}}^{\sigma,\tau}\left\Vert \sum_{p\in B_{F}^{c}}\sum_{r\in B_{F}}\overline{A_{p,q,r}}c_{p,\tau}^{\ast}c_{r,\sigma}^{\ast}\Psi\right\Vert ^{2}\nonumber \\
 & +\frac{1}{s}\sum_{r\in B_{F}}^{\sigma}\left\Vert \sum_{p'\in B_{F}^{c}}^{\tau'}\sum_{q'\in B_{F}}A_{p',q',r}c_{q',\tau'}^{\ast}c_{p',\tau'}\Psi\right\Vert ^{2}-\frac{1}{s}\sum_{p\in B_{F}^{c}}^{\sigma,\tau}\sum_{q\in B_{F}}\left\Vert \sum_{r\in B_{F}}\overline{A_{p,q,r}}c_{r,\sigma}^{\ast}\Psi\right\Vert ^{2}\\
 & \leq\frac{1}{s}\sum_{p\in B_{F}^{c}}^{\sigma,\tau}\mleft(\sum_{r\in B_{F}}\sqrt{\sum_{q\in B_{F}}\left|A_{p,q,r}\right|^{2}}\left\Vert c_{r,\sigma}^{\ast}\Psi\right\Vert \mright)^{2}+\frac{1}{s}\sum_{q\in B_{F}}^{\sigma,\tau}\mleft(\sum_{r\in B_{F}}\sqrt{\sum_{p\in B_{F}^{c}}\left|A_{p,q,r}\right|^{2}}\left\Vert c_{r,\sigma}^{\ast}\Psi\right\Vert \mright)^{2}\nonumber \\
 & +\frac{1}{s}\sum_{r\in B_{F}}^{\sigma}\mleft(\sum_{p'\in B_{F}^{c}}^{\tau'}\sqrt{\sum_{q'\in B_{F}}\left|A_{p',q',r}\right|^{2}}\left\Vert c_{p',\tau'}\Psi\right\Vert \mright)^{2}\nonumber \\
 & \leq\mleft(2+s\mright)\left\langle \Psi,\mathcal{N}_{E}\Psi\right\rangle \leq3s\left\langle \Psi,\mathcal{N}_{E}\Psi\right\rangle .\nonumber 
\end{align}
$\hfill\square$

Applying this to equation (\ref{eq:SumofQ1kAkEstimate}) we conclude
the desired bound:
\begin{align}
\sum_{k\in\mathbb{Z}_{\ast}^{3}}\left\langle \Psi,Q_{1}^{k}\mleft(A_{k}\mright)\Psi\right\rangle  & \leq\frac{1}{\sqrt{s}}\sqrt{3s\sum_{q\in B_{F}^{c}}\sum_{p'\in B_{F}^{c}}\sum_{q',r'\in B_{F}}\left|\sum_{k\in\mathbb{Z}_{\ast}^{3}}\sum_{p\in L_{k}}\delta_{p,p'}\delta_{p-k,q'}\delta_{q-k,r'}1_{L_{k}}\mleft(q\mright)\left\langle A_{k}e_{q},e_{p}\right\rangle \right|^{2}}\left\langle \Psi,\mathcal{N}_{E}\Psi\right\rangle \nonumber \\
 & =\sqrt{3\sum_{q\in B_{F}^{c}}\sum_{p'\in B_{F}^{c}}\sum_{q'\in B_{F}}\sum_{k\in\mathbb{Z}_{\ast}^{3}}\left|\sum_{p\in L_{k}}\delta_{p,p'}\delta_{p-k,q'}1_{L_{k}}\mleft(q\mright)\left\langle A_{k}e_{q},e_{p}\right\rangle \right|^{2}}\left\langle \Psi,\mathcal{N}_{E}\Psi\right\rangle \\
 & =\sqrt{3\sum_{q\in B_{F}^{c}}\sum_{p'\in B_{F}^{c}}\sum_{k\in\mathbb{Z}_{\ast}^{3}}\sum_{p\in L_{k}}\left|\delta_{p,p'}1_{L_{k}}\mleft(q\mright)\left\langle A_{k}e_{q},e_{p}\right\rangle \right|^{2}}\left\langle \Psi,\mathcal{N}_{E}\Psi\right\rangle \nonumber \\
 & =\sqrt{3\sum_{k\in\mathbb{Z}_{\ast}^{3}}\sum_{p,q\in L_{k}}\left|\left\langle e_{p},A_{k}e_{q}\right\rangle \right|^{2}}\left\langle \Psi,\mathcal{N}_{E}\Psi\right\rangle =\sqrt{3}\sqrt{\sum_{k\in\mathbb{Z}_{\ast}^{3}}\left\Vert A_{k}\right\Vert _{\mathrm{HS}}^{2}}\left\langle \Psi,\mathcal{N}_{E}\Psi\right\rangle .\nonumber 
\end{align}

\subsubsection*{Justification of the Transformation}

We can now justify the transformation. First note that the expression
we consider,
\begin{equation}
\sum_{k\in\mathbb{Z}_{\ast}^{3}}\mleft(2\,Q_{1}^{k}\mleft(P_{k}\mright)+Q_{2}^{k}\mleft(P_{k}\mright)\mright),
\end{equation}
defines a bounded operator as
\begin{equation}
\sqrt{\sum_{k\in\mathbb{Z}_{\ast}^{3}}\left\Vert P_{k}\right\Vert _{\mathrm{HS}}^{2}}=\sqrt{\sum_{k\in\mathbb{Z}_{\ast}^{3}}\left\Vert v_{k}\right\Vert ^{4}}=\sqrt{\sum_{k\in\mathbb{Z}_{\ast}^{3}}\mleft(\frac{s\hat{V}_{k}k_{F}^{-1}}{2\,\mleft(2\pi\mright)^{3}}\left|L_{k}\right|\mright)^{2}}\leq C\sqrt{\sum_{k\in\mathbb{Z}_{\ast}^{3}}\hat{V}_{k}^{2}}<\infty,
\end{equation}
where we simply estimate that $\left|L_{k}\right|\leq Ck_{F}^{3}$.

Now we note that the transformation rules of Proposition \ref{prop:TransformationofQuadraticTerms},
i.e.
\begin{align}
 & \;\,e^{\mathcal{K}}\mleft(2\,Q_{1}^{k}\mleft(T_{k}\mright)+2\,Q_{1}^{-k}\mleft(T_{-k}\mright)\mright)e^{-\mathcal{K}}=\mathrm{tr}\mleft(T_{k}^{1}\mleft(1\mright)-T_{k}\mright)+2\,Q_{1}^{k}\mleft(T_{k}^{1}\mleft(1\mright)\mright)+Q_{2}^{k}\mleft(T_{k}^{2}\mleft(1\mright)\mright)\\
 & +\int_{0}^{1}e^{\mleft(1-t\mright)\mathcal{K}}\mleft(\varepsilon_{k}\mleft(\left\{ K_{k},T_{k}^{2}\mleft(t\mright)\right\} \mright)+2\,\mathrm{Re}\mleft(\mathcal{E}_{k}^{1}\mleft(T_{k}^{1}\mleft(t\mright)\mright)\mright)+2\,\mathrm{Re}\mleft(\mathcal{E}_{k}^{2}\mleft(T_{k}^{2}\mleft(t\mright)\mright)\mright)\mright)e^{-\mleft(1-t\mright)\mathcal{K}}dt+\mleft(k\rightarrow-k\mright)\nonumber 
\end{align}
and
\begin{align}
 & \;\,e^{\mathcal{K}}\mleft(Q_{2}^{k}\mleft(T_{k}\mright)+Q_{2}^{-k}\mleft(T_{-k}\mright)\mright)e^{-\mathcal{K}}=\mathrm{tr}\mleft(T_{k}^{2}\mleft(1\mright)\mright)+2\,Q_{1}^{k}\mleft(T_{k}^{2}\mleft(1\mright)\mright)+Q_{2}^{k}\mleft(T_{k}^{1}\mleft(1\mright)\mright)\\
 & +\int_{0}^{1}e^{\mleft(1-t\mright)\mathcal{K}}\mleft(\varepsilon_{k}\mleft(\left\{ K_{k},T_{k}^{1}\mleft(t\mright)\right\} \mright)+2\,\mathrm{Re}\mleft(\mathcal{E}_{k}^{1}\mleft(T_{k}^{2}\mleft(t\mright)\mright)\mright)+2\,\mathrm{Re}\mleft(\mathcal{E}_{k}^{2}\mleft(T_{k}^{1}\mleft(t\mright)\mright)\mright)\mright)e^{-\mleft(1-t\mright)\mathcal{K}}dt+\mleft(k\rightarrow-k\mright)\nonumber 
\end{align}
for
\begin{align}
T_{k}^{1}\mleft(t\mright) & =\frac{1}{2}\mleft(e^{tK_{k}}T_{k}e^{tK_{k}}+e^{-tK_{k}}T_{k}e^{-tK_{k}}\mright)\\
T_{k}^{2}\mleft(t\mright) & =\frac{1}{2}\mleft(e^{tK_{k}}T_{k}e^{tK_{k}}-e^{-tK_{k}}T_{k}e^{-tK_{k}}\mright)\nonumber 
\end{align}
do actually hold without further justification by boundedness\footnote{Strictly speaking, as the $\mathcal{E}_{k}^{1}\mleft(A\mright)$ and
$\mathcal{E}_{k}^{2}\mleft(B\mright)$ operators are also defined as
infinite sums (due to the sum over $l$ in their definition), one
should also justify that these are bounded operators. This can be
done by considering limits of cut-offs in $l$ and the kind of estimation
we perform in Section \ref{sec:AnalysisofExchangeTerms} - we omit
the details.}, so it is the summation over $k\in\mathbb{Z}_{\ast}^{3}$ that must
be justified. Again we consider a cut-off: The above implies that
for any $R\in\mathbb{N}$ (taking the $Q_{1}^{k}$ case for definiteness)
\begin{align}
 & \qquad\quad\;e^{\mathcal{K}}\mleft(2\sum_{k\in\overline{B}\mleft(0,R\mright)\cap\mathbb{Z}_{\ast}^{3}}Q_{1}^{k}\mleft(T_{k}\mright)\mright)e^{-\mathcal{K}}\nonumber \\
 & =\sum_{k\in\overline{B}\mleft(0,R\mright)\cap\mathbb{Z}_{\ast}^{3}}\mathrm{tr}\mleft(T_{k}^{1}\mleft(1\mright)-T_{k}\mright)+2\sum_{k\in\overline{B}\mleft(0,R\mright)\cap\mathbb{Z}_{\ast}^{3}}Q_{1}^{k}\mleft(T_{k}^{1}\mleft(1\mright)\mright)+\sum_{k\in\overline{B}\mleft(0,R\mright)\cap\mathbb{Z}_{\ast}^{3}}Q_{2}^{k}\mleft(T_{k}^{2}\mleft(1\mright)\mright)\label{eq:AppendixFiniteQuadraticSumTransformation}\\
 & +\sum_{k\in\overline{B}\mleft(0,R\mright)\cap\mathbb{Z}_{\ast}^{3}}\int_{0}^{1}e^{\mleft(1-t\mright)\mathcal{K}}\mleft(\varepsilon_{k}\mleft(\left\{ K_{k},T_{k}^{2}\mleft(t\mright)\right\} \mright)+2\,\mathrm{Re}\mleft(\mathcal{E}_{k}^{1}\mleft(T_{k}^{1}\mleft(t\mright)\mright)\mright)+2\,\mathrm{Re}\mleft(\mathcal{E}_{k}^{2}\mleft(T_{k}^{2}\mleft(t\mright)\mright)\mright)\mright)e^{-\mleft(1-t\mright)\mathcal{K}}dt\nonumber 
\end{align}
and we must argue that the limit $R\rightarrow\infty$ is well-defined.
By Proposition \ref{prop:QuadraticSumBounds} and the estimates of
Section \ref{sec:AnalysisofExchangeTerms} this is assured if (for
$j=1,2$)
\begin{equation}
\sum_{k\in\mathbb{Z}_{\ast}^{3}}\left|\mathrm{tr}\mleft(T_{k}^{1}\mleft(1\mright)-T_{k}\mright)\right|,\sum_{k\in\mathbb{Z}_{\ast}^{3}}\Vert T_{k}^{j}\mleft(1\mright)\Vert_{\mathrm{HS}}^{2}<\infty
\end{equation}
and
\begin{equation}
\sum_{k\in\mathbb{Z}_{\ast}^{3}}\sup_{p\in L_{k}}\left|\left\langle e_{p},\left\{ K_{k},T_{k}^{j}\mleft(t\mright)\right\} e_{p}\right\rangle \right|,\,\sum_{k\in\mathbb{Z}_{\ast}^{3}}\sum_{p\in L_{k}}\max_{q\in L_{k}}\left|\left\langle e_{p},T_{k}^{j}\mleft(t\mright)e_{q}\right\rangle \right|^{2},\,\sum_{k\in\mathbb{Z}_{\ast}^{3}}\Vert T_{k}^{j}\mleft(t\mright)h_{k}^{-\frac{1}{2}}\Vert_{\mathrm{HS}}^{2}<\infty.
\end{equation}
In our particular case $T_{k}=P_{k}$ and $P_{k}^{1}\mleft(t\mright),\,P_{k}^{2}\mleft(t\mright)$
can be written as
\begin{equation}
P_{k}^{1}\mleft(t\mright)=P_{k}+\frac{1}{2}P_{k}^{+}\mleft(t\mright)+\frac{1}{2}P_{k}^{-}\mleft(t\mright),\quad P_{k}^{2}\mleft(t\mright)=\frac{1}{2}P_{k}^{+}\mleft(t\mright)-\frac{1}{2}P_{k}^{-}\mleft(t\mright),
\end{equation}
for
\begin{equation}
P_{k}^{\pm}=e^{\pm tK_{k}}P_{k}e^{\pm tK_{k}}-P_{k}.
\end{equation}
Arguing as in Proposition \ref{prop:etKPetK-PEstimate} (which really
concerns $P_{k}^{+}\mleft(t\mright)$) one can see that
\begin{equation}
\left|\left\langle e_{p},P_{k}^{\pm}\mleft(t\mright)e_{q}\right\rangle \right|\leq C\hat{V}_{k}^{2}
\end{equation}
independently of $t$, and naturally $\left|\left\langle e_{p},P_{k}e_{q}\right\rangle \right|\leq C\hat{V}_{k}$
which implies finiteness of the sums above.

In conclusion:
\begin{prop}
The expression $\sum_{k\in\mathbb{Z}_{\ast}^{3}}\mleft(2\,Q_{1}^{k}\mleft(P_{k}\mright)+Q_{2}^{k}\mleft(P_{k}\mright)\mright)$
defines a bounded operator on $\mathcal{H}_{N}$ and it holds that
\begin{align*}
 & \quad\;\;\;e^{\mathcal{K}}\mleft(\sum_{k\in\mathbb{Z}_{\ast}^{3}}\mleft(2\,Q_{1}^{k}\mleft(P_{k}\mright)+Q_{2}^{k}\mleft(P_{k}\mright)\mright)\mright)e^{-\mathcal{K}}\\
 & =\sum_{k\in\mathbb{Z}_{\ast}^{3}}\mathrm{tr}\mleft(P_{k}^{1}\mleft(1\mright)+P_{k}^{2}\mleft(1\mright)-P_{k}\mright)+\sum_{k\in\mathbb{Z}_{\ast}^{3}}\mleft(2\,Q_{1}^{k}\mleft(P_{k}^{1}\mleft(1\mright)+P_{k}^{2}\mleft(1\mright)\mright)+Q_{2}^{k}\mleft(P_{k}^{1}\mleft(1\mright)+P_{k}^{2}\mleft(1\mright)\mright)\mright)\\
 & +\sum_{k\in\mathbb{Z}_{\ast}^{3}}\int_{0}^{1}e^{\mleft(1-t\mright)\mathcal{K}}\mleft(\varepsilon_{k}\mleft(\left\{ K_{k},P_{k}^{1}\mleft(t\mright)+P_{k}^{2}\mleft(t\mright)\right\} \mright)+\mathcal{E}_{k}^{1}\mleft(P_{k}^{1}\mleft(t\mright)+P_{k}^{2}\mleft(t\mright)\mright)+\mathcal{E}_{k}^{2}\mleft(P_{k}^{1}\mleft(t\mright)+P_{k}^{2}\mleft(t\mright)\mright)\mright)e^{-\mleft(1-t\mright)\mathcal{K}}dt
\end{align*}
with the right-hand side likewise defining a bounded operator. 
\end{prop}

\subsection{Transformation of $H_{\mathrm{kin}}^{\prime}$}

We now come to $H_{\mathrm{kin}}^{\prime}$. As this is a proper unbounded
operator we must exercise more care in working with this than we did
with the quadratic operators.

To work with $H_{\mathrm{kin}}^{\prime}$ we will apply the following
general result, which we prove in appendix section \ref{subsec:TransformationofUnboundedOperators}:
\begin{prop*}[\ref{prop:ezKAe-zKProperties}]
Let $X$ be a Banach space, $A:D\mleft(A\mright)\rightarrow X$ be
a closed operator and let $K:X\rightarrow X$ be a bounded operator
which preserves $D\mleft(A\mright)$. Suppose that $AK:D\mleft(A\mright)\rightarrow X$
is $A$-bounded.

Then for every $z\in\mathbb{C}$ the operator $e^{zK}:X\rightarrow X$
likewise preserves $D\mleft(A\mright)$ and $e^{zK}Ae^{-zK}:D\mleft(A\mright)\rightarrow X$
is closed. If additionally $X$ is a Hilbert space, $A$ is self-adjoint
and $K$ is skew-symmetric then $e^{tK}Ae^{-tK}$ is self-adjoint
for all $t\in\mathbb{R}$.

Furthermore, for every $x\in D\mleft(A\mright)$ the mapping $z\mapsto e^{zK}Ae^{-zK}x$
is complex differentiable and $C^{1}$ with
\[
\frac{d}{dz}e^{zK}Ae^{-zK}x=e^{zK}\left[K,A\right]e^{-zK}x.
\]
\end{prop*}
To apply the result we must show that $\mathcal{K}$ preserves $D\mleft(H_{\mathrm{kin}}^{\prime}\mright)$
and that $H_{\mathrm{kin}}^{\prime}\mathcal{K}$ is $H_{\mathrm{kin}}^{\prime}$-bounded.
To do this we will work with the cut-off operators $\mathcal{K}_{R}$,
and obtain the corresponding results for $\mathcal{K}$ by the following
lemma:
\begin{lem}
\label{lemma:LimitOperatorCommutatorLemma}Let $X$ be a Banach space,
$A:D\mleft(A\mright)\rightarrow X$ be a closed operator and $\mleft(B_{k}\mright)_{k=1}^{\infty}\subset\mathcal{B}\mleft(X\mright)$
a collection of bounded operators such that $B_{k}\rightarrow B\in\mathcal{B}\mleft(X\mright)$
(in norm).

Suppose that all $B_{k}$ preserve $D\mleft(A\mright)$ and that the
commutators $\left[B_{k},A\right]:D\mleft(A\mright)\rightarrow X$ converge
pointwise to some $C:D\mleft(A\mright)\rightarrow X$.

Then $B$ also preserves $D\mleft(A\mright)$ and $\left[B,A\right]=C$.
\end{lem}

\textbf{Proof:} Let $x\in D\mleft(A\mright)$ be arbitrary. Then $B_{k}x\rightarrow Bx$
by assumption, and likewise
\begin{equation}
AB_{k}x=B_{k}Ax-\left[B_{k},A\right]x\rightarrow BAx-Cx.
\end{equation}
It follows by closedness of $A$ that $Bx\in D\mleft(A\mright)$, i.e.
that $B$ preserves $D\mleft(A\mright)$, and that
\begin{equation}
ABx=BAx-Cx
\end{equation}
i.e. $\left[B,A\right]=C$.

$\hfill\square$

We consider the operators $\mathcal{K}_{R}$. For this we require
another general result:
\begin{lem}
\label{lemma:DomainPreservationLemma}Let $A:D\mleft(A\mright)\rightarrow X$
be a closed operator with core $\mathcal{C}$ and let $K:X\rightarrow X$
be a bounded operator which maps $\mathcal{C}$ into $D\mleft(A\mright)$.
Suppose that $\left.AK\right\vert _{\mathcal{C}}:\mathcal{C}\rightarrow X$
is $\left.A\right\vert _{\mathcal{C}}$-bounded. Then $K$ preserves
$D\mleft(A\mright)$ and $AK:D\mleft(A\mright)\rightarrow X$ is $A$-bounded
(with the same relative bounds).
\end{lem}

\textbf{Proof:} Let $x\in D\mleft(A\mright)$ be arbitrary. As $\mathcal{C}$
is a core for $A$, there exists a sequence $\mleft(x_{k}\mright)_{k=1}^{\infty}\subset\mathcal{C}$
such that
\begin{equation}
x_{k}\rightarrow x\quad\text{and}\quad Ax_{k}\rightarrow Ax,\quad k\rightarrow\infty.
\end{equation}
Since $K$ is bounded, $Kx_{k}\rightarrow Kx$, and as $\left.AK\right\vert _{\mathcal{C}}$
is $\left.A\right\vert _{\mathcal{C}}$-bounded, the fact that $\mleft(Ax_{k}\mright)_{k=1}^{\infty}$
converges implies that $\mleft(AKx_{k}\mright)_{k=1}^{\infty}$ also
converges. By closedness of $A$ it then follows that $Kx\in D\mleft(A\mright)$
and $AKx_{k}\rightarrow AKx$. The first statement shows that $K$
indeed preserves $D\mleft(A\mright)$, while the second implies that
$AK:D\mleft(A\mright)\rightarrow X$ is $A$-bounded, since if $\left\Vert AKx'\right\Vert \leq a\left\Vert Ax'\right\Vert +b\left\Vert x'\right\Vert $
for $x'\in\mathcal{C}$ then also
\begin{equation}
\left\Vert AKx\right\Vert =\lim_{k\rightarrow\infty}\left\Vert AKx_{k}\right\Vert \leq\limsup_{k\rightarrow\infty}\mleft(a\left\Vert Ax_{k}\right\Vert +b\left\Vert x_{k}\right\Vert \mright)=a\left\Vert Ax\right\Vert +b\left\Vert x\right\Vert 
\end{equation}
for $x\in D\mleft(A\mright)$.

$\hfill\square$

We can now prove that $\left[\mathcal{K}_{R},H_{\mathrm{kin}}^{\prime}\right]$
behaves as expected:
\begin{prop}
For any $R\in\mathbb{N}$ it holds that $\mathcal{K}_{R}$ preserves
$D\mleft(H_{\mathrm{kin}}^{\prime}\mright)$ and
\[
\left[\mathcal{K}_{R},H_{\mathrm{kin}}^{\prime}\right]=\left.\sum_{k\in\overline{B}\mleft(0,R\mright)\cap\mathbb{Z}_{\ast}^{3}}Q_{2}^{k}\mleft(\left\{ K_{k},h_{k}\right\} \mright)\right\vert _{D\mleft(H_{\mathrm{kin}}^{\prime}\mright)}.
\]
\end{prop}

\textbf{Proof:} First we note that $\mathcal{K}_{R}$ maps $\bigwedge_{\mathrm{alg}}^{N}H^{2}\mleft(\mathbb{T}^{3};\mathbb{C}^{s}\mright)$,
which is a core for $H_{\mathrm{kin}}^{\prime}$, into $D\mleft(H_{\mathrm{kin}}^{\prime}\mright)$:
The operator $b_{k,p}$ can be written as
\begin{align}
b_{k,p} & =\frac{1}{\sqrt{s}}\sum_{\sigma=1}^{s}c_{p-k,\sigma}^{\ast}c_{p,\sigma}=\frac{1}{\sqrt{s}}\sum_{\sigma=1}^{s}\sum_{p',q'\in\mathbb{Z}^{3}}\delta_{p',p-k}\delta_{q',p}c_{p',\sigma}^{\ast}c_{q',\sigma}\nonumber \\
 & =\frac{1}{\sqrt{s}}\sum_{\sigma=1}^{s}\sum_{p',q'\in\mathbb{Z}^{3}}^{\tau,\tau'}\left\langle u_{p',\tau},u_{p-k,\sigma}\right\rangle \left\langle u_{p,\sigma},u_{q',\tau'}\right\rangle c_{p',\tau}^{\ast}c_{q',\tau'}\\
 & =\frac{1}{\sqrt{s}}\sum_{\sigma=1}^{s}\sum_{p',q'\in\mathbb{Z}^{3}}^{\tau,\tau'}\left\langle u_{p,\tau},P_{p-k,p}^{\mleft(\sigma\mright)}u_{q',\tau'}\right\rangle c_{p',\tau}^{\ast}c_{q',\tau'}=\frac{1}{\sqrt{s}}\sum_{\sigma=1}^{s}\mathrm{d}\Gamma\mleft(P_{p-k,p}^{\mleft(\sigma\mright)}\mright)\nonumber 
\end{align}
where $P_{p-k,p}^{\mleft(\sigma\mright)}=\left|u_{p-k,\sigma}\right\rangle \left\langle u_{p,\sigma}\right|$.
Now, $\mathrm{d}\Gamma\mleft(P_{p-k,p}^{\mleft(\sigma\mright)}\mright)$
preserves $\bigwedge_{\mathrm{alg}}^{N}H^{2}\mleft(\mathbb{T}^{3};\mathbb{C}^{s}\mright)$
for any $k,p\in\mathbb{Z}^{3}$ and $1\leq\sigma\leq s$, as $P_{p-k,p}^{\mleft(\sigma\mright)}$
simply takes an inner product and projects onto $u_{p,\sigma}\in H^{2}\mleft(\mathbb{T}^{3};\mathbb{C}^{s}\mright)$,
so $b_{k,p}$ likewise preserves $\bigwedge_{\mathrm{alg}}^{N}H^{2}\mleft(\mathbb{T}^{3};\mathbb{C}^{s}\mright)$.
The same argument applies to $b_{k,p}^{\ast}$, so as a finite sum
of products of operators which preserve $\bigwedge_{\mathrm{alg}}^{N}H^{2}\mleft(\mathbb{T}^{3};\mathbb{C}^{s}\mright)$,
$\mathcal{K}_{R}$ also preserves this, hence certainly maps it into
$D\mleft(H_{\mathrm{kin}}^{\prime}\mright)=D\mleft(H_{\mathrm{kin}}\mright)=\bigwedge^{N}H^{2}\mleft(\mathbb{T}^{3};\mathbb{C}^{s}\mright)$.

Having established that $H_{\mathrm{kin}}^{\prime}\mathcal{K}_{R}$
is well-defined on $\bigwedge_{\mathrm{alg}}^{N}H^{2}\mleft(\mathbb{T}^{3};\mathbb{C}^{s}\mright)$,
we note that the calculation we performed in Proposition \ref{prop:calKHkinCommutator}
shows that
\begin{equation}
\left[\mathcal{K}_{R},H_{\mathrm{kin}}^{\prime}\right]=\sum_{k\in\overline{B}\mleft(0,R\mright)\cap\mathbb{Z}_{\ast}^{3}}Q_{2}^{k}\mleft(\left\{ K_{k},h_{k}\right\} \mright),
\end{equation}
at least on this domain. It follows that $H_{\mathrm{kin}}^{\prime}\mathcal{K}_{R}$
is $H_{\mathrm{kin}}^{\prime}$-bounded here, since for any $\Psi\in\bigwedge_{\mathrm{alg}}^{N}H^{2}\mleft(\mathbb{T}^{3};\mathbb{C}^{s}\mright)$
\begin{equation}
\left\Vert H_{\mathrm{kin}}^{\prime}\mathcal{K}_{R}\Psi\right\Vert \leq\left\Vert \mathcal{K}_{R}H_{\mathrm{kin}}^{\prime}\Psi\right\Vert +\left\Vert \left[\mathcal{K}_{R},H_{\mathrm{kin}}^{\prime}\right]\Psi\right\Vert \leq\left\Vert \mathcal{K}_{R}\right\Vert _{\mathrm{Op}}\left\Vert H_{\mathrm{kin}}^{\prime}\Psi\right\Vert +\left\Vert \left[\mathcal{K}_{R},H_{\mathrm{kin}}^{\prime}\right]\right\Vert _{\mathrm{Op}}\left\Vert \Psi\right\Vert .
\end{equation}
Lemma \ref{lemma:DomainPreservationLemma} now implies that $\mathcal{K}_{R}$
in fact preserves all of $D\mleft(H_{\mathrm{kin}}^{\prime}\mright)$
and the commutator identity continues to hold.

$\hfill\square$

We can now extend this to $\mathcal{K}$ proper:
\begin{prop}
$\mathcal{K}$ preserves $D\mleft(H_{\mathrm{kin}}^{\prime}\mright)$,
the commutator
\[
\left[\mathcal{K},H_{\mathrm{kin}}^{\prime}\right]=\left.\sum_{k\in\mathbb{Z}_{\ast}^{3}}Q_{2}^{k}\mleft(\left\{ K_{k},h_{k}\right\} \mright)\right\vert _{D\mleft(H_{\mathrm{kin}}^{\prime}\mright)}
\]
extends to a bounded operator on all of $\mathcal{H}_{N}$, and $H_{\mathrm{kin}}^{\prime}\mathcal{K}$
is $H_{\mathrm{kin}}^{\prime}$-bounded.
\end{prop}

\textbf{Proof:} By Lemma \ref{lemma:LimitOperatorCommutatorLemma}
it only remains to be shown that $\lim_{R\rightarrow\infty}\sum_{k\in\overline{B}\mleft(0,R\mright)\cap\mathbb{Z}_{\ast}^{3}}Q_{2}^{k}\mleft(\left\{ K_{k},h_{k}\right\} \mright)$
exists on $D\mleft(H_{\mathrm{kin}}^{\prime}\mright)$. In fact this
exists everywhere, since Proposition \ref{prop:QuadraticSumBounds}
says that this is ensured if $\sum_{k\in\mathbb{Z}_{\ast}^{3}}\left\Vert \left\{ K_{k},h_{k}\right\} \right\Vert _{\mathrm{HS}}^{2}<\infty$,
and by the one-body operator estimates of Section \ref{sec:AnalysisofOne-BodyOperators},
\begin{align}
\left\Vert \left\{ K_{k},h_{k}\right\} \right\Vert _{\mathrm{HS}}^{2} & =\sum_{p,q\in L_{k}}\left|\left\langle e_{p},\left\{ K_{k},h_{k}\right\} e_{q}\right\rangle \right|^{2}=\sum_{p,q\in L_{k}}\left|\mleft(\lambda_{k,p}+\lambda_{k,q}\mright)\left\langle e_{p},K_{k}e_{q}\right\rangle \right|^{2}\nonumber \\
 & \leq\sum_{p,q\in L_{k}}\left|\mleft(\lambda_{k,p}+\lambda_{k,q}\mright)\frac{\left\langle e_{p},v_{k}\right\rangle \left\langle v_{k},e_{q}\right\rangle }{\lambda_{k,p}+\lambda_{k,q}}\right|^{2}=\mleft(\sum_{p\in L_{k}}\left|\left\langle e_{p},v_{k}\right\rangle \right|^{2}\mright)^{2}\\
 & =\mleft(\sum_{p\in L_{k}}\frac{s\hat{V}_{k}k_{F}^{-1}}{2\,\mleft(2\pi\mright)^{3}}\mright)^{2}\leq C\mleft(\hat{V}_{k}\left|L_{k}\right|\mright)^{2}\leq C'\hat{V}_{k}^{2},\quad k\in\mathbb{Z}_{\ast}^{3}.\nonumber 
\end{align}
$\hfill\square$

Proposition \ref{prop:ezKAe-zKProperties} now gives us the following:
\begin{cor}
The operator $e^{t\mathcal{K}}H_{\mathrm{kin}}^{\prime}e^{-t\mathcal{K}}:D\mleft(H_{\mathrm{kin}}^{\prime}\mright)\rightarrow\mathcal{H}_{N}$
is a well-defined, self-adjoint operator for all $t\in\mathbb{R}$,
and for any $\Psi\in D\mleft(H_{\mathrm{kin}}^{\prime}\mright)$ it
holds that
\[
\frac{d}{dt}e^{t\mathcal{K}}H_{\mathrm{kin}}^{\prime}e^{-t\mathcal{K}}\Psi=e^{t\mathcal{K}}\left[\mathcal{K},H_{\mathrm{kin}}^{\prime}\right]e^{-t\mathcal{K}}\Psi=\sum_{k\in\mathbb{Z}_{\ast}^{3}}e^{t\mathcal{K}}Q_{2}^{k}\mleft(\left\{ K_{k},h_{k}\right\} \mright)e^{-t\mathcal{K}}\Psi
\]
and this is continuous in $t$.
\end{cor}

We now have all the necessary prerequisites to carefully implement
Proposition \ref{prop:HKinTransformation}:
\begin{prop}
The statement of Proposition \ref{prop:HKinTransformation} holds
pointwise on $D\mleft(H_{\mathrm{kin}}^{\prime}\mright)$ and $e^{\mathcal{K}}H_{\mathrm{kin}}^{\prime}e^{-\mathcal{K}}-H_{\mathrm{kin}}^{\prime}$
extends continuously to all of $\mathcal{H}_{N}$.
\end{prop}

\textbf{Proof:} For any $R\in\mathbb{N}$, $\sum_{k\in\overline{B}\mleft(0,R\mright)\cap\mathbb{Z}_{\ast}^{3}}Q_{1}^{k}\mleft(h_{k}\mright)$
defines a bounded operator. Given $\Psi\in D\mleft(H_{\mathrm{kin}}^{\prime}\mright)$
we can then conclude by the corollary that
\begin{align}
 & \,\frac{d}{dt}e^{t\mathcal{K}}\mleft(H_{\mathrm{kin}}^{\prime}-2\sum_{k\in\overline{B}\mleft(0,R\mright)\cap\mathbb{Z}_{\ast}^{3}}Q_{1}^{k}\mleft(h_{k}\mright)\mright)e^{-t\mathcal{K}}\Psi\nonumber \\
 & =e^{t\mathcal{K}}\mleft(\sum_{k\in\mathbb{Z}_{\ast}^{3}}Q_{2}^{k}\mleft(\left\{ K_{k},h_{k}\right\} \mright)-\sum_{k\in\overline{B}\mleft(0,R\mright)\cap\mathbb{Z}_{\ast}^{3}}\left[\mathcal{K},Q_{1}^{k}\mleft(h_{k}\mright)\right]\mright)e^{-t\mathcal{K}}\Psi\\
 & =e^{t\mathcal{K}}\mleft(\sum_{k\in\mathbb{Z}_{\ast}^{3}\backslash\overline{B}\mleft(0,R\mright)}Q_{2}^{k}\mleft(\left\{ K_{k},h_{k}\right\} \mright)-\sum_{k\in\overline{B}\mleft(0,R\mright)\cap\mathbb{Z}_{\ast}^{3}}2\,\mathrm{Re}\mleft(\mathcal{E}_{k}^{1}\mleft(h_{k}\mright)\mright)\mright)e^{-t\mathcal{K}}\Psi,\nonumber 
\end{align}
which upon rearrangement reads
\begin{align}
\frac{d}{dt}e^{t\mathcal{K}}H_{\mathrm{kin}}^{\prime}e^{-t\mathcal{K}}\Psi & =\frac{d}{dt}e^{t\mathcal{K}}\mleft(2\sum_{k\in\overline{B}\mleft(0,R\mright)\cap\mathbb{Z}_{\ast}^{3}}Q_{1}^{k}\mleft(h_{k}\mright)\mright)e^{-t\mathcal{K}}\Psi\\
 & +e^{t\mathcal{K}}\mleft(\sum_{k\in\mathbb{Z}_{\ast}^{3}\backslash\overline{B}\mleft(0,R\mright)}Q_{2}^{k}\mleft(\left\{ K_{k},h_{k}\right\} \mright)-\sum_{k\in\overline{B}\mleft(0,R\mright)\cap\mathbb{Z}_{\ast}^{3}}2\,\mathrm{Re}\mleft(\mathcal{E}_{k}^{1}\mleft(h_{k}\mright)\mright)\mright)e^{-t\mathcal{K}}\Psi.\nonumber 
\end{align}
As the corollary also ensures that this is continuous in $t$, hence
Riemann integrable, the fundamental theorem of calculus together with
equation (\ref{eq:AppendixFiniteQuadraticSumTransformation}) shows
that
\begin{align}
e^{\mathcal{K}}H_{\mathrm{kin}}^{\prime}e^{-\mathcal{K}}\Psi & =H_{\mathrm{kin}}^{\prime}\Psi+e^{\mathcal{K}}\mleft(2\sum_{k\in\overline{B}\mleft(0,R\mright)\cap\mathbb{Z}_{\ast}^{3}}Q_{1}^{k}\mleft(h_{k}\mright)\mright)e^{-\mathcal{K}}\Psi-2\sum_{k\in\overline{B}\mleft(0,R\mright)\cap\mathbb{Z}_{\ast}^{3}}Q_{1}^{k}\mleft(h_{k}\mright)\Psi\nonumber \\
 & +\int_{0}^{1}e^{t\mathcal{K}}\mleft(\sum_{k\in\mathbb{Z}_{\ast}^{3}\backslash\overline{B}\mleft(0,R\mright)}Q_{2}^{k}\mleft(\left\{ K_{k},h_{k}\right\} \mright)-\sum_{k\in\overline{B}\mleft(0,R\mright)\cap\mathbb{Z}_{\ast}^{3}}2\,\mathrm{Re}\mleft(\mathcal{E}_{k}^{1}\mleft(h_{k}\mright)\mright)\mright)e^{-t\mathcal{K}}\Psi dt\\
 & =\sum_{k\in\overline{B}\mleft(0,R\mright)\backslash\mathbb{Z}_{\ast}^{3}}\mathrm{tr}\mleft(h_{k}^{1}\mleft(1\mright)-h_{k}\mright)\Psi+H_{\mathrm{kin}}^{\prime}\Psi+\sum_{k\in\overline{B}\mleft(0,R\mright)\backslash\mathbb{Z}_{\ast}^{3}}\mleft(2\,Q_{1}^{k}\mleft(h_{k}^{1}\mleft(1\mright)-h_{k}\mright)+Q_{2}^{k}\mleft(h_{k}^{2}\mleft(1\mright)\mright)\mright)\Psi\nonumber \\
 & +\sum_{k\in\overline{B}\mleft(0,R\mright)\backslash\mathbb{Z}_{\ast}^{3}}\int_{0}^{1}e^{\mleft(1-t\mright)\mathcal{K}}\mleft(\varepsilon_{k}\mleft(\left\{ K_{k},h_{k}^{2}\mleft(t\mright)\right\} \mright)+2\,\mathrm{Re}\mleft(\mathcal{E}_{k}^{1}\mleft(h_{k}^{1}\mleft(t\mright)-h_{k}\mright)\mright)+2\,\mathrm{Re}\mleft(\mathcal{E}_{k}^{2}\mleft(h_{k}^{2}\mleft(t\mright)\mright)\mright)\mright)e^{-\mleft(1-t\mright)\mathcal{K}}\Psi dt\nonumber \\
 & +\sum_{k\in\mathbb{Z}_{\ast}^{3}\backslash\overline{B}\mleft(0,R\mright)}\int_{0}^{1}e^{t\mathcal{K}}Q_{2}^{k}\mleft(\left\{ K_{k},h_{k}\right\} \mright)e^{-t\mathcal{K}}\Psi dt.\nonumber 
\end{align}
The formula of Proposition \ref{prop:HKinTransformation} now follows
provided we can take $R\rightarrow\infty$. As in the previous subsection,
this is possible if various sums involving the one-body operators
$h_{k}^{1}\mleft(t\mright)$ and $h_{k}^{2}\mleft(t\mright)$ are finite
- but with respect to the notation in Section \ref{sec:AnalysisofOne-BodyOperators},
\begin{equation}
h_{k}^{1}\mleft(t\mright)-h_{k}=A_{h_{k}}\mleft(t\mright),\quad h_{k}^{2}\mleft(t\mright)=B_{h_{k}}\mleft(t\mright)-tP_{v_{k}},
\end{equation}
and the bounds derived in that section for these operators yield the
desired estimates. The same bounds also imply the boundedness of $e^{\mathcal{K}}H_{\mathrm{kin}}^{\prime}e^{-\mathcal{K}}-H_{\mathrm{kin}}^{\prime}$
by the same argument.

$\hfill\square$

\subsection{\label{subsec:TransformationofcalQSR}Transformation of $\mathcal{Q}_{\mathrm{SR}}$}

For the short-range quartic terms
\begin{equation}
\mathcal{Q}_{\mathrm{SR}}=\frac{k_{F}^{-1}}{2\,\mleft(2\pi\mright)^{3}}\sum_{k\in\mathbb{Z}_{\ast}^{3}}\hat{V}_{k}\sum_{p,q\in B_{F}^{c}\cap\mleft(B_{F}^{c}-k\mright)}^{\sigma,\tau}c_{p+k,\sigma}^{\ast}c_{q,\tau}^{\ast}c_{q+k,\tau}c_{p,\sigma}
\end{equation}
we will switch our argument around and rather than cutting-off $\mathcal{K}$,
cut-off $\mathcal{Q}_{\mathrm{SR}}$ instead, and so consider for
$R\in\mathbb{N}$ the bounded operators
\begin{equation}
\mathcal{Q}_{\mathrm{SR}}^{\mleft(R\mright)}=\frac{k_{F}^{-1}}{2\,\mleft(2\pi\mright)^{3}}\sum_{k\in\overline{B}\mleft(0,R\mright)\cap\mathbb{Z}_{\ast}^{3}}\hat{V}_{k}\sum_{p,q\in B_{F}^{c}\cap\mleft(B_{F}^{c}-k\mright)}^{\sigma,\tau}c_{p+k,\sigma}c_{q,\tau}c_{q+k,\tau}c_{p,\sigma}.
\end{equation}
Now, we would like to say that $\mathcal{Q}_{\mathrm{SR}}\Psi=\lim_{R\rightarrow\infty}\mathcal{Q}_{\mathrm{SR}}^{\mleft(R\mright)}\Psi$
for any $\Psi\in D\mleft(H_{\mathrm{kin}}^{\prime}\mright)$, but here
arises a technical point: How is $\mathcal{Q}_{\mathrm{SR}}\Psi$
defined? We obtained $\mathcal{Q}_{\mathrm{SR}}$ by manipulating
the second-quantized form of $H_{N}$, but \textit{a priori} the action
of this representation need only be defined for elements of $\bigwedge_{\mathrm{alg}}^{N}H^{2}\mleft(\mathbb{T}^{3};\mathbb{C}^{s}\mright)$,
with the general action captured by extension arguments. Manipulating
such forms can therefore be a delicate issue (had we not included
the additional quadratic terms in our definition of $\mathcal{Q}$,
for instance, this would not be a well-defined operator, as an unavoidable
infinity then appears for unbounded $V$).

We must therefore clarify what we mean by $\mathcal{Q}_{\mathrm{SR}}$.
We note the following:
\begin{prop}
Let $\sum_{k\in\mathbb{Z}_{\ast}^{3}}\hat{V}_{k}^{2}<\infty$. Then
for any $\Psi\in D\mleft(H_{\mathrm{kin}}^{\prime}\mright)=D\mleft(H_{\mathrm{kin}}\mright)$
it holds that
\[
\left|\left\langle \Psi,\mathcal{Q}_{\mathrm{SR}}^{\mleft(R\mright)}\Psi\right\rangle \right|\leq C\mleft(\left\Vert \Psi\right\Vert ^{2}+\left\Vert H_{\mathrm{kin}}\Psi\right\Vert ^{2}\mright)
\]
for a $C>0$ independent of $R$.
\end{prop}

\textbf{Proof:} By Cauchy-Schwarz and the triangle inequality in the
form $\left|k\right|=\left|p+k-p\right|\leq\left|p+k\right|+\left|p\right|$
we can estimate
\begin{align}
\left|\left\langle \Psi,\mathcal{Q}_{\mathrm{SR}}^{\mleft(R\mright)}\Psi\right\rangle \right| & \leq\frac{k_{F}^{-1}}{2\,\mleft(2\pi\mright)^{3}}\sum_{k\in\overline{B}\mleft(0,R\mright)\cap\mathbb{Z}_{\ast}^{3}}\hat{V}_{k}\sum_{p,q\in B_{F}^{c}\cap\mleft(B_{F}^{c}-k\mright)}^{\sigma,\tau}\left\Vert c_{q,\tau}c_{p+k,\sigma}\Psi\right\Vert \left\Vert c_{q+k,\tau}c_{p,\sigma}\Psi\right\Vert \\
 & \leq C\sum_{k\in\overline{B}\mleft(0,R\mright)\cap\mathbb{Z}_{\ast}^{3}}\hat{V}_{k}\sum_{p,q\in B_{F}^{c}\cap\mleft(B_{F}^{c}-k\mright)}^{\sigma,\tau}\frac{\left|p\right|+\left|p+k\right|}{\left|k\right|}\frac{\left|q\right|+\left|q+k\right|}{\left|k\right|}\left\Vert c_{q,\tau}c_{p+k,\sigma}\Psi\right\Vert \left\Vert c_{q+k,\tau}c_{p,\sigma}\Psi\right\Vert \nonumber \\
 & \leq C\sum_{k\in\overline{B}\mleft(0,R\mright)\cap\mathbb{Z}_{\ast}^{3}}\frac{\hat{V}_{k}}{\left|k\right|^{2}}\sum_{p,q\in B_{F}^{c}\cap\mleft(B_{F}^{c}-k\mright)}^{\sigma,\tau}\mleft(\left|p\right|\left|q\right|+\left|p\right|\left|q+k\right|\mright)\left\Vert c_{q,\tau}c_{p+k,\sigma}\Psi\right\Vert \left\Vert c_{q+k,\tau}c_{p,\sigma}\Psi\right\Vert \nonumber 
\end{align}
where we apply the symmetry of the summations to reduce the consideration
of $\mleft(\left|p\right|+\left|p+k\right|\mright)\mleft(\left|q\right|+\left|q+k\right|\mright)$
to the two terms $\left|p\right|\left|q\right|$ and $\left|p\right|\left|q+k\right|$.
For the first kind of terms we bound as
\begin{align}
 & \qquad\;\sum_{p,q\in B_{F}^{c}\cap\mleft(B_{F}^{c}-k\mright)}^{\sigma,\tau}\left|p\right|\left|q\right|\left\Vert c_{q,\tau}c_{p+k,\sigma}\Psi\right\Vert \left\Vert c_{q+k,\tau}c_{p,\sigma}\Psi\right\Vert \nonumber \\
 & \leq\sqrt{\sum_{p,q\in B_{F}^{c}\cap\mleft(B_{F}^{c}-k\mright)}^{\sigma,\tau}\left|p\right|^{2}\left\Vert c_{q+k,\tau}c_{p,\sigma}\Psi\right\Vert ^{2}}\sqrt{\sum_{p,q\in B_{F}^{c}\cap\mleft(B_{F}^{c}-k\mright)}^{\sigma,\tau}\left|q\right|^{2}\left\Vert c_{q,\tau}c_{p+k,\sigma}\Psi\right\Vert ^{2}}\\
 & \leq\sum_{p\in B_{F}^{c}\cap\mleft(B_{F}^{c}-k\mright)}^{\sigma}\left|p\right|^{2}\Vert\mathcal{N}_{E}^{\frac{1}{2}}c_{p,\sigma}\Psi\Vert^{2}\leq C\sum_{p\in B_{F}^{c}\cap\mleft(B_{F}^{c}-k\mright)}^{\sigma}\left|p\right|^{2}\left\Vert c_{p,\sigma}\Psi\right\Vert ^{2}\nonumber \\
 & \leq C\Vert H_{\mathrm{kin}}^{\frac{1}{2}}\Psi\Vert^{2}\leq C\mleft(\left\Vert \Psi\right\Vert ^{2}+\left\Vert H_{\mathrm{kin}}\Psi\right\Vert ^{2}\mright).\nonumber 
\end{align}
For the second, observe that in the same manner one can show that
$\sum_{\sigma=1}^{s}\Vert\mathcal{N}_{E}^{\frac{1}{2}}c_{p,\sigma}\Psi\Vert^{2}\leq\sum_{\sigma=1}^{s}\Vert c_{p,\sigma}\mathcal{N}_{E}^{\frac{1}{2}}\Psi\Vert^{2}$,
as noted in equation (\ref{eq:NumberOperatorRearrangement}), it follows
that $\sum_{\sigma=1}^{s}\Vert H_{\mathrm{kin}}^{\frac{1}{2}}c_{p,\sigma}\Psi\Vert^{2}\leq\sum_{\sigma=1}^{s}\Vert c_{p,\sigma}H_{\mathrm{kin}}^{\frac{1}{2}}\Psi\Vert^{2}$.
We may then estimate
\begin{align}
 & \qquad\;\sum_{p,q\in B_{F}^{c}\cap\mleft(B_{F}^{c}-k\mright)}^{\sigma,\tau}\left|p\right|\left|q+k\right|\left\Vert c_{q,\tau}c_{p+k,\sigma}\Psi\right\Vert \left\Vert c_{q+k,\tau}c_{p,\sigma}\Psi\right\Vert \nonumber \\
 & \leq\sqrt{\sum_{p,q\in B_{F}^{c}\cap\mleft(B_{F}^{c}-k\mright)}^{\sigma,\tau}\left|p\right|^{2}\left|q+k\right|^{2}\left\Vert c_{q+k,\tau}c_{p,\sigma}\Psi\right\Vert ^{2}}\sqrt{\sum_{p,q\in B_{F}^{c}\cap\mleft(B_{F}^{c}-k\mright)}^{\sigma,\tau}\left\Vert c_{q,\tau}c_{p+k,\sigma}\Psi\right\Vert ^{2}}\\
 & \leq\sqrt{\sum_{p\in B_{F}^{c}\cap\mleft(B_{F}^{c}-k\mright)}^{\sigma}\left|p\right|^{2}\Vert H_{\mathrm{kin}}^{\frac{1}{2}}c_{p,\sigma}\Psi\Vert}\left\Vert \mathcal{N}_{E}\Psi\right\Vert \leq C\left\Vert \Psi\right\Vert \left\Vert H_{\mathrm{kin}}\Psi\right\Vert \leq C\mleft(\left\Vert \Psi\right\Vert ^{2}+\left\Vert H_{\mathrm{kin}}\Psi\right\Vert ^{2}\mright),\nonumber 
\end{align}
so in all
\begin{align}
\left|\left\langle \Psi,\mathcal{Q}_{\mathrm{SR}}\Psi\right\rangle \right| & \leq C\mleft(\sum_{k\in\overline{B}\mleft(0,R\mright)\cap\mathbb{Z}_{\ast}^{3}}\frac{\hat{V}_{k}}{\left|k\right|^{2}}\mright)\mleft(\left\Vert \Psi\right\Vert ^{2}+\left\Vert H_{\mathrm{kin}}\Psi\right\Vert ^{2}\mright)\\
 & \leq C\sqrt{\sum_{k\in\mathbb{Z}_{\ast}^{3}}\hat{V}_{k}^{2}}\sqrt{\sum_{k\in\mathbb{Z}_{\ast}^{3}}\left|k\right|^{-4}}\mleft(\left\Vert \Psi\right\Vert ^{2}+\left\Vert H_{\mathrm{kin}}\Psi\right\Vert ^{2}\mright)\leq C\mleft(\left\Vert \Psi\right\Vert ^{2}+\left\Vert H_{\mathrm{kin}}\Psi\right\Vert ^{2}\mright).\nonumber 
\end{align}
$\hfill\square$

By the proposition (or rather, its argument) it follows as we have
used repeatedly throughout this section that for any $\Psi\in D\mleft(H_{\mathrm{kin}}^{\prime}\mright)$,
the sequence $\mleft(\left\langle \Psi,\mathcal{Q}_{\mathrm{SR}}^{\mleft(R\mright)}\Psi\right\rangle \mright)_{R=1}^{\infty}$
is Cauchy, hence converges, and so we can define $\mathcal{Q}_{\mathrm{SR}}$
in quadratic form sense on all of $D\mleft(H_{\mathrm{kin}}^{\prime}\mright)$
by this limiting procedure\footnote{The cubic terms $\mathcal{C}$ arguably warrant a similar justification,
but this can be handled by the same kind of arguments we have just
used, so we omit this.}.

Having clarified $\mathcal{Q}_{\mathrm{SR}}$, the transformation
formula now follows by the calculations of the main text: For any
$R\in\mathbb{N}$ we have
\begin{equation}
e^{\mathcal{K}}\mathcal{Q}_{\mathrm{SR}}^{\mleft(R\mright)}e^{-\mathcal{K}}=\mathcal{Q}_{\mathrm{SR}}^{\mleft(R\mright)}+\int_{0}^{1}e^{t\mathcal{K}}\mleft(2\,\mathrm{Re}\mleft(\mathcal{G}^{\mleft(R\mright)}\mright)\mright)e^{t\mathcal{K}}dt
\end{equation}
where $\mathcal{G}^{\mleft(R\mright)}$ is given by
\begin{align}
\mathcal{G}^{\mleft(R\mright)} & =\frac{s^{-\frac{1}{2}}k_{F}^{-1}}{\mleft(2\pi\mright)^{3}}\sum_{k\in\overline{B}\mleft(0,R\mright)\cap\mathbb{Z}_{\ast}^{3}}\sum_{l\in\mathbb{Z}_{\ast}^{3}}\hat{V}_{k}\sum_{p,q\in B_{F}^{c}\cap\mleft(B_{F}^{c}+k\mright)}^{\sigma,\tau}1_{L_{l}}\mleft(q\mright)c_{p,\sigma}^{\ast}b_{l}\mleft(K_{l}e_{q}\mright)c_{-q+l,\tau}^{\ast}c_{-q+k,\tau}c_{p-k,\sigma}\\
 & +\frac{s^{-1}k_{F}^{-1}}{2\,\mleft(2\pi\mright)^{3}}\sum_{k\in\overline{B}\mleft(0,R\mright)\cap\mathbb{Z}_{\ast}^{3}}\sum_{l\in\mathbb{Z}_{\ast}^{3}}\hat{V}_{k}\sum_{p,q\in B_{F}^{c}\cap\mleft(B_{F}^{c}+k\mright)}^{\sigma,\tau}1_{L_{l}}\mleft(p\mright)1_{L_{l}}\mleft(q\mright)\left\langle K_{l}e_{q},e_{p}\right\rangle c_{p-l,\sigma}^{\ast}c_{-q+l,\tau}^{\ast}c_{-q+k,\tau}c_{p-k,\sigma}.\nonumber 
\end{align}
The same estimates used in Proposition \ref{prop:calGEstimate} now
apply to show $\mathcal{G}^{\mleft(R\mright)}\rightarrow\mathcal{G}$
in norm as $R\rightarrow\infty$, so for any $\Psi\in D\mleft(H_{\mathrm{kin}}^{\prime}\mright)$
\begin{align}
\left\langle \Psi,e^{\mathcal{K}}\mathcal{Q}_{\mathrm{SR}}e^{-\mathcal{K}}\Psi\right\rangle  & =\lim_{R\rightarrow\infty}\left\langle \Psi,e^{\mathcal{K}}\mathcal{Q}_{\mathrm{SR}}^{\mleft(R\mright)}e^{-\mathcal{K}}\Psi\right\rangle \nonumber \\
 & =\lim_{R\rightarrow\infty}\mleft(\left\langle \Psi,\mathcal{Q}_{\mathrm{SR}}^{\mleft(R\mright)}\Psi\right\rangle +\int_{0}^{1}\left\langle \Psi,e^{t\mathcal{K}}\mleft(2\,\mathrm{Re}\mleft(\mathcal{G}^{\mleft(R\mright)}\mright)\mright)e^{t\mathcal{K}}\Psi\right\rangle dt\mright)\\
 & =\left\langle \Psi,\mathcal{Q}_{\mathrm{SR}}\Psi\right\rangle +\int_{0}^{1}\left\langle \Psi,e^{t\mathcal{K}}\mleft(2\,\mathrm{Re}\mleft(\mathcal{G}\mright)\mright)e^{t\mathcal{K}}\Psi\right\rangle dt\nonumber 
\end{align}
which is the claim.


\begin{thebibliography}{10}

\bibitem{BohmPines-51}
David Bohm and David Pines,
"A Collective Description of Electron Interactions. I. Magnetic Interactions",
Phys. Rev. {\bf 82}, 625, 1951.

\bibitem{BohmPines-52}
David Pines and David Bohm,
"A Collective Description of Electron Interactions: II. Collective {\it vs} Individual Particle Aspects of the Interactions",
Phys. Rev. {\bf 85}, 338, 1952.

\bibitem{BohmPines-53}
David Bohm and David Pines,
"A Collective Description of Electron Interactions: III. Coulomb Interactions in a Degenerate Electron Gas",
Phys. Rev. {\bf 92}, 609, 1953.

\bibitem{Pines-53}
David Pines,
"A Collective Description of Electron Interactions: IV. Electron Interaction in Metals",
Phys. Rev. {\bf 92}, 626, 1953.

\bibitem{GellMannBrueckner-57}
Murray Gell-Mann and Keith A. Brueckner,
"Correlation Energy of an Electron Gas at High Density",
Phys. Rev. {\bf 106}, 364, 1957.

\bibitem{Sawada-57}
Katuro Sawada,
"Correlation Energy of an Electron Gas at High Density",
Phys. Rev. {\bf 106}, 372, 1957.

\bibitem{SawBruFukBro-57}
K. Sawada, K. A. Brueckner, N. Fukuda and R. Brout,
"Correlation Energy of an Electron Gas at High Density: Plasma Oscillations",
Phys. Rev. {\bf 108}, 507, 1957.



\bibitem{BenNamPorSchSei-20}
Niels Benedikter, Phan Th\`anh Nam, Marcello Porta, Benjamin Schlein and Robert Seiringer,
"Optimal Upper Bound for the Correlation Energy of a Fermi Gas in the Mean-Field Regime",
Commun. Math. Phys {\bf 374}, 2097, 2020.

\bibitem{BenNamPorSchSei-21}
Niels Benedikter, Phan Th\`anh Nam, Marcello Porta, Benjamin Schlein and Robert Seiringer,
"Correlation Energy of a Weakly Interacting Fermi Gas",
Invent. Math. {\bf 225}, 885, 2021.

\bibitem{BenPorSchSei-21}
Niels Benedikter, Marcello Porta, Benjamin Schlein and Robert Seiringer,
"Correlation Energy of a Weakly Interacting Fermi Gas with Large Interaction Potential",
Preprint 2021, arXiv:2106.13185.



\bibitem{ChrHaiNam-21}
Martin Ravn Christiansen, Christian Hainzl and Phan Th\`anh Nam,
"The Random Phase Approximation for Interacting Fermi Gases in the Mean-Field Regime",
Preprint 2021, arXiv:2106.11161.

\bibitem{ChrHaiNam-22a}
Martin Ravn Christiansen, Christian Hainzl and Phan Th\`anh Nam,
"On the Effective Quasi-Bosonic Hamiltonian of the Electron Gas: Collective Excitations and Plasmon Modes",
Preprint 2022, arXiv:2206.13073.

\bibitem{ChrHaiNam-22b}
Martin Ravn Christiansen, Christian Hainzl and Phan Th\`anh Nam,
"The Gell-Mann--Brueckner Formula for the Correlation Energy of the Electron Gas: A Rigorous Upper Bound in the Mean-Field Regime",
Preprint 2022, arXiv:2208.01581.



\bibitem{Bogolubov-47}
N. Bogolubov,
"On the theory of superfluidity",
J. Phys. (USSR), {\bf 11}, 23, 1947.

\bibitem{GrechSeiringer-13}
Philip Grech and Robert Seiringer,
"The Excitation Spectrum for Weakly Interacting Bosons in a Trap",
Commun. Math. Phys {\bf 322}, 559, 2013.

\bibitem{HaiPorRex-20}
Christian Hainzl, Marcello Porta and Felix Rexze,
"On the Correlation Energy of Interacting Fermionic Systems in the Mean-Field Regime",
Commun. Math. Phys {\bf 524}, 485, 2020.

\bibitem{Seiringer-11}
Robert Seiringer,
"The Excitation Spectrum for Weakly Interacting Bosons",
Commun. Math. Phys {\bf 306}, 565, 2011.

\bibitem{GelFondLinnik-66}
A. O. Gel'fond and Yu. V. Linnik,
"Elementary Methods in the Analytic Theory of Numbers",
translation by D. E. Brown and I. N. Sneddon,
Pergamon Press, 1966.

\end{thebibliography}
\end{document}